\documentclass[apj,twocolumn]{emulateapj_mod}
\usepackage{epsfig,apjfonts,mathptmx}
\usepackage{graphicx}
\usepackage{float}
\usepackage{placeins} 
\usepackage{amsfonts,amsmath,amssymb}
\usepackage{natbib}
\usepackage{url}
\usepackage[colorlinks,citecolor=blue]{hyperref} 
\usepackage{xcolor}
\usepackage{changes}
\usepackage[compatibility=false]{caption} 
\usepackage{subcaption} 

\newcommand{\Superdb}{{``Super-deblended'' }}
\newcommand{\superdb}{{``super-deblended'' }}
\newcommand{\crowdedness}{{\em crowdedness }}

\newcommand{\galfit}{\textit{galfit}}

\def\gtsima{$\; \buildrel > \over \sim \;$}
\def\ltsima{$\; \buildrel < \over \sim \;$}
\def\prosima{$\; \buildrel \propto \over \sim \;$}
\def\gsim{\lower.5ex\hbox{\gtsima}}
\def\lsim{\lower.5ex\hbox{\ltsima}}
\def\simgt{\lower.5ex\hbox{\gtsima}}
\def\simlt{\lower.5ex\hbox{\ltsima}}
\def\simpr{\lower.5ex\hbox{\prosima}}

\def\h1{$h^{-1}$}
\def\beq{\begin{equation}}
\def\eeq{\end{equation}}
\def\8mu{8\,$\mu{\rm m}$}
\def\16mu{16\,$\mu{\rm m}$}
\def\24mu{24\,$\mu{\rm m}$}
\def\70mu{70\,$\mu{\rm m}$}

\shorttitle{Super-deblended IR photometry of COSMOS star forming galaxies}
\shortauthors{S. Jin et al.}

\begin{document}

\title{\Superdb Dust Emission in Galaxies: II. Far-IR to (sub)millimeter photometry\\ and high redshift galaxy candidates in the full COSMOS field}

\author{%
	Shuowen Jin\altaffilmark{1, 2, 3}, 
	Emanuele Daddi\altaffilmark{2}, 
	Daizhong Liu\altaffilmark{4},
 	Vernesa Smol\v{c}i\'{c}\altaffilmark{5},
    Eva Schinnerer\altaffilmark{4},
    Antonello Calabr\`o\altaffilmark{2},
    Qiusheng Gu\altaffilmark{1,3},
    Jacinta Delhaize\altaffilmark{5}, 
    Ivan Delvecchio\altaffilmark{5},
    Yu Gao\altaffilmark{6},
    Mara Salvato\altaffilmark{7}, 
    Annagrazia Puglisi\altaffilmark{2},
    Mark Dickinson\altaffilmark{8},
    Frank Bertoldi\altaffilmark{9},
    Mark Sargent\altaffilmark{10},
    Mladen Novak\altaffilmark{5},
    Georgios Magdis\altaffilmark{11},
    Itziar Aretxaga\altaffilmark{12},
    Grant W. Wilson\altaffilmark{13},
    and Peter Capak\altaffilmark{14}
}
\altaffiltext{1}{School of Astronomy and Space Science, Nanjing University,
Nanjing 210093, China (shuowen.jin@gmail.com)}
\altaffiltext{2}{CEA, IRFU, DAp, AIM, Universit\'{e} Paris-Saclay, Universit\'{e} Paris Diderot, Sorbonne Paris Cit\'{e}, CNRS, F-91191 Gif-sur-Yvette, France}
\altaffiltext{3}{Key Laboratory of Modern Astronomy and Astrophysics, Nanjing University, Nanjing 210093, China}
\altaffiltext{4}{Max Planck Institute for Astronomy, K\"{o}nigstuhl 17, D-69117 Heidelberg, Germany}
\altaffiltext{5}{Department of Physics, Faculty of Science, University of Zagreb, Bijeni\v{c}ka cesta 32, 10000 Zagreb, Croatia}
\altaffiltext{6}{Purple Mountain Observatory/Key Laboratory for Radio Astronomy, Chinese Academy of Sciences, No. 2 West Bejing Road, Nanjing, China}
\altaffiltext{7}{Max-Planck-Institut f{\"{u}}r extraterrestrische Physik, Giessenbachstrasse 1, 85748, Garching, Bayern, Germany}
\altaffiltext{8}{National Optical Astronomy Observatory, Tucson, Arizona 85719, USA}
\altaffiltext{9}{Argelander Institute for Astronomy, University of Bonn, Auf dem Hugel 71, 53121 Bonn, Germany}
\altaffiltext{10}{Astronomy Centre, Department of Physics and Astronomy, University of Sussex, Brighton, BN1 9QH, UK}
\altaffiltext{11}{Dark Cosmology Centre, Niels Bohr Institute, University of Copenhagen, Juliane Maries Vej 30, DK-2100 Copenhagen, Denmark}
\altaffiltext{12}{Instituto Nacional de Astrof\'isica, \'Optica y Electr\'onica (INAOE), Luis Enrique Erro 1, Sta. Ma. Tonantzintla, 72840 Puebla, Mexico}
\altaffiltext{13}{Department of Astronomy, University of Massachusetts, Amherst, MA 01003, USA}
\altaffiltext{14}{Department of Astronomy, California Institute of Technology, MC 249-17, 1200 East California Blvd, Pasadena, CA 91125, USA}

\begin{abstract}
We present a ``super-deblended'' far-infrared to (sub)millimeter photometric catalog in the Cosmic Evolution Survey (COSMOS), prepared with the method recently developed by Liu et al. 2018, with key adaptations.
We obtain point spread function (PSF) fitting photometry at fixed prior positions including 88,008 galaxies detected in either VLA 1.4~GHz, 3~GHz and/or MIPS 24~$\mu$m images.
By adding a specifically carved mass-selected sample (with an evolving stellar mass limit), a highly complete prior sample of 194,428 galaxies is achieved for deblending FIR/(sub)mm images.
We performed ``active'' removal of non relevant priors at  FIR/(sub)mm bands using spectral energy distribution (SED) fitting and redshift information. 
In order to cope with the shallower COSMOS data
we subtract from the maps the flux of faint non-fitted priors and explicitly account for the uncertainty of this step.
The resulting photometry  (including data from Spitzer, Herschel, SCUBA2, AzTEC, MAMBO and NSF's Karl G. Jansky Very Large Array at 3~GHz and 1.4~GHz) displays well behaved quasi-Gaussian uncertainties,  calibrated from Monte Carlo simulations and tailored to observables (crowding, residual maps).
Comparison to ALMA photometry for hundreds of sources provide a remarkable validation of the technique.
We detect 11,220 galaxies over the 100--1200~$\mu$m range, extending to $z_{\rm phot}\sim7$.
We conservatively selected a sample of 85 $z>4$ high redshift candidates, significantly detected in the FIR/(sub)mm, often with secure radio and/or Spitzer/IRAC counterparts.
This provides a chance to  investigate the first generation of vigorous starburst galaxies (SFRs$\sim1000\mathrm{M}_\odot$~yr$^{-1}$).
The photometric and value added catalogs are publicly released. 
\end{abstract}


\section{Introduction}

\label{Section_Introduction}

Detailed studies of dust emission in galaxies, which peaks at far-infrared (FIR) to (sub)millimetre wavelengths as a function of redshift, have been revolutionizing our understanding of the formation and evolution of galaxies through cosmic epochs, providing an accurate bolometric tracer of their total {star formation rate \citep{Elbaz2017,Puglisi2017,Ciesla2014,Draine2007SINGS,Magdis2012SED,Tan2014,Aravena2016,Dunlop2017,Karim2013,Oteo2016ALMAsurvey,Cowie_2017,Aretxaga2011}}.
Such studies are now pushing well into the reionization era \citep{Riechers2013Nature,Strandet2017,Marrone2017Nature}
with abundant molecular gas detected [see, e.g.,the redshift record $z=7.5$ quasar in \cite{Venemans2017}].

The Cosmic Evolution Survey (COSMOS, PI: N. Scoville, \citealt{Scoville2007COSMOS}) field is one of the largest and most extensively observed blank deep fields with deep data at all wavebands. It has been surveyed in imaging to deep levels with the \textit{Herschel Space Observatory} \citep[hereafter \textit{Herschel};][]{Pilbratt2010} and other ground based (sub)mm telescopes (e.g., IRAM 30m and JCMT 15m telescopes).
Its nearly 2 square degrees sky coverage, as well as the wealth of deep FIR/(sub)mm imaging, potentially provide the largest dataset to select samples of dusty star forming galaxies in well-defined blank extragalactic fields. This in turn makes COSMOS an ideal survey to construct statistically meaningful samples to study the dust spectral energy distributions (SEDs) of galaxies and to search for dusty galaxies at $z>6-7$.

However, the very large beam sizes of FIR/(sub)mm detectors introduce heavy source confusion (blending), complicating photometric works by making fluxes of individual galaxies often difficult to measure, preventing us from precisely constraining their dusty SEDs and deriving their SFRs. Prior-extraction techniques, which make use of sources in high resolution images as priors to fit images with lower resolution, have been developed and applied to confused FIR/(sub)mm images in deep fields \citep{Roseboom2010,Elbaz2011, Lee2013, Hurley2016}. Nevertheless, despite the efforts mentioned above and many others, we are still far from a satisfactory resolution of the confusion problems, particularly in the full COSMOS field, where the completeness of the prior samples\footnote{{The completeness of the prior sample for galaxies that have intrinsic flux above some detectable threshold (e.g., $>3\sigma$) in an image under exam is defined as the fraction of them for which a prior is actually present in the prior sample.}}, the contribution of fainter sources to blending and the optimal selection of priors for the lowest resolution and most confused (SPIRE) images have not yet been exhaustively considered for attempting the highest quality deblending in $Herschel$ and other (sub)mm images.

\citet{Liu_DZ2017} (hereafter, L18)
developed a novel ``{Super-deblending}'' approach for obtaining prior-fitting multi-band photometry for FIR/(sub)mm data sets in the GOODS-North field. This work used galaxies detected in deep MIPS 24~$\mu$m and radio images as priors for deblending FIR/(sub)mm images, ensuring a high completeness of their prior sample even out to high redshifts. Based on SED fitting with photometry available for each prior at each step in wavelength, excessively faint priors are actively excluded in the fitting and their flux removed from the original maps. In this way the remaining priors could be fitted with less crowding ($\lesssim 1$ prior per beam) and the emission of fainter sources could also be better constrained. Also, as an indispensable feature of this technique, Monte Carlo simulations were used to precisely correct flux biases of various kinds and to obtain calibrated and quasi-Gaussian flux uncertainties for the photometry in all bands. Furthermore, sources extracted in the residual images were also included in the prior list and fitted, further improving the completeness of the prior sample. L18 used the GOODS-N catalog obtained in this way to study the star formation rate density (SFRD) of the Universe to $z=6$ and to search for high redshift galaxy candidates up to $z\sim 7$. 

Benefiting from the much larger area covered and its rich data sets, it would be highly desirable to have a similar ``Super-deblending'' technique applied to the COSMOS field to produce high quality de-confused FIR/(sub)mm photometry and thus much larger samples of dusty star forming galaxies at all redshifts with reliable dust SEDs. We have dealt with this endeavour in the work described in this paper.
Notwithstanding, there were crucial hurdles to solve to obtain our goal: when comparing to the GOODS-N field, the COSMOS field is shallower at MIPS 24~$\mu{\rm m}$, radio and PACS bands while it has comparable depth in the SPIRE and (sub)mm images. Galaxies only detected in 24~$\mu{\rm m}$ and/or radio images would thus make an incomplete set of priors for FIR/(sub)mm sources particularly at high redshifts. Therefore, finding alternative ways to complete the prior sample is crucial for the deblending work in COSMOS. At the same time, the shallower 24~$\mu{\rm m}$, radio 
and PACS data lead to larger uncertainties in SED fitting, where predicted fluxes of faint sources (an essential input for the ``Super-deblended'' approach) are much less well-constrained, introducing extra errors on the photometry. Hence further optimizations, with respect to the L18 work, in dealing with the subtraction of faint sources are required, and the Monte Carlo simulations have to be improved in order to account for the less reliable treatment of this part of the procedure.

In this paper, we apply the ``Super-deblending'' technique on the FIR/(sub)mm data sets available for the COSMOS field, based on a highly complete prior catalog that we have devised particularly for this work. 
We select $K_s$ sources from UltraVISTA catalogs and use radio detections from the VLA 3~GHz catalog as an initial prior sample, to fit MIPS 24~$\mu$m and radio images. Then we combine resulting detections from the 24~$\mu$m and/or radio fitting with a mass-limited sample of $K_s$ sources (with the actual mass limit changing with redshift) to fit the $Herschel$, SCUBA2, AzTEC and MAMBO images. 
The building of the prior catalog is described in Section~\ref{Section_Initial_general_priors_list}, including the fitting of 24~$\mu$m and radio images.
We optimize the subtraction of faint sources, which is shown in Section~\ref{Section_Faint_Prior}. 
We improve the Monte Carlo simulations by adding a new correction to account for subtracted fluxes, which is described in Section~\ref{section_simulations}.
The final photometry catalog and selection of high redshift candidates are shown in Section~\ref{result_catalog_candidates}.

We emphasize that the \superdb photometry technique is described in full extent in L18. We refer readers interested about the method to that crucial reference if required for a better understanding of technical parts in this paper, where we limit the detailed description only to specific variations with respect to L18 and particularities that apply and had to be adopted for the COSMOS field. For this reason, we also have decided to reproduce in the style and presentation most of the figures of the L18 paper with showing results from the COSMOS field, in order to allow for direct comparison. Notice that the limitations of this approach, as discussed in Sect.~7.6 of L18 do apply also to this work. We will explore the possibility for major improvements, namely to account for correlated photometric noise across bands and to attempt all-bands simultaneous processing, in future works.

{Finally, it is important to clarify that we do not embark in this paper in a direct study of the  completeness of our IR photometric catalog. Given the large PSFs in the FIR/(sub)mm bands, the probability to detect a galaxy of a given flux largely depends on the properties and densities of surrounding galaxies. 
Estimating this probability would thus require extended simulations to be performed simultaneously in all bands and over the whole COSMOS field. This is beyond the scope of this paper, and we defer the study of IR completeness to future works.}

We adopt $H_{0}=73$, $\Omega_{M}=0.27$, $\Lambda_{0}=0.73$, and a Chabrier IMF \citep{Chabrier2003} unless specified in text for specific comparisons to other works.

\section{Data sets}
\label{data_sets}

The imaging data sets on which measurements are performed in this paper are obtained from various surveys: MIPS 24~$\mu$m data is the GO3 image from the COSMOS-\textit{Spitzer} program \citep[PI: D. Sanders;][]{LeFloch2009}, \textit{Herschel}/PACS 100~$\mu$m \& 160~$\mu$m data are from PEP \citep[PI: D. Lutz;][]{Lutz2011} and CANDELS-$Herschel$ (PI: M. Dickinson) programs, while SPIRE 250~$\mu$m, 350~$\mu$m and 500~$\mu$m data are nested maps from the Herschel Multi-tiered Extragalactic Survey (HerMES, PI: S. Oliver). The SCUBA2 850~$\mu$m images are from the S2CLS program \citep{Geach2016,Cowie_2017}. The AzTEC 1.1~mm data are nested maps from \citet{Aretxaga2011} and MAMBO 1.2~mm images are from \citet{Bertoldi2007}. The deep radio data are VLA 3~GHz images from \citet{Smolcic2017} and 1.4~GHz images from \citet{Schinnerer2010}. 
The image products at each band are shown in Appendix~\ref{image_products}, and the detailed performance figures of the simulation-based correction recipes are presented in Appendix~\ref{Section_Simulation_Performance}.


\section{Setting up the prior catalogs}
\label{Section_Initial_general_priors_list}

Given the strong correlation between IR luminosity and the luminosity at 24~$\mu$m and radio bands \citep{Dale_Helou2002,Yun_M2001,Delhaize2017}, as well as the lower confusion in MIPS 24~$\mu$m and radio images, blind-extracted detections at 24~$\mu$m and/or radio bands are widely used as priors for $Herschel$ deblending works \citep{Lee2013, Hurley2016}. However, blind extractions {are restricted to large flux detection} and poorer performances (e.g., requiring $>5\sigma$ detections) in order to reduce spurious detections rates. This leaves many potentially detectable sources without being eventually extracted, and thus suppresses the completeness of the photometric sample. To identify and include fainter detections missing in the blind-extracted catalogs,
the prior-extraction method, in which one fits the PSF or other models at the positions of known sources, is an effective solution to obtain higher quality photometry with lower spurious detection rate. Thus this method allows us to obtain reliable detections to lower absolute significances.
This is crucial for our work in the COSMOS field, where blind catalogs at MIPS 24~$\mu$m
and radio were already obtained \citep[e.g.,][]{LeFloch2009,Schinnerer2010,Smolcic2017}.

At 24~$\mu$m, the COSMOS MIPS 24~$\mu$m image has formally an rms sensitivity of $\sigma=10~\mu$Jy, which is $\sim2$ times shallower than the GOODS-N 24~$\mu$m data
\footnote{As discussed later, we find evidence that the 24~$\mu$m photometry in COSMOS should be scaled up by a factor of $1.5-1.7$ in order to be consistent with the GOODS 24~$\mu$m to FIR flux ratios. 
This would imply that the actual rms sensitivity is $\sim15-17\mu$Jy at this band, and $\sim 3 \times$ shallower than
GOODS-N.}
(see Table.1 in L18).
At 1.4~GHz, the deepest central region of the VLA 1.4~GHz Deep Project map \citep{Schinnerer2010} reaches $\sigma \sim 12\mu$Jy, which is $>4$ times shallower than the combined VLA 1.4~GHz images ($\sigma \sim 2.74 \mu$Jy) \citep[Owen 2017]{Morrison2010} in the whole GOODS-N field (see Table.1 in L18). 
The deeper 3~GHz VLA map \citep{Smolcic2017} has $\sigma \sim 2.3 \mu$Jy, but it is still shallower than the 1.4~GHz images in GOODS-N by a factor of 1.5 (assuming that the radio flux density scales as $\lambda^{-0.7}$), and it has higher spatial resolution ($0.75''$ vs $2.0''$) that might result in more important flux losses for extended sources.

\begin{figure}
	\centering
	\includegraphics[width=0.5\textwidth, trim={0 1.6cm 0 0, clip}]{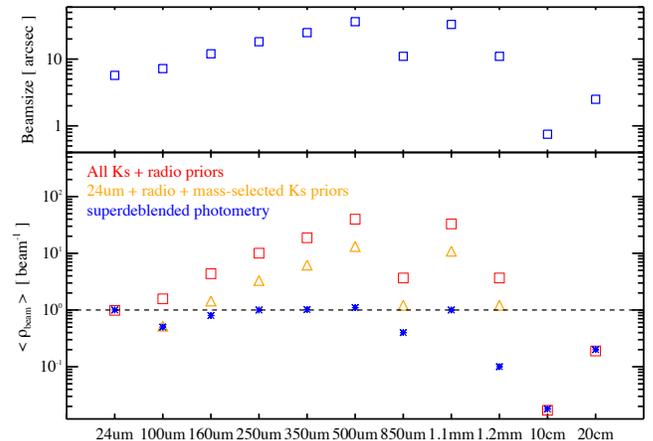}
	\caption{
		Analog to Fig.~1 in L18, but showing here source density properties for datasets in the COSMOS field.
        Upper panel: the beam sizes of the COSMOS images (Table 1).
		Bottom panel: the prior selection and the reduction of sources density ($\left<\rho_{\mathrm{beam}}\right>$) in the 1.7 $\deg^2$ UltraVISTA area. 
        The source densities of the full $K_s$+radio initial catalog are shown as the red squares, the 24 $\mu$m + radio + mass-selected sources as orange triangles. The \superdb prior sources that are actually fitted are shown as blue crosses.
	\label{Fig_Galsed_Plot_Number_per_Beam}
	}
\end{figure}

In the \superdb GOODS-N catalog, L18 selected Spitzer IRAC prior-based detections in deep 24~$\mu$m and/or radio images as the further starting subset of priors to be used for deblending FIR/(sub)mm images. 
{Although their deep 24~$\mu$m+radio catalog has higher completeness than what we can do in COSMOS, they still find $80$ additional sources (approaching one per arcmin$^2$) detected at FIR/(sub)mm bands (in the residual images) without detections at 24~$\mu$m and/or radio. Therefore, the shallower prior catalog that can be built from 24~$\mu$m+radio photometry in COSMOS is likely to be in even more substantial shortage of priors, having a lower completeness which is not ideally suited for high quality FIR/(sub)mm photometric work.}

In order to obtain a highly complete prior catalog for FIR/(sub)mm deblending in the COSMOS field we proceeded as following.
Firstly, we run prior-extraction in the MIPS 24~$\mu$m and VLA images on the positions of a set of $K_s$+radio priors (see Section~\ref{Section_prior_fit_24radio},~\ref{Section_Photometry_24} and ~\ref{Section_Photometry_radio}), in order to obtain more complete source detection lists than in blind-extracted catalogs.
Secondly, in order to further complete the prior sample including all relevant sources that cannot be detected in the available 24~$\mu$m and/or radio images, we select a supplementary set of priors using stellar masses, to eventually define a new 24~$\mu$m+radio+mass-selected prior catalog for the FIR/(sub)mm deblending work (see Section~\ref{Section_The_24_Radio_Mass_Catalog}).

\begin{figure*}
\centering
\includegraphics[width=0.33\textwidth]{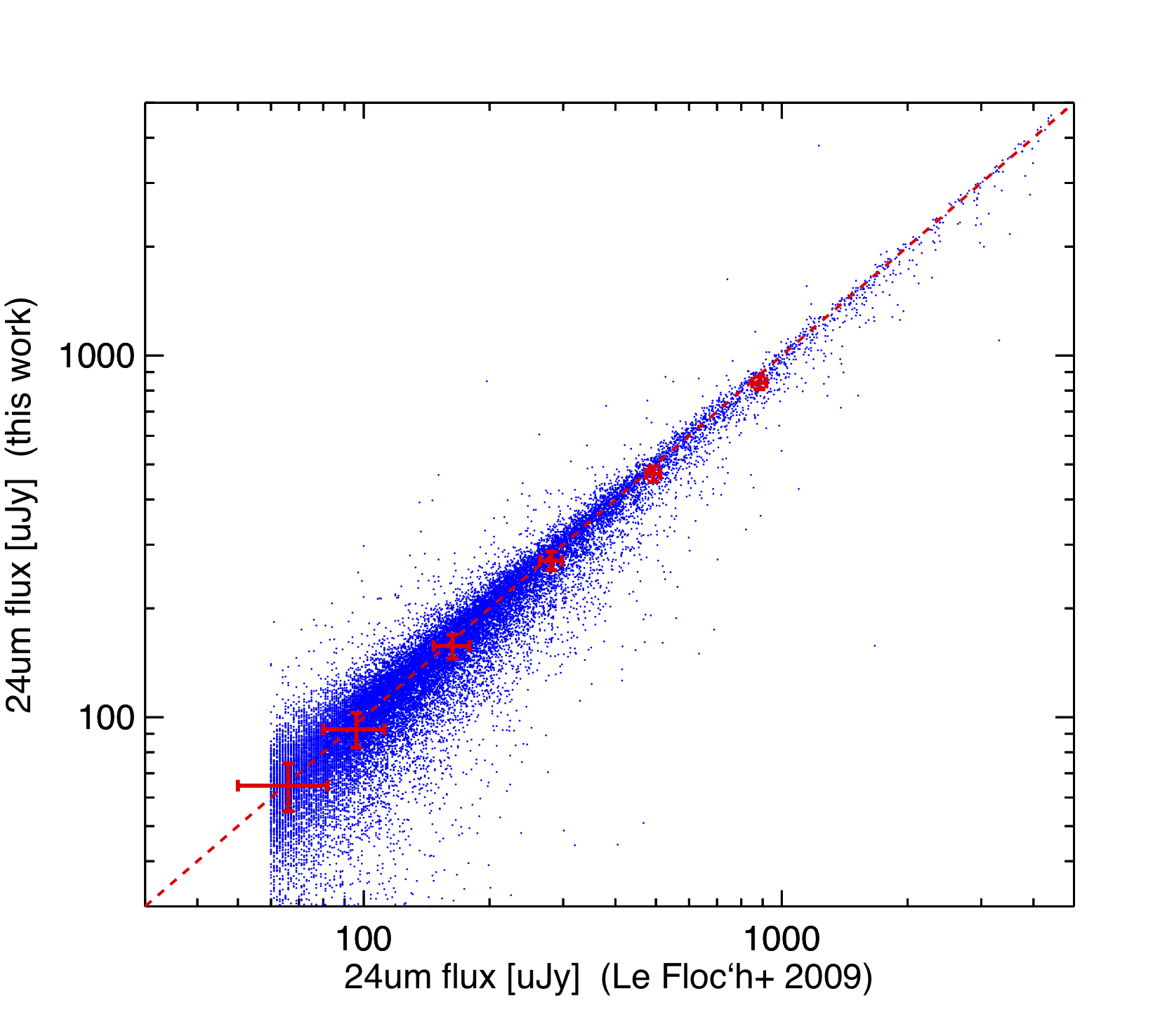}
\includegraphics[width=0.33\textwidth]{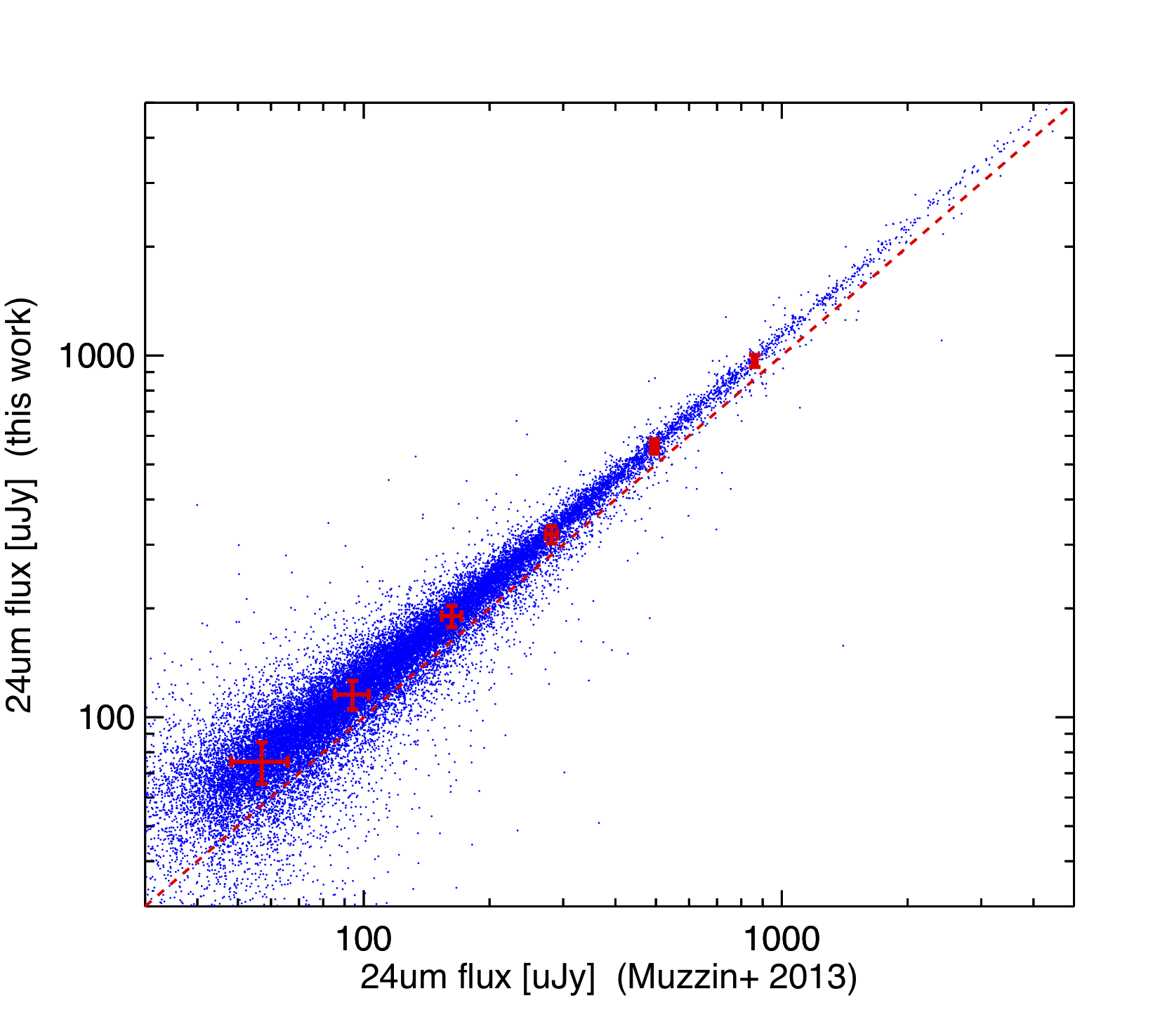}
\includegraphics[width=0.33\textwidth, trim={2cm -0.2cm 1.4cm 2.1cm}, clip]{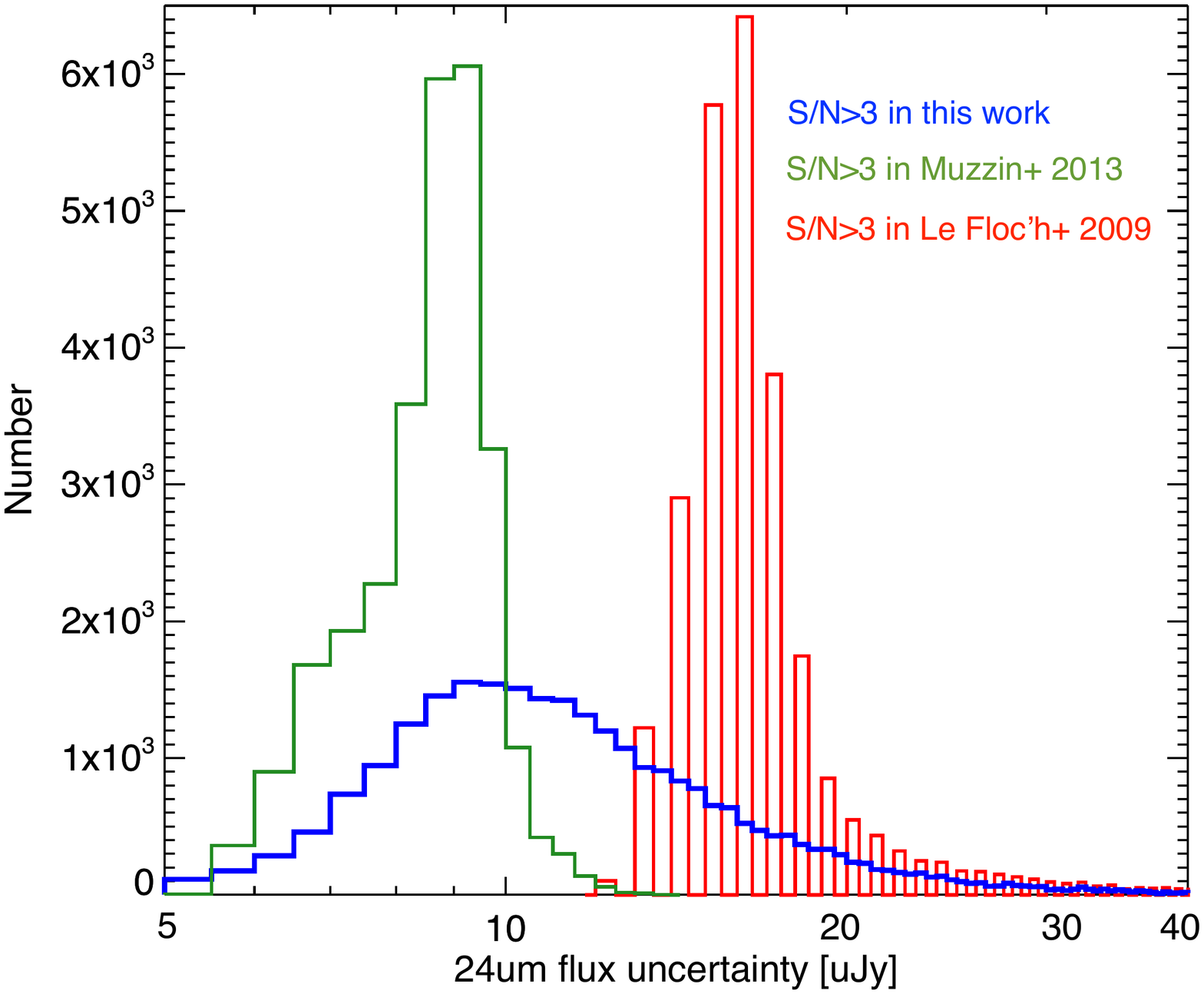}
\caption{%
	Our 24~$\mu$m photometry vs. measurements in the catalog of \citet{LeFloch2009} (left) and \citet{Muzzin2013} (center). The right panel shows the distribution of flux uncertainties for detected sources in each catalog. {Red points with error bars show median flux and flux uncertainty from the two catalogs for matched sources in several bins.}
\label{24um_result}%
}
\end{figure*}

\subsection{$K_s$(+radio) priors for 24~$\mu$m and radio images}
\label{Section_prior_fit_24radio}

In this section, we describe the creation of a $K_s$ catalog that we wish to use to perform prior-based fitting in the MIPS 24~$\mu$m, VLA 1.4~GHz and 3~GHz images. 
While in L18 we used IRAC as a parent catalog of massive galaxies from which to build prior samples, in the COSMOS field we prefer to start from a $K_s$ catalog because substantially larger amounts of effort were put in by the community to create value added catalogs for $K_s$-selected samples, as opposed to IRAC selected samples \footnote{The situation might change in the future from completion of the Spitzer Large Area Survey with Hyper-Suprime-Cam photometry (SPLASH, PI: P. Capak).}.

The COSMOS2015 catalog \citep{Laigle2016} from UltraVISTA-DR2 survey contains 1,182,108 $YJHK_s$ sources in total, and over half of them with well-measured photometric redshifts and stellar masses. 
We included 528,889 sources with photometric redshift and stellar mass in the COSMOS2015 catalog in our initial prior sample. 
The spectroscopic redshifts are taken from the new COSMOS master spectroscopic catalog (M. Salvato et al., in prep.) {that is based on a variety of  spectroscopic surveys published in the literature\footnote{We only use publicly available redshifts for this work, taken from: \citet[][zCOSMOS]{Lilly2007zCOSMOS,Lilly2009zCOSMOS}; \citet{Kartaltepe2010}; \citet[][DEIMOS]{Hasinger2018DEIMOS}; \citet[][DEIMOS-C3R2]{Masters2017C3R2}; \citet{Roseboom2012}; \citet[][FMOS]{Silverman2015FMOS}; \citet[][MMT]{Prescott2006}; \citet[][VUDS]{Tasca2017VIMOS}; \citet[][LEGA-C]{vanderWel2016_LEGA_C}, \citet[][SINFONI]{Perna2015SINFONI}; \citet[][MOSDEF]{Kriek2015MOSDEF}; \citet[][ZFIRE]{Nanayakkara2016_zFIRE}; \citet[][MOIRCS]{Onodera2012,Onodera2015}, \citet{Faure2011}; \citet[][GEMINI-S]{Balogh2014}; \citet[][FORS2]{Comparat2015} \citet[][PRIMUS]{Coil2011PRIMUS}; \citet[][NIRSPEC]{Marsan2017}; \citet{Marchesi2016a}; \citet[][2dF]{Colless2001_2dF}; \citet[][2MASS]{Skrutskie2006_2MASS}; \citet[][3D-HST]{Momcheva2016_3DHST}; \citet{Fu2011}; \citet{Yun2015AzTECC5} ; \citet[][SDSS DR14]{Abolfathi2018_SDSS_dr14}; and \citet[][IMACS]{Trump2007IMACS}.}.} 
{Sources with a reliable spectroscopic redshift (redshift quality $>3$ and $\lvert z_\mathrm{spec}-z_\mathrm{phot} \lvert <0.1\times(1+z_\mathrm{phot})$) are fitted at their fixed redshift in SED fitting.}
{Sources in the COSMOS2015 catalog that are flagged to be X-ray detected by $Chandra$ are also included in our prior catalog, but their photometric redshifts are not used (hence all redshift range explored) as they might be problematic.}
While objects located in the regions surrounding saturated optical stars are removed from this Laigle et al catalog, we fill up these blank regions by adding 57,862 $K_s$ sources from the UltraVISTA catalog of \citet{Muzzin2013}, ensuring a good coverage and evenness of prior distribution in the full UltraVISTA area. 
By matching the 3~GHz catalog of \citet{Smolcic2017}, we find 2,962 radio sources that are not present in the $K_s$ catalog. These radio sources lacking a near-IR counterpart (separation $>1''$ from any $K_s$ source), are also included in our initial prior catalog (mainly because some might be genuine high redshift galaxies).

In total, we obtained 589,713 priors in the $K_s$(+radio) prior catalog.
As can be seen from the red squares marked in Fig.~\ref{Fig_Galsed_Plot_Number_per_Beam}, this prior catalog has a source density of $\left<\rho_{\mathrm{beam}}\right>= 1$ source per beam in the MIPS 24~$\mu$m image, $\left<\rho_{\mathrm{beam}}\right>=0.2$ source per beam in the VLA 1.4~GHz image and $\left<\rho_{\mathrm{beam}}\right>=0.06$ source per beam in the VLA 3~GHz image, which is appropriate for prior fitting in all these bands. 
Comparing to the initial IRAC catalog used in L18, the surface density of the $K_s$+radio priors is compatible to the one in GOODS-N, with both showing 1 source per beam at MIPS 24~$\mu$m (i.e., 35 galaxies~arcmin$^{-2}$). 

{Note that the UltraVISTA imaging in COSMOS is not homogeneously deep: more sources are detected in the Ultra-deep strips. However, there is only a 2.5\% difference in prior density in our $K_s$(+radio) catalog. This difference gets to 0.9\% in the following 24~$\mu$m+radio+mass-selected prior catalog (see Section~\ref{Section_The_24_Radio_Mass_Catalog}). This is negligible and does not impact significantly our results.}

The 589,713 $K_s$(+radio) priors are used to fit MIPS 24~$\mu$m, VLA 3~GHz and 1.4~GHz images. 
Given that the $K_s$(+radio) prior catalog over COSMOS contains $56\times$ more sources than the 19437 IRAC priors in GOODS-N from L18, performing parallelized computations over dozens of CPUs was essential to bring the efficiency of the prior fitting to a manageable level.
The detailed image fitting in MIPS 24~$\mu$m and VLA images are described in Section~\ref{Section_Photometry_24} and ~\ref{Section_Photometry_radio}, respectively.

\subsection{Photometry at MIPS 24~$\mu$m on $K_s$-selected priors}
\label{Section_Photometry_24}
We obtain PSF-fitting photometry with \galfit{} \citep{Peng2002,Peng2010}, assuming that the intrinsic size of distant sources is negligible with respect to the size of the PSF, which is a good hypothesis for 24~$\mu$m except perhaps with rare cases of very low redshift galaxies that are not the main focus of this paper.

We perform PSF fitting at the positions of 589,713 $K_s$+radio priors in Spitzer/MIPS 24~$\mu$m GO3 images from the S-COSMOS team (PI: D. Sanders, \citealt{LeFloch2009}). 
{The PSF used for the MIPS 24~$\mu$m image is identical to the PSF profile used for the 24~$\mu$m image fitting of L18 in GOODS-N, which is consistent with the COSMOS one from \citet{LeFloch2009} but at much higher S/N. }
Based on the first-pass results with fixed RA-Dec positions, we run a second-pass PSF fitting allowing for up to 1 pixel
($1.2''$)
variations of prior source positions for those high $\mathrm{S/N}$ sources (\galfit{} $\mathrm{S/N}>10$). This improves the fitting for bright sources with returning a residual image that is cleaner than the one from the first-pass fitting.

After the \galfit{} PSF fitting, we run Monte Carlo simulations in the MIPS 24~$\mu$m map, then correct the \galfit{} outputs via the three-step correction recipes based on the simulations ($\sigma_{\galfit{}}$, ${S}_\mathrm{residual}$ and \crowdedness), {see Section~\ref{section_simulations} for details}.
The figures describing simulation performances are shown in Fig.~\ref{Simu_fig_24um} in the Appendix. The flux bias has also been calibrated and we obtain well behaved, quasi-Gaussian distributions of flux uncertainties, in a similar way to L18.

In Fig.~\ref{24um_result}, we compared our 24~$\mu$m photometry to the catalogs from \citet{LeFloch2009} and \citet{Muzzin2013}. Our flux measurements are in excellent agreement with the ones from \citet{LeFloch2009}, while our flux errors are significantly smaller (see panel 3 in Fig.~\ref{24um_result}), suggesting that our fitting is indeed more accurate and thus reaching quite deeper than this blindly extracted catalog. Note that some galaxies have significantly lower 24~$\mu$m fluxes in our catalog as compared to \citet{LeFloch2009}. We believe that the lower flux measurements are due to the resolution of blending of the 24~$\mu$m image from our method: fluxes for sources with close neighbors (e.g., with distance less than the beam size $5.7''$) are difficult to measure reliably with the blind extraction method.

\begin{figure*}
\centering
\includegraphics[width=0.48\textwidth]{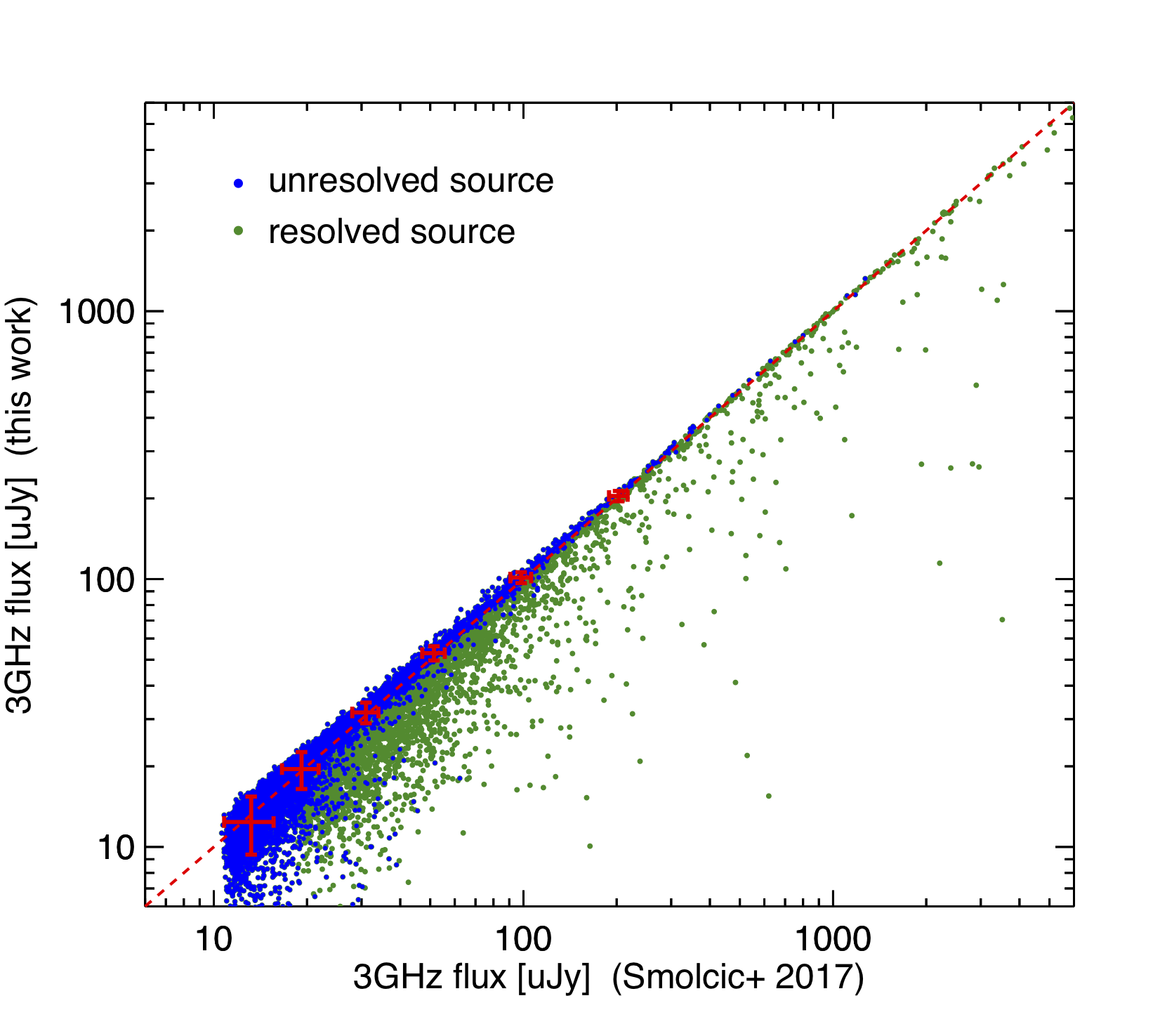}
\includegraphics[width=0.48\textwidth]{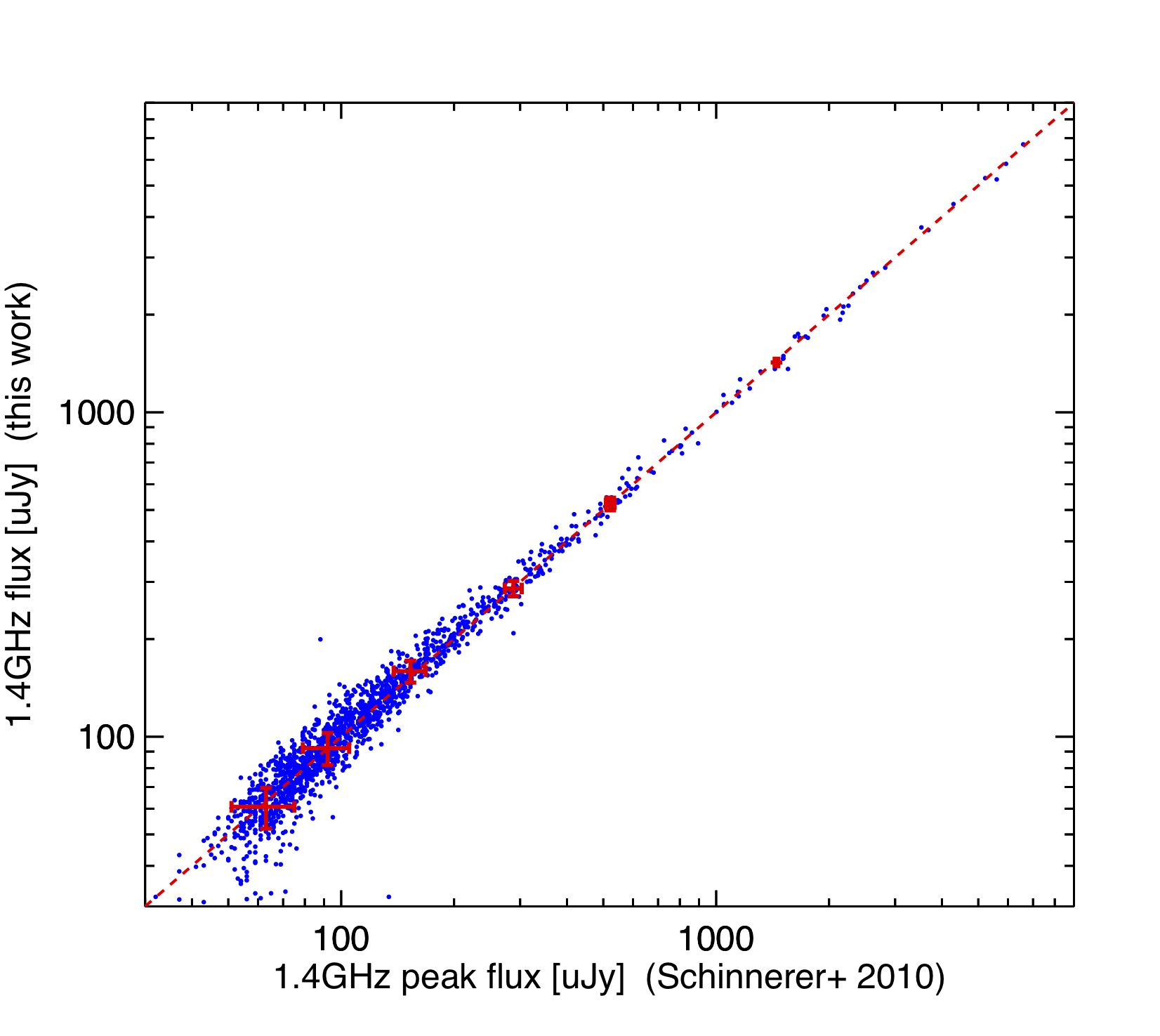}
\caption{%
	Our photometry at 3~GHz \& 1.4~GHz vs. the blind-extracted catalogs. Left panel: 3~GHz fluxes in our work versus the 5-sigma catalog of \citet{Smolcic2017}. Blue and green dots show unresolved and resolved sources, respectively, according to \citet{Smolcic2017}. {Red points with error bars show median flux and flux uncertainty from the two catalogs, for unresolved sources in several bins.}
    Our PSF fitted fluxes are consistent with fluxes of unresolved sources, while underestimating fluxes of resolved sources. We preferentially use 3~GHz photometry in the 5-sigma catalog of \citet{Smolcic2017} for matched sources. Right panel: our 1.4~GHz flux versus 1.4~GHz peak fluxes in the catalog of \citet{Schinnerer2010}.
\label{radio_result}%
}
\end{figure*}

The comparison with the 24~$\mu$m photometry reported in \citet{Muzzin2013}, that also adopts fitting at prior positions,
shows that our measured fluxes are systematically higher than those from \citet{Muzzin2013} by a factor of 1.2, which appears to be a calibration offset that applies consistently also with respect to the \citet{LeFloch2009} catalog.
Given that \citet{Muzzin2013} did not compare to the catalog of \citet{LeFloch2009}, the source of the observed difference is unclear. However, we notice that the scaling goes in the opposite direction as what is inferred elsewhere in our work, namely that the 24~$\mu$m photometry might need to be re-calibrated higher by some factor even with respect to our measurements (and those from \citealt{LeFloch2009}).
More details about the calibration of the 24~$\mu$m photometry are discussed in Appendix~\ref{24um_calibration}. Once accounting for this calibration difference, our results suggest that the photometric uncertainties of 24~$\mu$m fluxes from \citet{Muzzin2013} are often mildly underestimated.

\subsection{Photometry at VLA 1.4 and 3~GHz on $K_s$-selected priors}
\label{Section_Photometry_radio}

Radio continuum emission is also an important tracer of star formation.
\citet{Yun_M2001} found a strong correlation between radio and far-infrared, expressed in terms of the logarithmic ratio $q_\mathrm{IR}$ between IR and radio fluxes. Recently, some evolution of $q_\mathrm{IR}$ has also been revealed: $q_\mathrm{IR}$ appears to be decreasing with increasing redshifts \citep{Magnelli2015,Delhaize2017}. This helps for the detection of high redshift star forming galaxies in the radio, because their radio emission is expected to be brighter.

We use the 1.4~GHz Deep Project map \citep{Schinnerer2010} and the 3~GHz Large Project image \citep{Smolcic2017} from the VLA-COSMOS team.
The 1.4~GHz Deep Project map was combined with the existing data from the VLA-COSMOS 1.4~GHz Large map \citep{Schinnerer2010}. It covers 1.7 $\deg^2$ area with an angular resolution of $2.5''$, reaching $\sigma \approx 12 \mu$Jy in the central $50' \times 50'$ but shallower elsewhere.
The VLA-COSMOS 3 GHz Large Project \citep{Smolcic2017} is based on 384$h$ of VLA observations. The final mosaic reaches a median rms of 2.3 $\mu$Jy beam$^{-1}$ over the 2 $\deg^2$ at an angular resolution of 0.75$''$, corresponding to the best sensitivity and highest resolution of available radio surveys in COSMOS. \citet{Smolcic2017} presented a catalog of 10,830 blindly extracted $\mathrm{S/N}>5$ radio sources. With our prior-based PSF fitting technique we can push to deeper radio flux levels with high fidelity and completeness and with a very low spurious detection rate expected.

We use a circular Gaussian\footnote{A Gaussian PSF is a very good approximation of the true beam, especially given the fact that the large number of VLA antennas  does not result in significant sidelobes and that a Gaussian PSF is used for reconstruction during the CLEAN process \citep{Schinnerer2010,Smolcic2017}. 
On the other hand, the comparison of our measurements to the catalogs confirms this.
} PSF with FWHM=$2.5''$ and FWHM=$0.75''$, respectively, and run PSF fitting at the positions of 589,713 $K_s$+radio priors in the 1.4~GHz and 3~GHz images, {keeping also into account their RMS maps}. To improve the fitting, a second-pass fitting is performed in both images, allowing for variation of positions for bright sources (\galfit{} $\mathrm{S/N>20}$) of up to 2 pixels ($0.4''$ at 3~GHz, $0.7''$ at 1.4~GHz). 

Given the high resolution of the radio images we only ran 2-step correction recipes in the Monte Carlo simulations. Correction for crowding (see L18) is not required as priors in the image are basically never crowded ($crowdedness\sim1$ always). 
The simulation recipes return a typical effective rms sensitivity of 2.5--2.7 $\mu$Jy at 3~GHz with well behaved Gaussian-like uncertainties, which is close to the nominal 2.3$\mu$Jy rms noise, allowing for reliable $\mathrm{S/N}>3$ detections at 3~GHz down to $\sim8\mu$Jy. 
The 1.4~GHz fitting gives a median uncertainty of 10.22~$\mu$Jy (albeit better in the central area), shallower than the 3~GHz photometry but useful for SED fitting.

We compare our radio photometry to the blind-extracted catalogs of \citet{Smolcic2017} and \citet{Schinnerer2010}, as shown in Fig.~\ref{radio_result}. At 3~GHz, our flux measurements are consistent to the unresolved fluxes in the catalog of \citet{Smolcic2017}, while we underestimate the fluxes of resolved sources.
In the right panel of Fig.~\ref{radio_result}, we show the comparison with the catalog of \citet{Schinnerer2010}: our 1.4~GHz measurements are consistent to the peak fluxes in their catalog.
As a result of these measurements, we obtain additional 7,005 $\mathrm{S/N>3}$ radio sources which have a $K_s$ counterpart but were not reported in the catalog of \citet{Smolcic2017}. We consider most of these to be reliable radio detections given that our photometric uncertainties are quasi Gaussian and we would therefore expect only of order of 10\% of these extra sources to be the result of noise fluctuations (for purely Gaussian noise).
Including sources already detected in the \citet{Smolcic2017} catalog, for which we adopt their fluxes,
we have 15,645 reliable radio detections in total in the UltraVISTA area.

\begin{figure}[ht]
\centering
\includegraphics[width=0.48\textwidth]{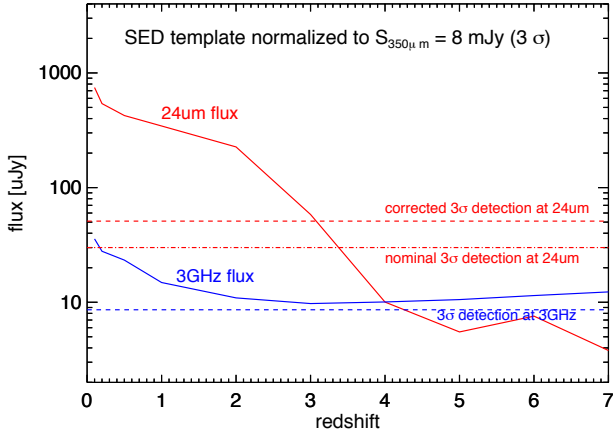}
\caption{%
	Analog to Fig.~3 in L18, showing the flux density expected at 24~$\mu$m (red) and 3 GHz (blue) in the COSMOS field as a function of redshift, for the faintest detectable sources at 350$\mu$m. Red and blue horizontal lines show the $3\sigma$ detection limits at 24~$\mu$m and 3~GHz respectively. The dot dashed line marks the nominal $3\sigma$ detection limit at 24~$\mu$m (i.e, $1\,\bar{\sigma}$ in Table~\ref{Table_1}), while the $1.7\times$ corrected limit is shown in red and dashed style.
	The SED template adopted to scale among different wavelengths is identical to the one used in L18, a main sequence template from \citet{Magdis2012SED} with redshift-evolving $q_\mathrm{IR}$ \citep{Magnelli2015}, normalized to a $3\sigma$ detection at SPIRE 350~$\mu$m ($S_{350\mu m}=8m$Jy).
	\label{Fig_SED_expected_flux_against_z}%
}
\end{figure}

To improve PSF fitting for resolved sources in the 3~GHz image, and attempting to recover the radio flux contributions that we might be resolving out, we perform prior-based fitting in Gaussian convolved 3~GHz images whose PSFs were increased to 1.5$''$ and 2$''$. Obviously the convolution increases the noise in the data, 
rising the median flux uncertainties to $\sigma=5.57\mu$Jy and $\sigma=7.66\mu$Jy in the fitting of convolved images with 1.5$''$ and 2$''$ PSFs, respectively.

On the other hand, this procedure might also introduce biases, e.g., for sources with radio lobes on the sizes of these beams, which could eventually contribute to the recovered radio flux.
Weighting the advantages and disadvantages of using fluxes from convolved radio images,
we decided to use the photometry from the original image fitting for the rest of this work because it is much deeper.
However, we publicly release the photometry from the convolved images as well, as it might be useful to the users from the community.

Early results from our radio photometry catalog constructed in this way were used for the work reported by \citet{Daddi2017}, which found that a strong overdensity of radio sources are hosted by the $z=2.5$ X-ray-detected cluster Cl~J1001, demonstrating
interesting prospects for future deep and wide-area radio surveys to discover large samples of the first generation of forming galaxy clusters.

\subsection{The effective depth of our 24~$\mu$m+radio prior catalog}
\label{smallsec}

In Fig.~\ref{Fig_SED_expected_flux_against_z} we plot the predicted flux at 24~$\mu$m and 3~GHz as a function of redshift for the faintest SPIRE sources that we could hope to be detected with our survey, using dust continuum SED templates from \citet{Magdis2012SED} and adopting the evolving FIR-radio correlation from \citet{Magnelli2015}, 
All SEDs are normalized to a common $S_{350\,{\mu}\mathrm{m}}=8.0\,mJy$, which is the posterior 3$\,\sigma_{350\,{\mu}\mathrm{m}}$ detection limit in our final catalog. 
The flux at 24~$\mu$m decreases with increasing redshift, getting below the detection threshold at $z>3$--3.5 (lower if the actual 24$\mu$m calibration has to be altered).
Meanwhile, as can be seen from the blue solid line in Fig.~\ref{Fig_SED_expected_flux_against_z}, the radio emission at 3~GHz is always above the 3$\,\sigma_{3\,\mathrm{GHz}}$ detection limit at redshift 0--7, suggesting that galaxies within reach of detection at 350~$\mu$m are always in principle also detectable at 3~GHz. 
Therefore, including detections at VLA 3~GHz, we can obtain a more complete prior catalog particularly for $z\gtrsim3$ galaxies, and improve the completeness of our prior sample for fitting PACS, SPIRE and (sub)mm photometry. However, in reality, given the scatter in the FIR-radio correlation, the effect of noise, and possible flux losses from the use of $0.75''$ data, our effective depth in the radio is so close to the actual required limit that we do expect substantial amounts of incompleteness in our priors at $z>3$, even when considering radio.

{We have performed PSF fitting in the negative 3~GHz image to test our procedure. We found 795 $\mathrm{S/N>3}$ (negative) detections out of the total 589,713 positions fitted. This is 0.13\%, very consistent with the expected one-tail Gaussian probability at 3-sigma. 
Comparing to the 15,645 (positive) $\mathrm{S/N>3}$ detections, the spurious detection rate in the 3GHz catalog is estimated to be at the level of $5.1\%$, pretty low: this will not spuriously alter our prior source density while ensuring that we are fitting the IR maps at truly detected positions in most cases.
The same test in the 24um negative image shows $\mathrm{S/N>3}$ spurious detection rate $0.07\%$ with respect to fitted positions and 0.5\% with respect to actual detections, a return rate of spurious even slightly lower than expected from Gaussian statistics.   
}

\subsection{Completing the prior catalog with stellar mass-selected galaxies}
\label{Section_The_24_Radio_Mass_Catalog}

\begin{figure}
\centering
\includegraphics[width=0.5\textwidth]{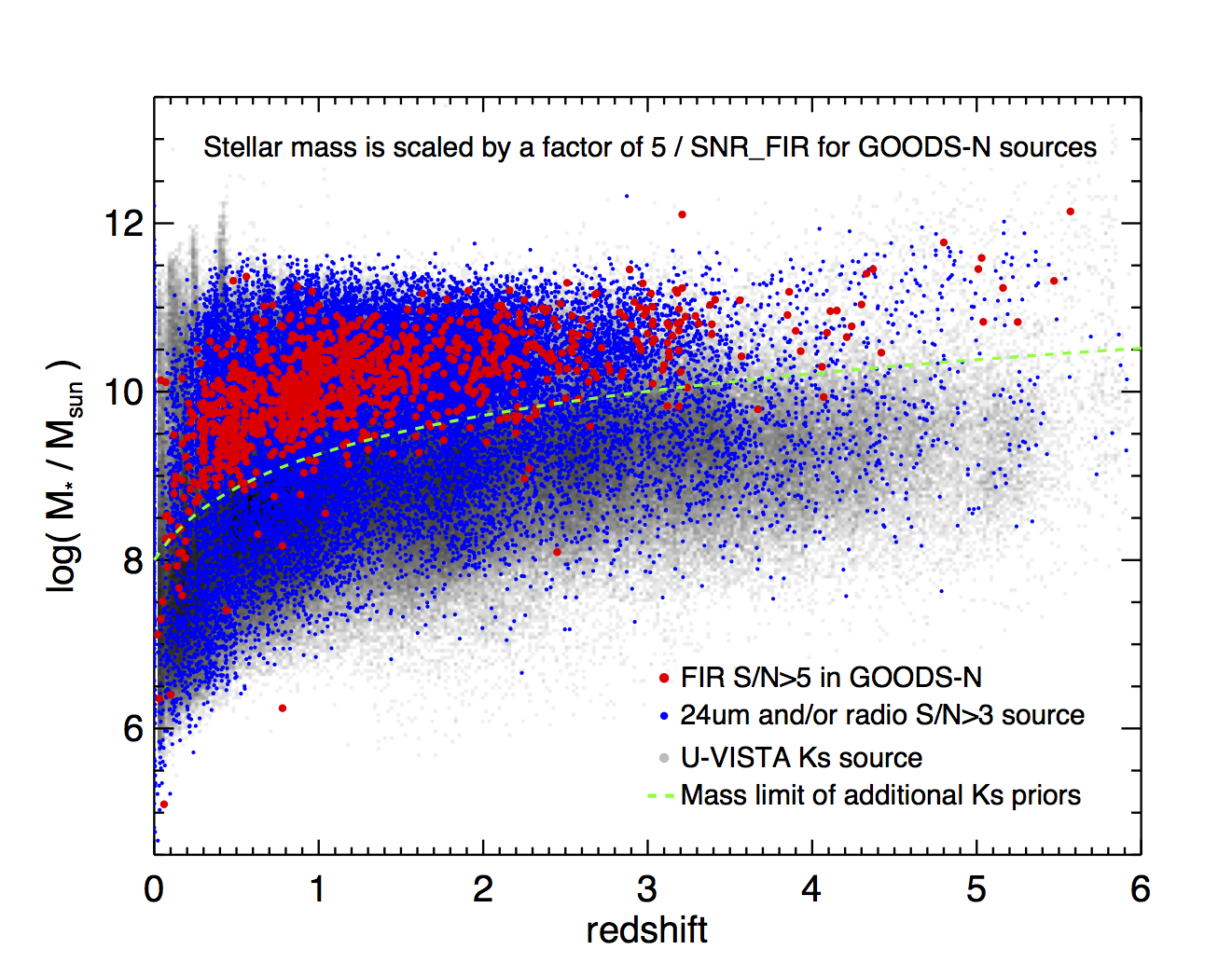}
\caption{%
	The selection of the 24~$\mu$m+radio+mass-selected prior catalog for FIR/(sub)mm bands. Red dots: the $\mathrm{S/N_\mathrm{FIR+mm}}>5$ sources from the \Superdb catalog in the GOODS-North field \citep{Liu_DZ2017}, whose stellar mass is multiplied by a factor of $\frac{5}{\mathrm{S/N_\mathrm{FIR+mm}}}$ to effectively re-normalize all these IR detected sources to the $5\sigma$ detection limit in GOODS-N. Blue dots: the 3$\sigma$ detections at 24~$\mu$m and/or radio bands in our catalog. Grey dots: the UltraVISTA $K_s$ sources \citep{Laigle2016,Muzzin2013}. Green dashed line: the stellar mass limit to select additional $K_s$ priors. Most of mass-scaled FIR detected galaxies (red dots) are above our stellar mass limit $\log M_{\ast} > 1.8 \times \log (1+4\times z)+8$ (green dashed line). 
	\label{Mass_selected_priors}%
	}
\end{figure}

From the prior-based fitting at positions of 589,713 $K_s$+radio priors in 24~$\mu$m and radio images, we obtained 88,008 galaxies with 24~$\mu$m and/or radio $\mathrm{S/N}>3$. They are shown as blue dots in Fig.~\ref{Mass_selected_priors}. 
As mentioned before, the full sample of 589,713 $K_s$+radio sources in COSMOS are too dense to be useful in the fitting of FIR/(sub)mm images, where the source density reaches 5-50 sources per beam at FIR/(sub)mm bands. 
The 88,008 detections at 24~$\mu$m and/or radio bands are preferentially included in the prior catalog for FIR/(sub)mm deblending, and have much more manageable sky densities, but are lacking in completeness as discussed in the previous sections.
Thanks to well-measured stellar masses and photometric redshift in the UltraVISTA catalogs, we can select a supplemented prior sample by stellar mass, exploiting once again the tight connection between stellar mass and SFR and thus IR luminosity.

In Fig.~\ref{Mass_selected_priors}, $\mathrm{S/N_\mathrm{FIR+mm}}>5$ sources from the GOODS-N \superdb catalog \citep{Liu_DZ2017} are shown as red dots. We scaled down their stellar masses multiplying their values by a factor of $\frac{5}{\mathrm{S/N_\mathrm{FIR+mm}}}$ to re-normalize the location of all sources to the 5$\sigma$ detection limit of the catalog.
There is a close connection between FIR luminosity and stellar masses in star forming galaxies {(as widely discussed in the literature in terms of the so-called star forming Main Sequence; \citealt{Elbaz2007,Noeske2007,Daddi2007,Pannella2009,Rodighiero2011,Karim2011,Schreiber2015})}, that can be seen by the relative small dispersion of $\sim0.3$~dex of the red dots along their average redshift trend.

The scaled GOODS-N sources statistically locate positions expected for galaxies detectable on average at $\mathrm{S/N_\mathrm{FIR}} = 5$ in this mass-redshift diagram. Scaling their trend a factor of 5 lower we obtained the stellar mass limit shown as a dashed line in Fig.~\ref{Mass_selected_priors}, which corresponds to the $\mathrm{S/N_\mathrm{FIR}} = 1$ limit, or to deviations of $\sim 2.5\sigma$ from the average trend.
{Therefore a prior catalog including galaxies down to this stellar mass limit will be able to include the vast majority of  FIR/(sub)mm detectable sources.}
The number of IR detected outliers in GOODS-N that would drop below our mass limit is only of 4\% at all redshift, and 7\% at $z>2$. Notice though that we still have 24$\mu$m and radio selected priors below this stellar mass limit.

We thus selected 106,420 $K_s$ sources, with stellar mass $\log \mathrm{M_\ast} > 1.8\times \log (1+4\times z)+8$, as a supplement for the 24~$\mu$m+radio priors. 
In total, we have 194,428 sources in this 24~$\mu$m+radio+mass-selected prior catalog. 
The green dashed line in Fig.~\ref{Mass_selected_priors} visually shows this stellar mass selection threshold.
{Some 96\% of all UltraVISTA sources with stellar mass $\mathrm{M_\ast>10^{9.5} M_\odot}$ are included in this prior catalog, and 68\% of sources with $\mathrm{M_\ast>10^{9.0} M_\odot}$ in the $z=0-4$ redshift range.}
This prior catalog has a source density $\left<\rho_{\mathrm{beam}}\right>\sim 0.5$ at PACS 100~$\mu$m (i.e., surface density $\sim 11$ sources per $arcmin^2$), $2\times$ the surface density of the 24~$\mu$m+radio prior catalog in L18 (i.e., $\sim 5$ sources per $arcmin^2$ in GOODS-N).

The remaining 395,285 $K_s$ sources are not included in our prior catalog. Similar to L18, we assume that their flux contributions to the PACS, SPIRE and (sub)mm images are negligible and do not consider them for the rest of this work. As discussed in L18, their presence will act as a background whose average level will be accounted for consistently by our procedure, while their possibly inhomogeneous distribution will also be accounted for by error bars in the finalized photometry.

We notice that, in principle, we might have supplemented stellar mass-selected priors only for $z>2$ as in general we expect our $24\mu$m catalog to be highly complete at $z<2$ at least for SPIRE bands (see Fig.~\ref{Fig_SED_expected_flux_against_z}).
However, we might be missing in this way silicate-dropout sources at $1<z<2$ \citep{Magdis2011Dropout24}, that are otherwise detectable in FIR, e.g., in PACS images. On the other hand, for the SPIRE bands our \superdb procedure effectively removes from the fitting pool most of the stellar mass selected extra priors at $z<2$, so their presence is not negatively impacting our results, while guaranteeing a higher completeness for priors at all redshifts and for the PACS bands.


\section{\Superdb photometry in the FIR/(sub)mm images}
\label{Section_Superdeblending}

In this section, we describe the ``Super-deblending'' process applied to images from PACS, SPIRE, SCUBA2, AzTEC and MAMBO.
Key differences with respect to the deblending work of L18 are listed in Section~\ref{deblending_difference}.

\begin{figure}
\centering
\includegraphics[width=0.37\textwidth, trim={1cm 17.8cm 3.65cm 3.3cm}, clip]{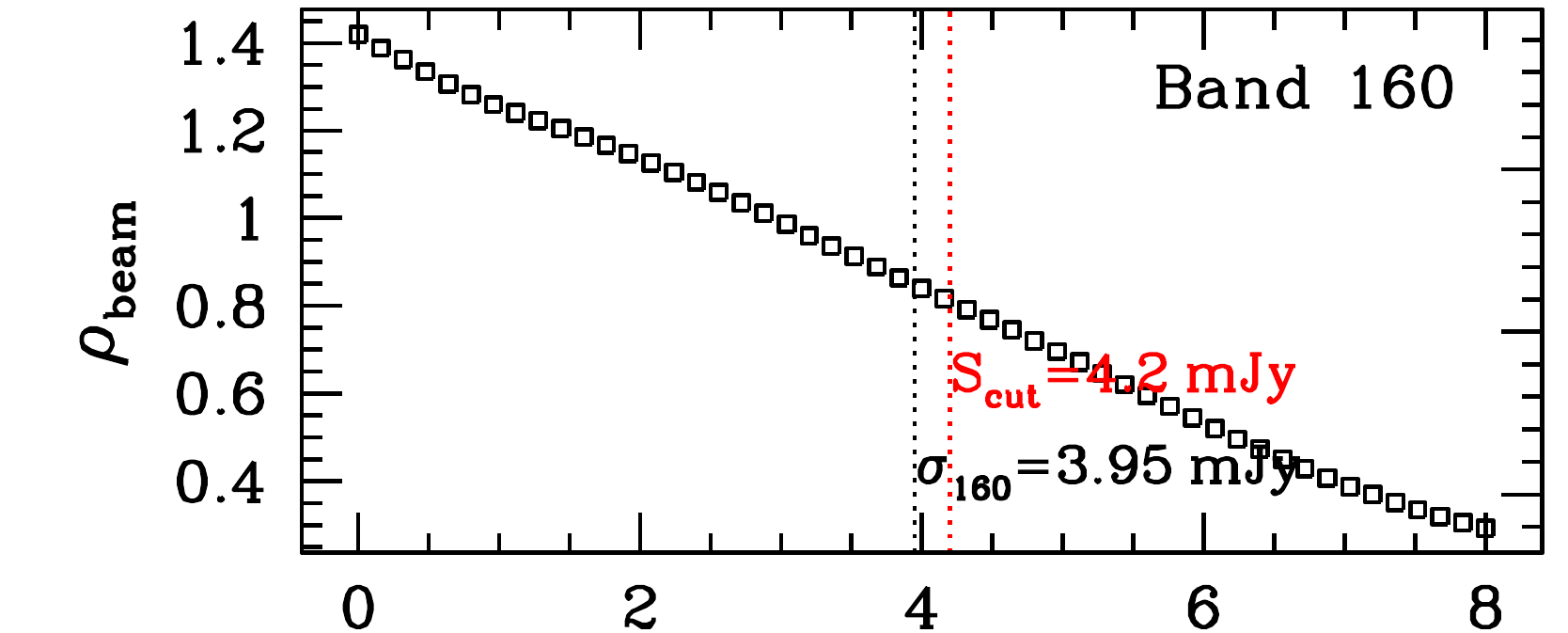}
\includegraphics[width=0.37\textwidth, trim={1cm 17.8cm 3.65cm 3.3cm}, clip]{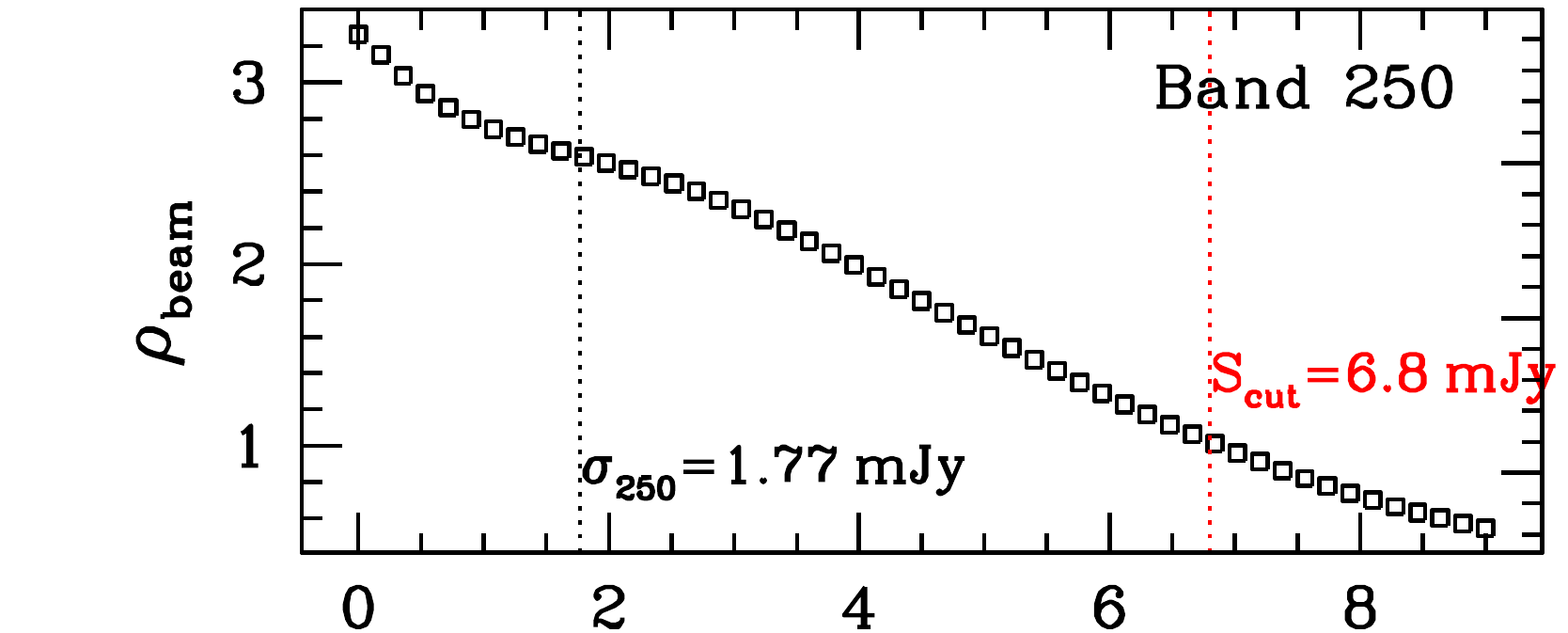}
\includegraphics[width=0.37\textwidth, trim={1cm 17.8cm 3.65cm 3.3cm}, clip]{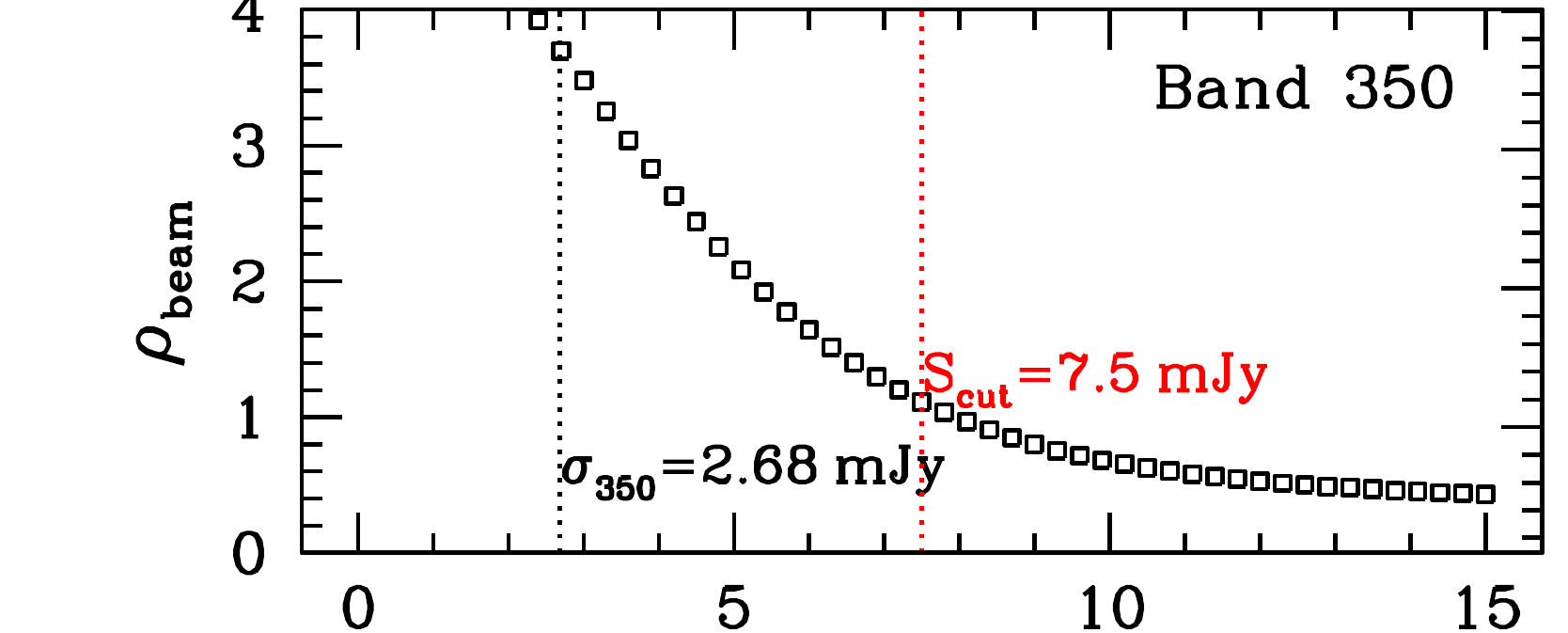}
\includegraphics[width=0.37\textwidth, trim={1cm 17.8cm 3.65cm 3.3cm}, clip]{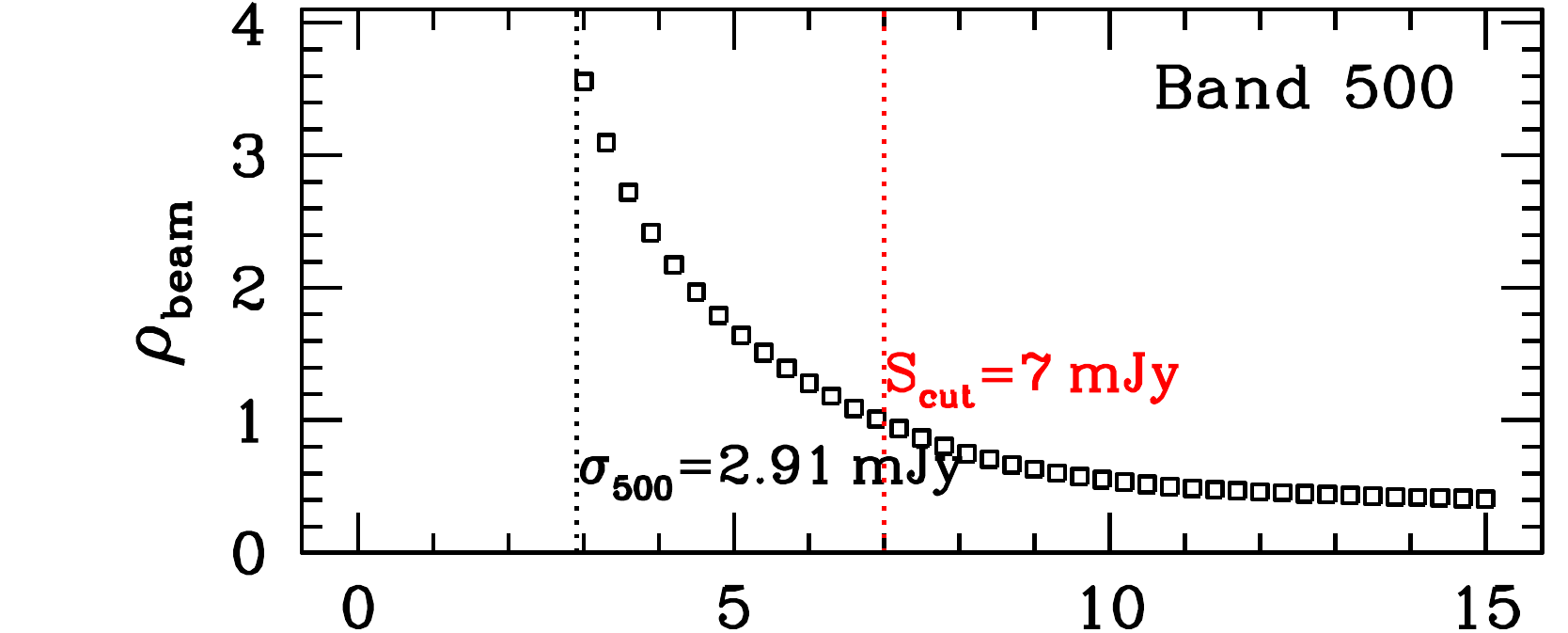}
\includegraphics[width=0.37\textwidth, trim={1cm 17.8cm 3.65cm 3.3cm}, clip]{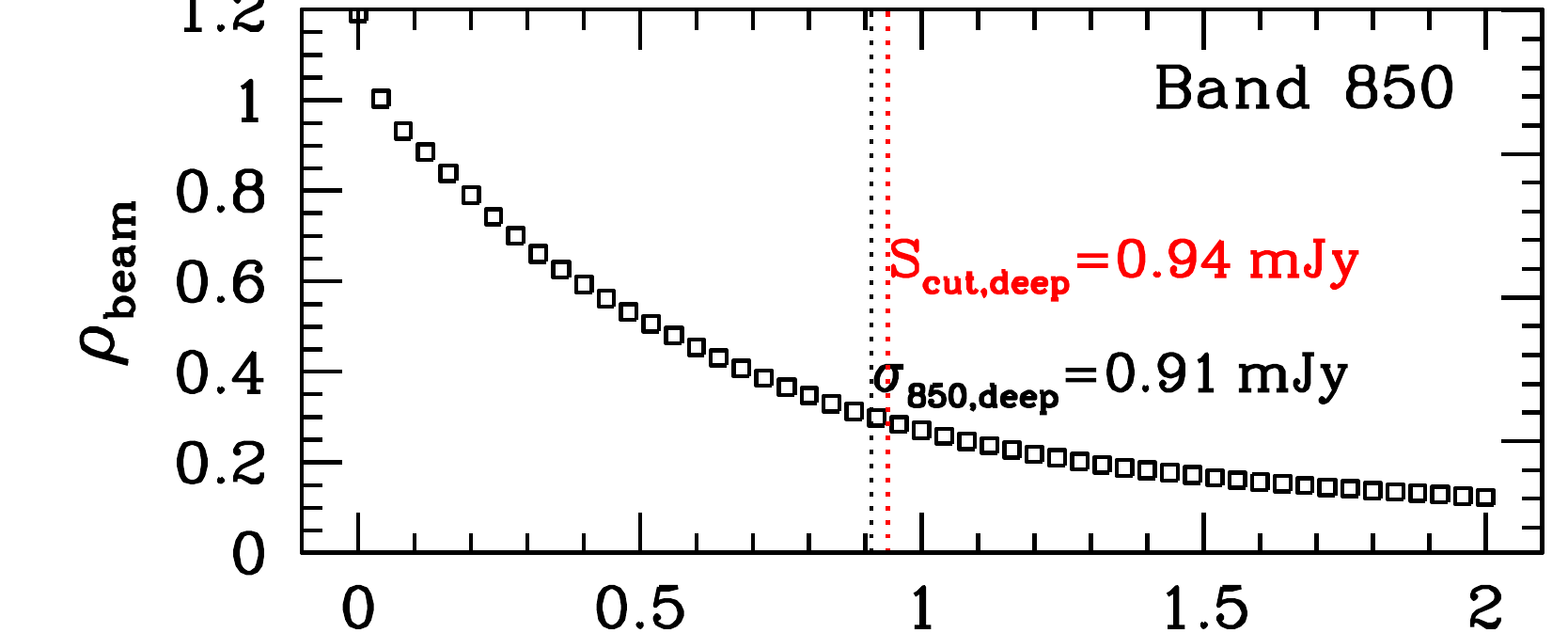}
\includegraphics[width=0.37\textwidth, trim={1cm 17.8cm 3.65cm 3.3cm}, clip]{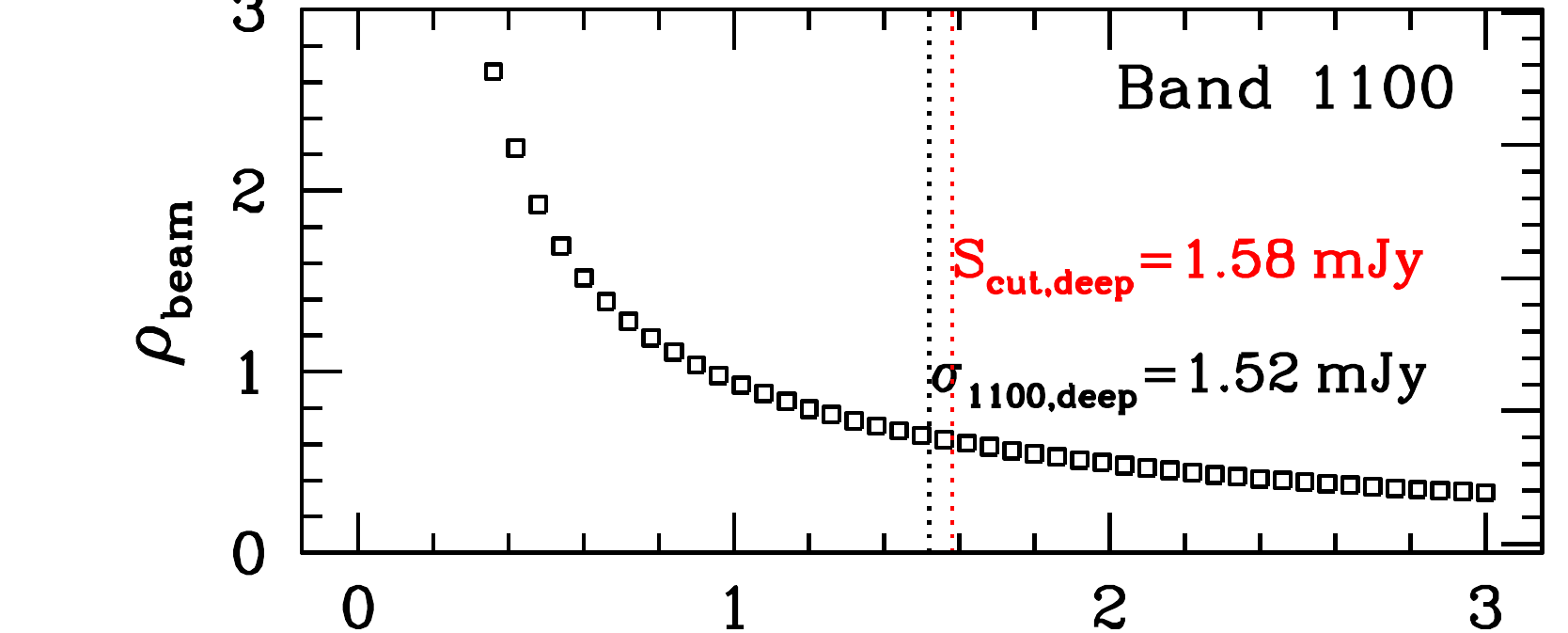}
\includegraphics[width=0.37\textwidth, trim={1cm 16.5cm 3.65cm 3.3cm}, clip]{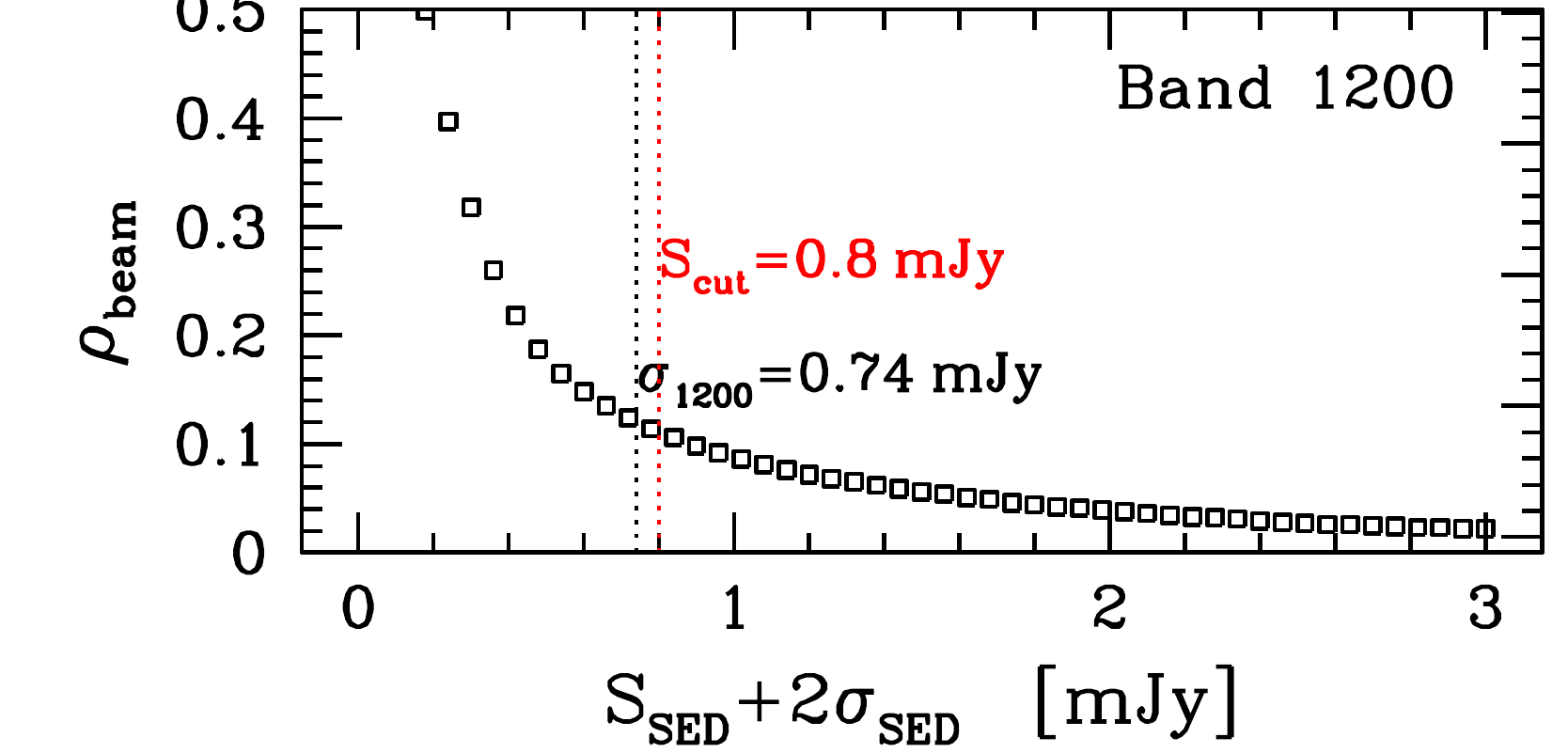}
\caption{%
	Analog to Fig.~5 in L18, showing the chosen flux limits for selection of \textit{excluded} and \textit{selected} sources at each band. Each panel shows the cumulative sources density $\rho_{\mathrm{beam}}$ of priors versus their expected flux. 
    At PACS, SPIRE and MAMBO band, sources with $S_{\mathrm{SED}}+2\,{\sigma}_{{\mathrm{SED}}} \ge S_{\mathrm{cut}}$ are selected for fitting, while the rest are excluded from the fitting.
    Fitted priors at SCUBA2 850~$\mu$m and AzTEC 1.1mm are selected via $(S_{\mathrm{SED}}+2\,{\sigma}_{{\mathrm{SED}}} ) / \sigma_{\mathrm{rms\,noise}} > 1$, and the corresponding $S_{\mathrm{cut,deep}}$ and sensitivity ${\sigma}_{deep}$ of their deepest region are shown in panels of 850~$\mu$m and 1.1~mm, where lesser priors are fitted in shallower regions.
}
\label{Fig_Galsed_cumulative_number_function}%
\end{figure}

We first run SED fitting to predict the flux of each source in each band, where the SED procedures and parameters in this work are identical to that in L18 (see Section 3 in L18).
Then we determine a critical flux value for selecting an actual prior source list to fit at each band by considering both the number density and the expected flux detection limit.
The selection of fitted priors is described in Section~\ref{Section_Faint_Prior}, and the critical fluxes at each band are presented in Fig.~\ref{Fig_Galsed_cumulative_number_function} and table~\ref{Table_1}.
{We apply the same sources density criteria to define prior lists for fitting at FIR/(sub)mm bands, as discussed in Section 7.4 of L18.}
In this way the number of fitted sources in each band can be kept to reasonable values of $\leqslant 1$ per PSF beam area. 

{At the next step, we perform the faint-source-subtraction and the actual flux measurement on retained priors, by running PSF fitting on the map with \galfit{}. These measurements  will be corrected for biases and reliable uncertainties determined, by applying the results of Monte Carlo simulations that are performed by inserting one source (of known flux) at a time in the image  (with its own crowding and blending). This is a crucial step to ensure the quality of measurements for both fluxes and errors.
The newly obtained photometry at the band under exam will then be appended to the catalog being constructed, and used in the SED fitting for predicting fluxes at next band.}

\begin{figure*}
\centering
\includegraphics[width=0.57\textwidth]{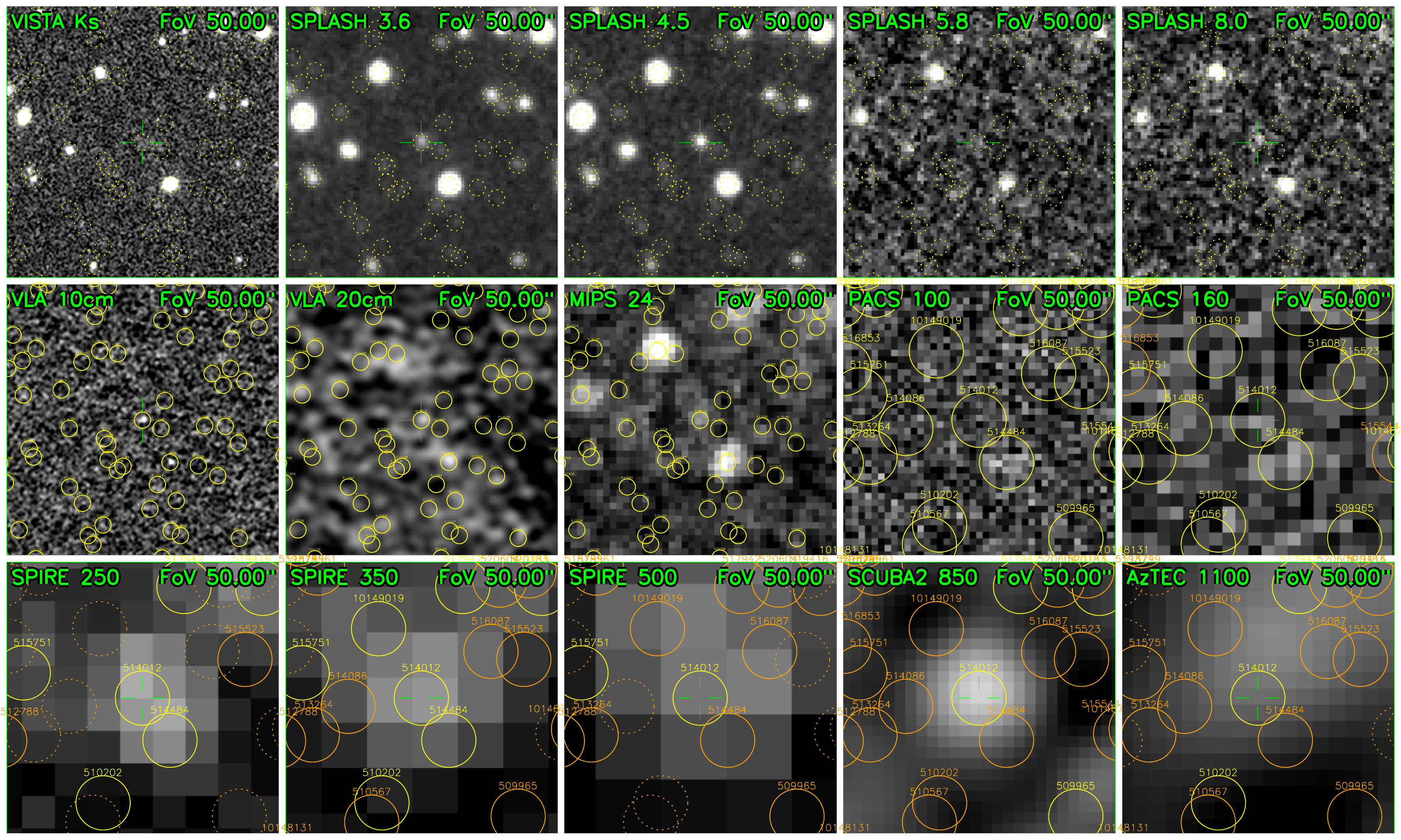}
\includegraphics[width=0.35\textwidth, trim={0.6cm 5cm 1cm 3.5cm}, clip]{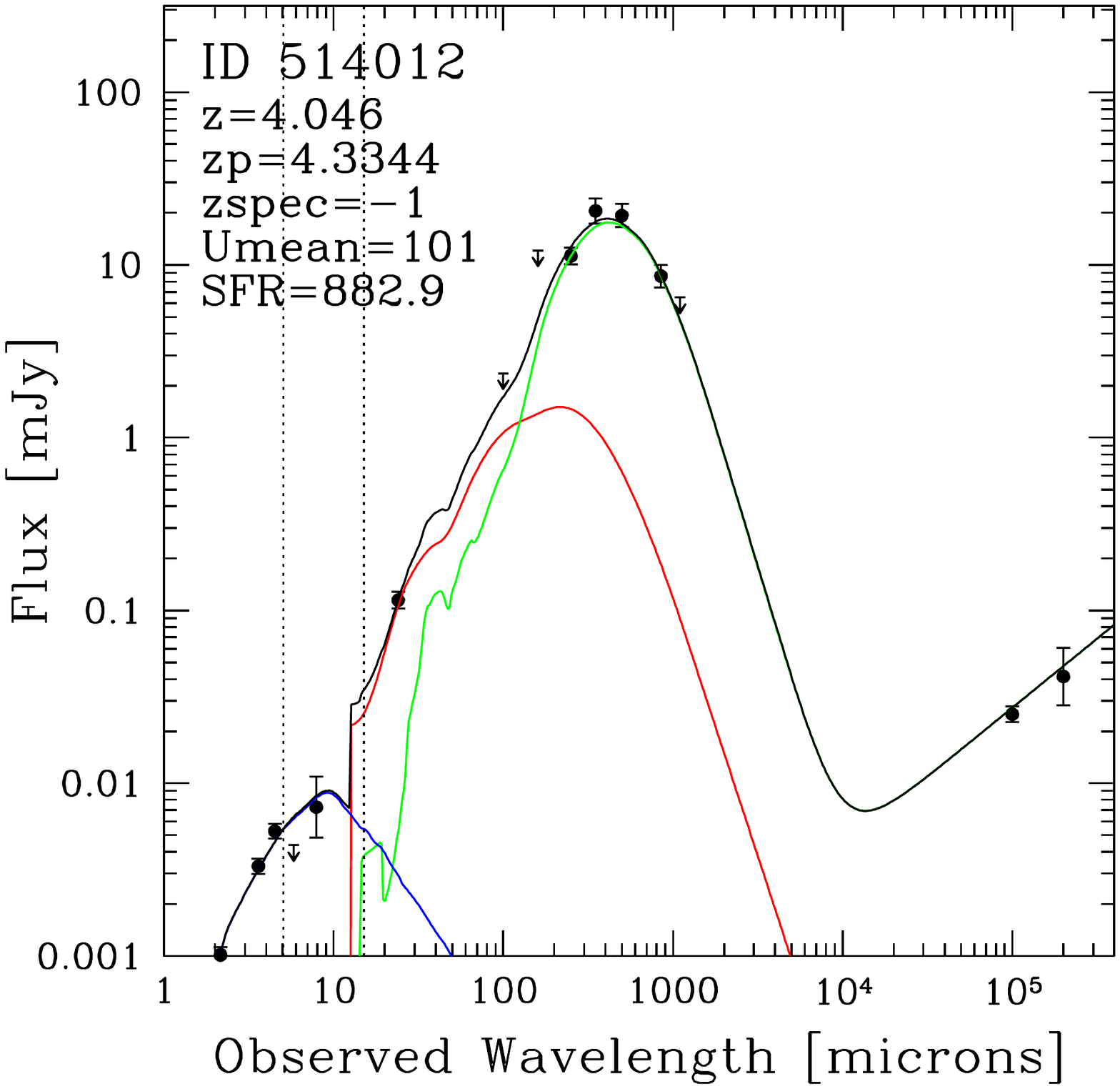}
\caption{%
  	An example of the ``Super-deblending'' process: we fit the source ID514012 together with other sources in an image box at each band and show its finalized SED in the right panel. 
    {Left panel:} Multi-band cutouts in a $50''\times 50''$ box. Green text in each cutout marks the data set and field of view (FoV). Yellow solid circles show the priors that are actually fitted in each image (yellow dashed circles show positions of all the prior sources in UltraVISTA $K_s$ and SPLASH images), while orange solid circles show faint sources which are excluded from fitting and subtracted from the map. Finally, orange dashed circles show \textit{excluded} sources neither fitted nor subtracted (because we could not accurately determine their flux, a part from the fact that they are too faint to deserve fitting).
    {Right panel:} The finalized SED fitting of source ID514012. Blue and red curves show the stellar component \citep{BC03} and the AGN torus emission \citep{Mullaney2011}, and the dust continuum emission is shown in green \citep{Magdis2012SED}. The $z_{\rm p}$ is the NIR photometric redshift from the COSMOS2015 catalog; Umean=101 is our code to mark that the source was fitted by a starburst-like template (GN20 template from \citealt{Magdis2012SED}). Downward arrow shows the 2-sigma upper limit at given wavelength.
	\label{example_SED}%
	}
\end{figure*}

Overall, for the \superdb work, SED fitting is ran for 194,428 prior sources for predicting their fluxes at FIR/(sub)mm bands for each of the 7 wavelength steps (see bands for which this is done in Fig.~\ref{Fig_Galsed_cumulative_number_function}),
and once more for finalizing the fitting with all photometry in the end. PSF fitting with \galfit{} at each band is performed twice (2 passes) for selected priors and additional residual sources. As in L18, \galfit{} is run on overlapping sub-image regions with the size of order 5-10 times the PSF in each band in order to allow convergence and keep computation time manageable. There are of order of $6\times 10^{3}$--$2\times 10^{5}$ fitting regions per band (depending on the band) over the COSMOS field. 
Given that all these steps require very large computation time, we have optimized our algorithm to run on parallel computing clusters with large numbers of CPUs, and we were able to use 60 CPUs on average from the CEA/Irfu clusters.
Globally, the full measurement procedure 
in COSMOS once the prior sample is in hand requires of order $\sim 60$ computation days to be carried out on the 60 CPU cores, the most time-consuming part being the SED fitting for flux prediction, requiring of order 5 days at each wavelength step.
The actual \galfit{} fitting takes on average two days per wavelength step
(longer on radio/24$\mu$m images where we run on all $K_s$ galaxies, hence on a much larger number of priors, shorter for SPIRE where fewer priors are fitted).

Examples of fitting images and a finalized SED are shown in Fig.~\ref{example_SED}.

\begin{figure}
\includegraphics[width=0.4\textwidth, trim={1cm 13.5cm 0cm 3.2cm}, clip]{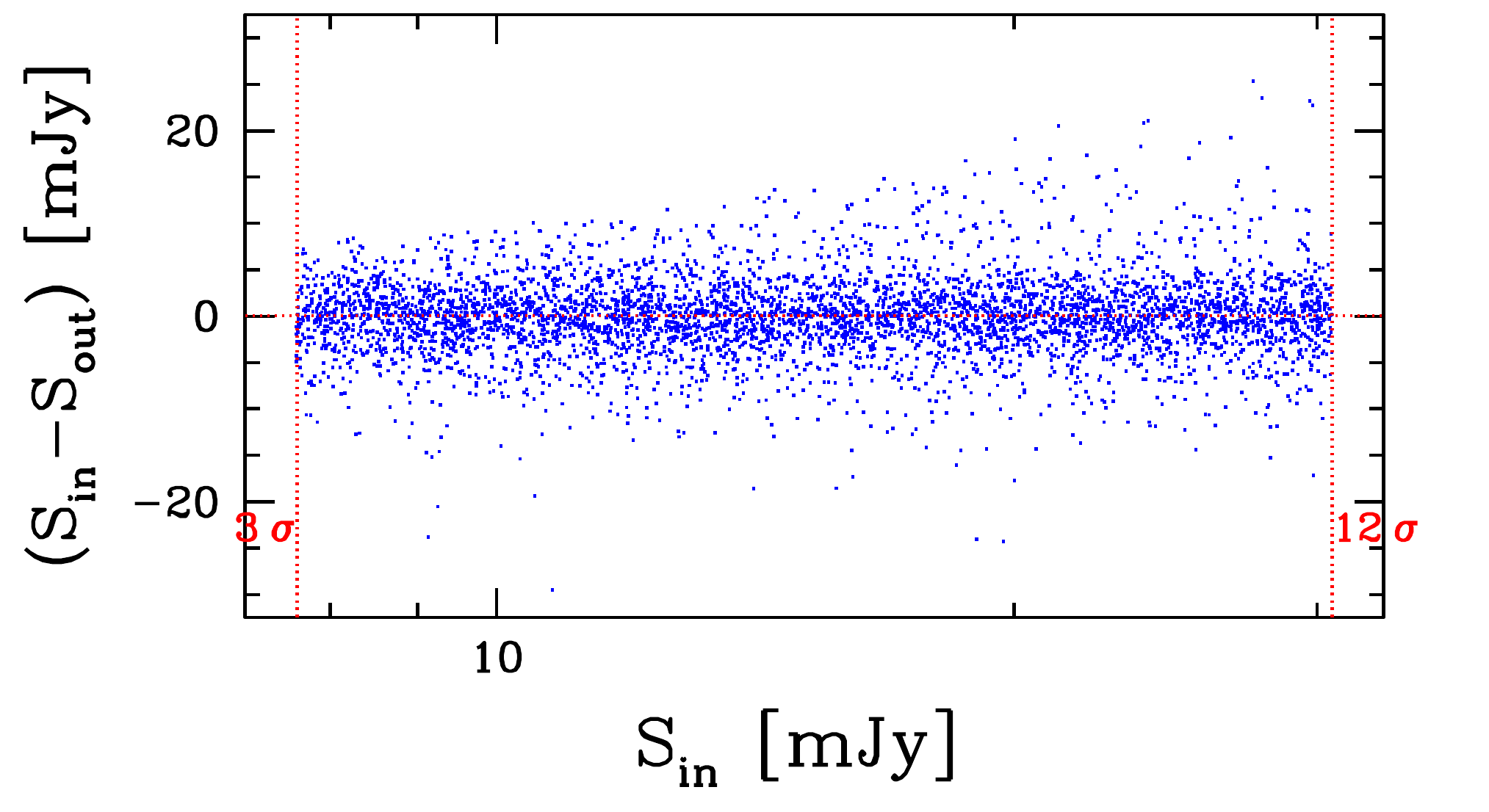}
\includegraphics[width=0.4\textwidth, trim={1cm 15cm 0cm 3.5cm}, clip]{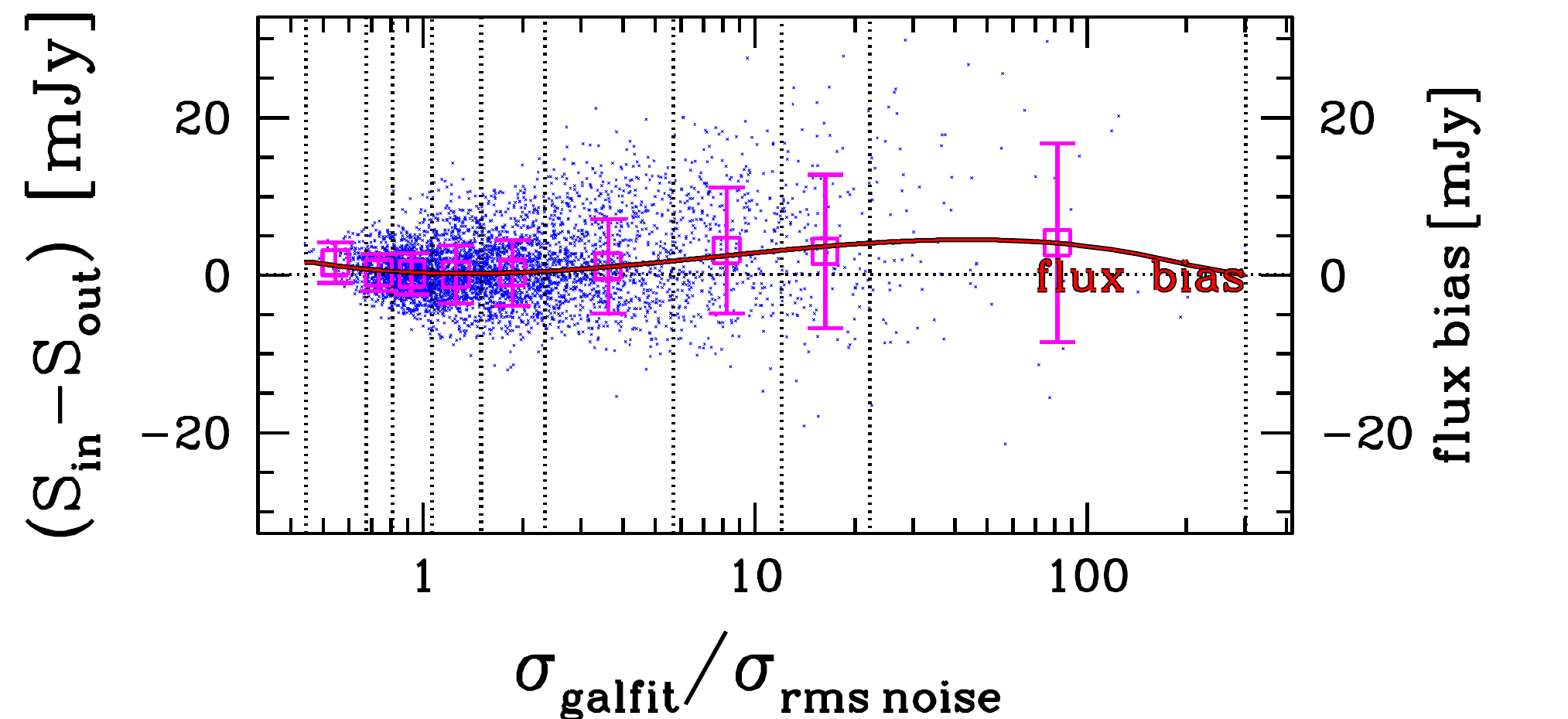}
\includegraphics[width=0.4\textwidth, trim={1cm 15cm 0cm 3.5cm}, clip]{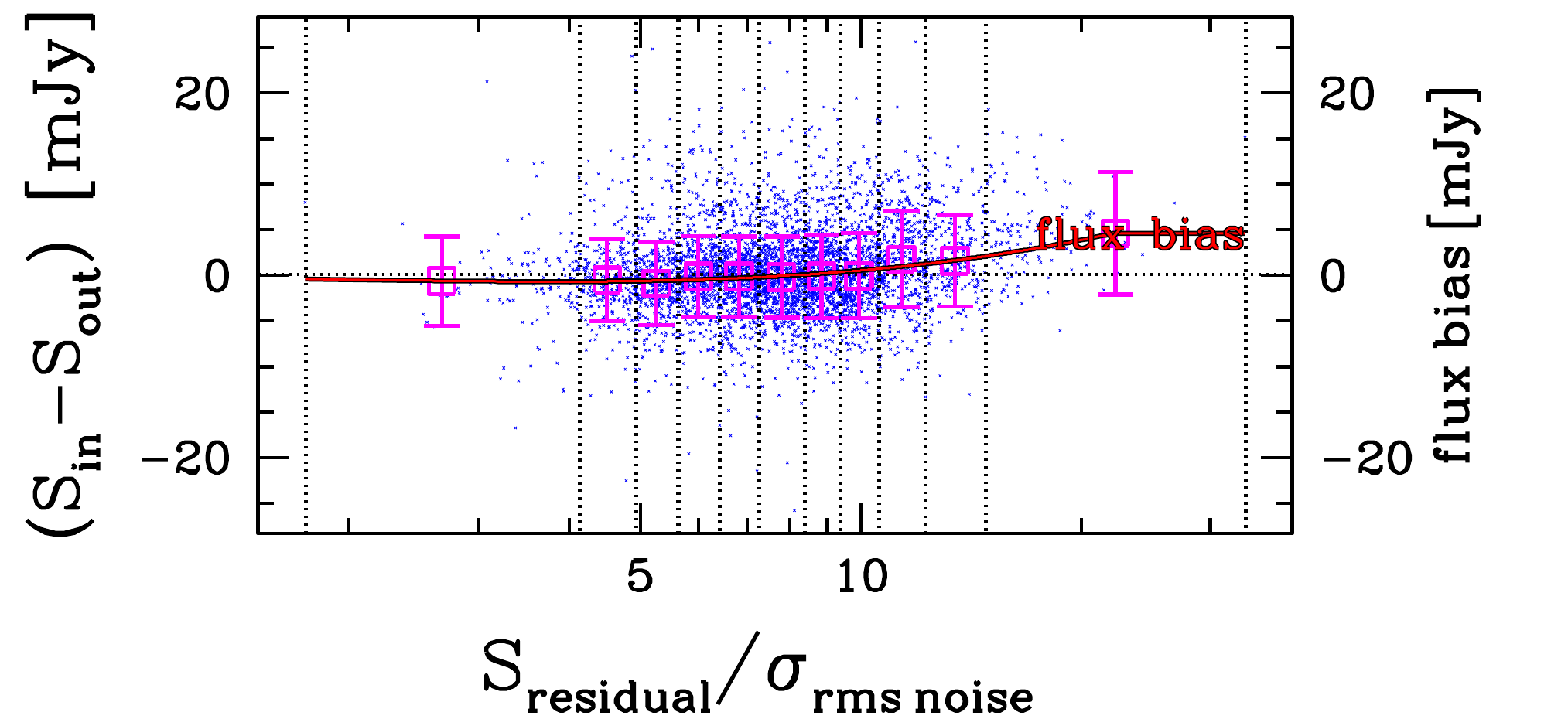}
\includegraphics[width=0.4\textwidth, trim={1cm 15cm 0cm 3.5cm}, clip]{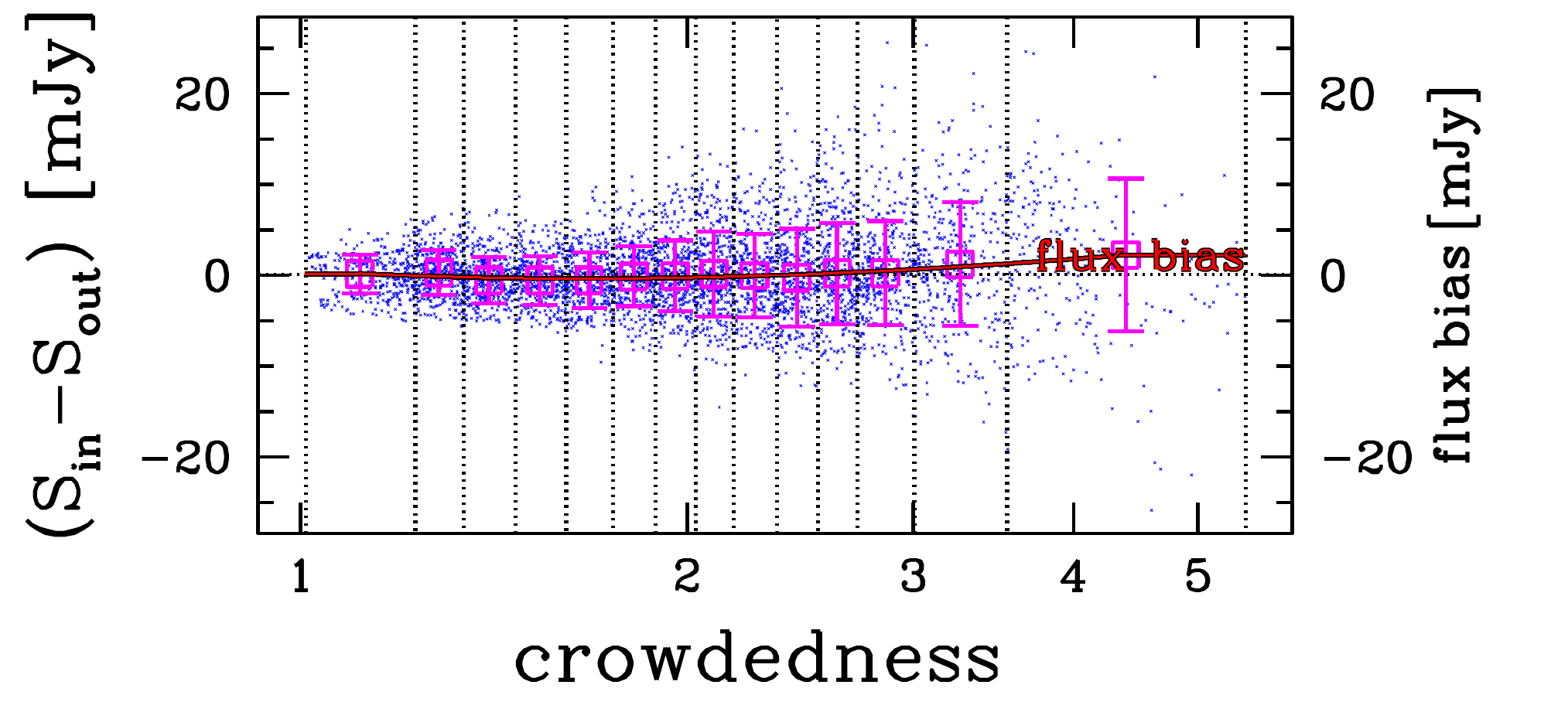}
\includegraphics[width=0.4\textwidth, trim={1cm 15cm 0cm 3.5cm}, clip]{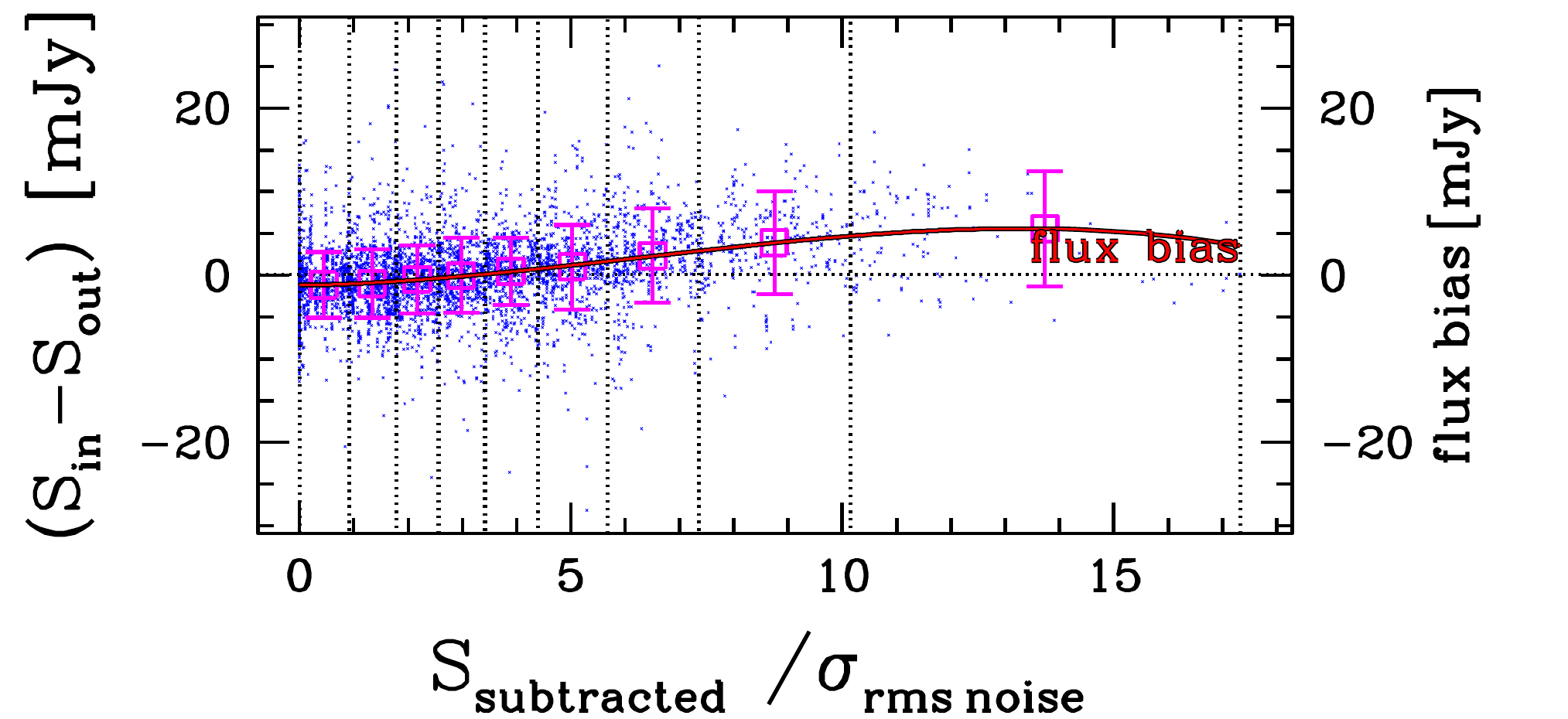}
\caption{%
	Analog to Fig.~10 in L18, we show the flux bias in SPIRE 350~$\mu$m simulations.  Magenta squares with error bars show the median flux bias and the dispersion of the data in each bin.
    Differently from L18, we consider the normalized subtracted flux (i.e., $S_\mathrm{subtracted} / \sigma_\mathrm{rms\ noise}$) as a fourth parameter for calibrating flux bias. This is shown in the bottom panel, where we analyze the variation of flux bias on the normalized subtracted flux, and fit the variation by a polynomial function (the red curve). 
	\label{Fig_Galsim_flux_bias_SPIRE}%
}
\end{figure}

\subsection{Differences in deblending work from L18}
\label{deblending_difference}

We list below all relevant differences with respect to the L18 GOODS-N work for the process of FIR/(sub)mm deblending (details of some of the most crucial differences are further provided in the following sections): 

(1) We do not subtract faint priors from the PACS 100~$\mu$m map, given that the 24~$\mu$m+radio+mass-selected prior catalog is less confused in this image ($\left<\rho_{\mathrm{beam}}\right> \sim0.5$ sources per beam). This step was almost irrelevant also in L18.

(2) We deblend the SCUBA2 850~$\mu$m image before fitting the SPIRE 350~$\mu$m map, given that the SCUBA2 850~$\mu$m image has a much higher spatial resolution (FWHM=11$''$ in the non-match-filtered image) and is very deep in about one fourth of the COSMOS field. This change of order provides better constraints on the SEDs and helps to improve flux predictions at 350~$\mu$m and 500~$\mu$m (particularly useful given that PACS imaging is generally shallower in COSMOS).

(3) The SCUBA2 850~$\mu$m deblending is performed in the non-match-filtered image, after removing hot pixels that appear to plague quite substantially the released COSMOS maps (both match-filtered and non-match-filtered). These hot pixels were identified as strong $>5\sigma$ outliers by median filtering the SCUBA images normalized by their RMS maps on scales of twice the beam (20 pixels squares), and replacing them with the actual median from $10\times10$ (PSF) filtering. 
The SCUBA2 PSF profile is obtained by stacking bright sources deemed to be reasonably isolated based on our modeling, and fitting the result with a 2D Gaussian.
{We note that we display the match-filtered version in the multi-wavelength cutouts (Fig.~\ref{example_SED}, Fig.~\ref{highz_4example} and Appendix~\ref{Section_highz}), for illustrative purposes.}

(4) Given that SCUBA2 and AzTEC images display significant depth variations over the COSMOS field, we adapted the flux threshold for accepting priors to the actual local depth, removing from the fitting pool all priors with predicted flux (plus twice the flux uncertainty, as in L18) below the $1\sigma$ depth (i.e., we exclude sources with $S_{\mathrm{SED}}+2\,{\sigma}_{{\mathrm{SED}}} < \sigma_{\mathrm{rms\,noise}}$ from the fitting, where $\sigma_{\mathrm{rms\,noise}}$ varies locally).

(5) Because the 24$\mu$m, radio and, most importantly, PACS imaging is shallower in COSMOS with respect to GOODS-N, the flux prediction process at each wavelength step displays considerably larger uncertainties.
This implies that for a large fraction of priors, while we can confidently infer that they are faint (below the threshold) and should not be considered for fitting, we cannot accurately determine their actual expected fluxes. This poses a problem to appropriately subtract their best/predicted flux from the images as they are ill-defined, and doing such might actually degrade the performances of our \superdb process rather than improving them.
We deal with this problem in Section~\ref{Section_Faint_Prior}.

(6) Meanwhile, in order to further cope with the limitation discussed above, we added a fourth step in our calibration recipes, analyzing results as a function of the total flux locally subtracted at each position, thus testing for flux biases and noise variations induced by this step (Section~\ref{section_simulations}).

Finally, we notice that 
the COSMOS field contains a sub-area observed by the ESA CANDELS-$Herschel$ key program (PI: M. Dickinson) with much deeper PACS imaging data \citep{Schreiber2015}.
We have fitted the deep PACS 100~$\mu$m \& 160~$\mu$m images in the CANDELS field, and 
used the results based on these deeper images for the CANDELS area.
Their fluxes and flux uncertainties have been calibrated by dedicated Monte Carlo simulations executed in the deep images. 

 Note that PACS fluxes measured from images need to be scaled up by a factor of 1.12$\times$ because of the flux losses from the high-pass filtering processing of PACS images \citep[e.g.][]{Popesso2012,Magnelli2013}. We have applied this factor in the SED fitting and also the released catalog, while this factor is not applied for the flux comparison in Fig~\ref{compare_PACS}.

\begin{figure*}
\begin{center}
\includegraphics[width=0.75\textwidth]{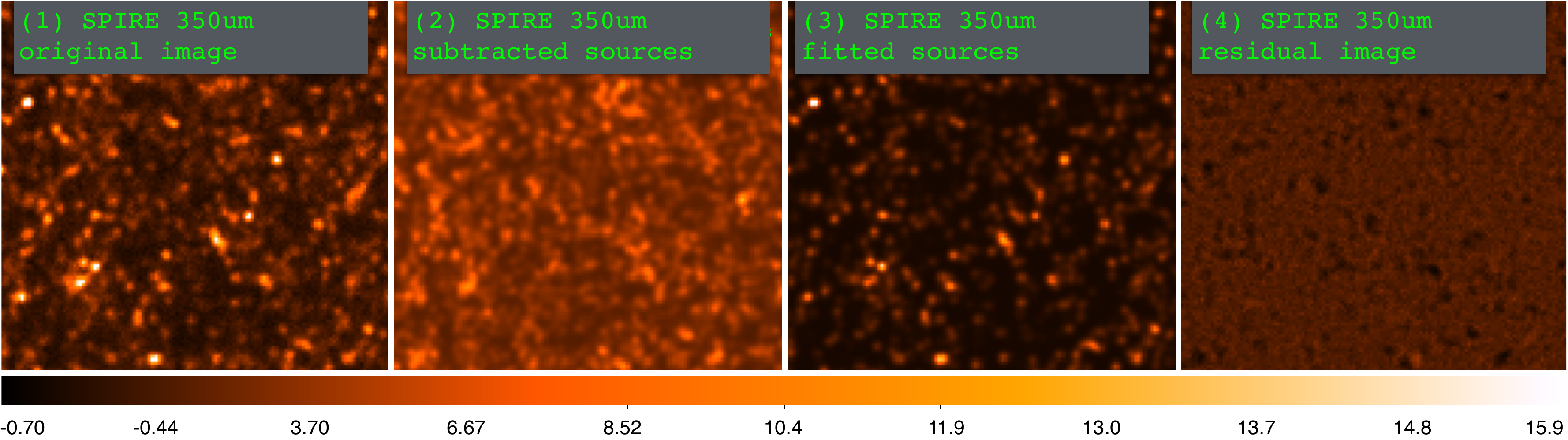}
\includegraphics[width=0.233\textwidth, trim={3cm 0cm 3cm 1.8cm},clip]{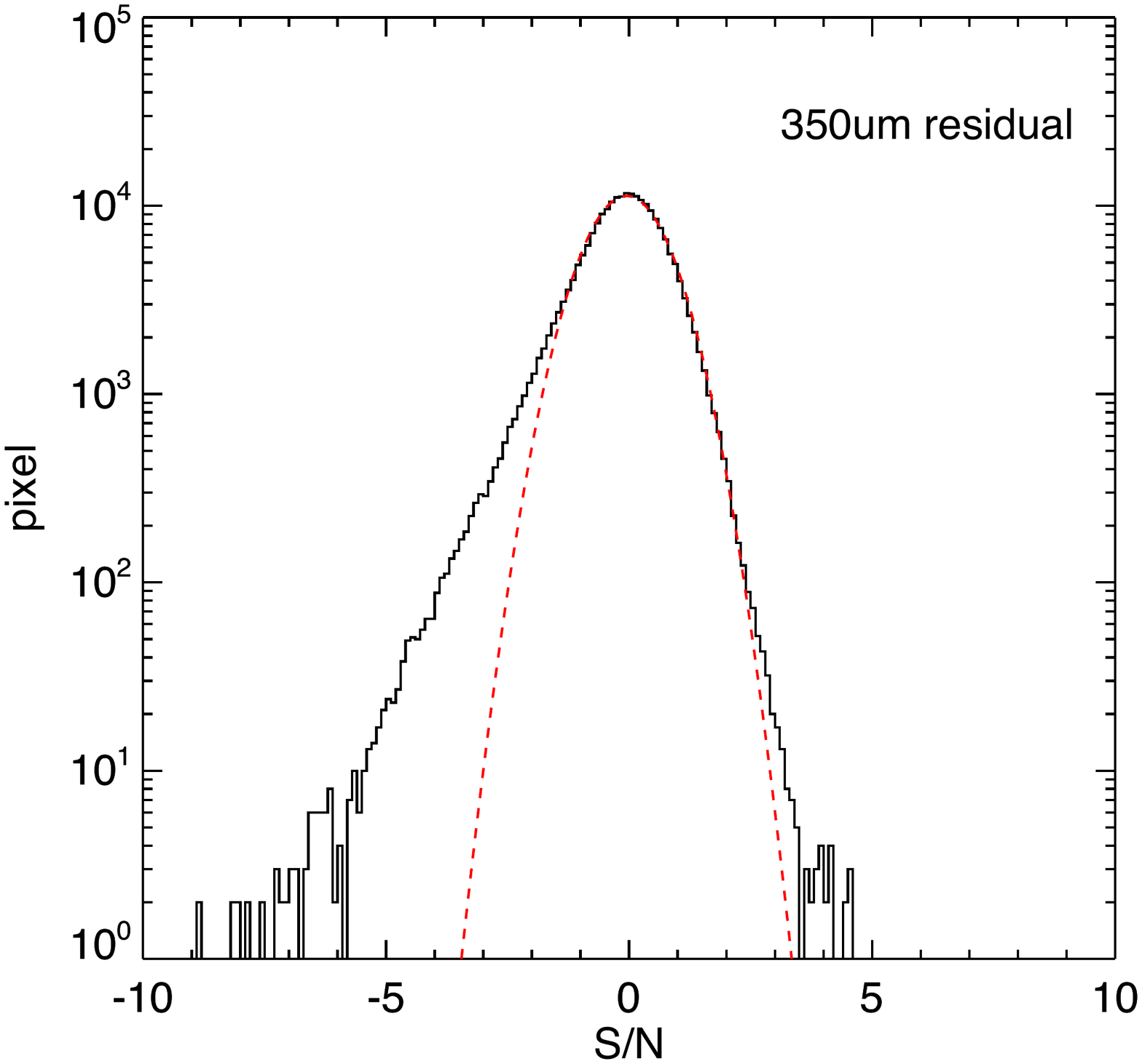}

\caption{%
	Analog to Fig.~6 in L18, we show
	the SPIRE 350~$\mu$m photometry images in our Super-deblending processes. Panel (1) is the original image of SPIRE 350~$\mu$m. Panel (2) is the modeled image of sources to be \textit{subtracted} (having $S_{\mathrm{SED}} / {\sigma}_{\mathrm{SED}} > 2$). Panel (3) is the best fitting model image of selected prior sources for fitting with \galfit{} (e.g., Section~\ref{Section_Photometry_24}), and panel (4) is the residual image. The image in panel (1) is the sum of panels (2), (3) and (4). Scales are the same for all panels as indicated by the bottom color bar and expressed in terms of $\mathrm{S/N}$ ratio (respect to the typical noise at the band, as detailed in Table~\ref{Table_1}). 
    The last panel show the histogram of pixel S/N values from the residual images, with a Gaussian fit overlaid.
    We note that the skewness of SPIRE residuals at $\mathrm{S/N<0}$ are due to over-subtraction of faint sources. This effect has been discussed in Section 7.5 of L18 and corrected at the fourth step $S_{subtracted}$ in the Monte Carlo simulations (Section~\ref{section_simulations}).
	\label{Fig_Photometry_Images}%
}
\end{center}
\end{figure*}

\subsection{SED fitting algorithm and parameters}
\label{SED_fitting}

{The SED fitting recipes and parameters are identical to those in L18. We recall them in some detail here for clarity. Four distinct SED components are used in the fitting procedure: 1) a stellar component \citep{BC03} with a Small Magellanic Cloud attenuation law (we recall that we fit only down to $K_s$-band); 2) a mid-infrared AGN torus component \citep{Mullaney2011}; 3) a dust continuum emission from the \citet{Magdis2012SED} library with the more updated $L_{IR}/M_{dust}$-redshift evolution taken from \citet{Bethermin2015} to fit galaxy SEDs and predict photometric redshift and FIR/mm fluxes; 4) a power-law radio continuum with an evolving $q_\mathrm{IR}=2.35\times (1+z)^{-0.12}+\mathrm{log(1.91)}$ \citep{Magnelli2015,Delhaize2017}.
We perform SED fitting at fixed redshift for sources with reliable spectroscopic redshifts, while we do allow redshift variations within $\pm10\%\times(1+z_{phot})$ for sources with an optical/near-IR photometric redshift. 
Using the newly fitted $\mathrm{SFR_{IR}}$s from our catalog (as updated at each wavelength step) and the stellar masses from \citet{Laigle2016}, we perform MS/SB classification by measuring the distance to the Main-sequence from \citet{Sargent2014}, at the fitted redshift. Sources are considered to be pure SBs if $log(SFR/SFR_{MS}) > 0.6$ dex and $SFR/\sigma_{SFR} > 3$, and to be pure MS if $log(SFR/SFR_{MS}) < 0.4$ dex and $SFR/\sigma_{SFR} > 3$, and fitted with the appropriate templates. When a clear MS/SB classification cannot be obtained, they are fitted with all SB+MS templates.
In case of radio-excess sources classified as radio-loud AGNs, we do not include the radio photometry in the fit. Our fairly conservative criterion requires observed radio fluxes $\times 2$ higher than the prediction from FIR-radio correlation with $>3\sigma$ significance. 
For sources with combined $\mathrm{S/N\geqslant 5}$ over the 100$\mu$m-1.1mm range we do not fit the 24~$\mu$m and radio photometry so to avoid being affected by the scatter of the FIR-radio correlation and by the variation of mid-IR features.
As already briefly mentioned, the SED fitting is performed before  the photometric measurement work at each band from 160--1200~$\mu$m, predicting FIR flux at each band in exam before actual measurements. The SED fitting is also performed a last time on the final catalog, once measurements in all bands have been made, to provide final phyical quantities released with the catalog.
}

\subsection{Faint sources subtraction in COSMOS}
\label{Section_Faint_Prior}

We show the adopted limits for
retaining sources for fitting at each band in Fig.~\ref{Fig_Galsed_cumulative_number_function}: priors with $S_{\mathrm{SED}}+2\,{\sigma}_{{\mathrm{SED}}}>S_{\mathrm{cut}}$ are maintained and fitted (hereafter \textit{selected} sources), while sources fainter than this limit are excluded from the fitting (hereafter \textit{excluded} sources). $S_{\mathrm{SED}}$ is the predicted flux based on SED fitting, and ${\sigma}_{\mathrm{SED}}$ its uncertainty based on the $\chi^2$ statistics.

L18 subtracted fluxes $S_{\mathrm{SED}}$ of all \textit{excluded} sources from the GOODS-N maps. However, this blind approach would be problematic in the photometric work in COSMOS. 
As mentioned in Section~\ref{Section_Initial_general_priors_list}, both the 24~$\mu$m, radio and especially PACS photometry are shallower than the ones in GOODS-N. 
Although we have included the 100~$\mu$m photometry in the first run of SED fitting to better constrain SEDs, the PACS 100~$\mu$m photometry (${\sigma}\sim 1.44$ mJy) in COSMOS is still a factor of 4.5 shallower than the one in GOODS-N (${\sigma}\sim 0.32$ mJy), while at PACS 160~$\mu$m the sensitivity ratio is {5.2 (3.55 vs 0.681 mJy)},
leaving the SEDs less well-constrained in COSMOS than in the GOODS-N field. 
Moreover, the 106,420 mass-selected $K_s$ priors do not have any 24~$\mu$m or radio detection (by construction, although they do have upper limits in these bands), leaving large uncertainty on their fluxes at FIR/(sub)mm bands.
The large uncertainty on flux predictions introduces errors and possible systematics on faint sources subtraction, which would imply in turn flux biases and increased photometric errors on measured fluxes.

{To minimize the uncertainty in faint source subtraction,
we decided to subtract SED-predicted fluxes only for sources with $S_{\mathrm{SED}} / {\sigma}_{\mathrm{SED}} > 2$ (hereafter, \textit{subtracted} sources) from the original map. }
This method is applied on PACS 160~$\mu$m, SPIRE and AzTEC images, where the numbers of \textit{subtracted} sources are listed in Table~\ref{Table_1}. Using simulations we verified that this criterion improves the overall performances and produces lower uncertainties for fitted sources. Also, we tested that a threshold around $2\sigma$ is optimal for this step, in terms of reducing the final noise with the procedure discussed here.
In the left panel of Fig.~\ref{example_SED}, the orange solid circles in each cutout mark the sources subtracted from the map, while the sources in orange dashed circles are neither subtracted nor fitted because their predicted flux is definitely faint { but uncertain}. Treating these sources differently makes a difference because there are obviously a lot more faint than brighter sources intrinsically so even if their flux is small, combining large numbers of them can have an impact on the result.

Secondly, we add one more step in the Monte Carlo simulations analysis to identify and correct the flux bias introduced from the subtraction, and to include the errors in the finalized photometry. 
In the bottom panel of Fig.~\ref{Fig_Galsim_flux_bias_SPIRE}, we can inspect the difference between the input and measured fluxes $S_{in}-S_{out}$ from the simulations, as a function of a parameter $S_{\rm subtracted}$ that is the sum of the total flux of subtracted galaxies convolved with the beam at the position of each specific source examined in the simulation. 
For positions where $S_{\rm subtracted}$ is high $S_{in}-S_{out}$ is also large, and the opposite is also true (the median bias is zero at this fourth processing step by construction). This implies that subtracted fluxes are too large, i.e., are overestimated. This is expected in a regime of low-accuracy predictions, given that fluxes are always positively defined.

Note that we do not apply the $2 \sigma$-requirement to subtract sources in the SCUBA2 and MAMBO images where we subtract all the \textit{excluded} sources, because 
we found from simulations that this did not make any difference in these bands. This might be owing to the smaller beam size ($\mathrm{FWHM}=11''$ for SCUBA2) with respect to SPIRE images, as well as the fact that most \textit{excluded} priors are at low redshift and have very low fluxes predicted at $\approx1$~mm.

In Fig.~\ref{Fig_Photometry_Images}, we present a portion of the SPIRE 350~$\mu$m image as an example of the overall procedure. Panel (2) in this figure shows the image combining all faint sources to be \textit{subtracted}. It is quite apparent how their crowding in certain regions (clustering) simulate brighter individual objects that are not really individual galaxies but just spurious superposition of many faint galaxies.
In Appendix~\ref{image_products}, 
we show the same set of images produced during these steps at each wavelength band.

\subsection{Flux bias and uncertainties calibration via improved Monte Carlo simulation}
\label{section_simulations}

Different from the the case in GOODS-N field, it is impractical and unnecessary to randomly distribute simulated galaxies over the whole COSMOS field. We thus chose an experimental and representative area of $10'\times 10'$ (similar to the size of the GOODS-N field) in the center of the COSMOS field, where the depth of the imaging data was typical/average for the whole map.  
Monte Carlo simulations are performed in the experimental area of the original images at 24~$\mu$m, 1.4~GHz, 3~GHz and 100~$\mu$m bands (as no sources are subtracted at these bands), while in same area but in the faint-source-subtracted images at other bands. We simulate $\sim$5000 galaxies per band, one at a time.

Following the method defined by L18, Monte Carlo simulations are performed simulating one source at a time in the actual image, with the following steps.

Firstly, the position of each simulated source is randomly generated within the experimental area. Its  flux, $S_{in}$, is drawn from a  uniform distribution in log space within a range of $\sim3\sigma$ to $\sim12\sigma$, where $\sigma$ is the median flux density uncertainty at each band. We model the source as a PSF and add it in the actual image (which include all the other,  real sources).
Secondly, we include the coordinates of the simulated source to the fitted prior list (i.e., selected sources, as well as additional sources from residual images), and perform our photometric measurements fitting simultaneously all priors together with the extra simulated source. From the \galfit{} output, the flux $S_{out}$ and the flux uncertainty ${\sigma}_{\mathit{galfit}}$ of the simulated source will be measured.
Finally, we repeated this process $\sim$5000 times. In this way, we obtain $\sim$5000 values of $S_{in}$, $S_{out}$ and ${\sigma}_{\mathit{galfit}}$, realistically representing the measurement process in the real images. Several additional properties are measured for each of the simulated sources: (1) the instrument rms noise value ${\sigma}_{rms\,noise}$ at the position of the simulated source; (2) the local flux in the absolute valued residual image (hereafter $S_{\mathrm{residual}}$), measured by considering the  absolute  values of the pixels of the residual image, within the PSF aperture; (3) and the \crowdedness parameter, defined by summing up the Gaussian weighed distances of all sources at position of source $i$ (and including it): 
\begin{equation}
 crowdedness \equiv \sum\limits_{j=1}^{N} e^{\left(-d_{j,i}^2/d_{\mathrm{PSF}}^2\right)}
\label{crowdedness}
\end{equation}
where $d_{j,i}$ is the angular distance in arcsec from source $j$ to source $i$ and $d_{\mathrm{PSF}}$ is the FWHM in arcsec of the PSF.
In this way, the \crowdedness is a weighted measure of the number of sources present within the beam, including the specific source under consideration.
These  parameters, measurable in the same way for all fitted priors, provide key information to check the expected quality of fitting and the actual local crowding (hence blending) of prior sources. 
{They will be used to calibrate the flux bias corrections and the flux uncertainties of each source.}

\begin{figure*}
\begin{center}
\includegraphics[width=0.4\textwidth, trim={1cm 15cm 0cm 2.5cm}, clip]{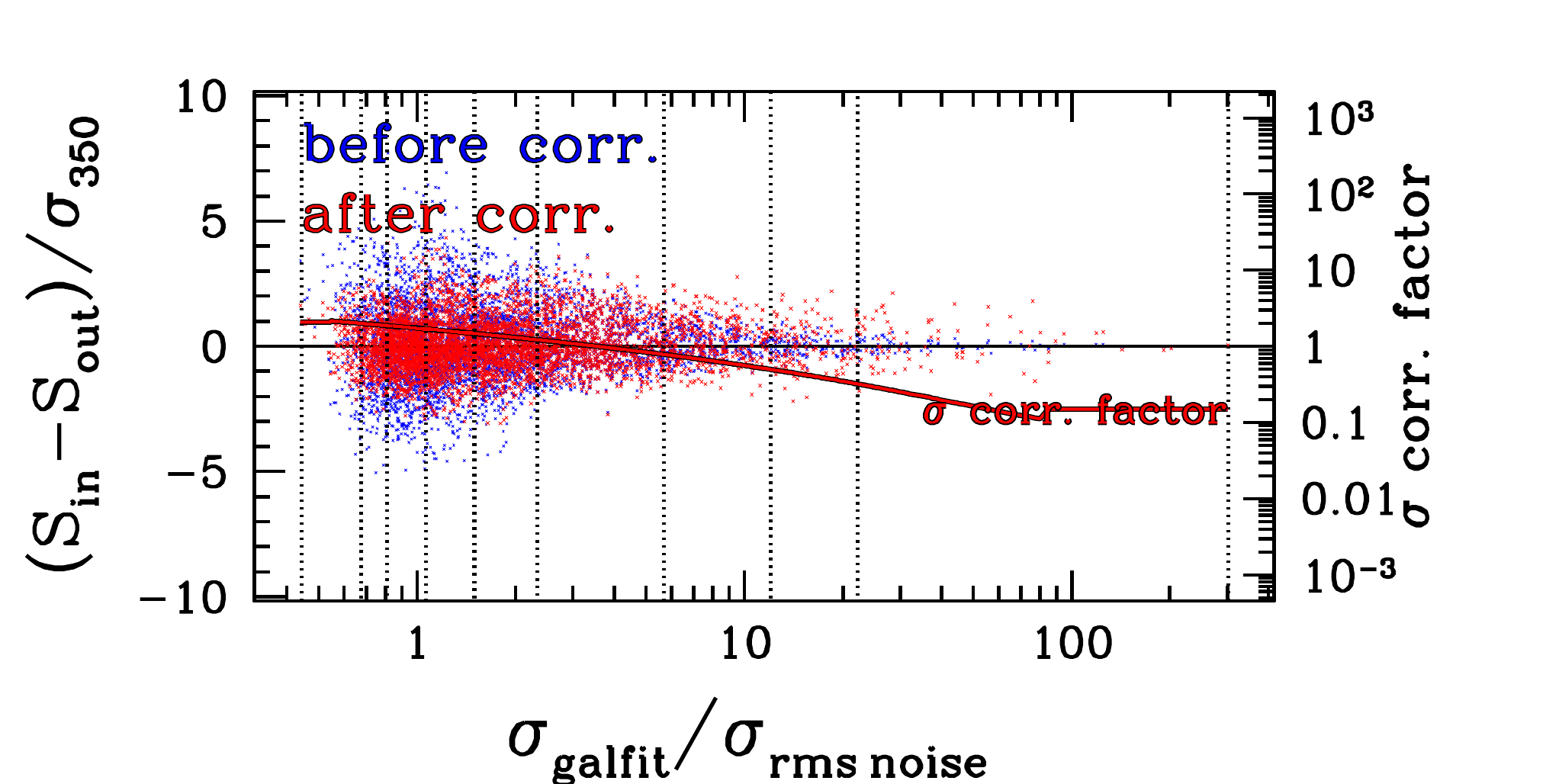}
\includegraphics[width=0.4\textwidth, trim={1cm 15cm 0cm 2.5cm}, clip]{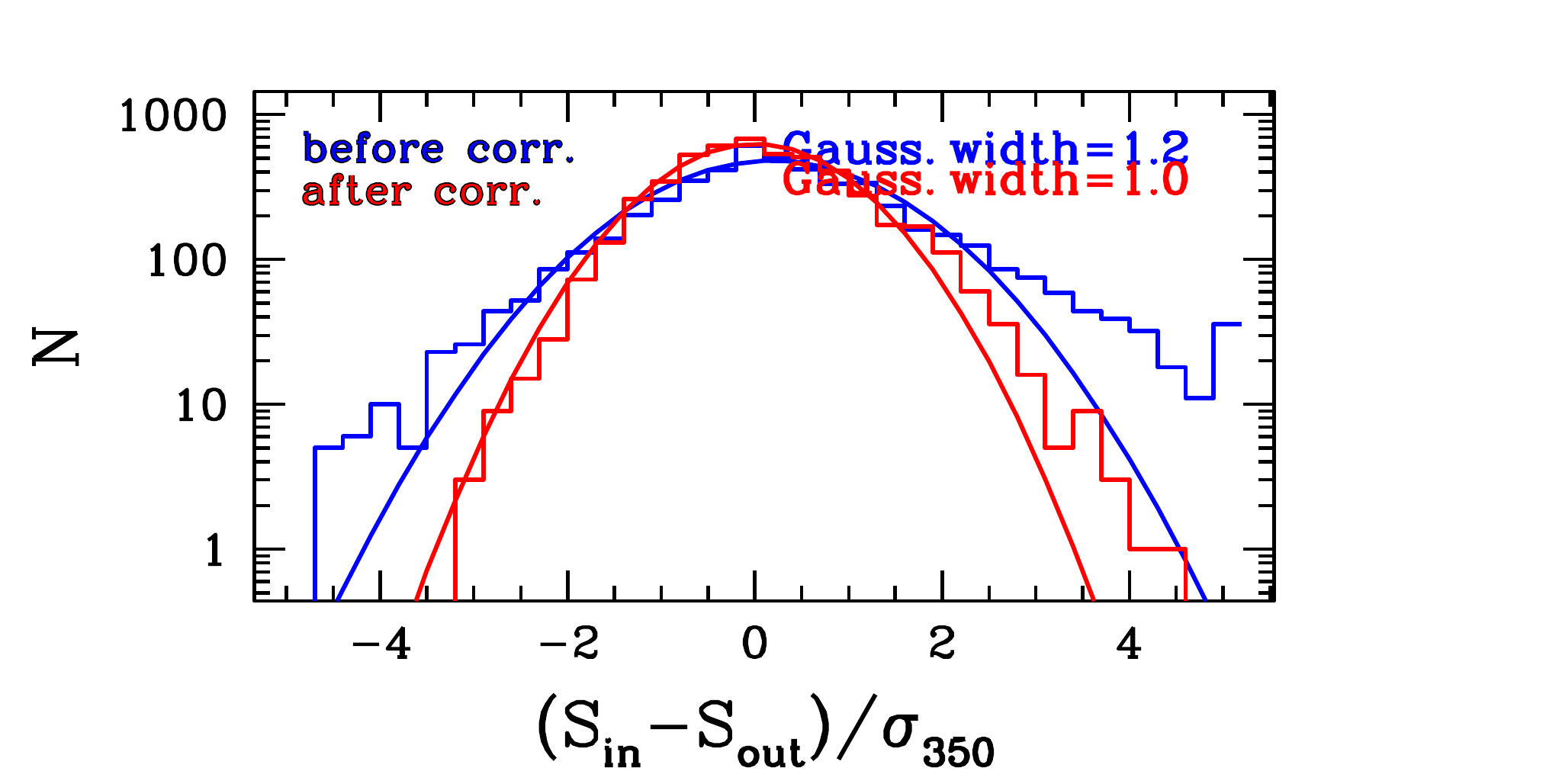}
\includegraphics[width=0.4\textwidth, trim={1cm 15cm 0cm 2.5cm}, clip]{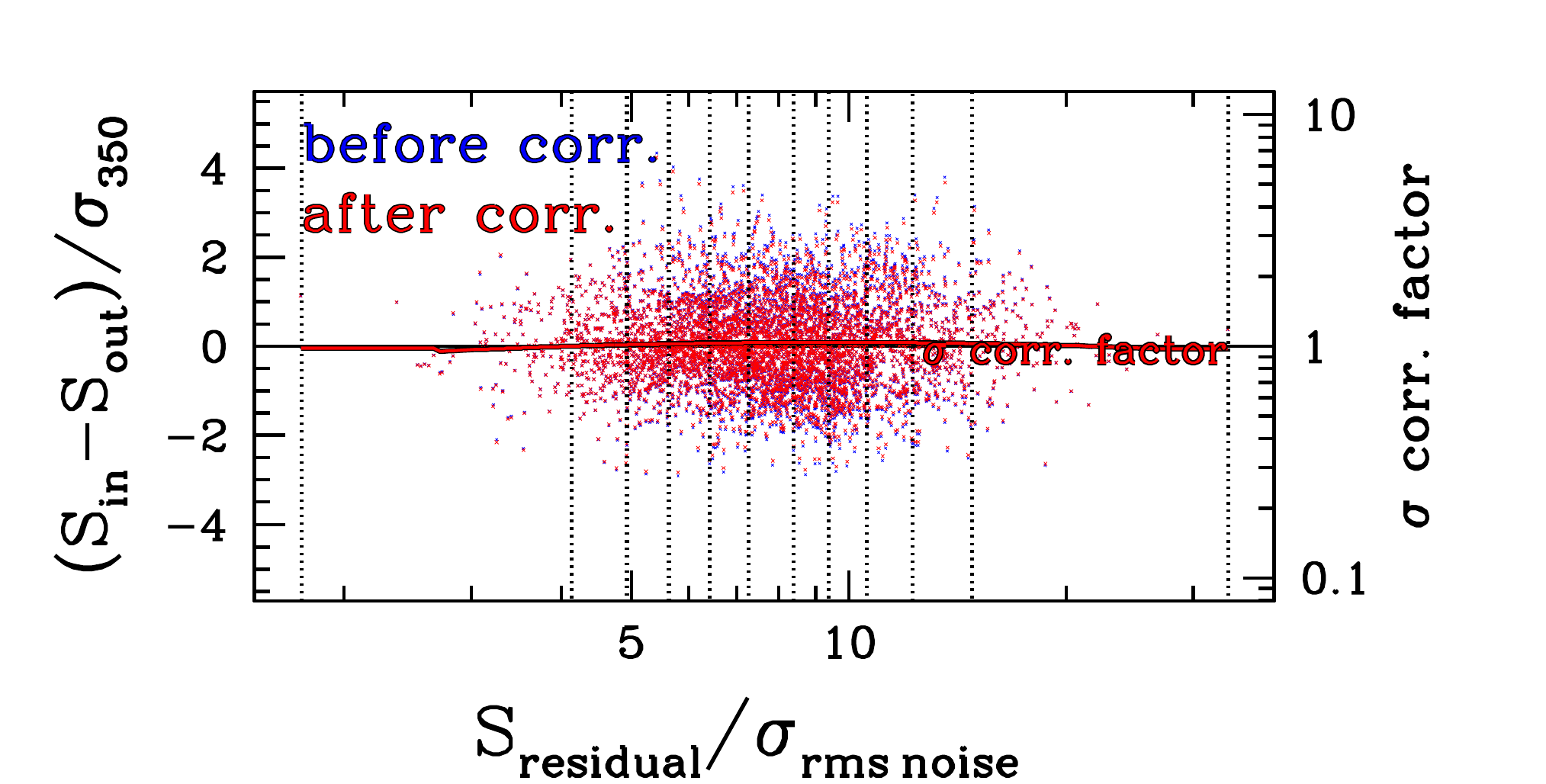}
\includegraphics[width=0.4\textwidth, trim={1cm 15cm 0cm 2.5cm}, clip]{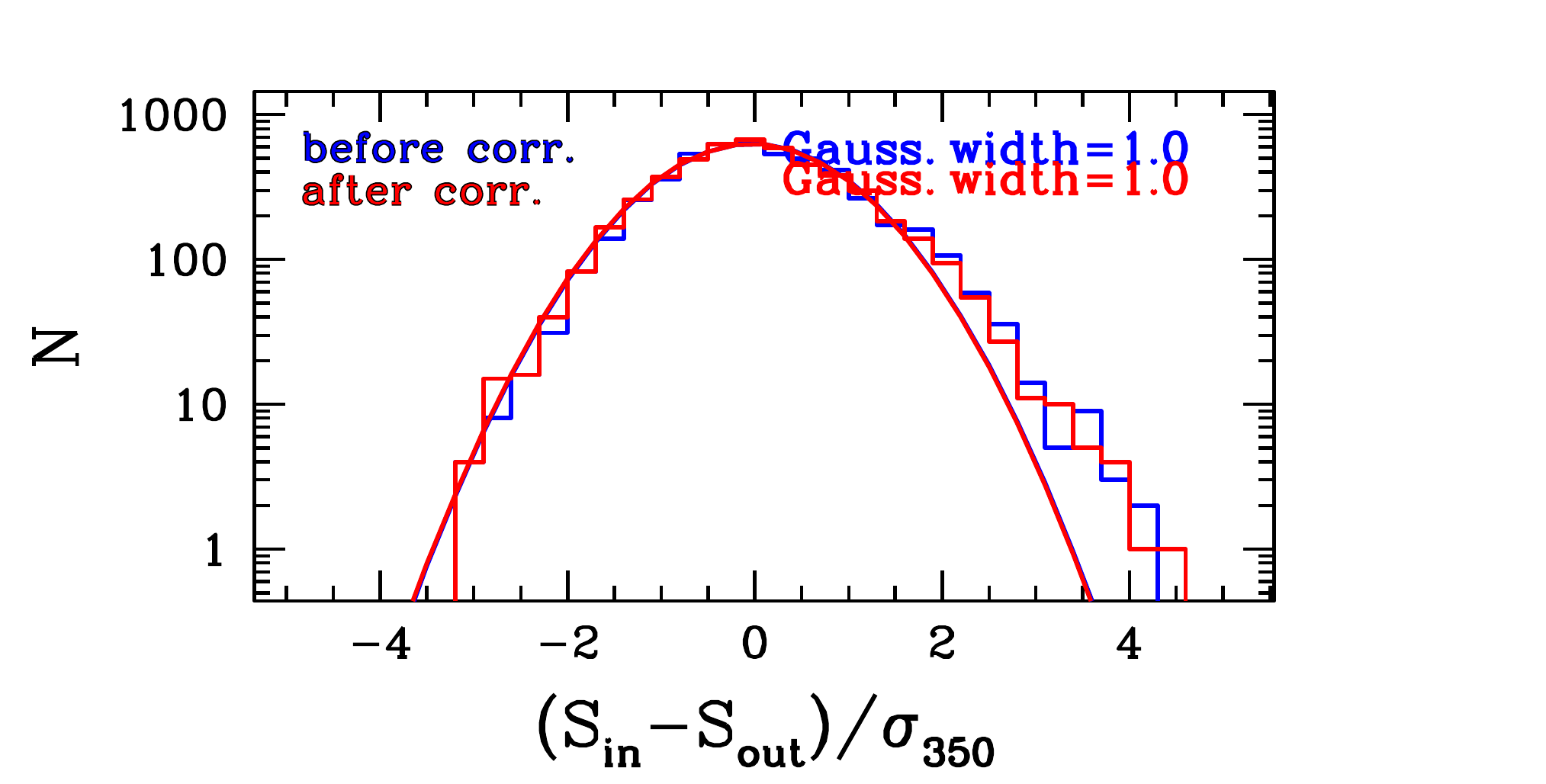}
\includegraphics[width=0.4\textwidth, trim={1cm 15cm 0cm 2.5cm}, clip]{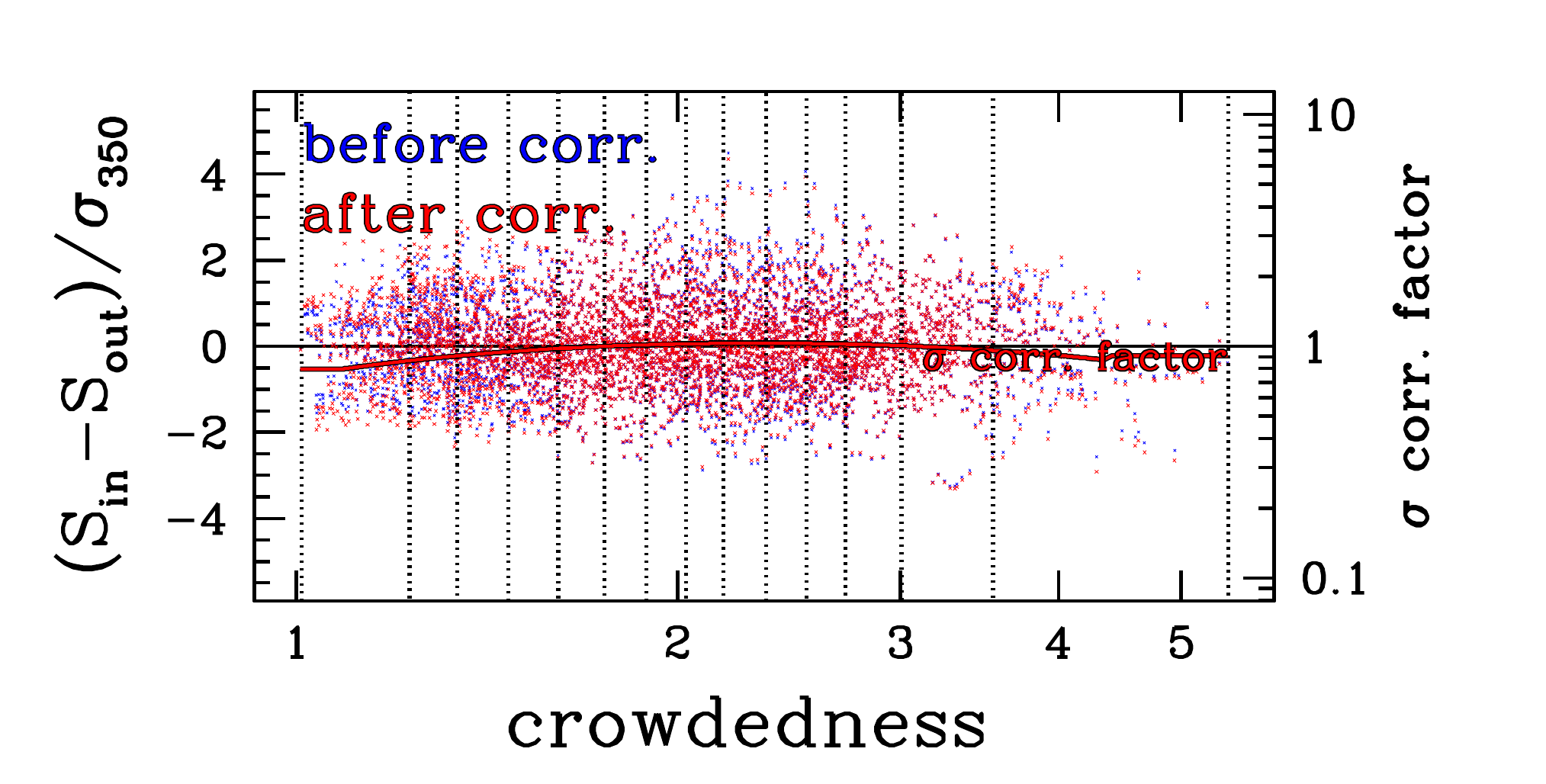}
\includegraphics[width=0.4\textwidth, trim={1cm 15cm 0cm 2.5cm}, clip]{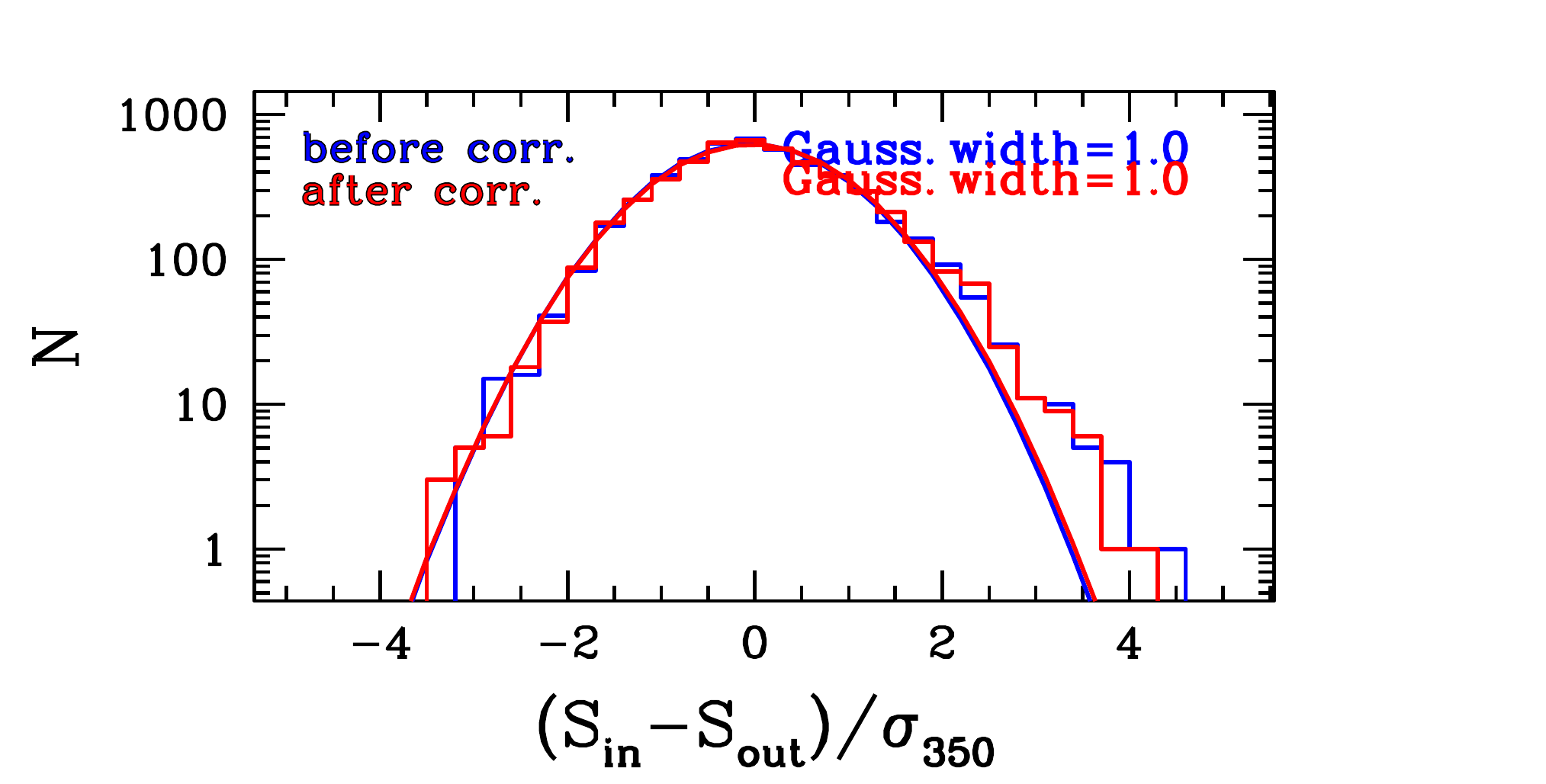}
\includegraphics[width=0.4\textwidth, trim={1cm 15cm 0cm 2.5cm}, clip]{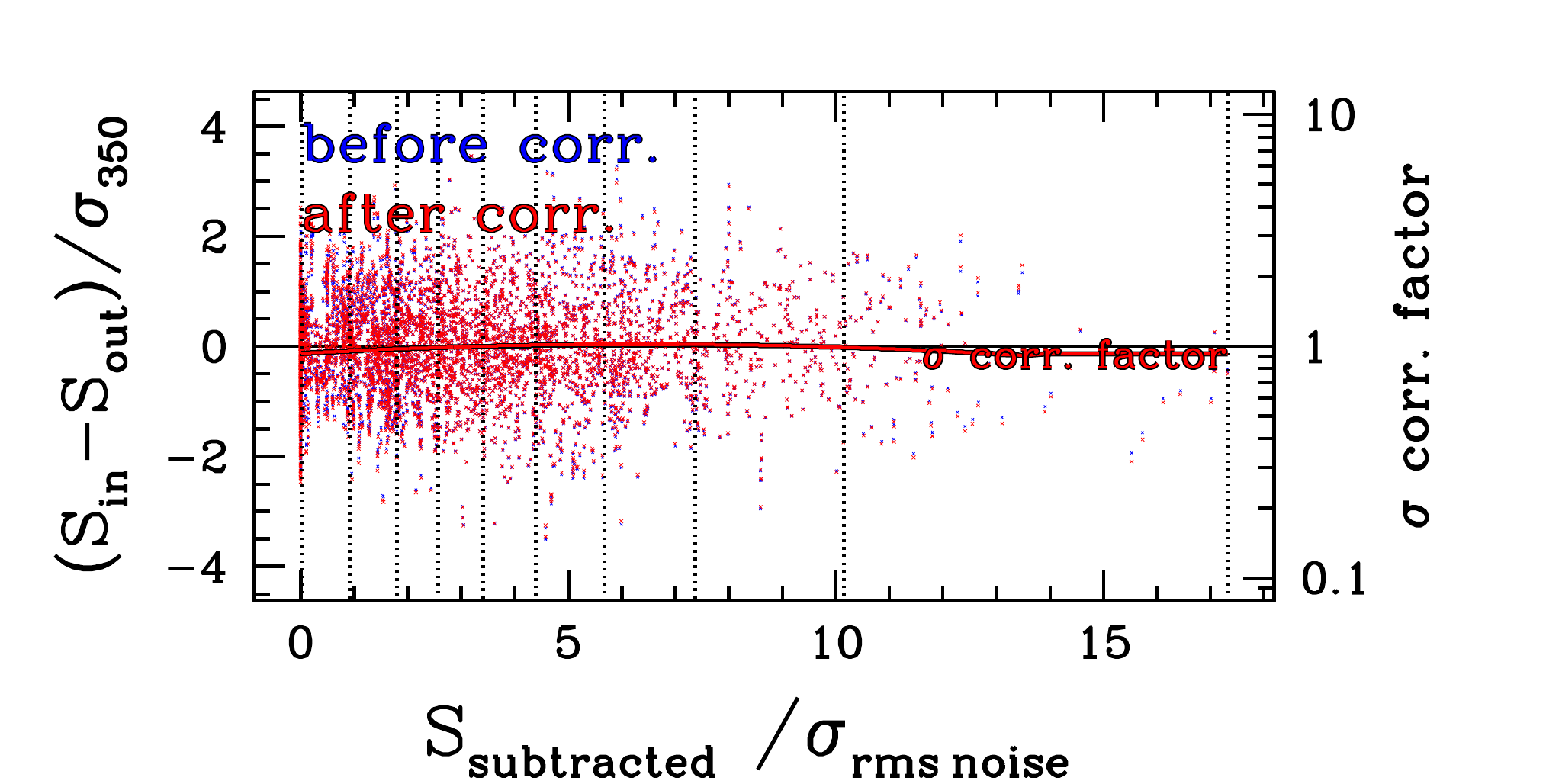}
\includegraphics[width=0.4\textwidth, trim={1cm 15cm 0cm 2.5cm}, clip]{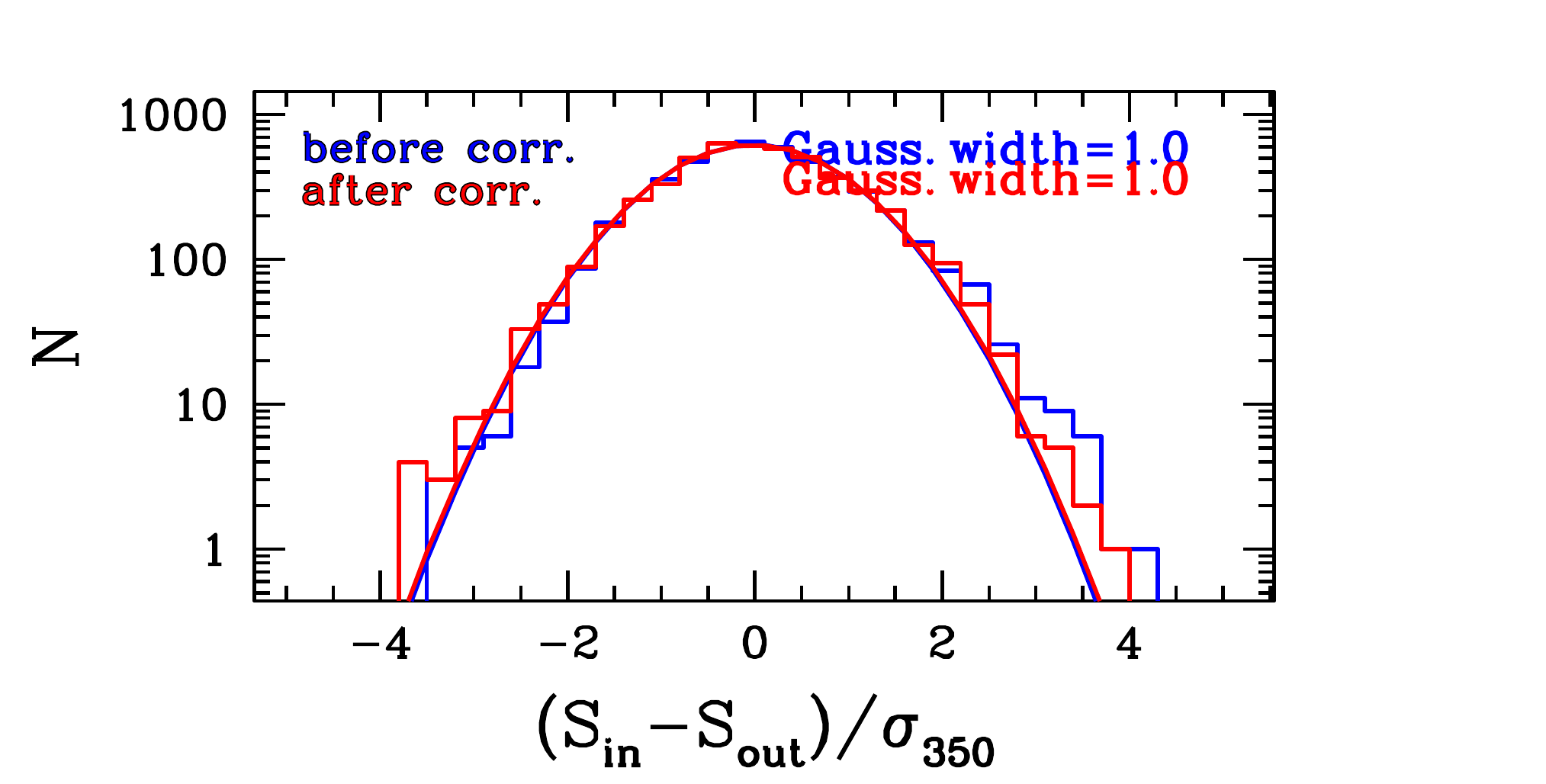}
\caption{%
Analog to Fig.~11 in L18, the corrections of flux bias and flux uncertainty in 350~$\mu$m Monte Carlo simulations.
\textbf{Left panels:} Normalized flux bias (i.e., $(S_{in}-S_{out})/{\sigma}$) versus four parameters: the normalized \galfit{} flux uncertainty $\sigma_\mathrm{galfit} / \sigma_\mathrm{rms\ noise}$, the normalized flux density in residual image $S_\mathrm{residual} / \sigma_\mathrm{rms\ noise}$, the $crowdedness$ and the normalized subtracted flux $S_\mathrm{subtracted} / \sigma_\mathrm{rms\ noise}$, while the right sides show the correction factors.
\textbf{Right panels:} The histograms of the $(S_{in}-S_{out})/{\sigma}$ in logarithm space, and with a solid line representing a best fitting Gaussian profile to the inner part of each histogram. 
From top to bottom, we analyze this quantity against four parameters as indicated by the x axis label. 
We bin the simulated objects by the vertical dashed lines and compute the rms in each bin for deriving the correction factors.
The data points before and after correction (i.e., $(S_{in}-S_{out})/({\sigma,\,\mathrm{uncorrected}})$ and $(S_{in}-S_{out})/({\sigma,\,\mathrm{corrected}})$) are shown in blue and red respectively. 
After the four-step corrections the histograms are well-behaved and generally well consistent with a Gaussian distribution. 
Similar figures of other bands are shown in Appendix~\ref{Section_Simulation_Performance}.
\label{Fig_Galsim_df_corr_SPIRE}%
}
\end{center}
\end{figure*}

We have analyzed the simulations following the method  detailed above, with some important changes with respect to L18.
First, we use the newly defined $S_\mathrm{subtracted}$ variable ({i.e., the sum of the subtracted fluxes from excluded sources within the PSF circle})
 in a  fourth step calibration  in the analysis of simulations (L18 had only 3 steps, similar to the first three shown in Fig.~\ref{Fig_Galsim_flux_bias_SPIRE} for flux biases). We introduced this new parameter because of the lower overall depth of COSMOS in many bands, so that  
 we expect that the actual noise will be appreciably higher in regions where more flux was subtracted from faint sources (see Fig.~\ref{Fig_Galsim_df_corr_SPIRE}). 

{Using the simulations, we derive the dependence of the flux bias $S_{in}-S_{out}$ on the four  adopted observational  parameters. As shown in Fig.~\ref{Fig_Galsim_flux_bias_SPIRE}, we fit the flux bias in each bin using a 3rd-order polynomial function.
Meanwhile, we determine a correction factor (i.e., the ``$\sigma$ corr. factor'' in Fig.~\ref{Fig_Galsim_df_corr_SPIRE}) on the flux uncertainty at each step, defined as the multiplicative factor that is required to scale the rms dispersion of $(S_{in}-S_{out})/\sigma$ in each bin to 1. 
These polynomial functions and correction factors are then applied to each real source for which the four measured observational parameters.}
As shown in the right column of Fig.~\ref{Fig_Galsim_df_corr_SPIRE}, the four-step correction significantly improves the performance of the deblended photometry with lowering the overall rms noise of the photometry and reducing (often removing) the prominent non-Gaussian tails in the flux uncertainties distributions. 

Also, we have to account in COSMOS for more substantial effects of depth variations of the data across the field. 
This is important often towards the edges of each dataset, but also in some cases across the datasets as for SCUBA2.
We find that we can rescale simulation results for calibration of flux uncertainties to areas with different depth in the same dataset by calibrating performances based on normalized observables (flux bias terms are not affected). 
Notably, we have normalized the parameters used in the simulations by the local rms noise, i.e., we consider quantities as $\sigma_\mathrm{galfit} / \sigma_\mathrm{rms\ noise}$, $S_\mathrm{residual} / \sigma_\mathrm{rms\ noise}$ and $S_\mathrm{subtracted} / \sigma_\mathrm{rms\ noise}$ (this is not necessary for {\em crowdedness} that does not depend on the noise in the dataset). 

We tested this scaling by directly performing additional simulation in data portions with different depths, and comparing the direct simulation results from those rescaled from a region with different depth. For example, for the SCUBA2 image we have executed our primary simulations in the deep area and we also carried out our independent simulations in a 3.6~$\times$ shallower area, then applied the scaled deep-area simulation results on the \galfit\ outputs to the shallow area. We find that the flux uncertainty scaled from the deep area simulation are slightly larger than those directly performed on the shallower data, but overall consistent (or at least conservative). 
As a further test, we have performed PSF fitting in the inverted SCUBA2 image and found 97 (negative) $\mathrm{S/N>3}$ detections, after calibration of the fluxes and uncertainties as for positive sources. 
The spurious fraction at $\mathrm{S/N>3}$ is 0.4\% with respect to the number of fitted priors.
This is close to Gaussian expectations, albeit a factor of 2 higher.
Compared to positive $\mathrm{S/N>3}$ detections, the SCUBA2 spurious detection rate is $9.5\%$.

Figures for all bands simulations are reported in the Appendix~\ref{Section_Simulation_Performance}.

\subsection{Selecting additional sources in the residual images}
\label{Section_Additional_Sources_In_Residual}

\begin{figure*}
\begin{center}
\includegraphics[width=\textwidth]{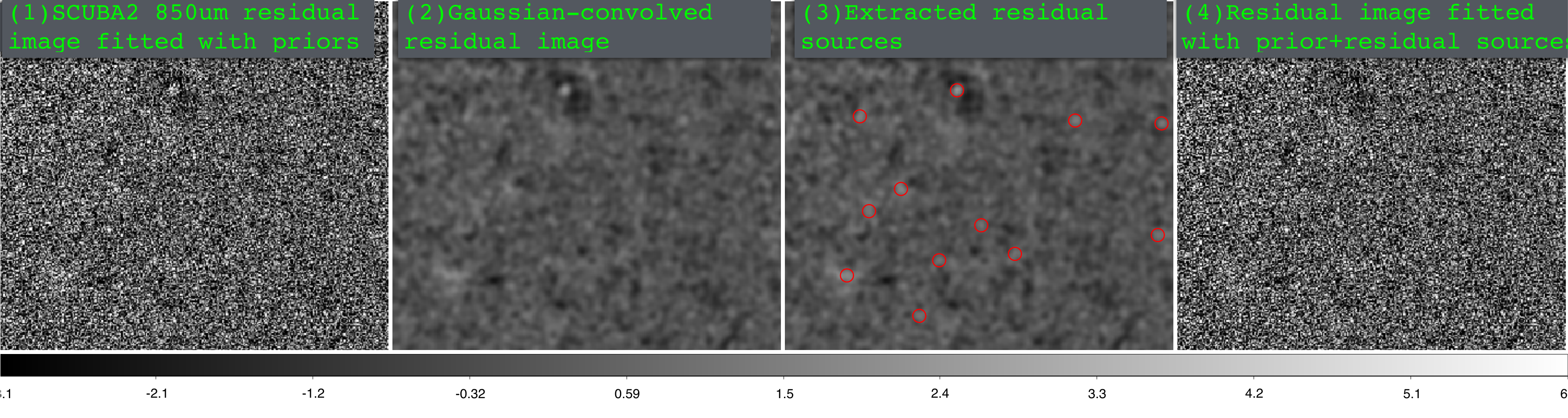}
\caption{%
	Analog to Fig.~7 in L18,
	the extraction of additional sources in the SCUBA2 residual image. Panel (1): the residual image fitted with {sources from the} prior catalog. Panel(2): the Gaussian-convolved images of panel 1. We extract bright sources in image (1) via SExtractor and show the residual sources in red circles in panel (3). We add the positions of residual sources in our prior catalog and fit the whole faint-source-subtracted image by \galfit, where panel (4) shows the final residual image fitted at the positions of prior+residual sources.
	\label{Fig_residual_extraction}%
}
\end{center}
\end{figure*}


Although the 24~$\mu$m+radio+mass-selected prior catalog is expected to have high completeness, some FIR emitters are still likely missing in this prior catalog. 
For example, in Fig.~\ref{Fig_SED_expected_flux_against_z}, there are several GOODS-N $\mathrm{S/N_{FIR+mm}} > 5$ galaxies below our stellar mass limit (red dots below the green dashed line), suggesting that galaxies with lower stellar masses could be detectable in the FIR imaging, e.g., if they are starburst.
On the other hand, $K_s$-undetected galaxies ($K_s$-dropouts) that are not already included in the \citet{Smolcic2017} 5-sigma radio catalog, are also missing from our prior selection.
These missing priors, which are probably very high-z galaxies or extreme low-mass starbursts, might have detectable fluxes at FIR/(sub)mm bands and emerge in the residual images of our photometric products (see Fig.~\ref{Fig_residual_extraction}).
Extracting these sources will improve the photometry of sources around them, 
in addition to providing potentially interesting high redshift candidates.

Similar to L18, we blindly extract residual sources with SExtractor \citep{Bertin1996} in the residual maps, at each wavelength. The positions of residual sources are then added in the \textit{selected} prior list and fitted together in the 2nd-pass. The 
\galfit{} outputs of the 2nd-pass fitting are then corrected via simulation recipes and archived in the finalized photometry catalog. 
We show an example of extraction of SCUBA2 residual sources in Fig.~\ref{Fig_residual_extraction}. 
We convolve the residual image (panel 1) with a Gaussian filter to improve the visibility and detectability (panel 2), and show the extracted sources in red circles in panel 3. The final residual image (panel 4) is cleaner after prior+residual sources fitting. 
Residual sources with $\mathrm{S/N_{850\mu m}} > 2.5$ are kept in the prior list and fitted in subsequent fitting of SPIRE, AzTEC and MAMBO images. 
The counts of residual sources at each band are listed as $N_\mathrm{add}$ in Table~\ref{Table_1}.

Note that we have included residual sources in the Monte Carlo simulations, although there is no obvious difference from masking residual sources in simulations. We do not extract any residual sources in PACS 100~$\mu$m and 160~$\mu$m images because of their shallow depths and clean residual maps. 
Also, the residual images of AzTEC and MAMBO are clean, no residual sources are extracted there.

\section{\Superdb photometry catalog}
\label{result_catalog_candidates}

\begin{figure}
\centering
\includegraphics[width=0.48\textwidth]{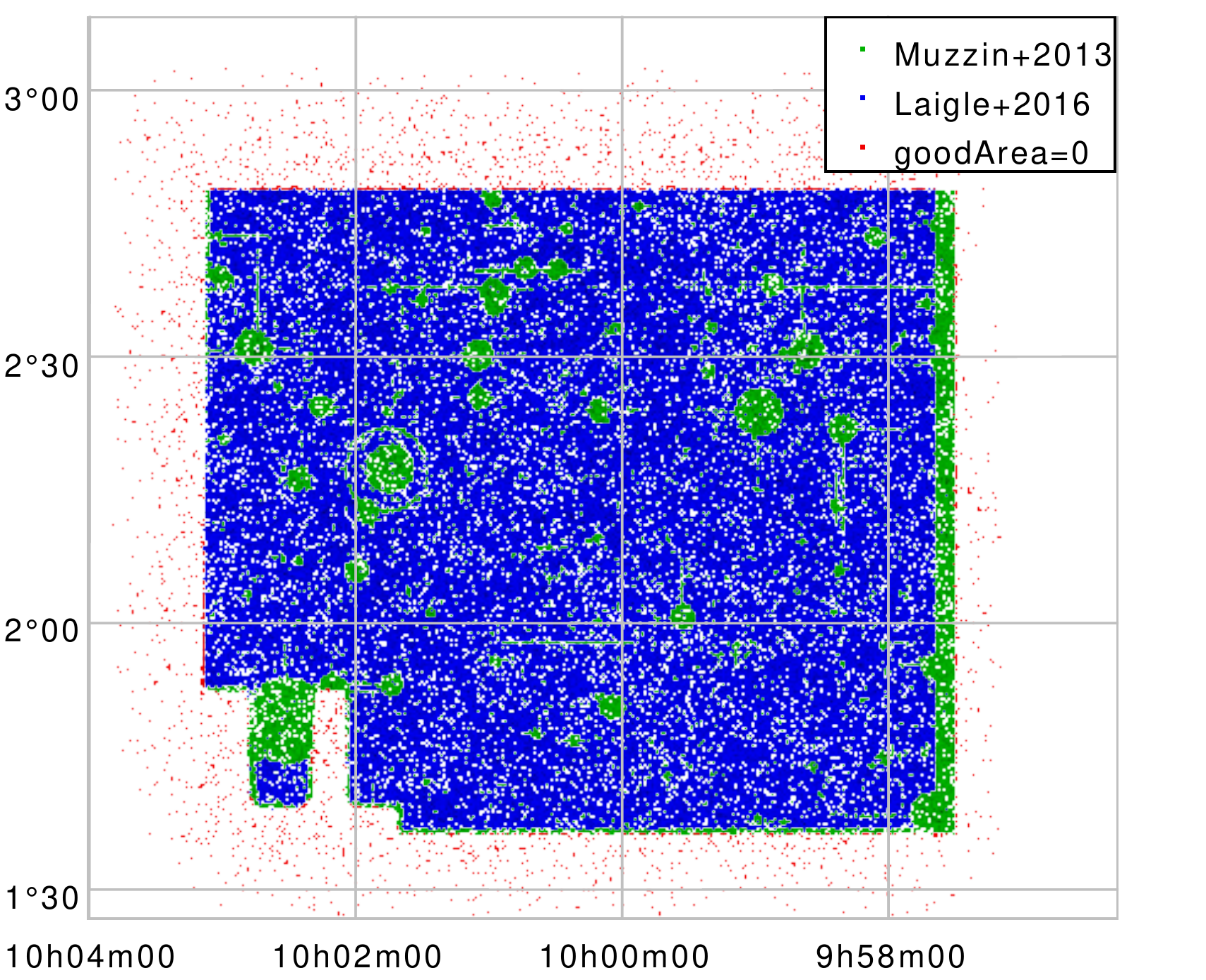}
\caption{
	The definition of $goodArea$: sources located in the UltraVISTA area \citep{Muzzin2013,Laigle2016} have $goodArea=1$. Red dots show sources with $goodArea=0$, which are mostly radio sources \citep{Smolcic2017} located outside of the UltraVISTA FoV and some UltraVISTA sources on the edge. 
\label{good_Area}
}
\end{figure}

As in L18, we define a $goodArea$ boolean parameter to describe regions with the best and most homogeneous prior coverage, corresponding in COSMOS to the UltraVISTA 1.7 $deg^2$ area.
The final catalog that we are releasing includes photometry (fluxes and uncertainties) from Spitzer, $Herschel$, SCUBA2, AzTEC, MAMBO and VLA (3~GHz and 1.4~GHz) for 191,624 prior galaxies within the $goodArea=1$ region (we also release measurements for the $goodArea=0$ regions but warn that they suffer from lack of complete prior coverage, non resolved blending, and flux uncertainties are likely underestimated).
In Fig.~\ref{good_Area}, sources with $goodArea=1$ are shown in blue and green, depending on whether they were taken from the catalog of \citet{Laigle2016} or \citet{Muzzin2013}, respectively. Sources with $goodArea=0$ are shown in red, which are mostly radio sources \citep{Smolcic2017} that are located outside or just on the edge of the UltraVISTA area.
We do include additional sources selected in the residual images in the released catalog, although we warn that the spurious contamination fraction among them is uncertain (see discussion in L18).
Some of them do appear to be solid detections with well established unique counterparts, as discussed in Section~\ref{high_z_candi}.
The number of detections within $goodArea$ and the median rms uncertainty (hence sensitivity) at each band are listed in Table~\ref{Table_1}.
Note that we use identical IDs for sources selected from the COSMOS2015 catalog \citep{Laigle2016}, while we designedly set $\mathrm{ID = ID_{Muzzin} + 1E8}$ for sources supplemented from \citet{Muzzin2013} catalog, and $\mathrm{ID = ID_{Smolcic} + 2E8}$ for radio sources supplemented from the \citet{Smolcic2017} 3~GHz catalog in order to avoid ID duplications. Sources selected in the residual images are marked by the wavelength combined with the ID from SExtractor output, e.g., ID=85000019 is a source extracted in SCUBA2 850~$\mu$m residual image with SExtractor ID=19.

\begin{figure*}
\centering
\includegraphics[width=0.98\textwidth]{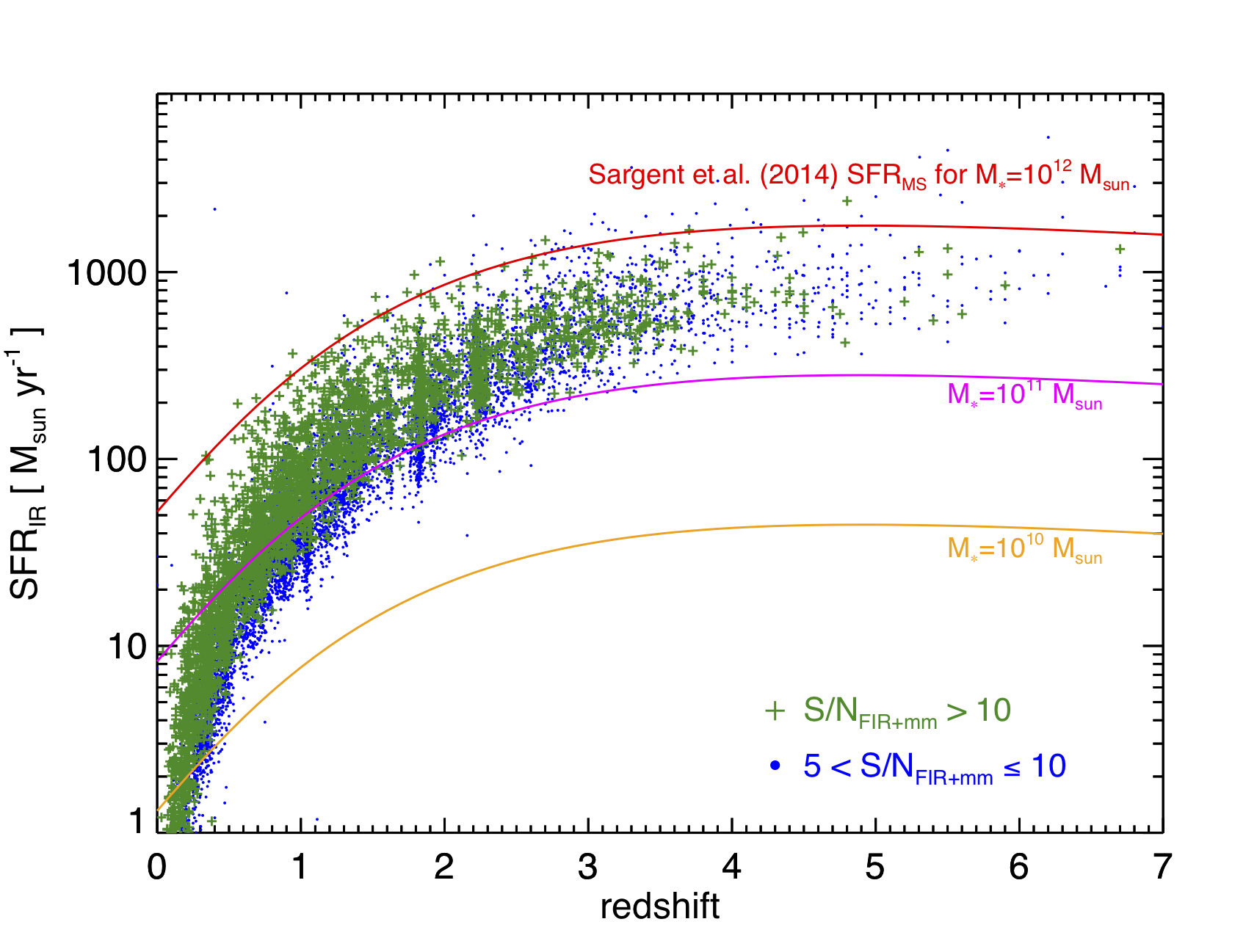}
\caption{%
	Star formation rate (SFR) versus redshift for $\mathrm{S/N}_{\mathrm{FIR+mm}}>5$ sources. 
   SFRs are computed from the integrated 8-1000~$\mu$m infrared luminosities derived from FIR+mm SED fitting, assuming a Chabrier IMF \citep{Chabrier2003}. Colors indicate the combined S/N over the FIR+mm bands.
   The colored curves show the empirical tracks of the MS galaxy SFR as a function of redshift at three Main-Sequences in specific stellar masses \citep{Sargent2014}.
\label{plot_SFR_z}%
}
\end{figure*}

In this catalog, there are 85,171 sources with $\mathrm{S/N}>3$ at 24~$\mu$m and/or radio bands, 25 times larger than the 24~$\mu$m+radio detections in the GOODS-N \superdb catalog.
Similar to L18, we adopt a combined $\mathrm{S/N}_{\mathrm{FIR+mm}}$ over PACS, SPIRE, 850~$\mu$m, 1.1~mm and 1.2~mm bands:

\begin{equation}
\label{Equation_SNR}
\begin{split}
\mathrm{({S/N})}^{2}_\textnormal{FIR+mm}
= \mathrm{({S/N})}^{2}_{100\,{\mu}\mathrm{m}}
+ \mathrm{({S/N})}^{2}_{160\,{\mu}\mathrm{m}}
+ \mathrm{({S/N})}^{2}_{250\,{\mu}\mathrm{m}}
+ \\
\mathrm{({S/N})}^{2}_{350\,{\mu}\mathrm{m}}
+ \mathrm{({S/N})}^{2}_{500\,{\mu}\mathrm{m}}
+ \mathrm{({S/N})}^{2}_{850\,{\mu}\mathrm{m}}
+ \\
\mathrm{({S/N})}^{2}_{1.1\,\mathrm{mm}}
+ \mathrm{({S/N})}^{2}_{1.2\,\mathrm{mm}}
\end{split}
\end{equation}

This results in the detection of {11,220} galaxies with $\mathrm{S/N}_{\textnormal{FIR+mm}}>5$, extending all the way to possibly $z\sim7$ (Fig.~\ref{plot_SFR_z}).
Taking at face value the \citet{Rodighiero2011} criterion classifying starbursts as objects with sSFR a factor of 4 above the MS, we find 1,769 starbursts in the sample at $0<z<4$, for a fraction of {15.6\% of all the IR detections}, higher of course than \citet{Rodighiero2011} because of the IR selection, and also higher than in L18 given our somewhat shallower photometry that digs less deeper in the MS. 
There are {63} galaxies with SFR$_{\rm IR}>1000M_\odot$~yr$^{-1}$,
but the highest luminosity sources in the field do not reach much beyond that of GN20 ($L_{IR}\sim2\times10^{13}L_\odot$; \citealt{Daddi2009GN20,Tan2014}), showing again that GN20-like galaxies are rare and finding one in GOODS-N was a lucky occurrence (see also \citealt{Pope2006}).

Comparing the detection limits at the FIR/(sub)mm bands, i.e., the $1\bar\sigma$ values shown in Table~\ref{Table_1} in this paper and Table~1 in L18, the detection limits at PACS are $\sim$5 times shallower in COSMOS than the ones in GOODS-N. 
The detection limits in the SPIRE bands are comparable, albeit slightly shallower here than in GOODS-N. 
The sensitivity in SCUBA2 and MAMBO photometry is also comparable to the results in the GOODS-N catalog. It seems therefore that we reached our goal of exploiting the depth of the SPIRE and (sub)mm images in COSMOS, despite the shallower supporting data as discussed earlier.

A sample of {11,220} sources with $\mathrm{S/N}_{\mathrm{FIR+mm}}>5$ is contained in this catalog. This is 10 times larger than the FIR/(sub)mm sample in the \superdb GOODS-N catalog. 
Among the FIR/(sub)mm detections, there are {770} sources at $z\geqslant 3$, which is a 11 times larger sample than the 71 $z\geqslant 3$ sources detected in the \superdb GOODS-N catalog. 
The consistent $\sim 10\times$ factors above suggest that this catalog does not present detection biases with respect to the GOODS-N catalog between low and high redshifts. However, COSMOS has an area $\sim50\times$ larger than GOODS-N, showing that we are not reaching as deep as in the latter, as expected especially given the much different depths at PACS bands.

\begin{table*}
	\centering
\caption{COSMOS {``Super-deblended''} Photometry Results
	\label{Table_1} 
	}
\begin{tabular*}{0.85\textwidth}{ @{\extracolsep{\fill}} c c c c c c c c c c }
	\hline
	        Band         &       Instrument  & Beam FWHM \tablenotemark{a} %
                                         & $S_{\mathrm{cut}}$ \tablenotemark %
                                         & $\rho_{\mathrm{fit}}$ \tablenotemark{b} %
                                         & $N_{\mathrm{fit}}$ \tablenotemark{c} %
                                         & $N_{\mathrm{excl.}}$ \tablenotemark{d} %
                                         & $N_{\mathrm{S/N}>3}$ \tablenotemark{e} %
                                         & $N_{\mathrm{add.}}$ \tablenotemark{f} %
                                         & $1\,\bar{{\sigma}}$ \tablenotemark{g} \\
                     &                   & arcsec %
                                          & mJy
                                         & beam$^{-1}$ %
                                         & %
                                         & %
                                         & %
                                         & %
                                         & mJy \\
			\hline
			24$\mu$m	 & 	Spitzer/MIPS 		& 	$5.7$  & --	 	&	 $1.0$ 			  		& 	$589,713$ 	      & 0 				& 81,551			&0			&	$10.00\times 10^{-3}$ \\
			1.4~GHz			&	 VLA 		    		& 	$2.5$   & --	&	 $0.2$ 			 	 &	    $589,713$	      & 0 				& 4,311			&0			&	$10.22\times 10^{-3}$ \\
			3~GHz 			& 	VLA 					& 	$0.75$  & --	& 	$0.06$ 			 	 &	    $589,713$ 	      & 0 				& 15,645		&0			&	$2.89\times 10^{-3}$ \\
			100$\mu$m 	& 	Herschel/PACS   & 	$7.2$   & --	& 	$0.5$ 		 		 	&	   $191,624$	    & 0 				& 9,541		&0				&	1.44 \\
			160$\mu$m 	& Herschel/PACS 	& 	$12.0$  & 4.2	& 	$0.8$ 				  & 	$109,366$		  & 58,558 			& 6,106		&0			&	3.55 \\
			250$\mu$m 	& Herschel/SPIRE 	& 	$18.2$  & 6.8	& 	$1.0$ 				  & 	$59,371$ 		  & 20,637 			& 10,311	&111		&	1.77 \\
			350$\mu$m 	& Herschel/SPIRE 	& 	$24.9$  & 7.5	& 	$1.1$ 				  & 	$36,781$ 		  & 109,773 		& 4,874		&6		&	2.68 \\
			500$\mu$m 	& Herschel/SPIRE 	& 	$36.3$  & 7.0	& 	$1.0$ 				  & 	$16,333$ 		  & 123,355 		& 2,588		&24			&	2.91 \\
			850$\mu$m 	& JCMT/SCUBA2 	  &  $11.0$  & 0.94	& 	$\leqslant0.4$	& 	$23,868$ 		  & 170,560 		& 536		&484		&	1.37 \\
			1.1mm 		& ASTE/AzTEC 	    	& 	$33.0$  & 1.58	& 	$\leqslant0.8$			& 	$7,024$ 		  & 111,507 		& 137		&0				&	1.58 \\
			1.2mm 		& IRAM/MAMBO 	   	 &	$11.0$  	& 0.8	& 	$0.1$ 		  			& 	$2,501$ 		  & 25,730 				& 50		&0				&	0.74 \\ 
			\hline
		\end{tabular*}
		
\begin{minipage}{0.82\textwidth}
    
\flushleft

$^\mathrm{a}$ 
    Beam FWHM is the full width at half maximum of the circular-Gaussian-approximation point spread function of each image data. %

$^\mathrm{b}$ 
    $\rho_{\mathrm{fit}}$ is the number density of prior sources fitted at each band, normalized by the Gaussian-approximation beam area. %

$^\mathrm{c}$ 
    $N_{\mathrm{fit}}$ is the number of prior sources fitted at each band. %

$^\mathrm{d}$ 
    $N_{\mathrm{excl.}}$ is the number of prior sources excluded from fitting at each band. These sources are subtracted from original image with their spectral energy distribution predicted flux at each band. %

$^\mathrm{e}$ 
    $N_{\mathrm{S/N}>3}$ is the number of prior sources with $\mathrm{S/N}\ge3$ (i.e. detected) at each band. %

$^\mathrm{f}$ 
    $N_{\mathrm{add.}}$ is the number of $\mathrm{S/N}\ge3$ additional sources that are not in the prior source catalog but blindly-extracted from the intermediate residual image product at each band (see Section~\ref{Section_Additional_Sources_In_Residual}). %

$^\mathrm{g}$ 
    $1\,\bar{{\sigma}}$ is the detection limit computed as the median of the flux error of all detected sources at each band.

\end{minipage}


\end{table*}

\begin{figure}
\centering
\includegraphics[width=0.48\textwidth,trim={1cm 0cm 0cm 1.5cm},clip]{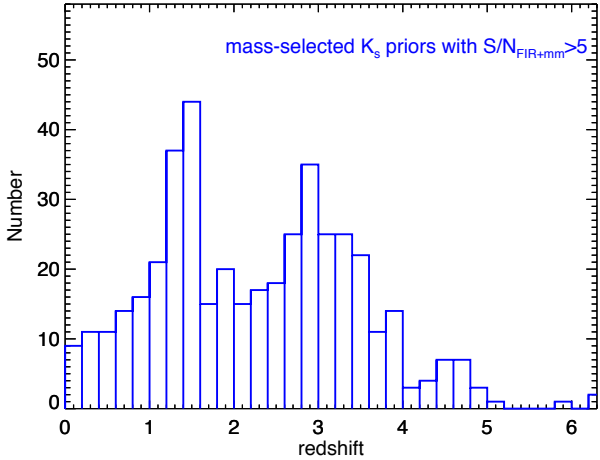}
\caption{%
	Redshift histogram of mass-selected $K_s$ priors detected in the FIR/(sub)mm with $\mathrm{S/N}_{\mathrm{FIR+mm}}>5$. These priors have no detection at 24~$\mu$m and radio bands. The redshifts shown here were derived from FIR+(sub)mm SED fitting.
\label{hist_mass_only}%
}
\end{figure}

Furthermore, we have {434} detections with $\mathrm{S/N}_{\mathrm{FIR+mm}}>5$ from the mass-selected $K_s$ priors without detection at 24~$\mu$m or radio bands. 
As shown in Fig.~\ref{hist_mass_only}, these $K_s$ priors show a double-peak distribution in redshift with a valley at $z\sim2$. As mentioned in Section~\ref{Section_The_24_Radio_Mass_Catalog}, these $z<2$ detections could be (rare) silicate-dropout sources at $1<z<2$ \citep{Magdis2011Dropout24}, or sources that were lost because of high $crowdedness$ in the 24~$\mu$m map (recall that the radio does not reach quite as deep as 24$\mu$m at $z<2$, see Fig.~\ref{Fig_SED_expected_flux_against_z}). While only $\sim0.6$\% of the additional stellar mass-selected priors were eventually IR-detected (recall we did not limit to those at $z>2$--3 on purpose, so this low number is misleading), their presence enhances the sample of $z>3$ detected galaxies by more than 20\%.

We publicly release the photometric catalog for all $K_s$+radio+mass-selected sources and the additional sources extracted from the residual images\footnote{\href{https://drive.google.com/open?id=18iknbRBUJSqU3Tc5Fh6DRVJFrnLcWcl4}{https://drive.google.com/open?id=18iknbRBUJSqU3Tc5Fh6DRVJFrn\\LcWcl4} as an extended online version of Table~\ref{Table_all_candidates}}.

\section{Comparison to catalogs from the literature}
\label{Section_Compare_Measurements}

\begin{figure*}
\centering
\includegraphics[width=0.285\textwidth]{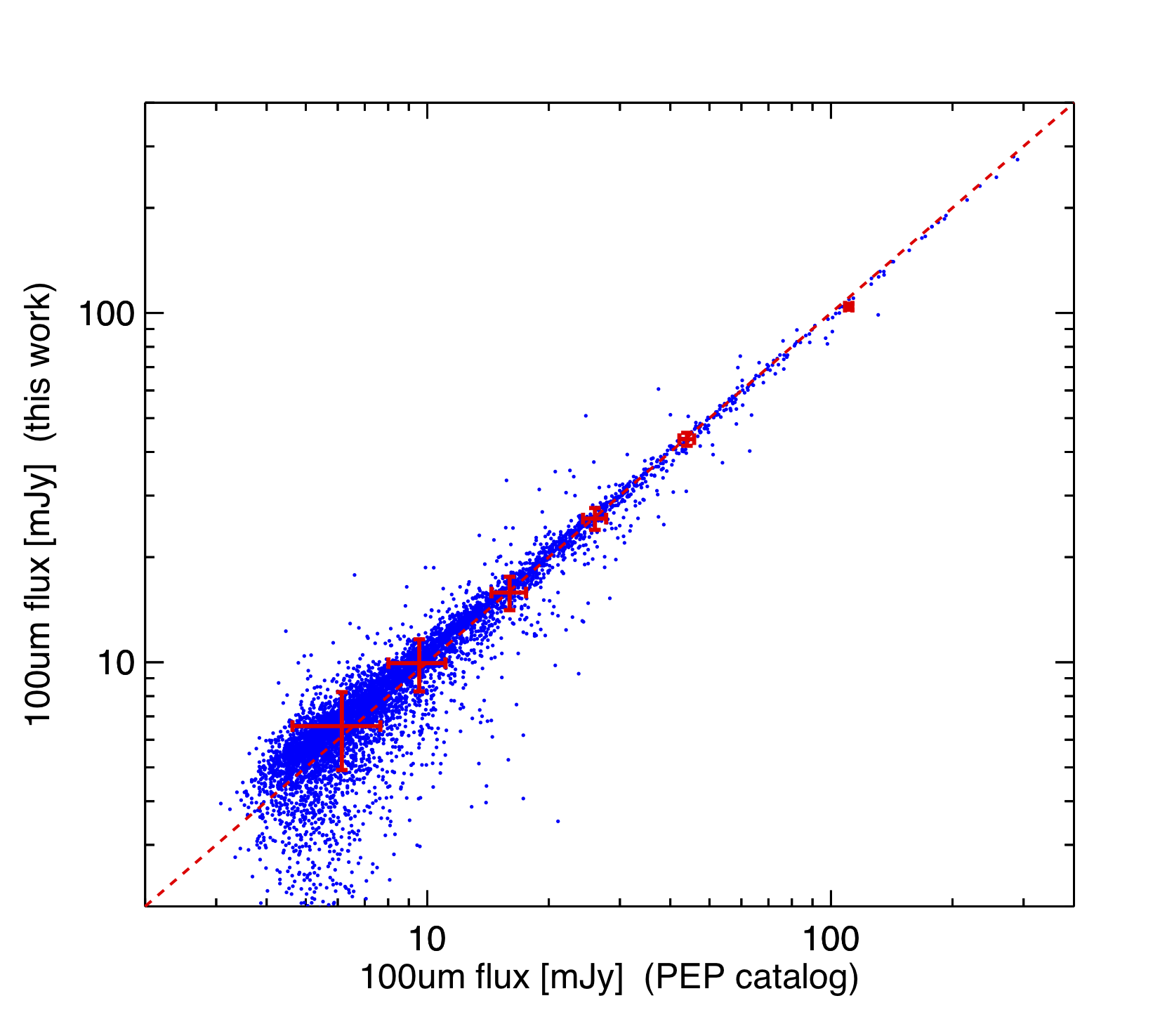}
\includegraphics[width=0.285\textwidth]{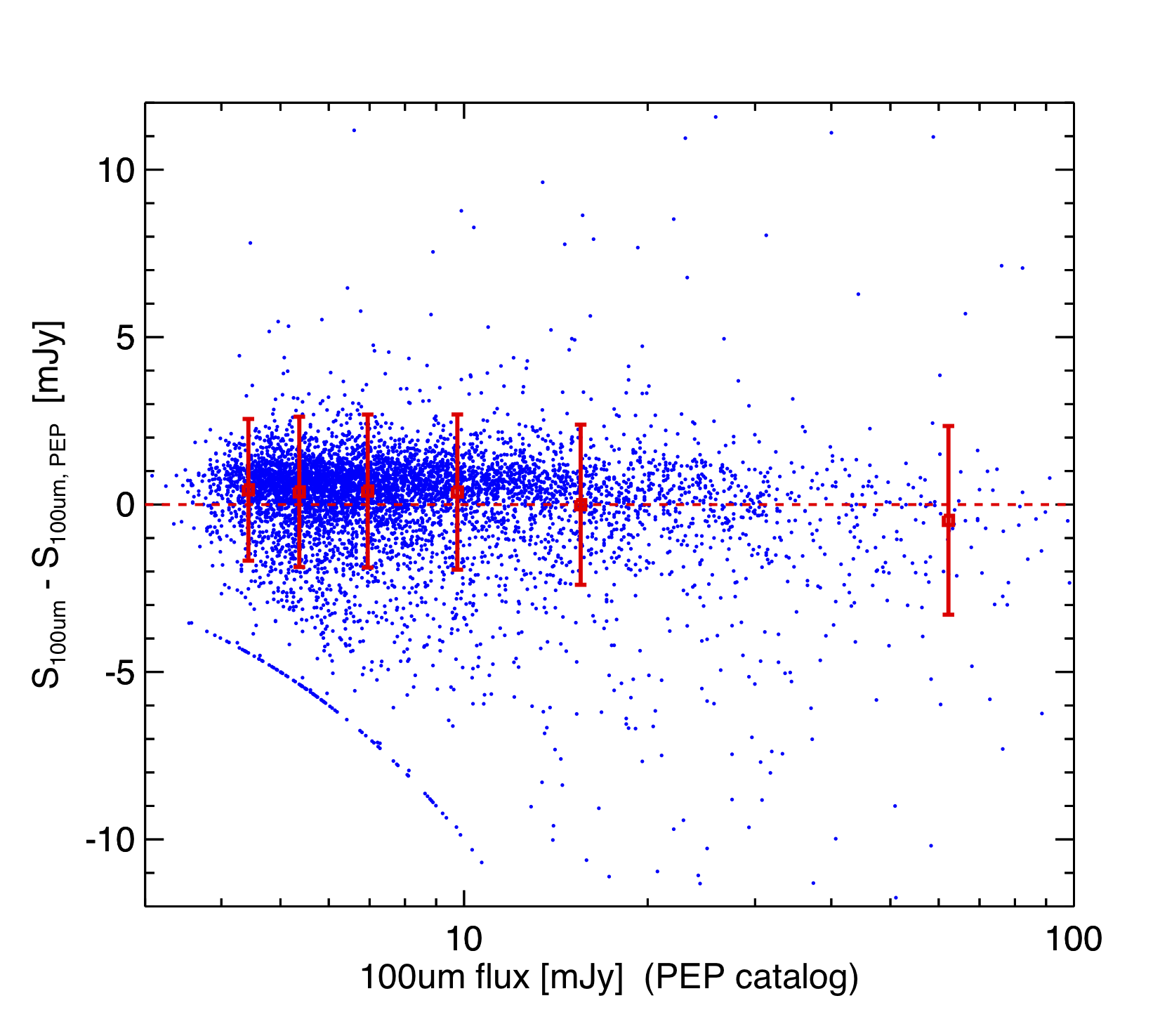}
\includegraphics[width=0.33\textwidth]{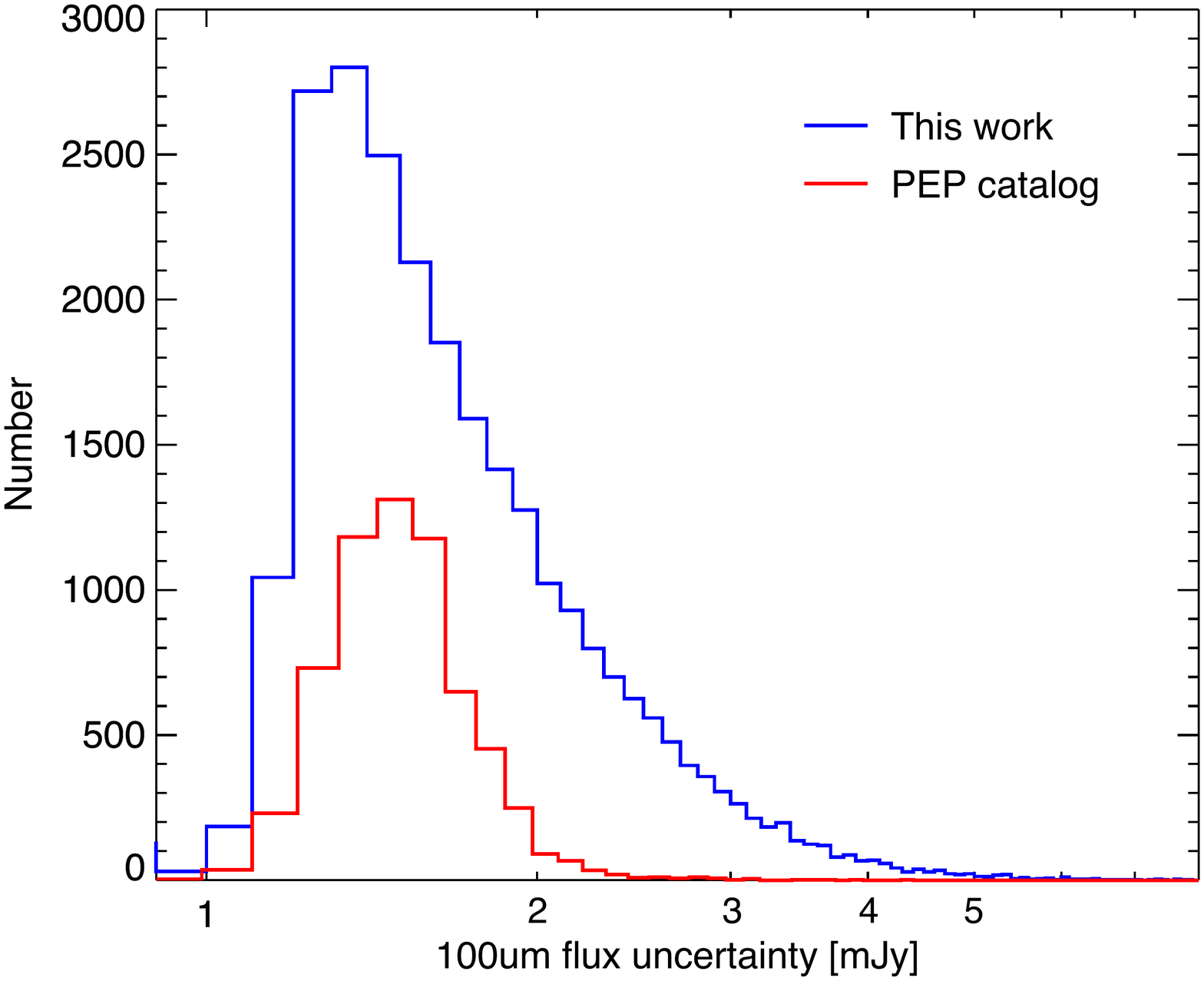}
\includegraphics[width=0.285\textwidth]{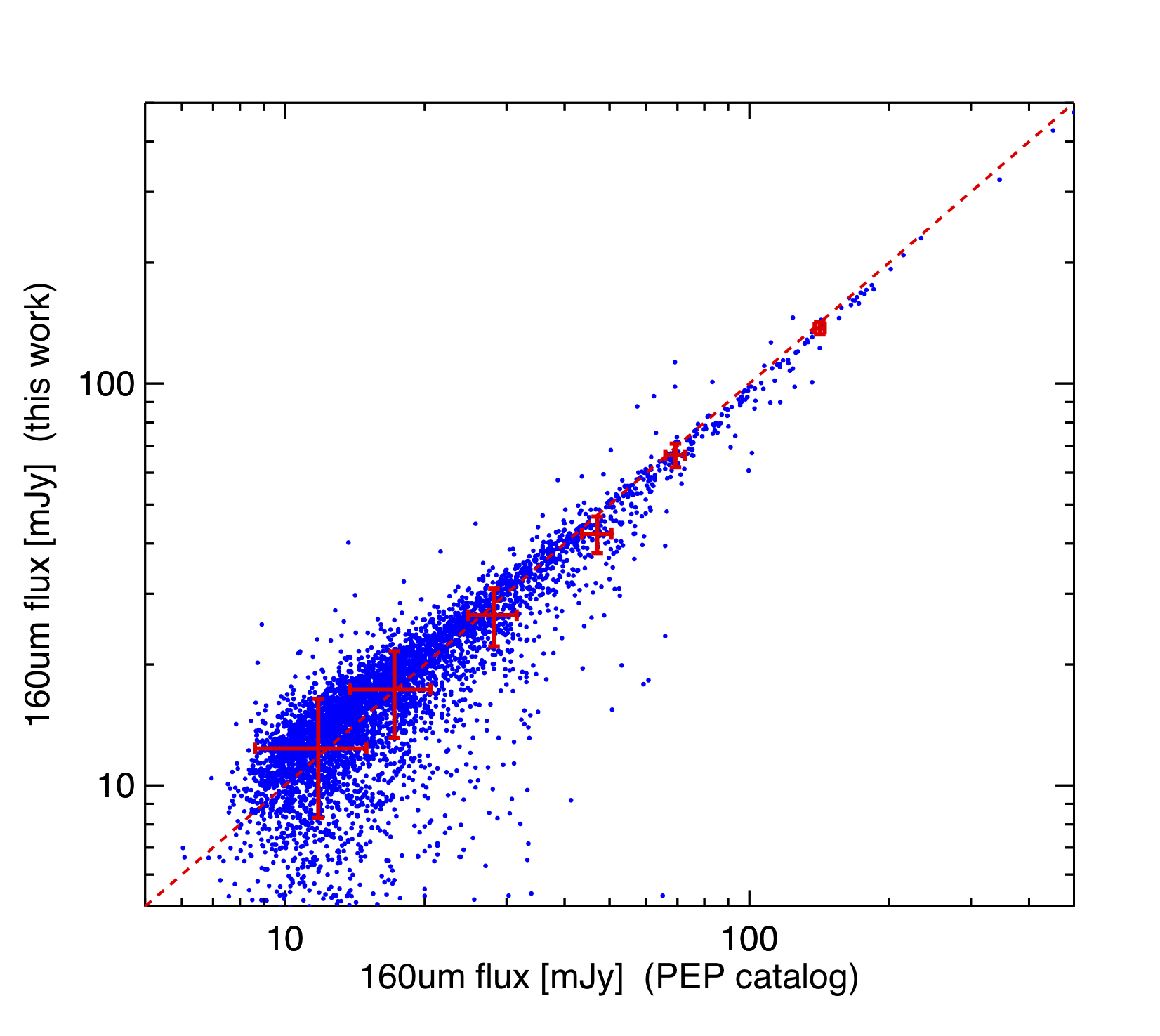}
\includegraphics[width=0.285\textwidth]{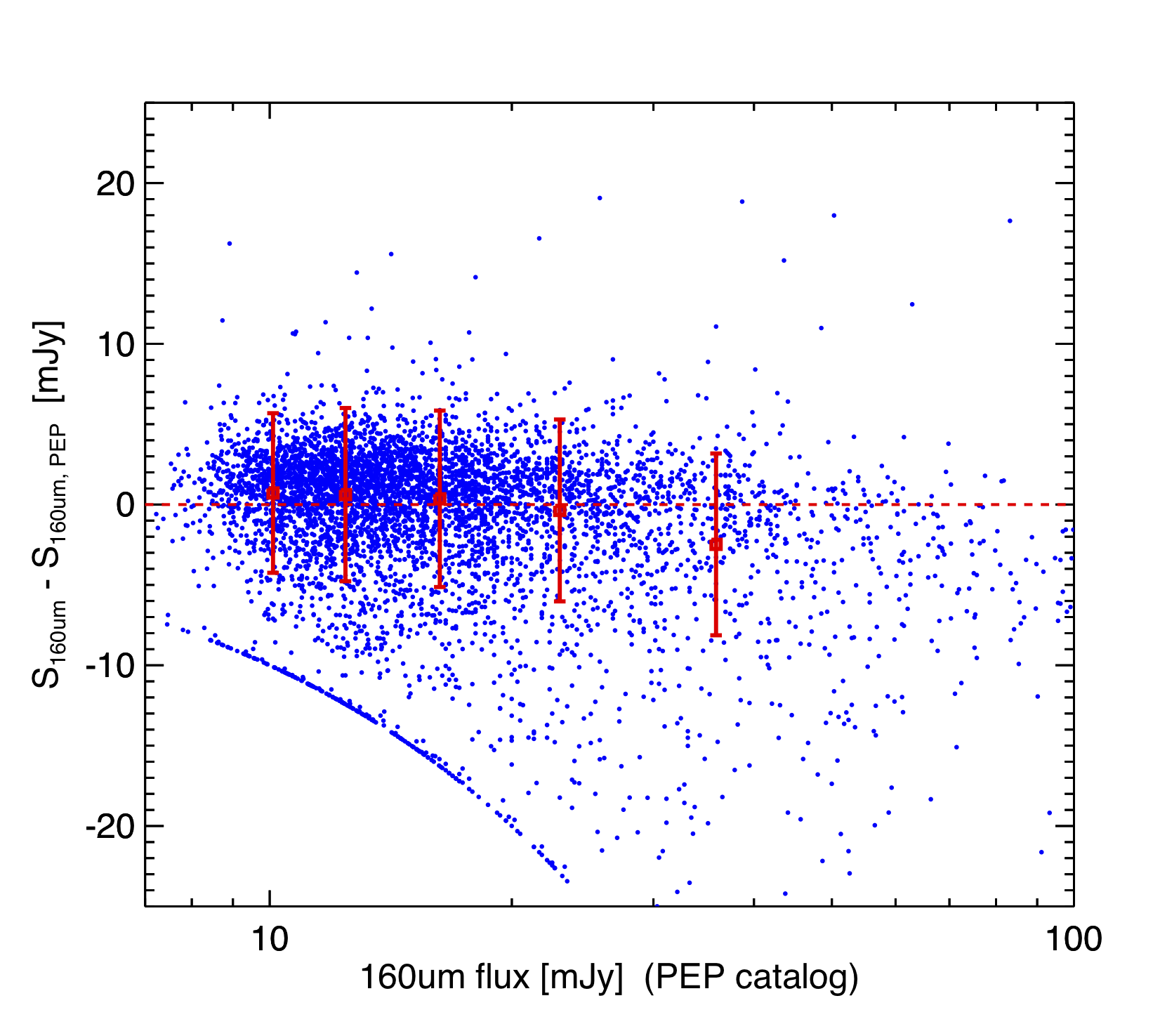}
\includegraphics[width=0.33\textwidth]{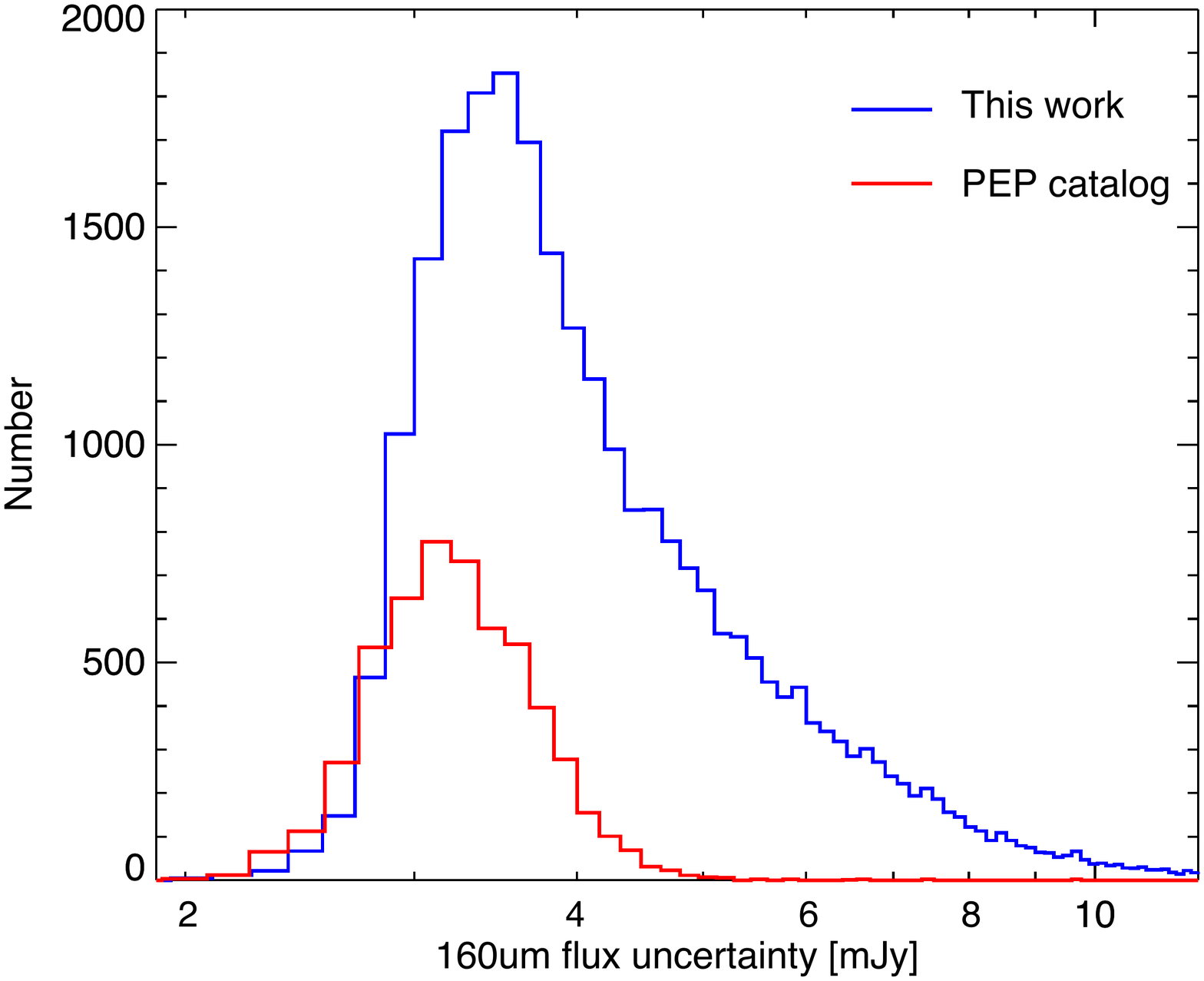}
\caption{
	Comparison of the PACS 100~$\mu$m \& 160~$\mu$m photometry with the PEP catalog \citep{Lutz2011}. Fluxes here are those directly measured, without correcting them for flux losses from the high-pass filtering processing of PACS images \citep{Magnelli2013,Popesso2012}. 
    Left panels show the flux and uncertainty comparison for the 100~$\mu$m (top) and 160~$\mu$m (bottom) measurements. 
    {Red points with error bars show median flux and flux uncertainty from the two catalogs for matched sources in several bins.}
    Middle panels show the flux difference {and combined uncertainty (error bars)} between our work and the PEP catalog. Right panels show distributions of flux uncertainty.
\label{compare_PACS}%
}
\end{figure*}

\begin{figure*}
\centering
\includegraphics[width=0.31\textwidth, trim={0cm 1cm 0cm 1.5cm}, clip]{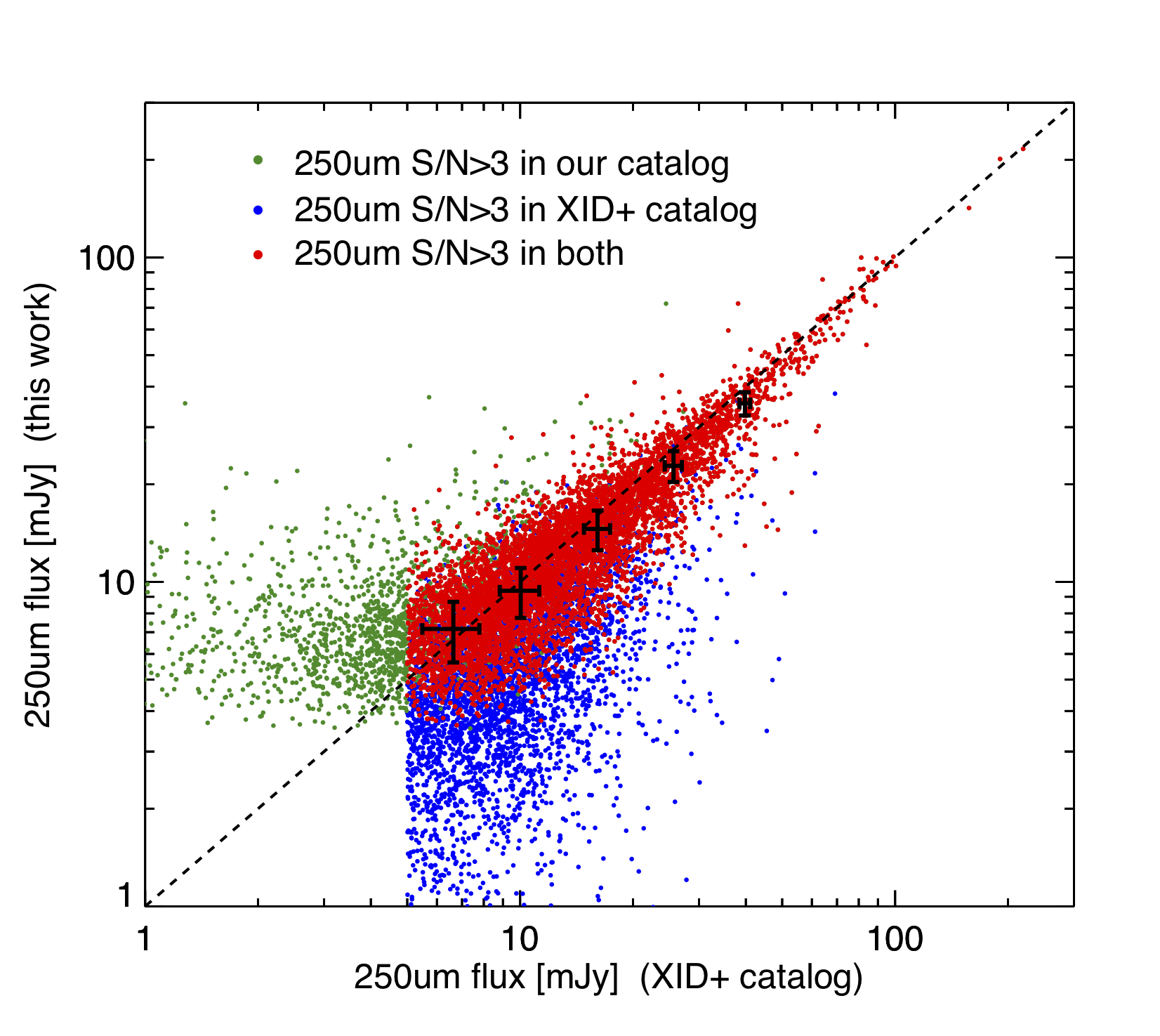}
\includegraphics[width=0.31\textwidth, trim={0cm 1cm 0cm 1.5cm}, clip]{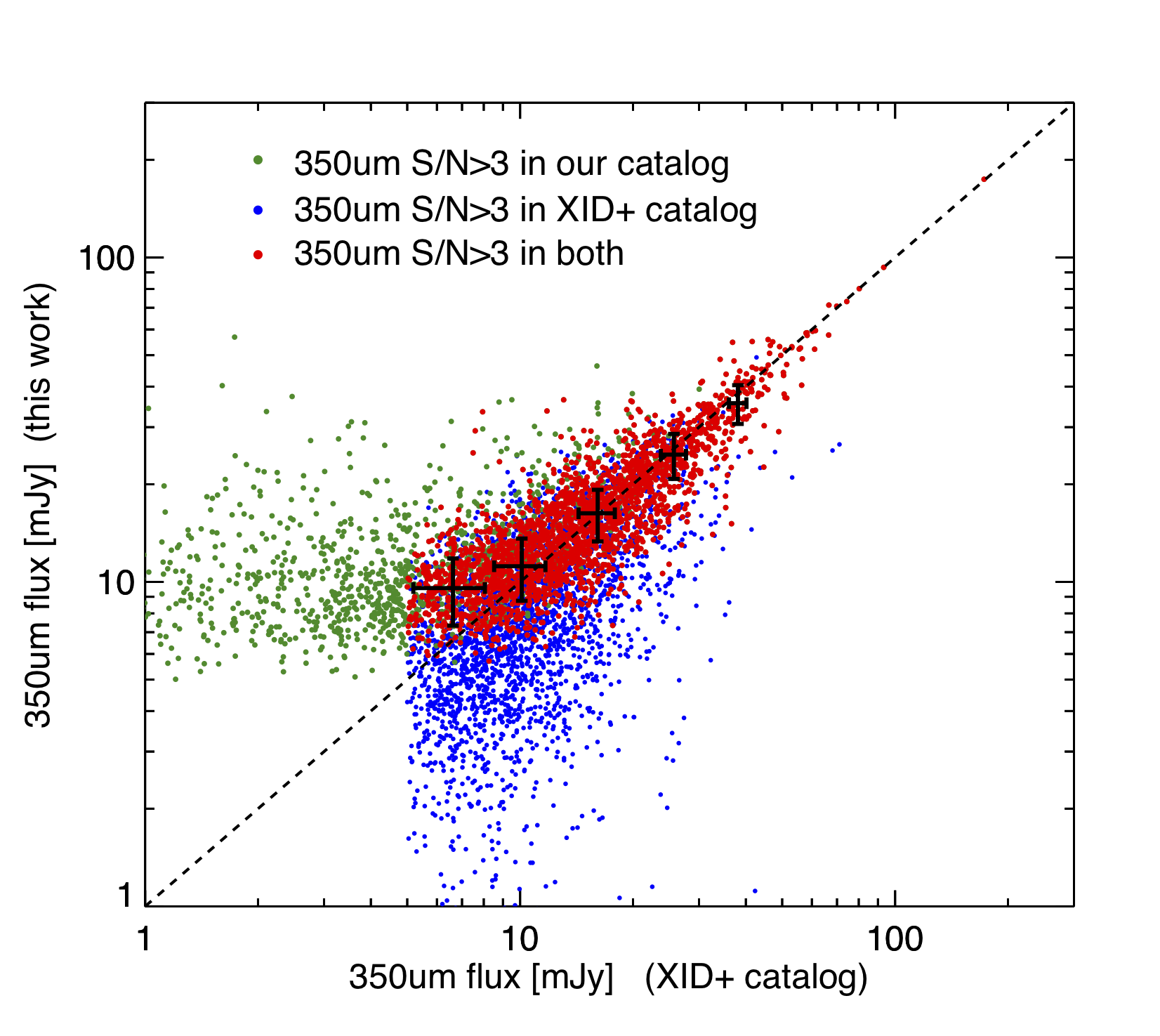}
\includegraphics[width=0.31\textwidth, trim={0cm 1cm 0cm 1.5cm}, clip]{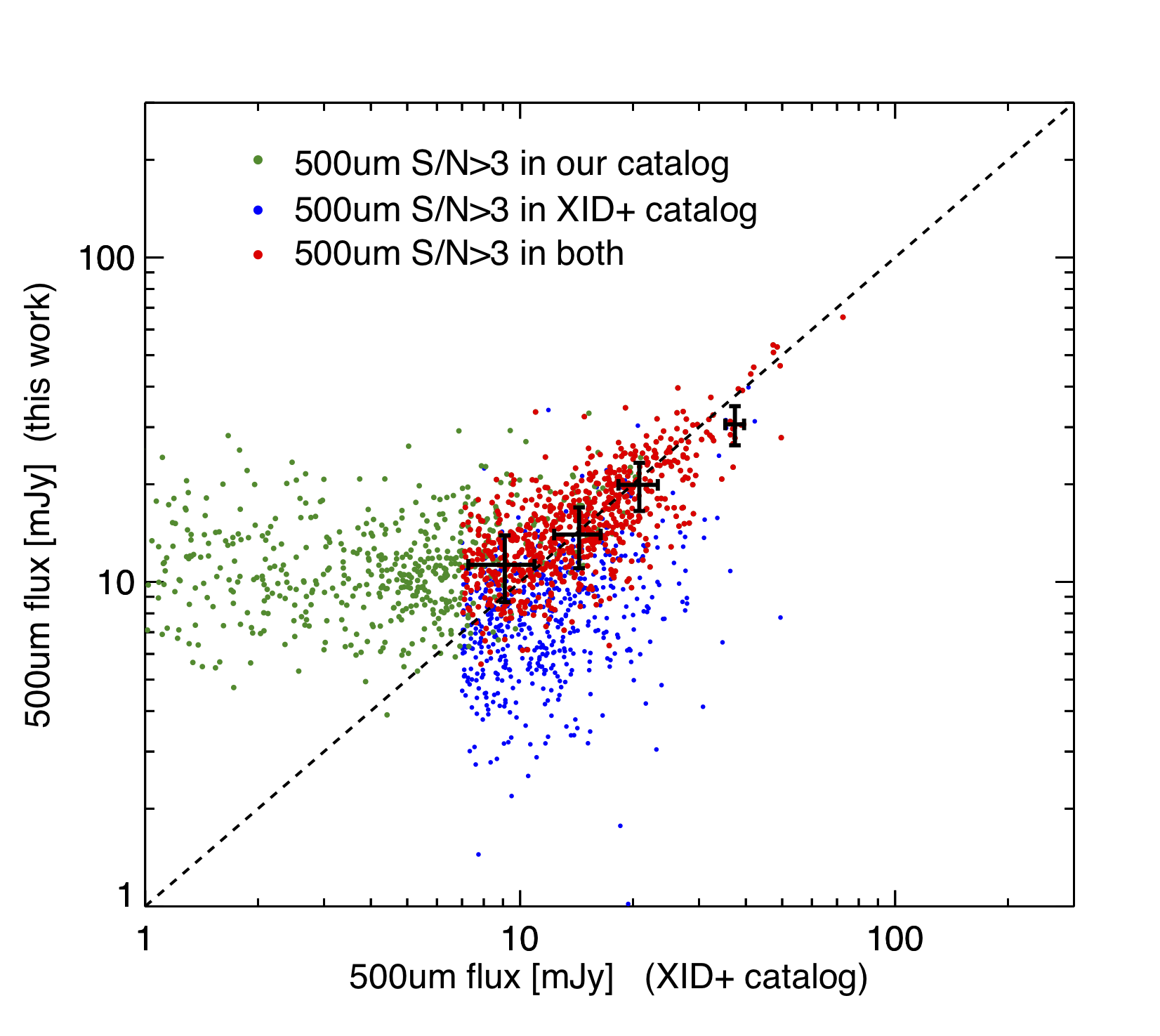}
\includegraphics[width=0.31\textwidth, trim={1.75cm 1cm 1.75cm 1.5cm}, clip]{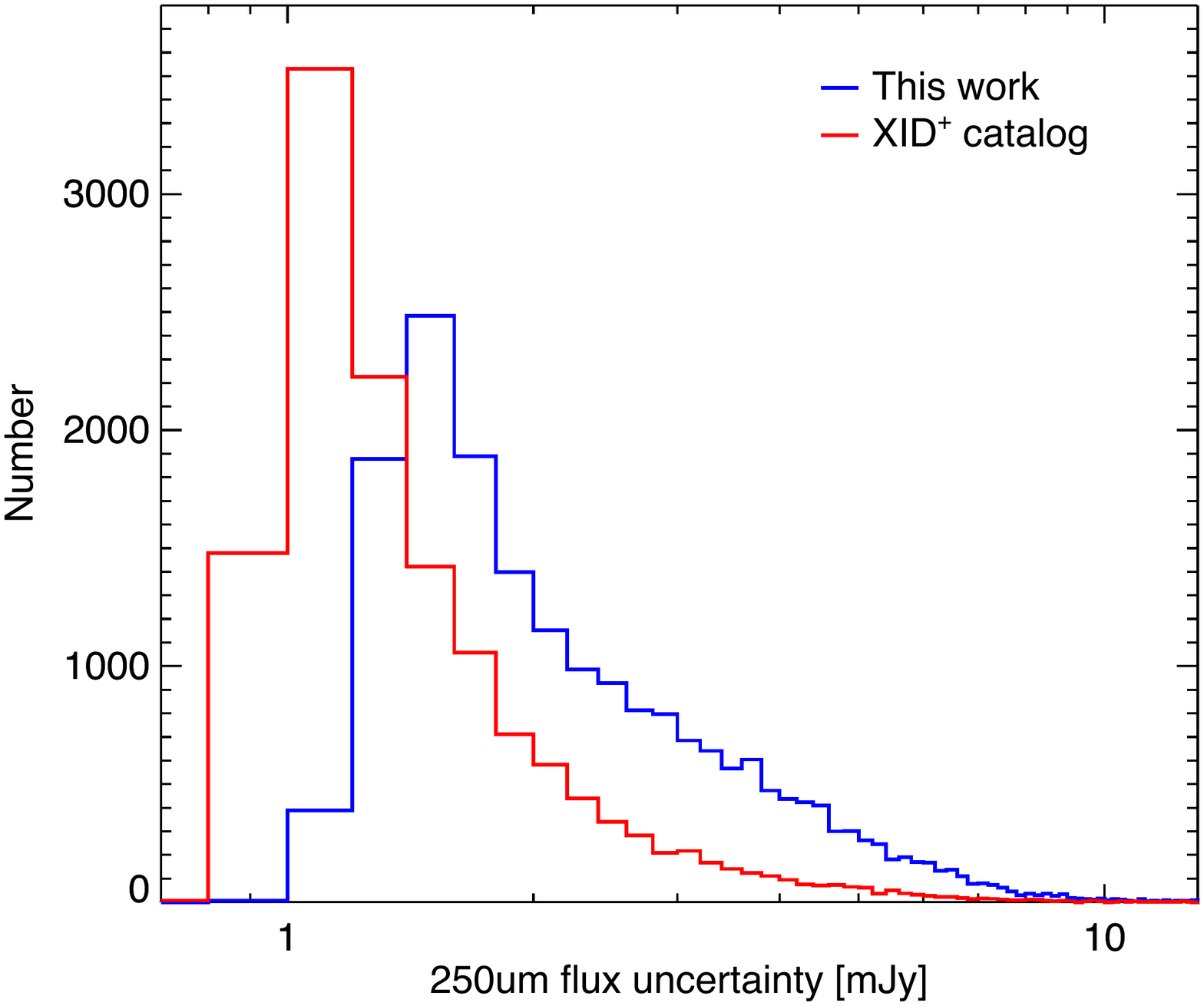}
\includegraphics[width=0.31\textwidth, trim={1.75cm 1cm 1.75cm 1.5cm}, clip]{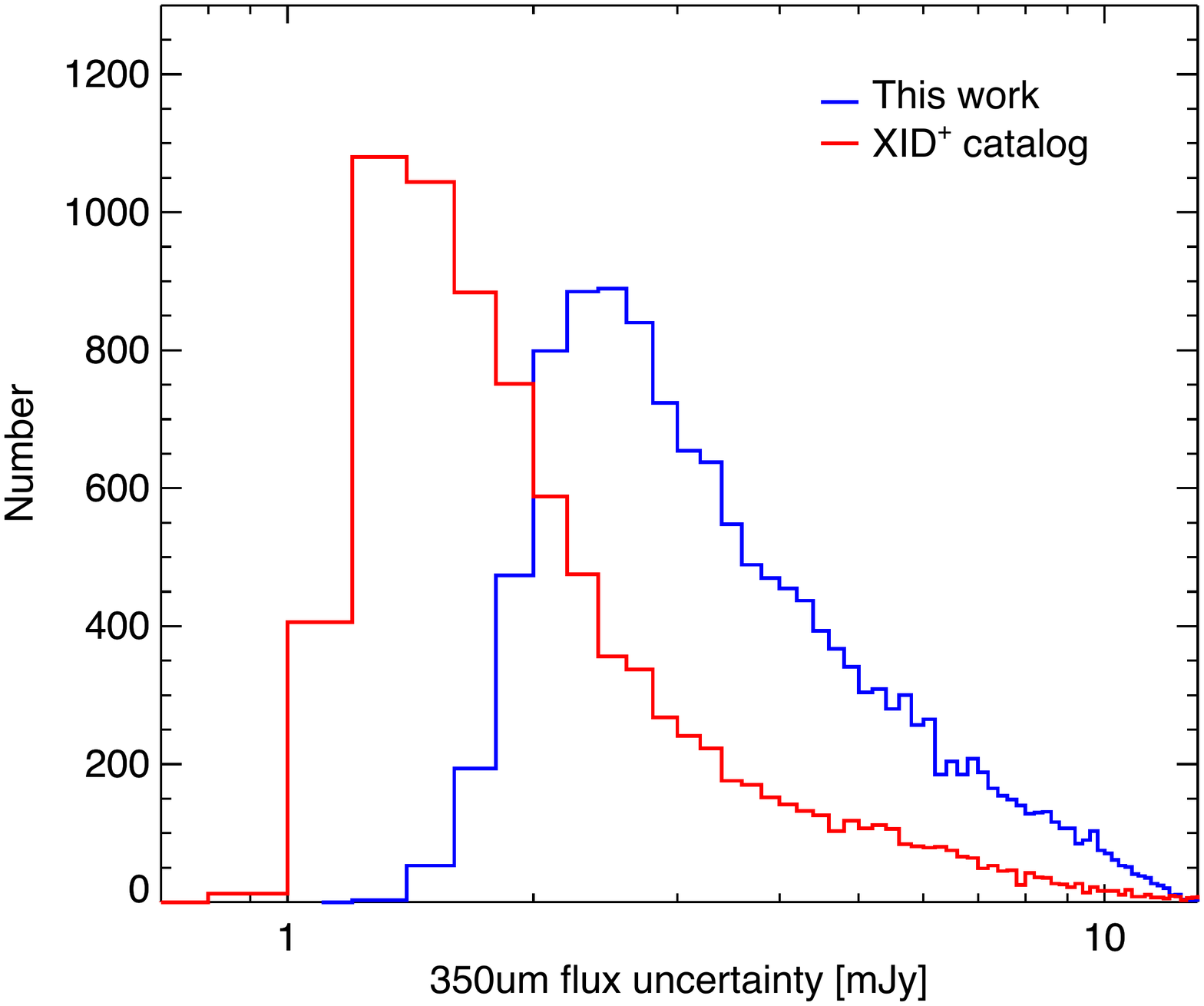}
\includegraphics[width=0.31\textwidth, trim={1.75cm 1cm 1.75cm 1.5cm}, clip]{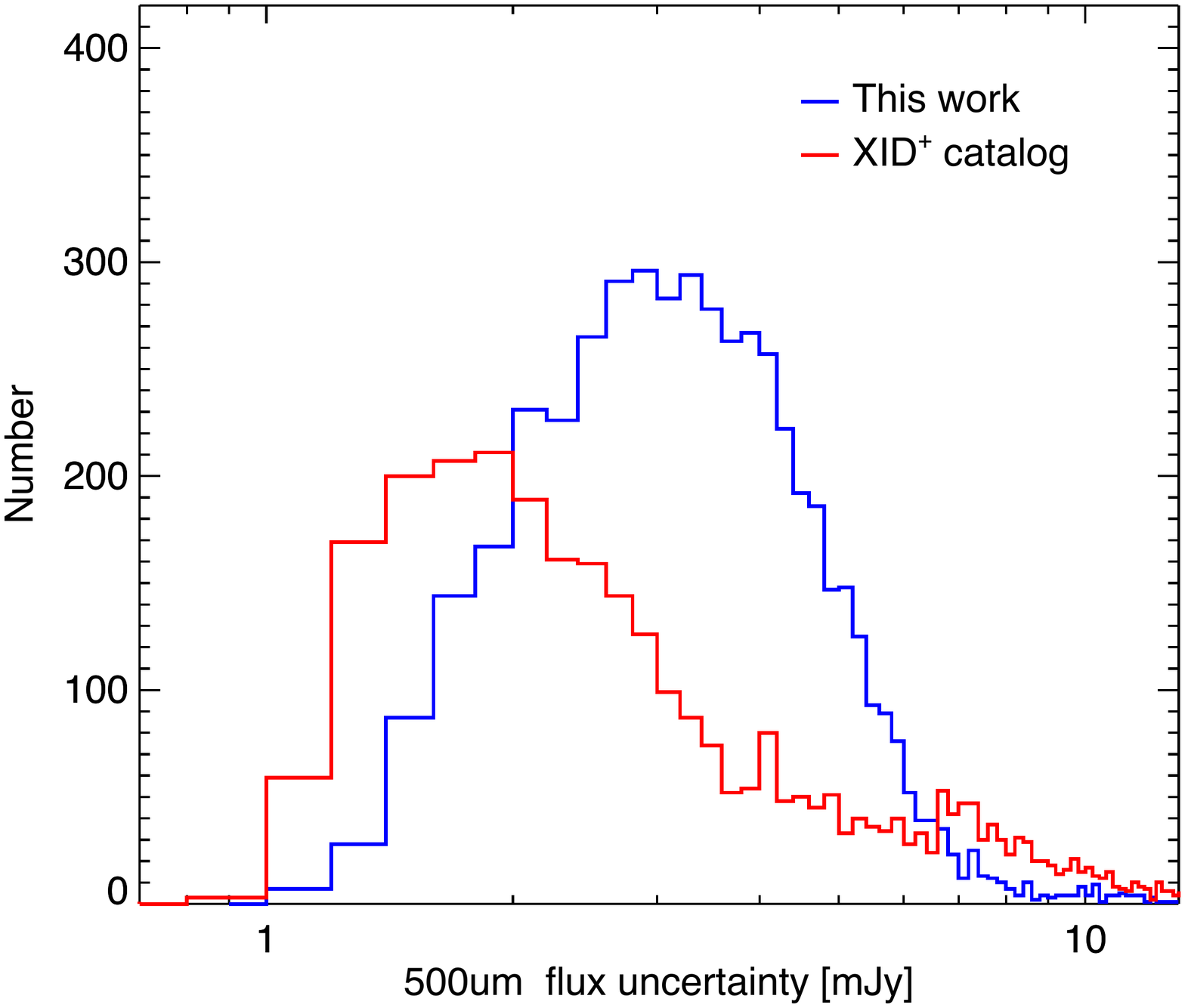}
\caption{%
	Comparisons between our \superdb catalog and the XID$^{+}$ catalog for SPIRE 250~$\mu$m, 350~$\mu$m and 500~$\mu$m bands. 
    Upper panels: Red dots show 3-sigma detections in both catalogs. Green and blue dots show sources that are only detected in our catalog and in the XID$^{+}$ catalog, respectively. {Black points with error bars show median flux and flux uncertainty from two catalogs for matched sources in several bins.}
    {Bottom panels}: histogram of flux uncertainty at each SPIRE band.
\label{compare_spire}%
}
\end{figure*}

In this section, we compare our \superdb photometry to public photometry catalogs for the COSMOS field: the PEP catalog at PACS bands \citep{Lutz2011}, the $\mathrm{XID}^{+}$ catalog at SPIRE bands \citep{Hurley2016} and the SCUBA2 catalog of \citet{Geach2016}. The comparisons are shown in Figs.~\ref{compare_PACS}, \ref{compare_spire} and \ref{compare_scuba2}.

\subsection{PEP catalogs}

At PACS 100~$\mu$m \& 160~$\mu$m we matched our catalog to the PEP catalog \citep{Lutz2011} with a tolerance of 1$''$. 
In Fig.~\ref{compare_PACS}, our flux measurements are generally consistent with the measurements in the PEP catalog, although showing a tail of sources for which we suggest systematically lower fluxes than PEP. 
The PEP catalog is produced by fitting 47,437 priors selected at 24~$\mu$m from \citet{LeFloch2009}, while $5\times$ more priors are fitted in our work. 
We thus believe that the lower flux measurements found in our work are due to sources blending in the PEP catalog.
There is also a systematic effect affecting fluxes of all sources, as 
can be seen by plotting flux differences (Fig.~\ref{compare_PACS}). 
There is a constant difference of 0.8~mJy at 100~$\mu$m and 1.8~mJy at 160~$\mu$m, which are both of the order of 0.2$\times$ the rms noise at each band.
We attribute these constant offsets to a small background difference applied in the two works, although we advocate that we could measure the background to better than this accuracy with our simulations, hence we tend to believe that our measurements are correct. Our uncertainties agree well with those in the PEP catalog, as shown by the right panels of Fig.~\ref{compare_PACS}.

\subsection{SPIRE $\mathrm{XID}^{+}$ catalogs}

At SPIRE bands we compare our results to the $\mathrm{XID}^{+}$ catalog \citep{Hurley2016}, that deblend SPIRE images via a MCMC-based prior-extraction method on 52,092 priors selected as 24~$\mu$m detections. 
The $\mathrm{XID}^{+}$ catalog covers an area of 2.27 deg$^2$, we limit the comparison to sources with $goodArea=1$ and obtain 30,372 matches with a tolerance of $1''$.
Apart from the matched sources, there are 161,252 priors in our catalog that are not listed in the XID$^{+}$ catalog, and 8,200 XID$^{+}$ priors missing from our list. The first is due to our deeper 24$\mu$m photometry, the use of radio and especially of mass priors. The latter is likely mis-associations of fluxes at 24$\mu$m\footnote{For these 8,200 XID$^{+}$ 24$\mu$m-detected priors missing from our work, we tend to believe they come from mismatches between our accurate positions from our $K_s$ and radio catalogs and the blind-extracted 24~$\mu$m sources in XID$^{+}$}. These are $\sim83\%$ and $\sim4\%$ of the 24~$\mu$m+radio+mass-selected priors respectively.
In the 161,252 priors missed by XID$^{+}$, we have 2,198 detections at 250~$\mu$m, 1,175 detections at 350~$\mu$m and 1,097 detections at 500~$\mu$m respectively.
Hence we obtain more detections benefiting from the higher completeness of our prior catalog. Also, the lack of these extra priors in XID$^{+}$ likely exacerbate blending and flux boosting problems, leading to flux discrepancies also with sources in common among the two works, as discussed below.

In Fig.~\ref{compare_spire}, we show the flux comparison of matched sources that are detected with $\mathrm{S/N>3}$ either in our \superdb catalog or in the $\mathrm{XID}^{+}$ catalog. 
We find that our measurements are generally consistent to the measurements in the $\mathrm{XID}^{+}$ catalog at high fluxes while there are significant discrepancies towards faint fluxes.
We highlight two populations with significant flux discrepancies, which are shown in blue and green in each panel.
The blue dots show sources that have $\mathrm{S/N>3}$ only from $\mathrm{XID}^{+}$, and lower fluxes and S/N in our catalog.
By inspecting priors around these sources, we find that they are located in crowded regions. 
We are generally de-blending the signal among different galaxies based on our {\em physical} rather than {\em statistical} approach.
Hence we believe that $\mathrm{XID}^{+}$ often overestimated their fluxes.

Meanwhile, there are some sources that are only detected in this work (the green dots in each panel), and not by $\mathrm{XID}^{+}$. We have visually checked these sources on the original and the faint-source-subtracted images. We find most of them have visible signals in both maps. We do not know why $\mathrm{XID}^{+}$ catalogs do not retain these sources that appear to be significant according to our work. 
Quantitatively, in the first panel of Fig.~\ref{compare_spire},
there are 12,947 sources in total, 1,360 of them have $\mathrm{S}_\mathrm{250\mu m,us}/\mathrm{S}_\mathrm{250\mu m,XID+}>1.5$, while 4,479 sources have $\mathrm{S}_\mathrm{250\mu m,XID+}/\mathrm{S}_\mathrm{250\mu m,us}>1.5$.
In the second panel, there are 6,887 sources in total, 1,388 of them have $\mathrm{S}_\mathrm{350\mu m,us}/\mathrm{S}_\mathrm{350\mu m,XID+}>1.5$, while 2,014 sources have $\mathrm{S}_\mathrm{350\mu m,XID+}/\mathrm{S}_\mathrm{350\mu m,us}>1.5$.
In the third panel, there are 2,498 sources in total, 978 of them have $\mathrm{S}_\mathrm{250\mu m,us}/\mathrm{S}_\mathrm{250\mu m,XID+}>1.5$, while 444 sources have $\mathrm{S}_\mathrm{500\mu m,XID+}/\mathrm{S}_\mathrm{500\mu m,us}>1.5$. 
So it seems that at the shortest SPIRE wavelength we are deblending into smaller/weaker sources many of the $\mathrm{XID}^{+}$ detections but this is getting less imbalanced at 350$\mu$m and reversed at 500$\mu$m, probably because of the rapidly decreasing fraction of sources seen by $\mathrm{XID}^{+}$ at these longest wavelengths.

We show the flux uncertainties for each band from our and the $\mathrm{XID}^{+}$ catalog in the bottom panels of Fig.~\ref{compare_spire}. The $\mathrm{XID}^{+}$ catalog has lower uncertainties in each SPIRE band. Since flux uncertainties in our catalog have been carefully calibrated by simulations and display well-behaved Gaussian-like uncertainties, we believe that $\mathrm{XID}^{+}$ generally underestimates the flux uncertainties in the SPIRE bands.

\begin{figure}
\centering
\includegraphics[width=0.48\textwidth, trim={1.5cm 0cm 2.8cm 1cm}, clip]{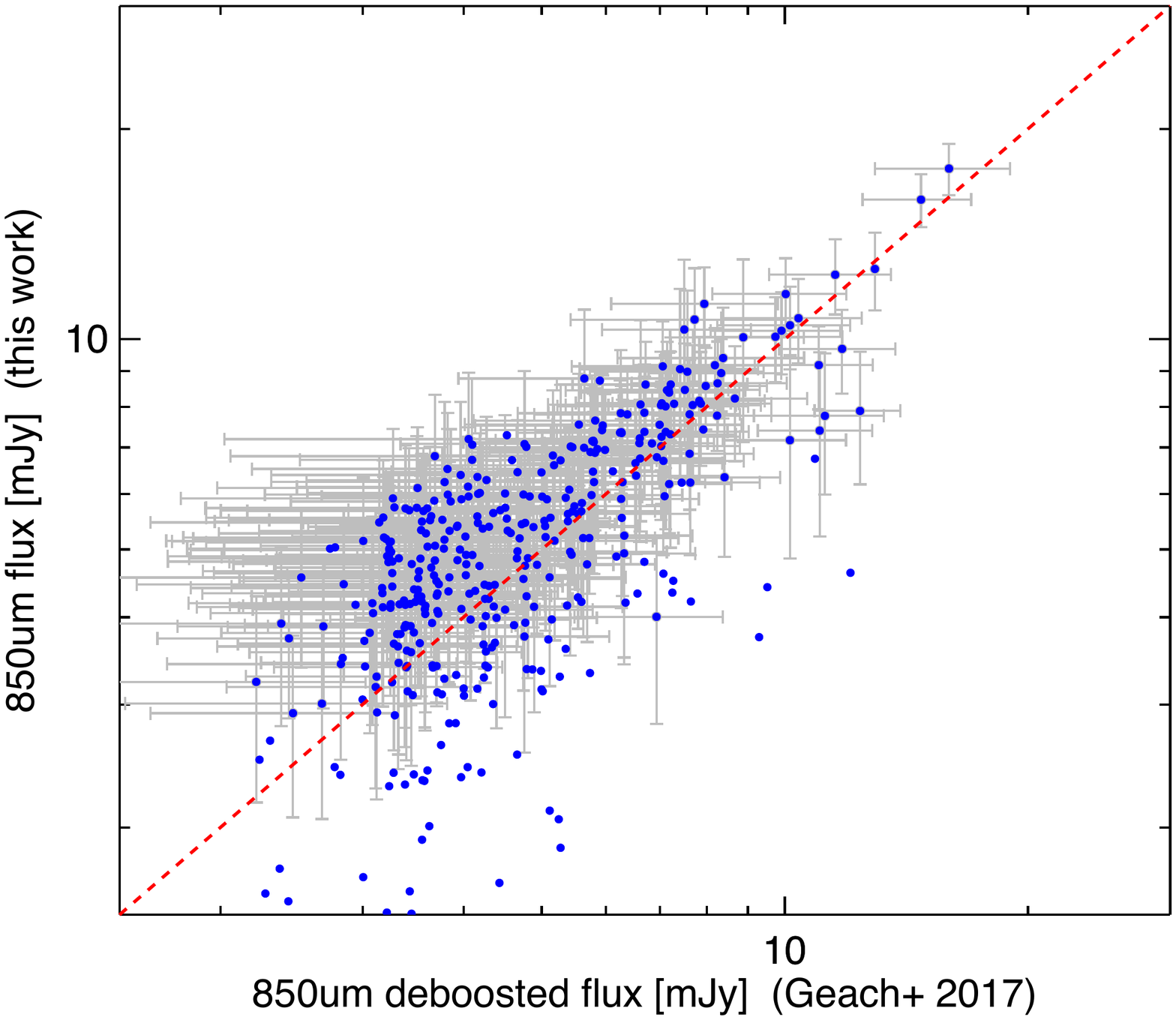}
\caption{%
	SCUBA2 850~$\mu$m fluxes in our work comparing to the de-boosted fluxes in \citet{Geach2016}. 
\label{compare_scuba2}%
}
\end{figure}

\begin{figure}
\centering
\includegraphics[width=0.48\textwidth]{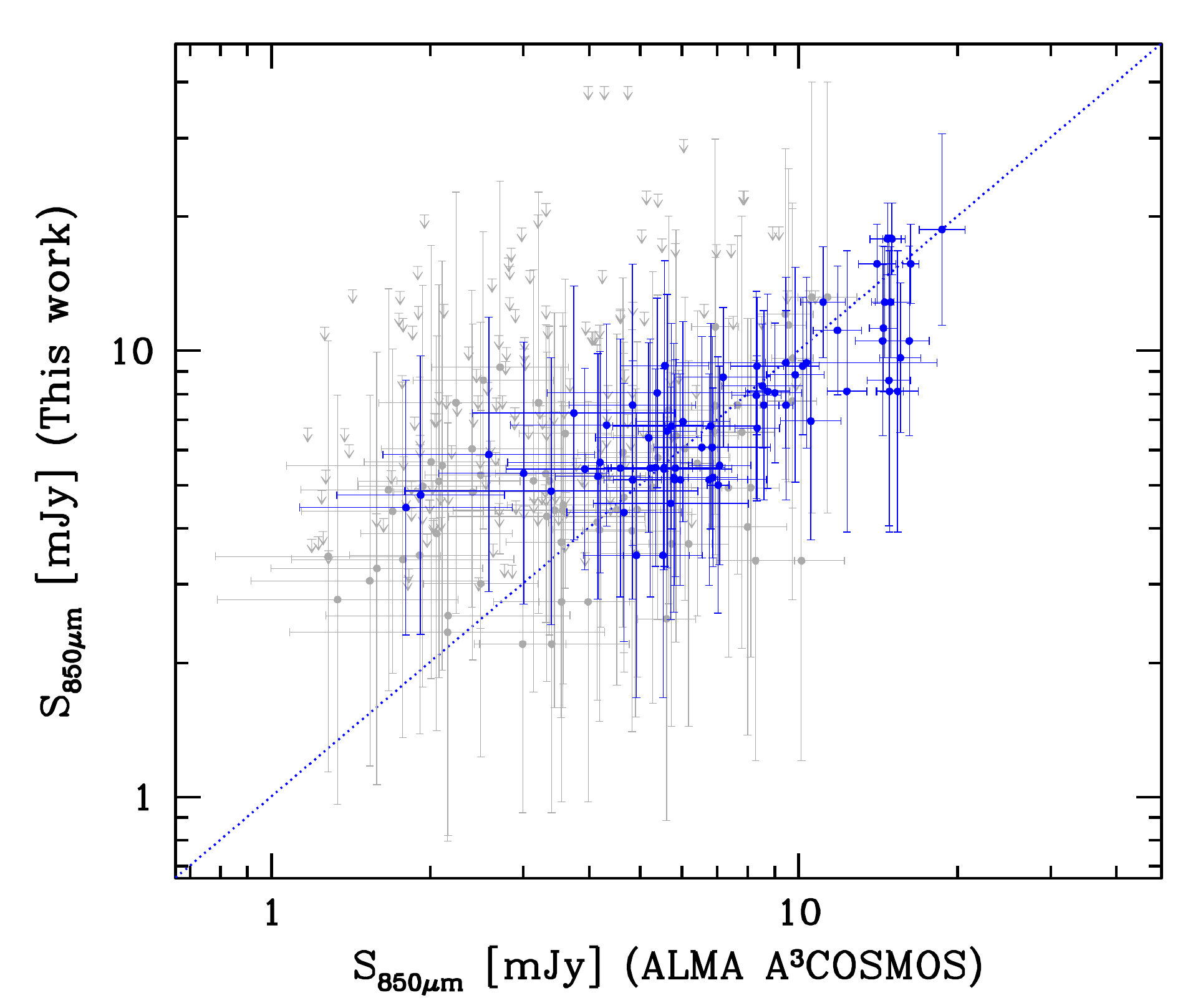}
\caption{%
Deblended SCUBA2 850~$\mu$m fluxes in this work comparing to the ALMA archival photometry from A$^3$COSMOS (see Sect.~\ref{Section_A3COSMOS}). Blue data points are sources commonly detected above $3\,\sigma$ in both works while gray data points are sources detected ($>3\,\sigma$) in A$^3$COSMOS but marginally detected ($\sim2$--$3\,\sigma$) in this work. Gray arrows are $3\,\sigma$ upper limits which are detected ($>3\,\sigma$) in A$^3$COSMOS but non-detected ($2\,\sigma$) in this work. 
\label{compare_alma}%
}
\end{figure}

\begin{figure}
\centering
\includegraphics[width=0.48\textwidth]{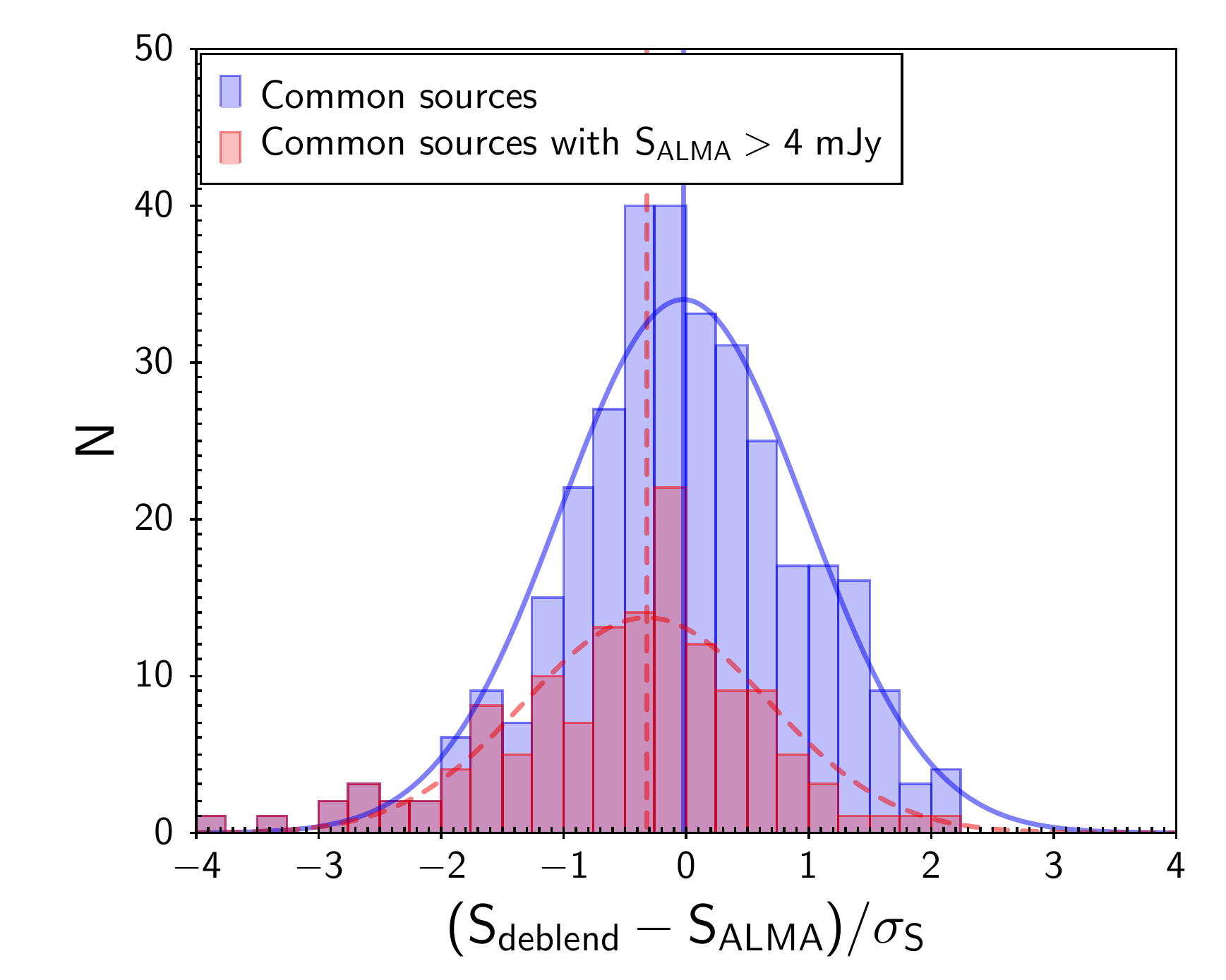}
\caption{%
Histogram of the flux difference normalized by total errors between this work and the ALMA archival photometry from A$^3$COSMOS (see Sect.~\ref{Section_A3COSMOS}). Blue histogram corresponds to all data points in Fig.~\ref{compare_alma} (including low $\mathrm{S/N}$ and upper limits). {Red histogram refer to the $>4$~mJy ALMA sample. A Gaussian fit with sigma fixed to 1 to the distribution is overlaid to both samples histograms. }
\label{compare_alma_histogram}%
}
\end{figure}

\subsection{SCUBA2 catalogs}

In Fig.~\ref{compare_scuba2}, we cross-match common sources (within a limiting separation of 6'' because of the large SCUBA2 PSF) and compare our SCUBA2 850~$\mu$m measurements to the de-boosted fluxes in \citet{Geach2016}. 
Our \superdb fluxes are consistent with the de-boosted ones.
Flux uncertainties are also fairly consistent with the de-boosted uncertainties in \citet{Geach2016}.
We have 1,020 galaxies (536 priors and 484 additional sources) detected with $\mathrm{S/N_{850\mu m}>3}$ from SCUBA2. This is a $3.3\times$ larger sample than the 306 detections reported by \citet{Geach2016} in the COSMOS field.

\subsection{Comparison to ALMA archival photometry}
\label{Section_A3COSMOS}
%
We further use 1000+ public ALMA archival data at submm-wavelengths in the COSMOS field to verify our deblended 850~$\mu$m photometry. These ALMA data were reduced, imaged and analyzed in the on-going ALMA archive mining project A$^3$COSMOS (D. Liu et al., in prep.)\footnote{\url{https://sites.google.com/view/a3cosmos}}. The A$^3$COSMOS team has processed all public ALMA archive in the COSMOS field and obtained accurate (sub)mm continuum photometry with both blind source extraction (mainly with the code PyBDSM) and prior source fitting. The catalogs are already available and verified with extensive Monte-Carlo simulations, and will be publicly released after the publication of their first paper (D. Liu et al., in prep.). Here we use their catalogs in advance (priv. comm.), benefiting from the exquisite $\approx1''$ ALMA resolution which resolves all blending issues, to verify our super-deblended flux measurements. 

Fig.~\ref{compare_alma} shows the flux comparison for common sources between this work and the A$^3$COSMOS prior source fitting catalog, {including all prior sources within the ALMA primary beam.}
Both works used optical+near-infrared+radio priors, therefore common sources can be easily cross-matched without confusion/multiplicity issue. 
This figure shows good agreements between the two catalogs without systematic bias. The error bars from this work are in general larger than those of the ALMA photometry, as expected (while ALMA, on the other hand, currently only covers a very small fraction of the COSMOS field).

In Fig.~\ref{compare_alma_histogram}, we further show the histogram of the flux difference between the two works normalized by the total uncertainty (quadratic sum of both errors from this work and A$^3$COSMOS). This comparison is very similar to the histograms shown for simulations for all bands in Appendix~\ref{Section_Simulation_Performance} figures.
All cross-matched sources, including low $\mathrm{S/N}$ and non-detections, are included in this histogram. The median of the distribution is -0.14 and the sigma (scatter) is 0.9, which is very close to 1, indicating that the scatter is consistent with the photometric error.  A Gaussian fit with sigma fixed to 1 is overlaid on the histogram in Fig.~\ref{compare_alma_histogram}. 
{If we limit this comparison to the 136 ALMA sources brighter than 4~mJy we find a small (0.3$\sigma$) average flux overestimate by ALMA, and the (uncertainty normalized) scatter rises only to 1.05. There are two $>3\sigma$ outliers, out of 136 galaxies (1.5\%). The ALMA pre-selection is slightly biasing the comparison towards brighter ALMA fluxes for this subsample, by construction.}
This allows us to conclude  that our fluxes and flux uncertainties at 850~$\mu$m from the ``Super-deblending'' technique are well defined and correctly derived.

\section{High redshift dusty star forming galaxies candidates}
\label{high_z_candi}

FIR/(sub)mm data are widely used for searching dusty star forming galaxies at very high redshifts \citep[e.g.][]{Riechers2013Nature,Riechers2017,Zavala2017}.
Although hundreds of square degrees have been mapped at FIR/(sub)mm wavelengths, where the photometry allows detection of these sources with roughly fixed SFR threshold up to $z\sim 10$ if they existed, only a handful of sources have been spectroscopically confirmed to lie at $z>5$ \citep{Capak2011Nature,Walter2012,Riechers2017,Smolcic2015} and only three at $z>6$ \citep{Riechers2013Nature,Strandet2017,Zavala2017}.
The sparsity of these very high-z samples is likely not only due to the intrinsic rarity of massive dusty galaxies in the very early Universe, but also to missing detections at lower fluxes in heavily blended FIR/(sub)mm images. 
Thus, it is of interest to inspect our \superdb FIR/(sub)mm photometry to search for candidates of dusty star forming galaxies at the highest redshifts and comparatively lower luminosities.

\begin{figure}
\centering
\includegraphics[width=0.47\textwidth]{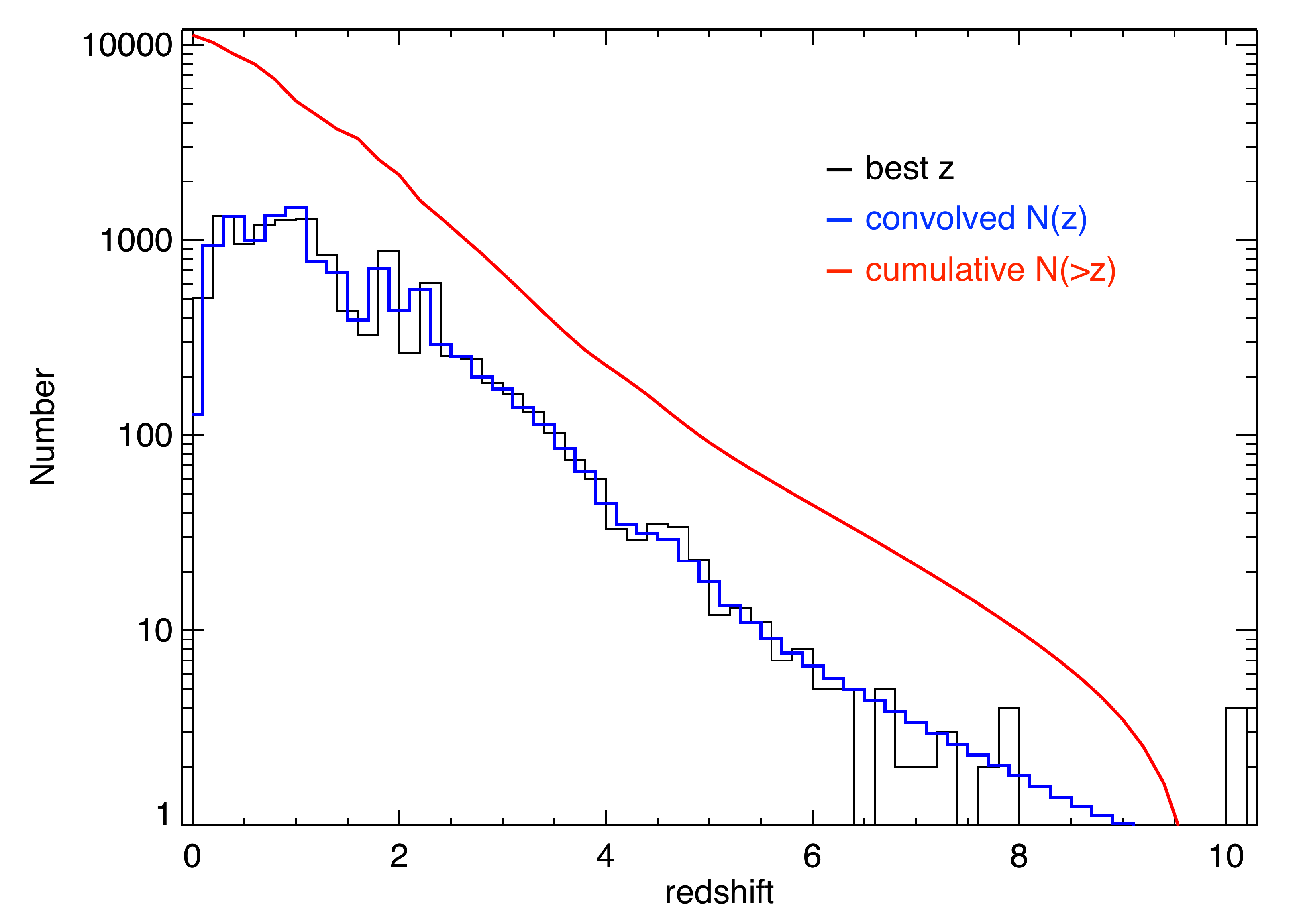}
\caption{%
	Redshift distribution of $\mathrm{S/N}_{\mathrm{FIR+mm}}>5$ sources. The black histogram shows the distribution of best fit redshifts from SED fitting. The uncertainty-convolved redshift distribution is shown as the blue histogram, while its cumulative distribution N(>$z$) is shown in red. 
\label{Nz_plot}%
}
\end{figure}

\begin{figure}
\centering
\includegraphics[width=0.46\textwidth]{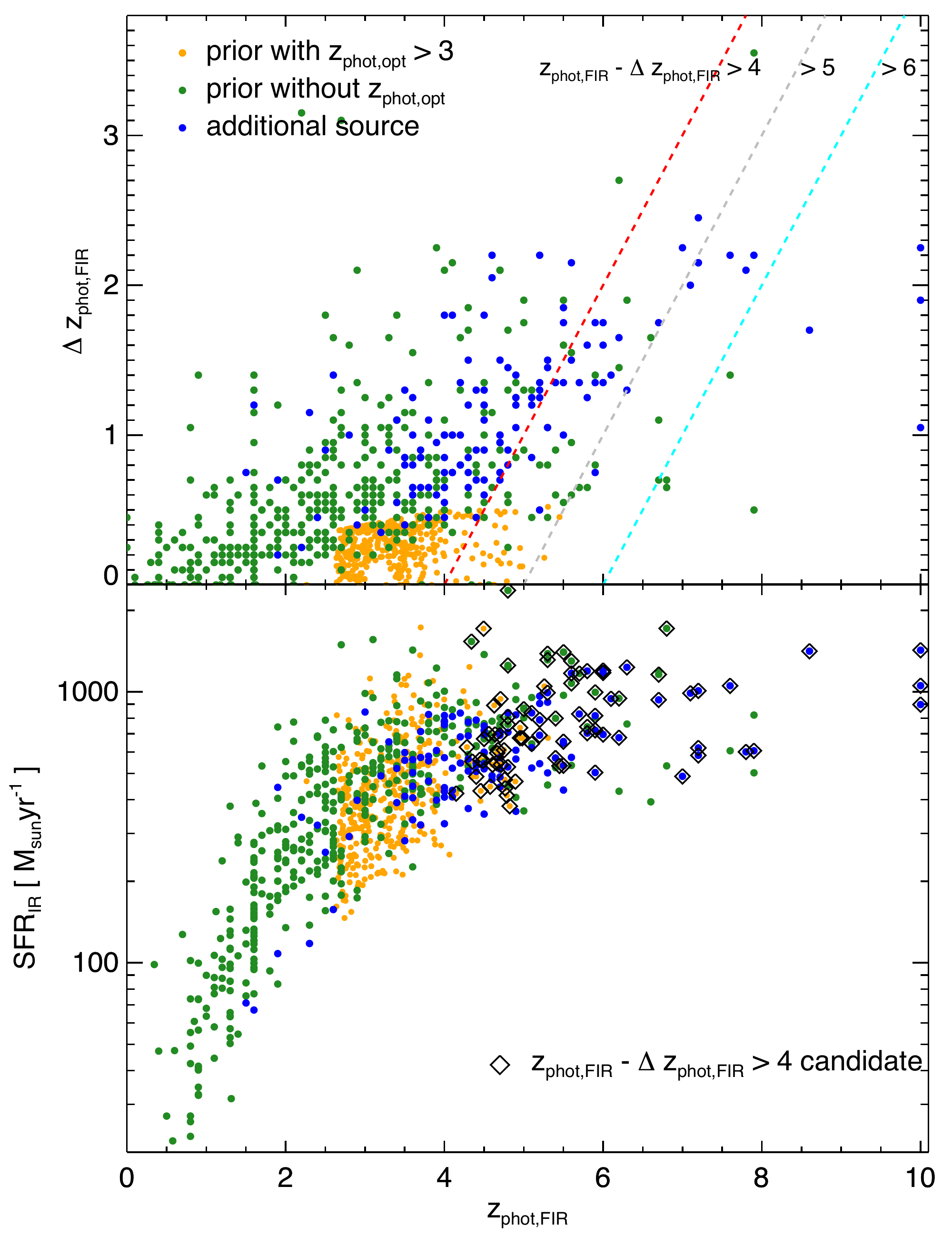}
\caption{%
	The selection of high redshift candidates, and SFR-redshift diagram. Sources below the red dashed line (i.e., $z_\mathrm{phot,FIR} - \Delta z_\mathrm{phot,FIR}>4$) are selected as our high redshift candidates.
\label{highz_selection}%
}
\end{figure}

In Figure~\ref{Nz_plot}, we present the redshift distribution of $\mathrm{S/N}_{\mathrm{FIR+mm}}>5$ detections in our catalog. 
There is a broad redshift peak at $z\sim 0.5$--1, and a very rapid, exponential-like decrease after $z>2$. Only a small tail of sources is reported as having $z>4$--6. Redshifts at $z>4$ are largely photometric, with a few exceptions discussed below. As shown by the red curve with cumulative numbers in Figure~\ref{Nz_plot}, only 2--3\% of all FIR/(sub)mm detections in this catalog are likely at redshift $z>4$ and possibly 0.5\% at $z>6$, while there are still few detections extending up to possibly even $z\sim 10$. Of course, as we get to these possibly highest redshifts the photometric redshift estimates have very large uncertainties, as discussed below. Hence the reality of this sample has to be better investigated. We concentrate in the {remainder} of the section on $z>4$ candidates among sources with $\mathrm{S/N}_{\mathrm{FIR+mm}}>5$.

\subsection{Candidates selection at $z>4$}

There are 3 main classes of candidate $z>4$ galaxies: 1) galaxies in our prior sample, and with an existing optically based photometric redshift: as in L18 we eventually re-derive a redshift from FIR SED fitting, constrained to be within 10\% of the optical $(1+z)$ to avoid catastrophic failures; 2) galaxies in our prior sample for which no optical photometric redshift exists, mostly because they were selected from radio and lack an obvious match to the $K_s$-selected catalogs: we only have a FIR redshift here; 3) additional sources added from the residual maps after the first-pass photometry -- also here we only have a FIR redshift, and some of these are possibly spurious sources or unresolved blends (see also L18). 

The FIR-based photometric redshifts from SED fitting are derived by considering the $\chi^2$ of the fit as a function of redshift, and applying a $\Delta\chi^2=2.5$ criterion (corresponding to the 90\% probability confidence) to determine the uncertainty of the redshift estimate following \citet{Avni1976}. 
High-z candidates are fitted using the \citet{Magdis2012SED} SB template for the dusty SF component, which in practice is GN20. This is appropriate regardless of their nature of SB or MS galaxies, given that even MS have fairly warm SEDs at $z>4$ \citep{Bethermin2015,Schreiber2017z4MS}. Still, the intrinsic SED shape as driven by dust temperature might have substantial variations among high-z galaxies. However, our approach is likely conservative as GN20 at $z=4$ has a relatively low dust temperature ($T_\mathrm{dust}=35\mathrm{K}$) compared to other $z>4$ dusty star forming galaxies \citep[e.g.][]{Smolcic2015,Riechers2017}. If using some of the latter as templates the redshifts would have been higher by up to 20-25\% (or about 5\%$\times(1+z)$, depending on the template). 
We were also careful of using full AGN \citep{Mullaney2011} and SF \citep{Magdis2012SED} decomposition of the SED only in presence of sufficient S/N in the photometry for allowing this, e.g., good detections in the mid-IR as well as FIR/(sub)mm ranges. When such a condition was lacking we considered only the dusty SF component, to avoid cases in which the hot AGN SED would spuriously produce very high redshift solutions, with low $\chi^2$, by dominating the SED fit. Further details of the dust temperature-redshift degeneracies are discussed in Section~\ref{z_dust_degeneracy}.

Fig.~\ref{highz_selection}-top 
shows the distribution of redshift errors for the three classes. It is obvious that the redshift uncertainty can grow to be fairly large for objects where only the FIR SED is used to constrain it. As a result, we decided to limit the discussion of high-redshift candidates to those objects satisfying $z-\Delta z>4$. 
Consequently, among significantly FIR-detected sources with existing $z_\mathrm{phot,opt}$
we obtain {31} sources with $z_\mathrm{phot,FIR} - \Delta z_\mathrm{phot,FIR}>4$ to class 1) above.
Among {420} (mainly radio) sources lacking an optical photometric redshift, we select {20} sources with $z_\mathrm{phot,FIR} - \Delta z_\mathrm{phot,FIR}>4$ to class 2) above. 
There are {126} additional sources from the SCUBA2 residual image that reach $\mathrm{S/N}_{\mathrm{FIR+mm}}>5$ (see Section~\ref{Section_Additional_Sources_In_Residual}), and {34} of them have $z_\mathrm{phot,FIR} - \Delta z_\mathrm{phot,FIR}>4$ and are in class 3). 
In total, we have
{85} sources with $z_\mathrm{phot,FIR} - \Delta z_\mathrm{phot,FIR}>4$ as our final sample of high redshift candidates.
In panel (1)--(3) of Fig.~\ref{highz_4example}, we present examples that are representing the different classes of candidates: (1) prior with $z_{\rm phot, NIR}$, (2) radio prior without $z_{\rm phot, NIR}$, (3) additional source with IRAC counterpart.

Note that no additional candidates with $z_\mathrm{phot,FIR} - \Delta z_\mathrm{phot,FIR}>4$ are selected from the SPIRE residual image, which are typically at a lower redshifts. The relative sensitivity of the SCUBA2 850~$\mu$m map is substantially higher than that of any SPIRE band for $z>4$ star forming galaxies. Also, no further residual sources candidates at $z>4$ are found from AzTEC and MAMBO maps. These bands were fitted after the SCUBA2 residual sources were already added. They generally support the reality of the SCUBA2 residual sources.

\subsection{Redshift-dust temperature degeneracy}
\label{z_dust_degeneracy}

Using the far-IR colors of a galaxy to determine its redshift (in absence of $z_{\rm spec}$) is known to suffer from the so called $T_{dust}-$redshift degeneracy. In particular the same colors could be fit by a colder template placed at lower redshift or by a warmer template placed at higher-z. To quantify this degeneracy, but also to determine how the uncertainties in optical $z_{\rm phot}$ affect the derivation of the dust temperature, or similarly of the mean radiation field, $<U> = L_{IR}/M_{dust}$, we perform the following simulation. We build DL07 models \citep{Draine2007SED} with representative $\gamma$, $q_{\rm PAH}$, $U_{\rm min}$ and $<U>$ parameters following \citet{Magdis2012SED}, and calculate flux densities at MIPS 24~$\mu$m, PACS, SPIRE and SCUBA2 850~$\mu$m bands by placing the template at a wide range of fixed redshifts ($z_{\rm or} =0-$6 with a step of 0.01). We then performed SED fitting using the full suit of DL07 models fixing the redshift first at $z_{\rm max}= z_{\rm or} + Dz\times (1+z_{\rm or})$ and then at $z_{\rm min} = z_{\rm or}-Dz \times (1+z_{\rm or})$ respectively, where $Dz = 0.03$, corresponding to the average photo$-z$ uncertainty in the COSMOS field \citep{Laigle2016}. Thus, at each redshift we derive three $<U>$ values: $U_{z_{or}}$, $U_{z_{min}}$ and $U_{z_{max}}$, and quantify the impact of the photo$-z$ uncertainty in $<U>$ by considering $DU = ( U_{z_{or}} - U_{z_{min}} ) / U_ {z_{or}}$ and $( U_{z_{or}} - U_{z_{max}} ) / U_{z_{or}}$. 
Our analysis yields an offset between the input and the extracted $<U>$ with the case of $z_{\rm min}$ ($z_{\rm max})$ systematically overestimating (underestimating) the true $<U>$ by 15\%. 
{We conclude that the uncertainty in $z_{\rm phot,FIR}$ introduces an extra small ( at least for the case of $\Delta z=0.03\times (1+z)$ ) uncertainty of 15\% in the determination of $<U>$.}
Repeating the process for the case of a MBB with a single $T_{dust}$ and fixed $\beta=1.8$, yields an uncertainty in the derived $T_{dust}$ of $\sim5$\%.

Furthermore, both MS and SB galaxies at any redshift are expected to exhibit a range of intrinsic $<U>$ values ($\Delta U$). Consequently, $\Delta U$, is expected to introduce an uncertainty in the far-IR based $z_{\rm phot,FIR}$ ( $\Delta z_{phot,FIR} =  z_{\rm spec} - z_{\rm phot,FIR} $). In order to quantify $\Delta z_{phot,FIR}$, we first need to adopt a reasonable value for $\Delta U$. Since $<U>$ $\propto$ $L_{\rm IR}/ M_{dust} \propto L_{\rm IR}/ (M_{gas} \times Z) \propto SFE/Z$ (e.g., \citealt{Magdis2012SED}), where Z is the gas phase metallicity, we can estimate the intrinsic range of $<U>$ in terms of variations of SFE and Z within and outside the MS. 
Assuming the various relations reported in the literature between SFE, distance from the Main Sequence ($\Delta \rm MS$) and Z (e.g., \citealt{Magdis2012SED}, \citealt{Sargent2014}, \citealt{Tacconi2017}), we estimate an intrinsic variation of $\Delta U$ $\approx$ 0.1-0.2$ dex$ for MS galaxies. This is agreement with the empirically determined 0.2$dex$ variation reported by \citet{Magdis2012SED} for local normal galaxies. Performing simulations as outlined above, we find that $\Delta U$= 0.1$-$0.2~dex corresponds to $\Delta z_{phot,FIR} \approx 0.06 - 0.1\times(1+z)$. We note that a similar estimate is reached for SBs at any redshift, assuming a range in gas depletion time scale of 50 to 200 Myrs. Thus, we conclude that the intrinsic variation of the shape of the far-IR SED of both MS and SB galaxies at any redshift introduces an intrinsic uncertainty of $\Delta z_{\rm phot,FIR}$ $\approx$0.06$-$0.1$\times(1+z)$, that should also be taken into account along with the uncertainties introduced by the photometric errors and the varying photometric coverage.

\subsection{An interesting case of possible lensing}

\begin{figure*}
\centering
\includegraphics[width=0.55\textwidth]{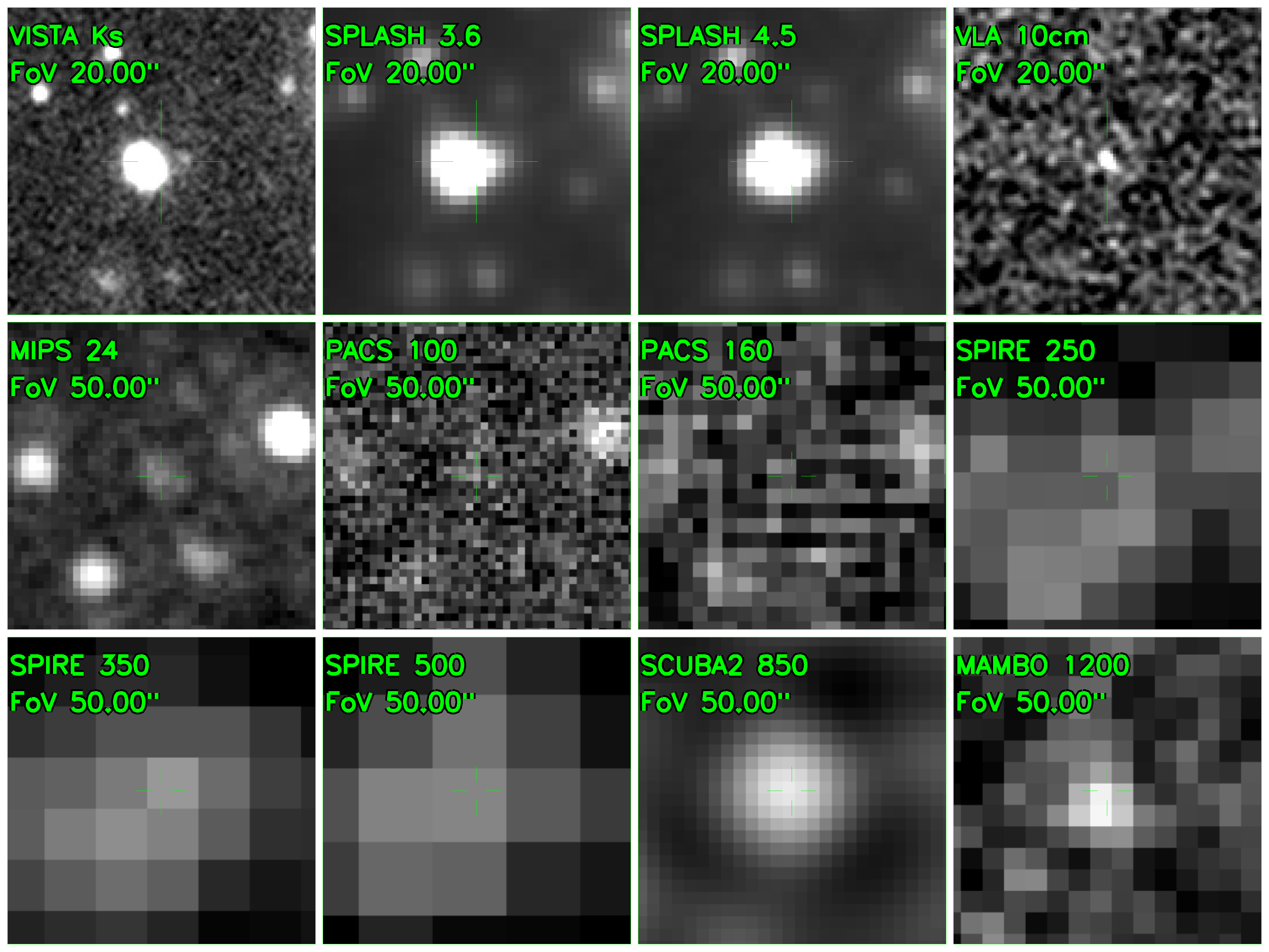}
\includegraphics[width=0.41\textwidth]{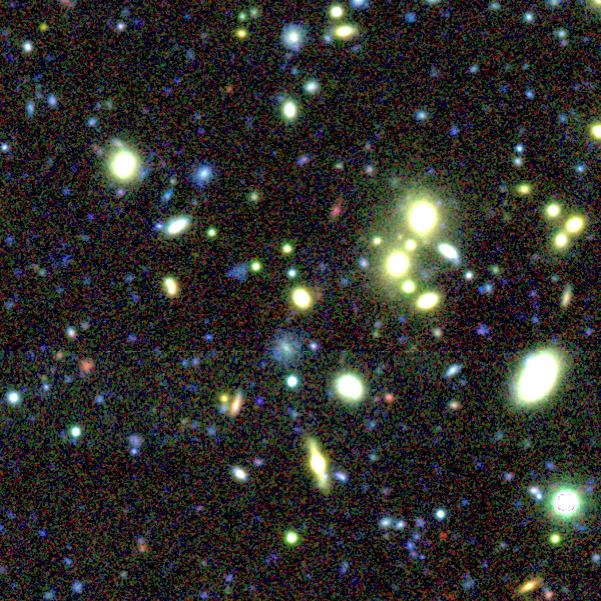}
\includegraphics[width=0.45\textwidth, trim={0.6cm 5cm 1cm 3.5cm}, clip]{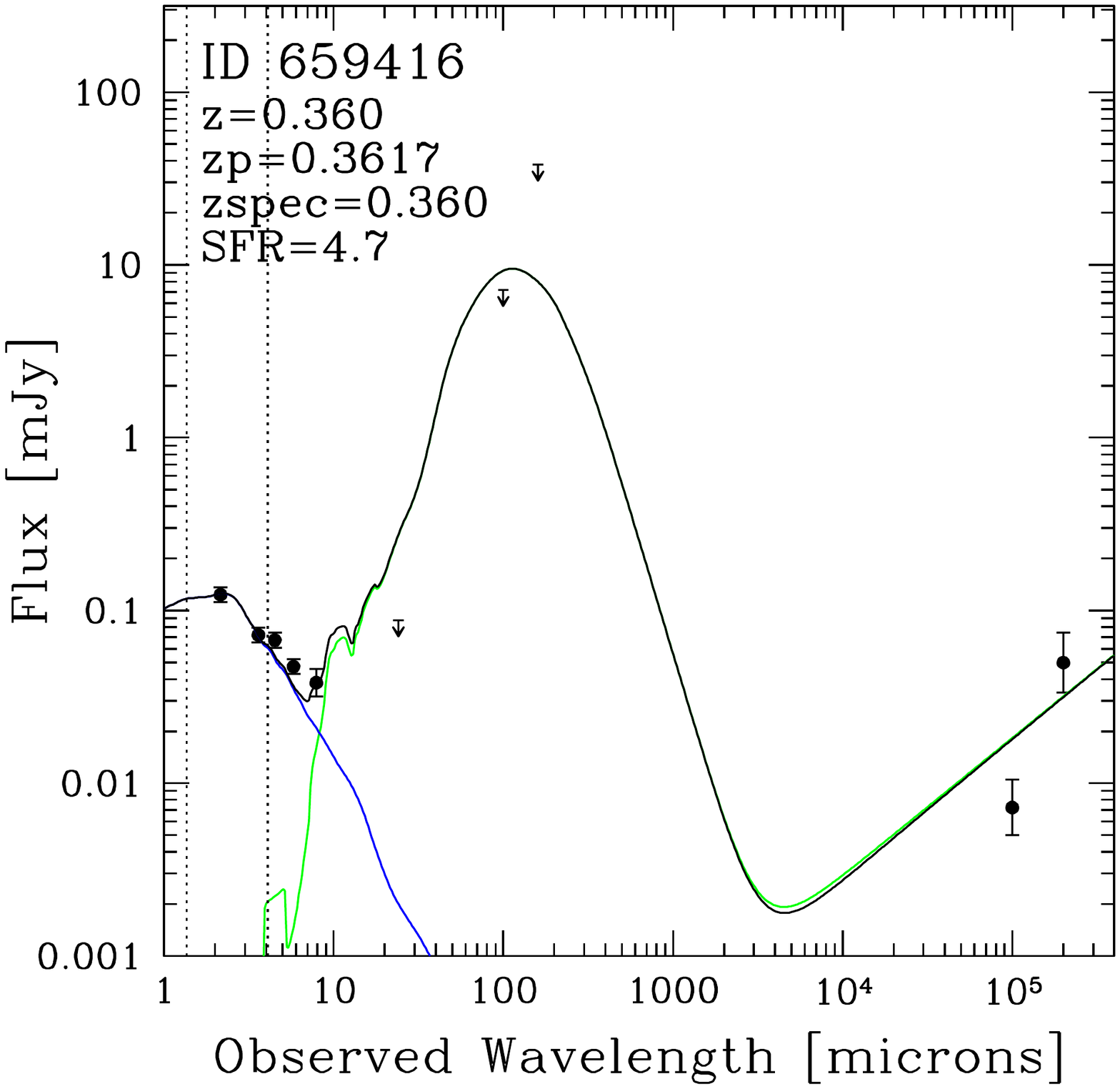}
\includegraphics[width=0.45\textwidth, trim={0.6cm 5cm 1cm 3.5cm}, clip]{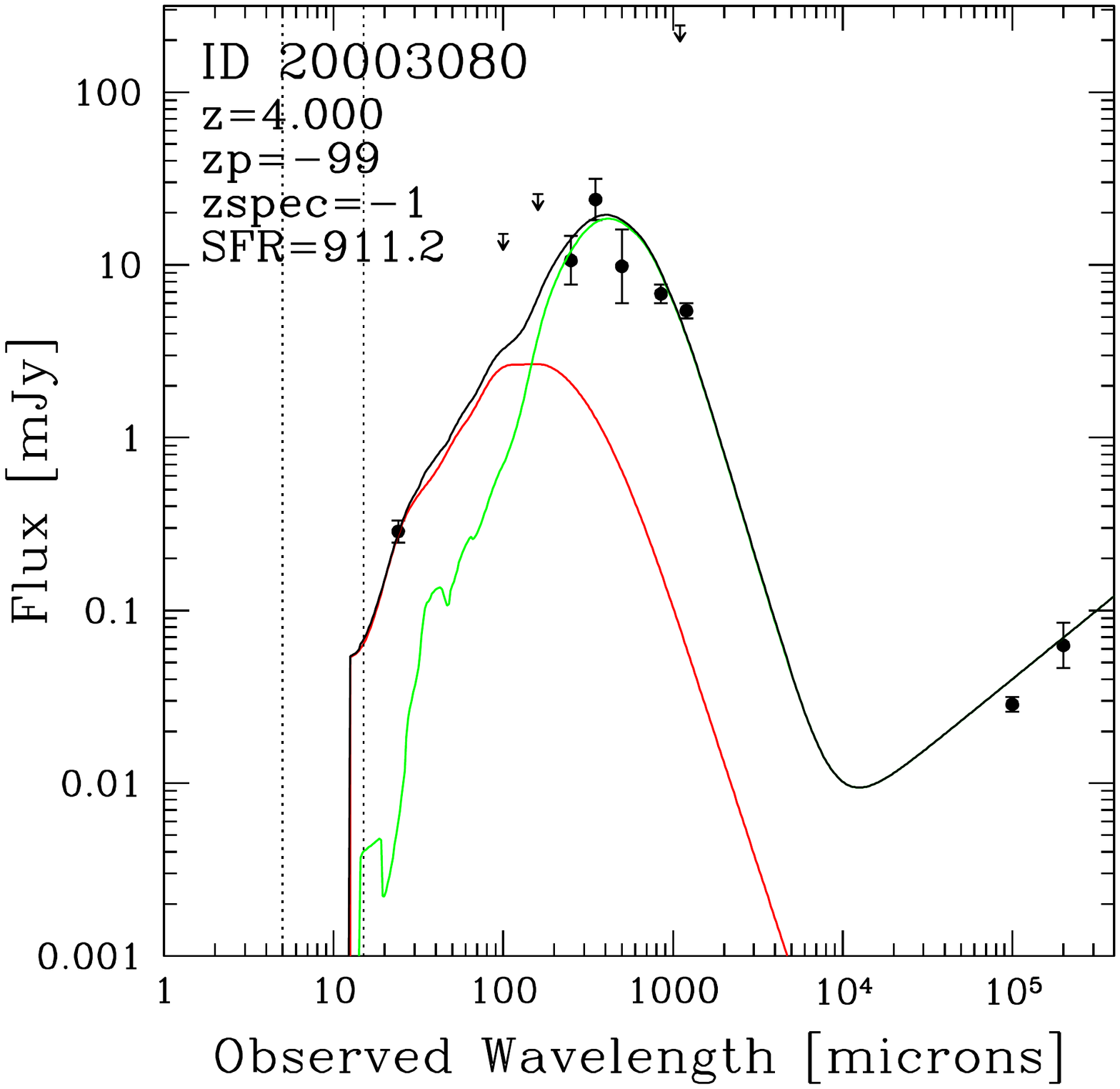}
\caption{%
	A likely case of a gravitationally lensed dusty galaxy ID20003080 (also known as Cosbo-7 in \citealt{Bertoldi2007}, AzTEC/C160 in \citealt{Aretxaga2011}).
    We show the multi-band cutouts centered on this source in panel (1), and a $90''\times90''$ color image (B, z, and $K_s$-bands) in panel (2).
    Panel (3)-(4) show SEDs of the $z=0.36$ elliptical ID659416 and the lensed ID20003080.
\label{lensing}%
}
\end{figure*}

In some cases the high-redshift candidates are interestingly coincident with previously existing low-redshift priors in our sample that were discarded for fitting at some longer wavelengths, but where a strong detection was later found in our procedure (often a residual source). 
This is similar to the case of source ID20003080 (AzTEC/C160) shown in Fig.~\ref{lensing}. 
This source is coincident with an early type galaxy with known spectroscopic redshift $z=0.36$ that is associated to a mass-selected prior (ID659416), which is fitted at PACS bands while excluded at longer wavelengths because it is obviously too faint. Thanks to its radio detection, we already knew in this case of the presence of the very close (1.2$''$ separation)
 radio source ID20003080 (i.e., Cosbo-7 in \citealt{Bertoldi2007}, AzTEC/C160 in \citealt{Aretxaga2011}) that is afterwards solidly detected at SCUBA2 850~$\mu$m ($\mathrm{S/N_{\rm 850\mu m}=8.2}$) and MAMBO 1.2~mm ($\mathrm{S/N_{\rm 1.2mm}=7.4}$). 
ID20003080 appears to be a high redshift galaxy that is aligned with (and perhaps gravitationally lensed by) an early type galaxy at $z=0.36$, likely part of a group or poor cluster of galaxies at the same redshift (see RBG image in Fig.~\ref{lensing} top-right panel).
We obtain a photometric redshift $z=4.0\pm 0.6$ by fitting 24~$\mu$m to radio photometry.

Other very similar cases, with the background sources at possibly even higher redshift, are ID85004261 and ID20003117 (see Table~\ref{Table_all_candidates} and Appendix~\ref{Section_highz}).
Investigating their nature in detail is left to future works.

\begin{figure*}
	\centering
	\includegraphics[width=0.28\textwidth]{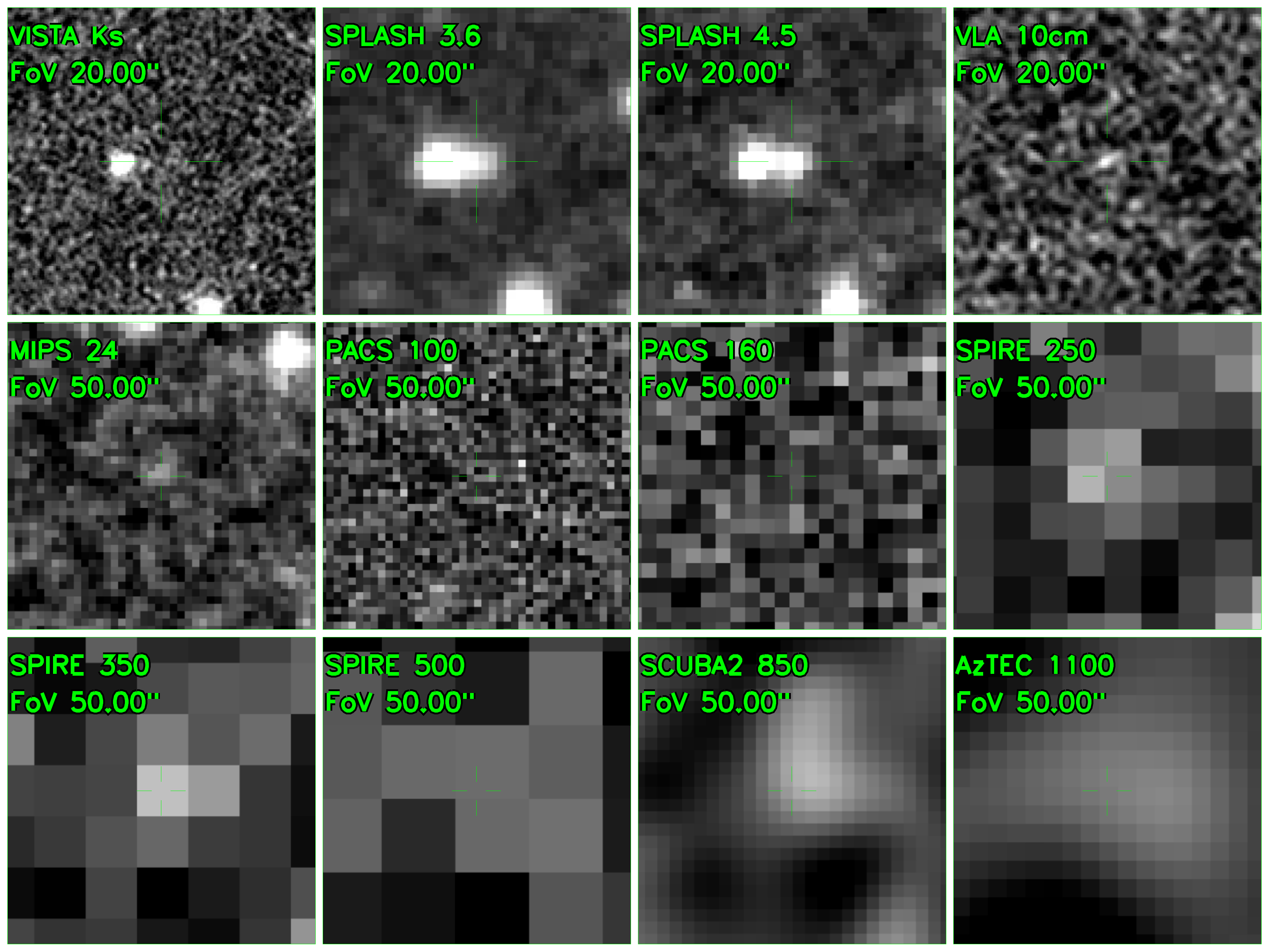}
	\includegraphics[width=0.21\textwidth, trim={0.6cm 5cm 1cm 3.5cm}, clip]{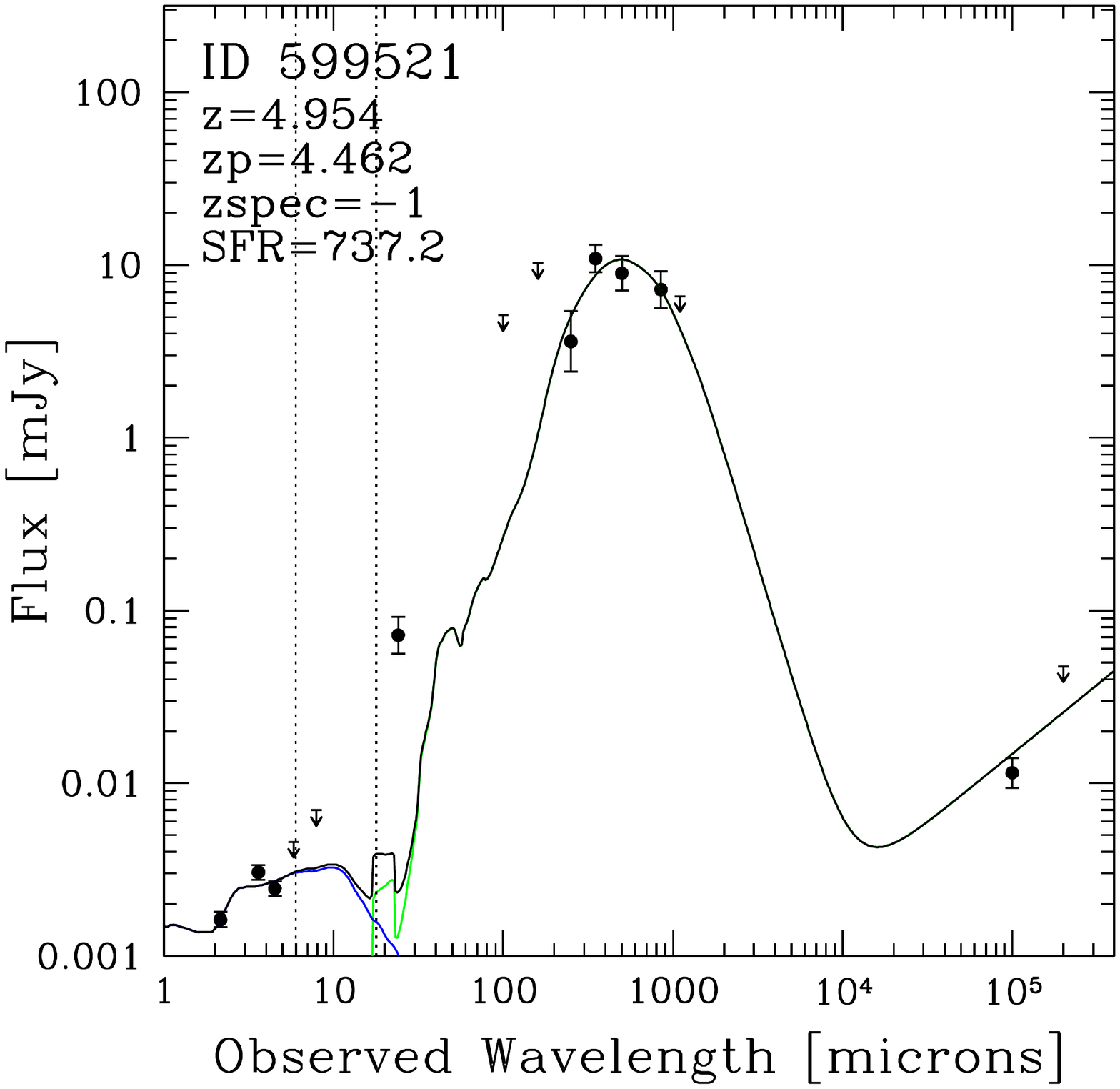}
    \includegraphics[width=0.28\textwidth]{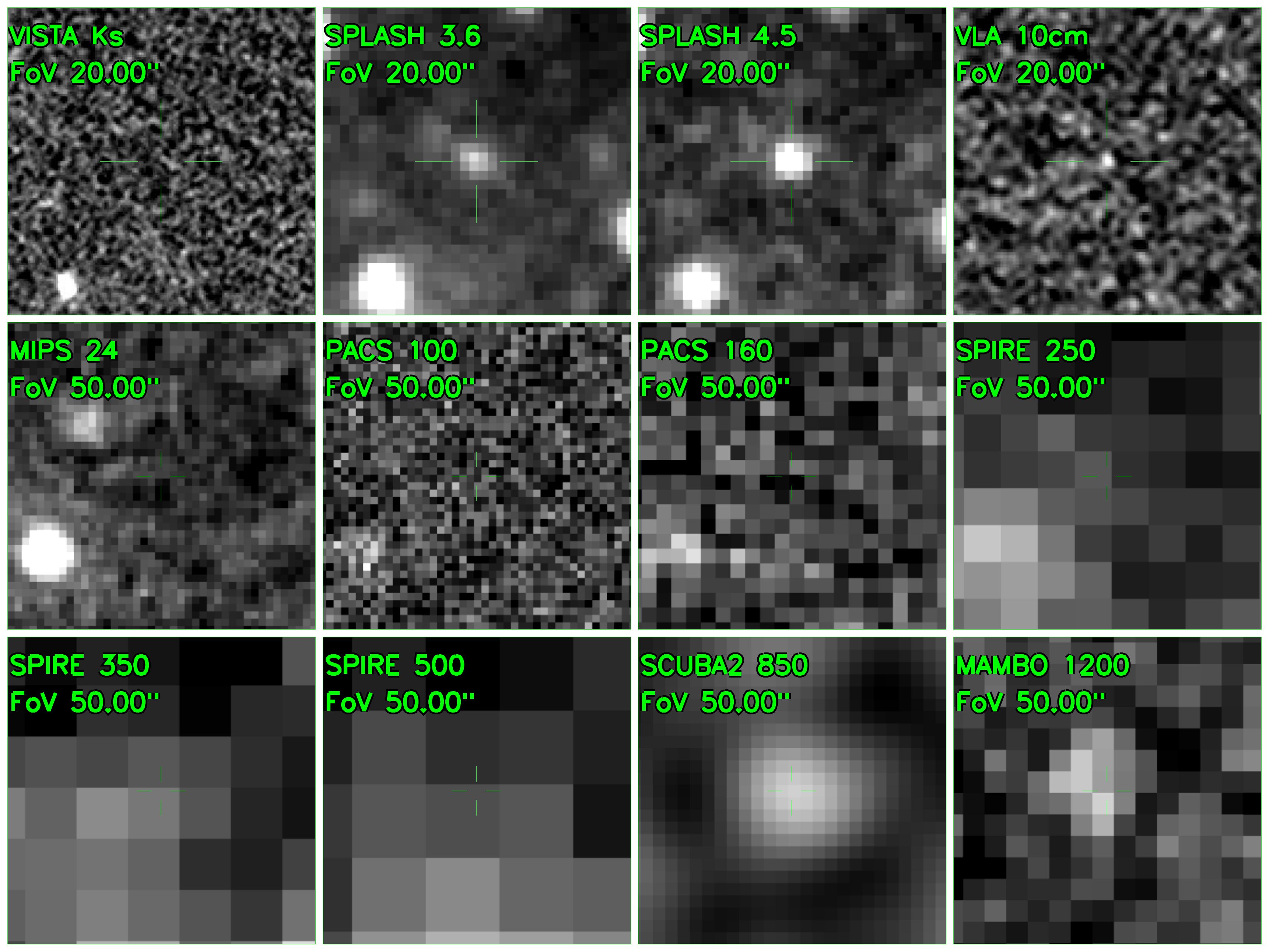}
	\includegraphics[width=0.21\textwidth, trim={0.6cm 5cm 1cm 3.5cm}, clip]{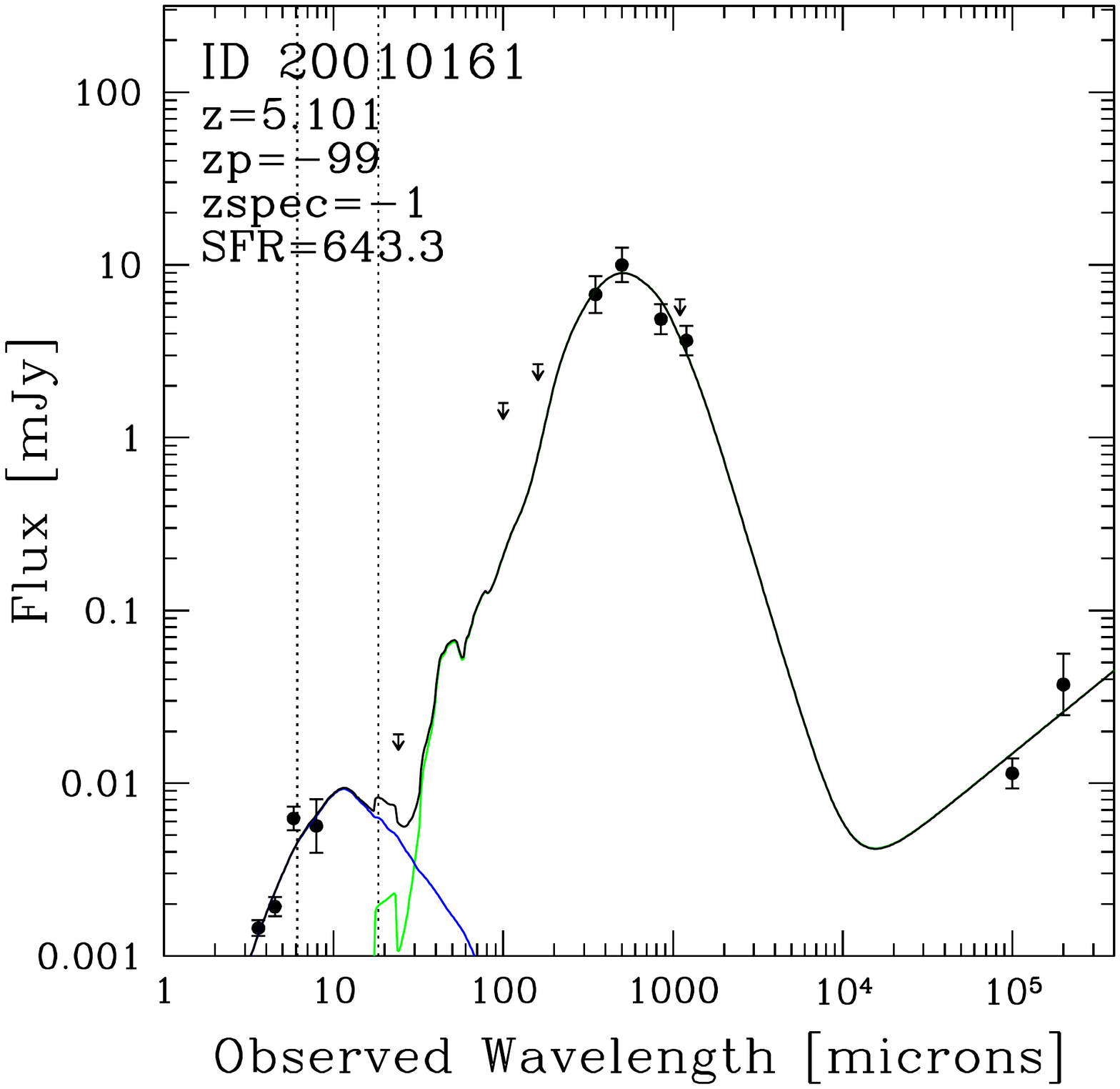}
    
    \includegraphics[width=0.28\textwidth]{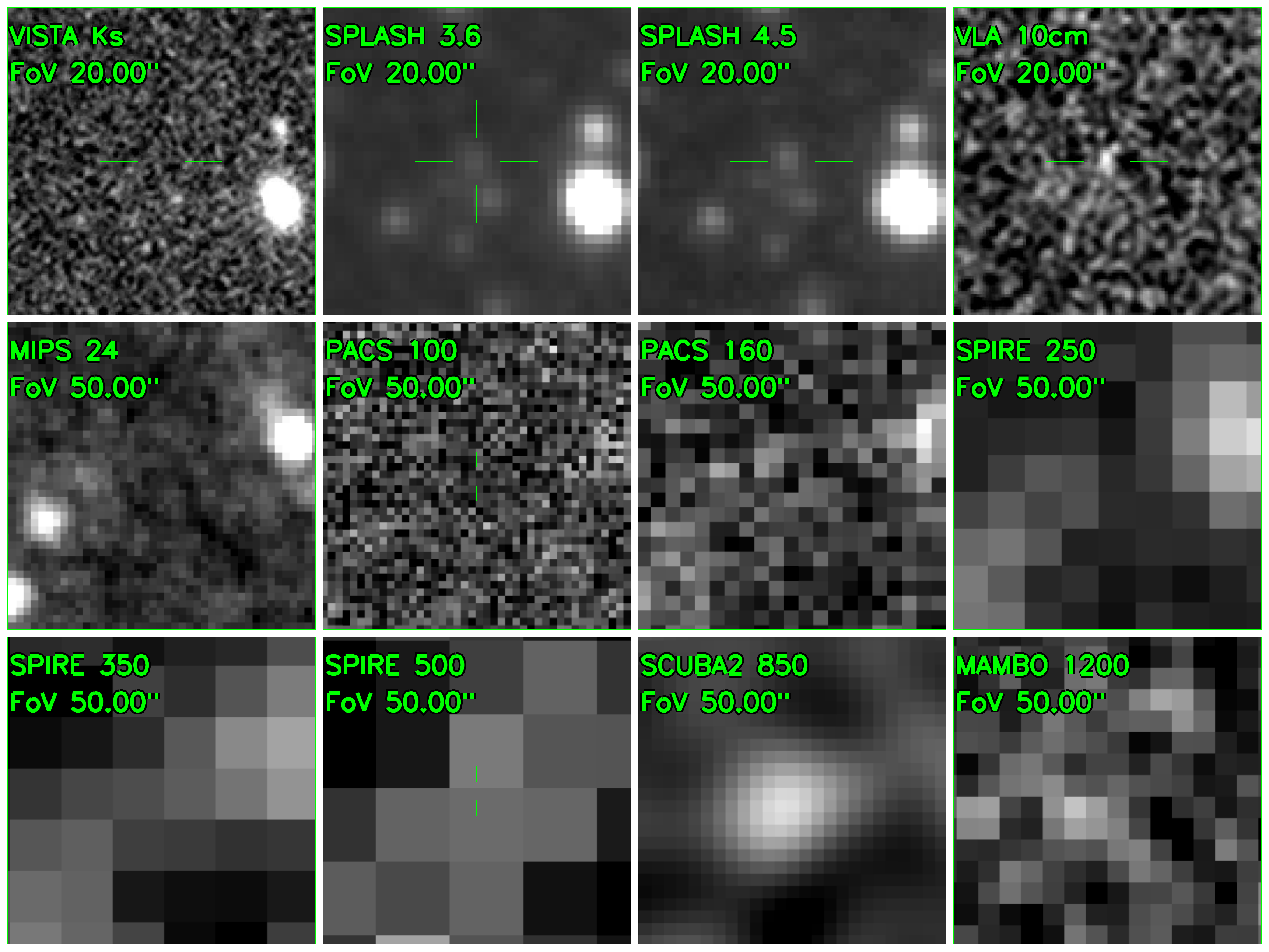}
	\includegraphics[width=0.21\textwidth, trim={0.6cm 5cm 1cm 3.5cm}, clip]{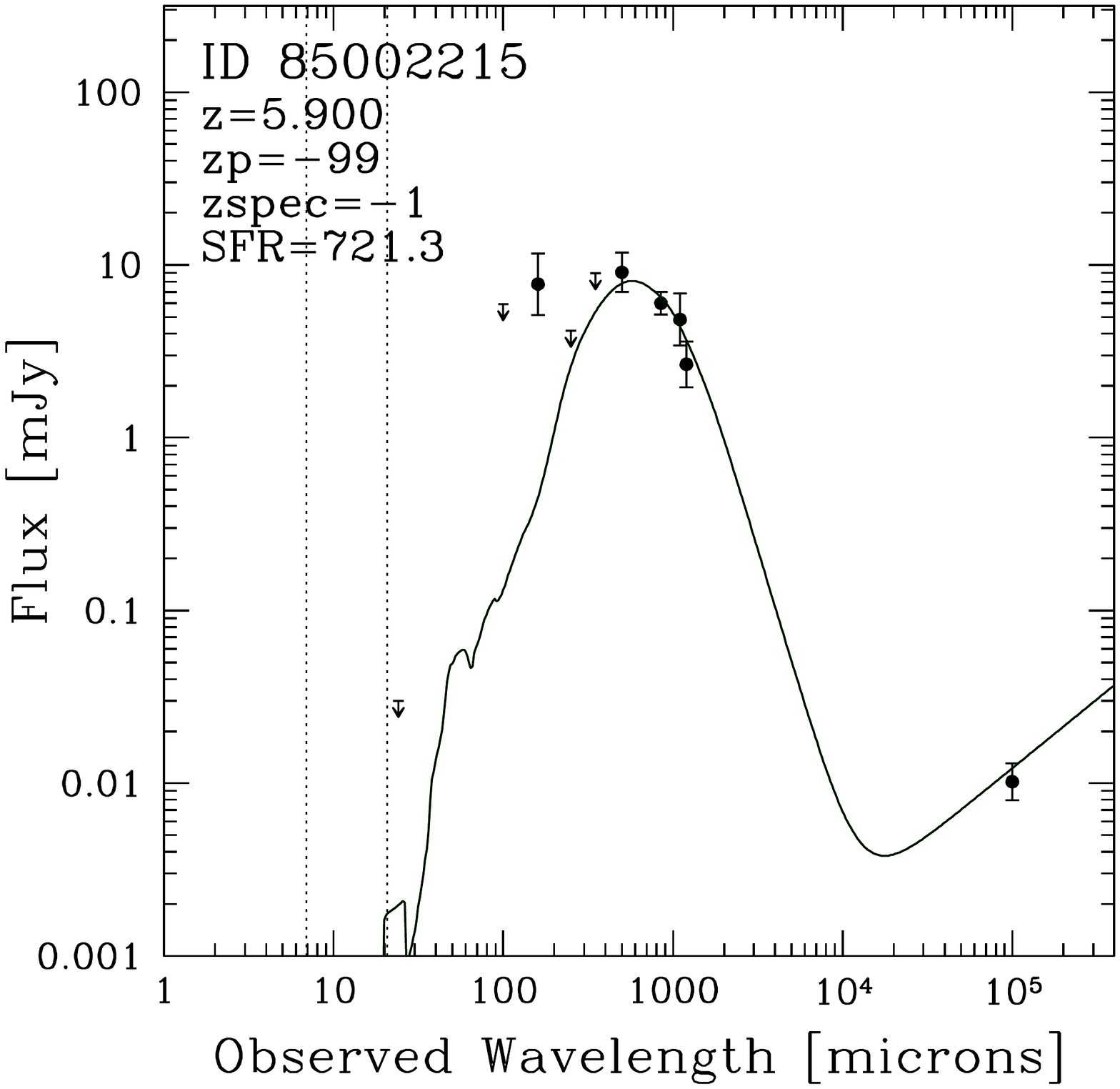}
     \includegraphics[width=0.28\textwidth]{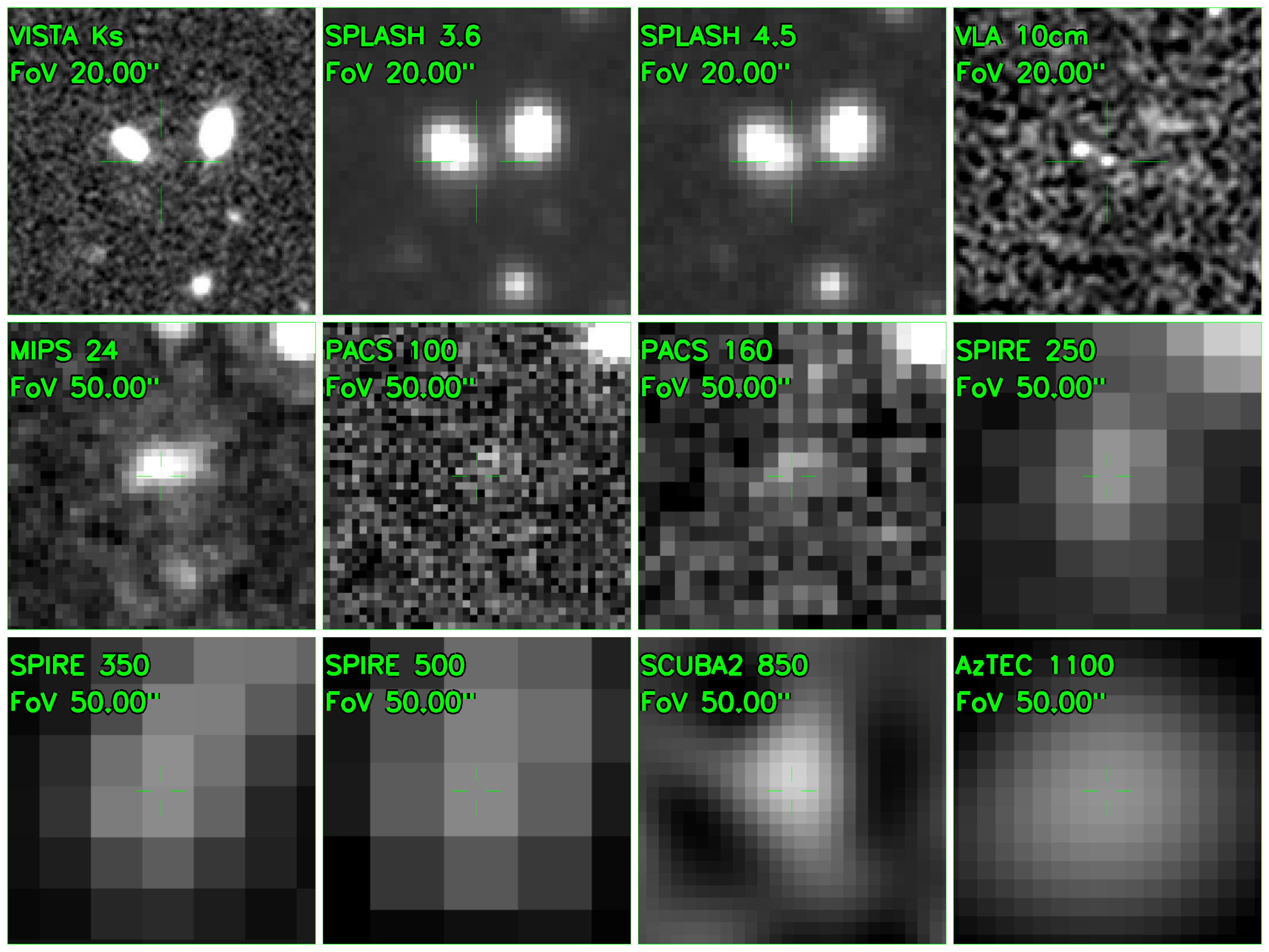}	        \includegraphics[width=0.21\textwidth, trim={0.6cm 5cm 1cm 3.5cm}, clip]{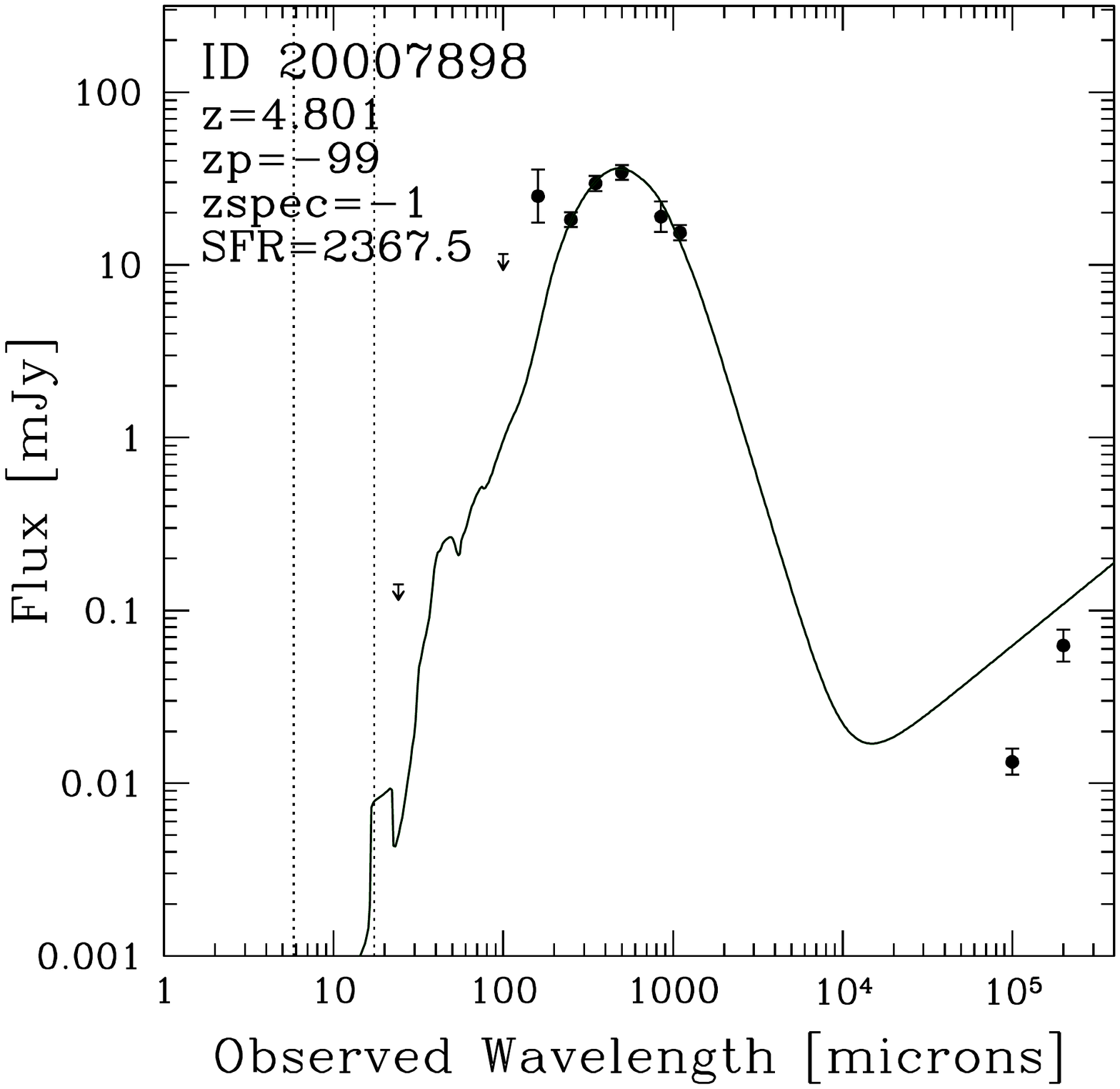}
	\caption{%
		Four examples of $z>4$ candidates. We show the multi-band cutouts on the left accompanying SED on the right. 
        The instrument used, wavelength (in unit of $\mu$m) and field of view (FoV) are shown in green text in each cutout. SEDs are fitted with a starburst-like template \citep[green curve,][]{Magdis2012SED} and a stellar component \citep[blue curve,][]{BC03}.
		\label{highz_4example}
		}
\end{figure*}

\subsection{AGN components at $z>4$}

We find {17} sources among our {\em robust} sample at $z>4$ that require an AGN component for their SED fitting (see Table~\ref{AGN_list}). 
The criterion adopted to conclude this requires that the AGN flux normalization divided by its total uncertainty range ($S_{\rm AGN}/\sigma_{\rm AGN}$) is larger than two. This is supported by visual inspection, and appear to be reliably returning reasonable results.
{In this AGN sample, there are 8 sources with $S_{\rm AGN}/\sigma_{\rm AGN}>3$ that we consider to be solid candidates, and the remaining 9 sources with $2<S_{\rm AGN}/\sigma_{\rm AGN}<3$ to be tentatives.}
The bolometric luminosities of the AGN is ranging over quite remarkable $L_{bol}^{\rm AGN}\sim10^{46-47}$~erg~s$^{-1}$, fully in the QSO regime. 
We show their multi-band cutouts and SEDs in Appendix~\ref{Section_highz}, where the AGN component is marked by the red curve in the SED panel.

For the {85} $z>4$ candidates, the sum of {SFR is $7.1\times10^4{\rm M_\odot~yr^{-1}}$}. The bolometric AGN luminosity contained among AGN detections adds up to $L_{bol}^{\rm AGN}=10^{48}$~erg~s$^{-1}$, corresponding to an integrated black hole accretion rate of $\mathrm{160~M_\odot~yr^{-1}}$.
The ratio of BH accretion rate to SFR is thus $2\times10^{-3}$ close but higher than the universal ratio discussed in \citet{Mullaney2012ApJL} despite the IR-selection that should favor star formation. This might suggest some evolution towards stronger AGN activity in these sources, that should be confirmed with further studies.
The number density of the 17 objects in COSMOS (assuming volume within $4<z<5$) is  $\phi \sim 8\times 10^{-7}$ Mpc$^{-3}$dex$^{-1}$, which matches reasonably well with the X-ray luminosity function extrapolations from \citet{Vito2018}, assuming a bolometric correction of $L_{bol} = 25 \times L_{X,2-10keV}$. 
Our AGN detections are predicted to have X-ray fluxes 2.9--31.1 $\times 10^{-15}$ erg cm$^{-2}$ s$^{-1}$ at 0.5--2~keV, in case of no obscuration, which would make them all detectable with the limiting depth of available X-ray imaging data ($2.2\times 10^-16$ erg cm$^{-2}$ s$^{-1}$ at 0.5--2~keV) in the Chandra-COSMOS legacy survey \citep{Civano2016}. 
However, only two sources ID10008348 and ID10137954 are cross-matched with the point source catalog of \citet{Civano2016}. 
These  X-ray sources show very high obscuration with $L_{bol}/L_{X,2-10keV}\sim 300$, even higher than the X-ray bolometric correction factors of Compton-Thick AGNs in \citet{Brightman2017}. 
The low X-ray detection rate and large $L_{bol}/L_X$ ratio clearly suggest that our IR-selected AGNs are heavily obscured in the X-rays, consistent with current understanding of high redshift populations \citep{Vito2018}.

\begin{table}
	\centering
		\caption{High redshift candidates that fitted with a significant AGN component}
        \label{AGN_list} 
\begin{tabular}{ccccccc}
\hline
  {ID} &
  {$z_\mathrm{phot,opt}$} &
  {$z_\mathrm{spec}$} &
  {$\mathrm{S/N_{FIR+mm}}$} &
  {$z_\mathrm{phot,FIR}$} &
  {$L_\mathrm{IR, SF}$} &
  {$L_\mathrm{bol, AGN}$} \\
  &  
  &
  &
  &
  &
   {$\times10^{12}~L_{\odot}$} &
   {$\times10^{12}~L_{\odot}$}  \\
\hline
  223720 & 4.19 & -- & 6.9 & $4.66 \pm 0.28$ & $5.5 \pm 0.3$ & $7.8 \pm 2.4$\\
  339785 & 4.92 & -- & 5.2 & $5.45 \pm 0.45$ & $5.3 \pm 0.2$ & $54.6 \pm 17.3$\\
  347052 & 4.34 & -- & 5.6 & $4.82 \pm 0.14$ & $3.8 \pm 0.4$ & $21.3 \pm 7.9$\\
  551174 & 4.21 & -- & 5.3 & $4.68 \pm 0.16$ & $6.1 \pm 0.3$ & $6.9 \pm 2.7$\\
  556890 & 4.49 & 4.63 & 9.6 & 4.63 & $8.9 \pm 0.1$ & $16.1 \pm 1.6$\\
  578482 & 4.32 & -- & 10.7 & $4.75 \pm 0.22$ & $6.1 \pm 0.4$ & $8.8 \pm 3.3$\\
  599184 & 4.41 & -- & 7.6 & $4.89 \pm 0.49$ & $4.7 \pm 0.4$ & $8.5 \pm 4.1$\\
  613818 & 4.47 & -- & 6.8 & $4.67 \pm 0.49$ & $5.9 \pm 0.8$ & $12.5 \pm 4.3$\\
  632541 & 4.29 & -- & 6.5 & $4.77 \pm 0.38$ & $4.4 \pm 0.5$ & $12.0 \pm 5.4$\\
  695002 & 4.31 & -- & 13.7 & $4.79 \pm 0.02$ & $4.1 \pm 0.2$ & $48.8 \pm 4.9$\\
  739920 & 4.42 & -- & 7.7 & $4.76 \pm 0.49$ & $4.7 \pm 0.3$ & $13.6 \pm 6.7$\\
  786213 & -- 	& 4.34 & 32.6 & 4.34 & $15.3 \pm 0.1$ & $5.1 \pm 0.7$\\
  965647 & 4.34 & -- & 6.2 & $4.58 \pm 0.19$ & $4.5 \pm 0.4$ & $5.6 \pm 2.7$\\
  10008348 & 4.16 & 4.45 & 5.2 & 4.45 & $5.6 \pm 0.1$ & $8.4 \pm 1.2$\\
  10015010 & 3.91 & -- & 8.4 & $4.35 \pm 0.18$ & $5.5 \pm 0.3$ & $11.0 \pm 2.9$\\
  10137954 & --   & 4.64 & 8.0 & 4.64 & $0.2 \pm 0.2$ & $34.9 \pm 2.4$\\
  10213589 & 4.66 & -- & 7.1 & $4.15 \pm 0.03$ & $4.2 \pm 0.2$ & $23.6 \pm 4.9$\\
\hline\end{tabular}

\end{table}

\subsection{Counterparts for candidates found in residual images}

\begin{table*}
	\centering
		\caption{High redshift candidates found among additional sources from the SCUBA2 residual map. We report here the result for our search of Spitzer IRAC and radio counterparts.}
        \label{Table_all_candidates} 
\begin{tabular}{|c|c|c|c|c|c|r|r|r|r|r|r|}
\hline
  \multicolumn{1}{|c|}{ID} &
  \multicolumn{1}{c|}{RA (J2000)} &
  \multicolumn{1}{c|}{Dec (J2000)} &
  {counterpart}\tablenotemark{a}  &
  {NAME}\tablenotemark{b}  &
  {Distance}\tablenotemark{c} &
  \multicolumn{1}{c|}{$\mathrm{S/N_{FIR+mm}}$} &
  \multicolumn{1}{c|}{$\mathrm{S/N_{850\mu m}}$} &
  {SFR}\tablenotemark{d} &
  {$\mathrm{\Delta SFR}$}\tablenotemark{e} &
  {$z_\mathrm{phot,FIR}$}\tablenotemark{f} &
  {$\Delta z_\mathrm{phot,FIR}$}\tablenotemark{g} \\
\hline
  85000019 & 150.24172 & 1.60795 & SPLASH, 3GHz & -- & $3.2''$ & 6.8 & 3.3 & 1194 & 309 & 5.8 & 1.3\\
  85000436 & 149.40348 & 1.73843 & 3GHz 				& -- & $1.9''$ & 5.7 & 4.2 & 981 & 205 & 5.4 & 0.9\\
  85000496 & 149.49761 & 1.77057 & -- 						& -- & -- & 5.3 & 4.3 & 702 & 324 & 6.0 & 1.8\\
  85000552 & 150.04723 & 1.78372 & -- 						& -- & -- & 5.5 & 3.3 & 534 & 104 & 5.5 & 1.4\\
  85000762 & 149.53400 & 1.85461 & -- 						& -- & -- & 6.4 & 3.8 & 528 & 131 & 4.8 & 0.8\\
  85000922 & 150.49660 & 1.90440 & SPLASH 			& -- & $2.2''$ & 6.2 & 3.8 & 1232 & 343 & 6.3 & 1.3\\
  85001050 & 149.98028 & 1.93565 & -- 						& -- & -- & 5.1 & 4.1 & 704 & 119 & 5.8 & 1.6\\
  85001505 & 149.46728 & 2.09288 & SPLASH 			& -- & $2.3''$ & 6.4 & 4.9 & 816 & 205 & 5.9 & 1.4\\
  85001571 & 150.02743 & 2.11300 & SPLASH 			& -- & $0.4''$ & 5.0 & 2.8 & 488 & 24  & 7.0 & 2.3\\
  85001674 & 150.30121 & 2.14766 & SPLASH, 3GHz & -- & $1.6''$ & 6.1 & 4.2 & 836 & 128 & 5.1 & 0.9\\
  85001756 & 149.87579 & 2.17826 & -- 						& -- & -- & 5.8 & 4.1 & 621 & 151 & 7.2 & 2.2\\
  85001929 & 150.10971 & 2.25753 & SPLASH 			& -- & $0.8''$ & 12.9& 6.8 & 692 & 151  & 5.2 & 0.5\\
  85001969 & 149.97064 & 2.31366 & ALMA & AzTEC/C71 & $3.0''$ & 5.5 & 4.5 & 606 & 103 & 7.9 & 2.2\\
  85002134 & 150.24795 & 2.38802 & -- 						& -- & -- & 5.7 & 3.3 & 850 & 107 & 10.0 & 1.1\\
  85002215 & 150.10063 & 2.33484 & SPLASH, 3GHz & AzTEC/C114 & $3.5''$ & 9.6 & 6.7 & 721 & 47  & 5.9 & 0.8\\
  85002966 & 149.42140 & 2.57688 & -- 						& -- & -- & 7.6 & 5.2 & 692 & 112 & 4.7 & 0.7\\
  85003151 & 150.02104 & 2.63323 & -- & AzTEC/C132 & -- & 5.5 & 3.3 & 678 & 154 & 6.2 & 1.7\\
  85004261 & 150.05635 & 2.57327 & ALMA, 3GHz 	& AzTEC/C10 & $2.6''$ & 8.3 & 5.8 & 1009 & 267 & 7.2 & 2.2\\
  85005253 & 150.41618 & 1.90769 & -- 						& -- & -- & 5.2 & 4.3 & 1413 & 223 & 8.6 & 1.7\\
  85005277 & 150.57793 & 1.93306 & -- 						& -- & -- & 5.0 & 3.3 & 1425 & 316 & 10.0 & 2.3\\
  85005285 & 150.68263 & 1.95067 & 3GHz 				& -- & $4.9''$ & 5.1 & 3.1 & 1053 & 509 & 7.6 & 2.2\\
  85005338 & 149.96455 & 1.98236 & -- 						& -- & -- & 5.2 & 3.3 & 898 & 144 & 10.0 & 1.9\\
  85005422 & 149.98861 & 2.09413 & -- 						& -- & -- & 5.1 & 3.4 & 601 & 57  & 7.8 & 2.1\\
  85005464 & 149.68279 & 2.09387 & -- 						& -- & -- & 5.5 & 2.7 & 570 & 152 & 5.4 & 1.4\\
  85005517 & 150.64576 & 2.14272 & -- 						& -- & -- & 5.0 & 2.5 & 934 & 272 & 6.7 & 1.8\\
  85005620 & 150.70247 & 2.25888 & -- 						& -- & -- & 5.1 & 2.6 & 989 & 312 & 7.1 & 2.0\\
  85005670 & 150.60669 & 2.31664 & -- 						& -- & -- & 5.3 & 2.9 & 828 & 136 & 5.7 & 1.4\\
  85005722 & 150.61282 & 2.41777 & -- 						& -- & -- & 5.1 & 2.9 & 1171 & 267 & 5.6 & 1.5\\
  85005759 & 149.78323 & 2.40500. & -- 						& -- & -- & 5.1 & 3.2 & 504. & 93 & 5.9 & 1.8\\
  85005769 & 150.36338 & 2.41572 & -- 						& -- & -- & 5.9 & 3.2 & 993 & 287 & 5.3 & 1.1\\
  85005926 & 149.95379 & 2.56544 & SPLASH, 3GHz & -- & $2.0''$ & 7.0 & 2.6 & 656 & 114 & 5.5 & 1.0\\
  85005933 & 149.75172 & 2.58143 & -- 						& -- & -- & 5.2 & 2.7 & 582 & 168 & 7.2 & 2.5\\
  85005963 & 150.69766 & 2.60405 & -- 						& -- & -- & 7.5 & 2.7 & 1176 & 574 & 6.0 & 1.6\\
  85006141 & 150.45836 & 2.81048 & 3GHz 			& -- & $1.8''$ & 8.3 & 2.5 & 1199 & 473 & 6.0 & 1.4\\
\hline\end{tabular}

\begin{minipage}{\textwidth}
    
\flushleft

$^\mathrm{a}$ 
    Counterpart image. SPLASH: P. Capak; 3~GHz: \citet{Smolcic2017}; ALMA: 1.2~mm continuum image (M. Aravena et al. in prep).%

$^\mathrm{b}$ 
    Reference name of the counterpart \citep{Aretxaga2011}.%

$^\mathrm{c}$ 
    Distance of additional source to its counterpart. %
    
$^\mathrm{d, e}$ 
    SFR \& $\mathrm{\Delta SFR}$: median fit and uncertainty of SFR based on FIR+(sub)mm SED fitting, Chabrier IMF \citep{Chabrier2003}. %
    
$^\mathrm{f, g}$ 
    {$z_\mathrm{phot,FIR}$} \& {$\Delta z_\mathrm{phot,FIR}$}: the photometric redshift and uncertainty based on FIR+(sub)mm SED fitting. %

\end{minipage}

\end{table*}

We list the {34} additional high-z candidates in Table~\ref{Table_all_candidates}.
Given that the positions of these additional sources are blindly extracted from the SCUBA2 residual image, that has a beam $\mathrm{FWHM=11''}$, we visually searched for counterparts in SPLASH, VLA 3~GHz and ALMA 1.2~mm (Aravena et al. in prep.) images with a tolerance of $5''$, and set coordinates of their counterparts as their final positions. 
We find that {13} additional sources have a well-defined counterpart in SPLASH and/or VLA and/or ALMA images. The counterparts are presented in cutouts in Appendix~\ref{Section_highz} and listed in Table~\ref{Table_all_candidates}. These associations are unlikely to happen by chance, and appear robust: 
{by cross-matching to the COSMOS2015 catalog for SPLASH and to our VLA catalog, we would expect only 0.03 chance coincidences per $5''$ radius aperture,
down to the flux limits reached in these probes. This suggests at most $\sim 1$ spurious association among the {34} additional candidates. The identification of these {13} sources with counterparts and solid detections at multiple FIR/(sub)mm wavelengths suggest that the spurious fraction among the additional candidates is quite contained.}

Like for the other candidates, we fit their photometry at 100--1200$\mu$m and 3~GHz using the GN20 template and an evolving $q_\mathrm{IR}$ without an AGN component. %
The 3~GHz photometry is taken from their counterparts in the VLA 3~GHz image, while for candidates without any counterpart we do not fit the 3~GHz photometry.
Note that there are {3} additional sources that result in a best fit $z=10$, the highest value that we allow. These {3} sources are only detected at (sub)mm wavelengths while their SEDs are weakly constrained by $Herschel$. Given the {redshift errors $\Delta z_\mathrm{phot,FIR}=1.1-2.3$}, it is quite plausible that these sources are at more reasonable, lower redshifts.
On the other hand, we notice that no counterpart is found for these sources, so that they might also possibly be spurious detections in the SCUBA2 residual image, in some cases. 

\subsection{General sample and final considerations on redshift estimates}

The cosmic volume sampled by COSMOS, adopting a generous redshift range $4<z<8$ is $7\times10^7$~Mpc$^3$, and the star formation rate density from our population in this volume is about $\mathrm{1\times10^{-3} M_\odot yr^{-1} Mpc^3}$. This seems quite low with respect to the total SFRD at these redshifts \citep{Madau2014a,Liu_DZ2017}, as expected due to the shallower depths reached. We are likely still sampling the high end tail of the luminosity function, albeit now with pretty fair statistics.

\citet{Ivison2016} selected a sample of 109 dusty star forming galaxies with $Hershcel$ colors ($S_{500um}\geq 30$mJy, $i.e., S_{500um}/S_{250um}\geq 1.5$ and $S_{500um}/S_{250um}\geq 0.85$) in a 600 deg$^2$ survey, estimated 32\% of that sample to have $z=$4--6 and reported a number density of $\rho_{z>4}\approx 6\times 10^{-7}$ Mpc$^{-3}$.
The sources of \citet{Ivison2016} have a median $LIR\sim 1.3\times 10^{13} L\odot$, while our $z>4$ sample reaches to fainter levels with a median $LIR\sim 8.0\times 10^12 L\odot$ and down to $LIR\sim 3.6\times10^12 L\odot$. 
We detected 64 dusty star forming galaxies at $z=$4--6 ($\sigma_{\rm 500\mu m}\sim 3$ mJy), implying a much higher space density than that in \citet{Ivison2016}. 
Given that the completeness of our $z>4$ sample is not yet constrained, detailed values for the space densities will be reported in our future paper.

\begin{figure}
\centering
\includegraphics[width=0.48\textwidth]{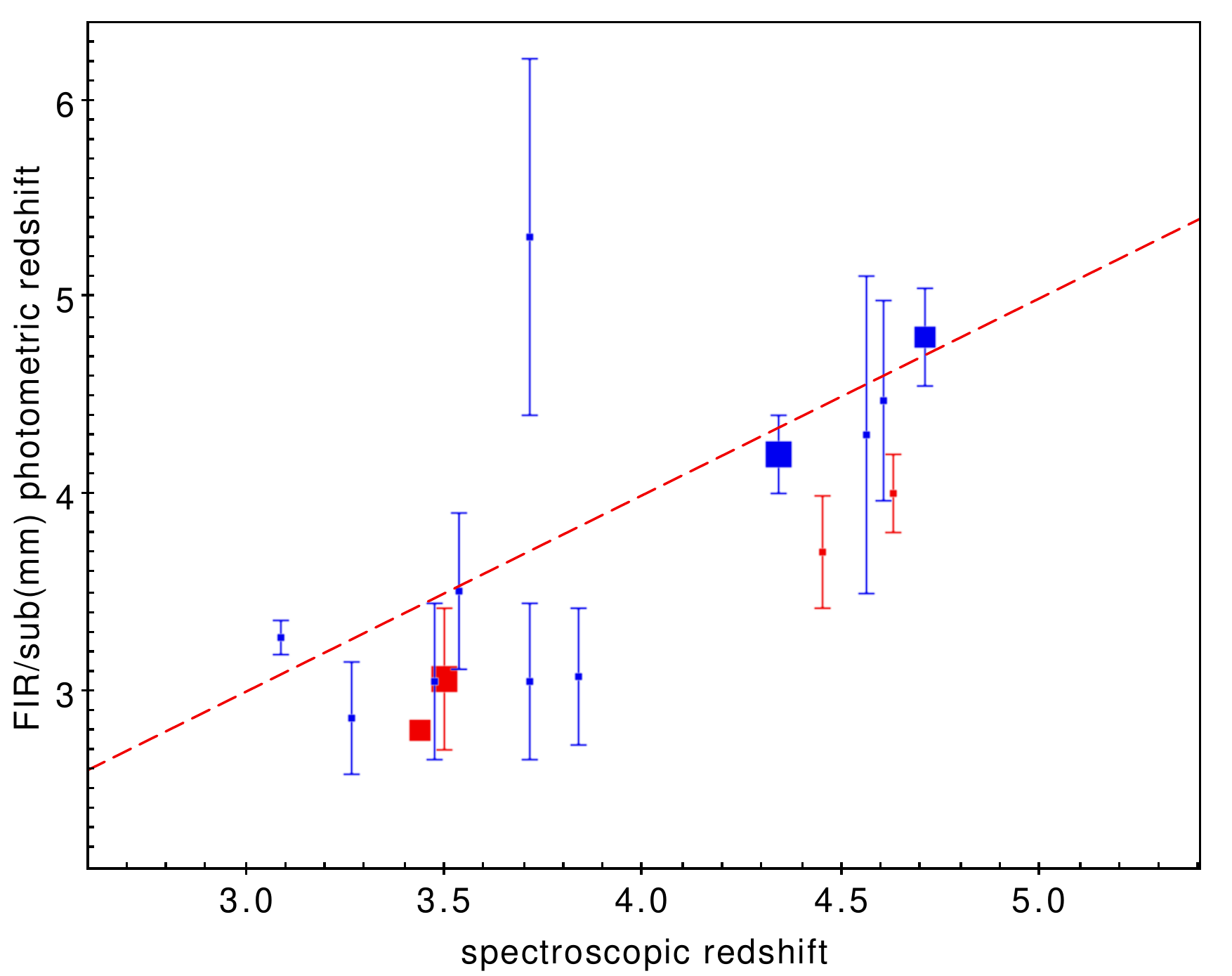}
\caption{%
	The comparison of photometric redshift and spectroscopic redshift of $z>3$ sources. 
The size of symbol is scaled by $\mathrm{S/N_{FIR+mm}}$. The red data points show sources fitted with AGN component. 
\label{compare_z}%
}
\end{figure}

Among the {85} candidates, there are {6} sources with confirmed spectroscopic redshift $z>4$.
Four of them have been listed in Table~\ref{AGN_list}, as they are fitted with an AGN component, while {5} of them are shown in Appendix~\ref{Section_highz} with their actual ``zspec" marked in the SED panels.
The last source with spectroscopic redshift is the source ID20007898 (i.e., AzTEC/C1 in \citealt{Smolcic2012AzTECC1}), as shown in panel (4) of Fig.~\ref{highz_4example}, whose spectroscopic redshift $z_{\rm spec}=4.7$ has been reported by \citet{Brisbin2017}.
As a test, we ignored the redshift of ID20007898 and fit its SED over $z=0-10$ to derive a photometric {$z_\mathrm{phot,FIR}=4.8\pm0.25$}. This agrees well with the $z_{\rm spec}$,
suggesting that our SED procedure is reasonable for giving photometric redshift on $z>4$ dusty galaxies. 
We have ran the same test on IR-detected sources with $z_{\rm spec}>3$ and shown the redshift comparison in Fig.~\ref{compare_z}. Our IR SED-driven photometric redshift is in good agreement with the spectroscopic redshift for galaxies with significant FIR/(sub)mm detections, while perhaps somewhat underestimated in case of galaxies with important AGN torus emission. The relevance of this hint is limited of course by low number statistics. 
We find that in order to obtain a fair comparison between photometric and spectroscopic redshifts the redshift uncertainties must be increased by adding in quadrature an extra component of about 10\% of (1+z). This is of order of the systematic effect expected from dust temperature variations, as discussed in Section~\ref{z_dust_degeneracy}. The redshift errors associated to our sources currently only reflect the accuracy of their SEDs and do not include such systematic uncertainties. Once weighting by the (total) redshift uncertainties, the comparison in Fig.~\ref{compare_z} suggest that our FIR photometric redshifts are underestimated by 6\%$\times(1+z)$ ($\Delta z\sim0.4$) on average, confirming the idea that the FIR SED of GN20 is somewhat colder than the high-z average. Larger samples are required to bring these results to firmer footings.

The multi-band cutouts and the SEDs of all candidates are presented in Appendix~\ref{Section_highz}.
Redshift confirmation of as many as possible of these candidates is clearly needed, and we are planning observations with NOEMA, ALMA and hopefully JWST in the future.
Such further studies of these high-z candidates will be needed to finally validate their selection, and understand critical issues like completeness, spurious fraction, actual redshift distribution (hence number densities), and others. Our photometric catalog in COSMOS will be the base for such explorations.

\section{Summary}

In this work we obtain (and publicly release) detailed \superdb photometry for far-infrared to (sub-)millimeter imaging datasets in the COSMOS full 2~deg$^2$ field, with the most accurate photometry information lying where complete prior information is available, i.e., in the 1.7~deg$^2$ UltraVISTA area. In order to overcome the heavy blending problems introduced by the large beam size at $Herschel$ SPIRE and (sub)mm bands, we adopted the ``Super-deblending'' technique which has been pioneered in the GOODS-North field by L18, and critically adapted it to the COSMOS field where data at PACS, Spitzer/MIPS and radio are shallower than in the GOODS-N field.

We selected a highly complete set of 194,428 priors for deblending the FIR to mm images. This prior catalog contains 88,008 detections from MIPS 24~$\mu$m and radio fitting, and $\sim1\times10^5$ mass-selected priors from UltraVISTA catalogs.

In the deblending of the FIR/(sub)mm images we improved the faint sources subtraction by only subtracting fluxes of galaxies with a reliable determination as predicted by the SED fitting. We calibrated and removed biases from this subtraction step using Monte Carlo simulations. This returned flux uncertainties with well-behaved Gaussian-like statistics.

A total of {11,220} galaxies are individually detected with a combined $\mathrm{S/N}>5$ over the FIR/(sub)mm wavelengths, including {770} detections at $z \geqslant 3$ (mostly photometric). Comparing with photometric catalogs in the literature, the \superdb photometry shows good agreement for bright sources and generally improved de-blending at the faint end.

Quite notably, the \superdb 850~$\mu$m photometry agrees remarkably well with  ALMA archival data photometry, with a scatter that is consistent with the \superdb photometric error, demonstrating that the \superdb photometry is correctly derived and errors are statistically well-defined.

Finally, we conservatively selected {85} {robust} high redshift candidates with solid detections at FIR/(sub)mm wavelengths
and requiring $z>4$ (we use $z-z_{\rm error}>4$).
These candidates have often well-determined counterparts in IRAC and/or radio images, weak or no detection at $K_s$ band. Their SEDs suggest redshifts over $z\sim$4--7, including possibly some of the most distant galaxies known. Confirmation of redshifts by future observations is needed. This unique sample will allow us to statistically investigate the first generation of vigorous star formation in the early Universe.

\acknowledgments

We are grateful to the full COSMOS team for their contributions in the build-up of such a rich multi-wavelength dataset. We thank the referee for useful comments and suggestions that helped improving the paper.
We thank B. Magnelli, P. Lang and the rest of the A$^3$COSMOS team for providing the ALMA photometry for comparison. 
SJ acknowledges funding from the China Scholarship Council. 
SJ and QG acknowledge supports from the National Key Research and Development Program of China (No. 2017YFA0402703) the National Natural Science Foundation of China (No. 11733002).
VS, JD and ID acknowledge support from the European Union's Seventh Frame-work program under grant agreement 337595 (ERC Starting Grant, "CoSMass''). 
ES and DL acknowledge funding from the European Research Council (ERC) under the European Union's Horizon 2020 research and innovation programme (grant agreement No. 694343). 
The National Radio Astronomy Observatory is a facility of the National Science Foundation operated under cooperative agreement by Associated Universities, Inc. ALMA is a partnership of ESO (representing its member states), NSF (USA) and NINS (Japan), together with NRC (Canada), MOST and ASIAA (Taiwan), and KASI (Republic of Korea), in cooperation with the Republic of Chile. The Joint ALMA Observatory is operated by ESO, AUI/NRAO and NAOJ.

\bibstyle{apj}
\bibliography{biblio}

\clearpage

\appendix

\section{MIPS 24~$\mu$\lowercase{m} calibration factor}
\label{24um_calibration}

\begin{figure}
\centering
\includegraphics[width=0.48\textwidth, trim={1.75cm 0cm 1.75cm 1.5cm}, clip]{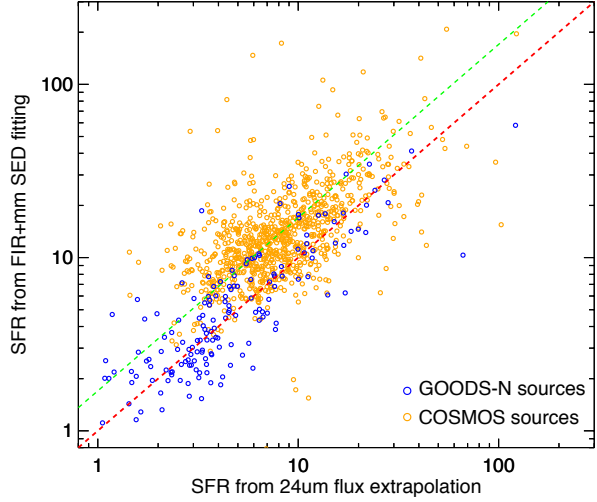}
\caption{%
	SFR from FIR+mm SED fitting vs. SFR from 24~$\mu$m extrapolation \citep{Magdis2012SED}. 
    All sources have redshift $z=0.4-0.58$ with $\mathrm{S/N}_{24\,{\mu}\mathrm{m}}>5$ and $\mathrm{S/N}_{\textnormal{FIR+mm}}>5$. 
    Blue circles show sources in the GOODS-North field, while orange circles show sources in the COSMOS field. The identity line is shown in red, while the median linear fit of COSMOS sources in green (i.e., $1.7\times$ the identity line).
    \label{24um_cal_factor}%
}
\end{figure}

We use a sample of significant IR detections at redshift $0.4<z<0.58$, a redshift range where starburst and main-sequence galaxies have identical ratios of 24~$\mu$m fluxes to bolometric FIR fluxes, based on the \citet{Magdis2012SED} templates.
In Fig.~\ref{24um_cal_factor} we compare the total SFR from the FIR SED fitting in our work with the SFR estimated only from the 24~$\mu$m flux on the basis of templates from \citet{Magdis2012SED}.
We only show sources with reliable detections: $\mathrm{S/N}_{24\,{\mu}\mathrm{m}}>5$ and $\mathrm{S/N}_{\textnormal{FIR+mm}}>5$ and from both COSMOS (this work) and GOODS-N (L18) fields. 
We find that in the COSMOS field, SFRs directly obtained from the FIR photometry (100~$\mu$m-1.2mm) are higher than the 24~$\mu$m extrapolated ones, by a factor of about 1.7, while this issue is not seen in the GOODS-North \superdb catalog.
The agreement in the GOODS-N field is, of course, by construction, given that the GOODS-N (and GOODS-S) data were used to construct the \citep{Magdis2012SED} templates, including the photometry at mid-IR bands from Spitzer MIPS and IRS.

We have also checked that this effect remains if we were using the SFRs only based on 100~$\mu$m and 160~$\mu$m fluxes in the \superdb catalog, which are subjected to much less blending than, e.g., SPIRE bands.

A similar effect is also reported by \citet{Ilbert2015} (see their Fig.~1), where they found evidence for the need of a factor of 1.5 up-scaling of 24$\mu$m fluxes in COSMOS when using the template of \citet{Magdis2012SED}, although reduced to a factor of 1.2 with \citet{Dale_Helou2002} templates. They suggest this difference might be due to the use of different data reduction pipelines, as well as the combined uncertainties in the absolute calibration of MIPS and $Herschel$ data.

Overall, it is difficult to conclude if this is really a problem of the COSMOS photometry, as it might be instead a problem with the GOODS photometry, or a mix there-of. Also, it might perhaps apply at least in part to the overall PACS and SPIRE photometry in COSMOS vs GOODS. 
What appears to be more reliable is that the interpretation of COSMOS 24~$\mu$m to FIR SEDs with \citet{Magdis2012SED} templates 
(or, more in general, their comparison to GOODS-N fluxes) 
requires some re-scaling. A possible solution to make the comparison consistent is the multiplication of COSMOS 24$\mu$m fluxes by a factor of 1.5--1.7, which is the range where our current analysis and that of Ilbert et al. are converging.

\section{Photometry Image Products}
\label{image_products}

\begin{figure}
	\centering
	\includegraphics[width=0.71\textwidth]{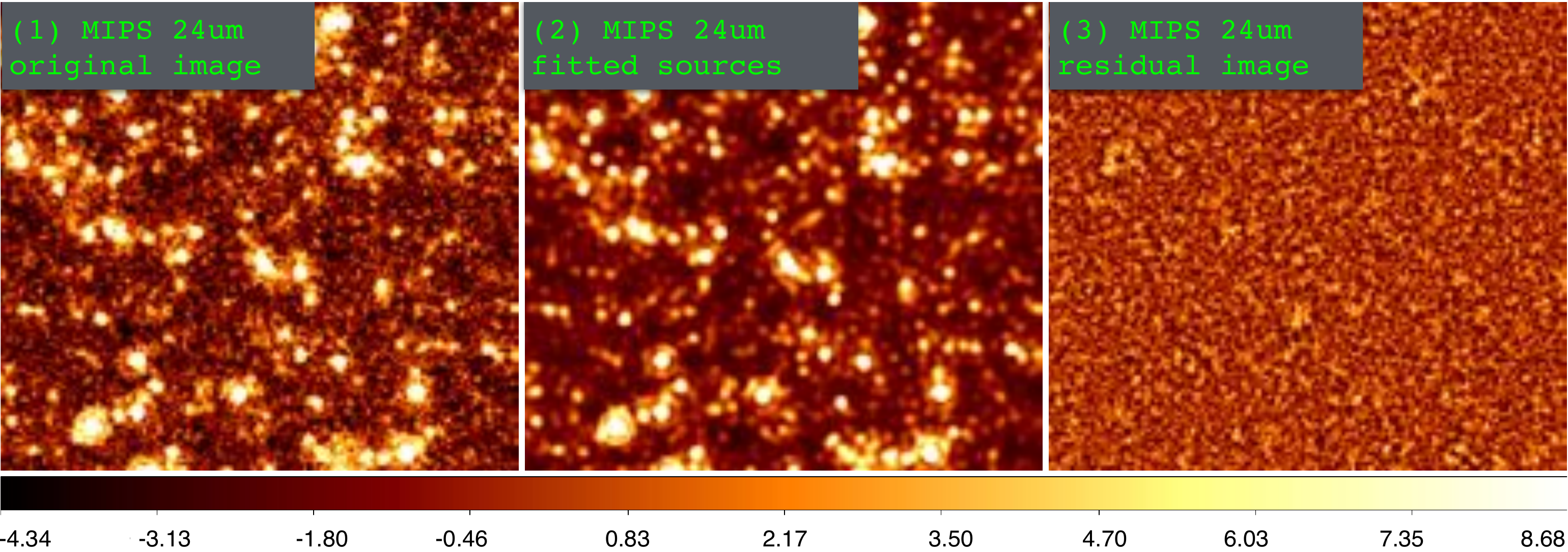}
    \includegraphics[width=0.28\textwidth,trim={3cm 0cm 3cm 1.8cm},clip]{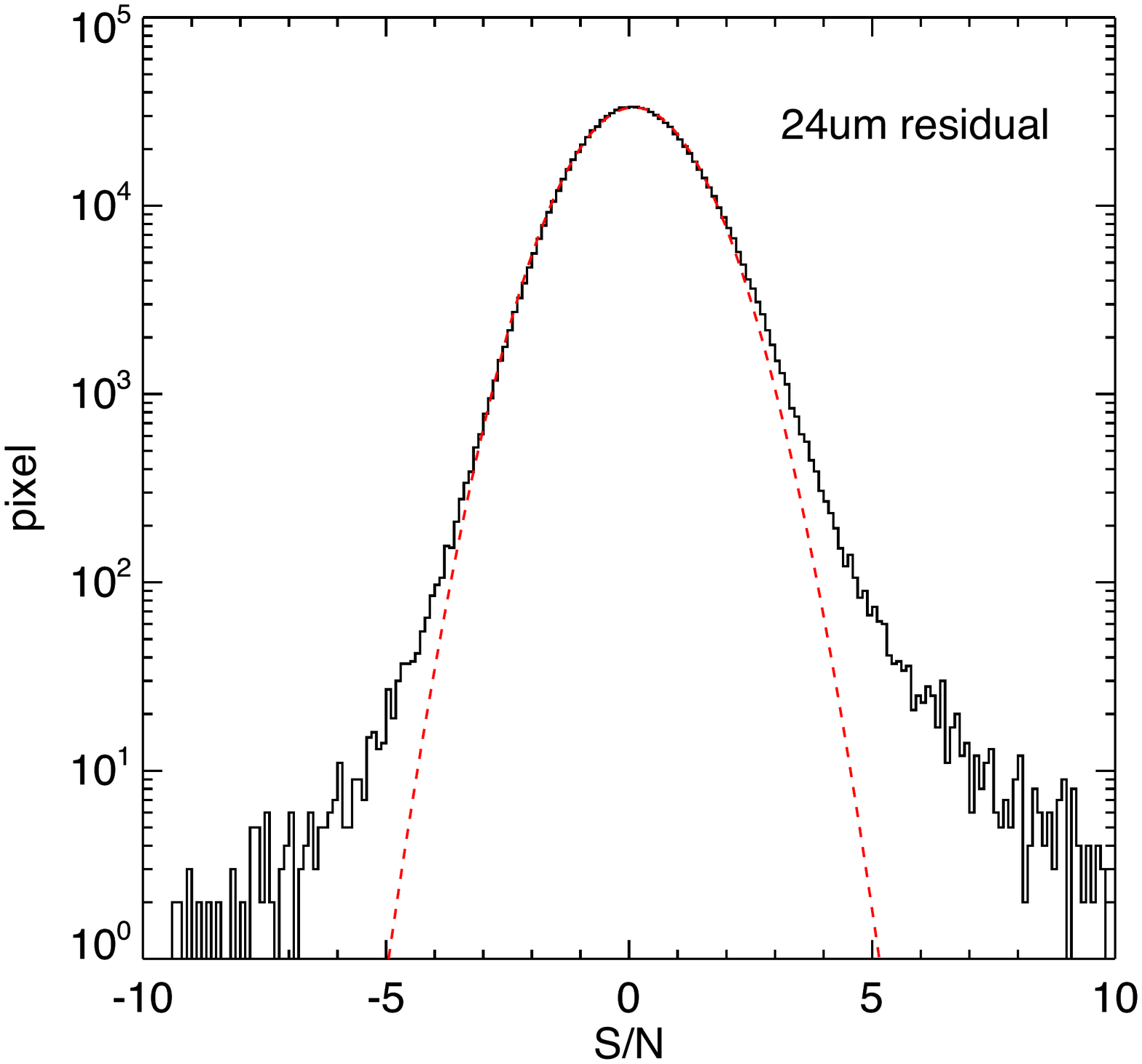}
	\includegraphics[width=0.71\textwidth]{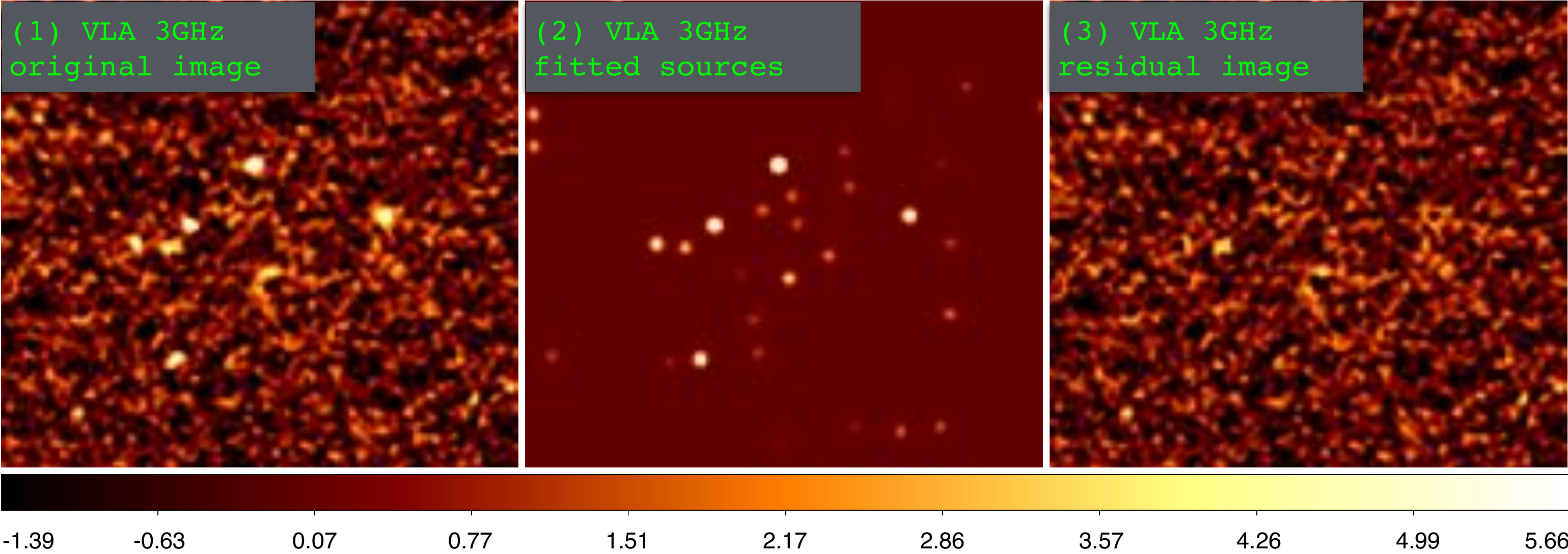}
    \includegraphics[width=0.28\textwidth, trim={3cm 0cm 3cm 1.8cm},clip]{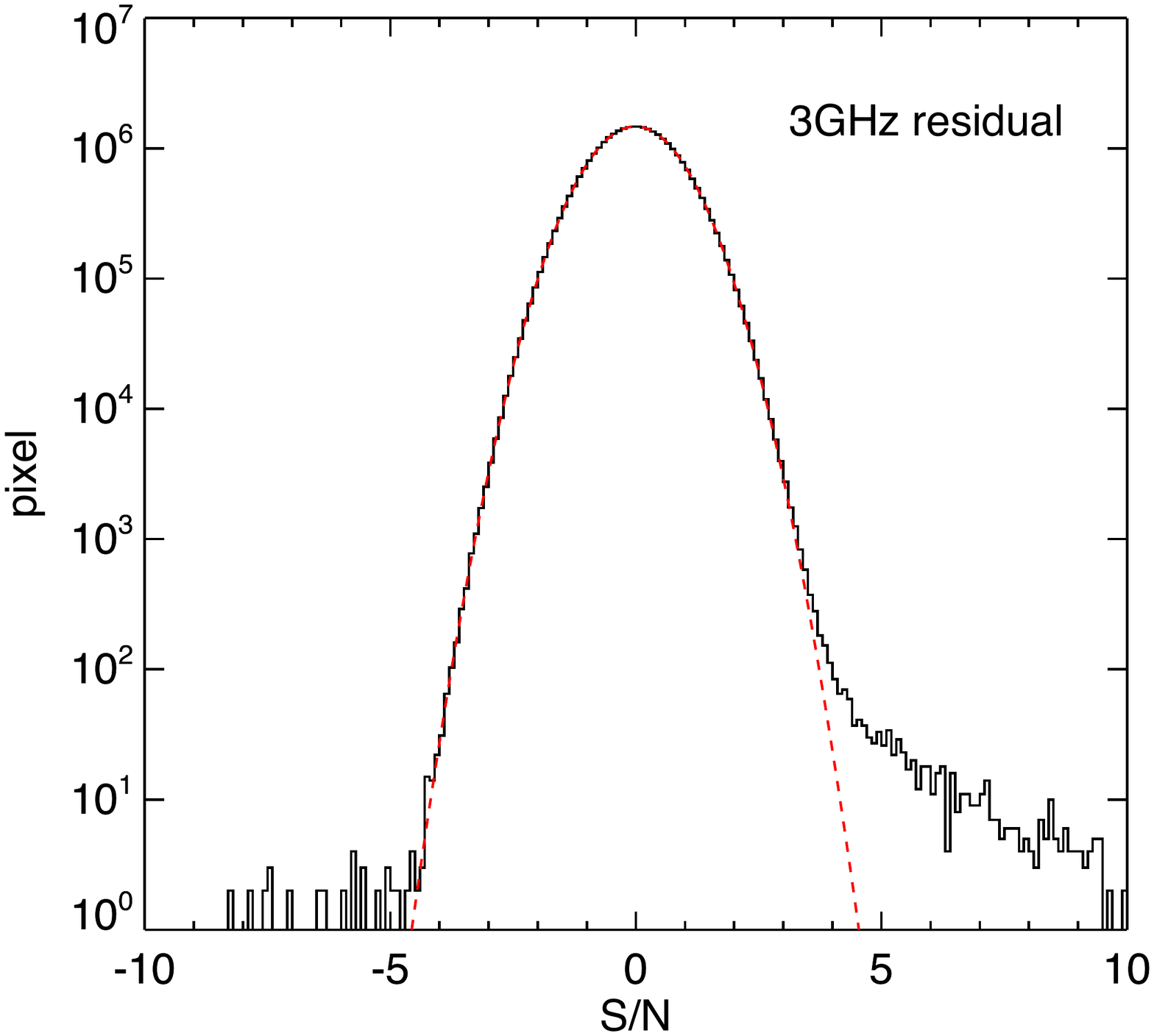}
	\includegraphics[width=0.71\textwidth]{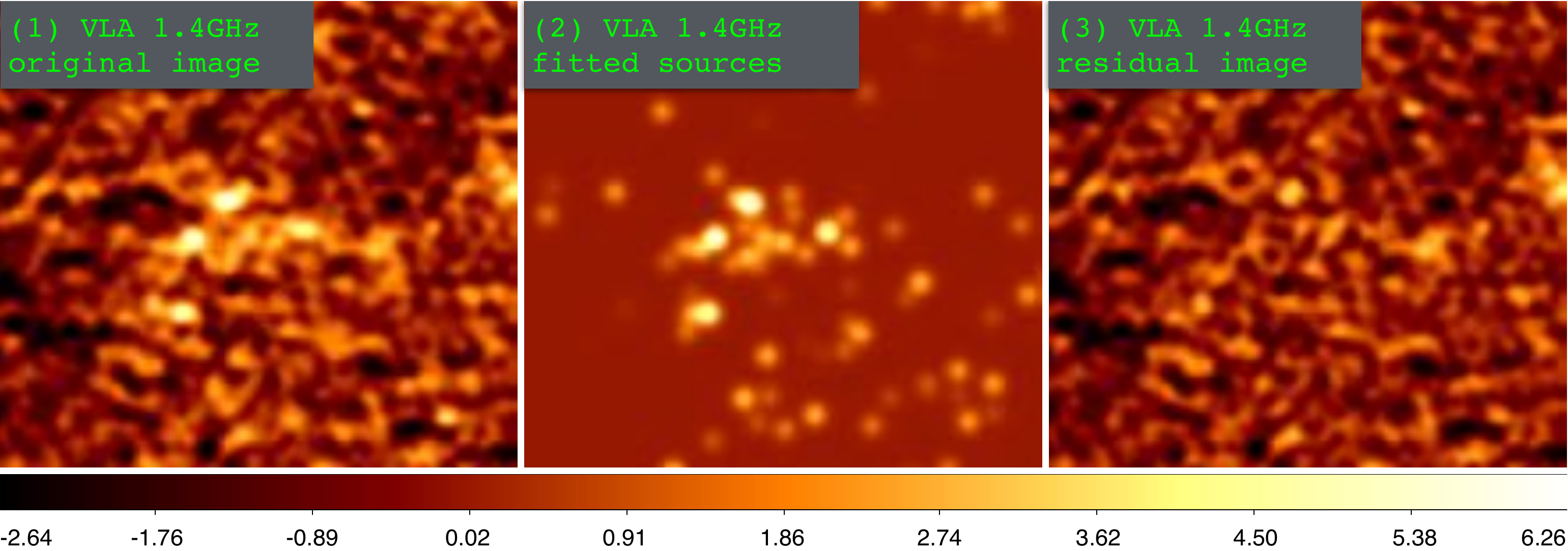}
    \includegraphics[width=0.28\textwidth, trim={3cm 0cm 3cm 1.8cm},clip]{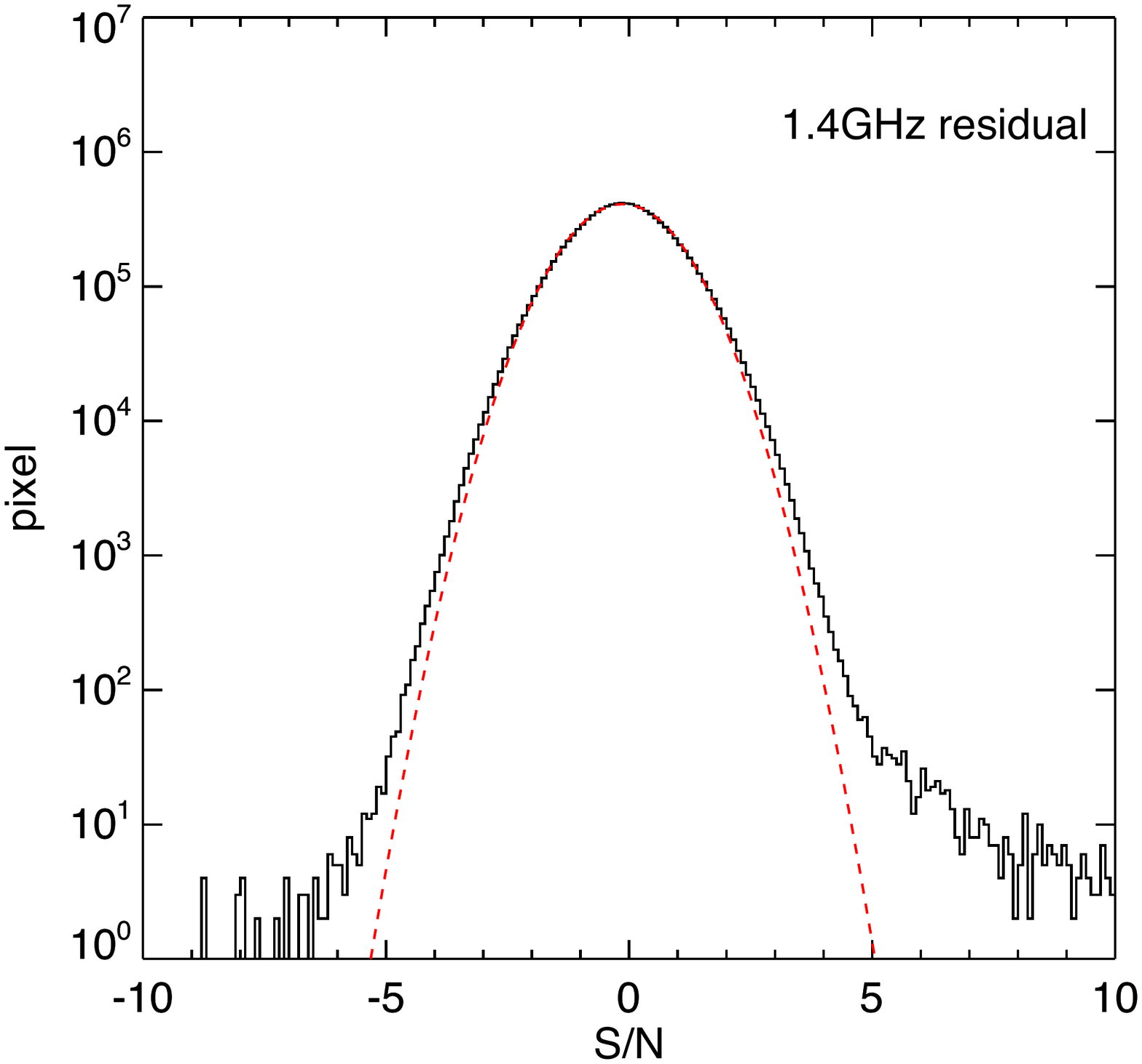}
    \includegraphics[width=0.71\textwidth]{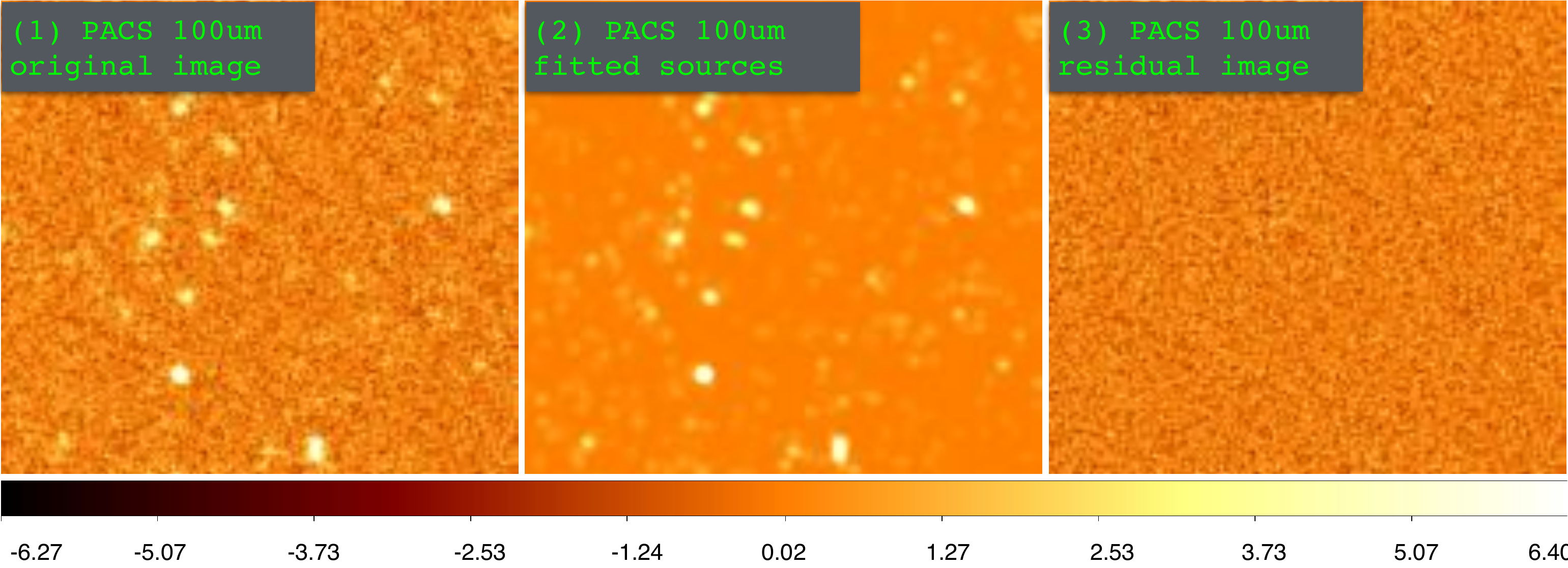}
    \includegraphics[width=0.28\textwidth, trim={3cm 0cm 3cm 1.8cm},clip]{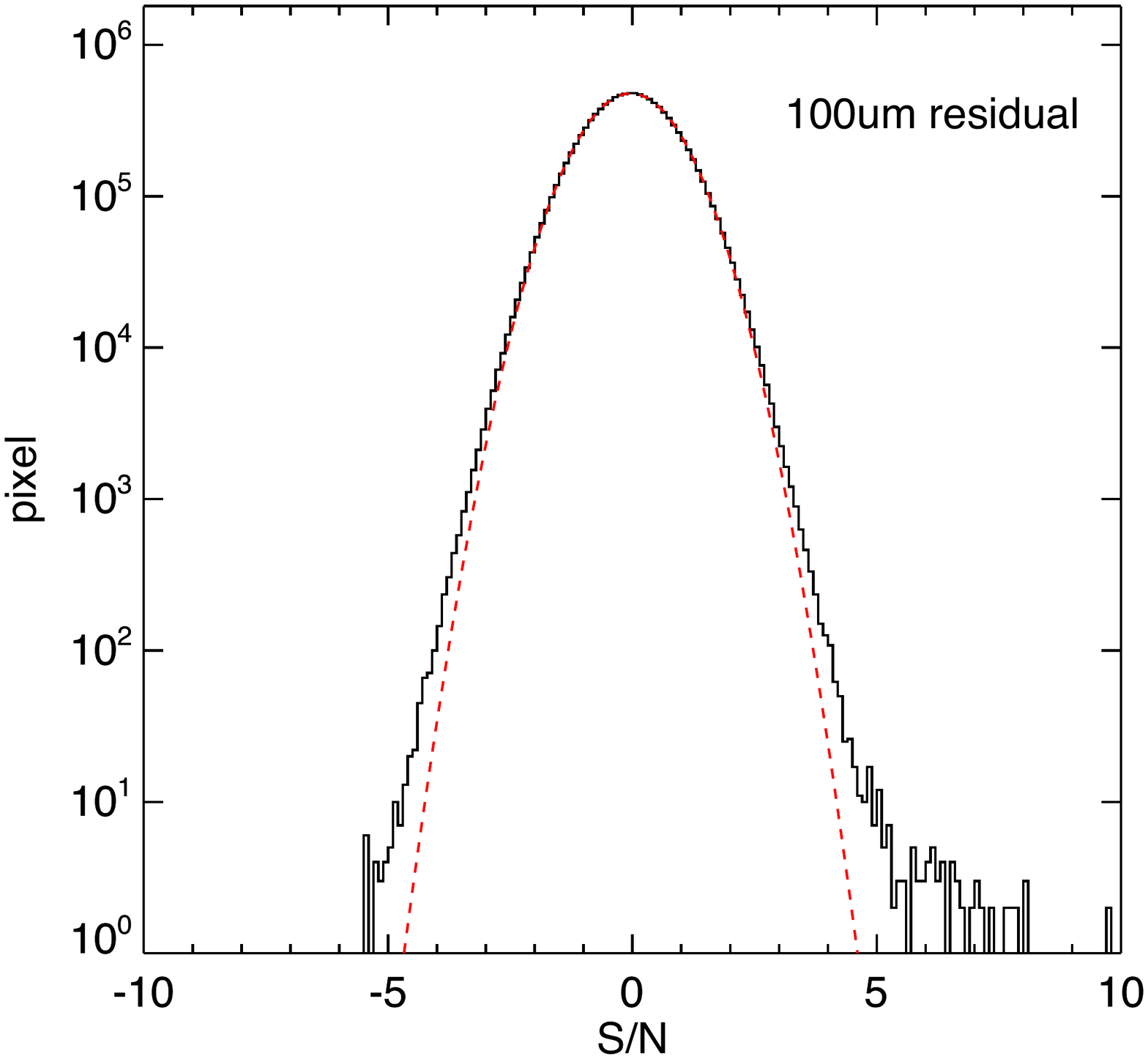}
	\caption{%
		Photometry image products at MIPS 24~$\mu$m, VLA 3~GHz \& 1.4~GHz and PACS 100~$\mu$m. Panel (1) is a portion of the original map of each band. Panel (2) shows the \galfit best fitting model image of fitted priors, and panel (3) is the residual image of panel (1) and (2). Image values and histograms are expressed in terms of S/N (see also the caption of Fig.~\ref{Fig_Photometry_Images}).
		\label{galfit_24_to_100_Map}
	}
\end{figure}

We present the photometry image products at each band here, all images have the same scaling and share the colorbar in terms of S/N ratio, following Fig.~\ref{Fig_Photometry_Images} (see also the caption there).
Fig.~\ref{galfit_24_to_100_Map} show image products in MIPS 24$\mu$m, VLA 3~GHz \& 1.4~GHz and PACS 100~$\mu$m images, where no faint source is subtracted from their original images. 

\begin{figure}
	\centering
	\includegraphics[width=0.75\textwidth]{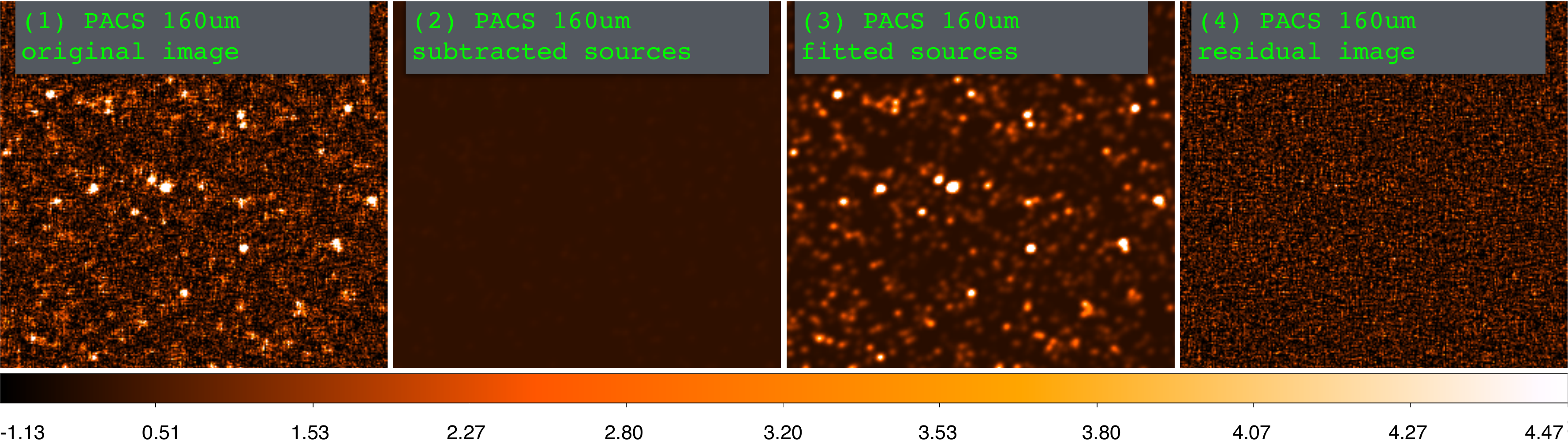}
    \includegraphics[width=0.233\textwidth, trim={3cm 0cm 3cm 1.8cm},clip]{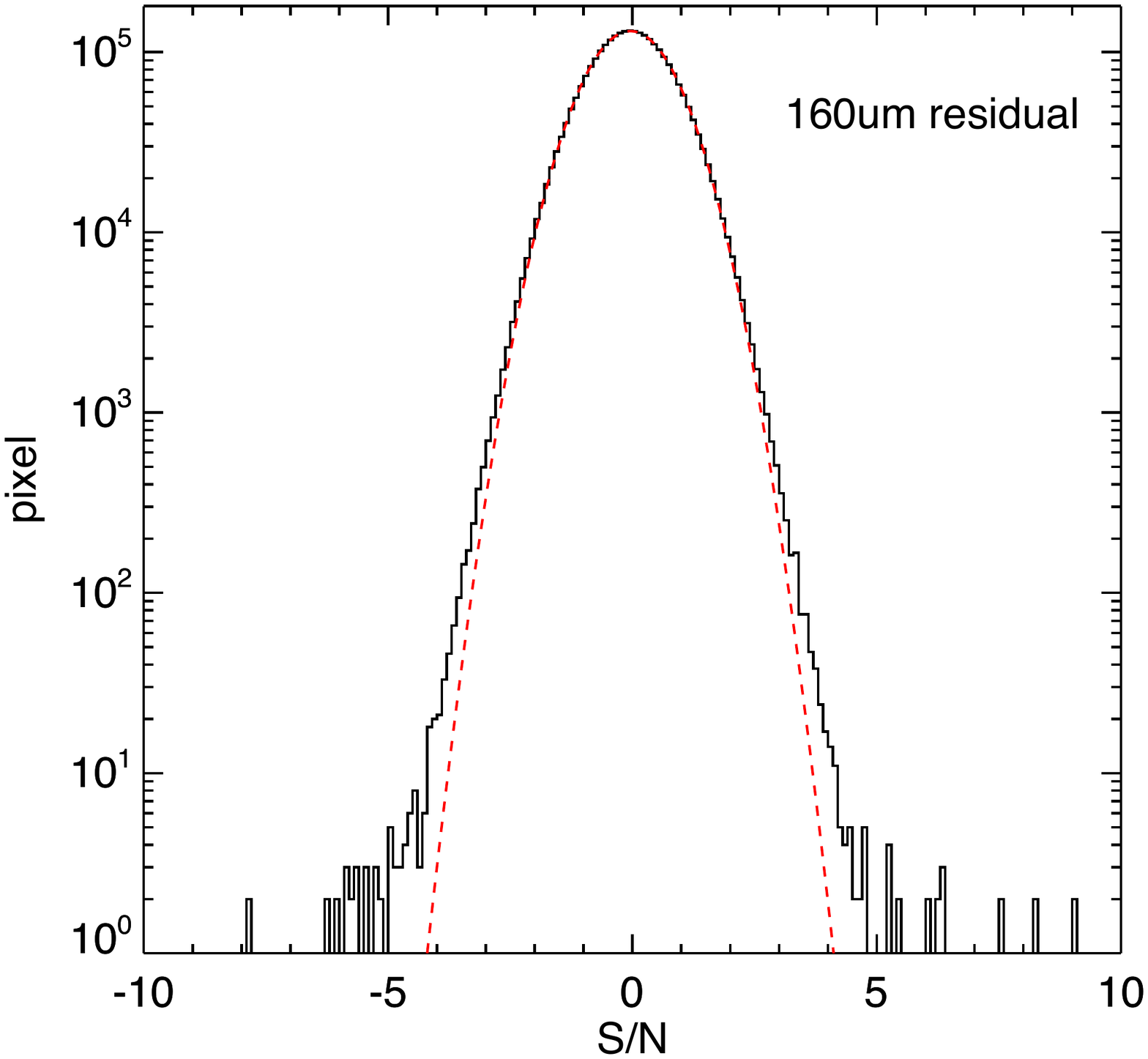}
	\includegraphics[width=0.75\textwidth]{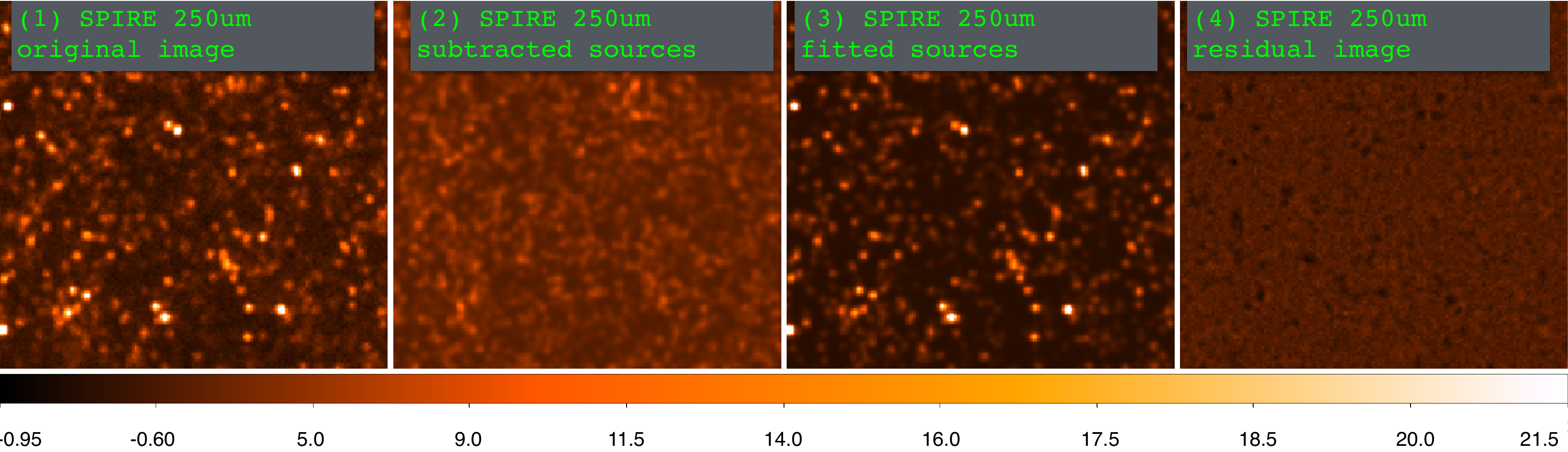}
    \includegraphics[width=0.233\textwidth, trim={3cm 0cm 3cm 1.8cm},clip]{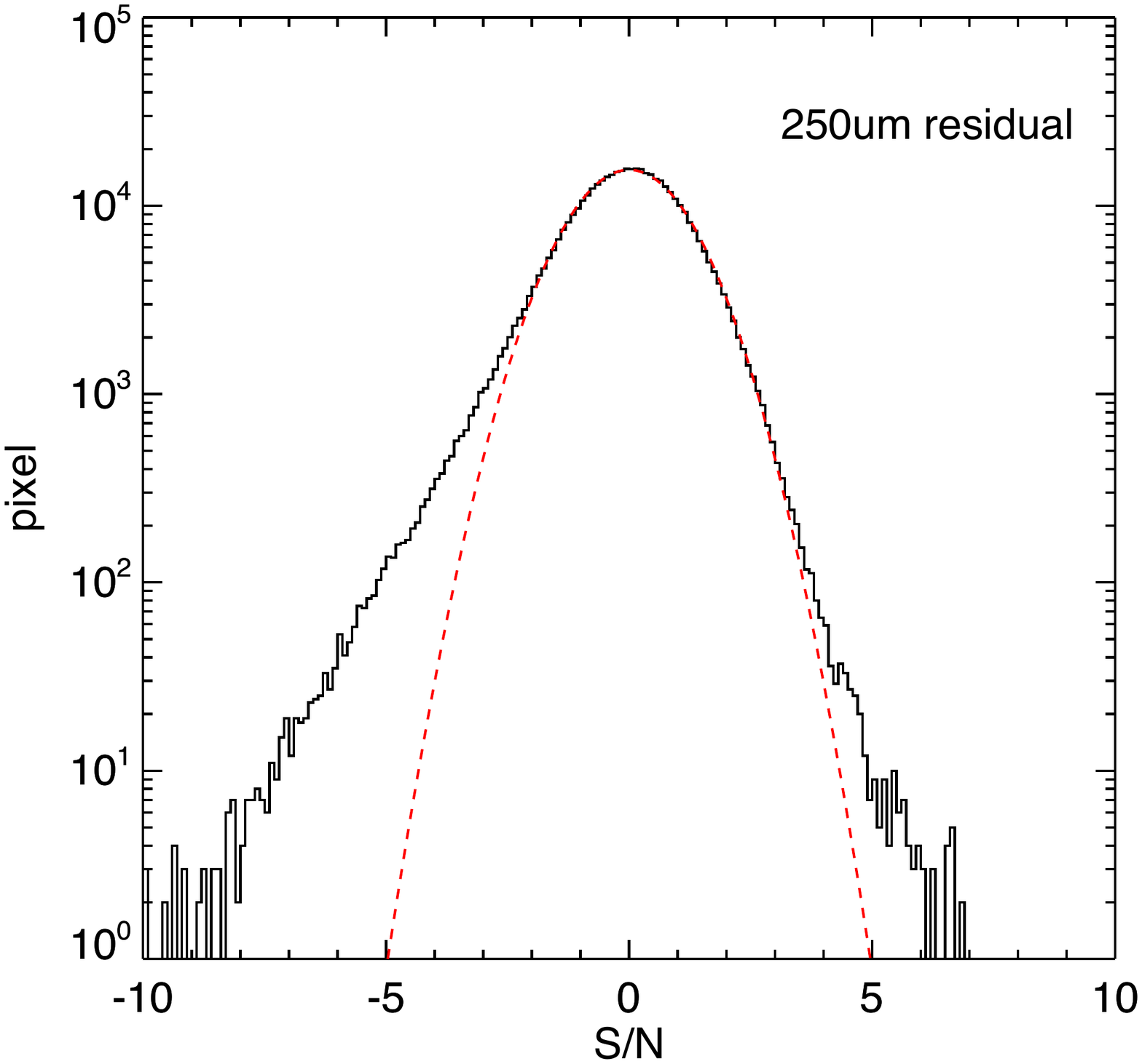}
	\includegraphics[width=0.75\textwidth]{350um_deblend_images}
    \includegraphics[width=0.233\textwidth, trim={3cm 0cm 3cm 1.8cm},clip]{hist_SN_350um.pdf}
	\includegraphics[width=0.75\textwidth]{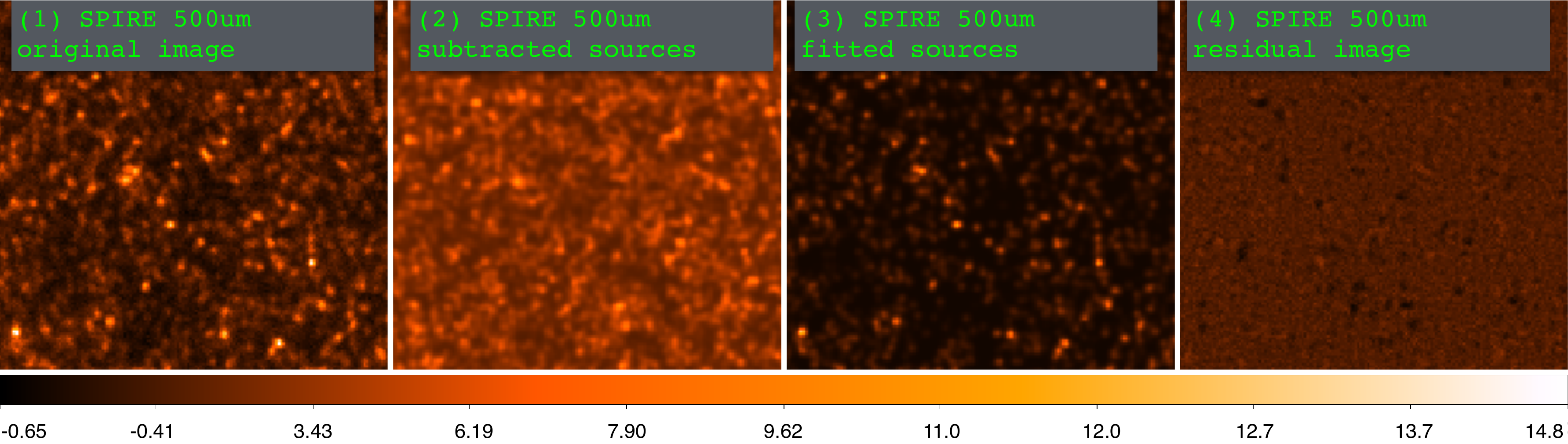}
    \includegraphics[width=0.233\textwidth, trim={3cm 0cm 3cm 1.8cm},clip]{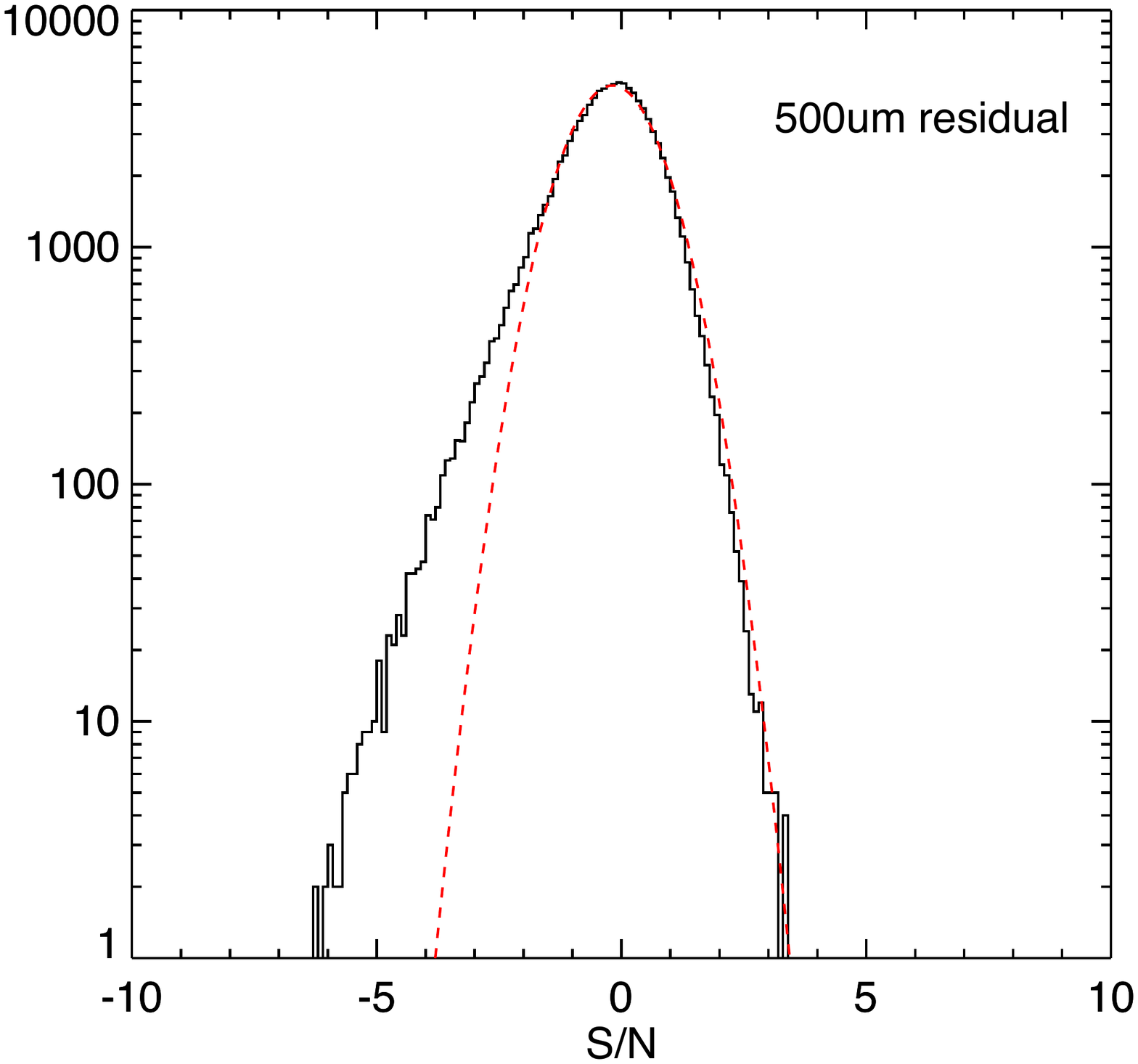}
	\caption{%
		Photometry image products at PACS 160~$\mu$m, SPIRE 250~$\mu$m, 350~$\mu$m and 500~$\mu$m. See descriptions in text. 
		\label{galfit_160_to_500_Map}
	}
\end{figure}

\begin{figure}
	\centering
	\includegraphics[width=0.75\textwidth]{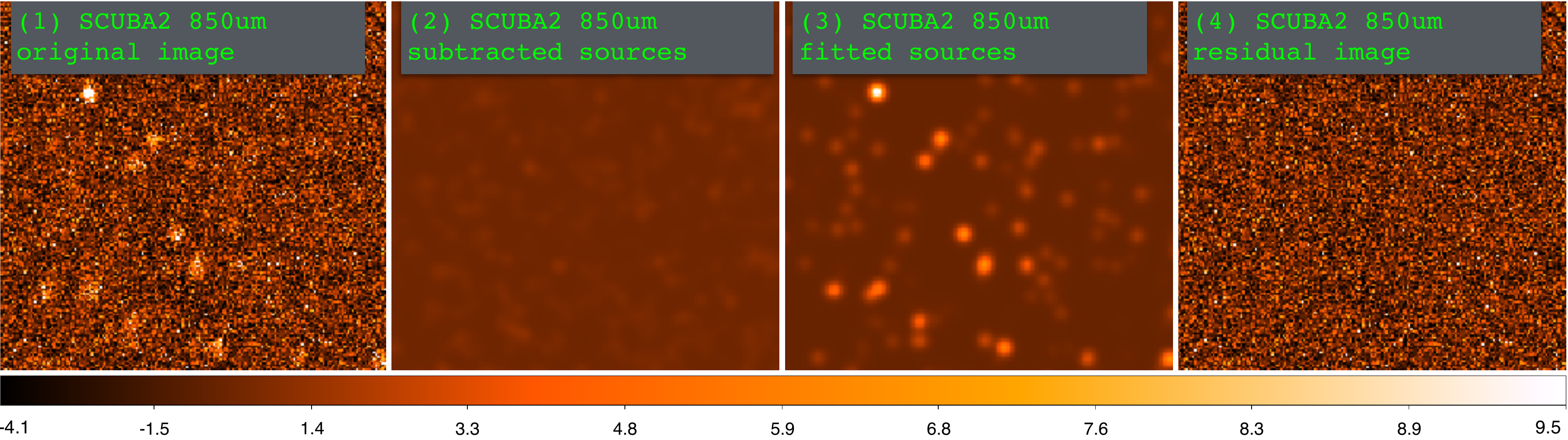}
    \includegraphics[width=0.233\textwidth, trim={3cm 0cm 3cm 1.8cm},clip]{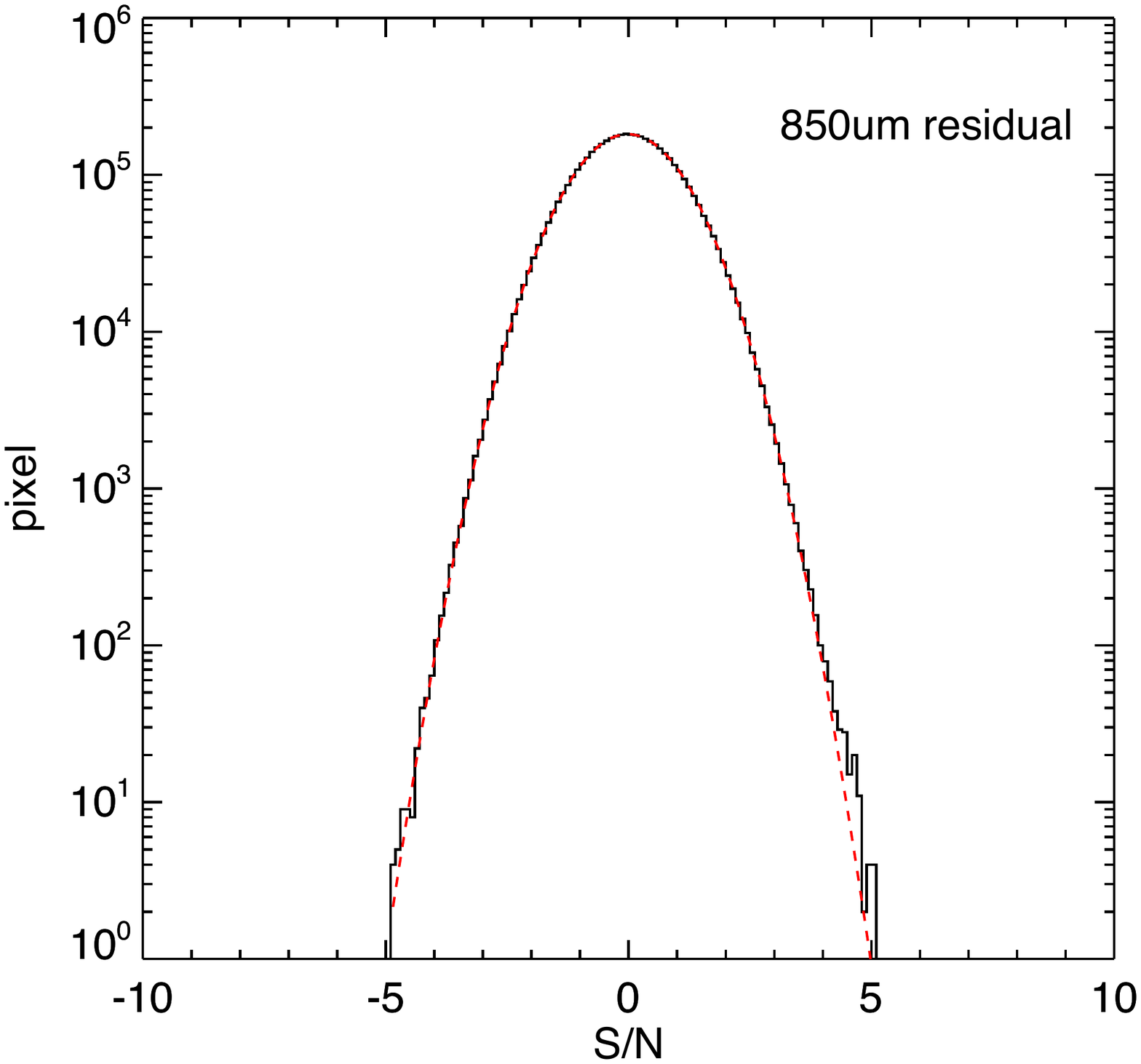}
	\includegraphics[width=0.75\textwidth]{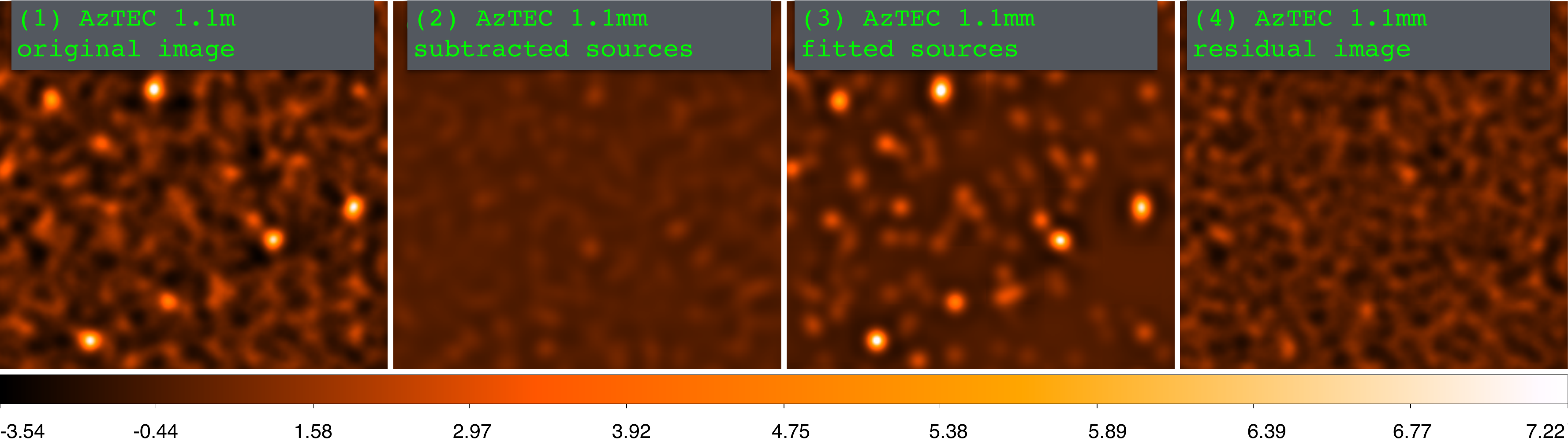}
    \includegraphics[width=0.233\textwidth, trim={3cm 0cm 3cm 1.8cm},clip]{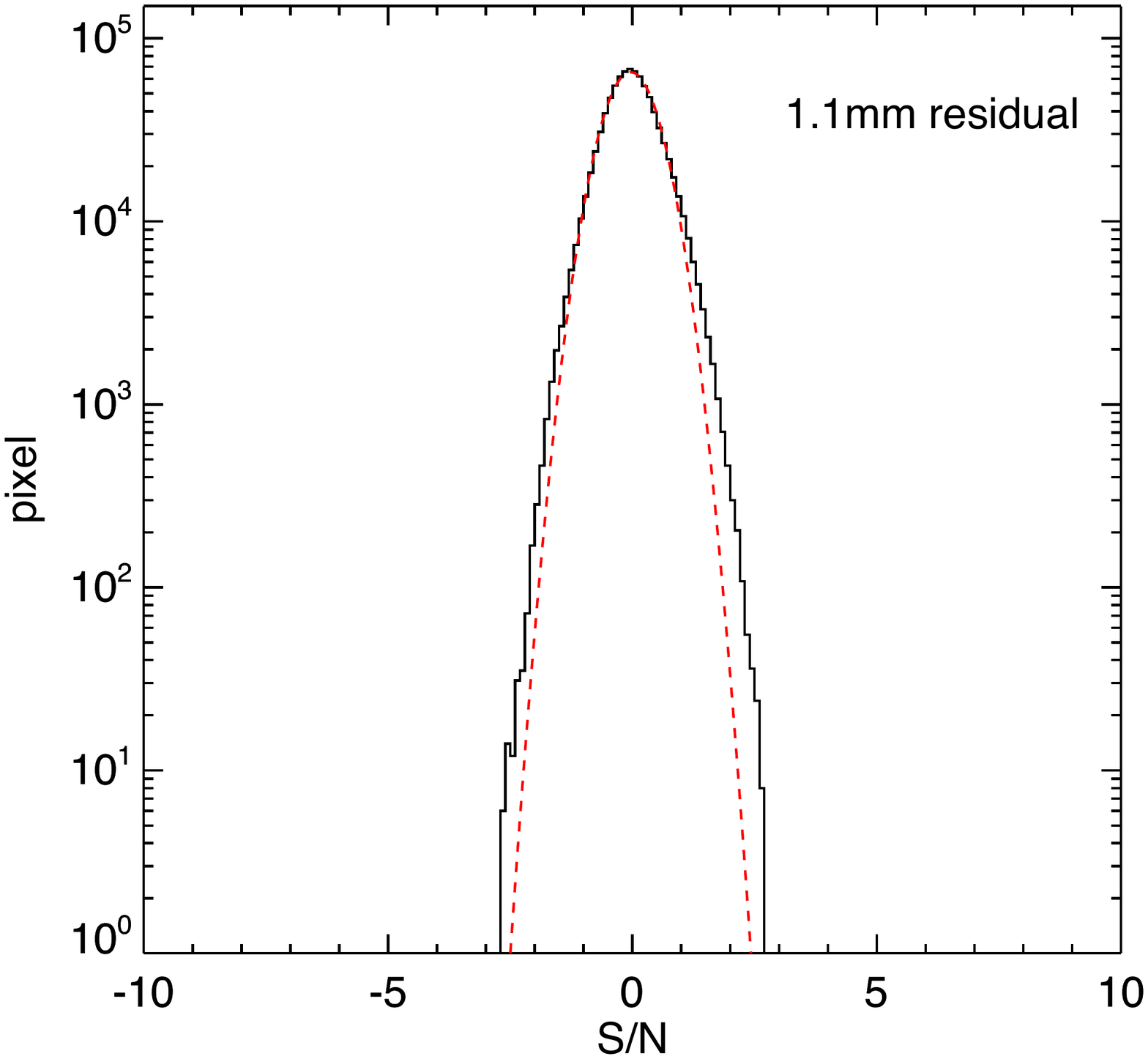}
	\includegraphics[width=0.75\textwidth]{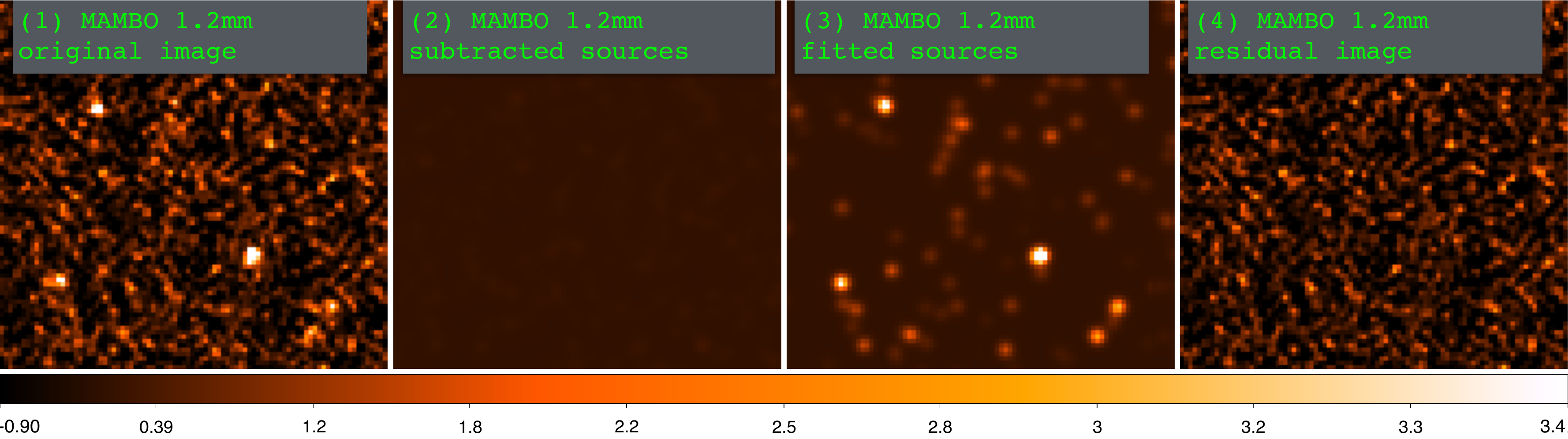}
    \includegraphics[width=0.233\textwidth, trim={3cm 0cm 3cm 1.8cm},clip]{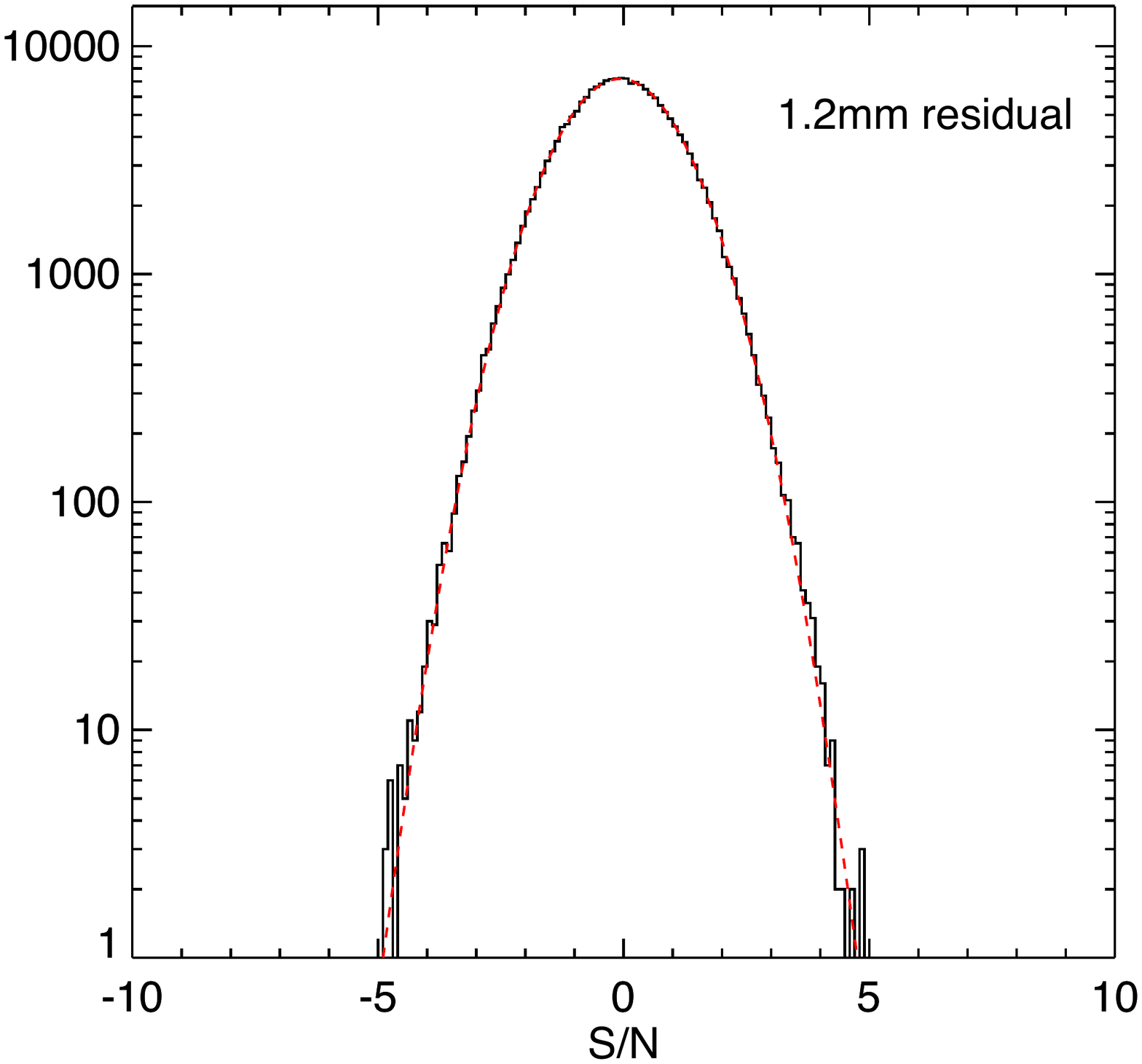}
	\caption{%
		Photometry image products at SCUBA2 850~$\mu$m, AzTEC 1.1mm and MAMBO 1.2mm. See descriptions in text. 
		\label{galfit_160_FIT_goodsn_160_Map}
	}
\end{figure}

\clearpage

\section{Simulation Correction Analyses}
\label{Section_Simulation_Performance}

Analog to Appendix B in L18, we present here the figures of our simulation-based correction recipes at each band. For example, in Fig.~\ref{Figure_galsim_250_bin}, simulation data points are binned by four measurable parameters: the \galfit{} flux uncertainty normalized by the local rms noise at the source position (${\sigma}_\mathit{galfit}/{\sigma}_{\mathrm{rms\,noise}}$) (the first column), the residual flux within one PSF beam area in the residual image normalized also by the local rms noise (${S}_{\mathrm{residual}}/{\sigma}_{\mathrm{rms\,noise}}$) (the second column), the \crowdedness parameter (the third column), and the normalized subtraction $S_\mathrm{subtracted}/{\sigma}_{\mathrm{rms\,noise}}$ (the forth column). Bins are indicated by the dashed vertical lines in the first and second row images. 

From left to right, the panels are in the same four-parameter order.
{\bf The first row:} the difference between the input and output flux of each simulated source ($S_{in}-S_{out}$) (i.e., the flux bias) vs the four parameters, which is fitted by a 3-order polynomial function (the red curve). 
{\bf The second row:} the flux difference divided by the flux uncertainty $(S_{in}-S_{out})/{\sigma}$ vs the four parameters. The scatter in each bin is considered as the correction factor that needs to be applied to ${\sigma}$. Uncorrected and corrected data are shown in Blue and red respectively. After correction, the scatter of $(S_{in}-S_{out})/{\sigma}$ in each bin is very close to 1.0, indicating that the corrected ${\sigma}$ is statistically consistent with the scatter of $S_{in}-S_{out}$. We fit a 3-order polynomial function to the bin-averaged flux uncertainty correction factor on each parameter, which is shown as the red curve. The right axis indicates its value. 
{\bf The third row:} the histogram of $(S_{in}-S_{out})/{\sigma}$ before and after correction of each parameter. Its shape, after correction (i.e., the red histogram), becomes well-behaved Gaussian distribution (i.e., symmetric and has a Gaussian width of 1.0), and is much better than the uncorrected one (i.e., the blue histogram). 
{\bf The fourth row:} the histogram of flux. {\bf The fifth row:} the histogram of flux uncertainty.

\begin{figure}[hb]
	\centering
    
    \begin{subfigure}[b]{\textwidth}\centering
	\includegraphics[width=0.3\textwidth, trim={1cm 15cm 0cm 2.5cm}, clip]{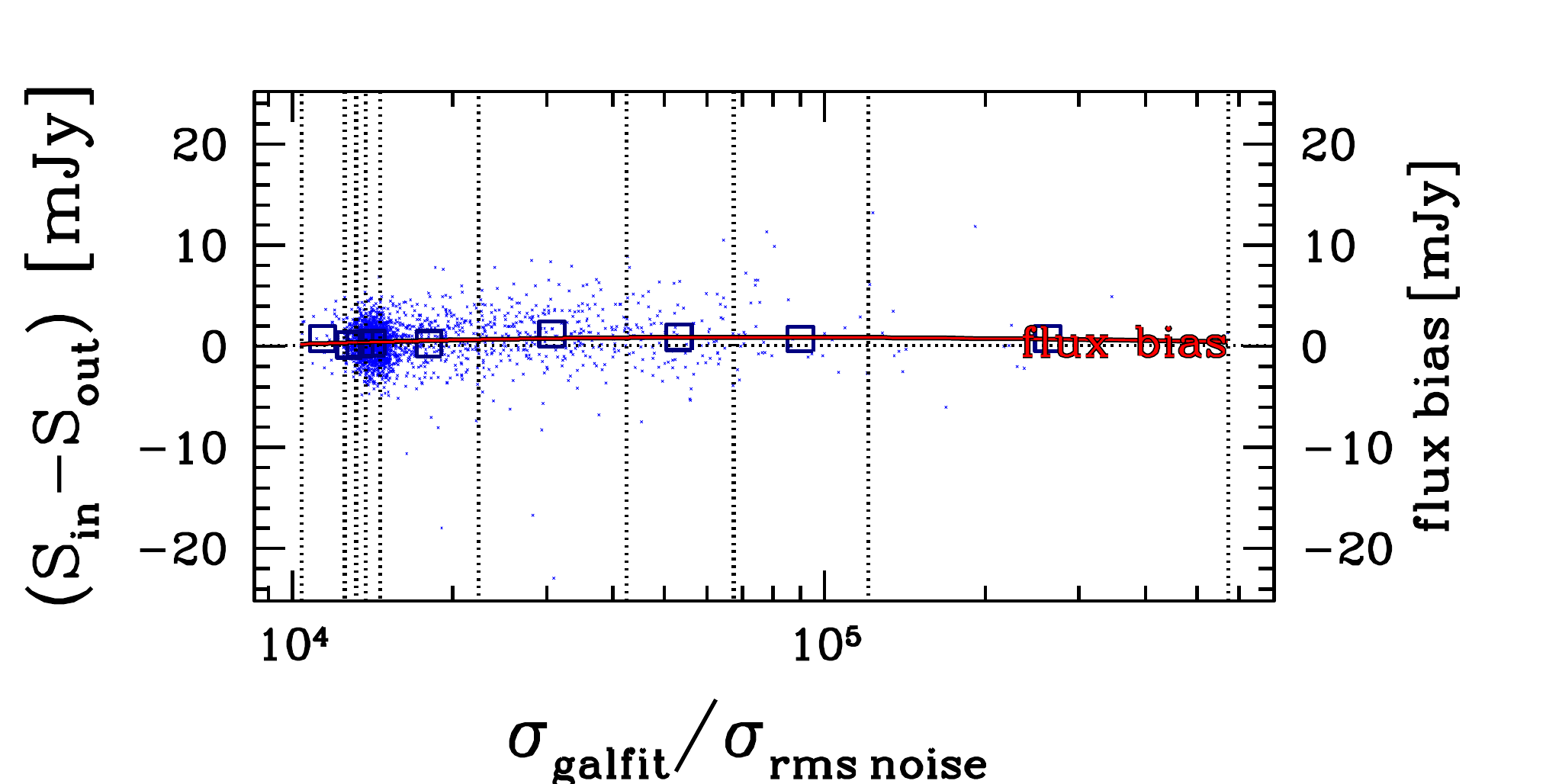}
	\includegraphics[width=0.3\textwidth, trim={1cm 15cm 0cm 2.5cm}, clip]{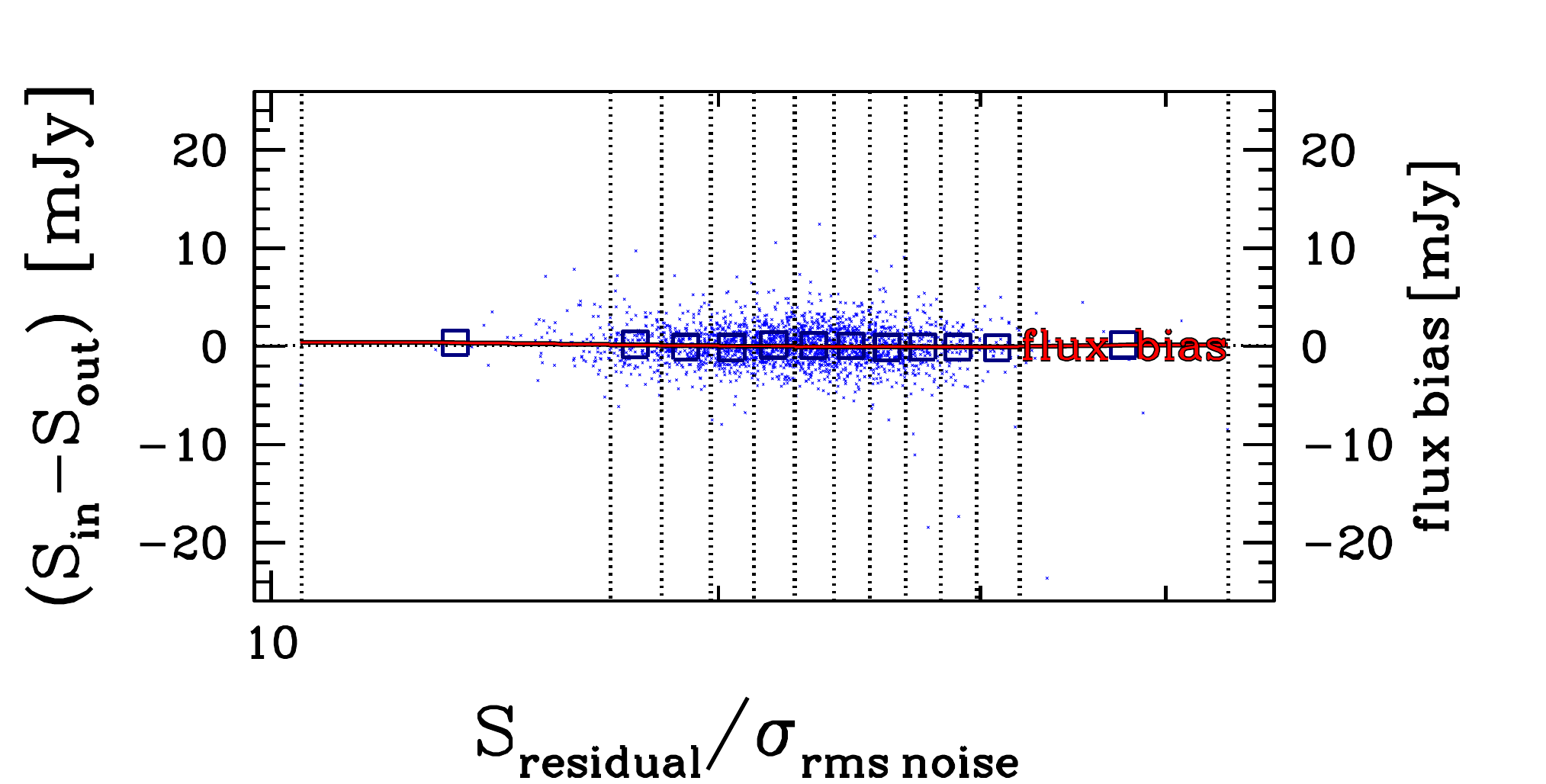}
	\includegraphics[width=0.3\textwidth, trim={1cm 15cm 0cm 2.5cm}, clip]{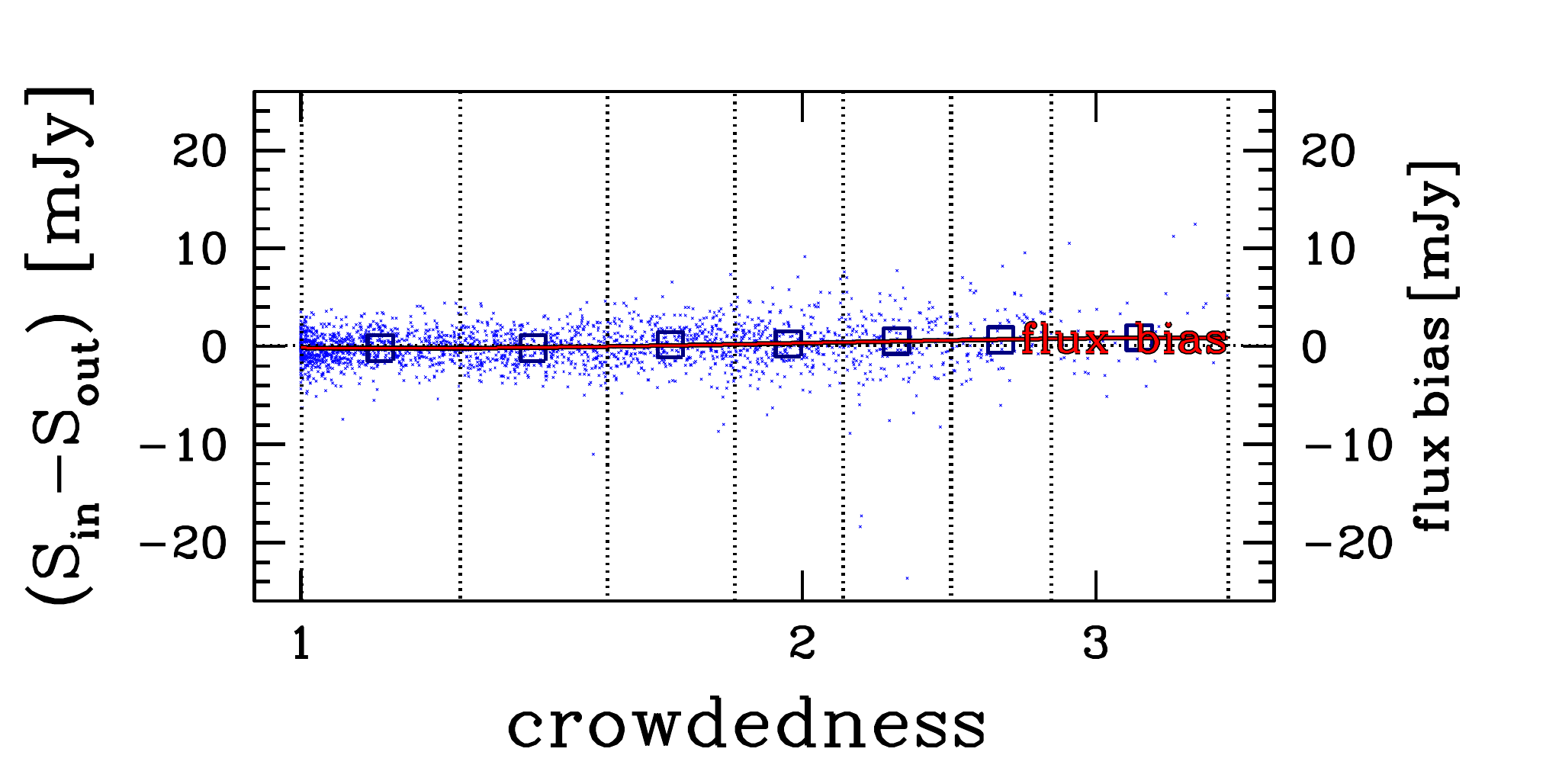}
    \end{subfigure}
    
    \begin{subfigure}[b]{\textwidth}\centering
	\includegraphics[width=0.3\textwidth, trim={1cm 15cm 0cm 2.5cm}, clip]{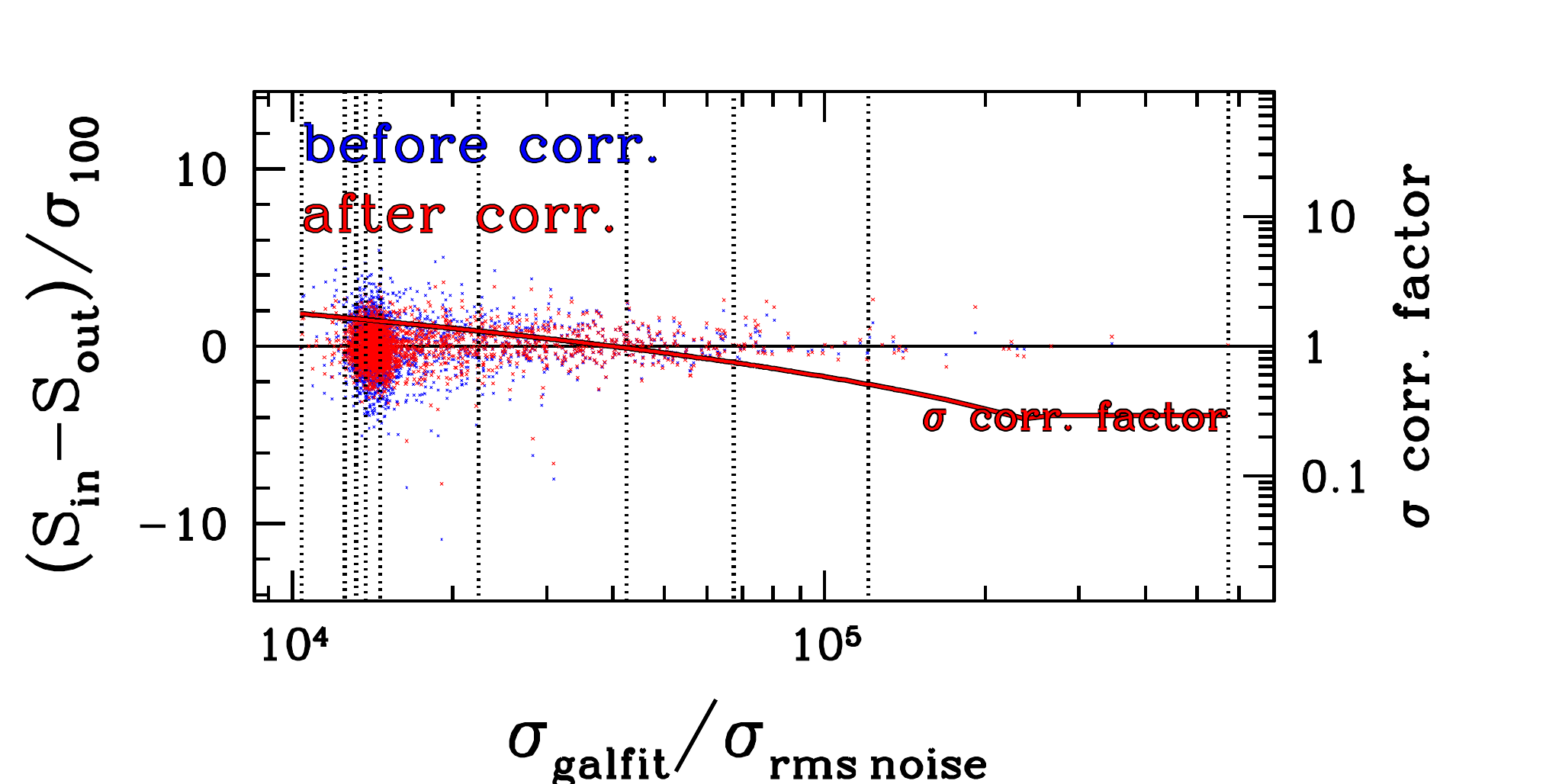}
	\includegraphics[width=0.3\textwidth, trim={1cm 15cm 0cm 2.5cm}, clip]{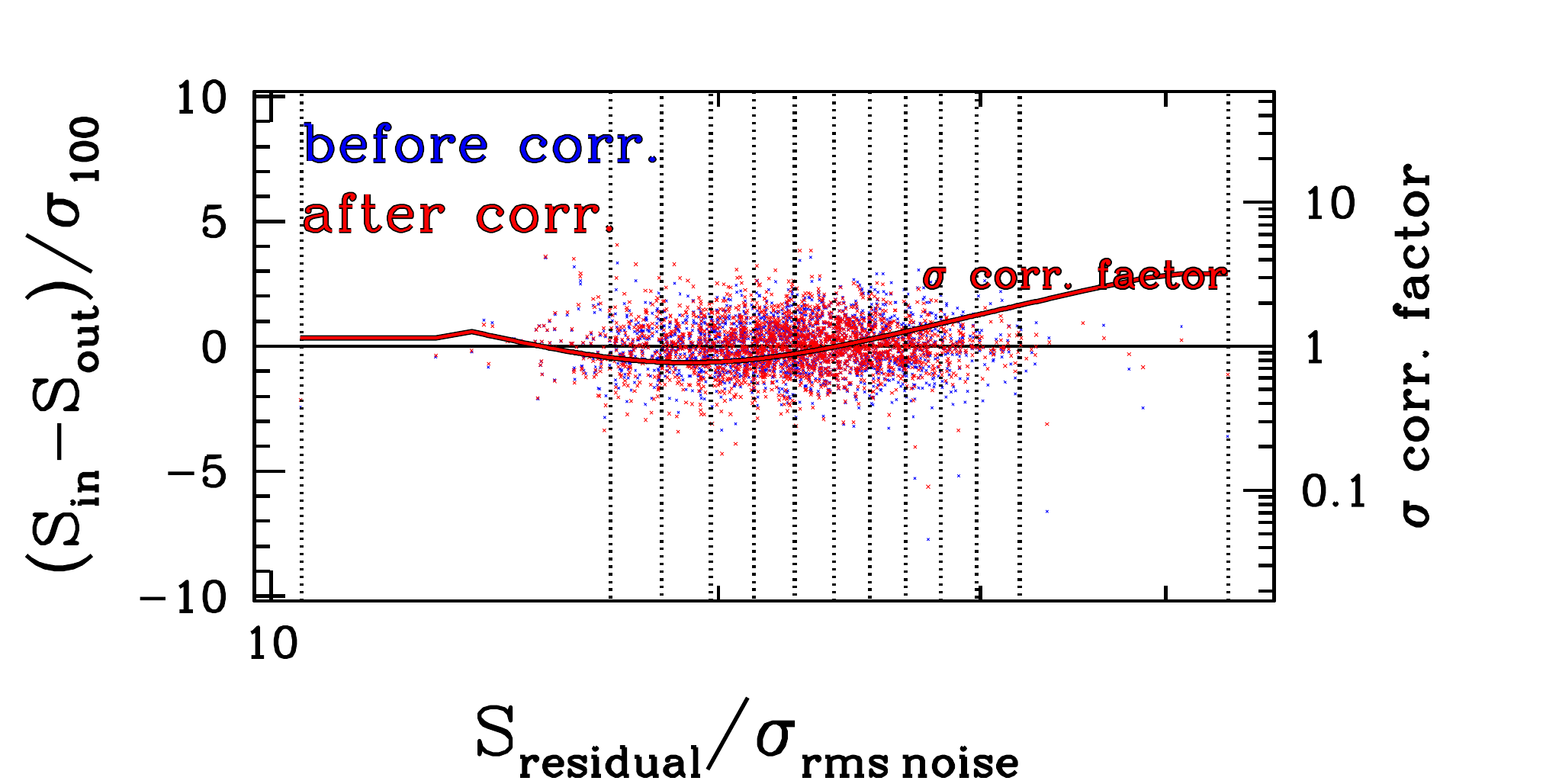}
	\includegraphics[width=0.3\textwidth, trim={1cm 15cm 0cm 2.5cm}, clip]{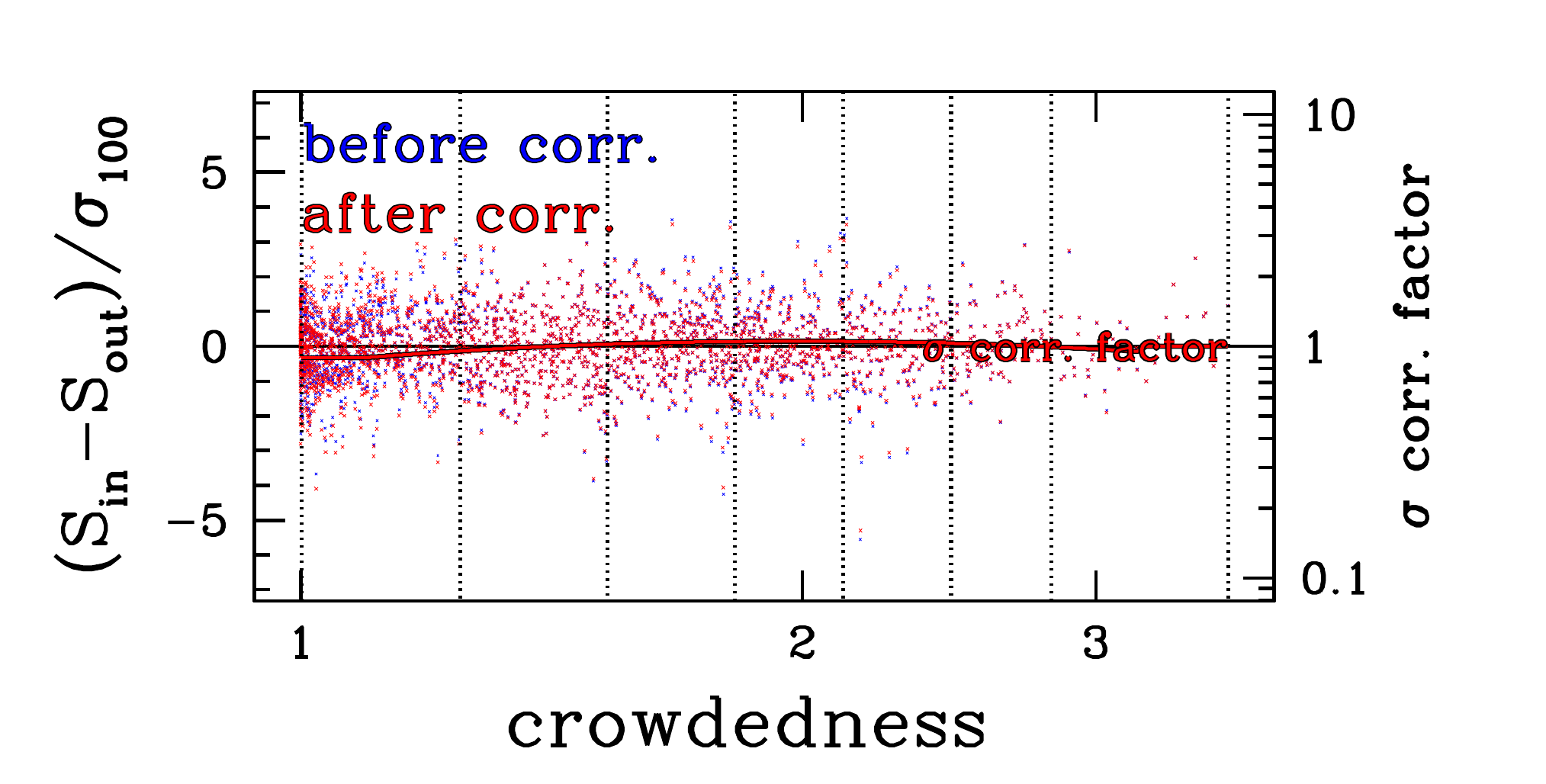}
    \end{subfigure}
    
    \begin{subfigure}[b]{\textwidth}\centering
	\includegraphics[width=0.3\textwidth, trim={1cm 15cm 0cm 2.5cm}, clip]{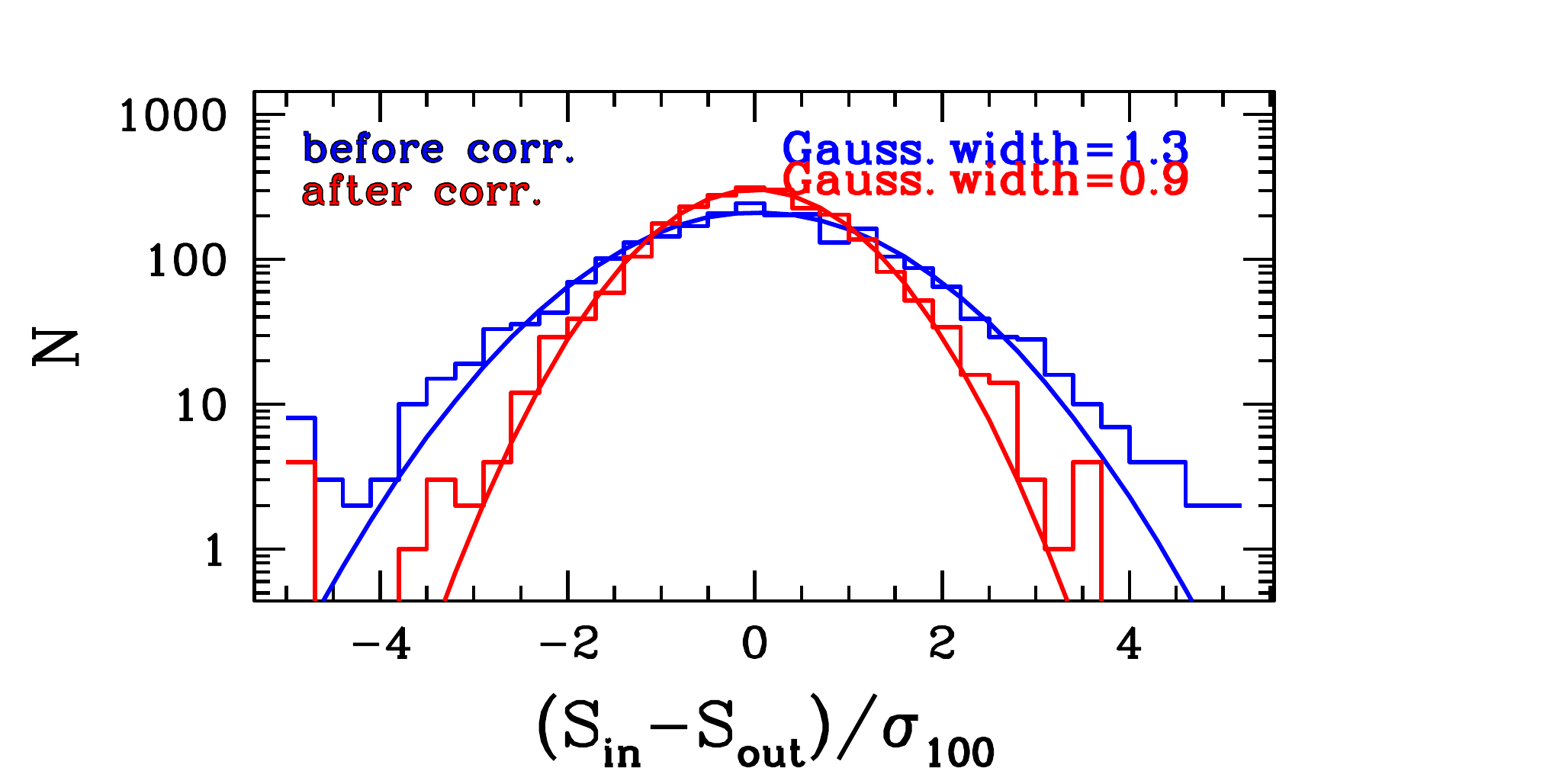}
	\includegraphics[width=0.3\textwidth, trim={1cm 15cm 0cm 2.5cm}, clip]{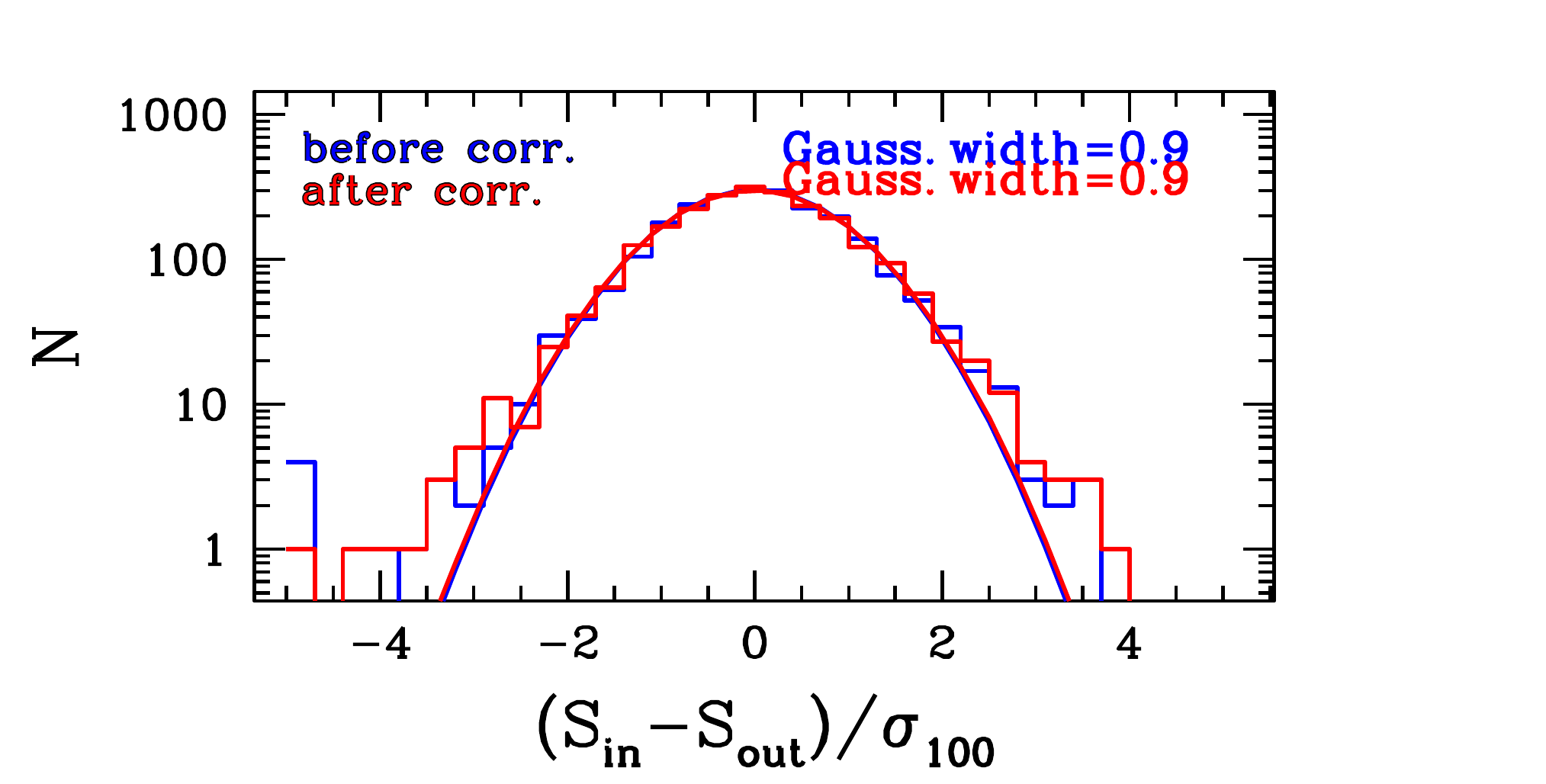}
	\includegraphics[width=0.3\textwidth, trim={1cm 15cm 0cm 2.5cm}, clip]{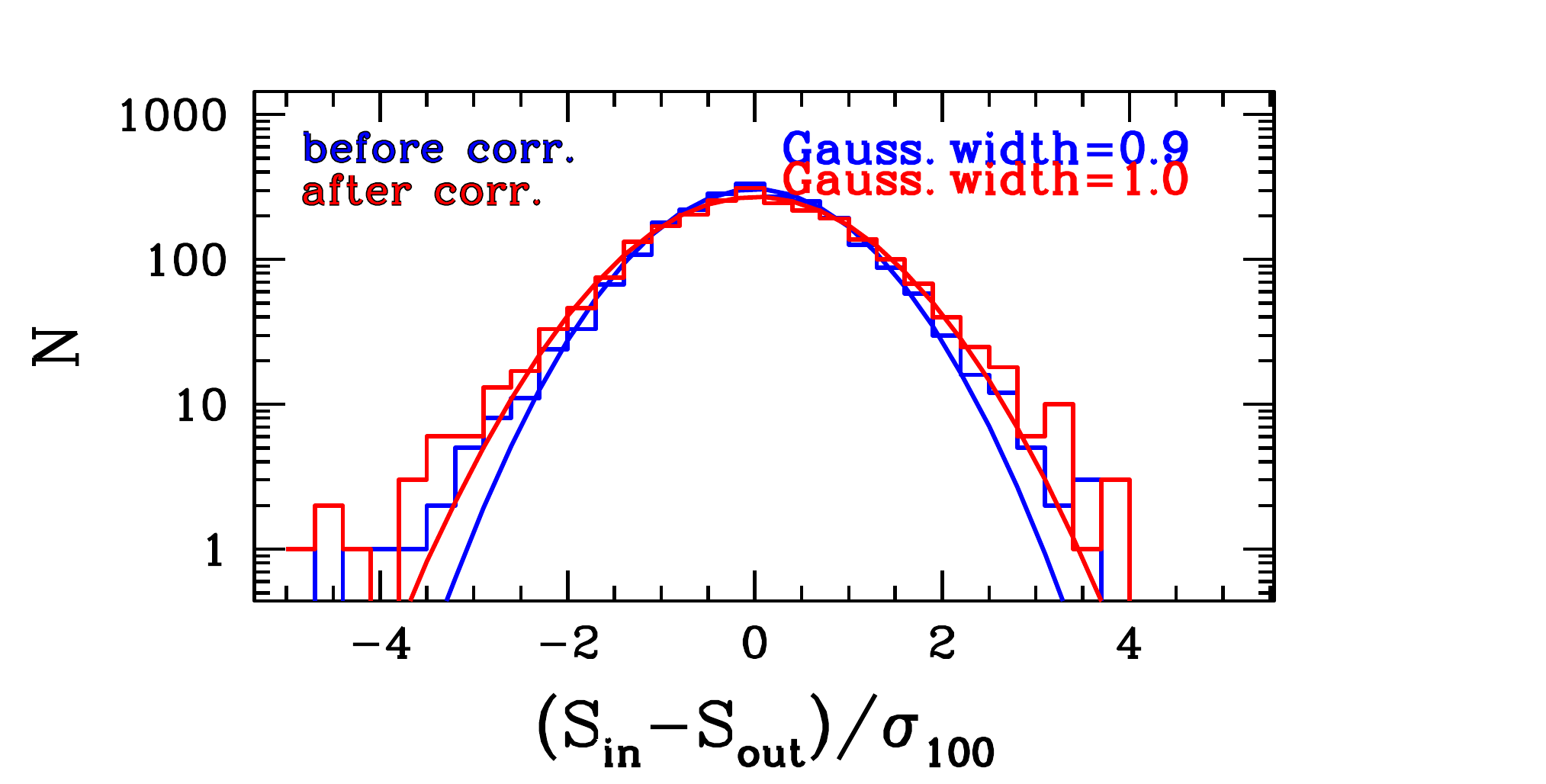}
    \end{subfigure}
    
    \begin{subfigure}[b]{\textwidth}\centering
	\includegraphics[width=0.3\textwidth, trim={1cm 15cm 0cm 2.5cm}, clip]{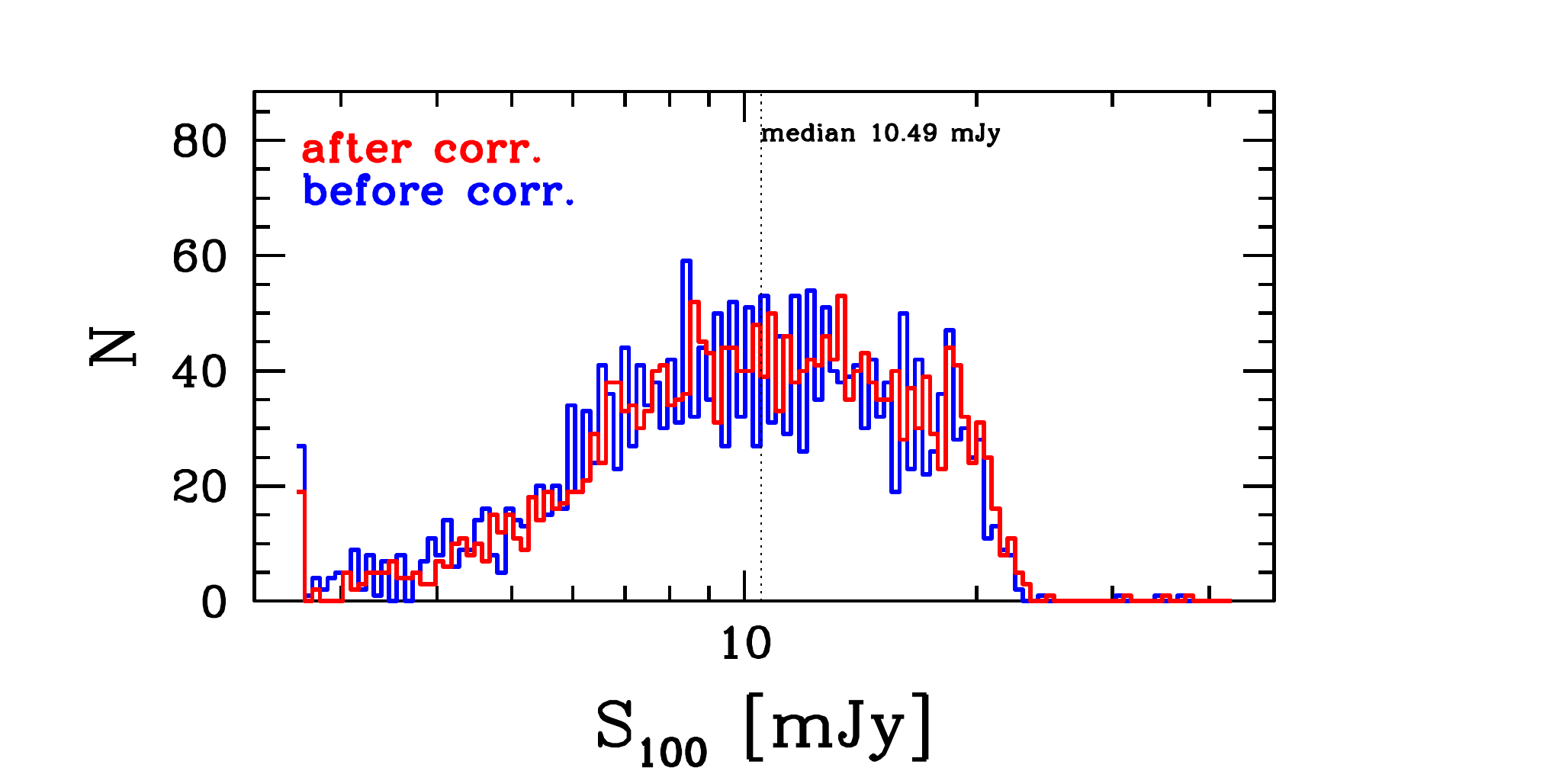}
	\includegraphics[width=0.3\textwidth, trim={1cm 15cm 0cm 2.5cm}, clip]{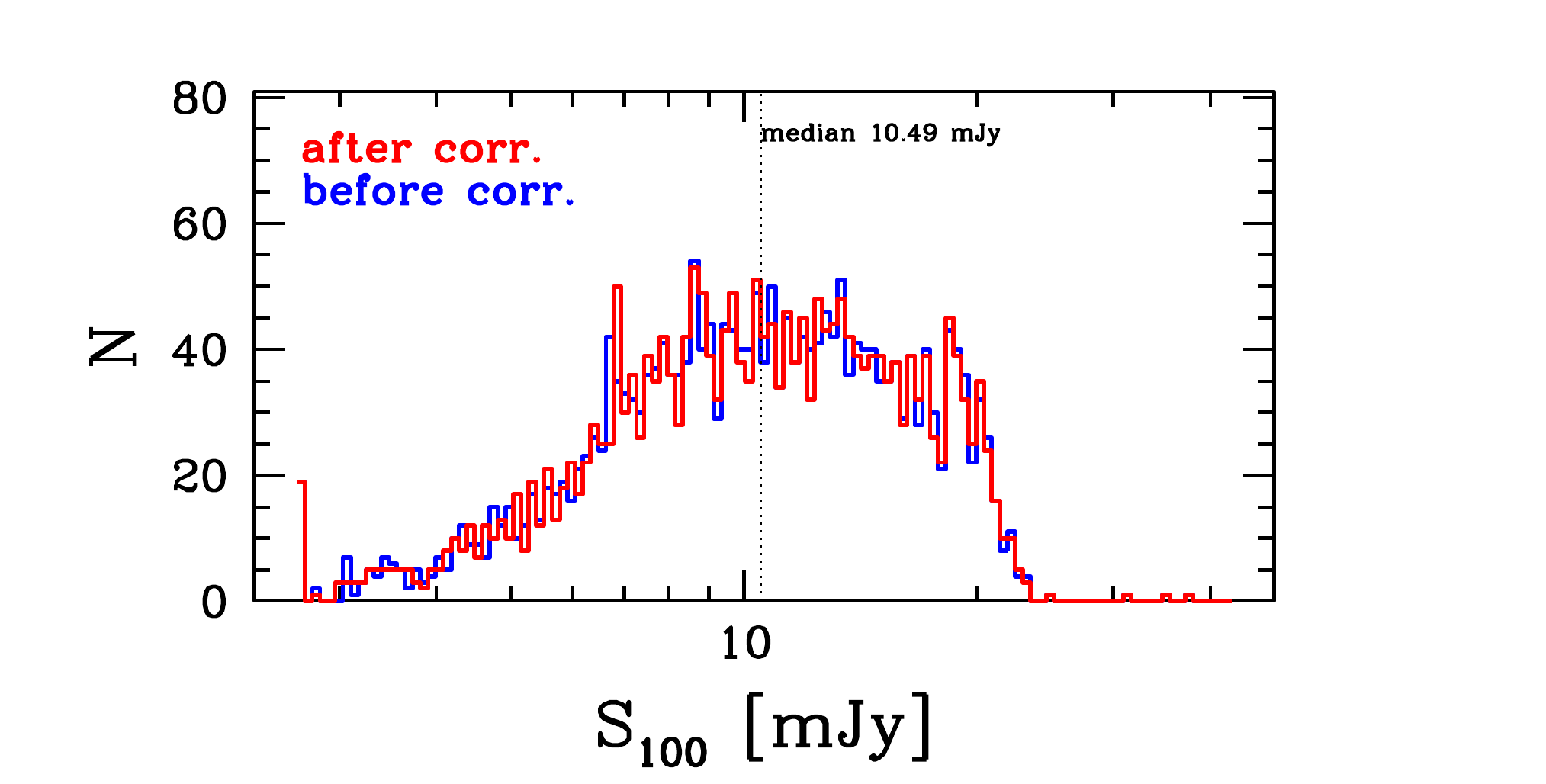}
	\includegraphics[width=0.3\textwidth, trim={1cm 15cm 0cm 2.5cm}, clip]{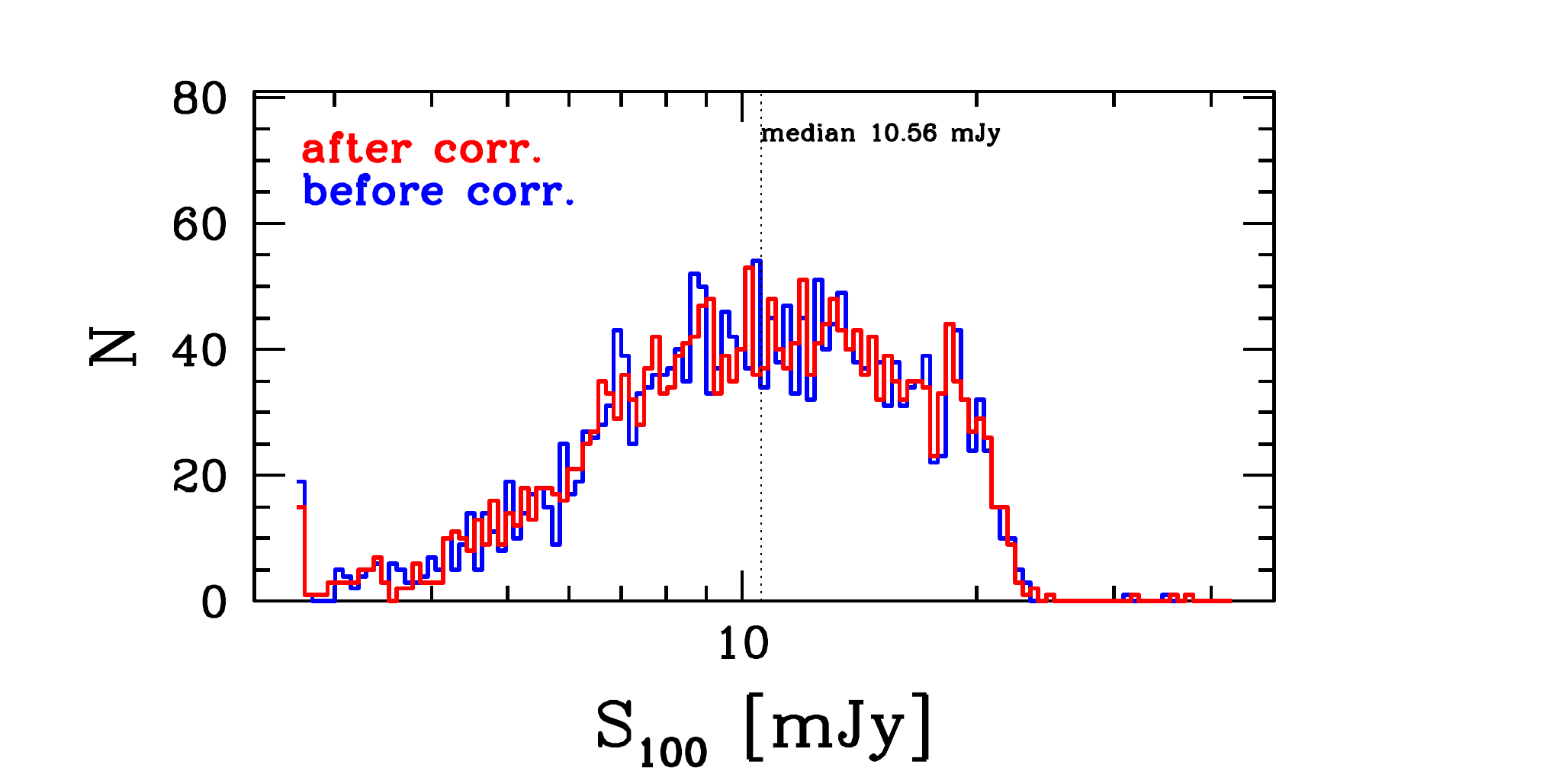}
    \end{subfigure}
    
    \begin{subfigure}[b]{\textwidth}\centering
	\includegraphics[width=0.3\textwidth, trim={1cm 15cm 0cm 2.5cm}, clip]{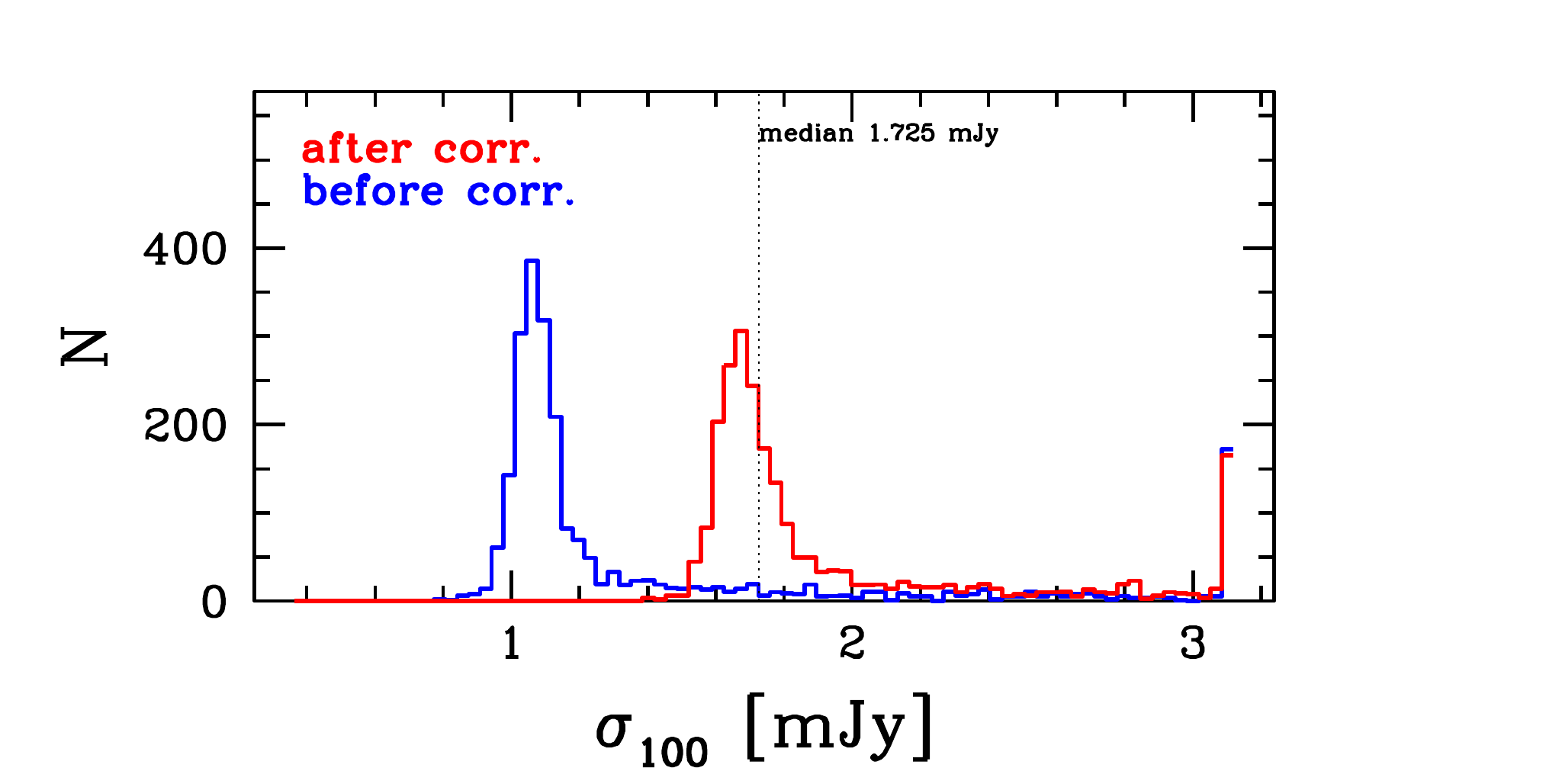}
	\includegraphics[width=0.3\textwidth, trim={1cm 15cm 0cm 2.5cm}, clip]{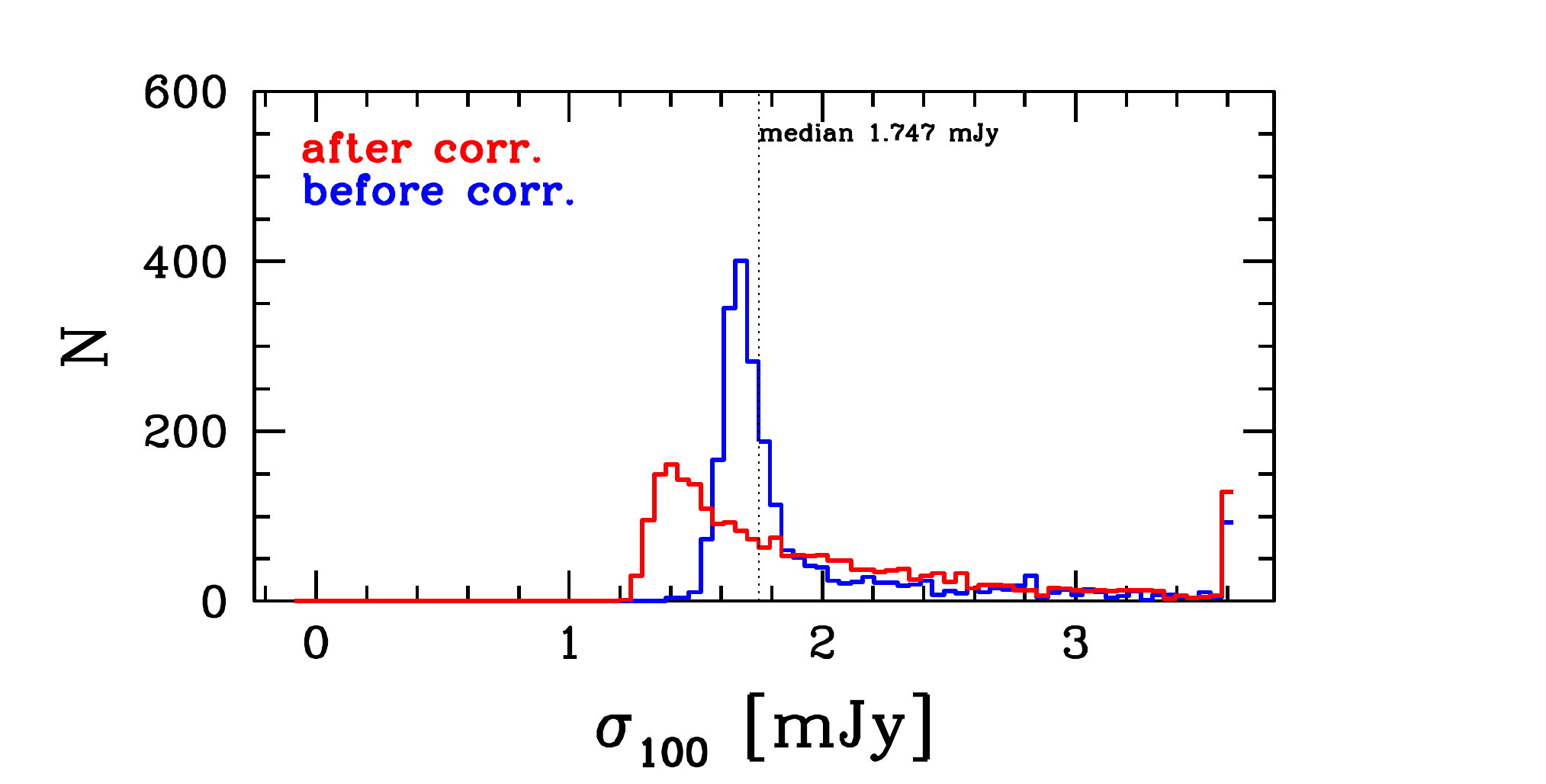}
	\includegraphics[width=0.3\textwidth, trim={1cm 15cm 0cm 2.5cm}, clip]{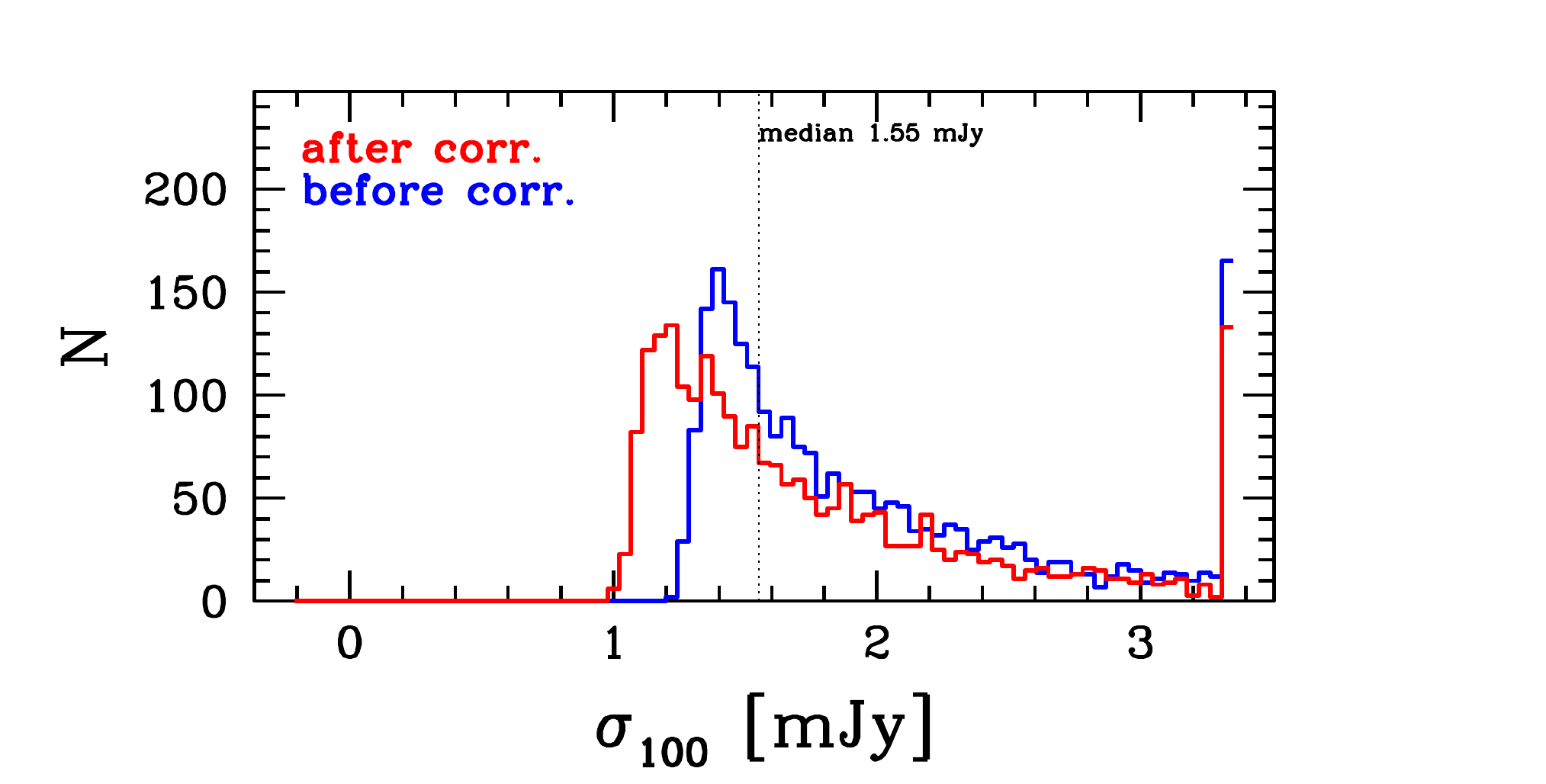}
    \end{subfigure}
    
	\caption{
		Simulation correction analyses at PACS 100~$\mu$m. See descriptions in text. 
		\label{Figure_galsim_100_bin}
	}
\end{figure}

\begin{figure}
	\centering
    
    \begin{subfigure}[b]{\textwidth}\centering
	\includegraphics[width=0.3\textwidth, trim={1cm 15cm 0cm 2.5cm}, clip]{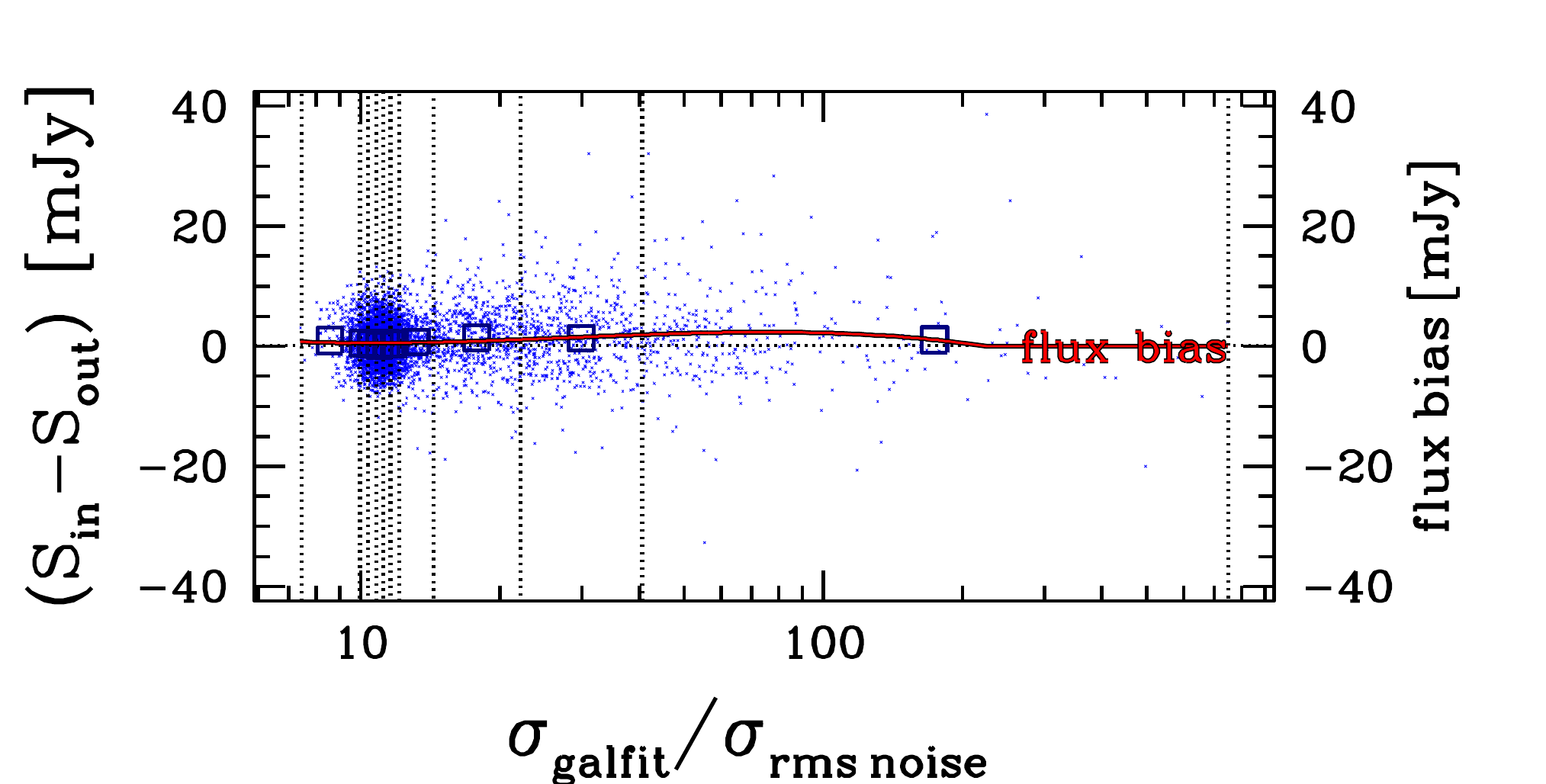}
	\includegraphics[width=0.3\textwidth, trim={1cm 15cm 0cm 2.5cm}, clip]{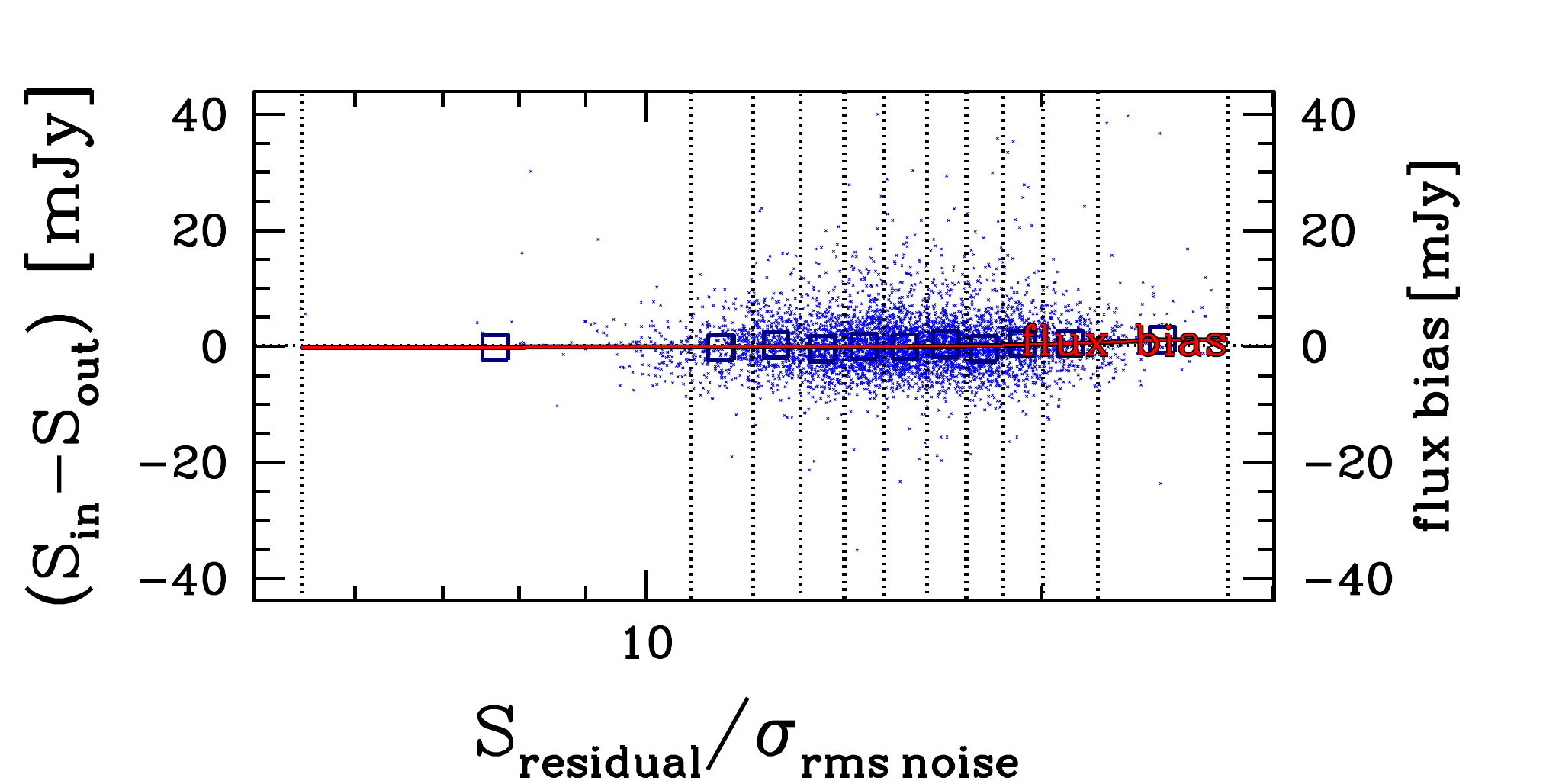}
	\includegraphics[width=0.3\textwidth, trim={1cm 15cm 0cm 2.5cm}, clip]{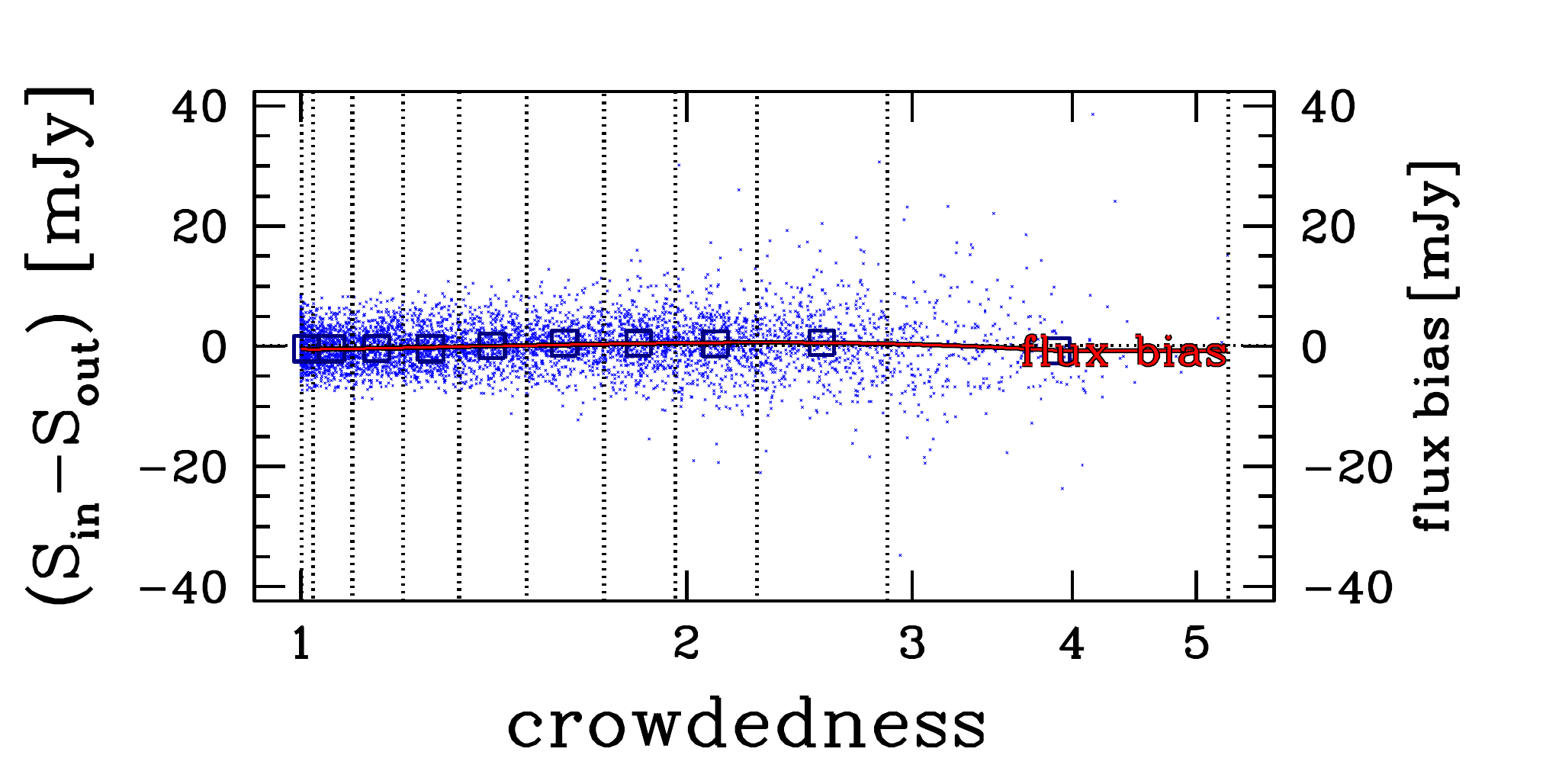}
    \end{subfigure}
    
    \begin{subfigure}[b]{\textwidth}\centering
	\includegraphics[width=0.3\textwidth, trim={1cm 15cm 0cm 2.5cm}, clip]{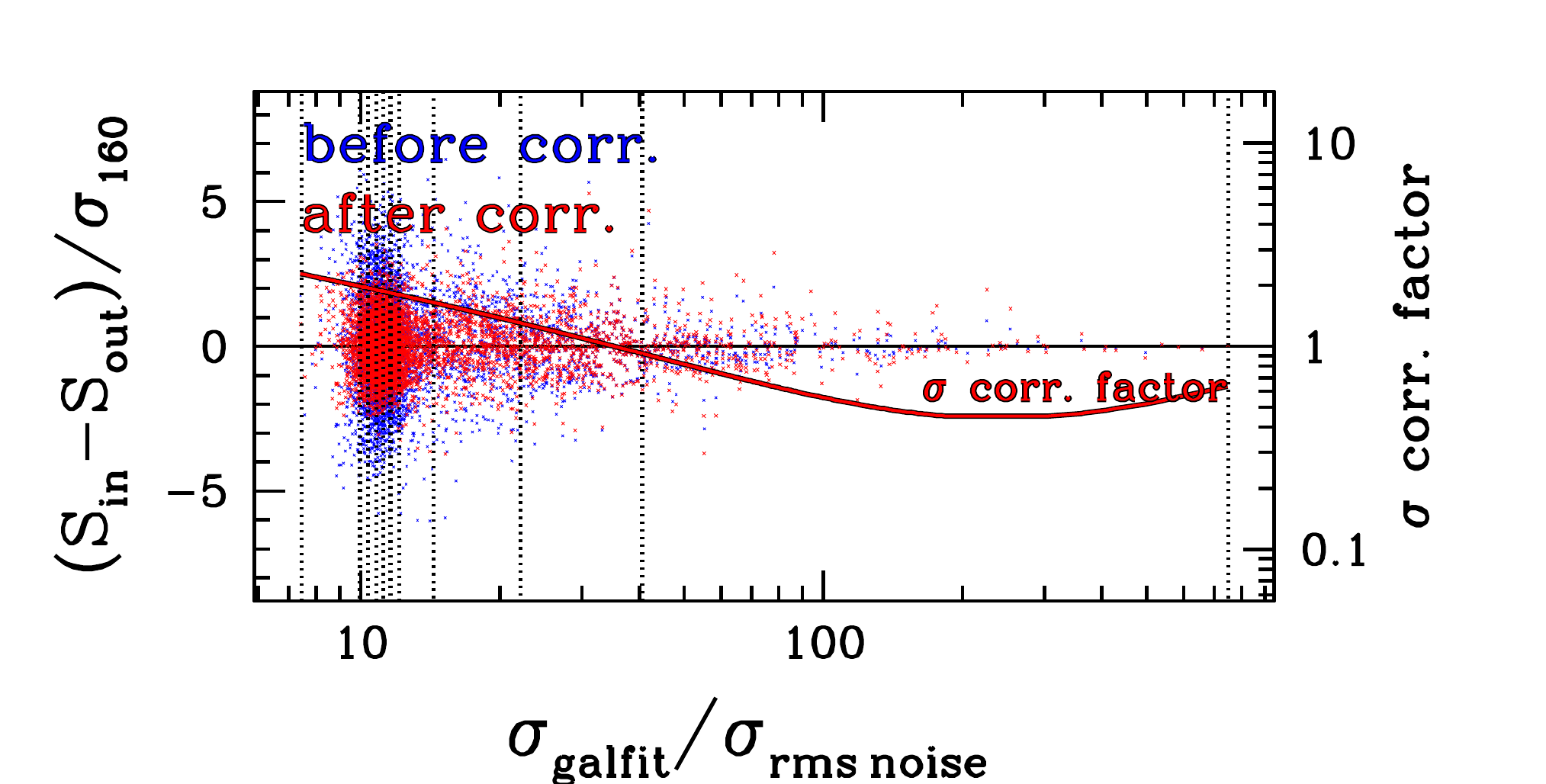}
	\includegraphics[width=0.3\textwidth, trim={1cm 15cm 0cm 2.5cm}, clip]{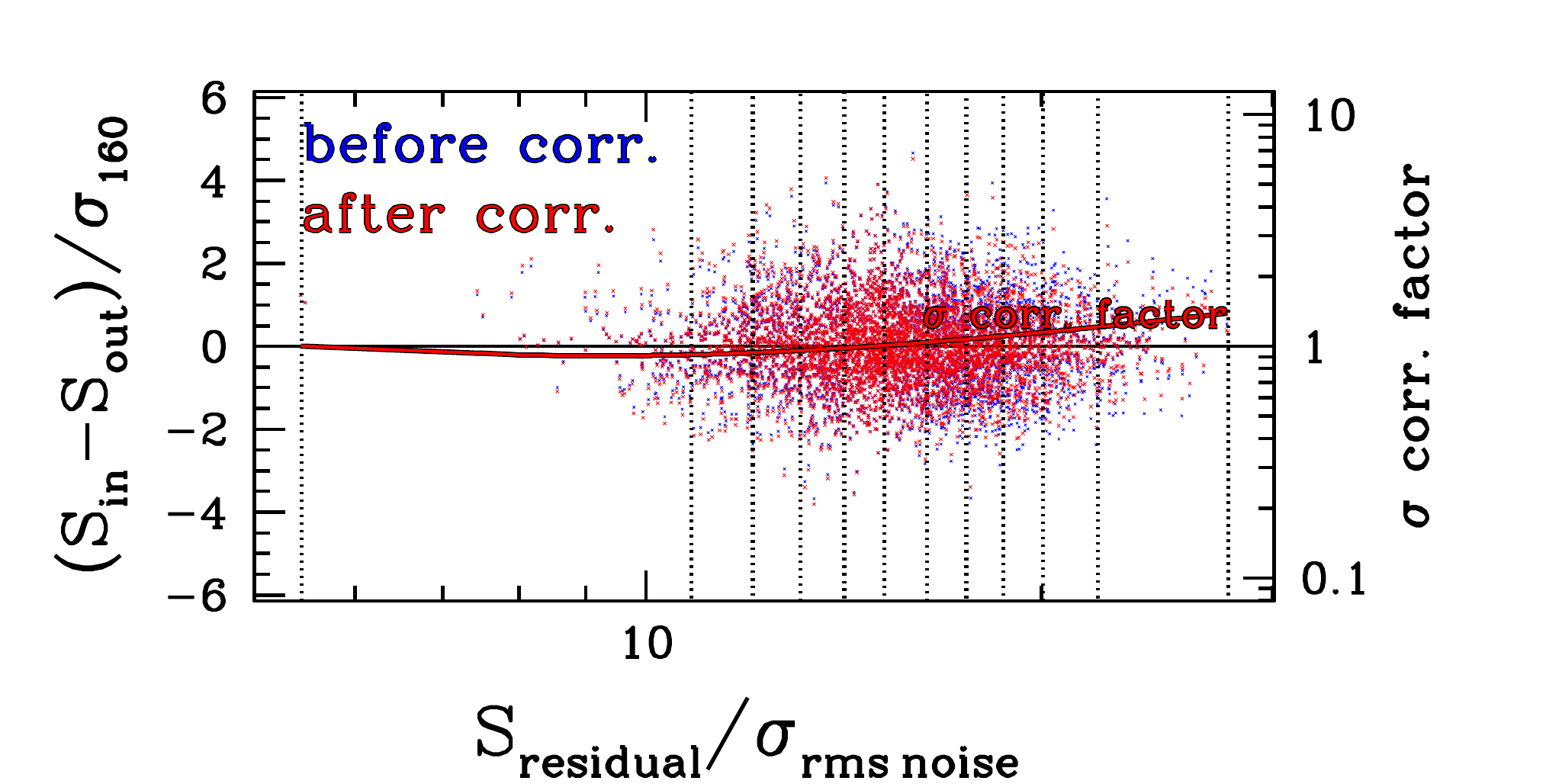}
	\includegraphics[width=0.3\textwidth, trim={1cm 15cm 0cm 2.5cm}, clip]{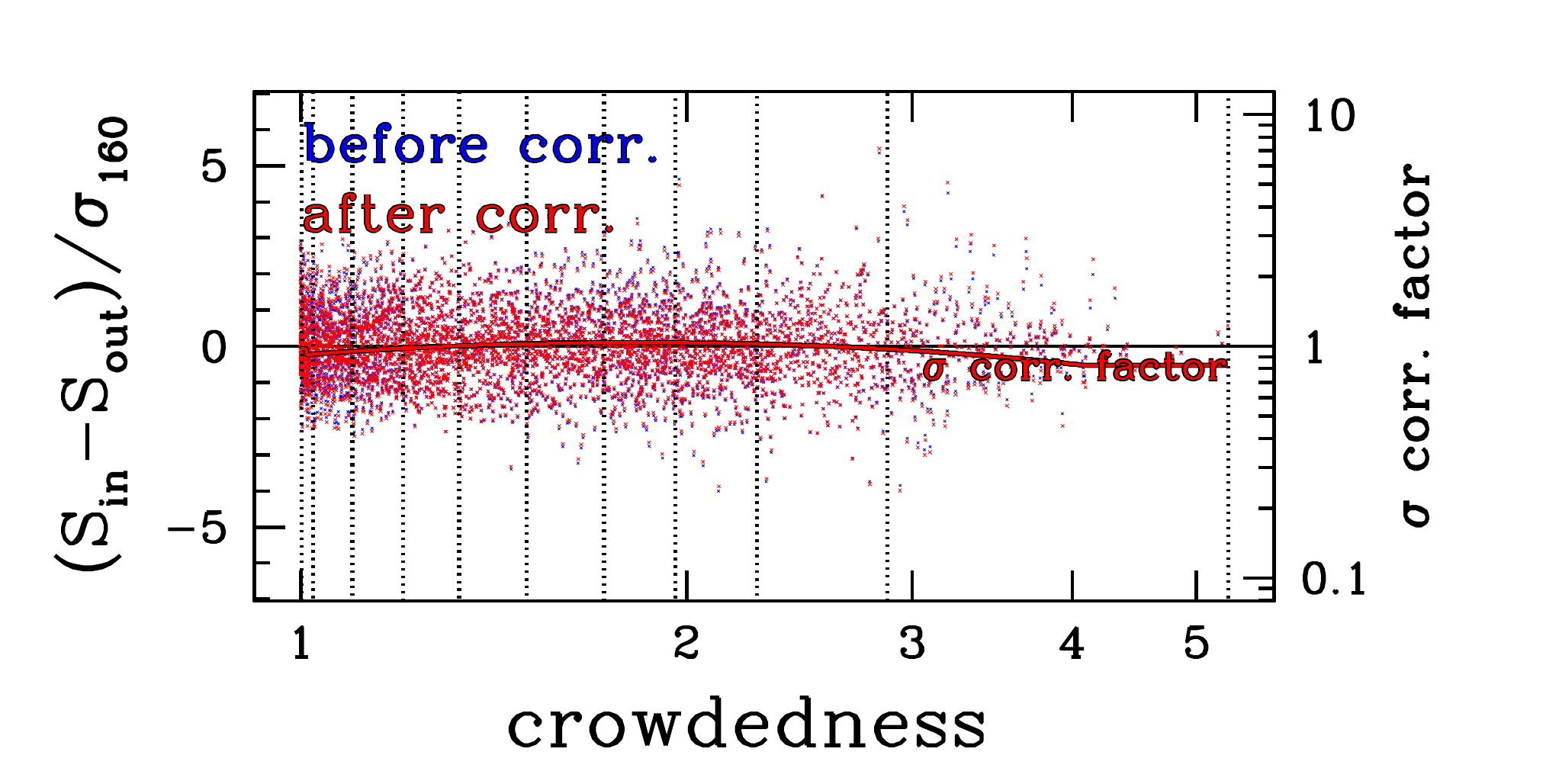}
    \end{subfigure}
	
    \begin{subfigure}[b]{\textwidth}\centering
	\includegraphics[width=0.3\textwidth, trim={1cm 15cm 0cm 2.5cm}, clip]{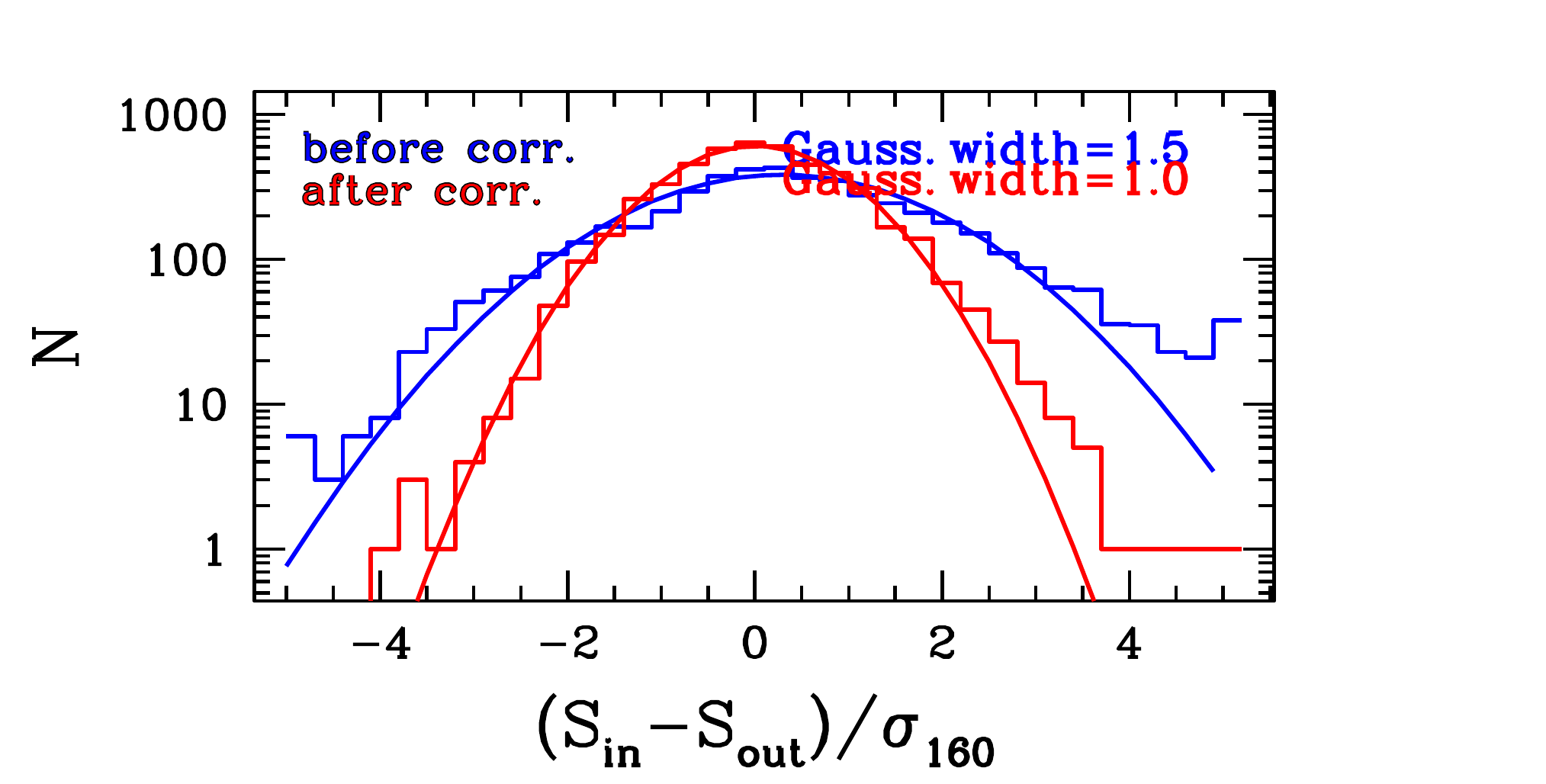}
	\includegraphics[width=0.3\textwidth, trim={1cm 15cm 0cm 2.5cm}, clip]{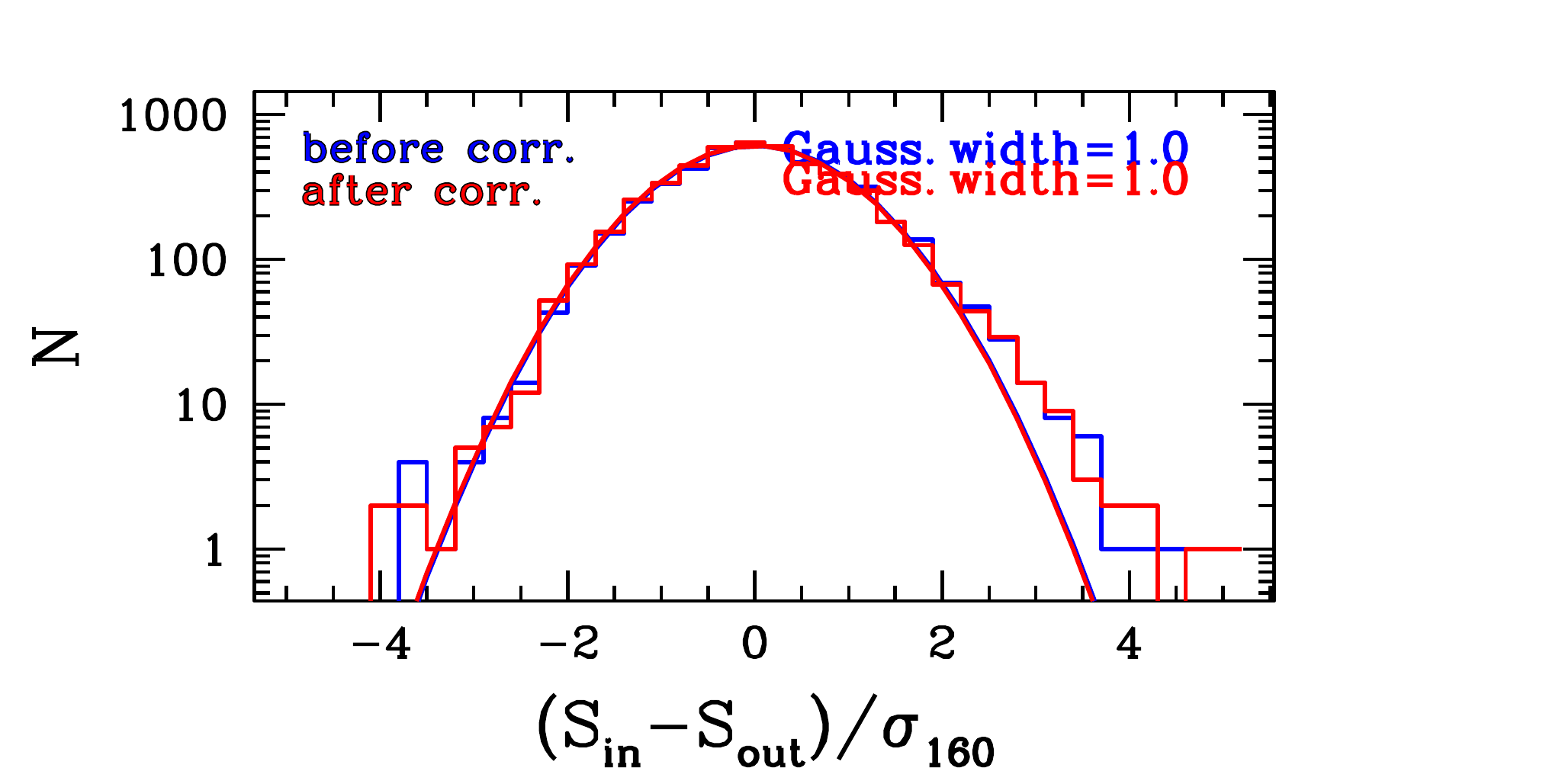}
	\includegraphics[width=0.3\textwidth, trim={1cm 15cm 0cm 2.5cm}, clip]{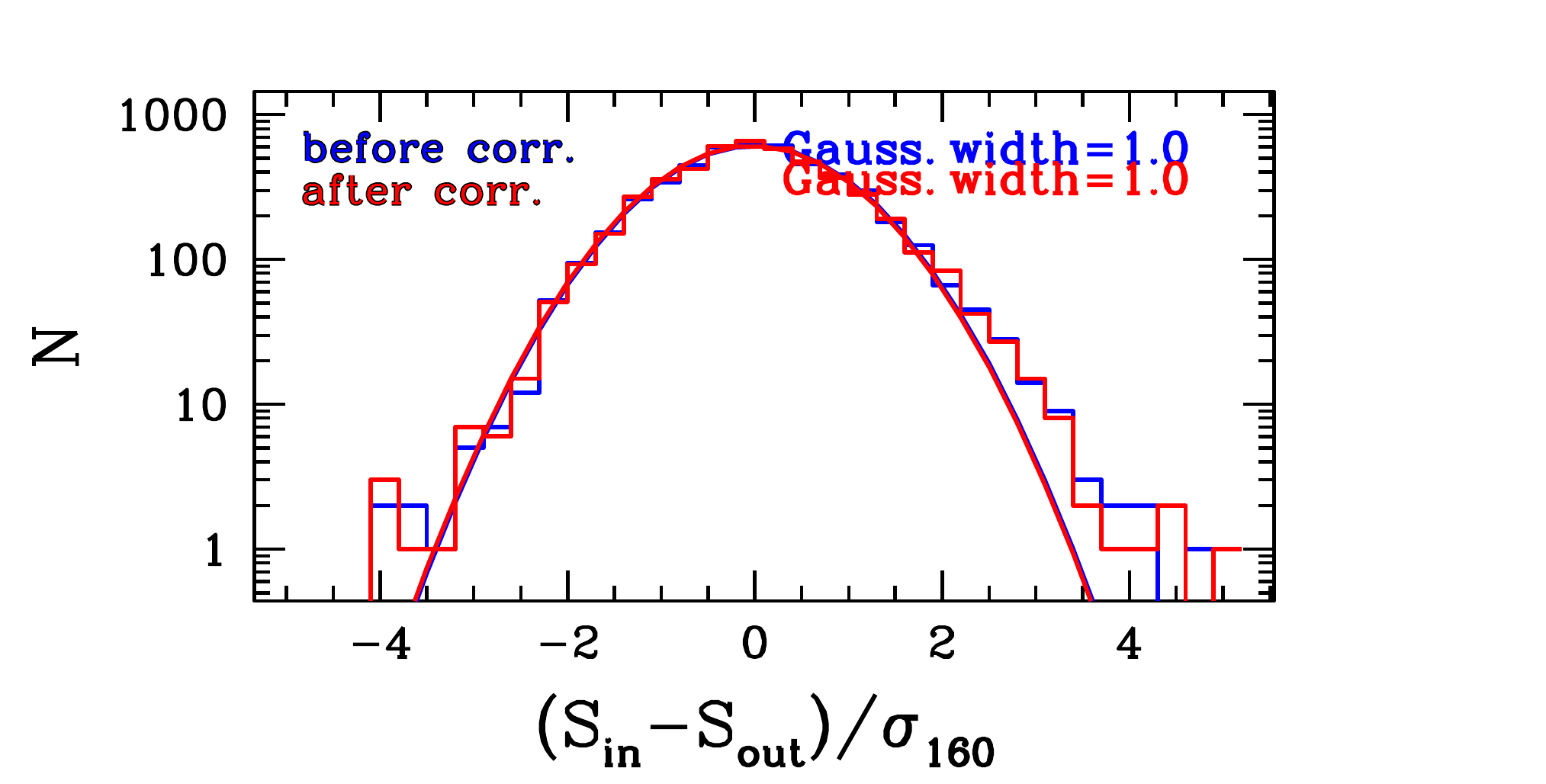}
    \end{subfigure}
    
    \begin{subfigure}[b]{\textwidth}\centering
	\includegraphics[width=0.3\textwidth, trim={1cm 15cm 0cm 2.5cm}, clip]{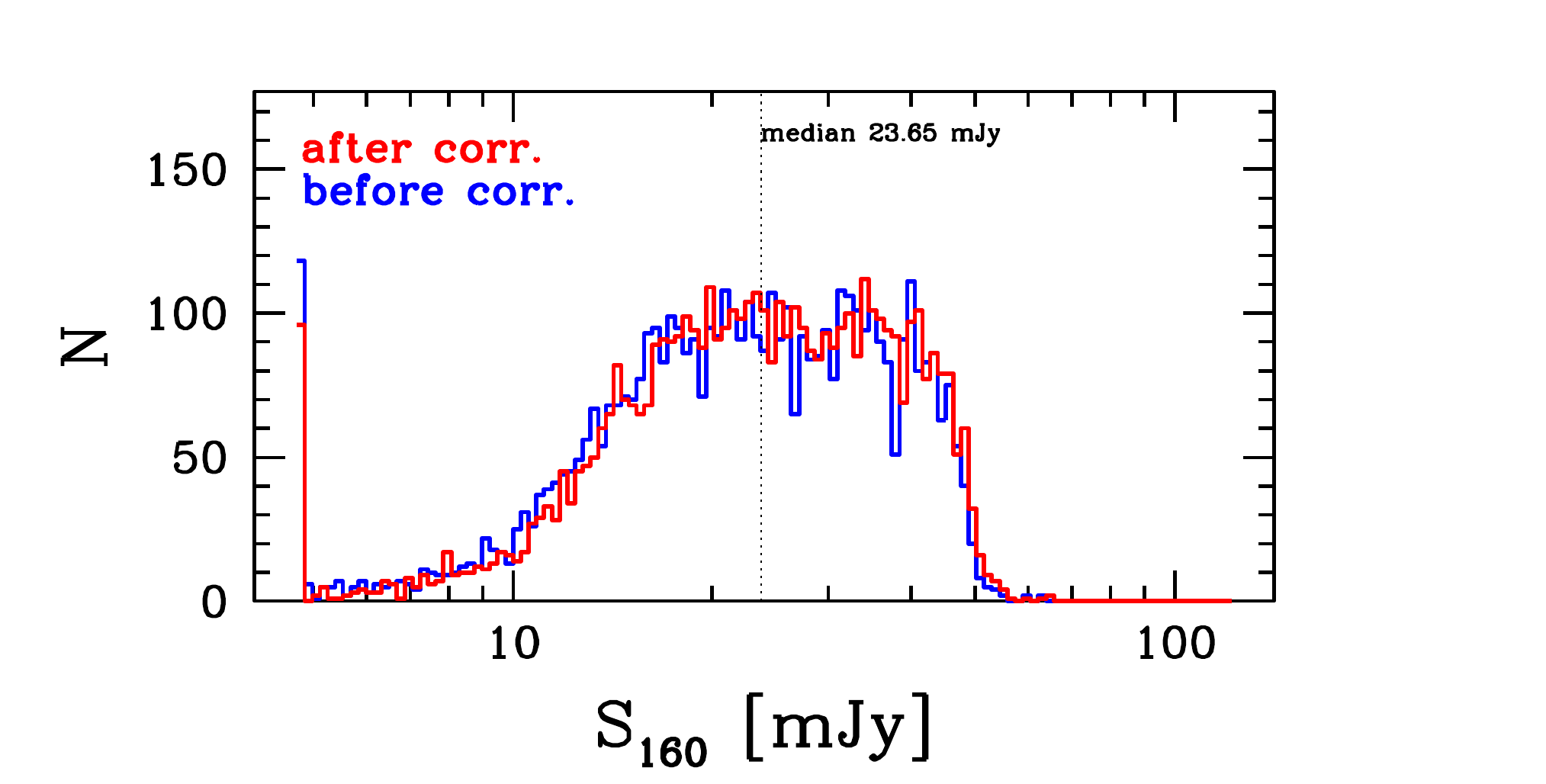}
	\includegraphics[width=0.3\textwidth, trim={1cm 15cm 0cm 2.5cm}, clip]{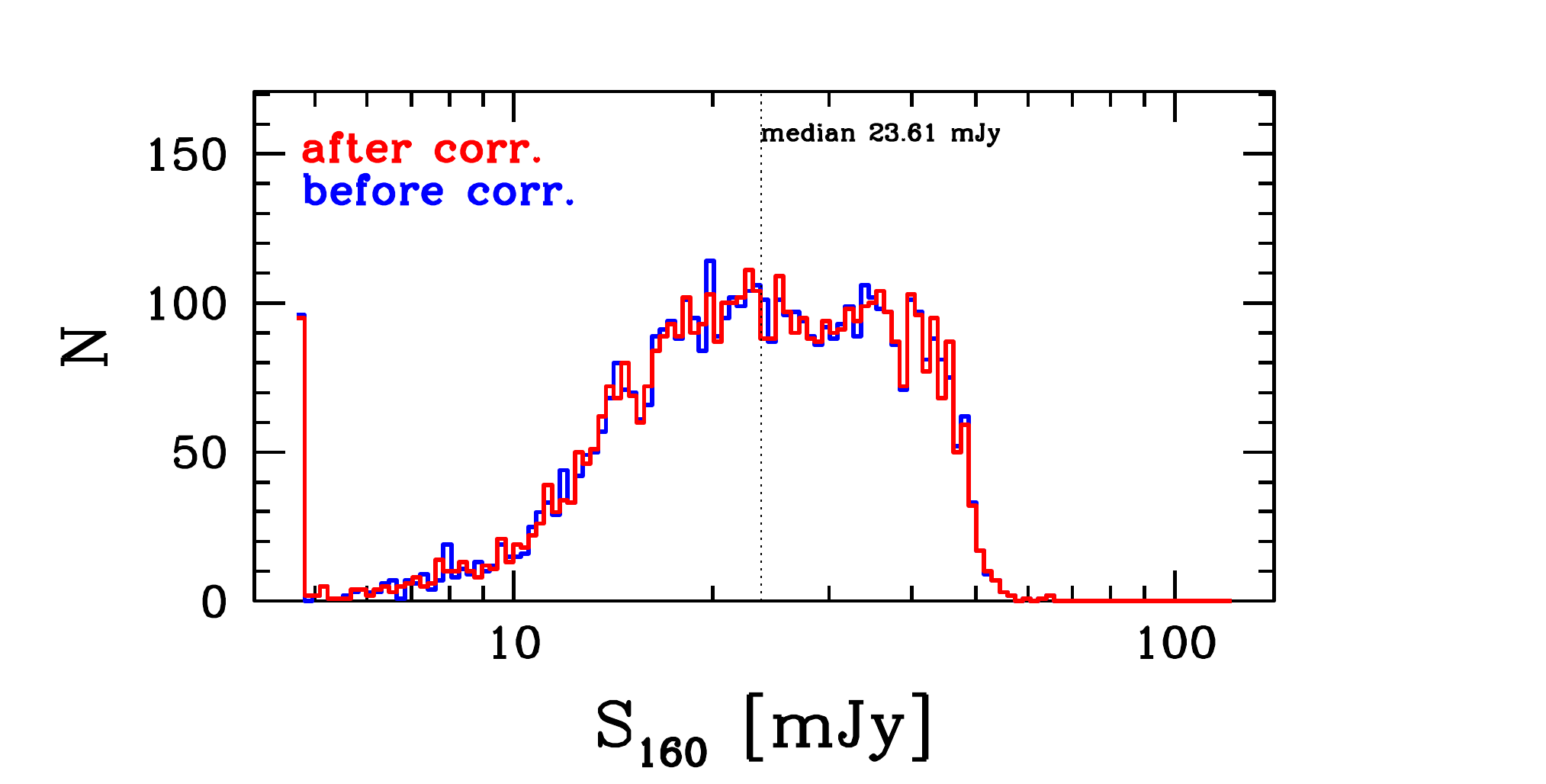}
	\includegraphics[width=0.3\textwidth, trim={1cm 15cm 0cm 2.5cm}, clip]{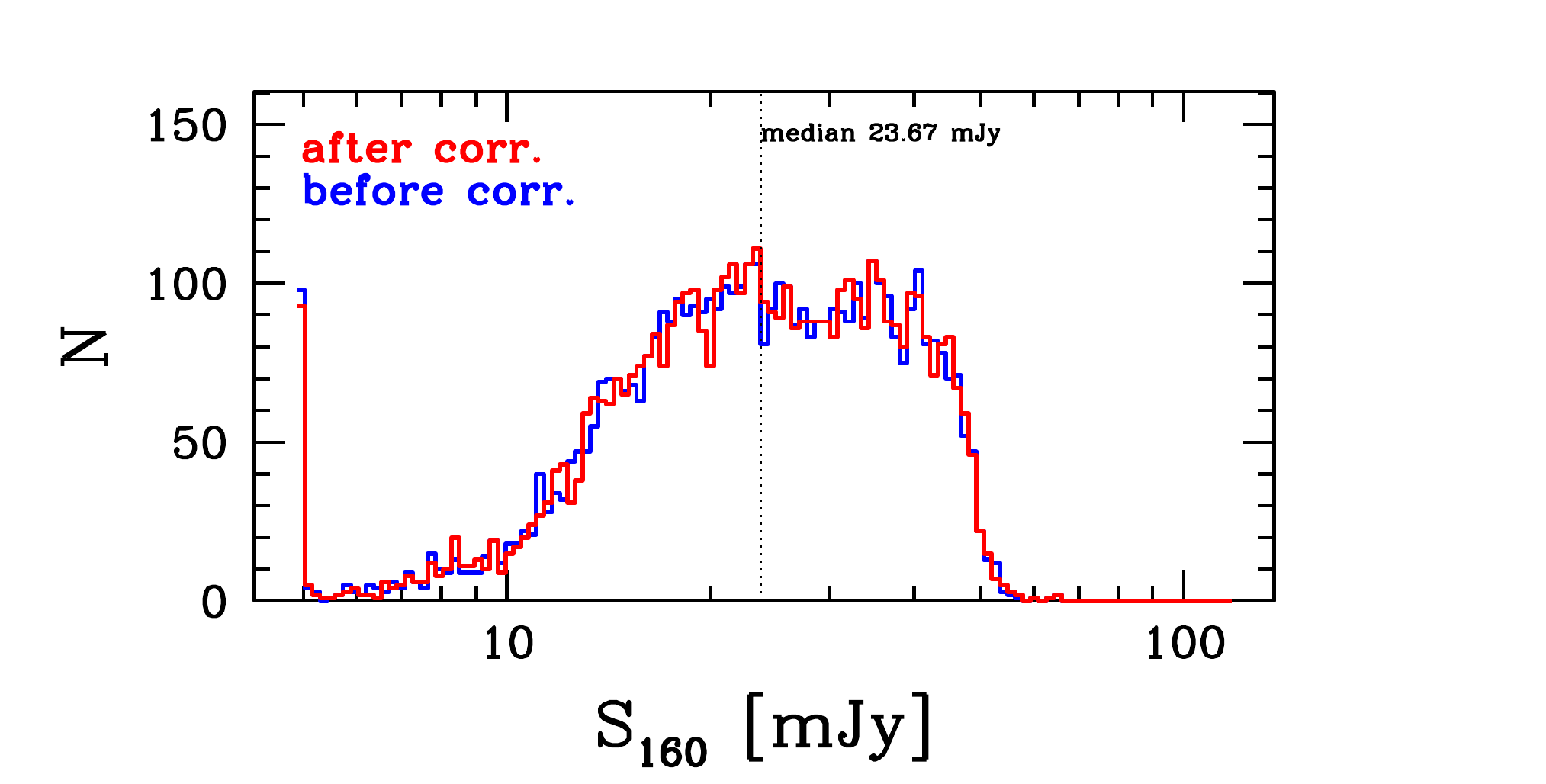}
    \end{subfigure}
    
    \begin{subfigure}[b]{\textwidth}\centering
	\includegraphics[width=0.3\textwidth, trim={1cm 15cm 0cm 2.5cm}, clip]{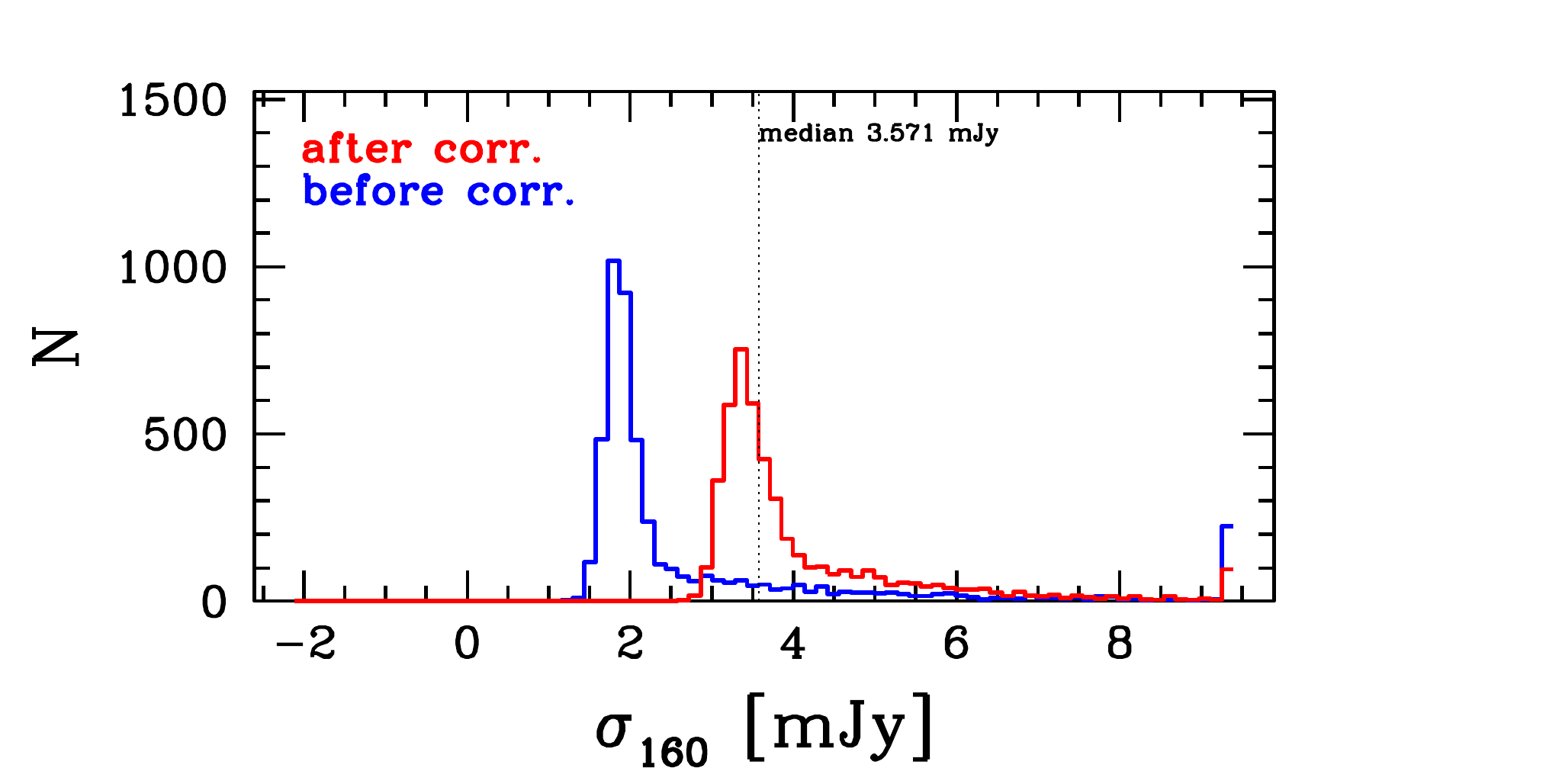}
	\includegraphics[width=0.3\textwidth, trim={1cm 15cm 0cm 2.5cm}, clip]{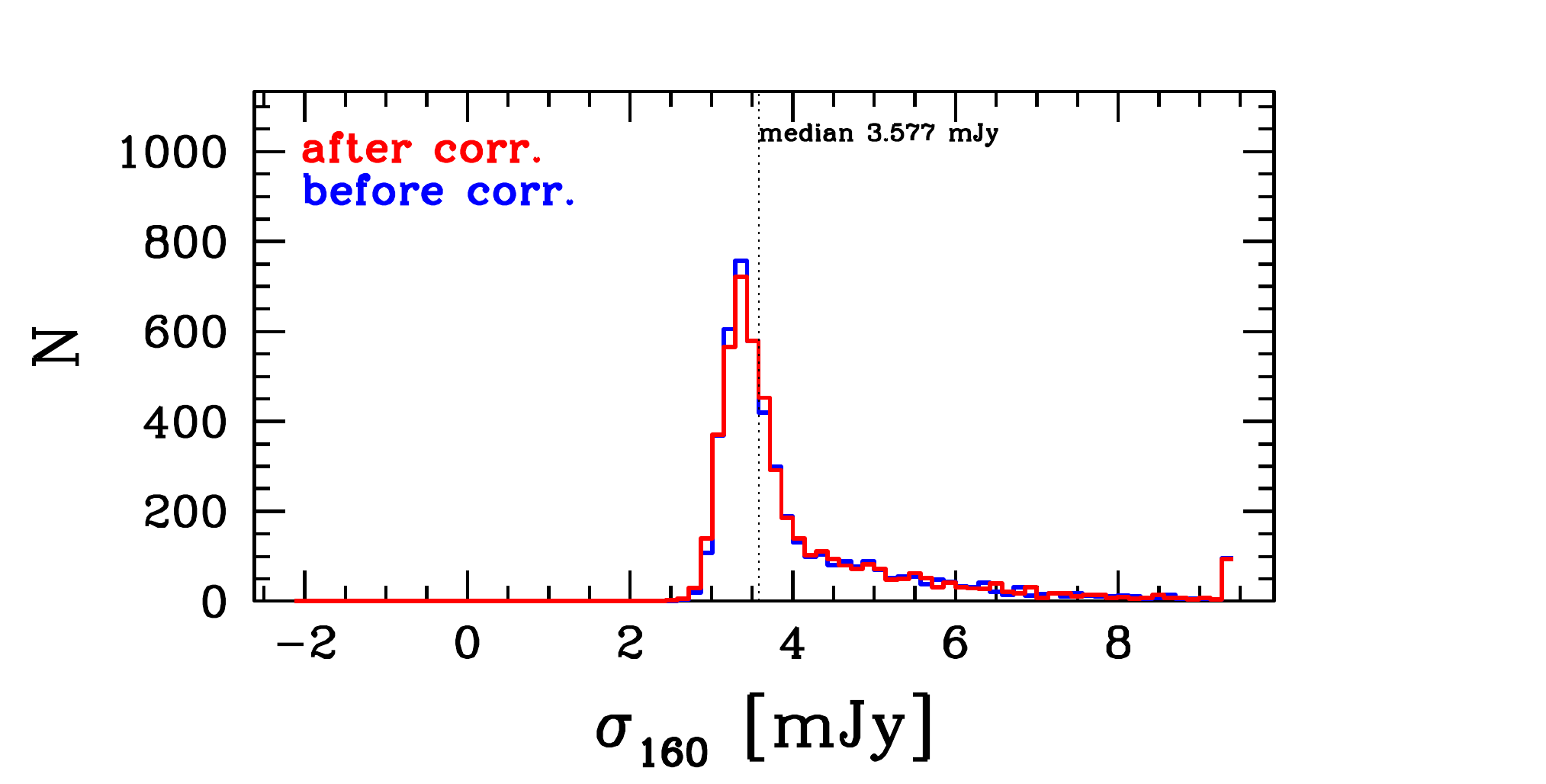}
	\includegraphics[width=0.3\textwidth, trim={1cm 15cm 0cm 2.5cm}, clip]{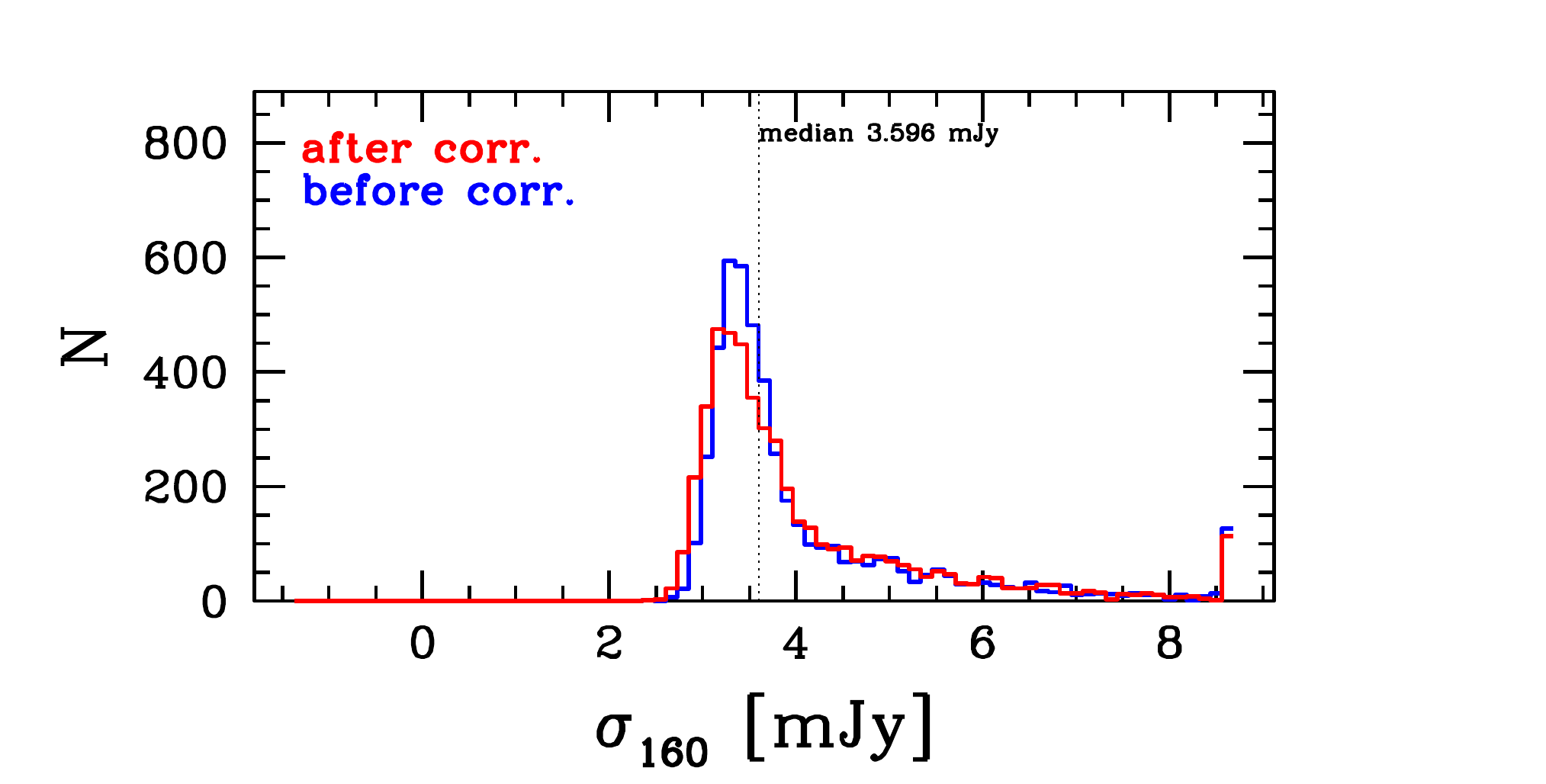}
    \end{subfigure}
    
	\caption{%
		Simulation correction analyses at PACS 160~$\mu$m. See descriptions in text. 
        \label{Figure_galsim_160_bin}
	}
\end{figure}

\begin{figure}
	\centering
    
    \begin{subfigure}[b]{\textwidth}\centering
	\includegraphics[width=0.23\textwidth, trim={1cm 15cm 0cm 2.5cm}, clip]{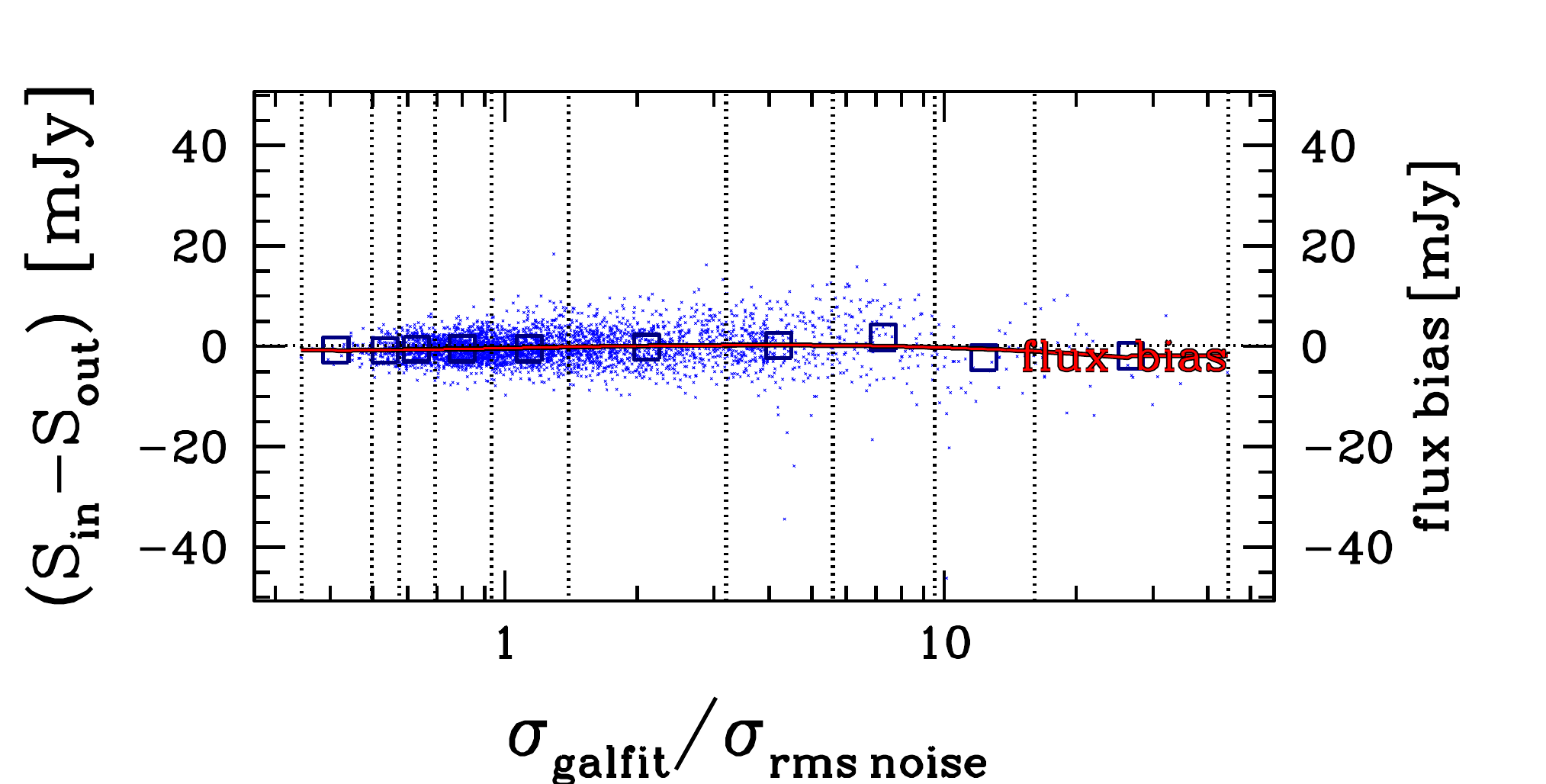}
	\includegraphics[width=0.23\textwidth, trim={1cm 15cm 0cm 2.5cm}, clip]{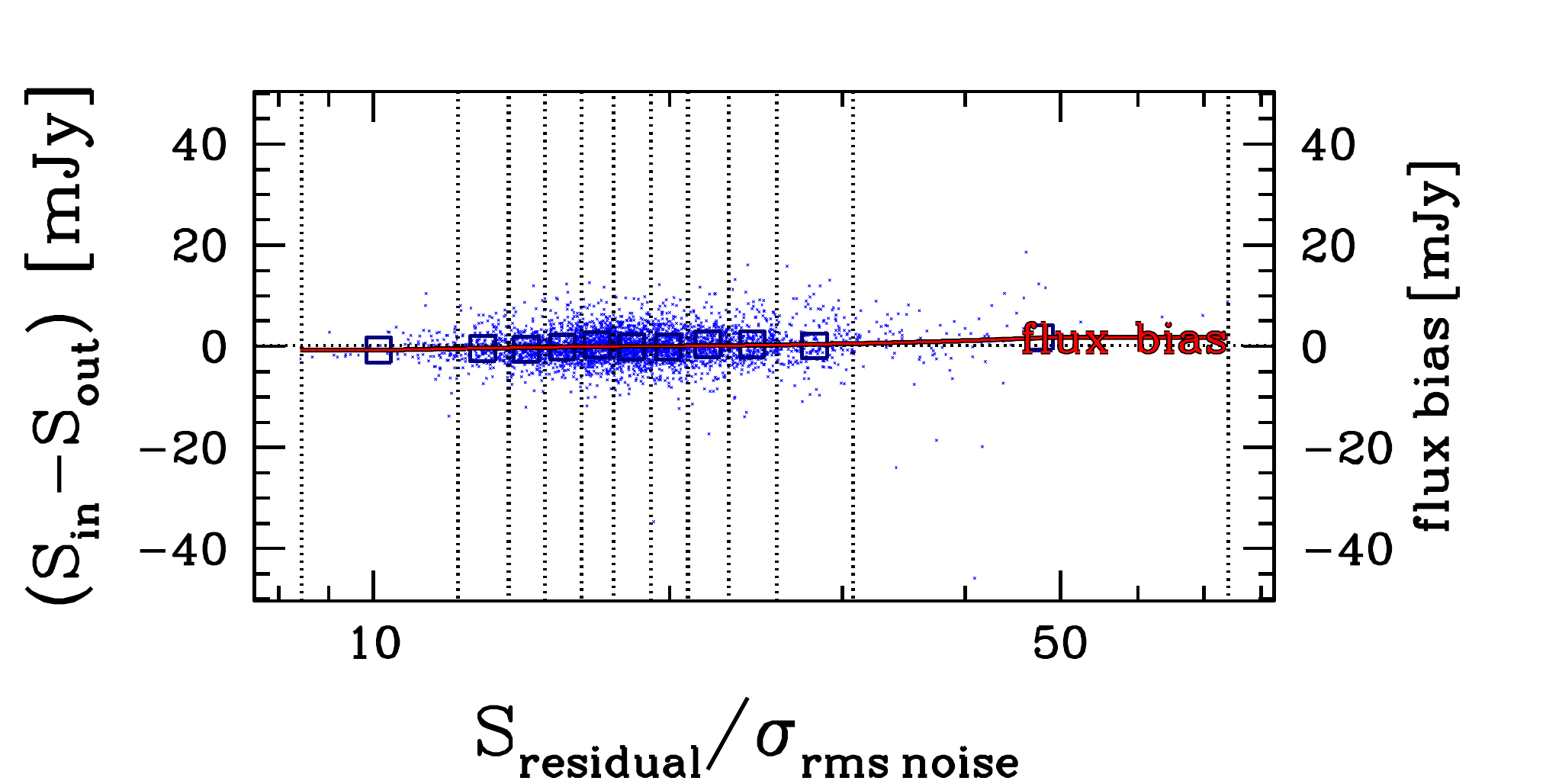}
	\includegraphics[width=0.23\textwidth, trim={1cm 15cm 0cm 2.5cm}, clip]{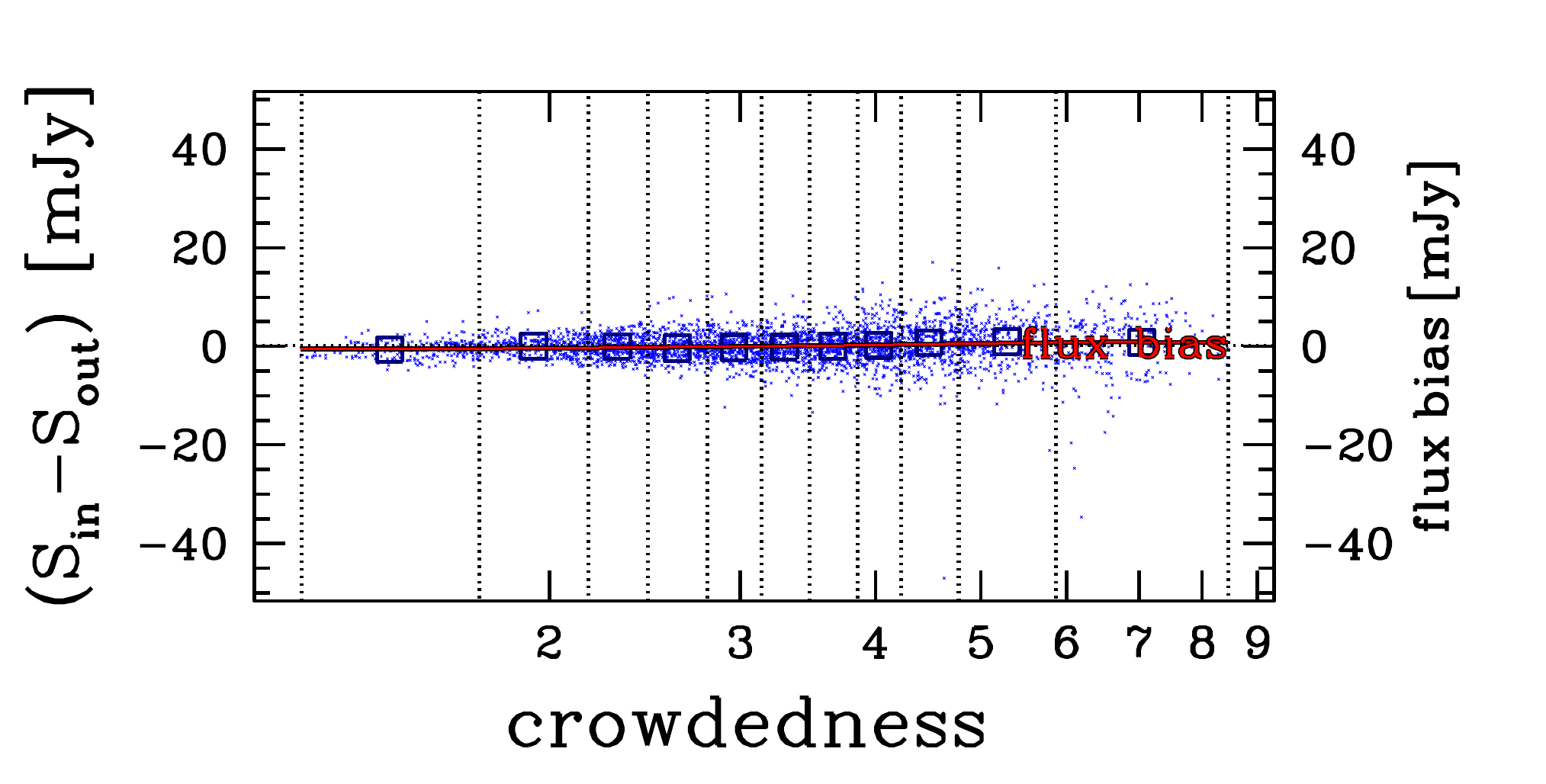}
	\includegraphics[width=0.23\textwidth, trim={1cm 15cm 0cm 2.5cm}, clip]{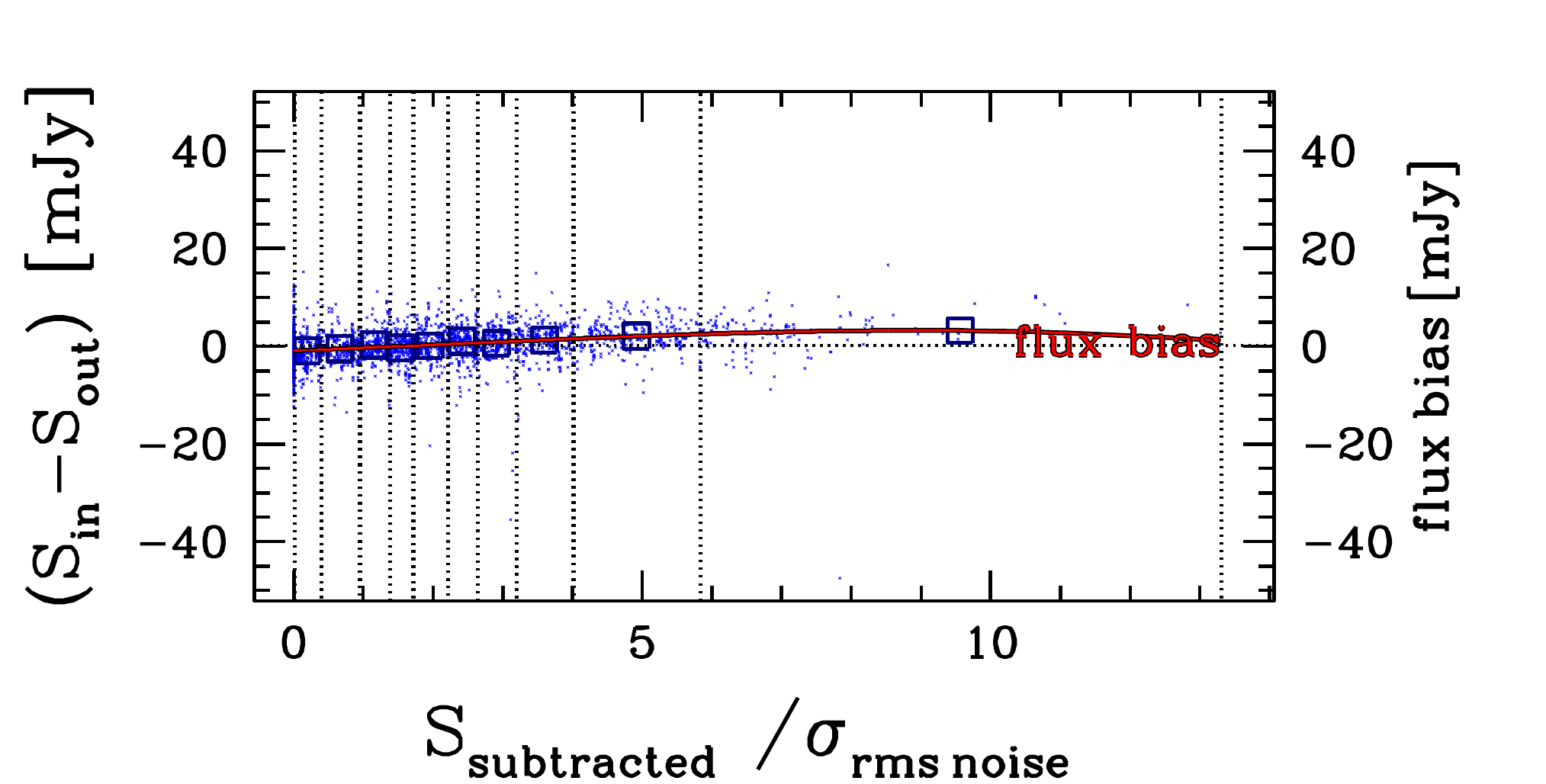}	
    \end{subfigure}
    
    \begin{subfigure}[b]{\textwidth}\centering
	\includegraphics[width=0.23\textwidth, trim={1cm 15cm 0cm 2.5cm}, clip]{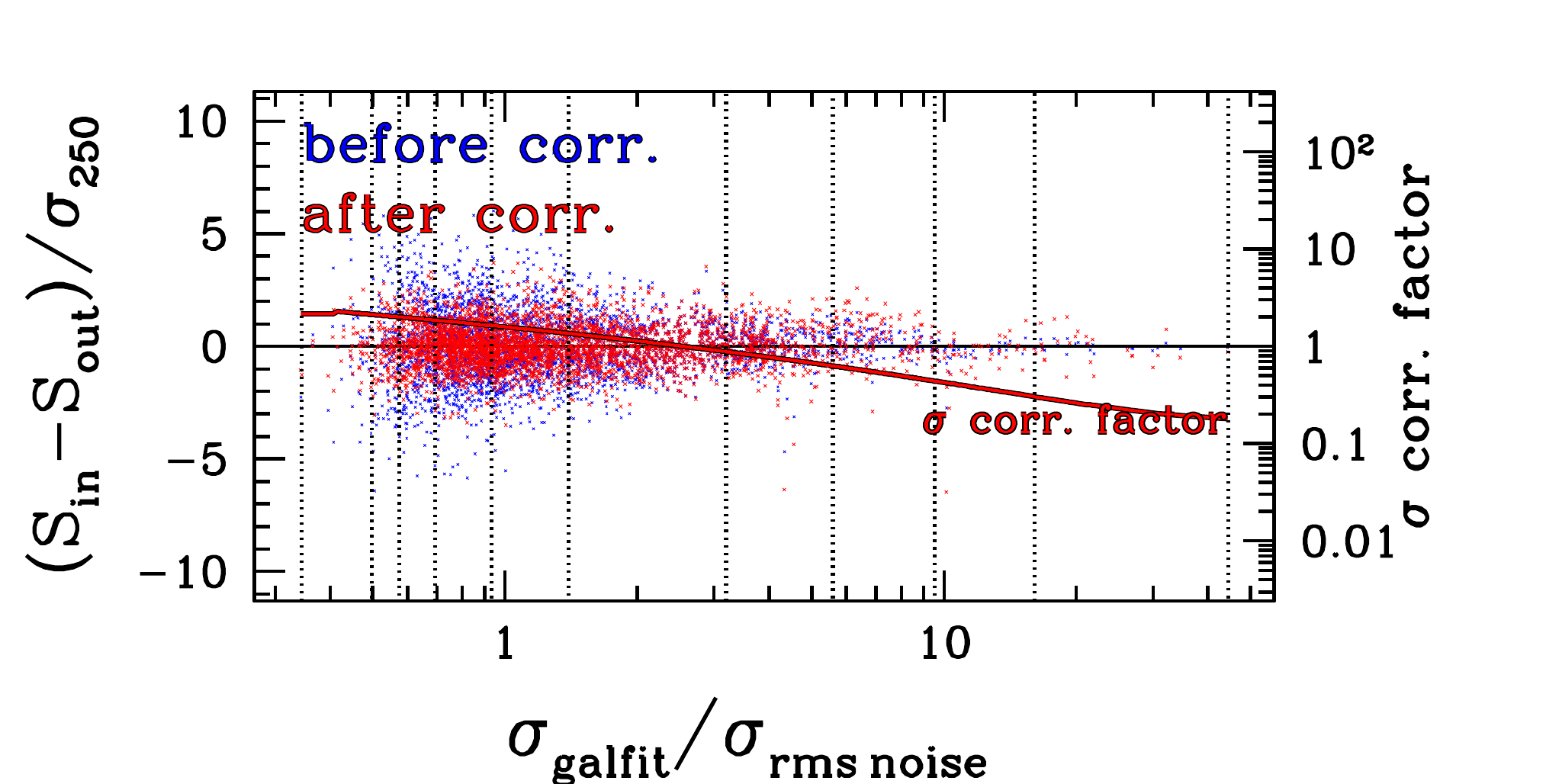}
	\includegraphics[width=0.23\textwidth, trim={1cm 15cm 0cm 2.5cm}, clip]{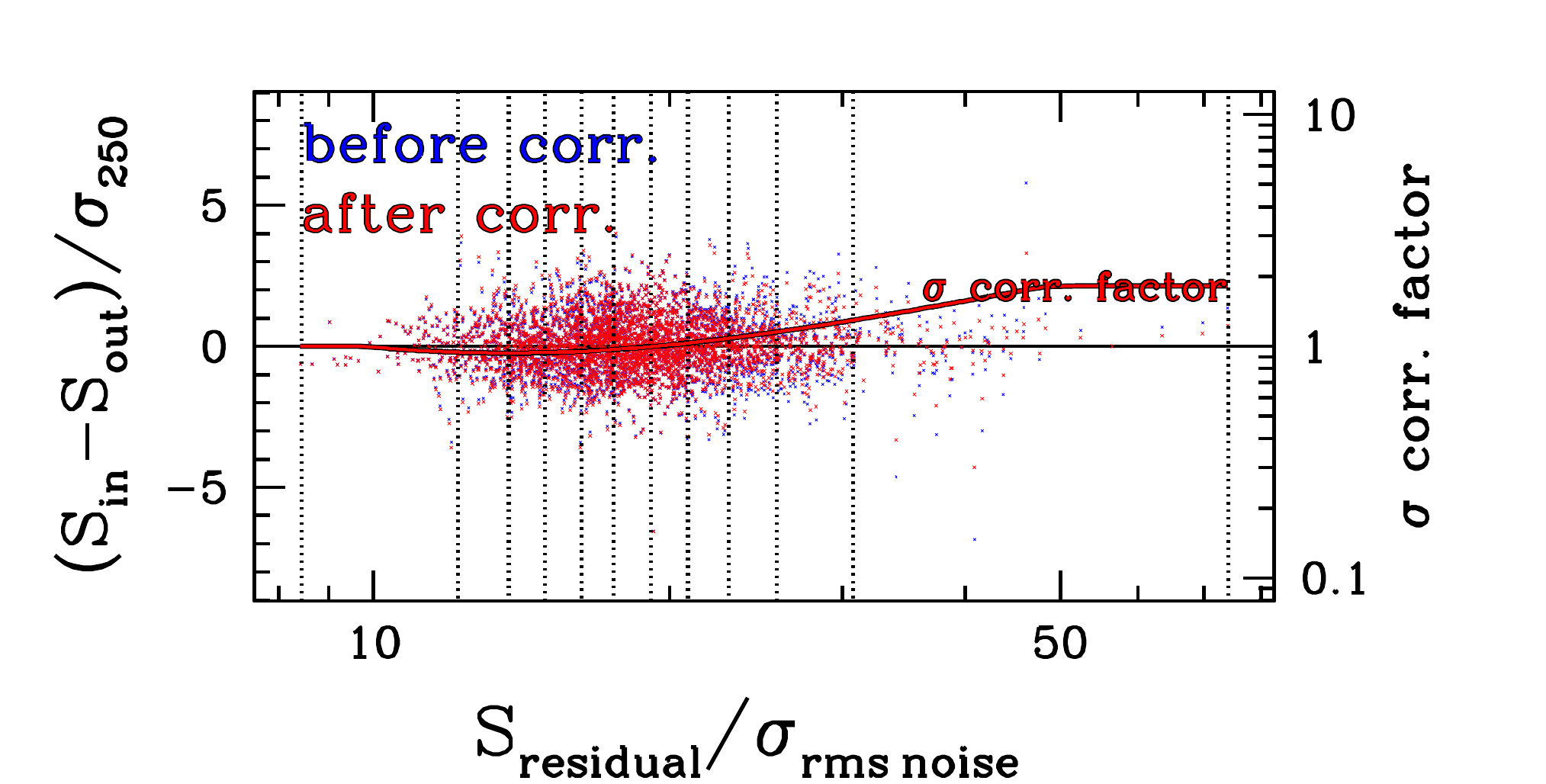}
	\includegraphics[width=0.23\textwidth, trim={1cm 15cm 0cm 2.5cm}, clip]{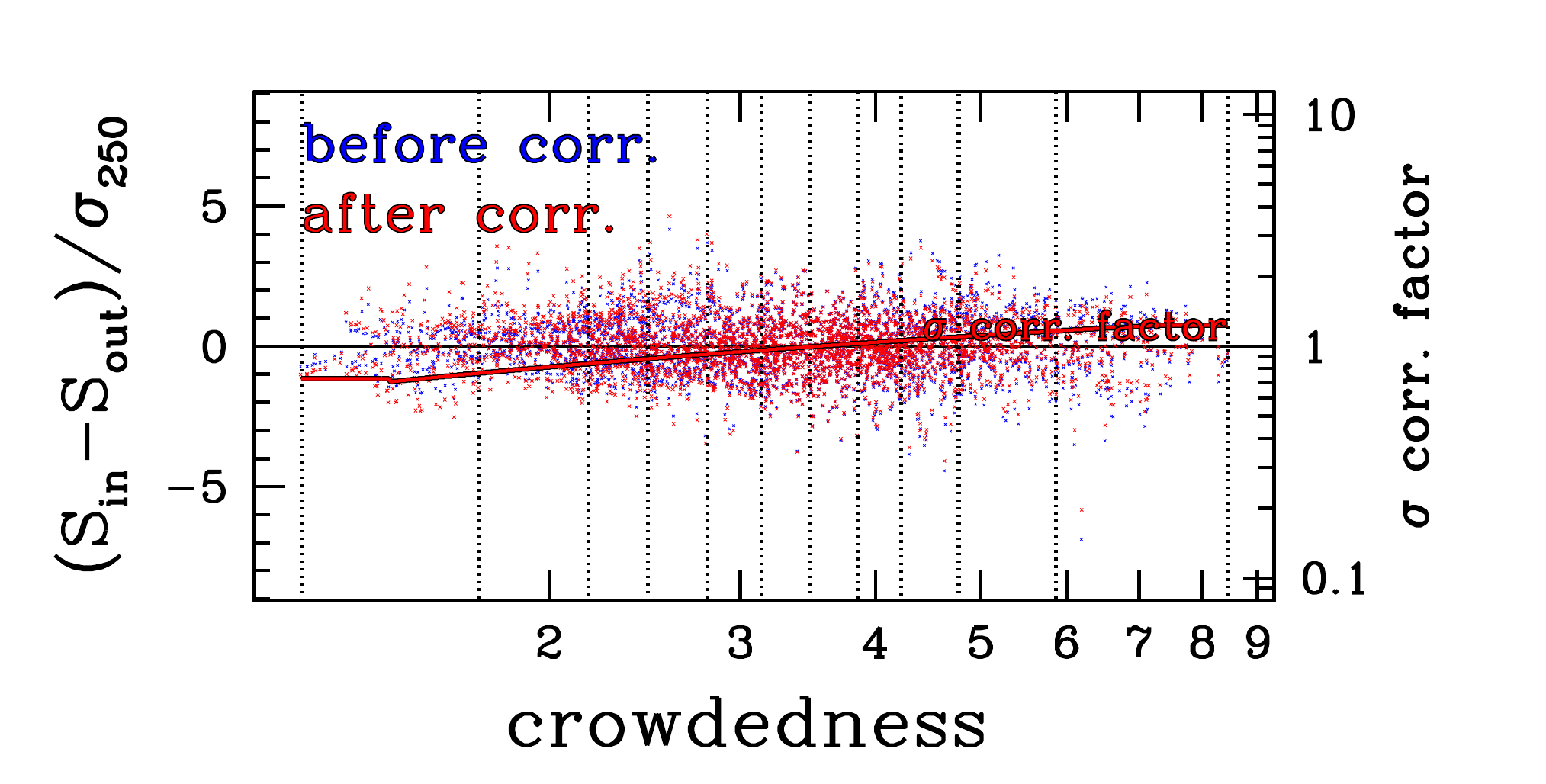}
	\includegraphics[width=0.23\textwidth, trim={1cm 15cm 0cm 2.5cm}, clip]{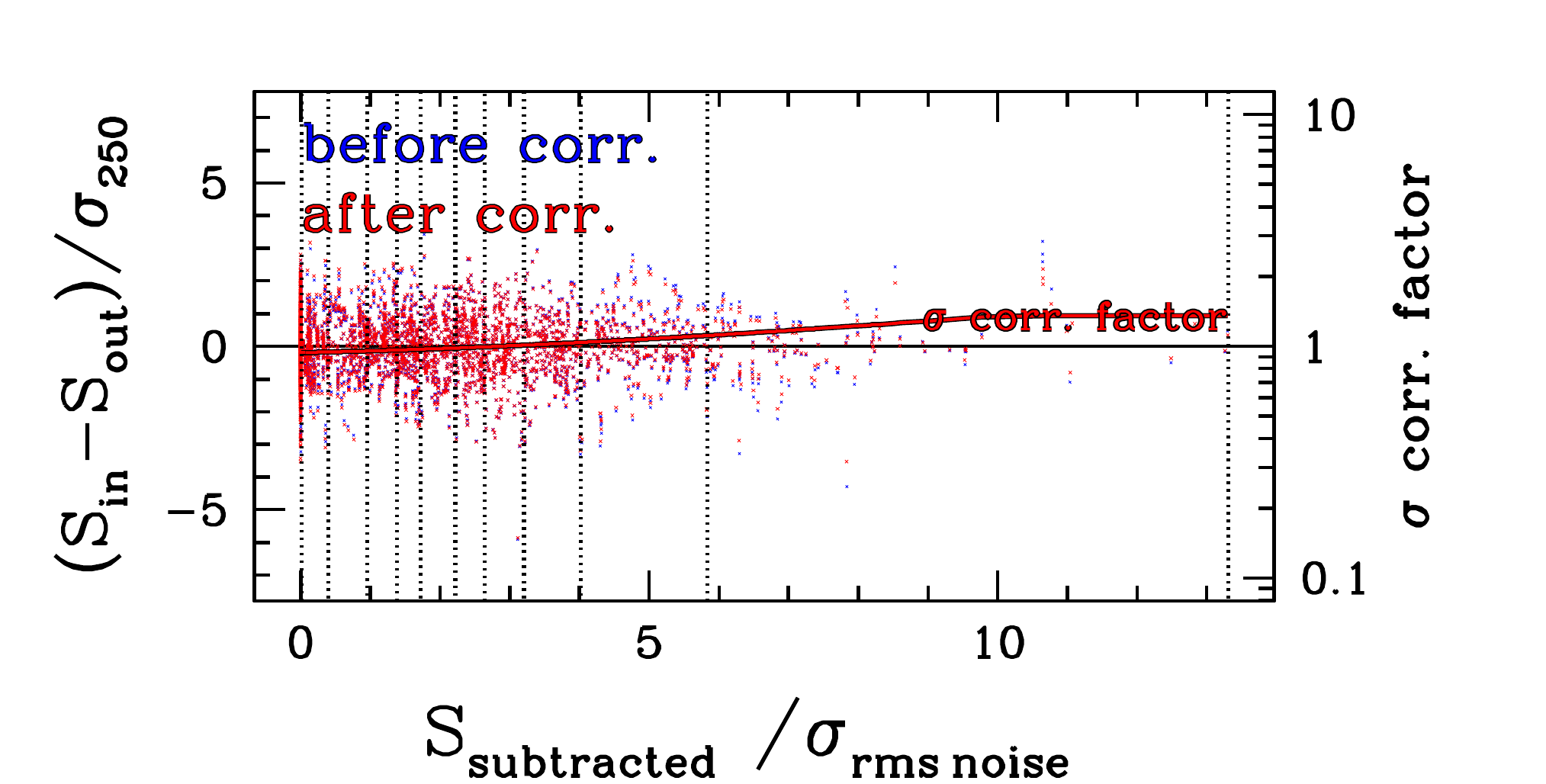}
    \end{subfigure}
    
    
    \begin{subfigure}[b]{\textwidth}\centering
	\includegraphics[width=0.23\textwidth, trim={1cm 15cm 0cm 2.5cm}, clip]{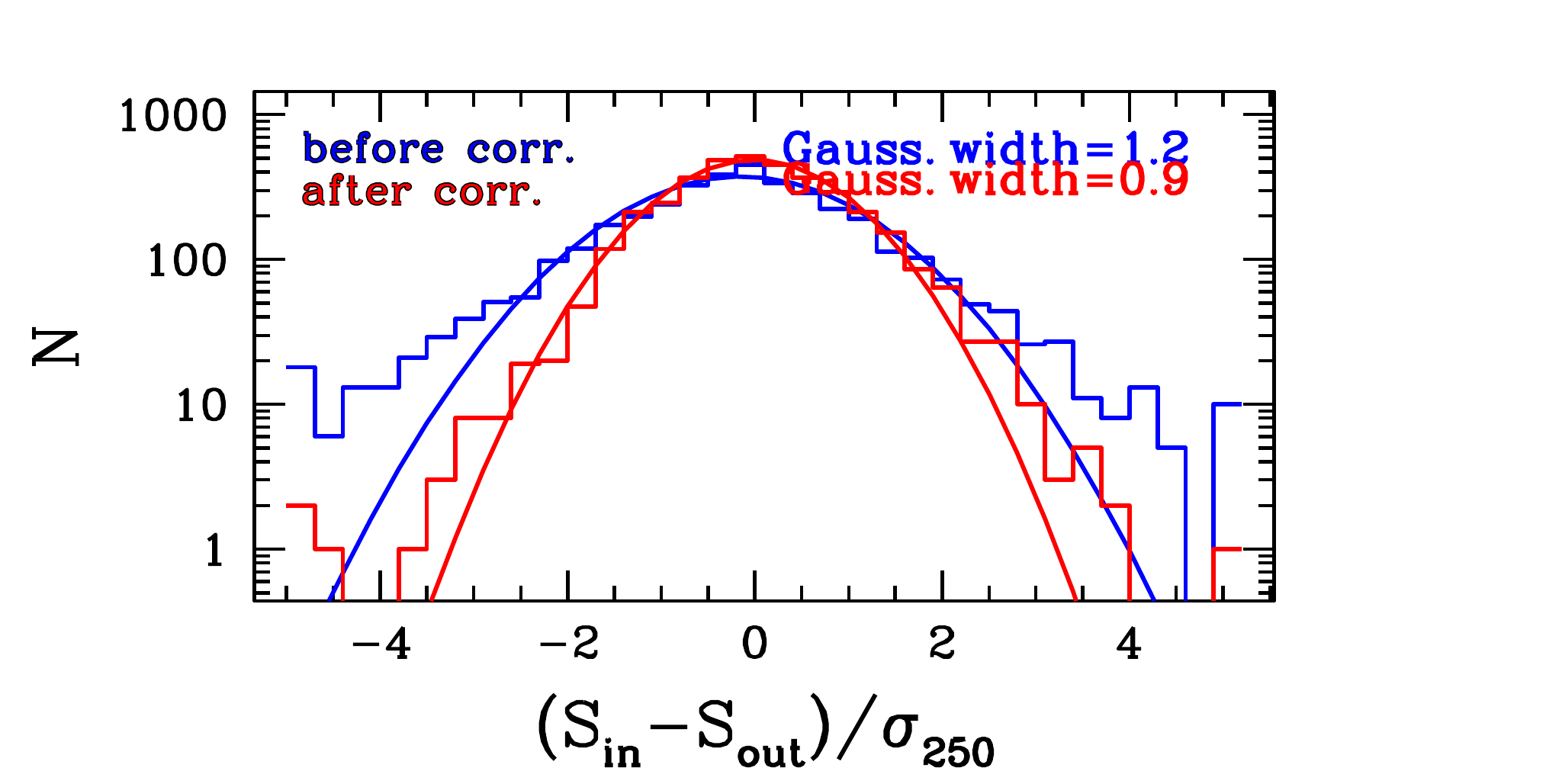}
	\includegraphics[width=0.23\textwidth, trim={1cm 15cm 0cm 2.5cm}, clip]{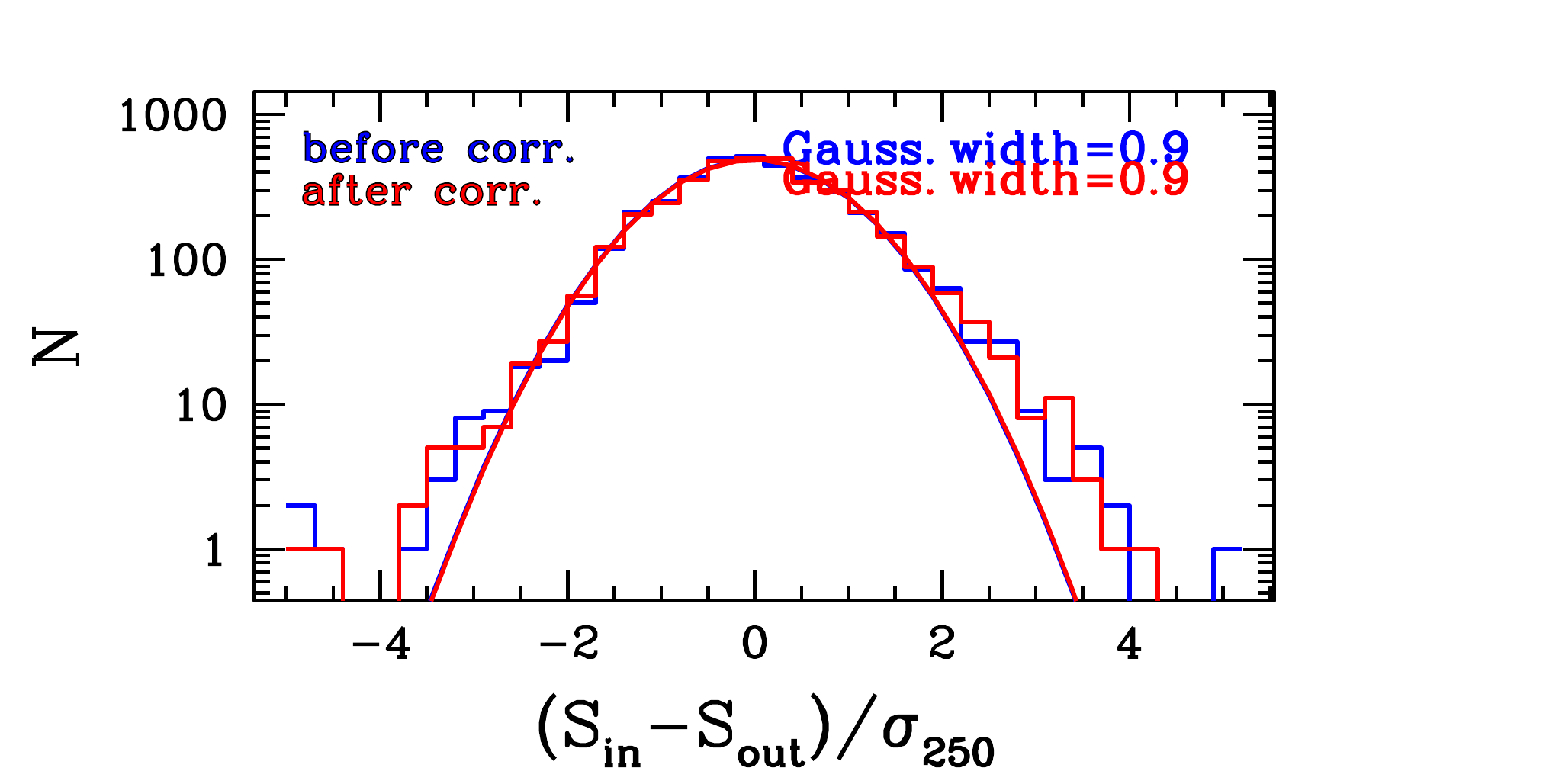}
	\includegraphics[width=0.23\textwidth, trim={1cm 15cm 0cm 2.5cm}, clip]{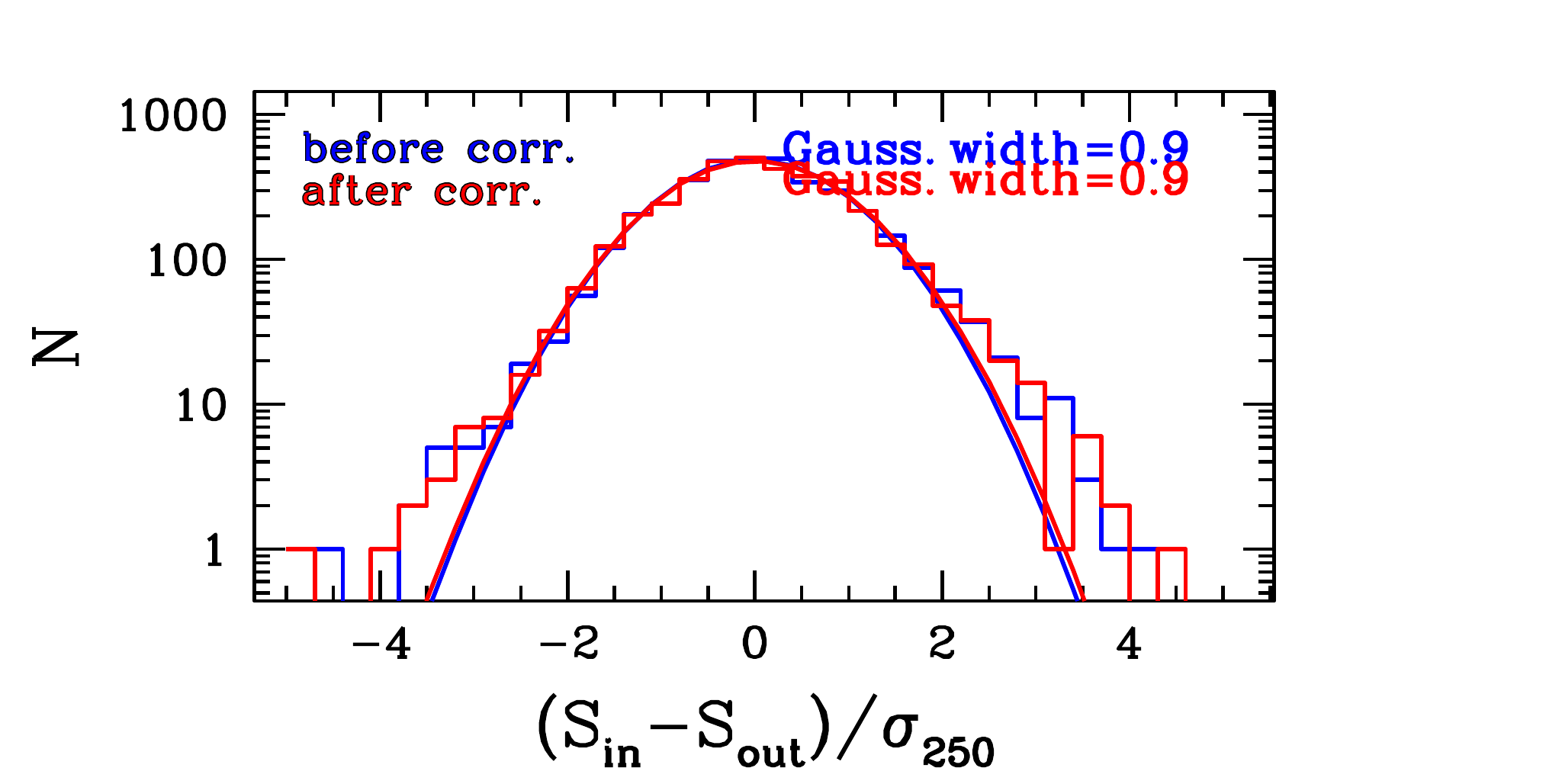}
	\includegraphics[width=0.23\textwidth, trim={1cm 15cm 0cm 2.5cm}, clip]{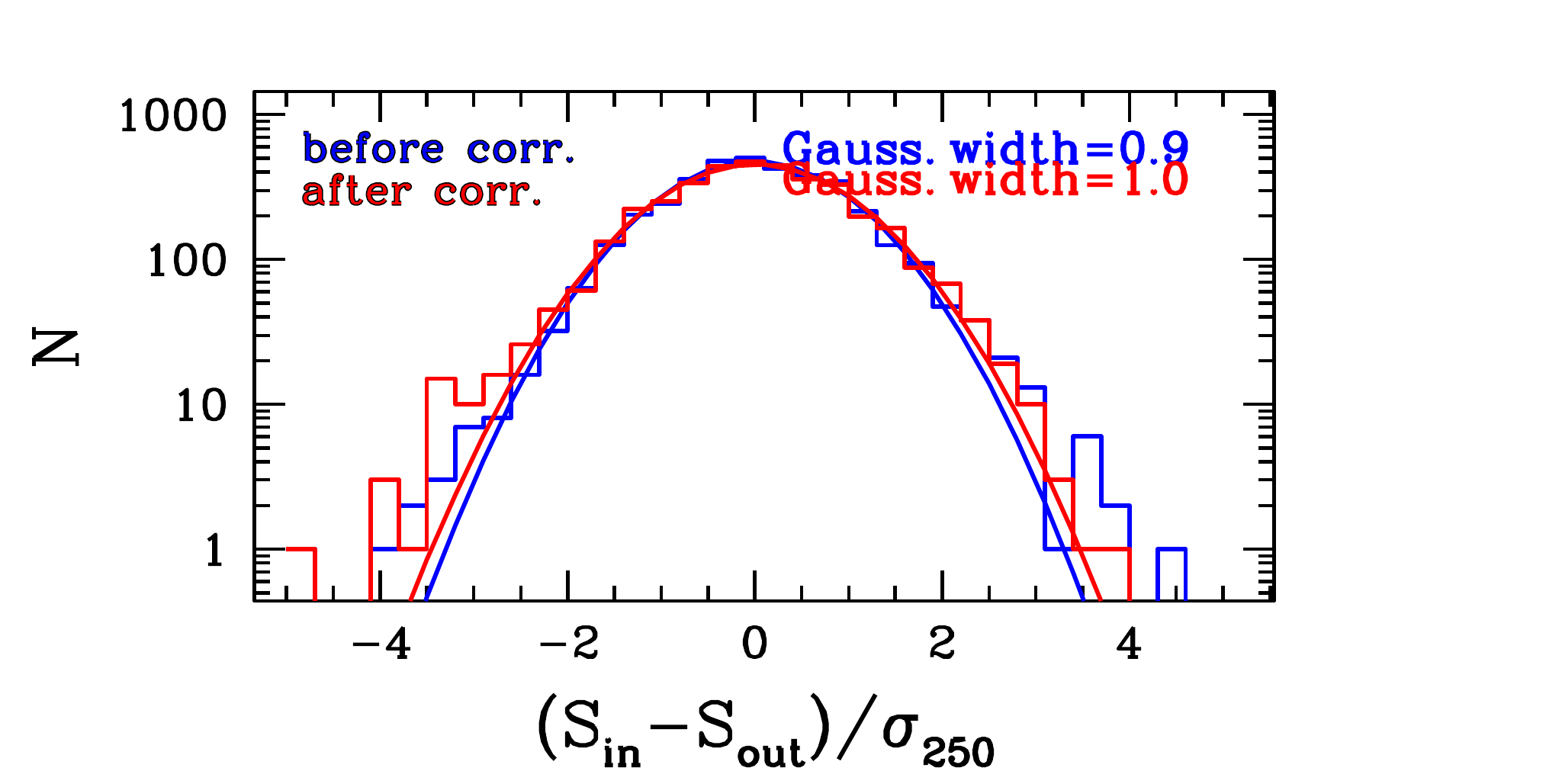}
    \end{subfigure}
    
    \begin{subfigure}[b]{\textwidth}\centering
	\includegraphics[width=0.23\textwidth, trim={1cm 15cm 0cm 2.5cm}, clip]{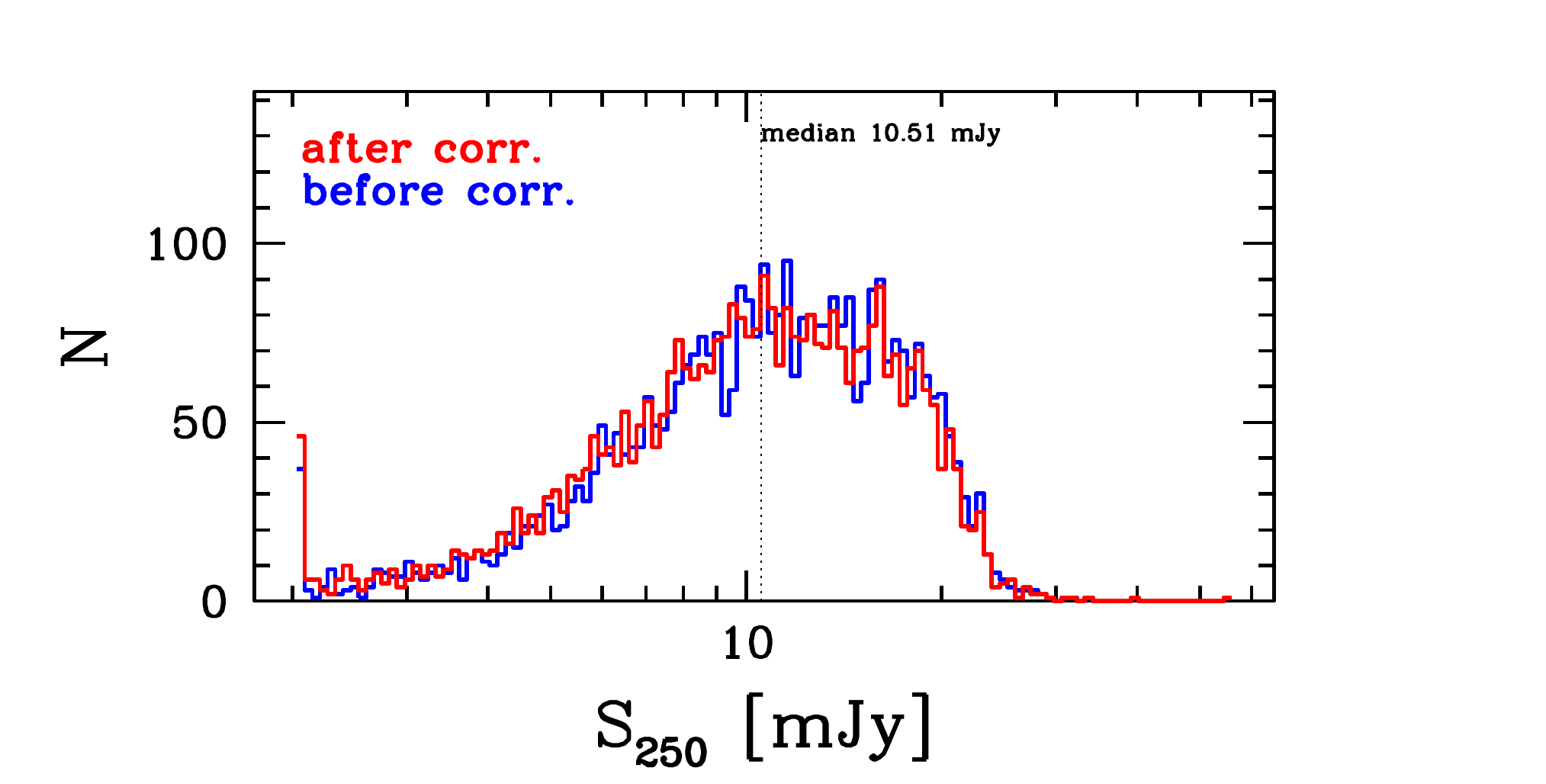}
	\includegraphics[width=0.23\textwidth, trim={1cm 15cm 0cm 2.5cm}, clip]{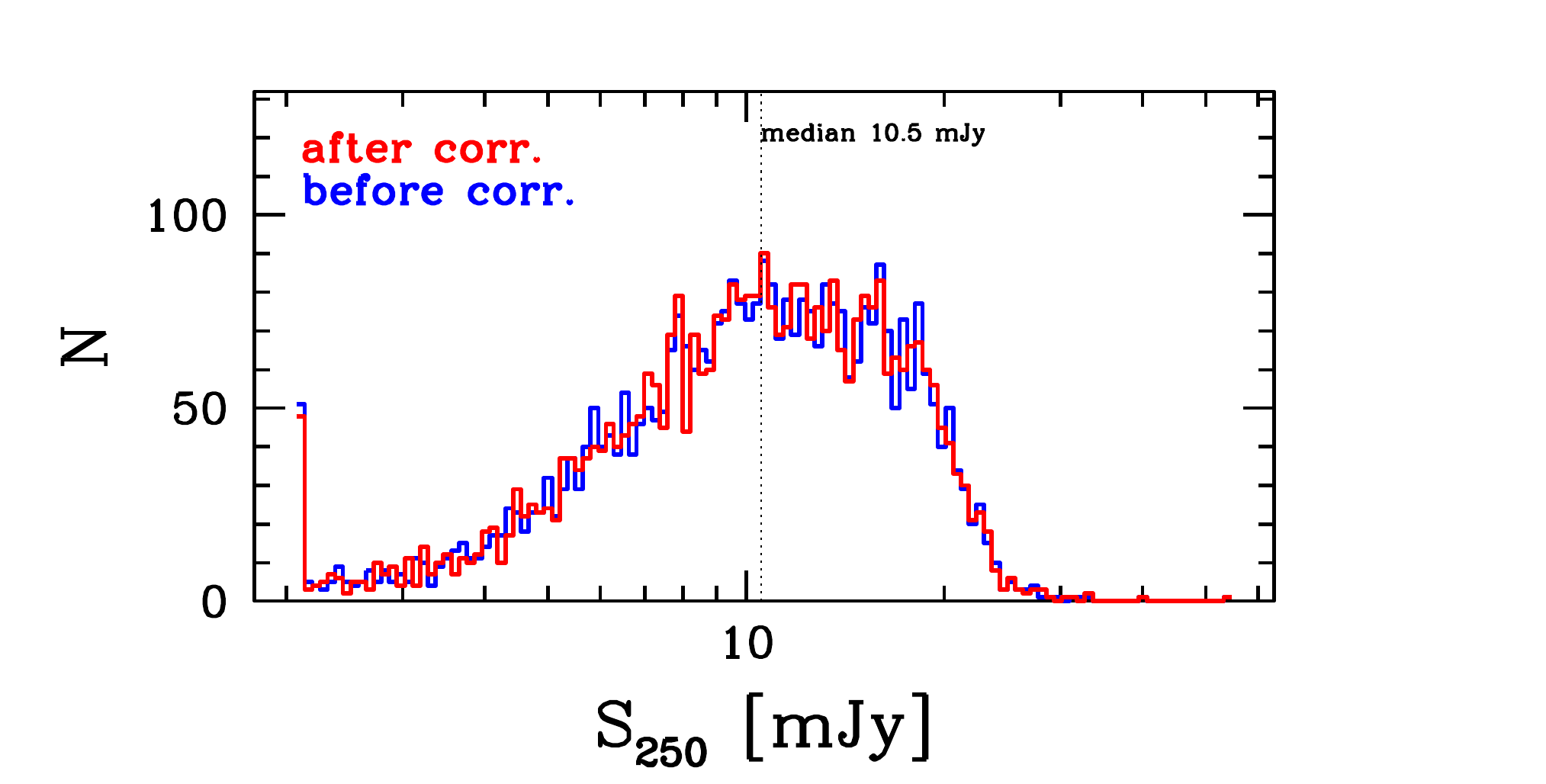}
	\includegraphics[width=0.23\textwidth, trim={1cm 15cm 0cm 2.5cm}, clip]{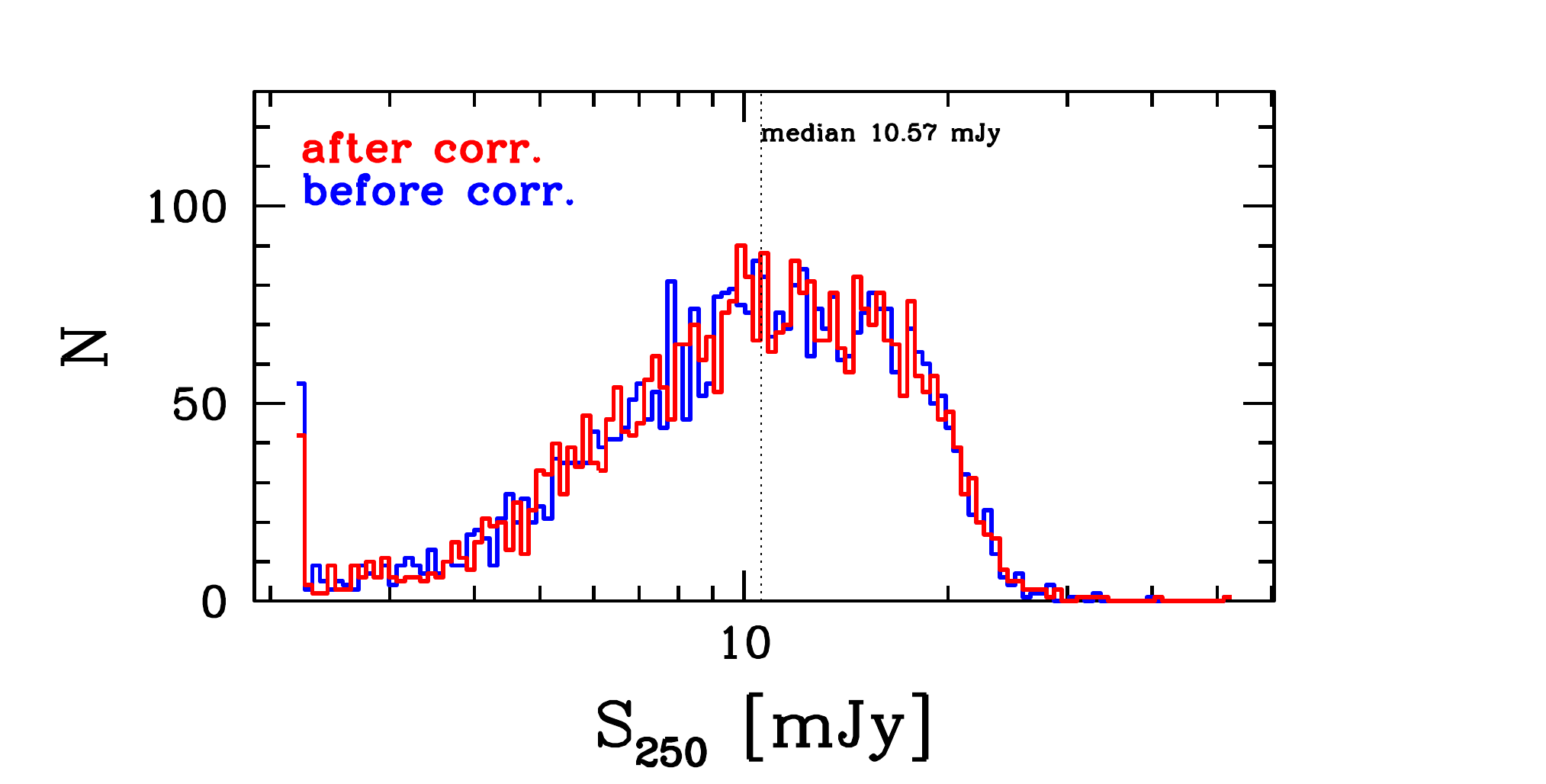}
	\includegraphics[width=0.23\textwidth, trim={1cm 15cm 0cm 2.5cm}, clip]{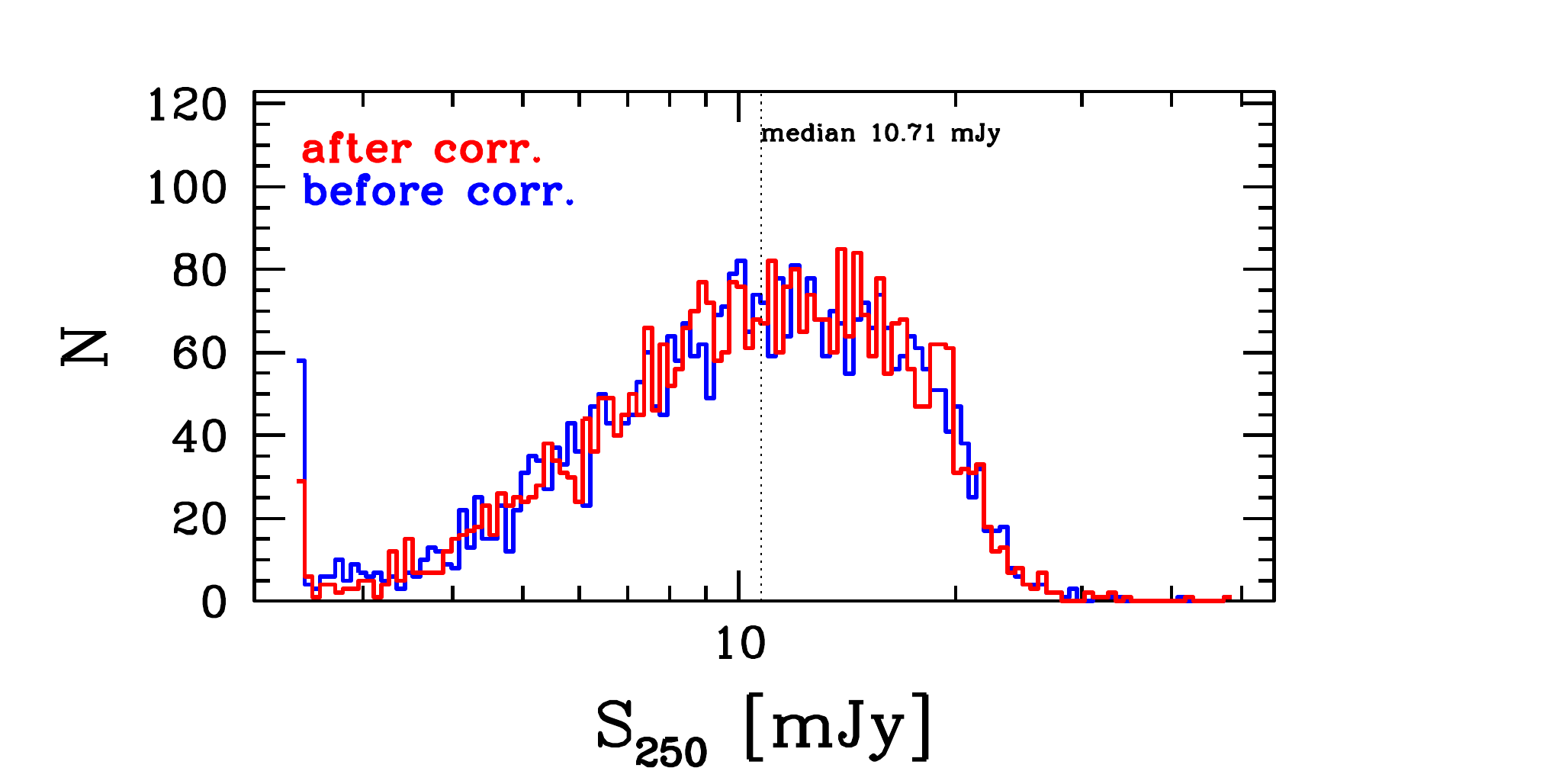}
    \end{subfigure}
    
    \begin{subfigure}[b]{\textwidth}\centering
	\includegraphics[width=0.23\textwidth, trim={1cm 15cm 0cm 2.5cm}, clip]{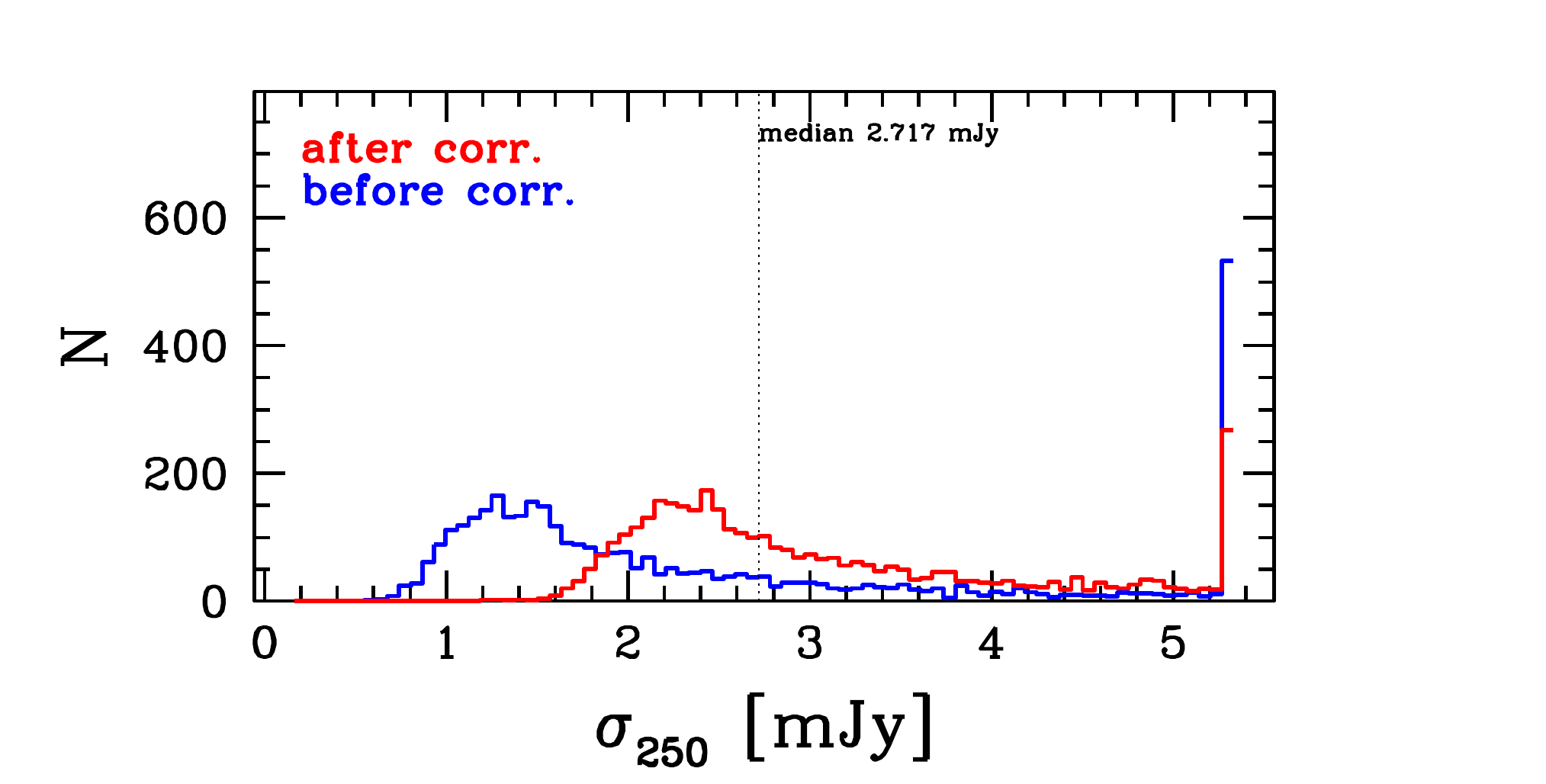}
	\includegraphics[width=0.23\textwidth, trim={1cm 15cm 0cm 2.5cm}, clip]{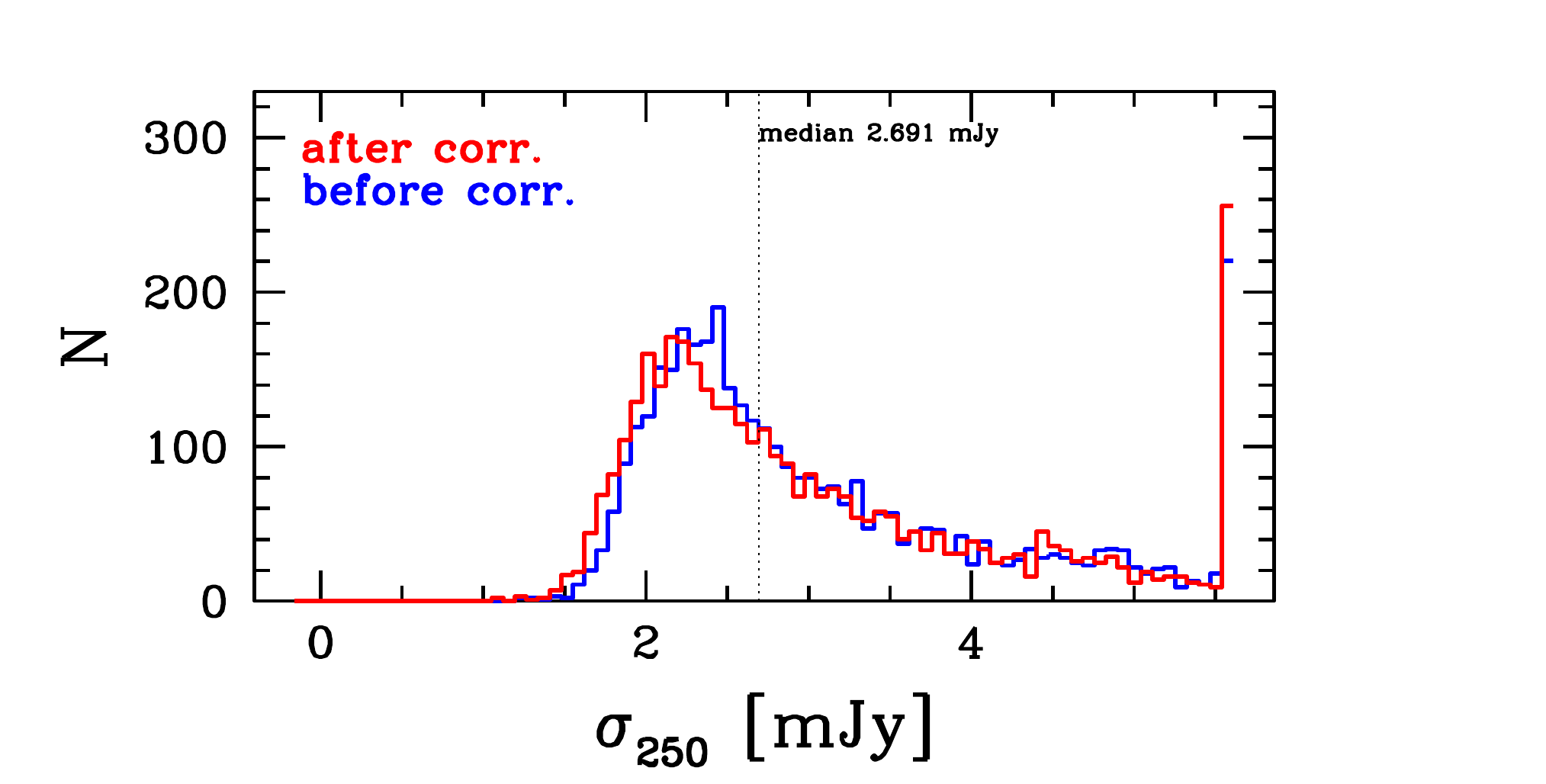}
	\includegraphics[width=0.23\textwidth, trim={1cm 15cm 0cm 2.5cm}, clip]{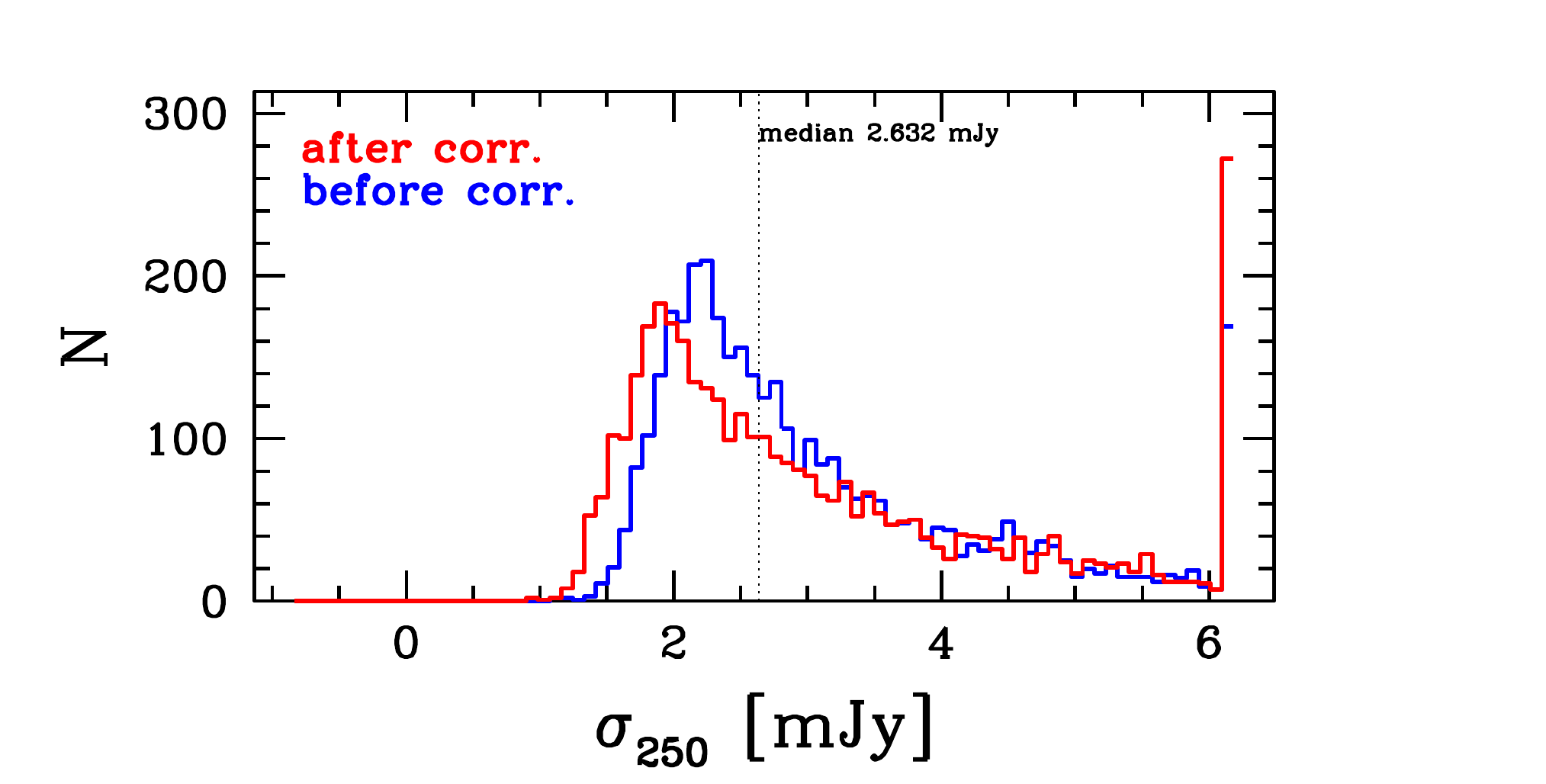}
	\includegraphics[width=0.23\textwidth, trim={1cm 15cm 0cm 2.5cm}, clip]{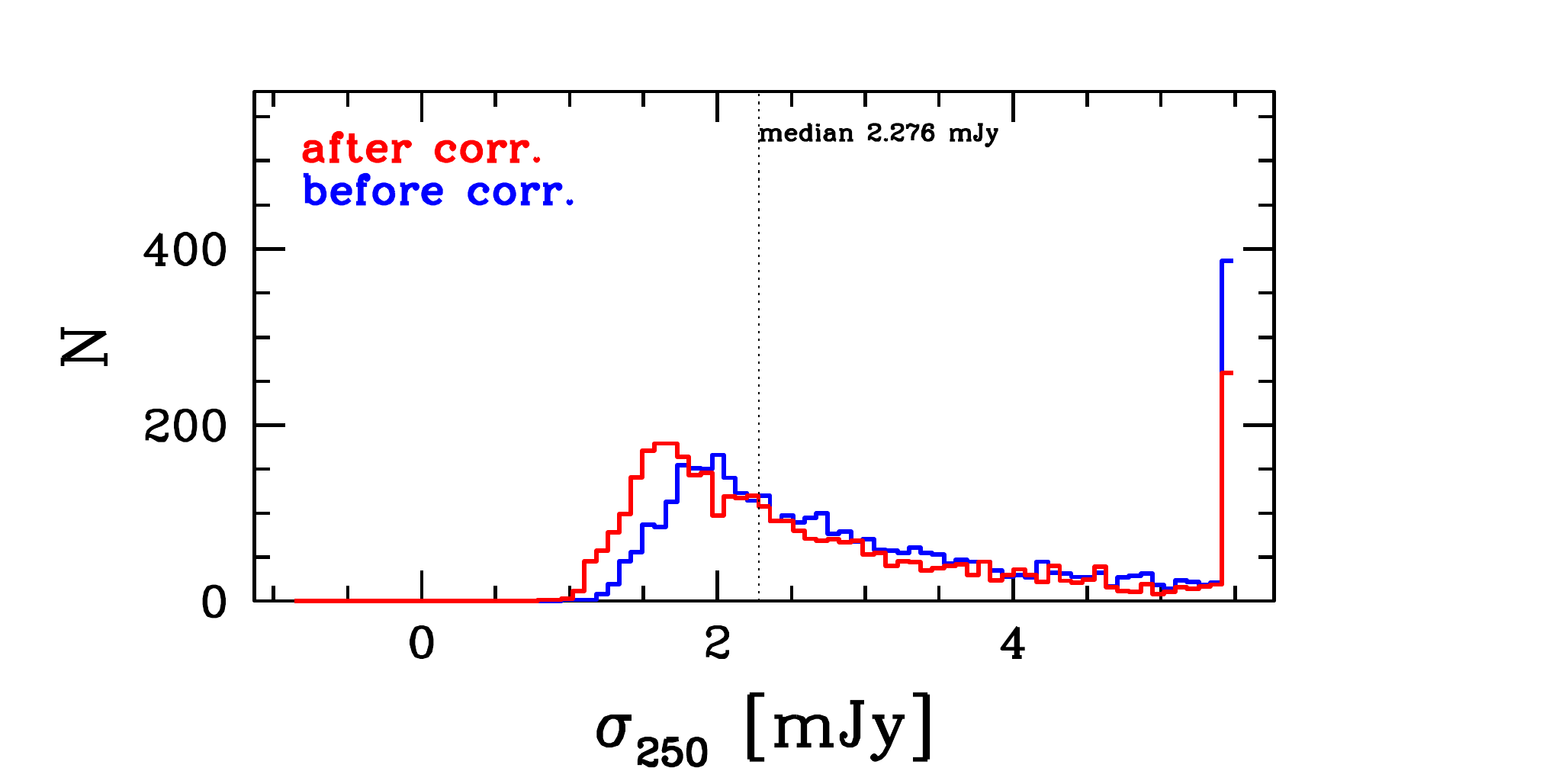}
    \end{subfigure}
    
	\caption{%
		Simulation correction analyses at SPIRE 250~$\mu$m. See descriptions in text. 
        \label{Figure_galsim_250_bin}
	}
\end{figure}

\begin{figure}
	\centering
    
    \begin{subfigure}[b]{\textwidth}\centering
	\includegraphics[width=0.23\textwidth, trim={1cm 15cm 0cm 2.5cm}, clip]{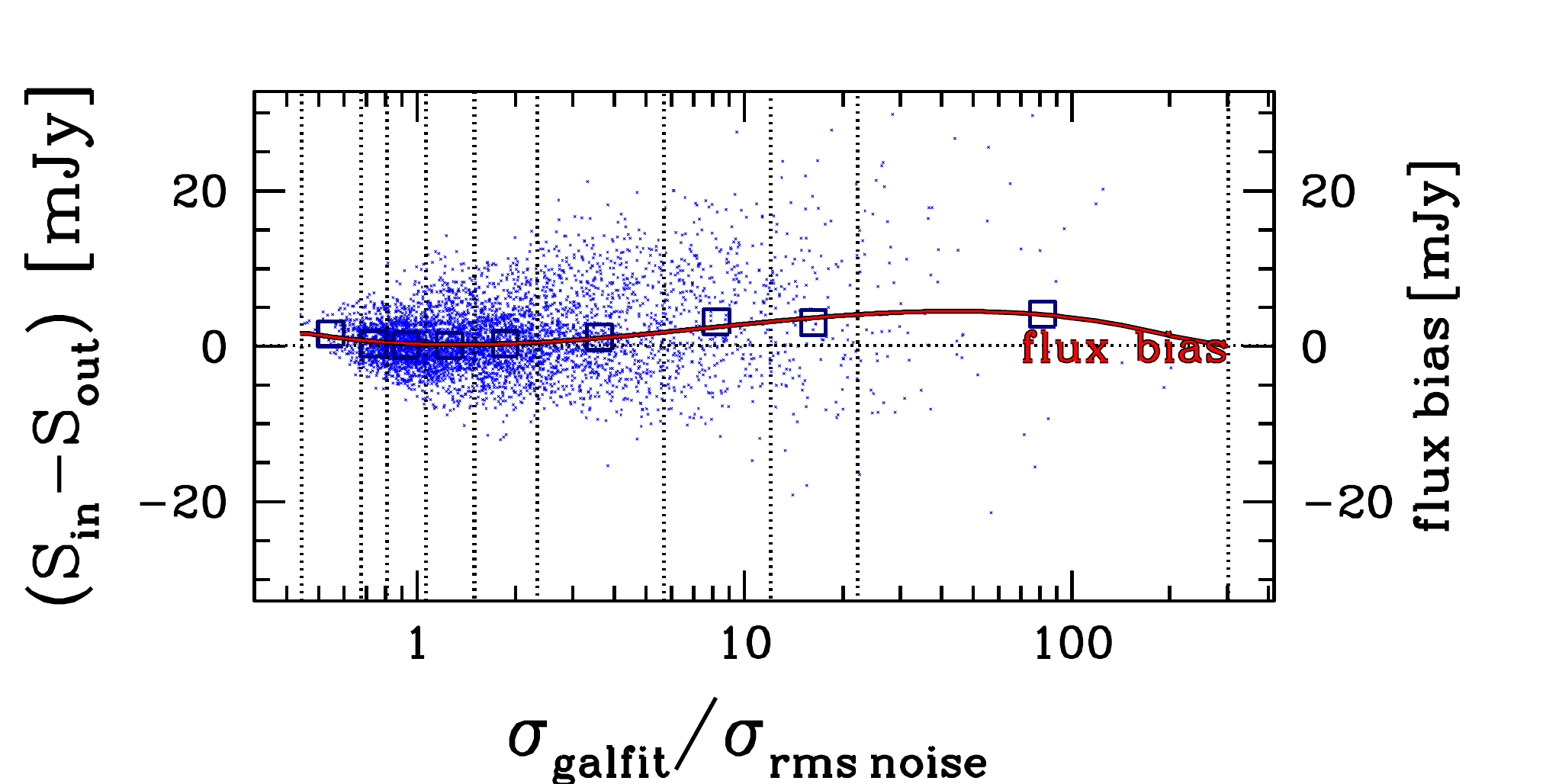}
	\includegraphics[width=0.23\textwidth, trim={1cm 15cm 0cm 2.5cm}, clip]{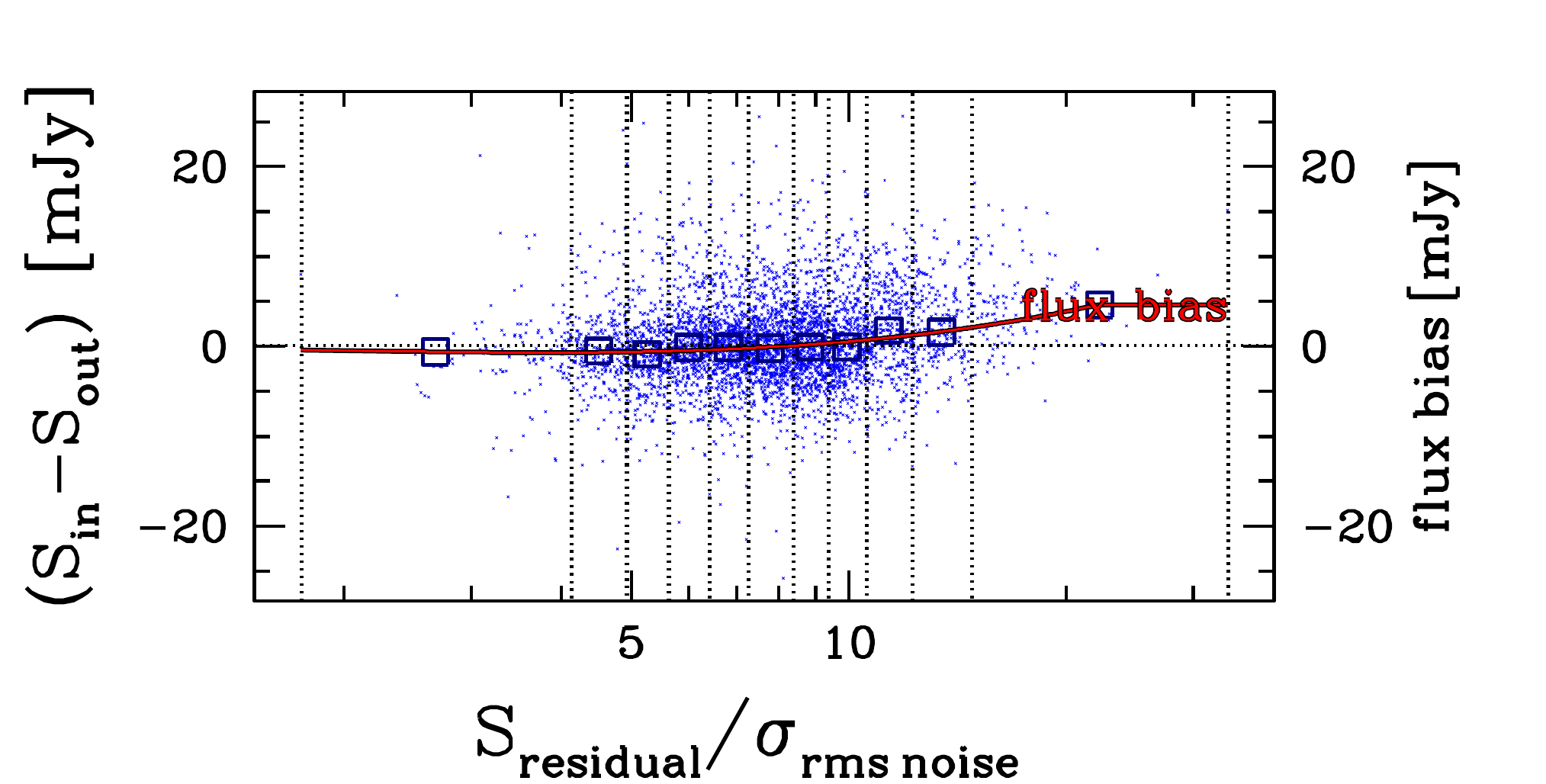}
	\includegraphics[width=0.23\textwidth, trim={1cm 15cm 0cm 2.5cm}, clip]{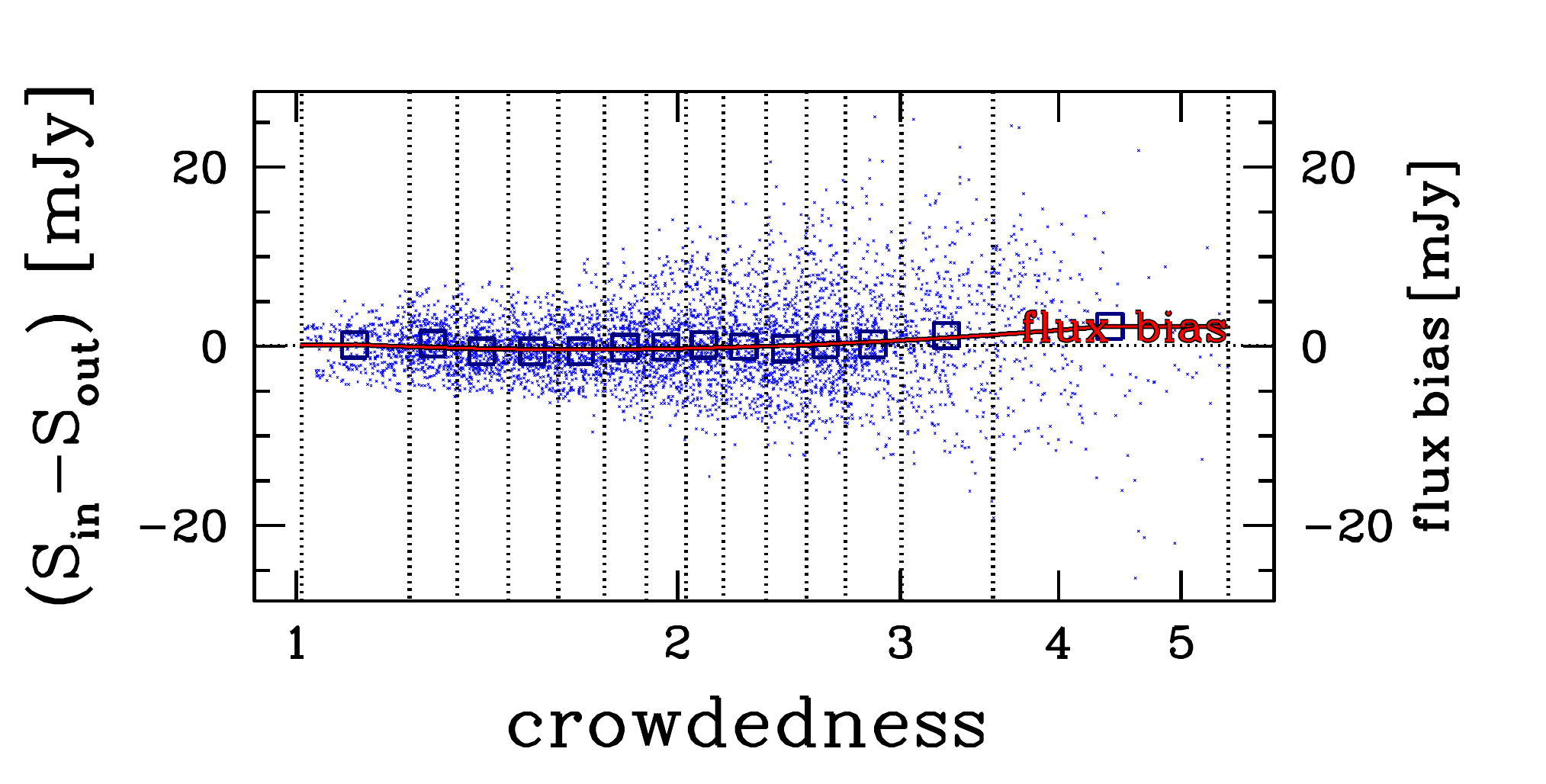}
	\includegraphics[width=0.23\textwidth, trim={1cm 15cm 0cm 2.5cm}, clip]{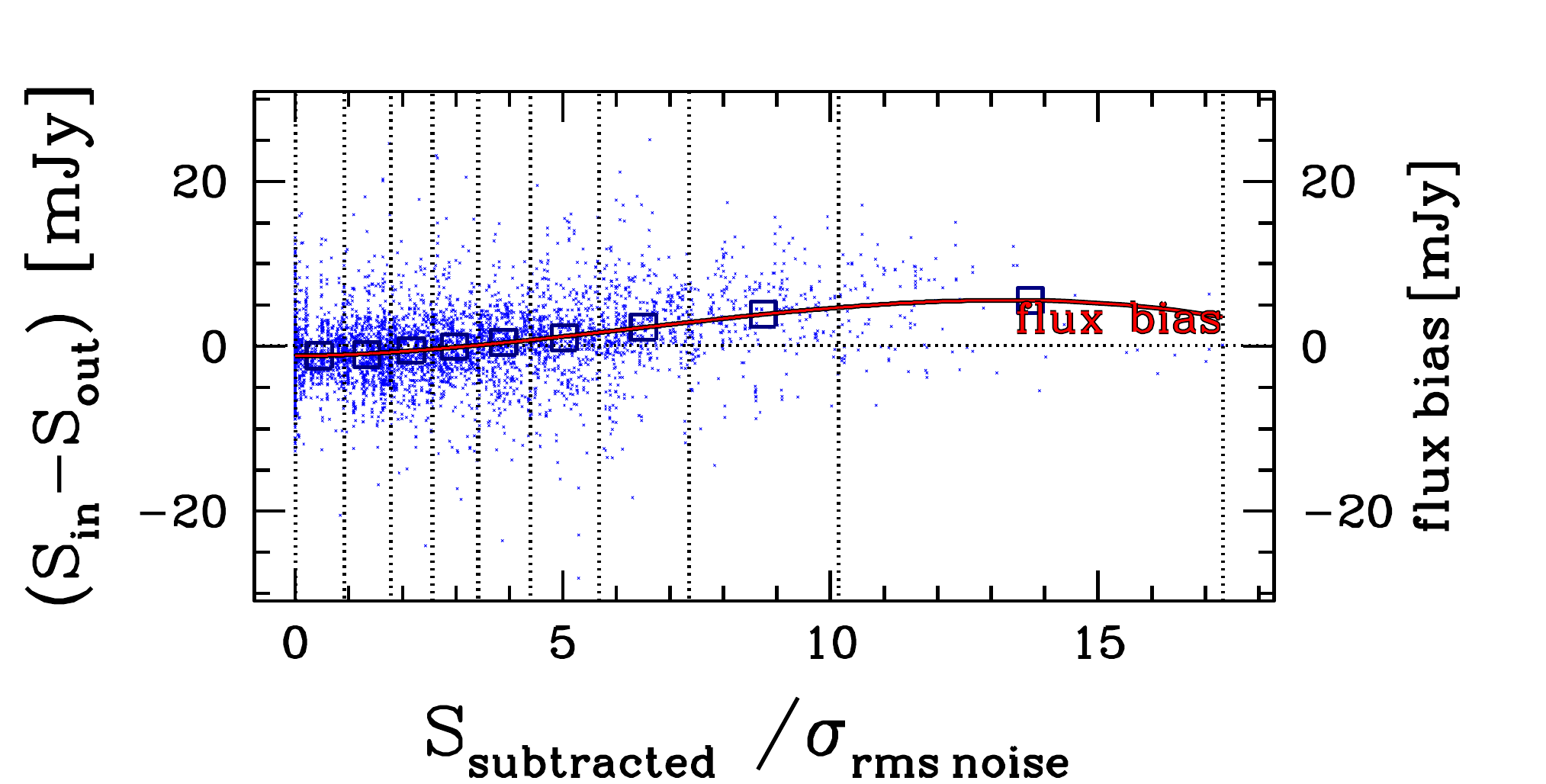}	
    \end{subfigure}
    
    \begin{subfigure}[b]{\textwidth}\centering
	\includegraphics[width=0.23\textwidth, trim={1cm 15cm 0cm 2.5cm}, clip]{galsim_350_bin_dfcorr_1}
	\includegraphics[width=0.23\textwidth, trim={1cm 15cm 0cm 2.5cm}, clip]{galsim_350_bin_dfcorr_2}
	\includegraphics[width=0.23\textwidth, trim={1cm 15cm 0cm 2.5cm}, clip]{galsim_350_bin_dfcorr_3}
	\includegraphics[width=0.23\textwidth, trim={1cm 15cm 0cm 2.5cm}, clip]{galsim_350_bin_dfcorr_4}
    \end{subfigure}
    
    
    \begin{subfigure}[b]{\textwidth}\centering
	\includegraphics[width=0.23\textwidth, trim={1cm 15cm 0cm 2.5cm}, clip]{galsim_350_hist_dfcorr_1_ylog}
	\includegraphics[width=0.23\textwidth, trim={1cm 15cm 0cm 2.5cm}, clip]{galsim_350_hist_dfcorr_2_ylog}
	\includegraphics[width=0.23\textwidth, trim={1cm 15cm 0cm 2.5cm}, clip]{galsim_350_hist_dfcorr_3_ylog}
	\includegraphics[width=0.23\textwidth, trim={1cm 15cm 0cm 2.5cm}, clip]{galsim_350_hist_dfcorr_4_ylog}
    \end{subfigure}
    
    \begin{subfigure}[b]{\textwidth}\centering
	\includegraphics[width=0.23\textwidth, trim={1cm 15cm 0cm 2.5cm}, clip]{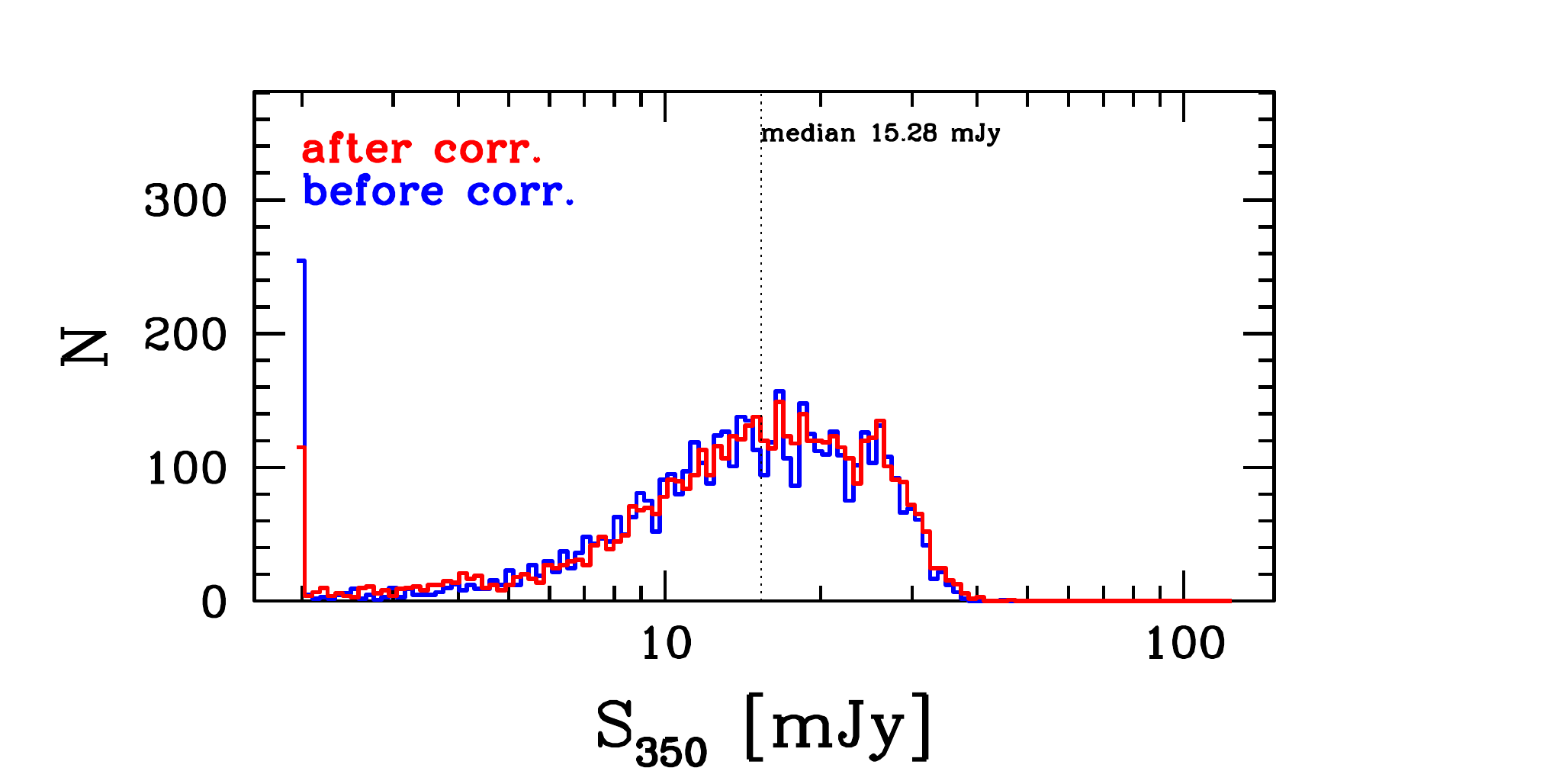}
	\includegraphics[width=0.23\textwidth, trim={1cm 15cm 0cm 2.5cm}, clip]{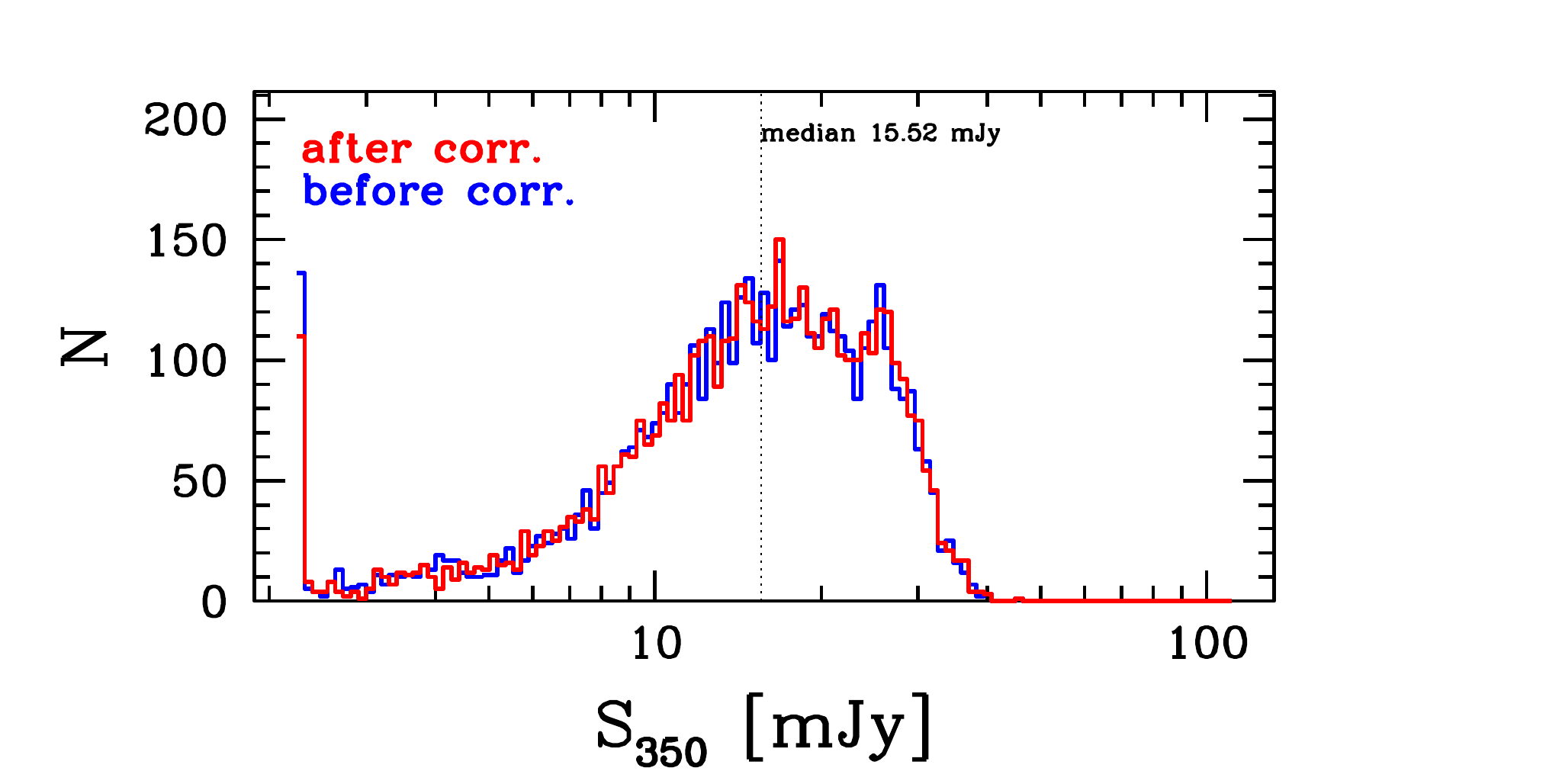}
	\includegraphics[width=0.23\textwidth, trim={1cm 15cm 0cm 2.5cm}, clip]{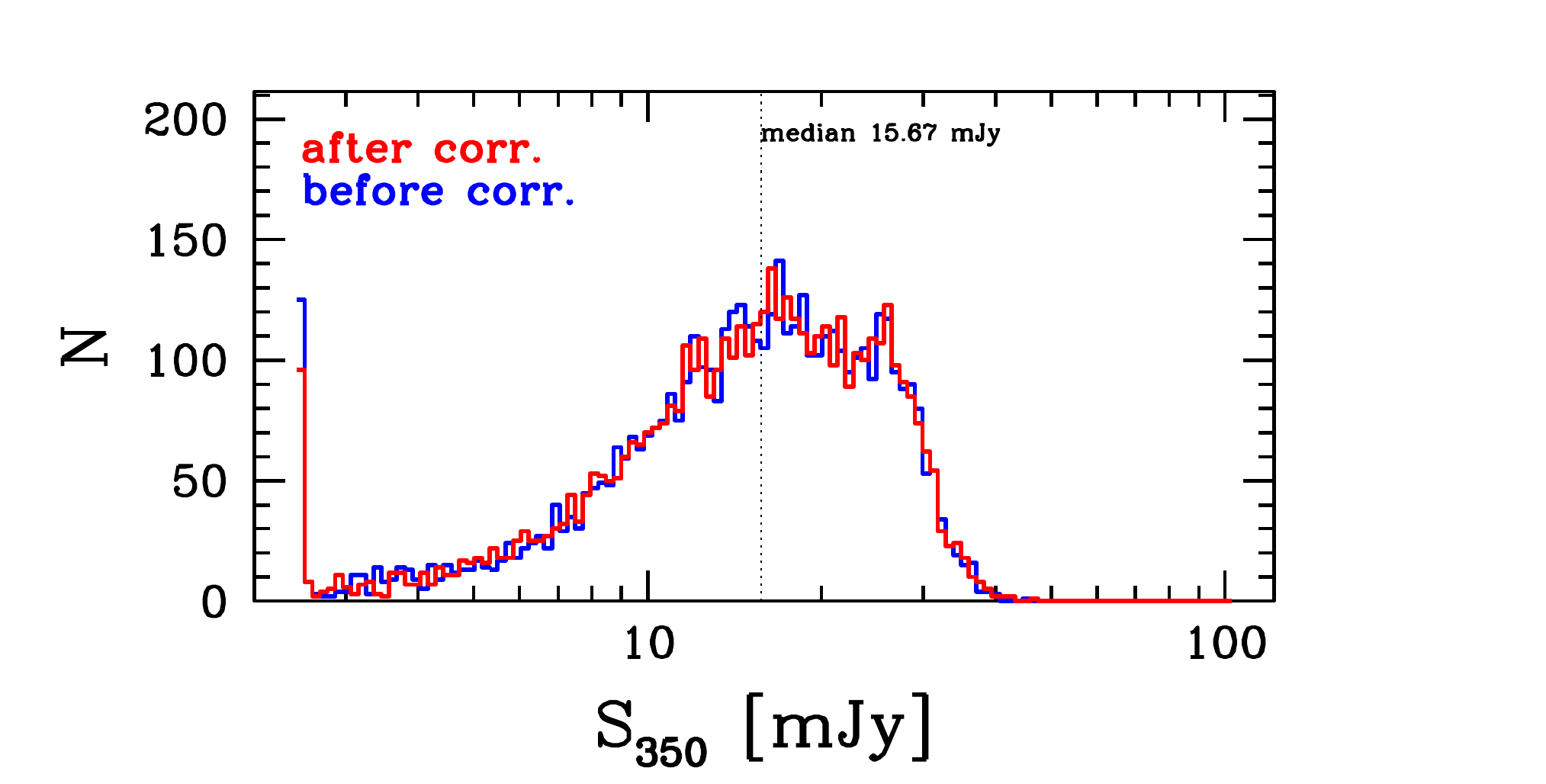}
	\includegraphics[width=0.23\textwidth, trim={1cm 15cm 0cm 2.5cm}, clip]{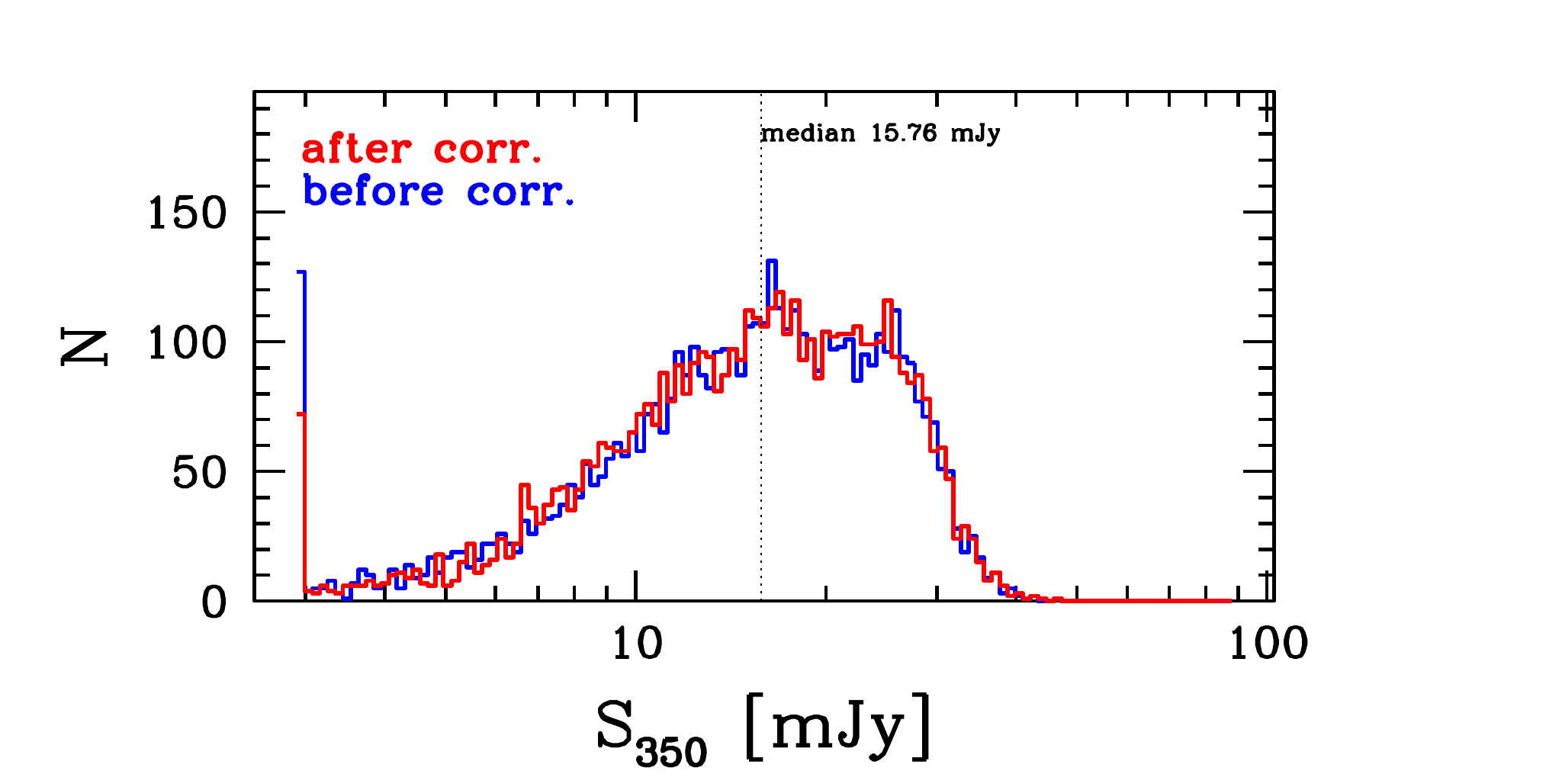}
    \end{subfigure}
    
    \begin{subfigure}[b]{\textwidth}\centering
	\includegraphics[width=0.23\textwidth, trim={1cm 15cm 0cm 2.5cm}, clip]{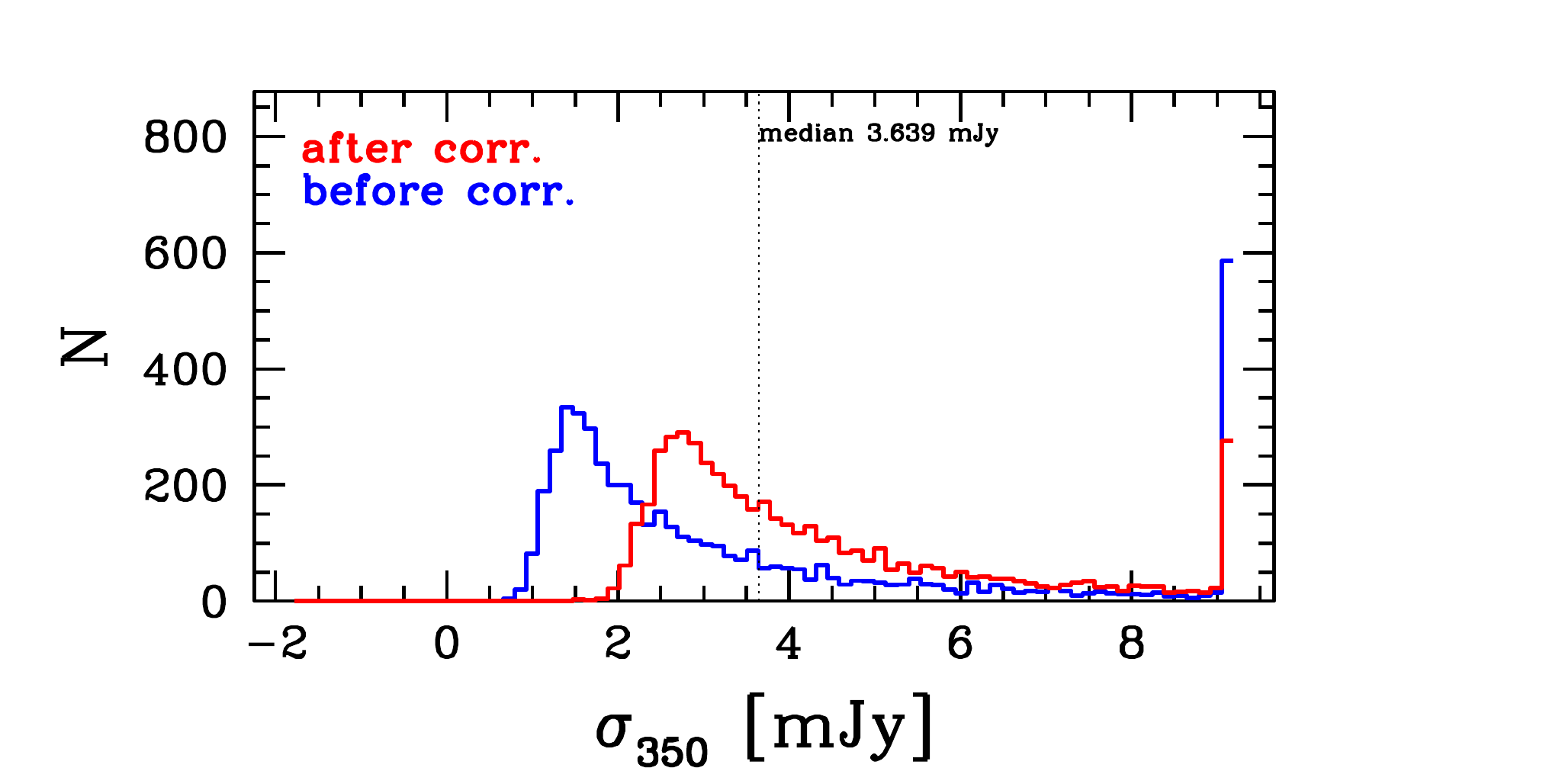}
	\includegraphics[width=0.23\textwidth, trim={1cm 15cm 0cm 2.5cm}, clip]{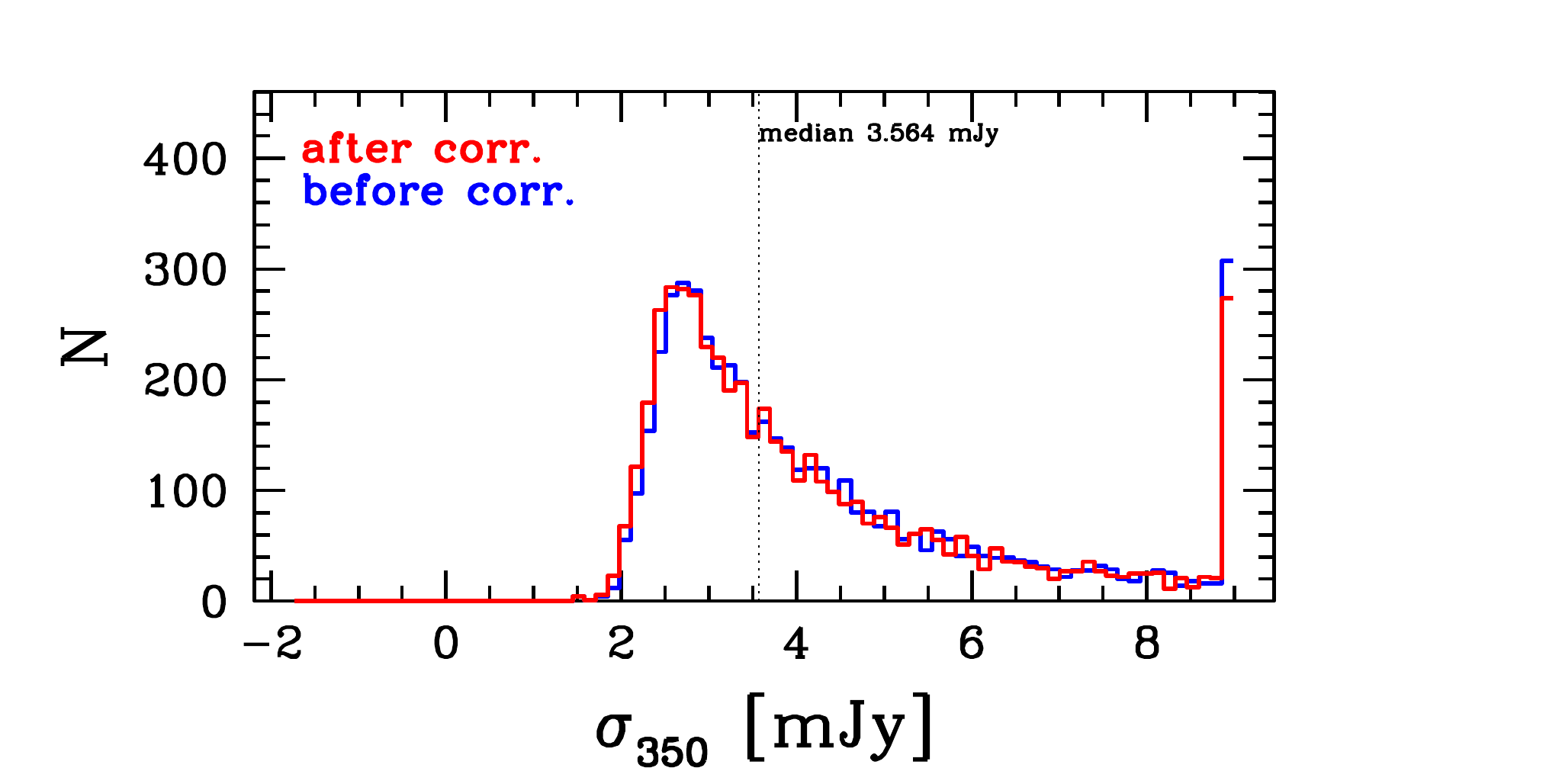}
	\includegraphics[width=0.23\textwidth, trim={1cm 15cm 0cm 2.5cm}, clip]{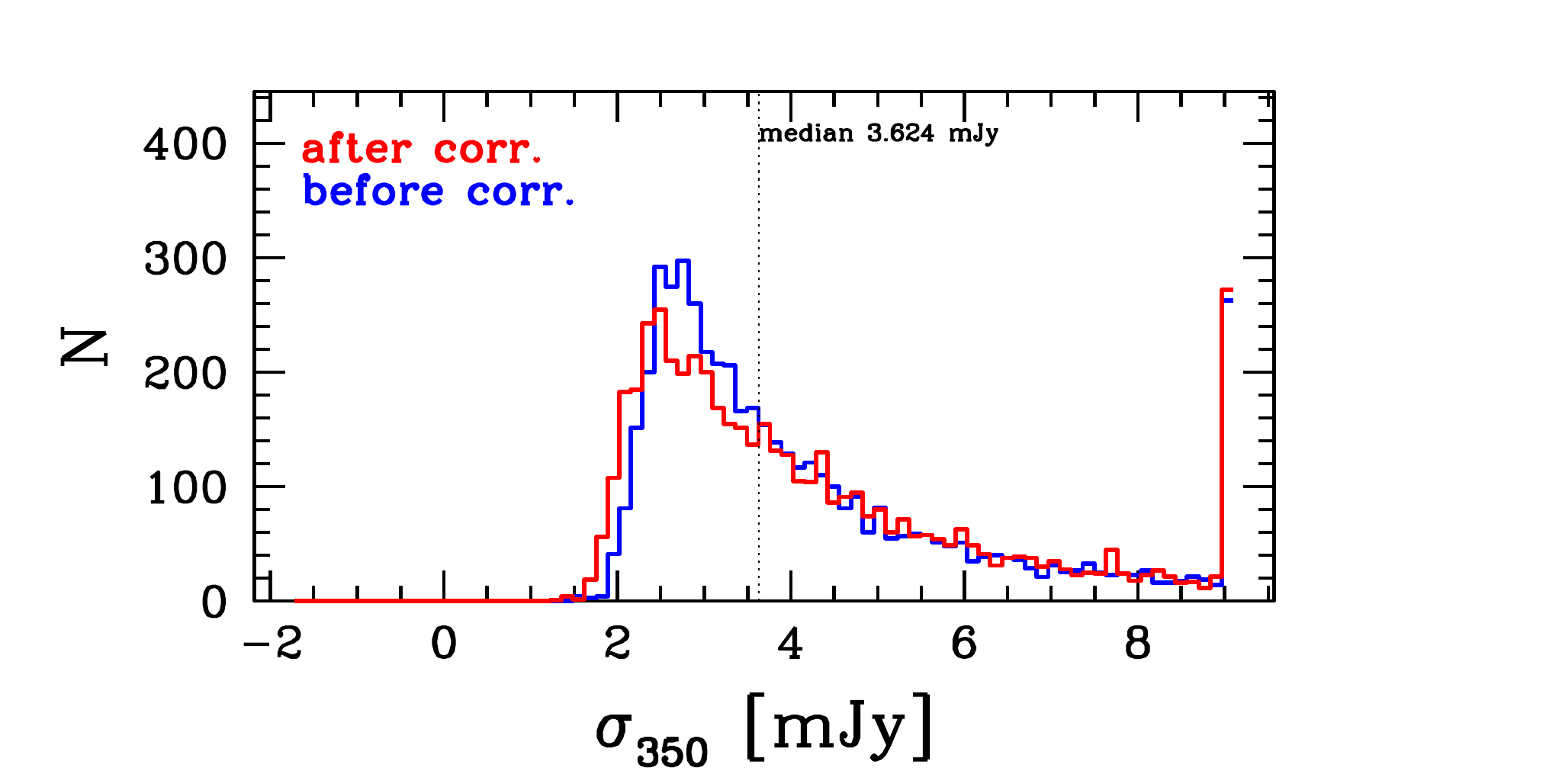}
	\includegraphics[width=0.23\textwidth, trim={1cm 15cm 0cm 2.5cm}, clip]{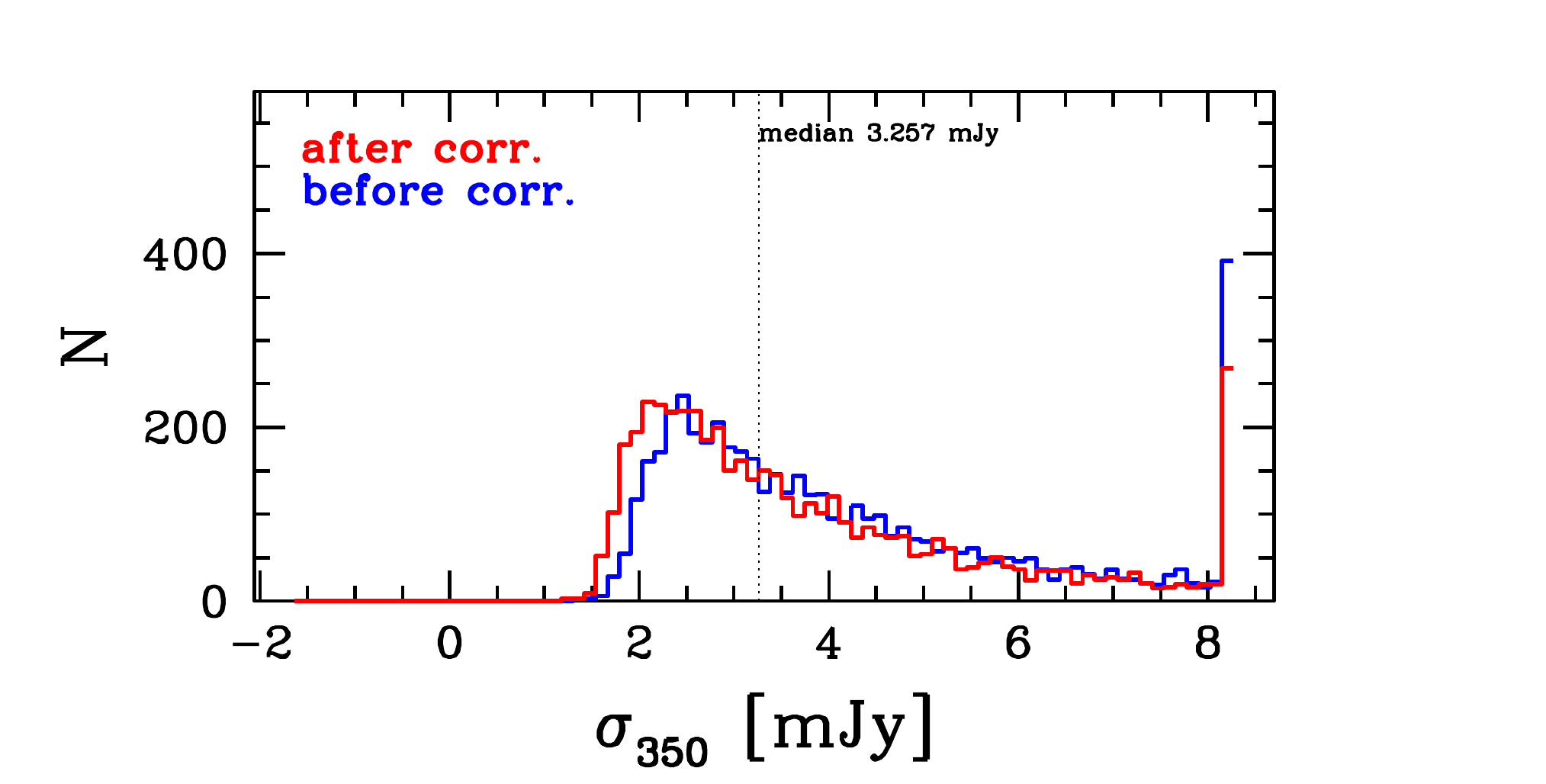}
    \end{subfigure}
    
	\caption{%
		Simulation correction analyses at SPIRE 350~$\mu$m. See descriptions in text. 
        \label{Figure_galsim_350_bin}
	}
\end{figure}

\begin{figure}
	\centering
    
    \begin{subfigure}[b]{\textwidth}\centering
	\includegraphics[width=0.23\textwidth, trim={1cm 15cm 0cm 2.5cm}, clip]{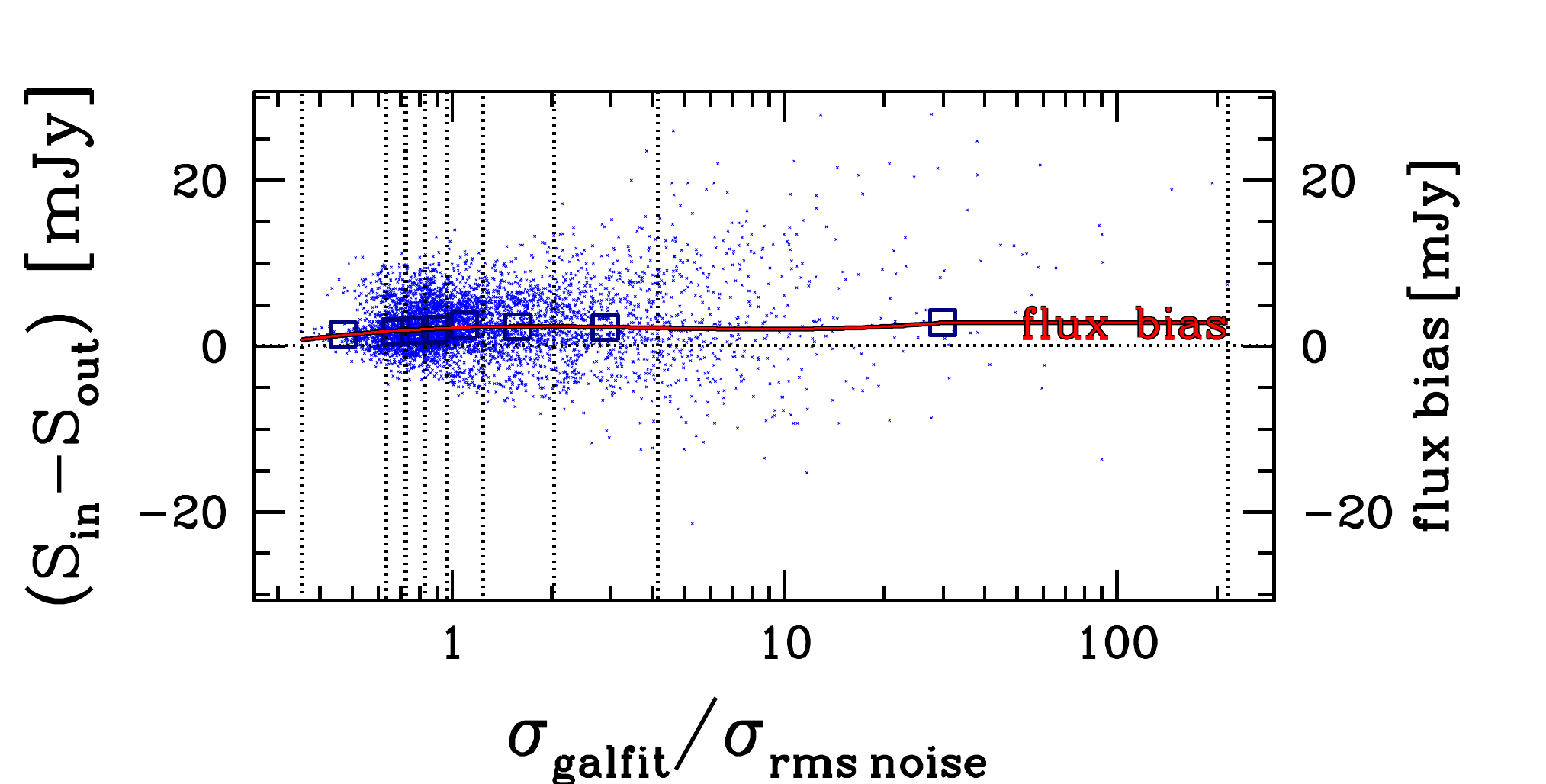}
	\includegraphics[width=0.23\textwidth, trim={1cm 15cm 0cm 2.5cm}, clip]{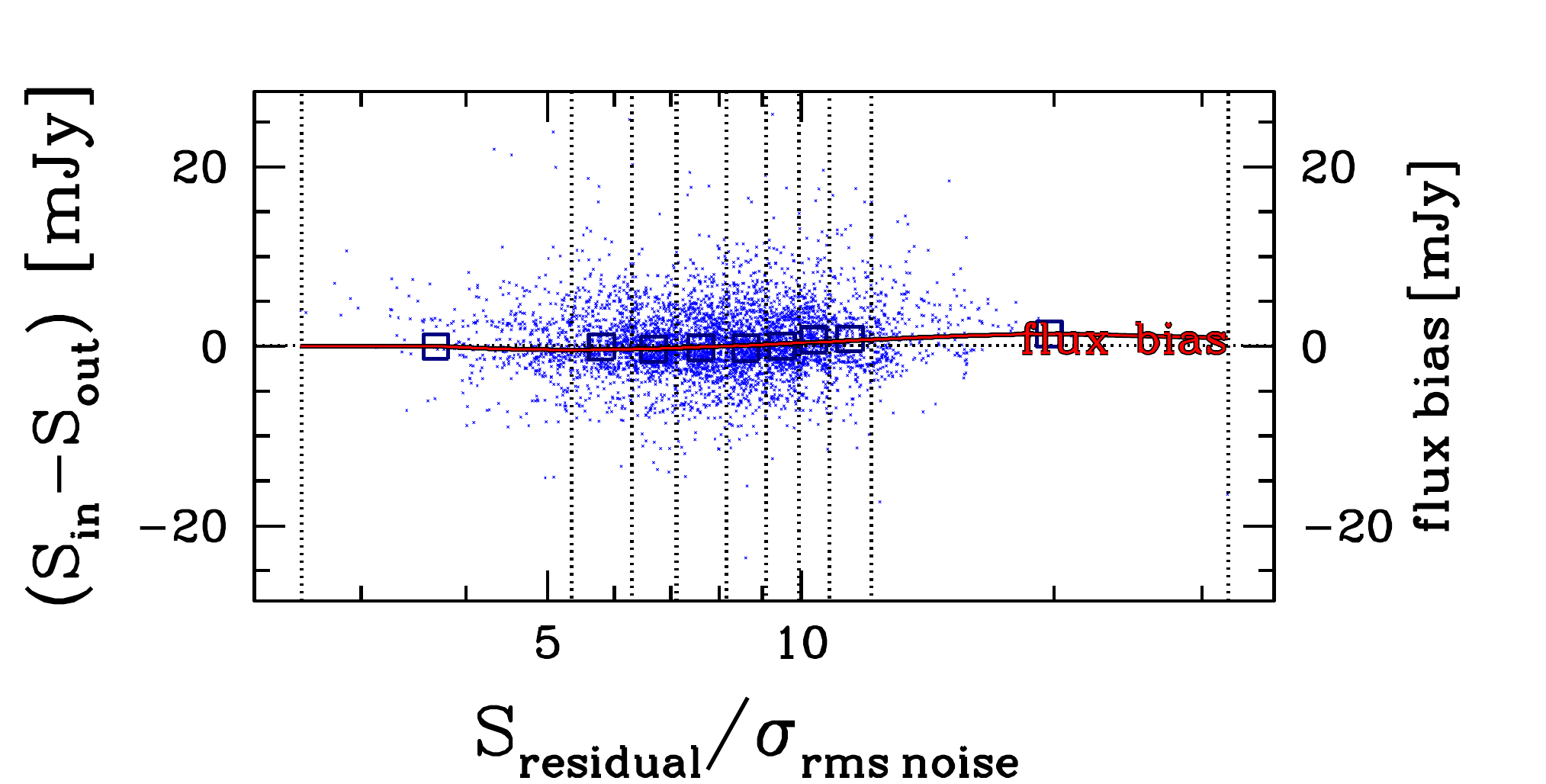}
	\includegraphics[width=0.23\textwidth, trim={1cm 15cm 0cm 2.5cm}, clip]{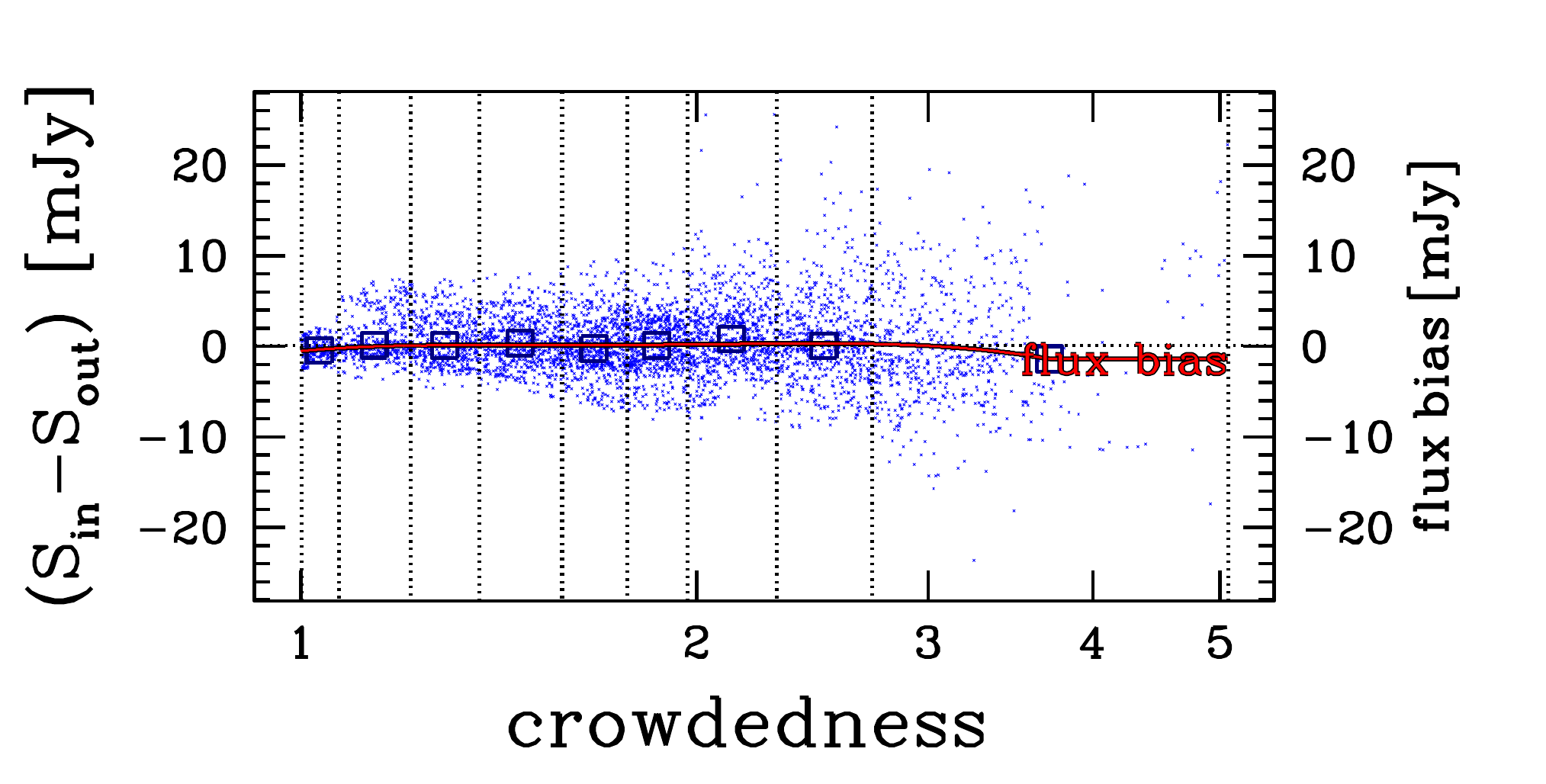}
	\includegraphics[width=0.23\textwidth, trim={1cm 15cm 0cm 2.5cm}, clip]{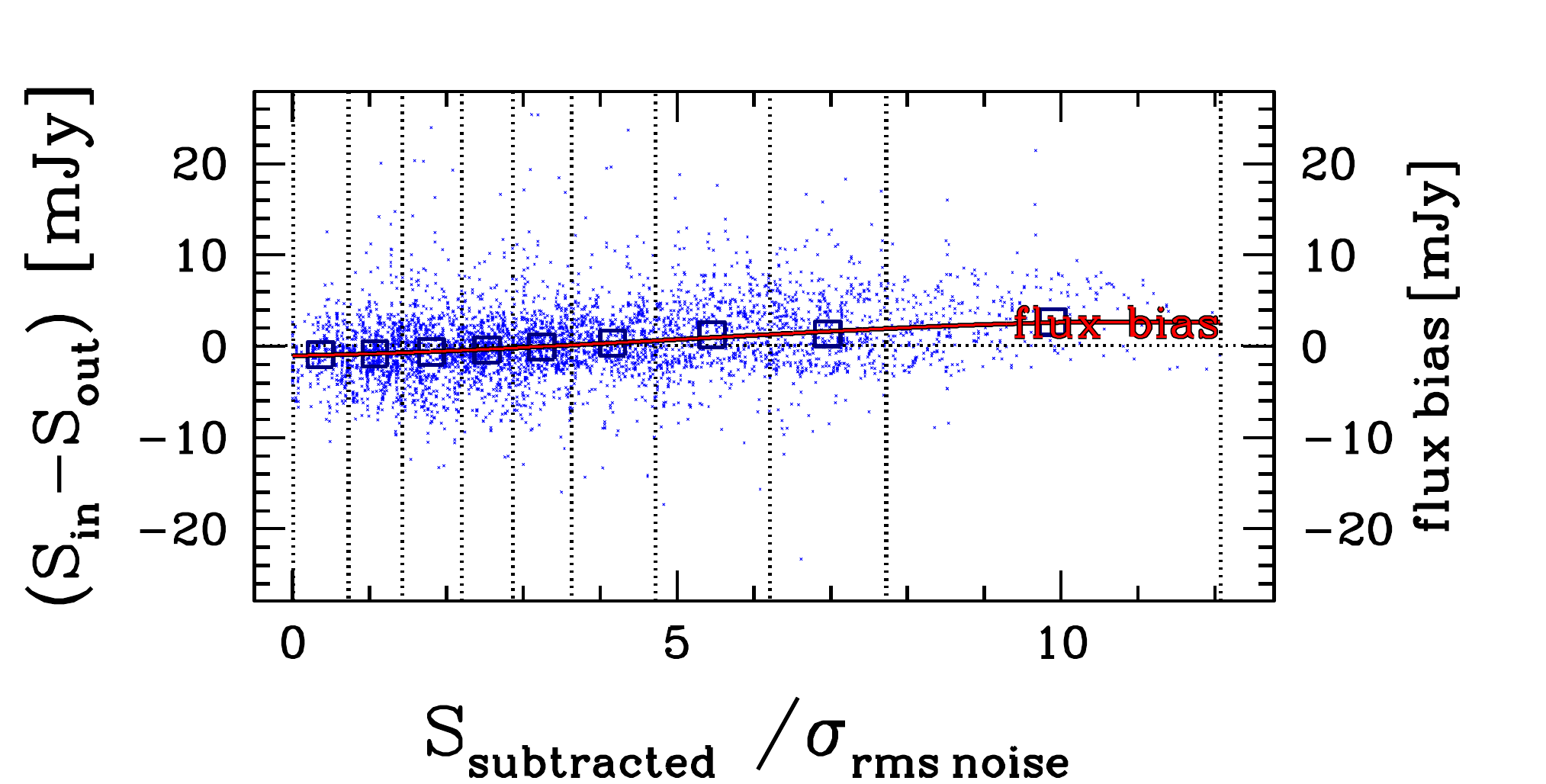}	
    \end{subfigure}
    
    \begin{subfigure}[b]{\textwidth}\centering
	\includegraphics[width=0.23\textwidth, trim={1cm 15cm 0cm 2.5cm}, clip]{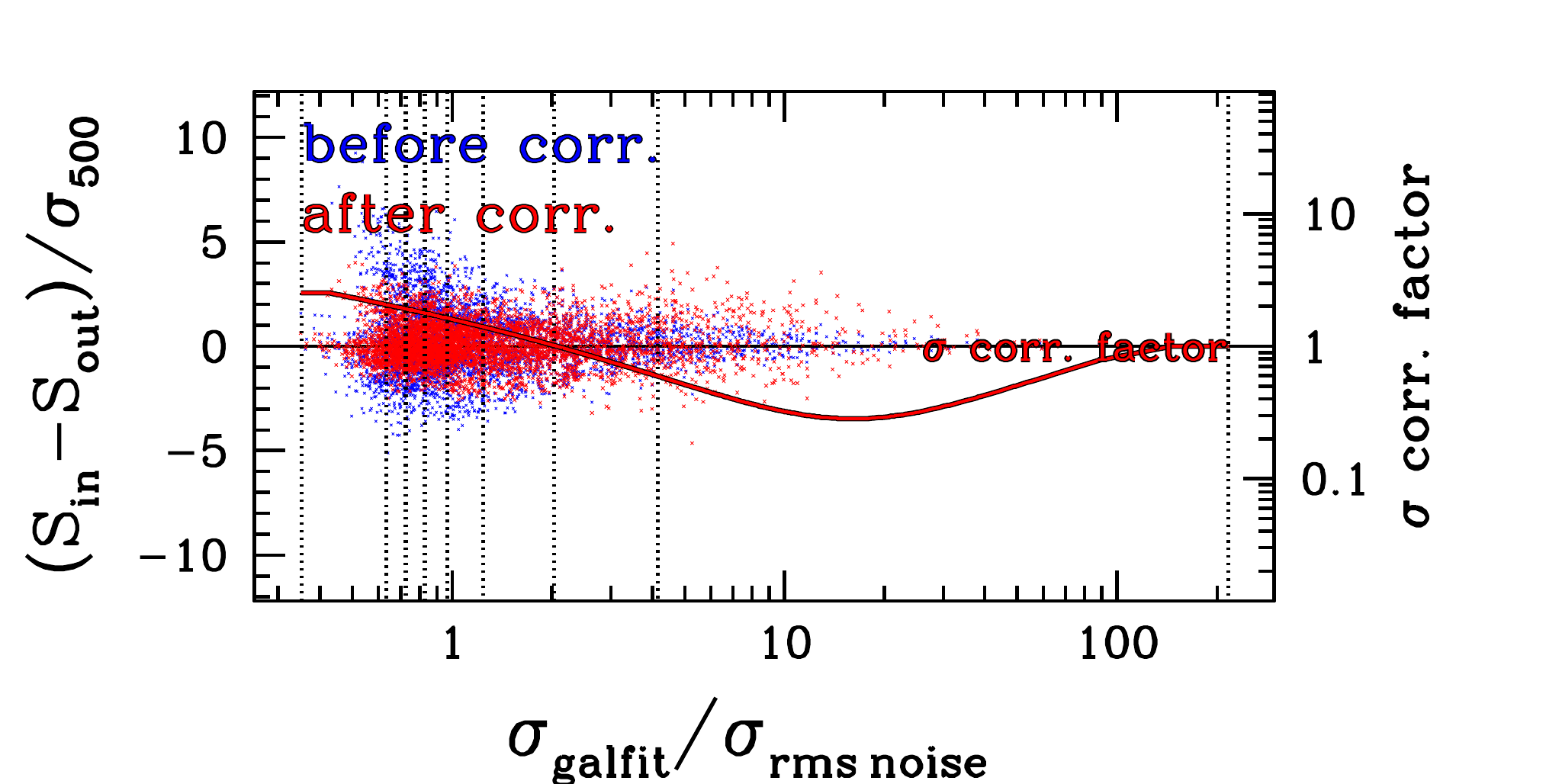}
	\includegraphics[width=0.23\textwidth, trim={1cm 15cm 0cm 2.5cm}, clip]{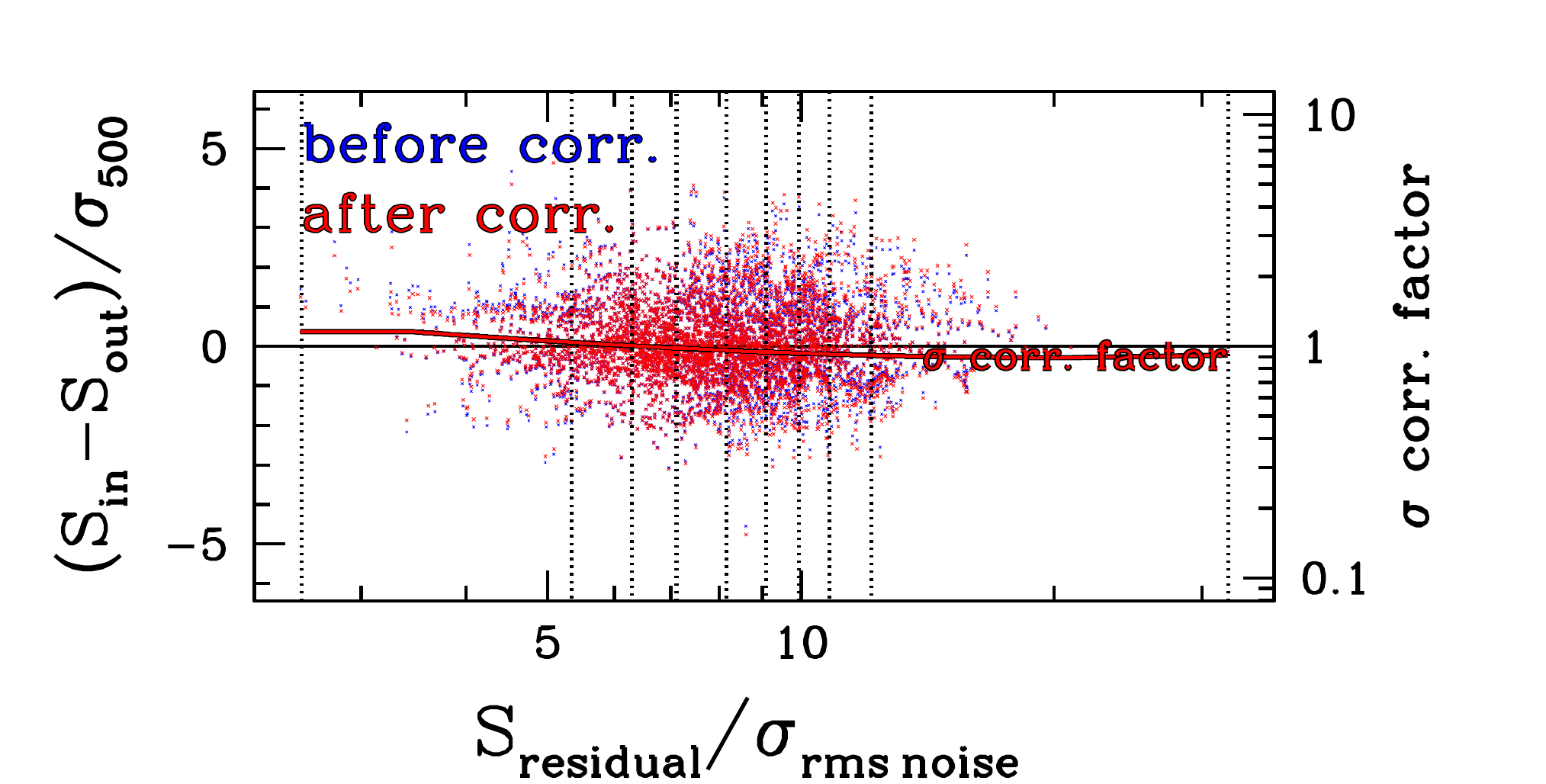}
	\includegraphics[width=0.23\textwidth, trim={1cm 15cm 0cm 2.5cm}, clip]{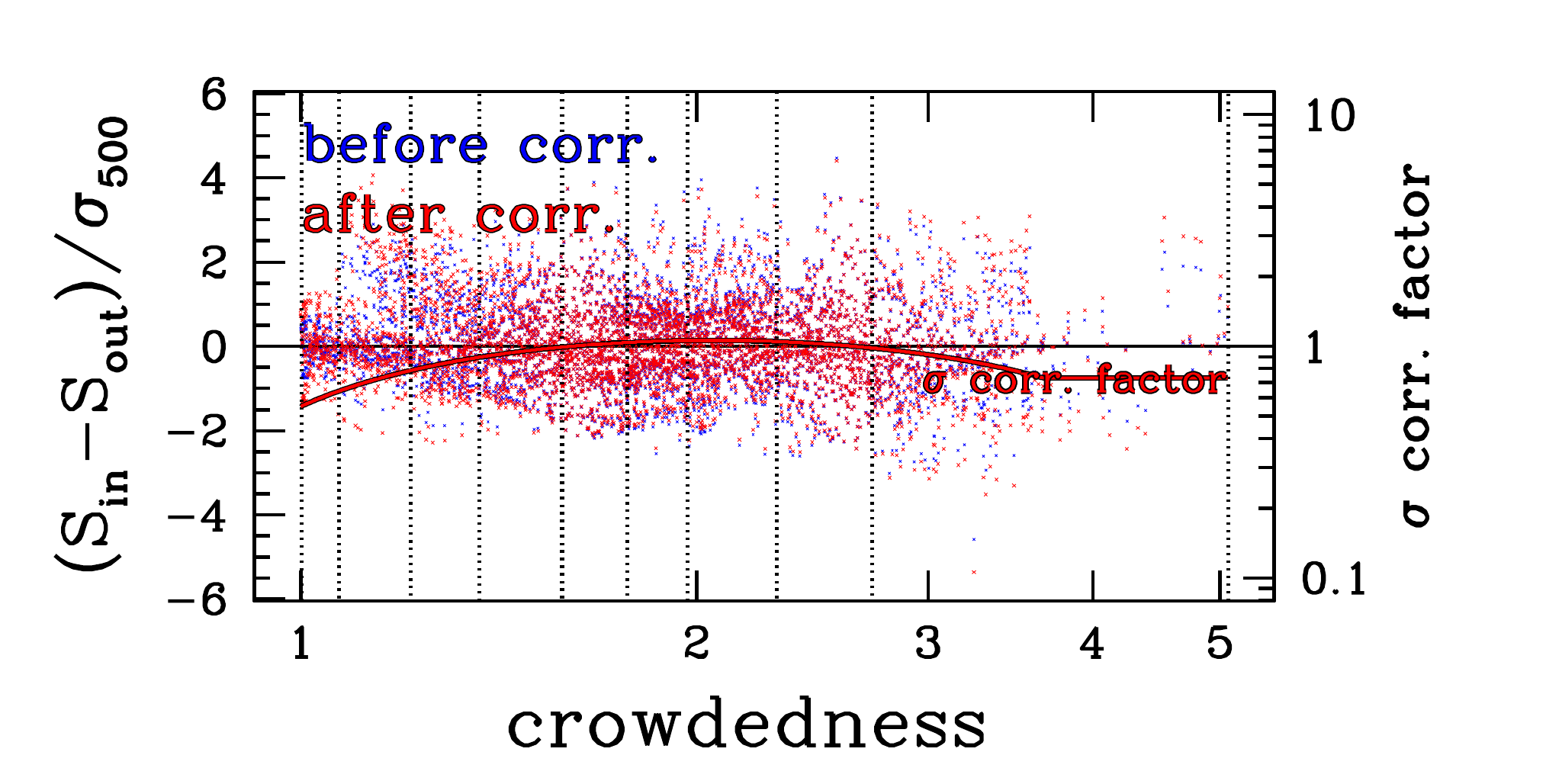}
	\includegraphics[width=0.23\textwidth, trim={1cm 15cm 0cm 2.5cm}, clip]{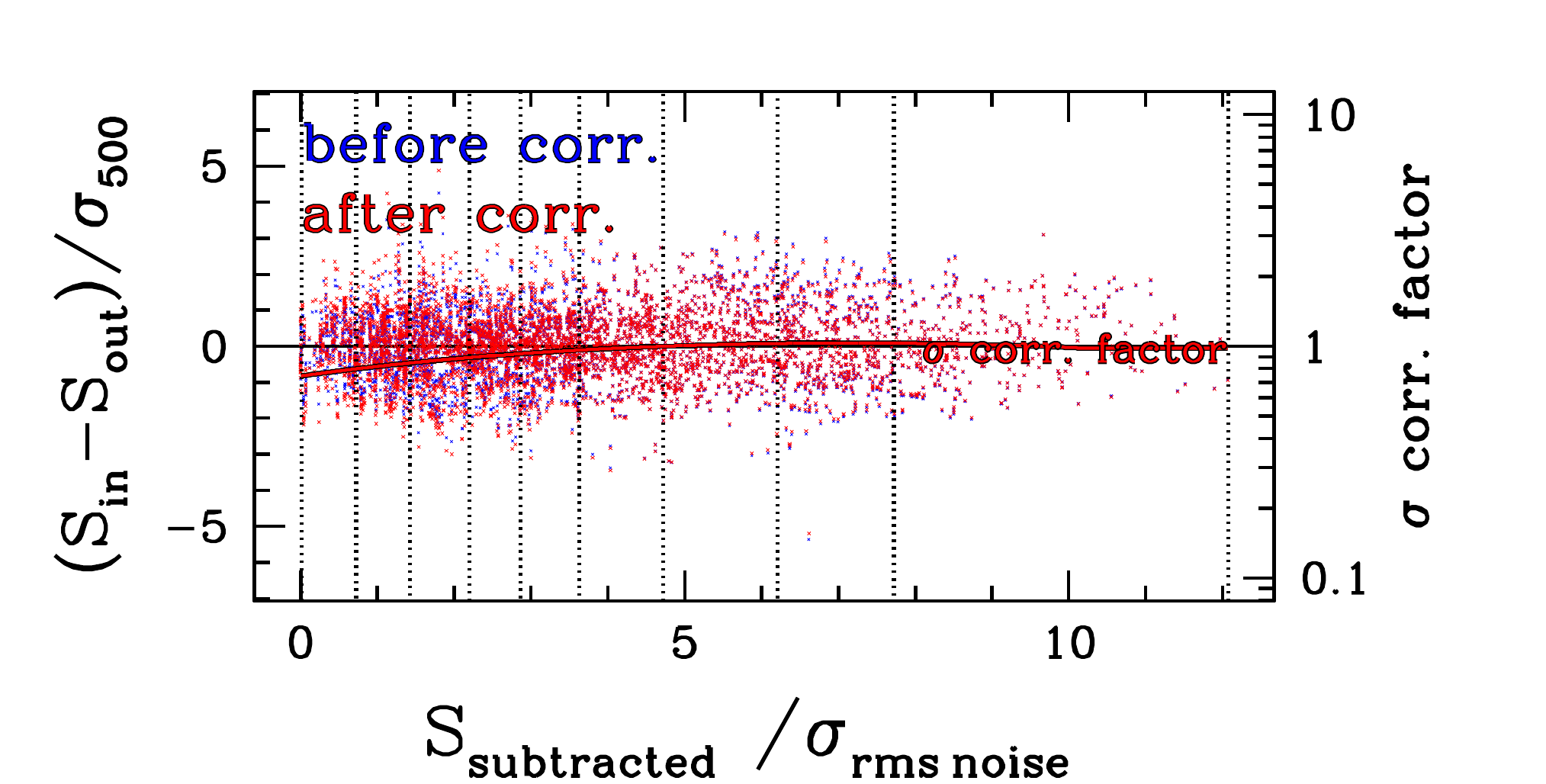}
    \end{subfigure}
    
    
    \begin{subfigure}[b]{\textwidth}\centering
	\includegraphics[width=0.23\textwidth, trim={1cm 15cm 0cm 2.5cm}, clip]{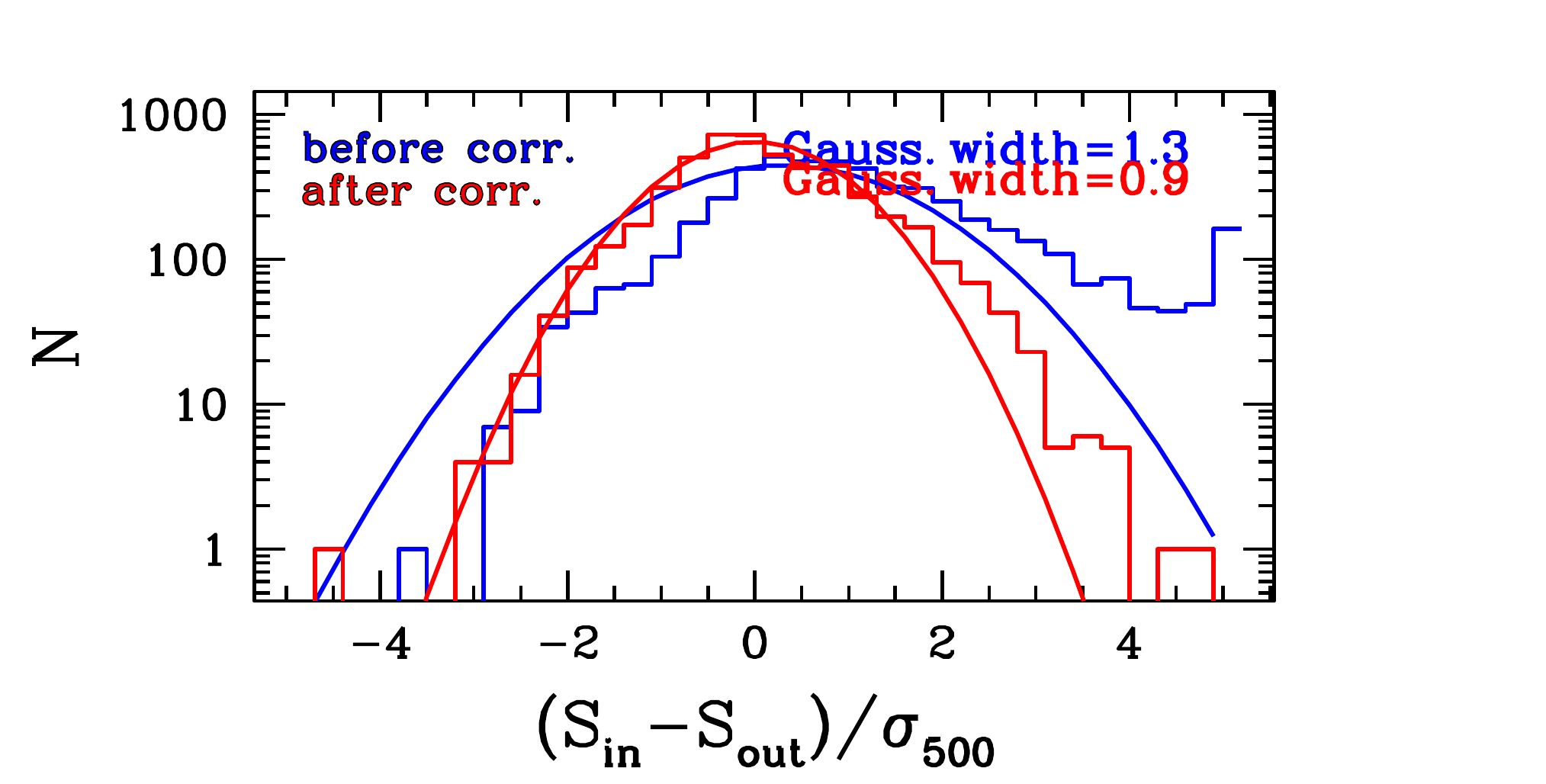}
	\includegraphics[width=0.23\textwidth, trim={1cm 15cm 0cm 2.5cm}, clip]{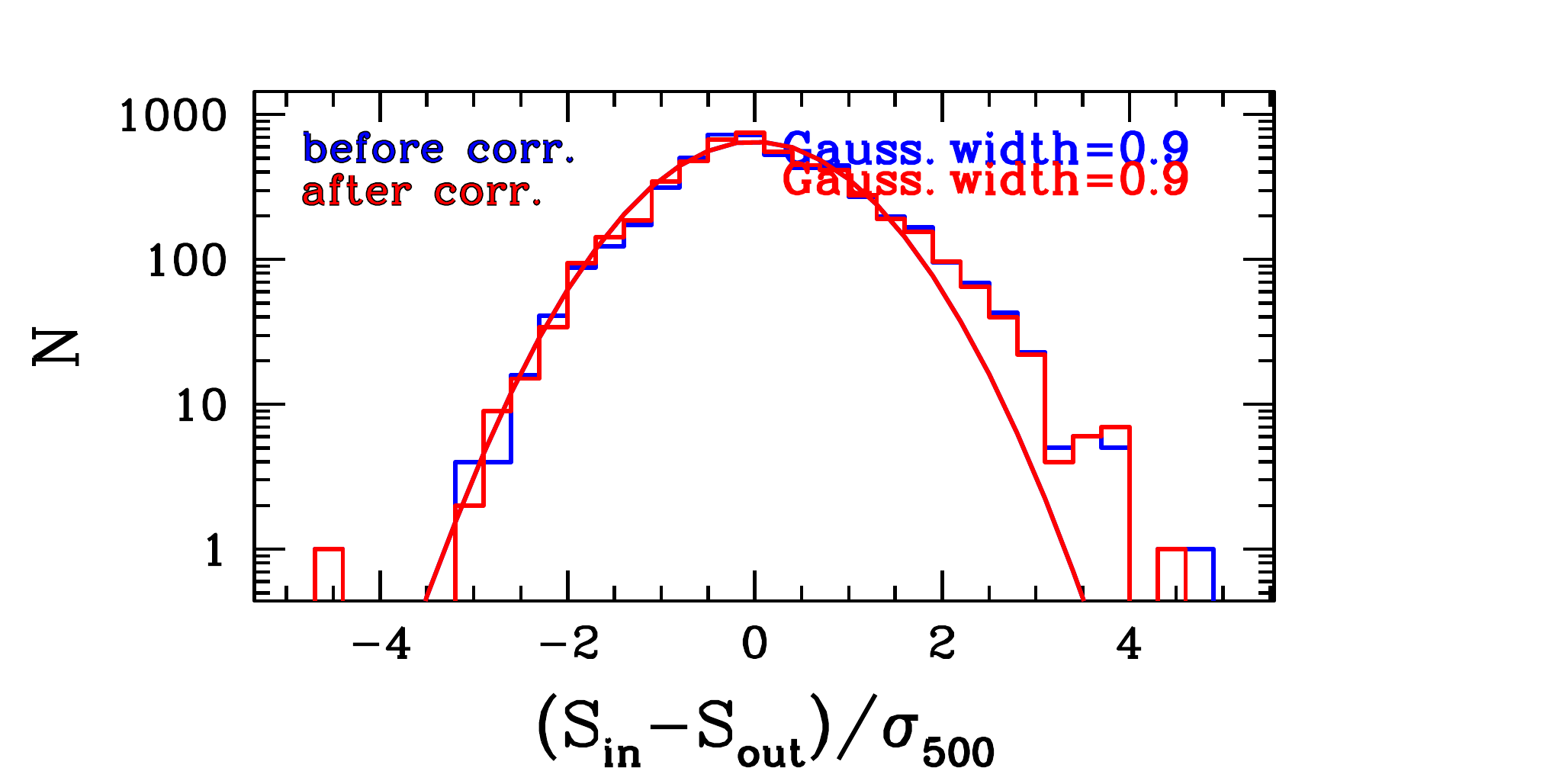}
	\includegraphics[width=0.23\textwidth, trim={1cm 15cm 0cm 2.5cm}, clip]{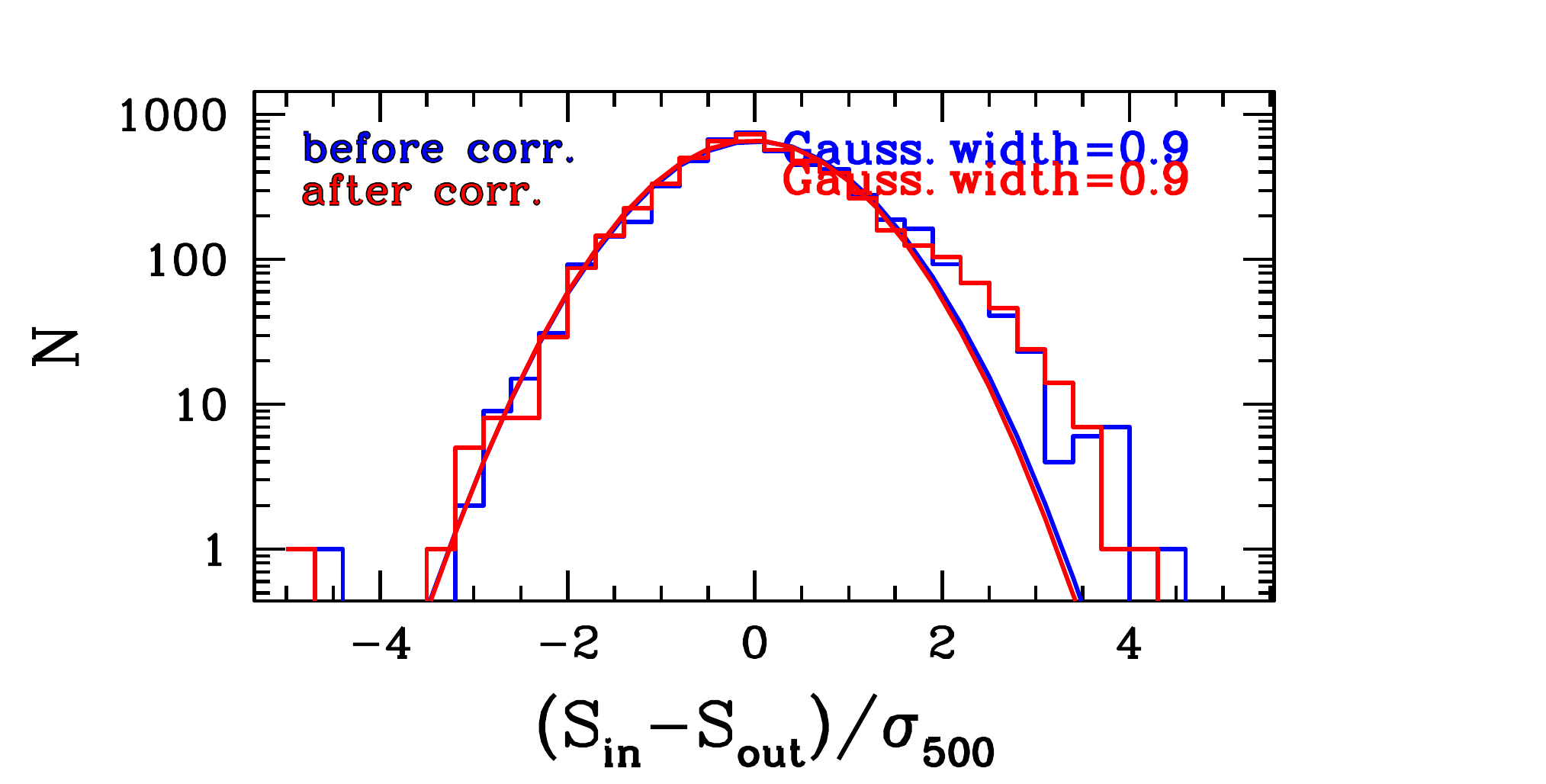}
	\includegraphics[width=0.23\textwidth, trim={1cm 15cm 0cm 2.5cm}, clip]{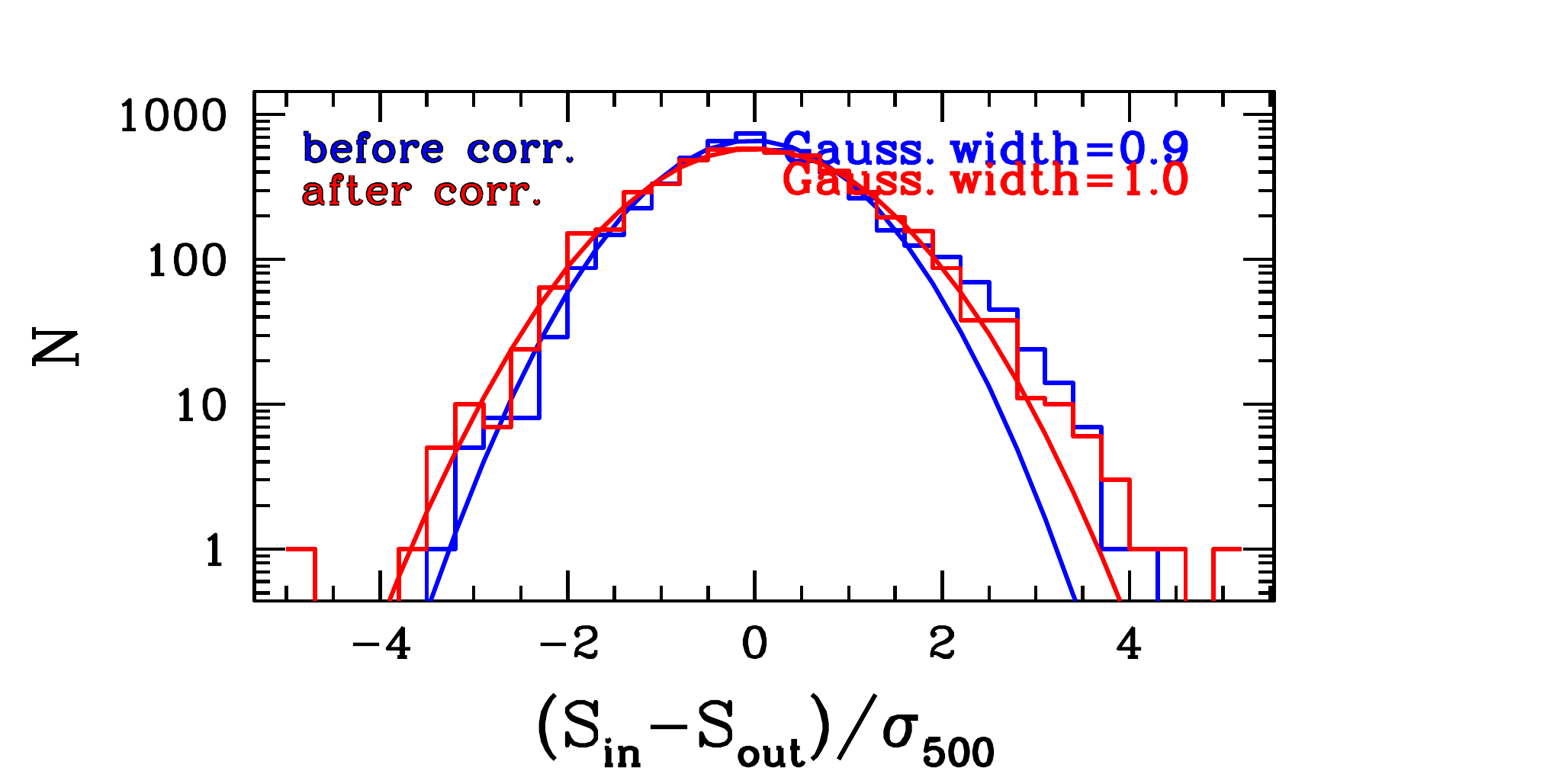}
    \end{subfigure}
    
    \begin{subfigure}[b]{\textwidth}\centering
	\includegraphics[width=0.23\textwidth, trim={1cm 15cm 0cm 2.5cm}, clip]{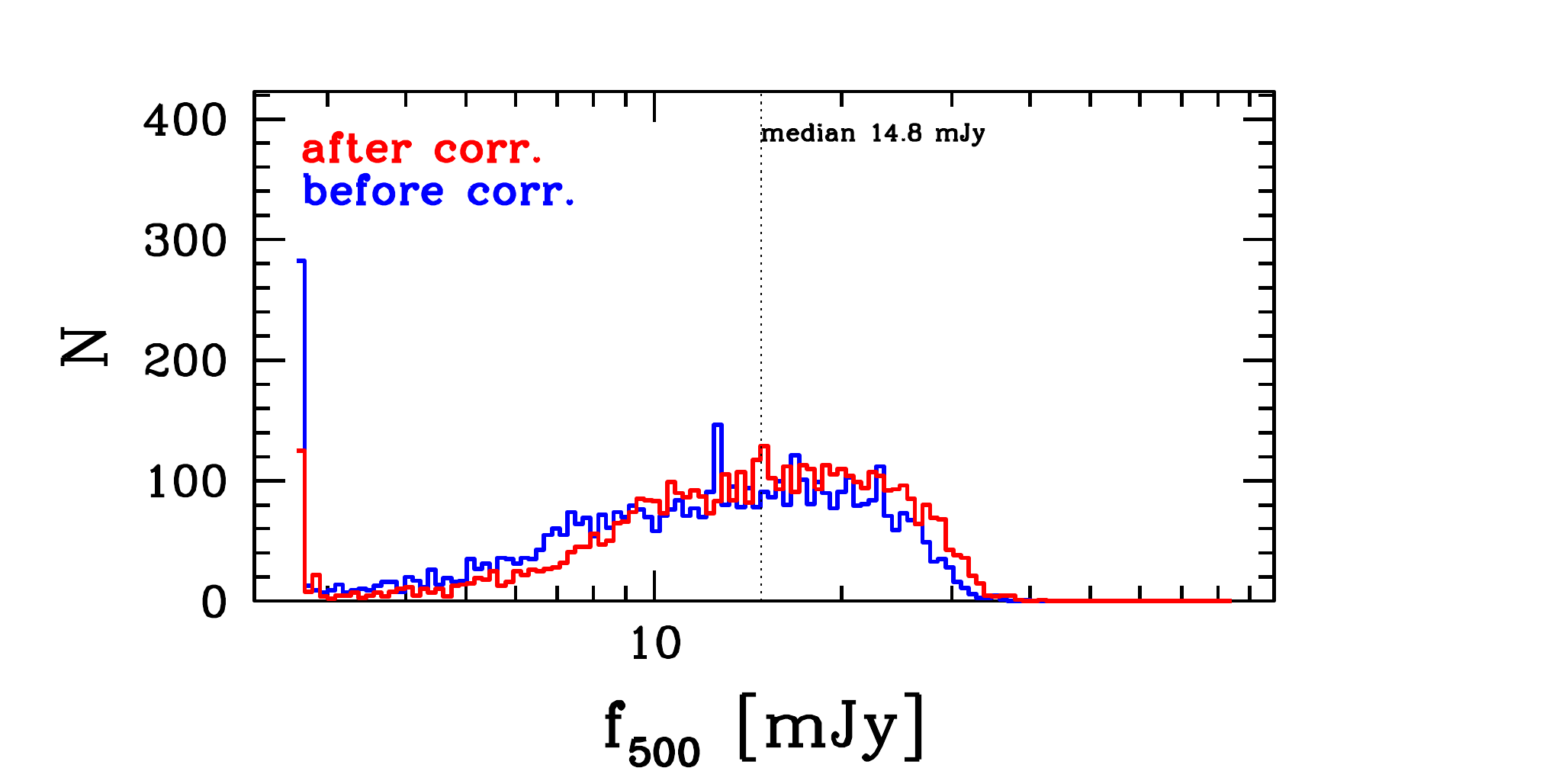}
	\includegraphics[width=0.23\textwidth, trim={1cm 15cm 0cm 2.5cm}, clip]{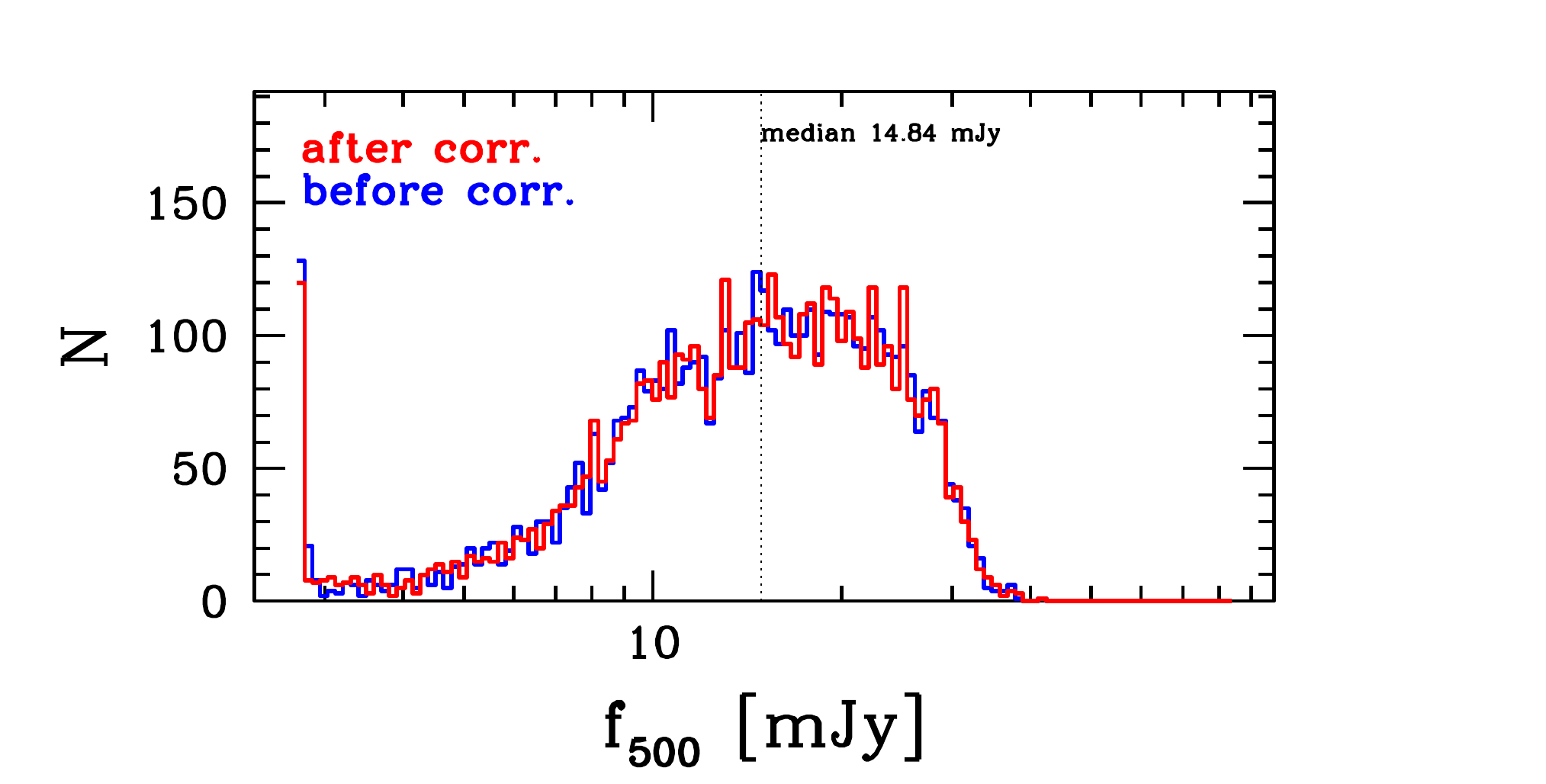}
	\includegraphics[width=0.23\textwidth, trim={1cm 15cm 0cm 2.5cm}, clip]{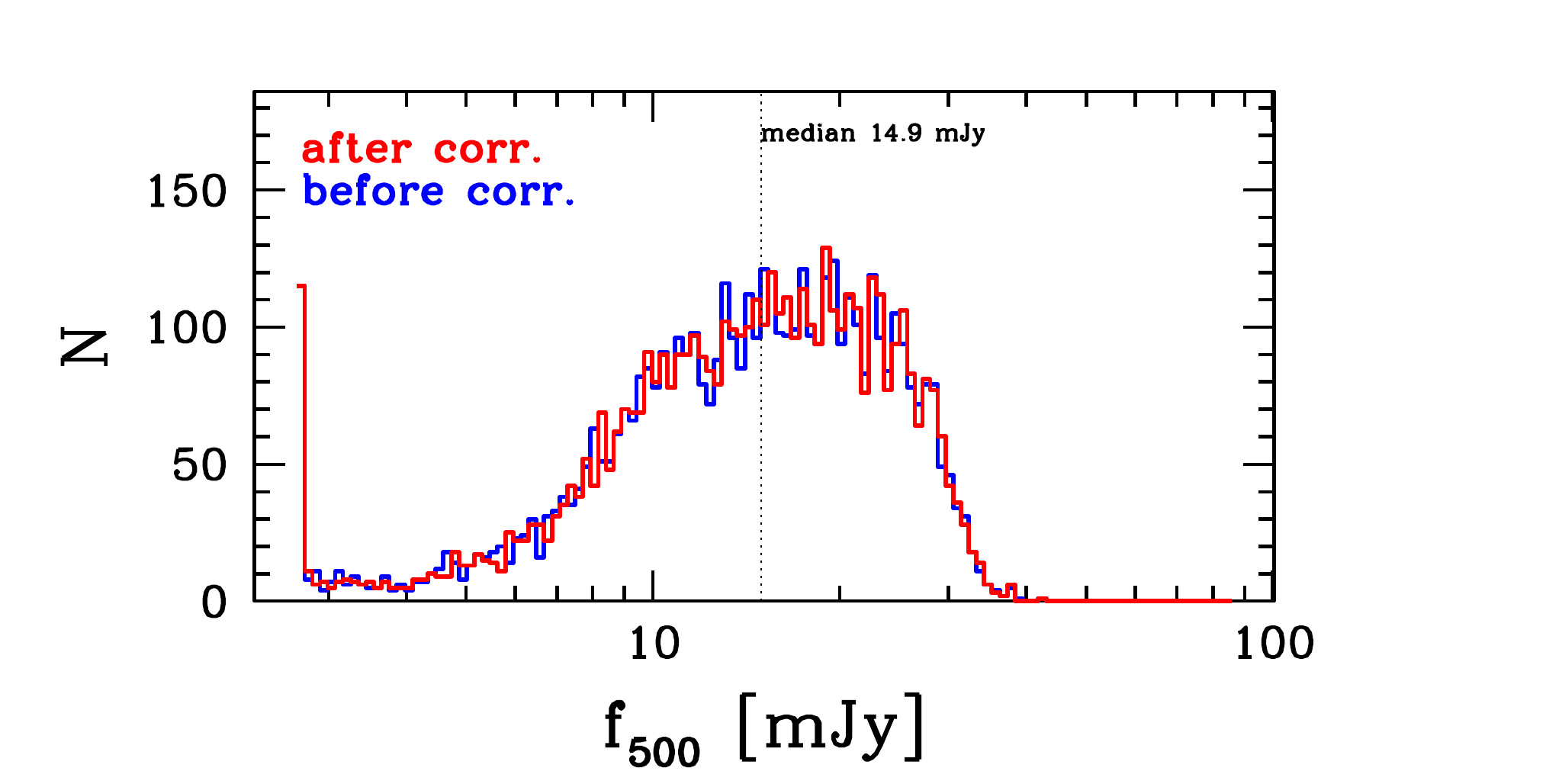}
	\includegraphics[width=0.23\textwidth, trim={1cm 15cm 0cm 2.5cm}, clip]{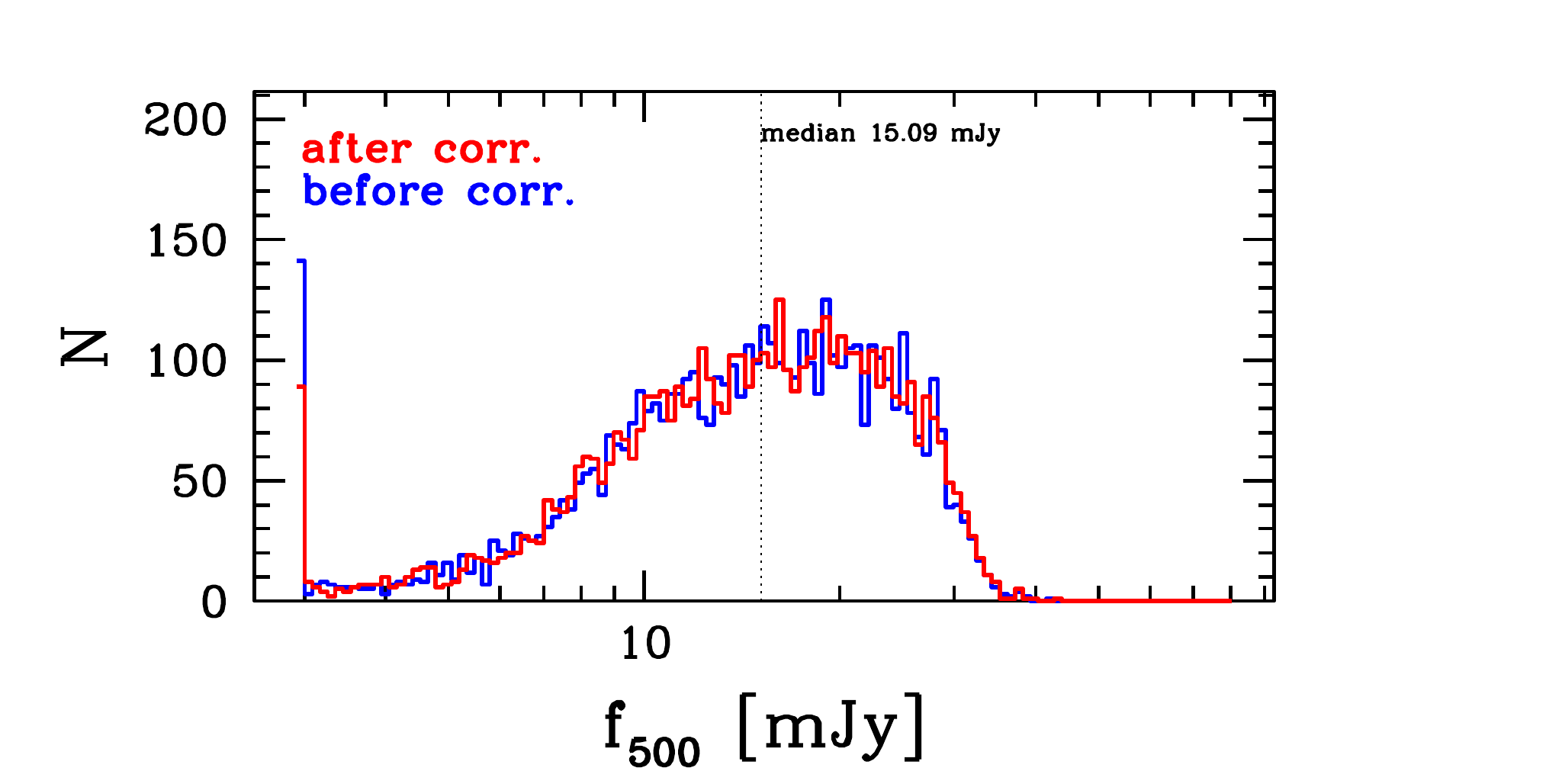}
    \end{subfigure}
    
    \begin{subfigure}[b]{\textwidth}\centering
	\includegraphics[width=0.23\textwidth, trim={1cm 15cm 0cm 2.5cm}, clip]{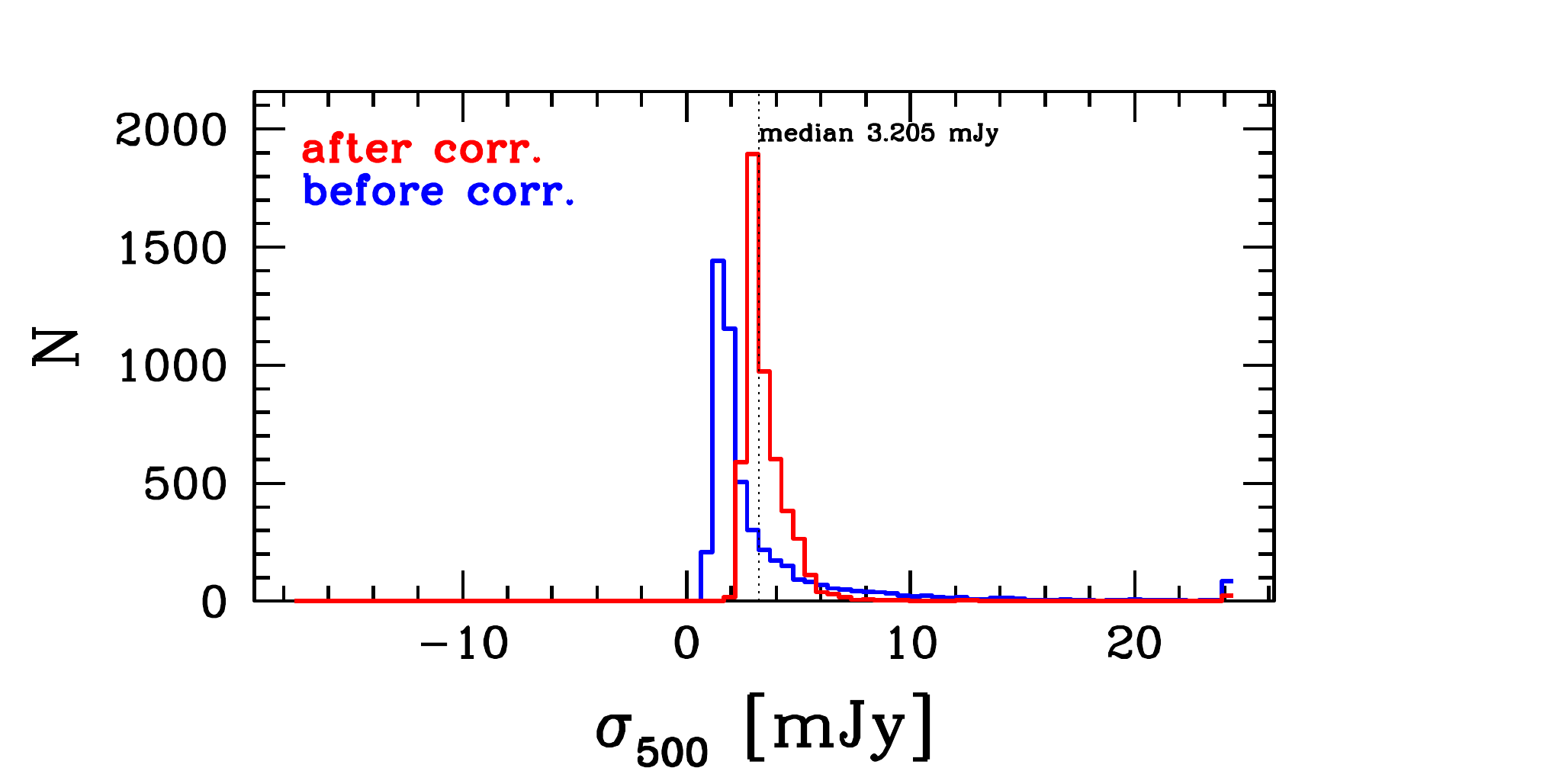}
	\includegraphics[width=0.23\textwidth, trim={1cm 15cm 0cm 2.5cm}, clip]{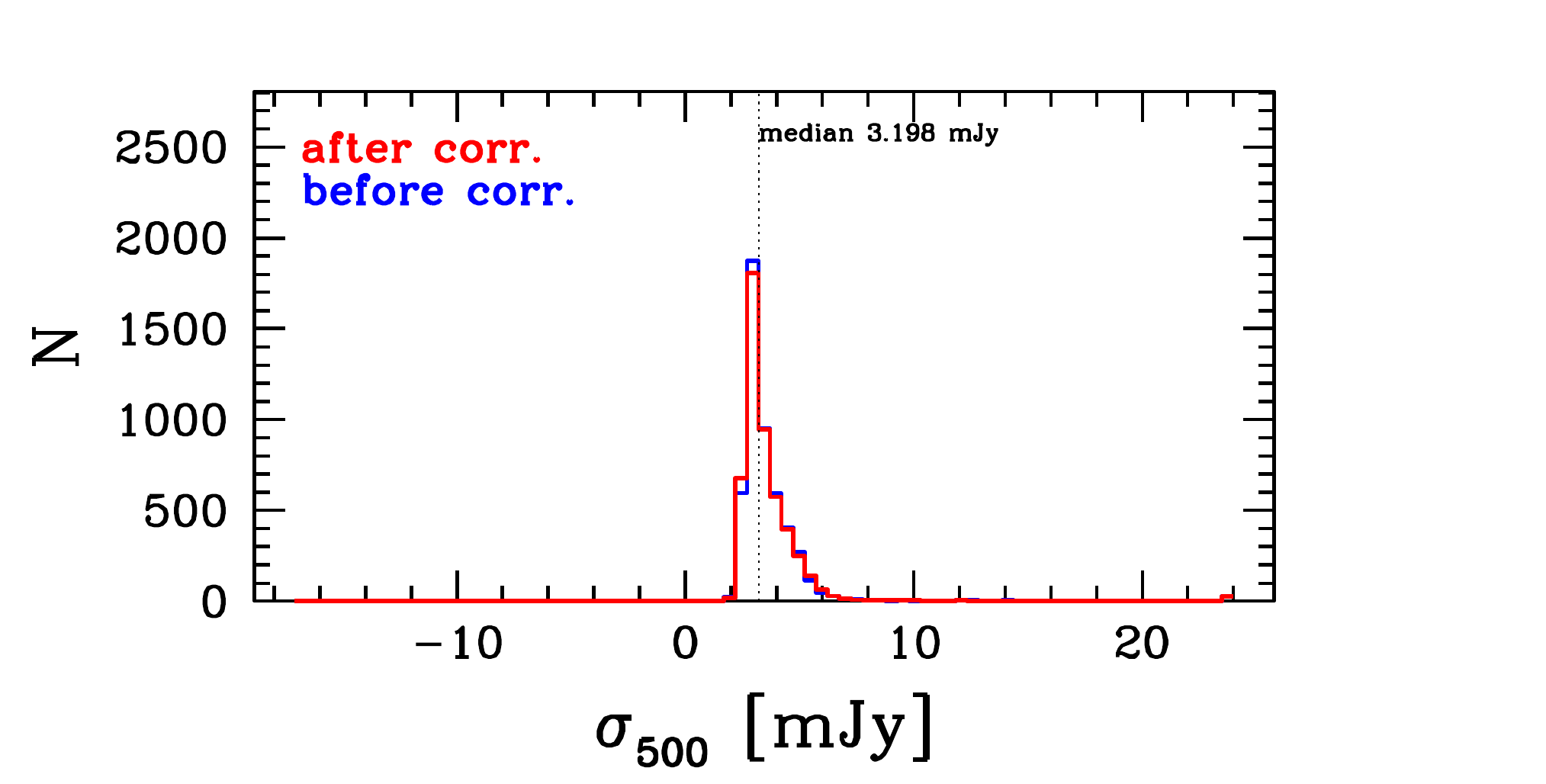}
	\includegraphics[width=0.23\textwidth, trim={1cm 15cm 0cm 2.5cm}, clip]{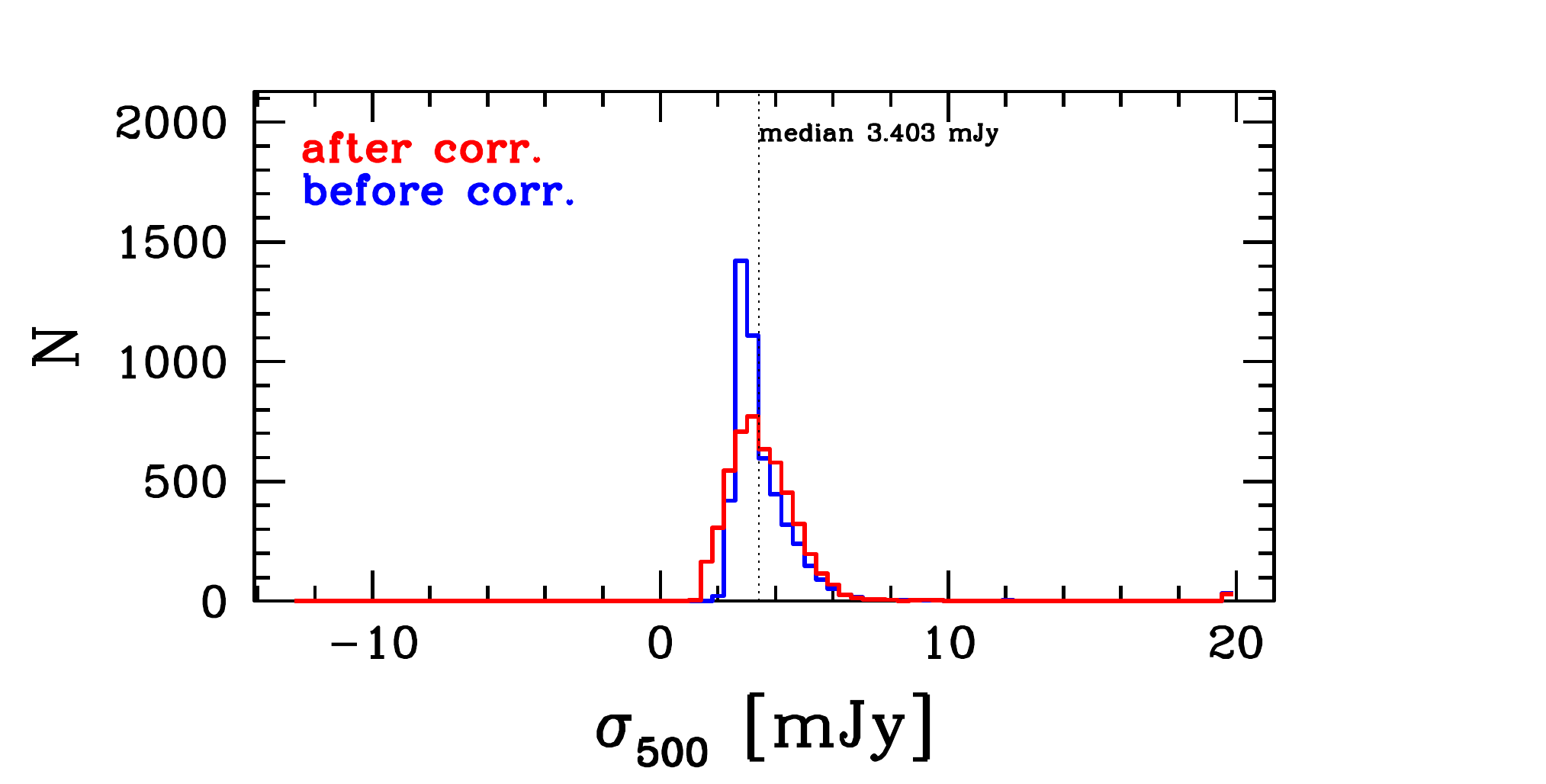}
	\includegraphics[width=0.23\textwidth, trim={1cm 15cm 0cm 2.5cm}, clip]{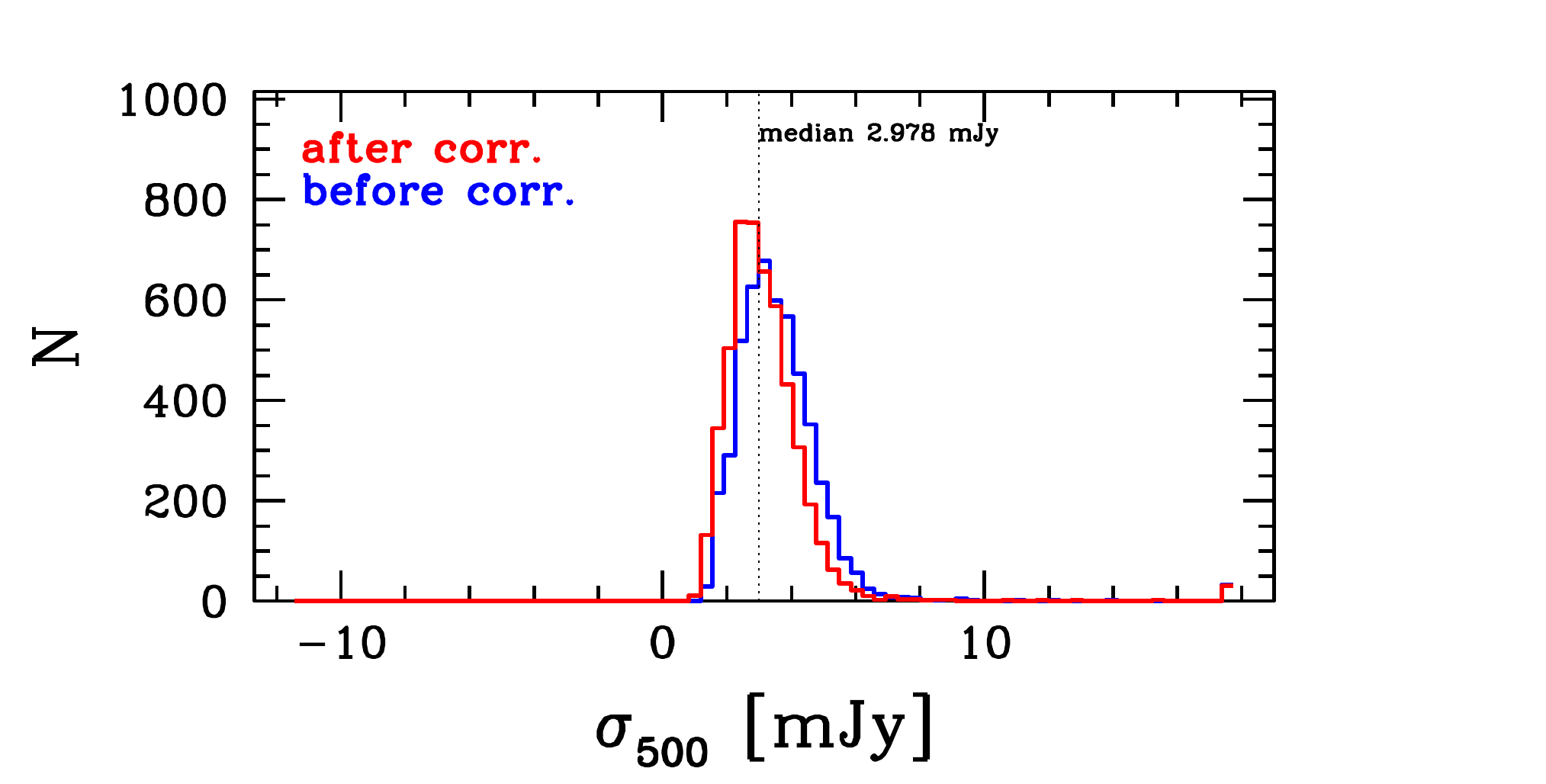}
    \end{subfigure}
    
	\caption{%
		Simulation correction analyses at SPIRE 500~$\mu$m. See descriptions in text. 
        \label{Figure_galsim_500_bin}
	}
\end{figure}

\begin{figure}
\centering

	\begin{subfigure}[b]{\textwidth}\centering
	\includegraphics[width=0.3\textwidth, trim={1cm 15cm 0cm 2.5cm}, clip]{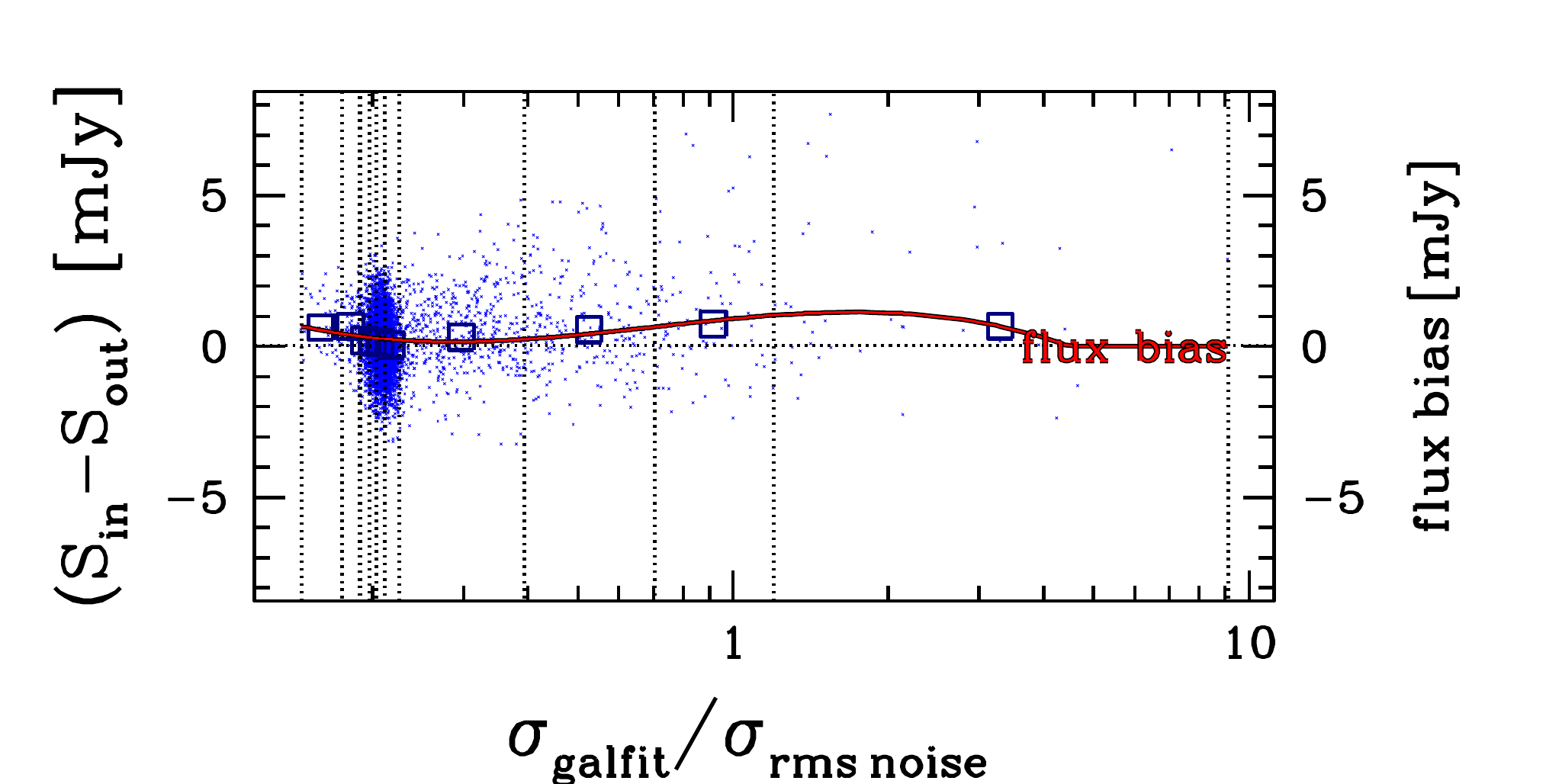}
	\includegraphics[width=0.3\textwidth, trim={1cm 15cm 0cm 2.5cm}, clip]{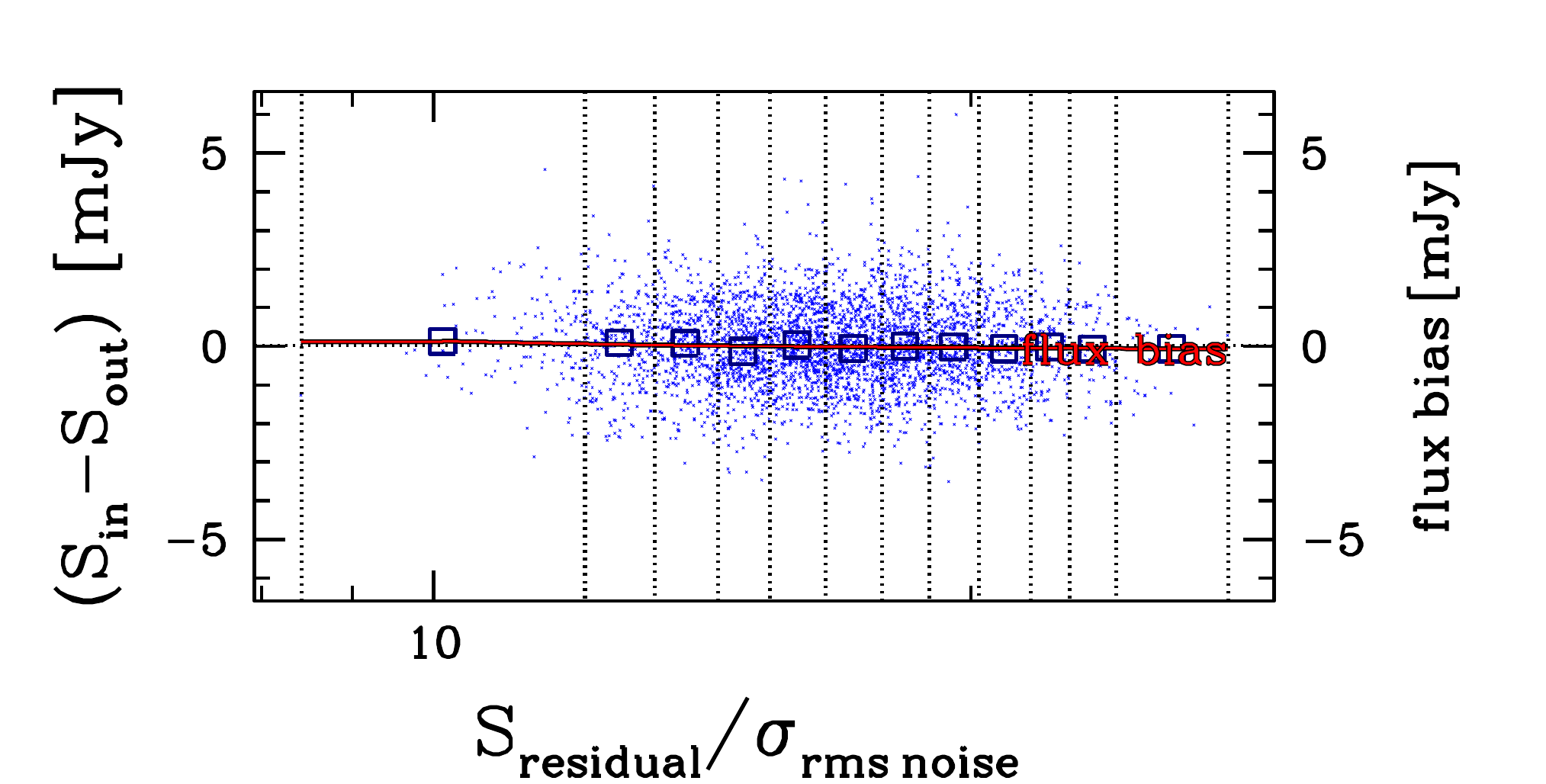}
	\includegraphics[width=0.3\textwidth, trim={1cm 15cm 0cm 2.5cm}, clip]{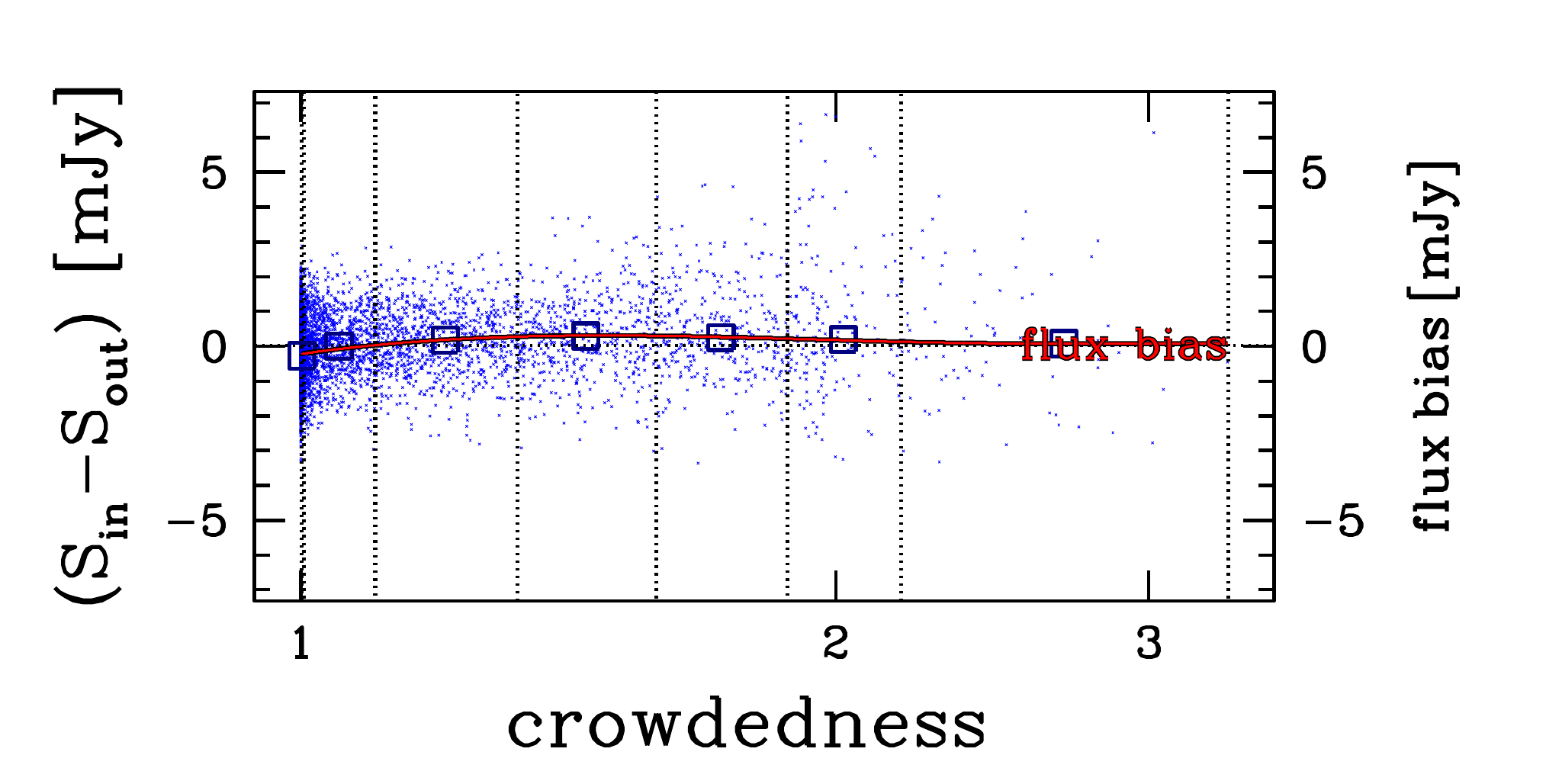}
	\end{subfigure}

	\begin{subfigure}[b]{\textwidth}\centering
	\includegraphics[width=0.3\textwidth, trim={1cm 15cm 0cm 2.5cm}, clip]{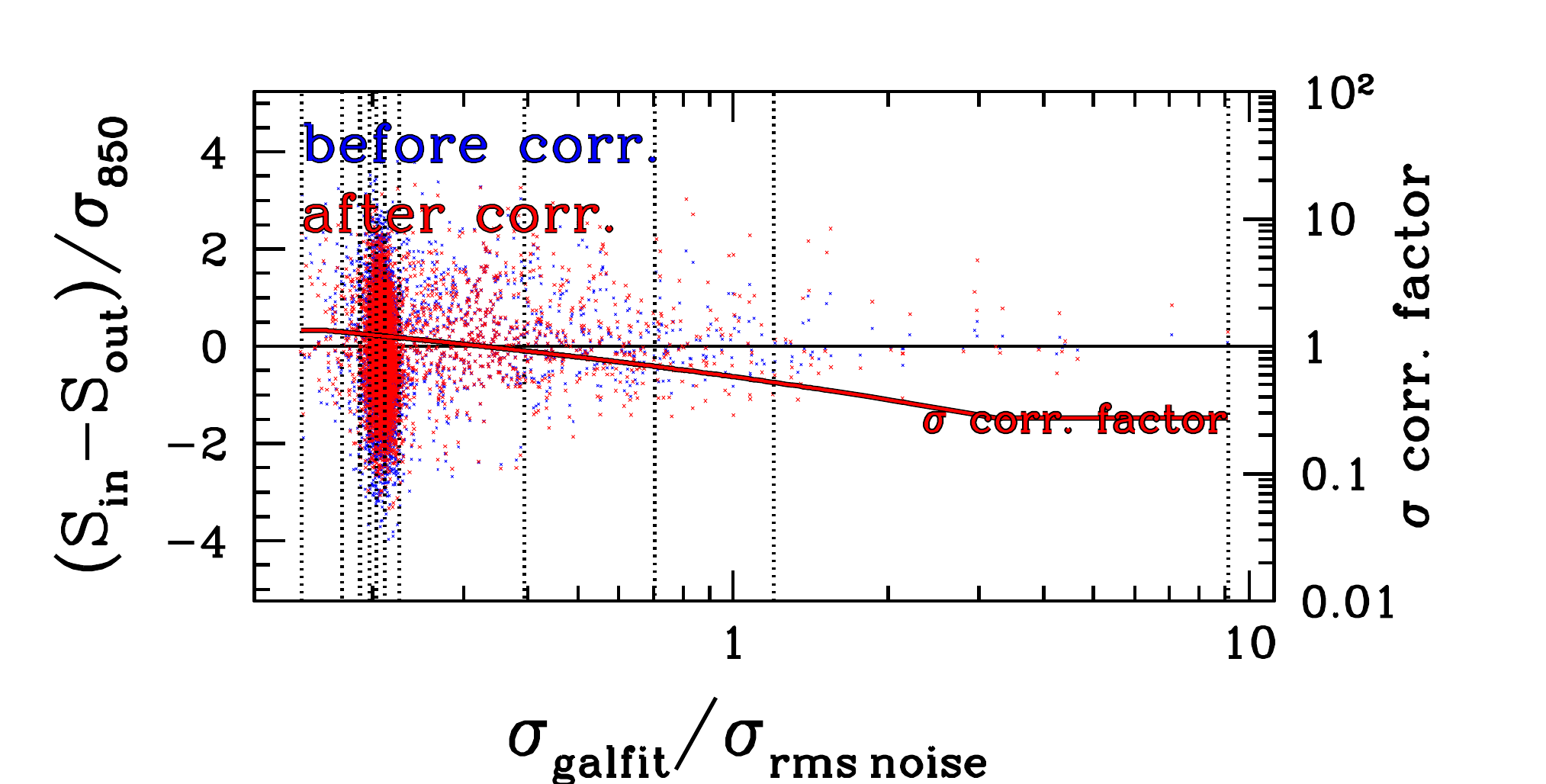}
	\includegraphics[width=0.3\textwidth, trim={1cm 15cm 0cm 2.5cm}, clip]{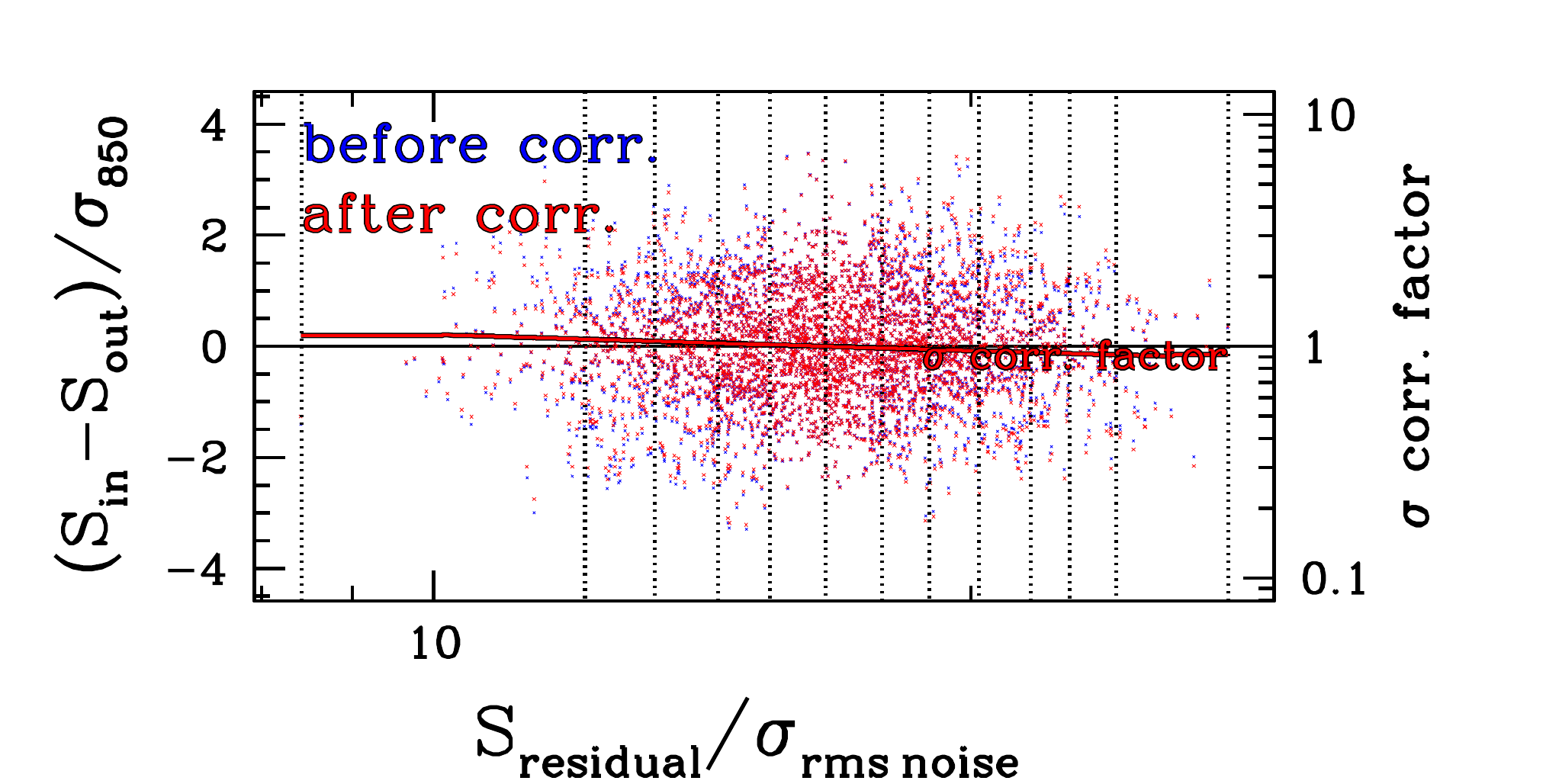}
	\includegraphics[width=0.3\textwidth, trim={1cm 15cm 0cm 2.5cm}, clip]{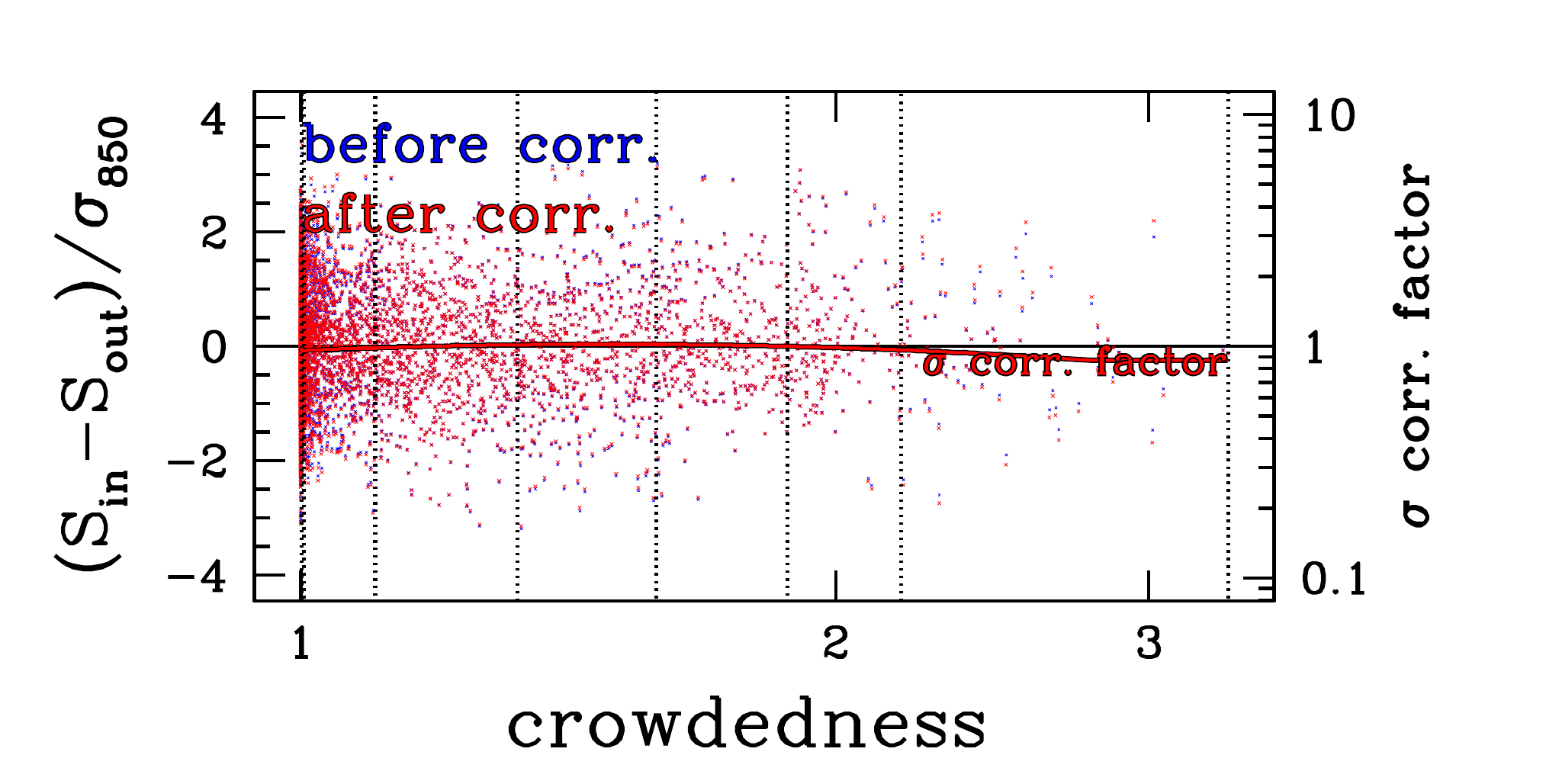}
	\end{subfigure}

	\begin{subfigure}[b]{\textwidth}\centering
	\includegraphics[width=0.3\textwidth, trim={1cm 15cm 0cm 2.5cm}, clip]{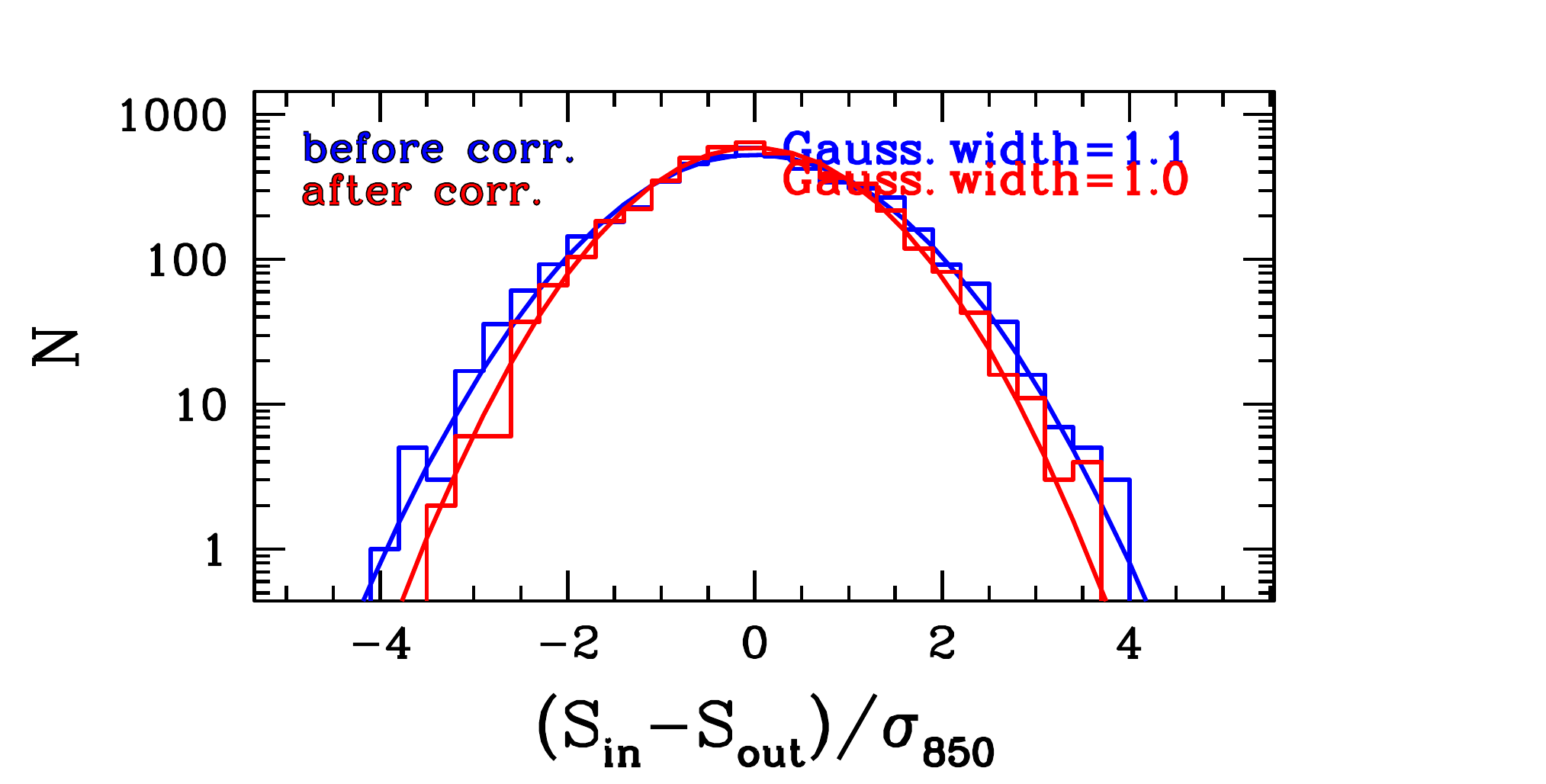}
	\includegraphics[width=0.3\textwidth, trim={1cm 15cm 0cm 2.5cm}, clip]{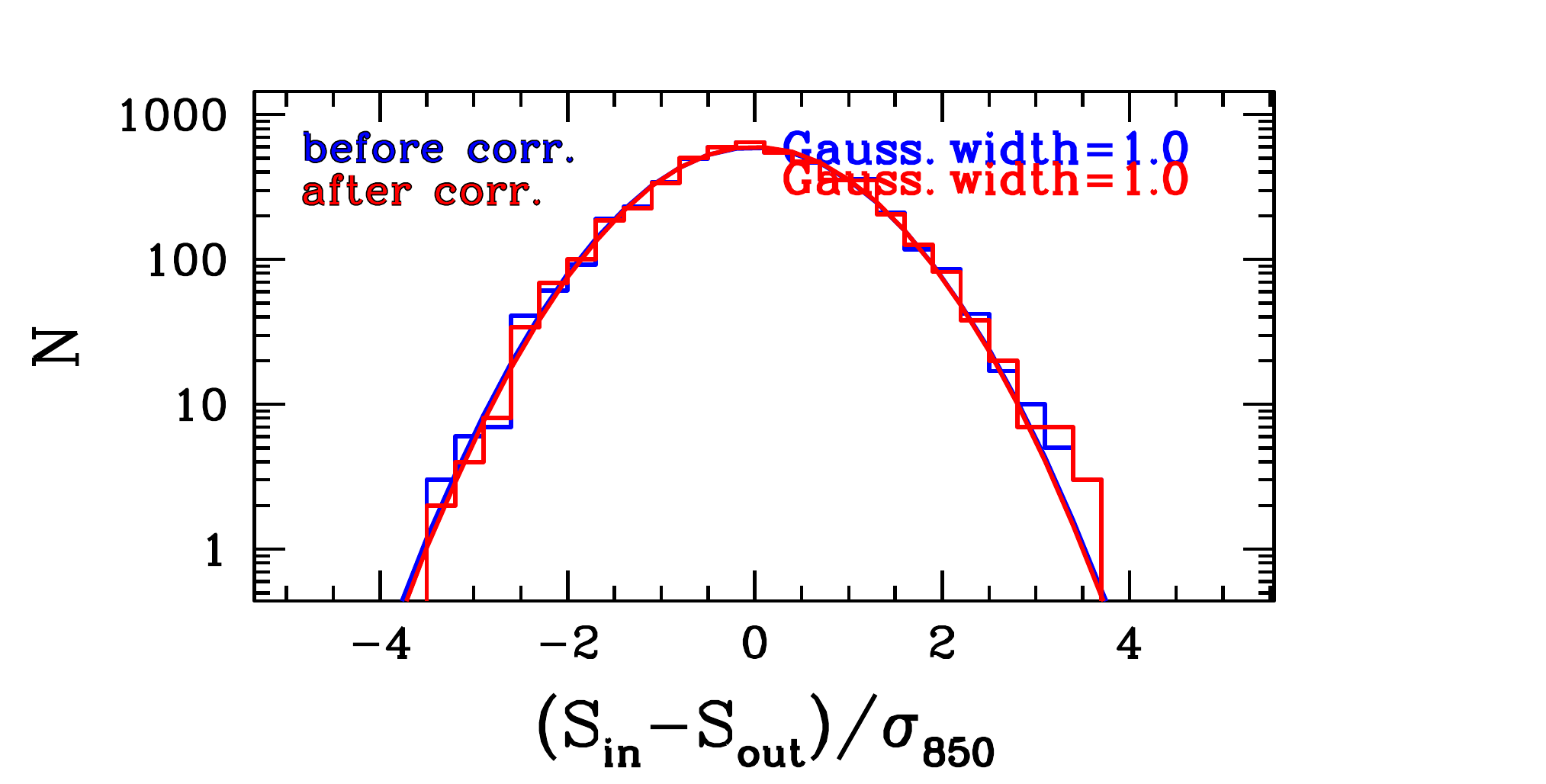}
	\includegraphics[width=0.3\textwidth, trim={1cm 15cm 0cm 2.5cm}, clip]{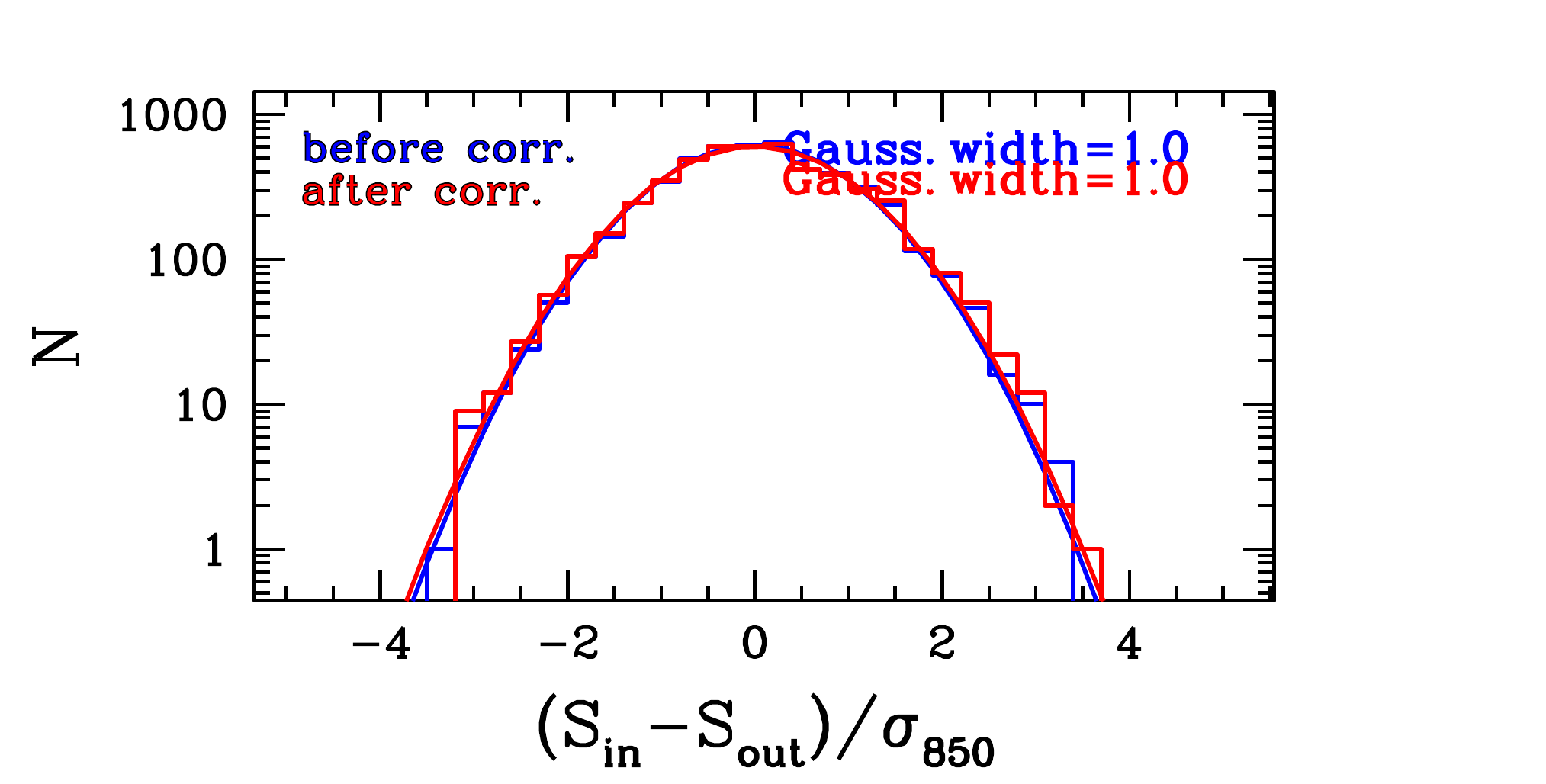}
	\end{subfigure}

	\begin{subfigure}[b]{\textwidth}\centering
	\includegraphics[width=0.3\textwidth, trim={1cm 15cm 0cm 2.5cm}, clip]{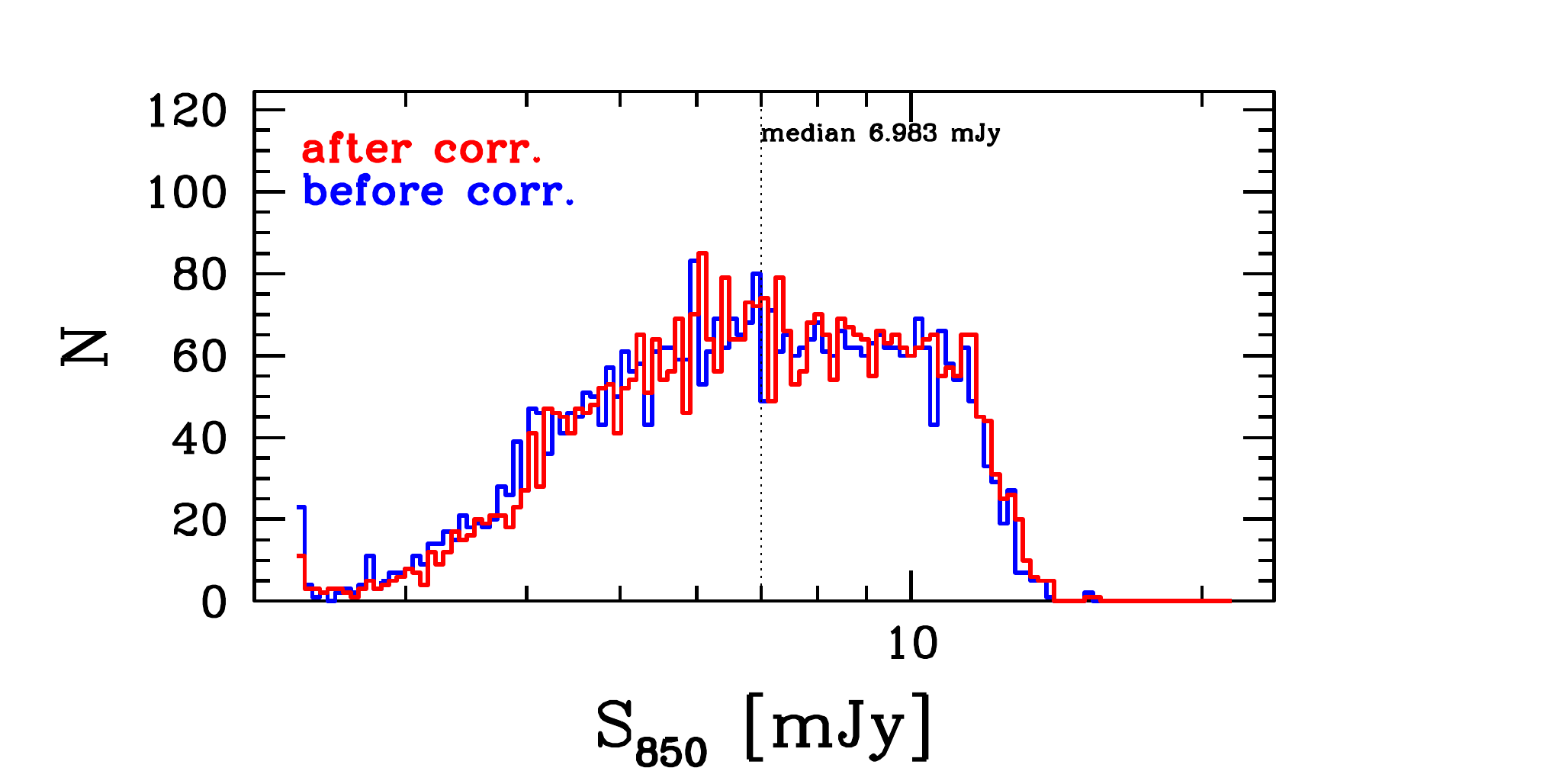}
	\includegraphics[width=0.3\textwidth, trim={1cm 15cm 0cm 2.5cm}, clip]{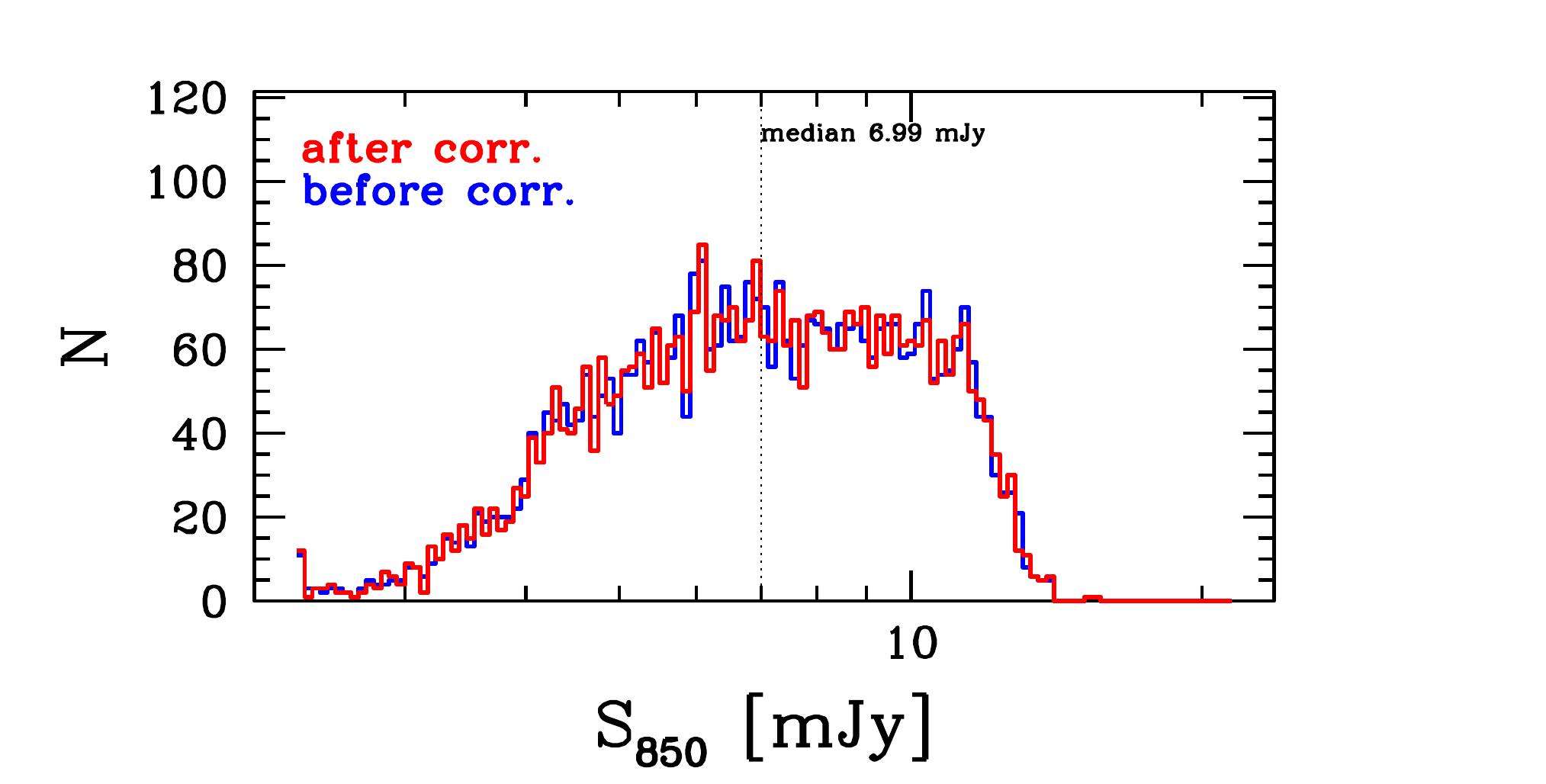}
	\includegraphics[width=0.3\textwidth, trim={1cm 15cm 0cm 2.5cm}, clip]{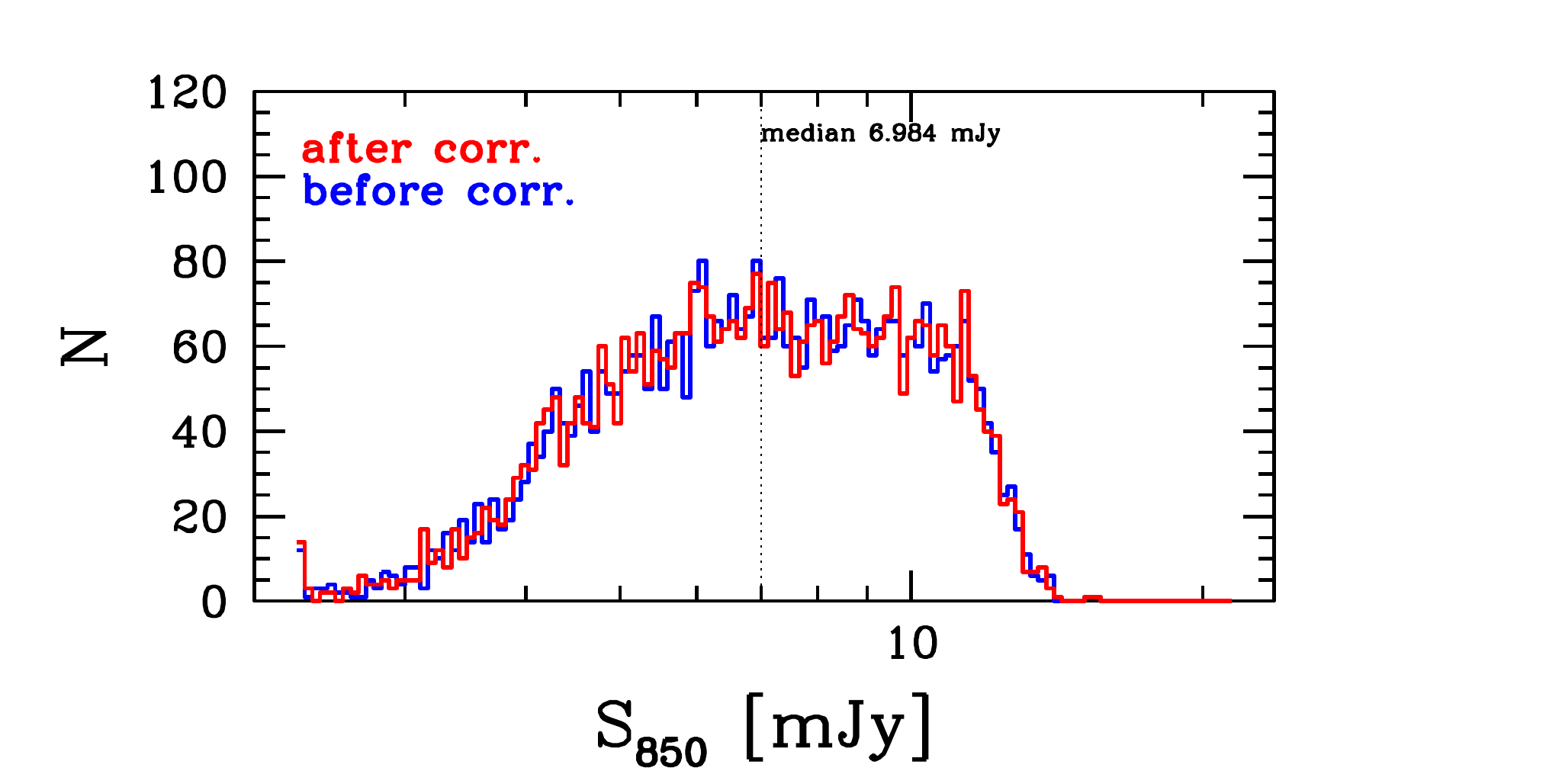}
	\end{subfigure}

	\begin{subfigure}[b]{\textwidth}\centering
	\includegraphics[width=0.3\textwidth, trim={1cm 15cm 0cm 2.5cm}, clip]{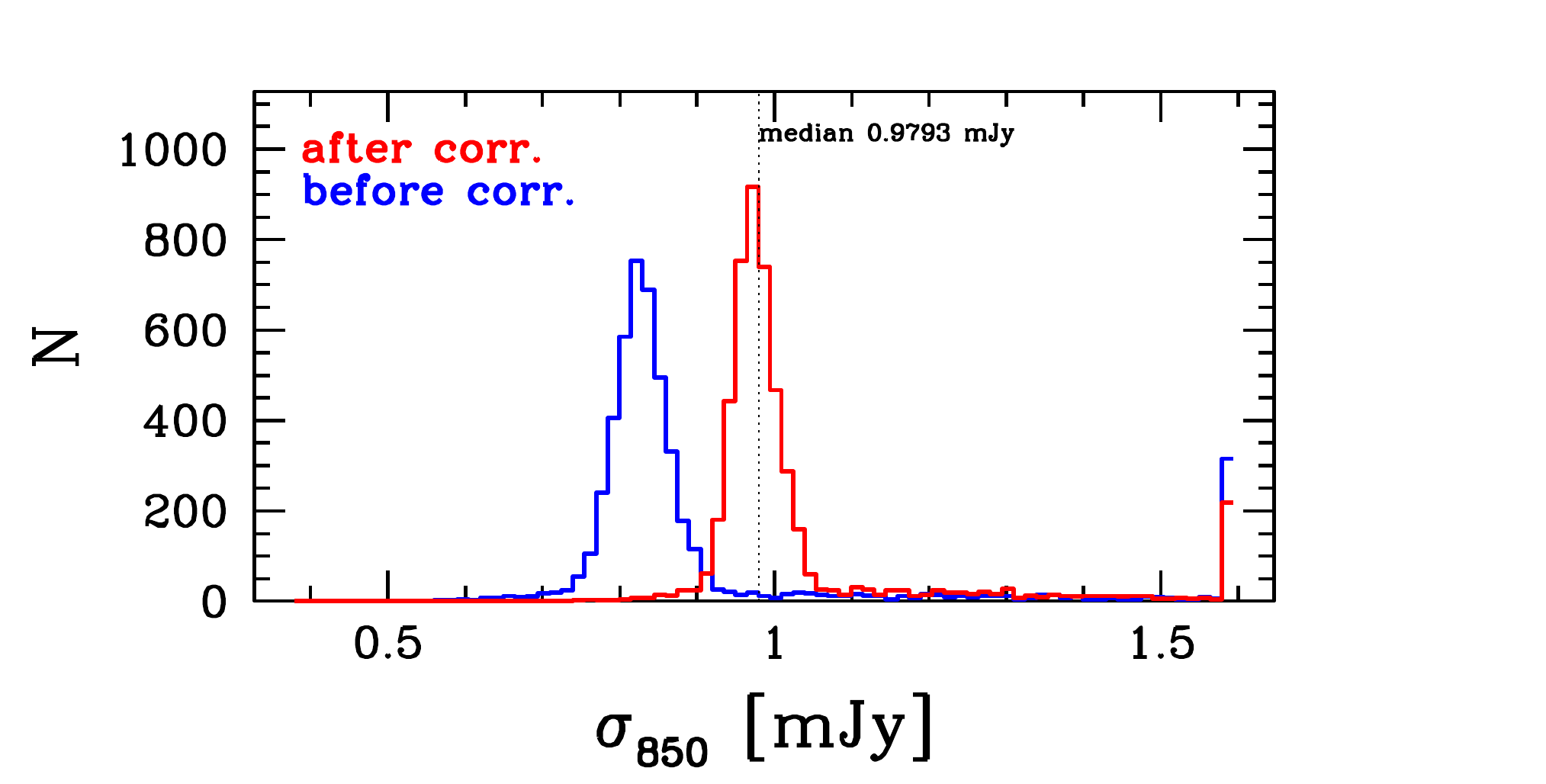}
	\includegraphics[width=0.3\textwidth, trim={1cm 15cm 0cm 2.5cm}, clip]{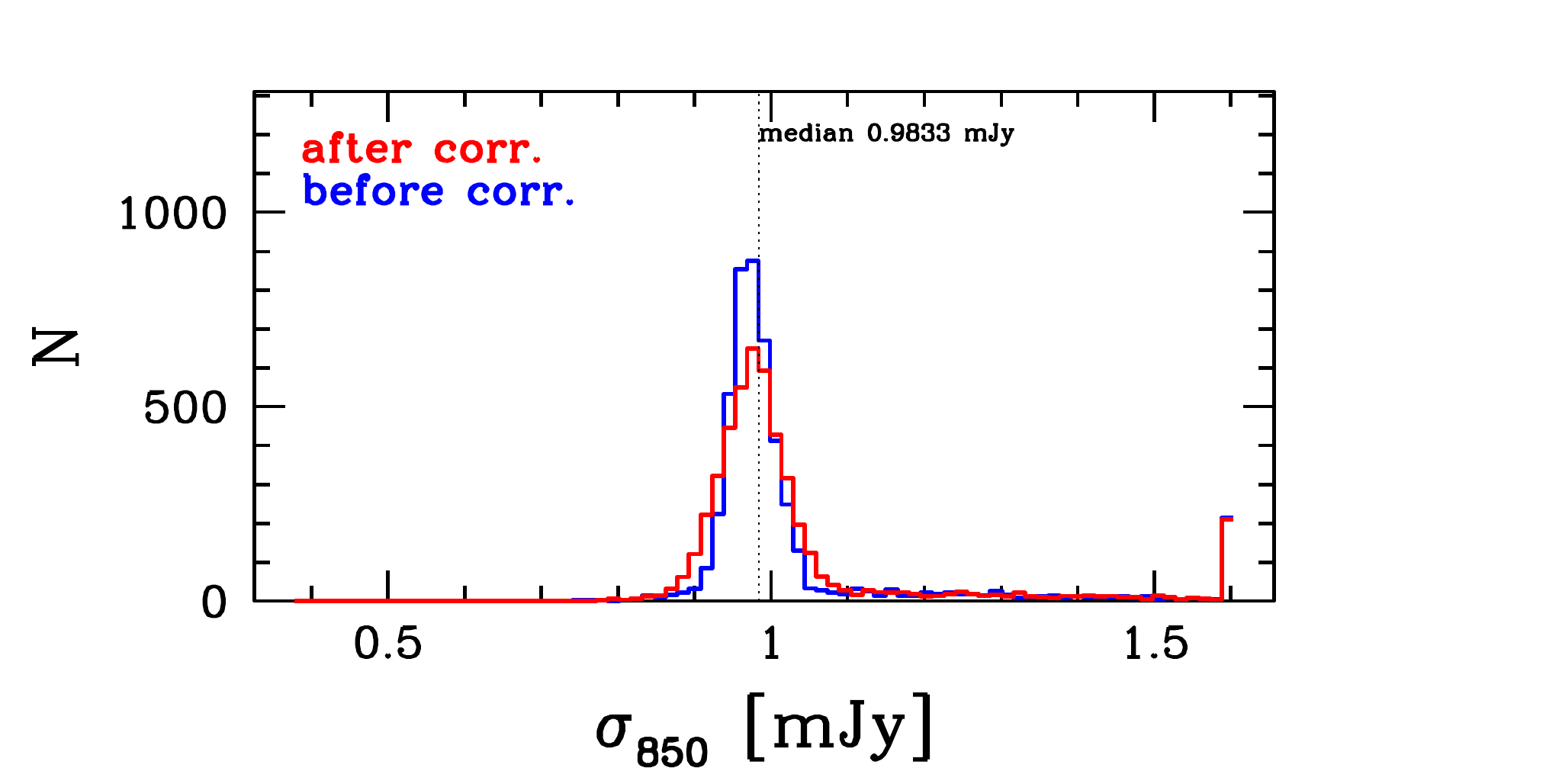}
	\includegraphics[width=0.3\textwidth, trim={1cm 15cm 0cm 2.5cm}, clip]{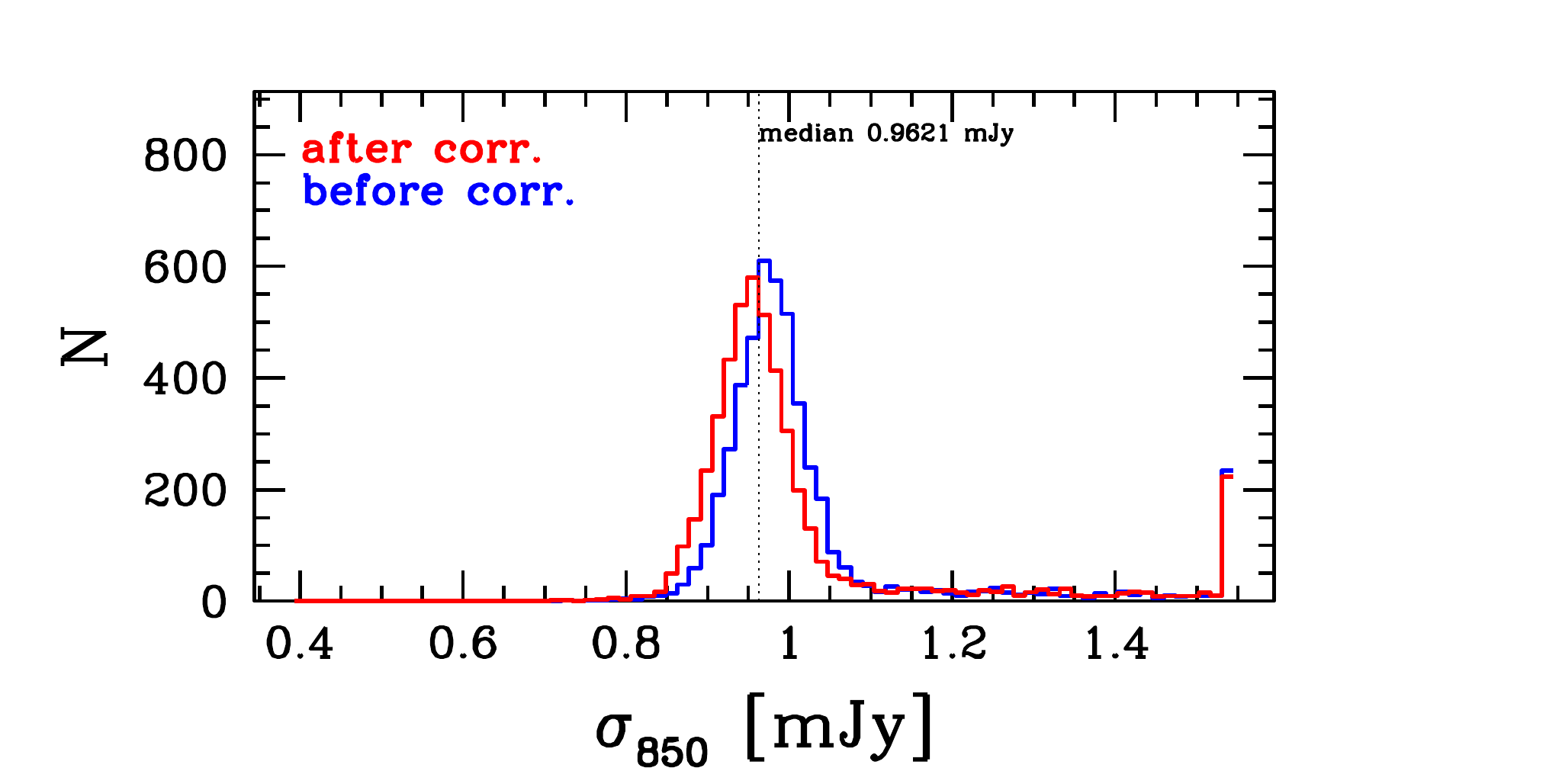}
	\end{subfigure}

\caption{%
	Simulation correction analyses at SCUBA2 850~$\mu$m. See descriptions in text. 
	\label{Figure_galsim_850_bin}
}
\end{figure}

\begin{figure}
	\centering
    
    \begin{subfigure}[b]{\textwidth}\centering
	\includegraphics[width=0.23\textwidth, trim={1cm 15cm 0cm 2.5cm}, clip]{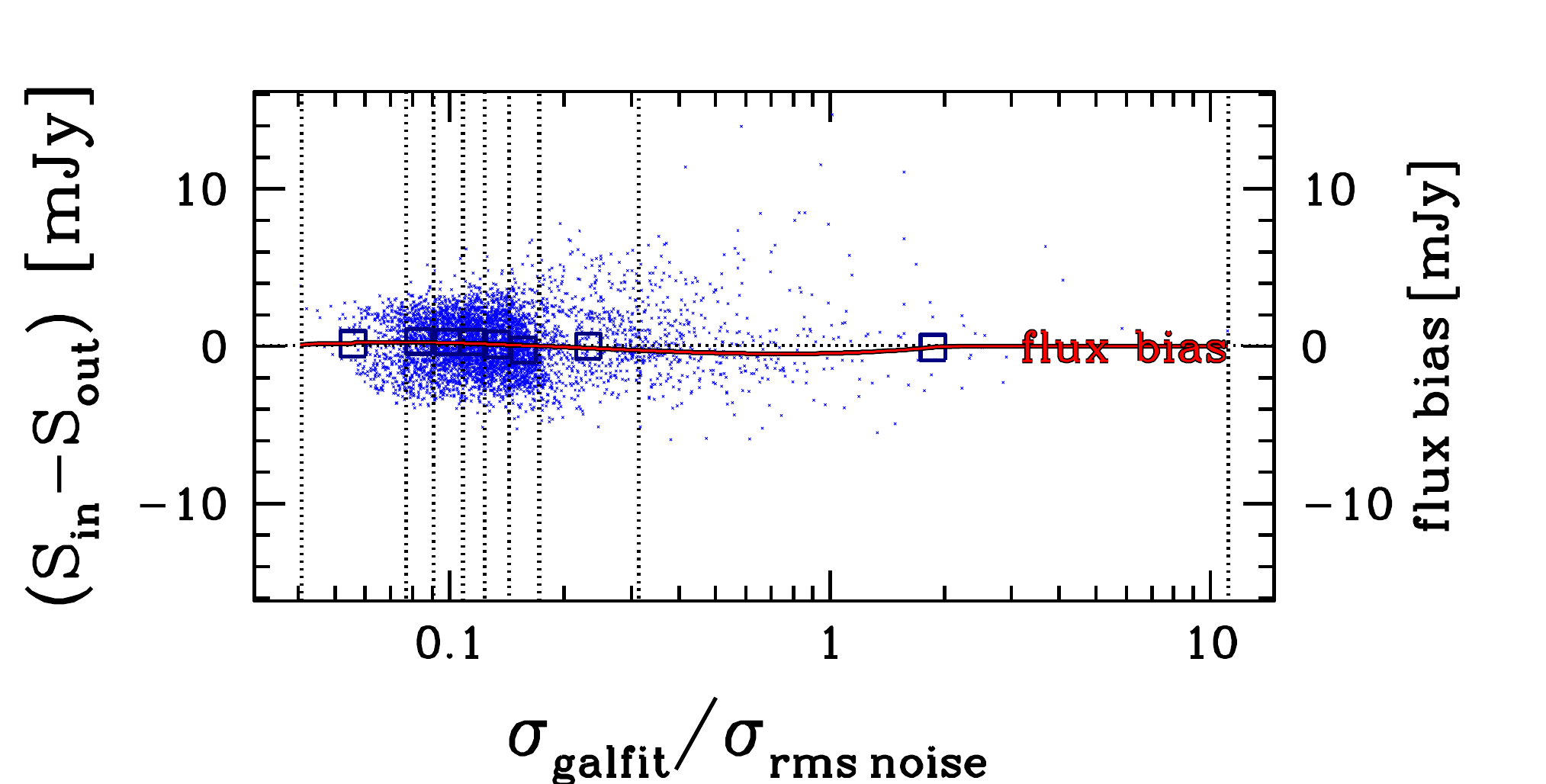}
	\includegraphics[width=0.23\textwidth, trim={1cm 15cm 0cm 2.5cm}, clip]{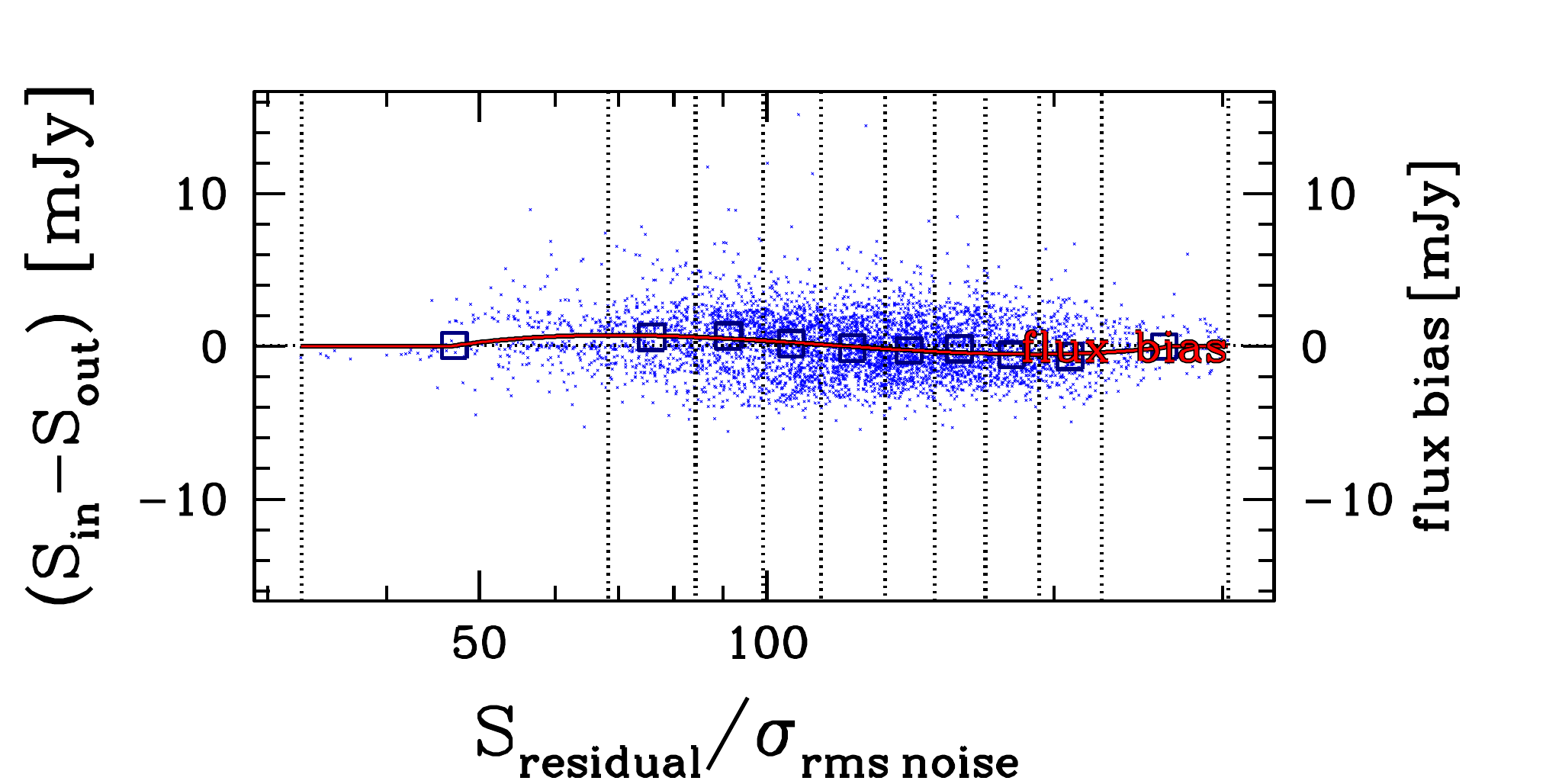}
	\includegraphics[width=0.23\textwidth, trim={1cm 15cm 0cm 2.5cm}, clip]{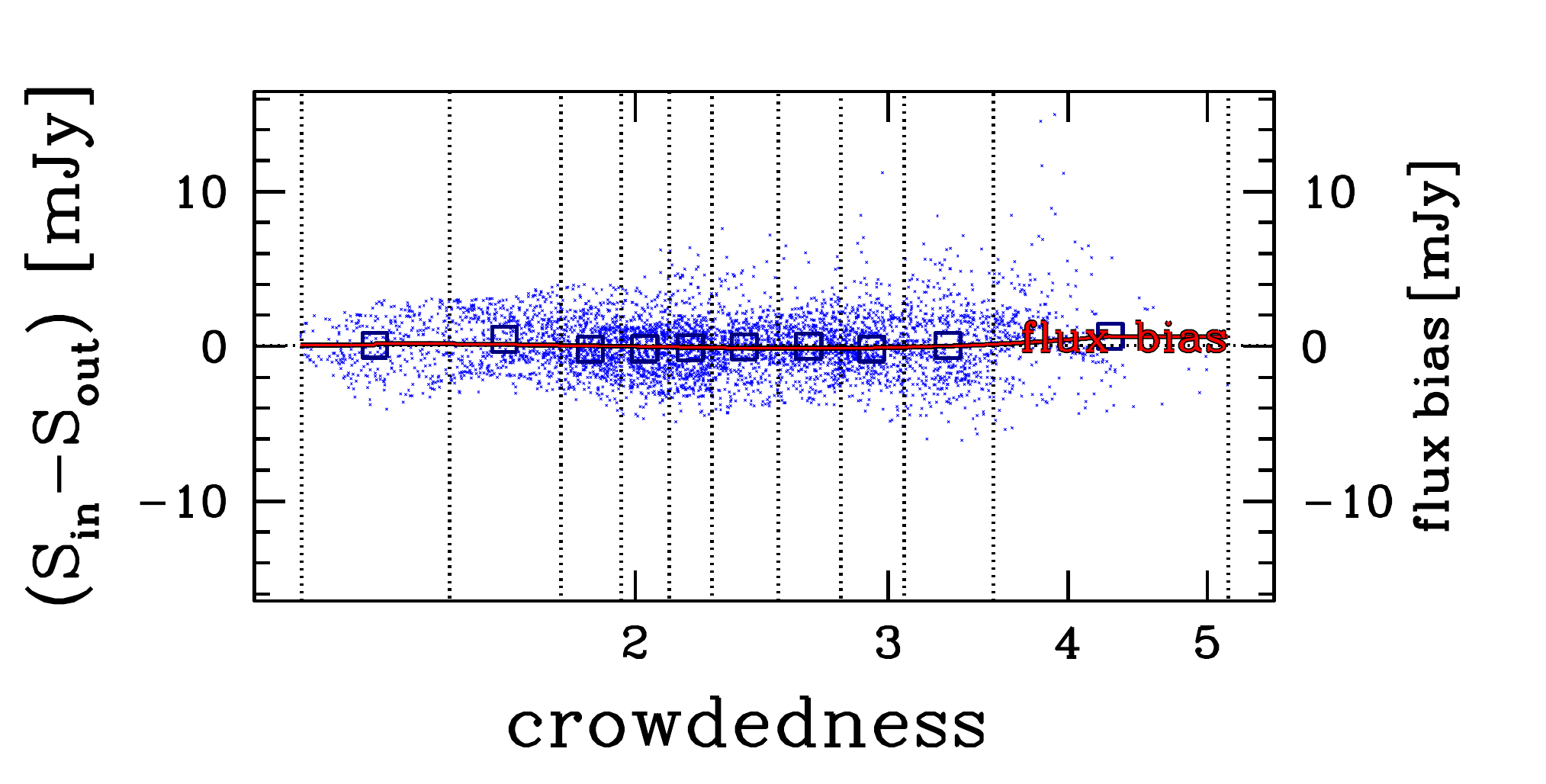}
	\includegraphics[width=0.23\textwidth, trim={1cm 15cm 0cm 2.5cm}, clip]{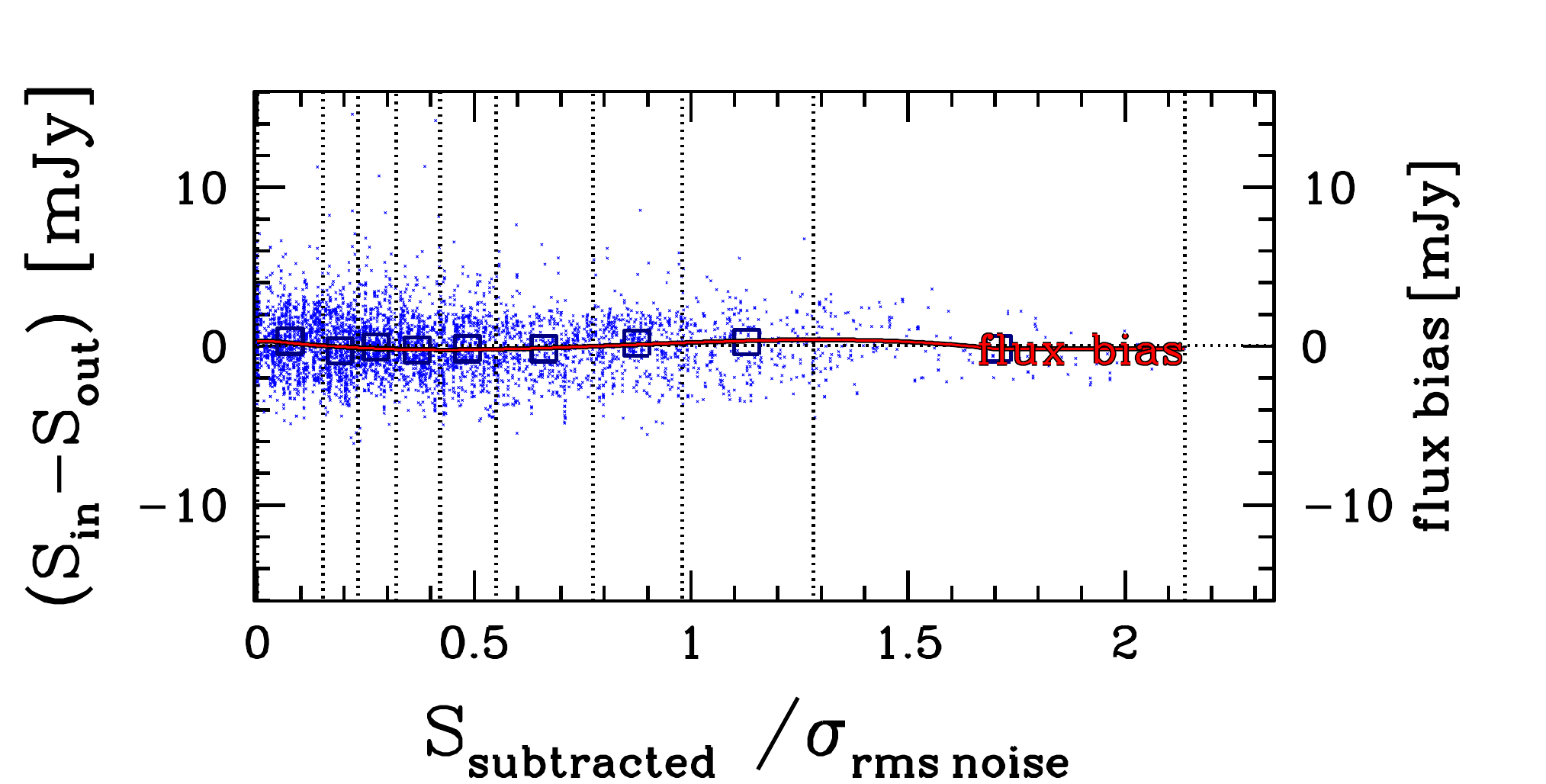}
    \end{subfigure}

    \begin{subfigure}[b]{\textwidth}\centering
	\includegraphics[width=0.23\textwidth, trim={1cm 15cm 0cm 2.5cm}, clip]{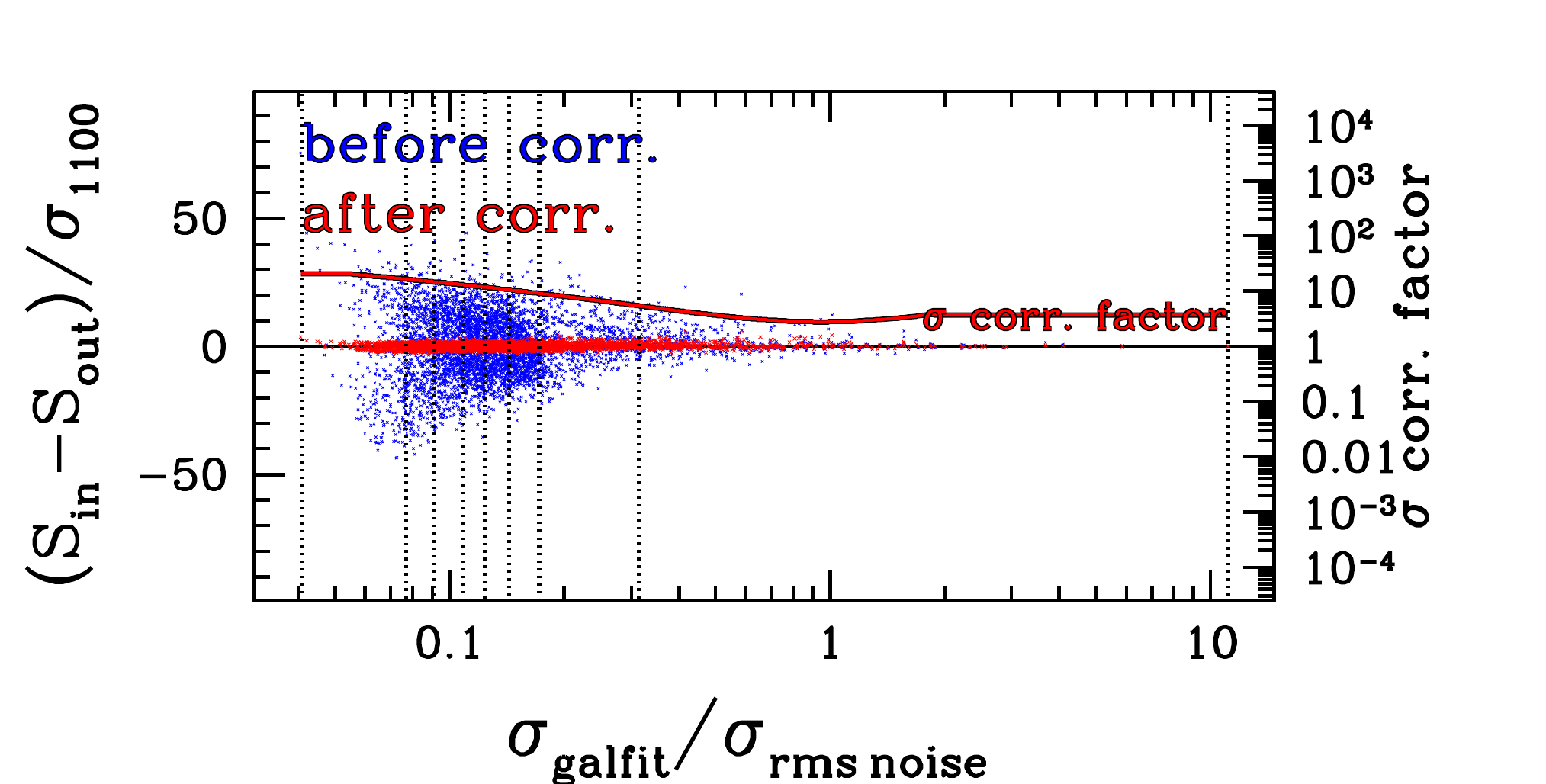}
	\includegraphics[width=0.23\textwidth, trim={1cm 15cm 0cm 2.5cm}, clip]{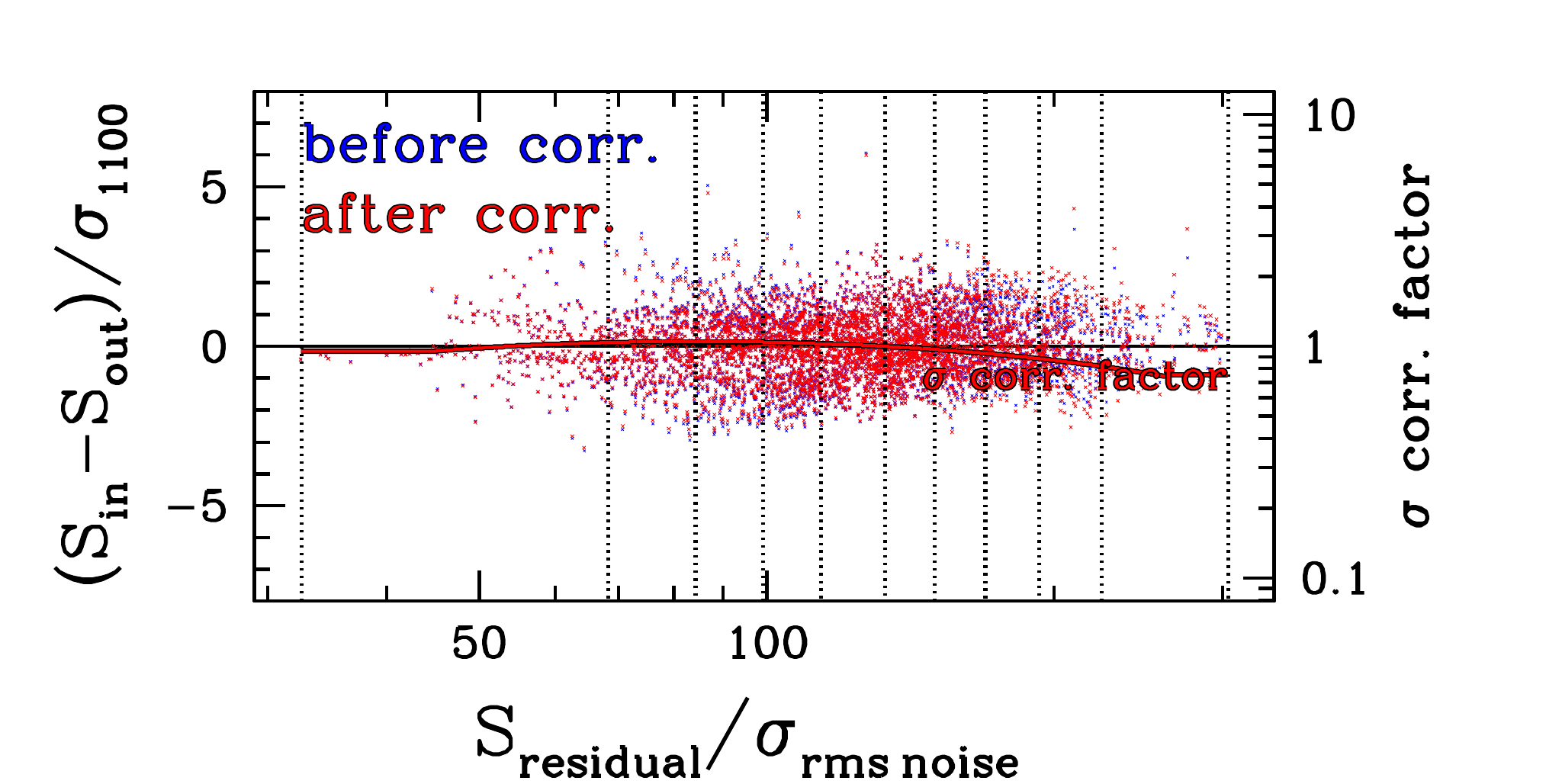}
	\includegraphics[width=0.23\textwidth, trim={1cm 15cm 0cm 2.5cm}, clip]{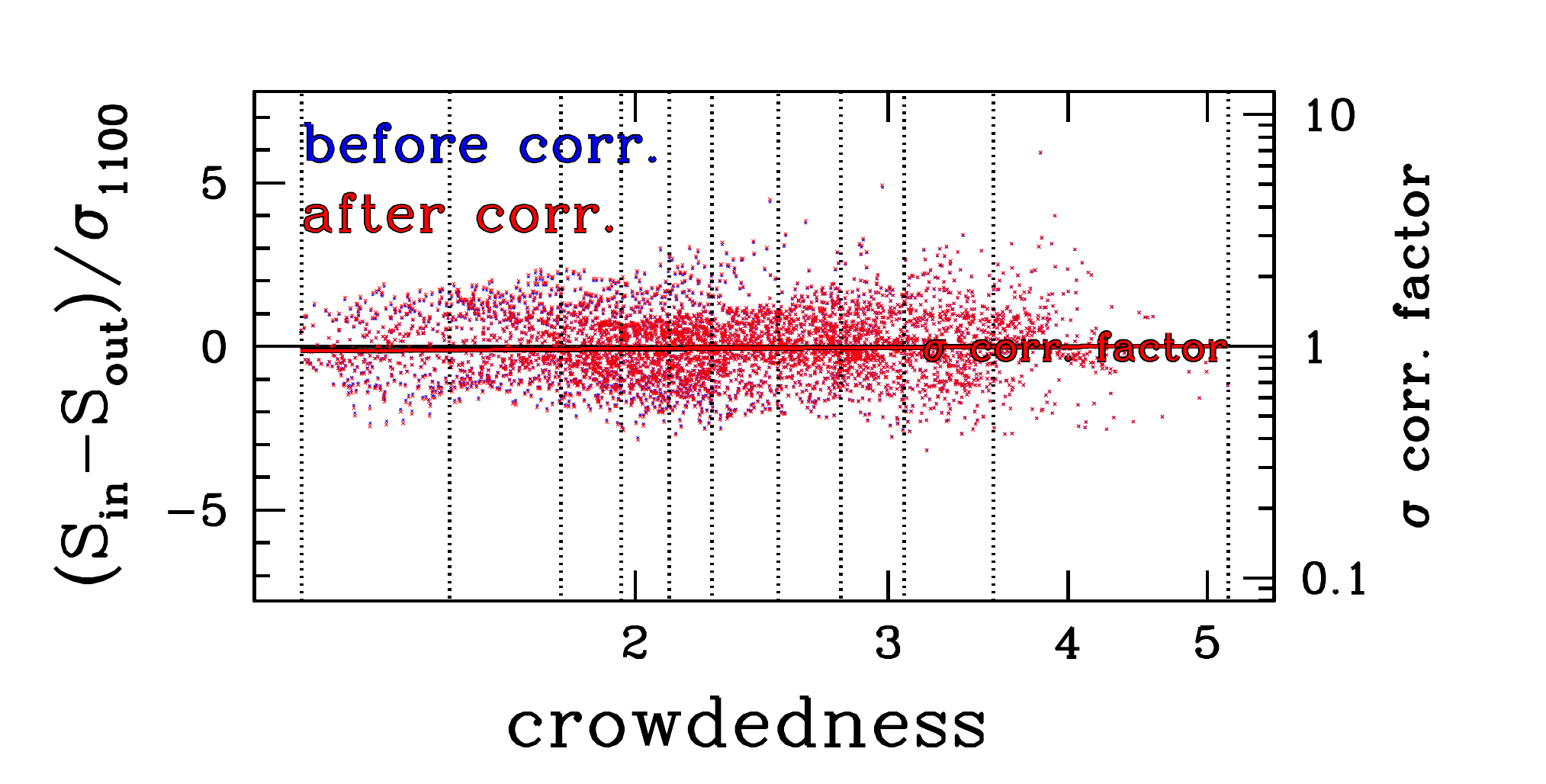}
	\includegraphics[width=0.23\textwidth, trim={1cm 15cm 0cm 2.5cm}, clip]{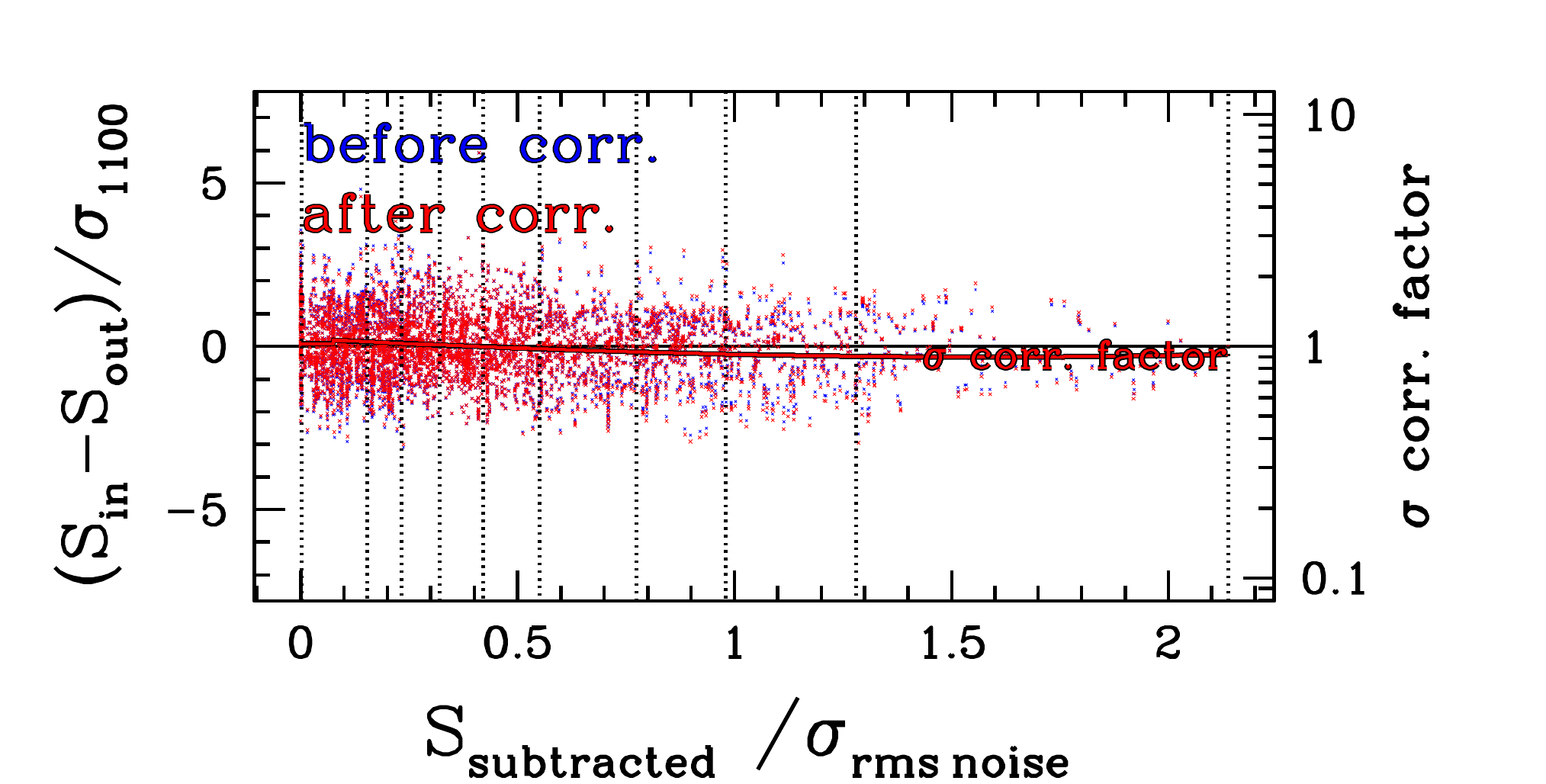}
    \end{subfigure}
    
    \begin{subfigure}[b]{\textwidth}\centering
	\includegraphics[width=0.23\textwidth, trim={1cm 15cm 0cm 2.5cm}, clip]{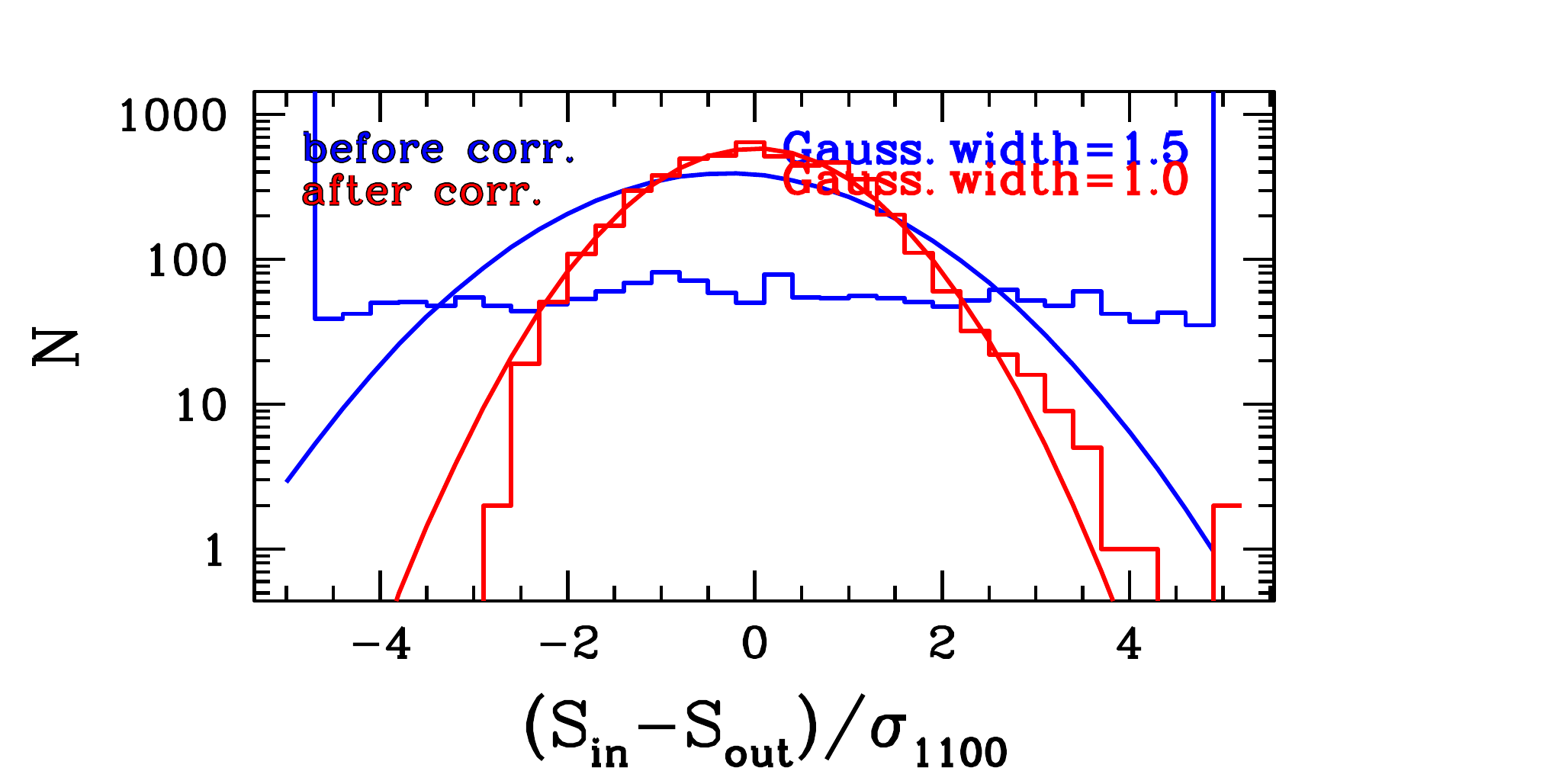}
	\includegraphics[width=0.23\textwidth, trim={1cm 15cm 0cm 2.5cm}, clip]{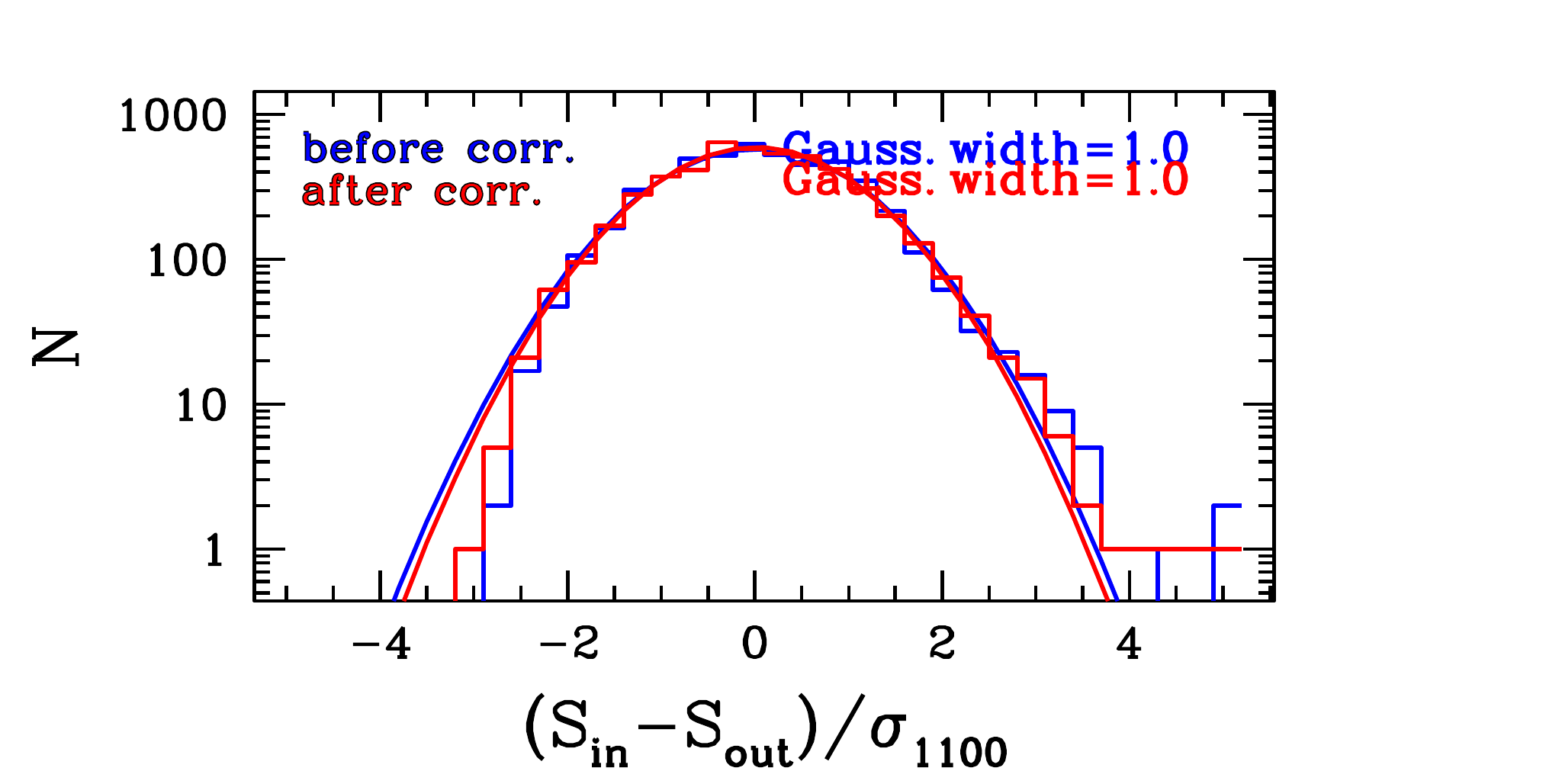}
	\includegraphics[width=0.23\textwidth, trim={1cm 15cm 0cm 2.5cm}, clip]{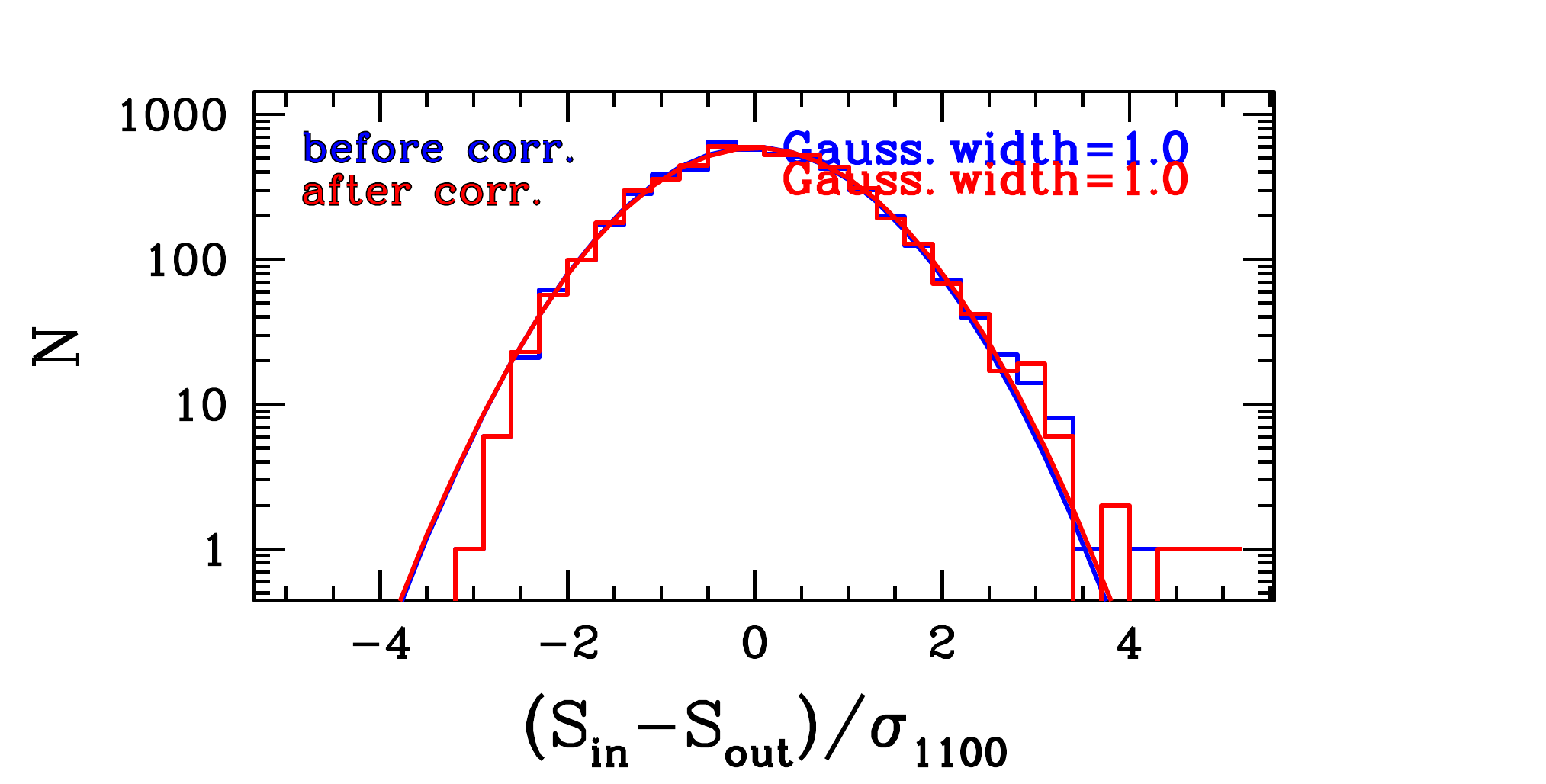}
	\includegraphics[width=0.23\textwidth, trim={1cm 15cm 0cm 2.5cm}, clip]{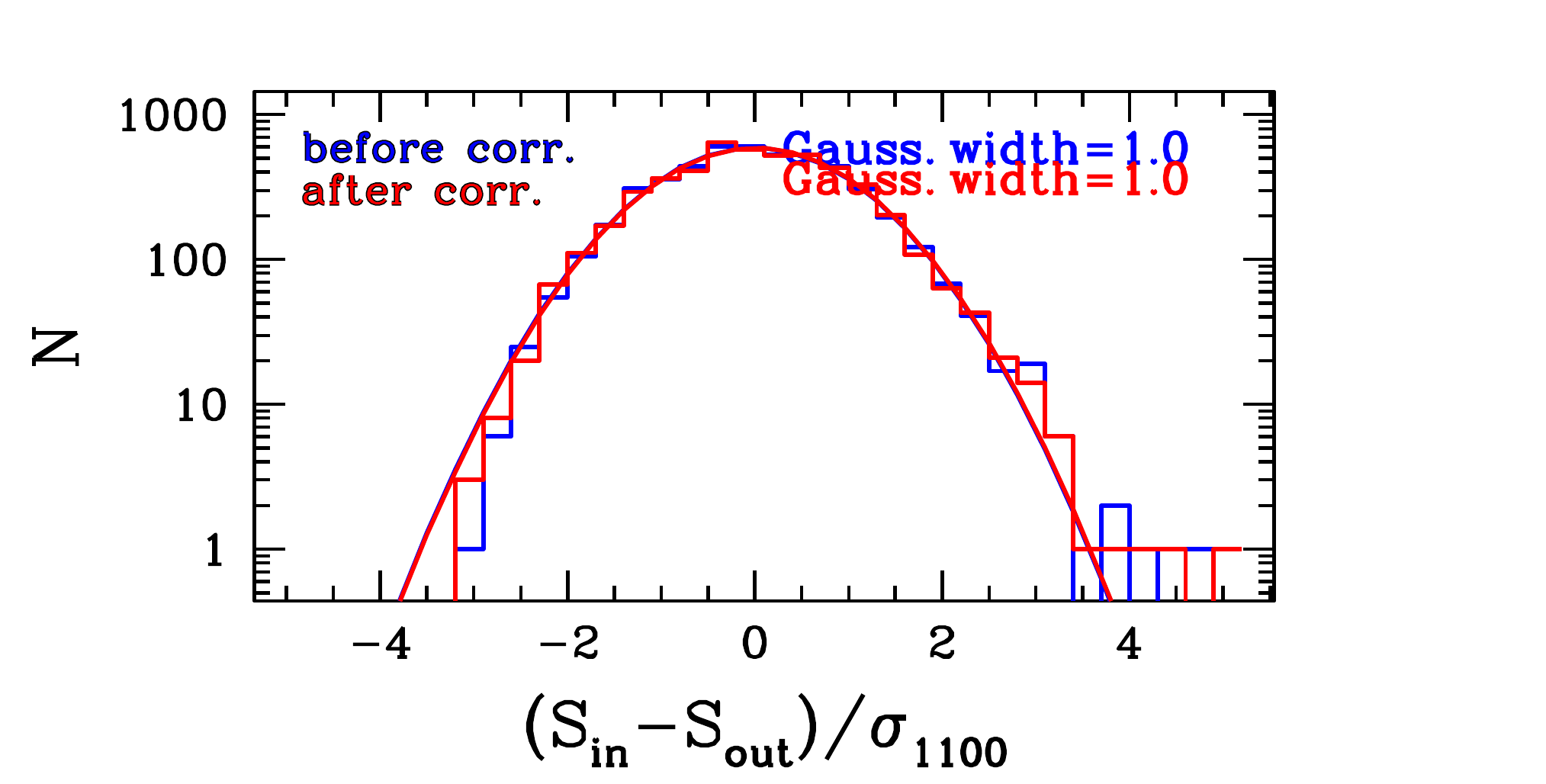}
    \end{subfigure}
    
    \begin{subfigure}[b]{\textwidth}\centering
	\includegraphics[width=0.23\textwidth, trim={1cm 15cm 0cm 2.5cm}, clip]{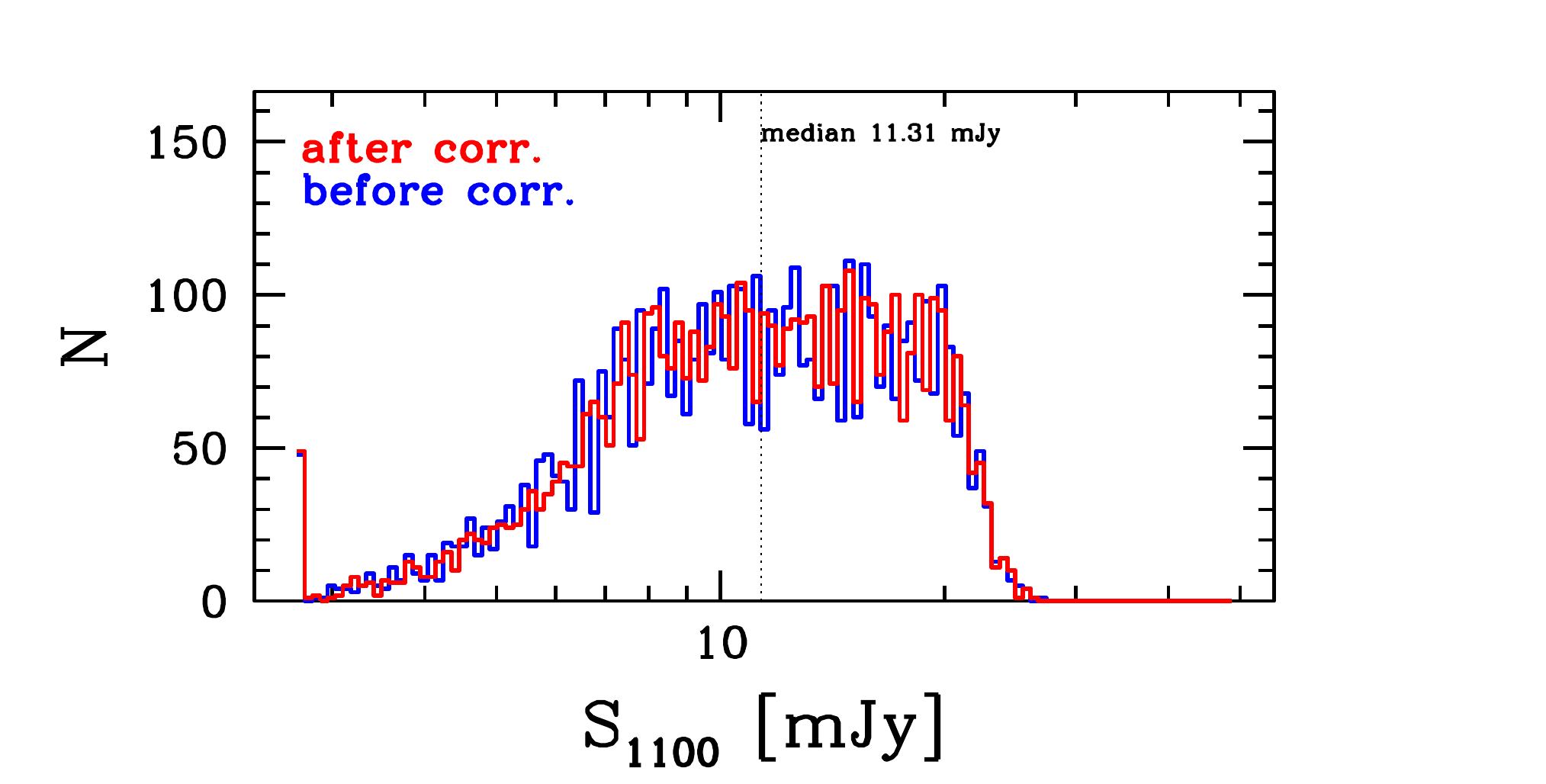}
	\includegraphics[width=0.23\textwidth, trim={1cm 15cm 0cm 2.5cm}, clip]{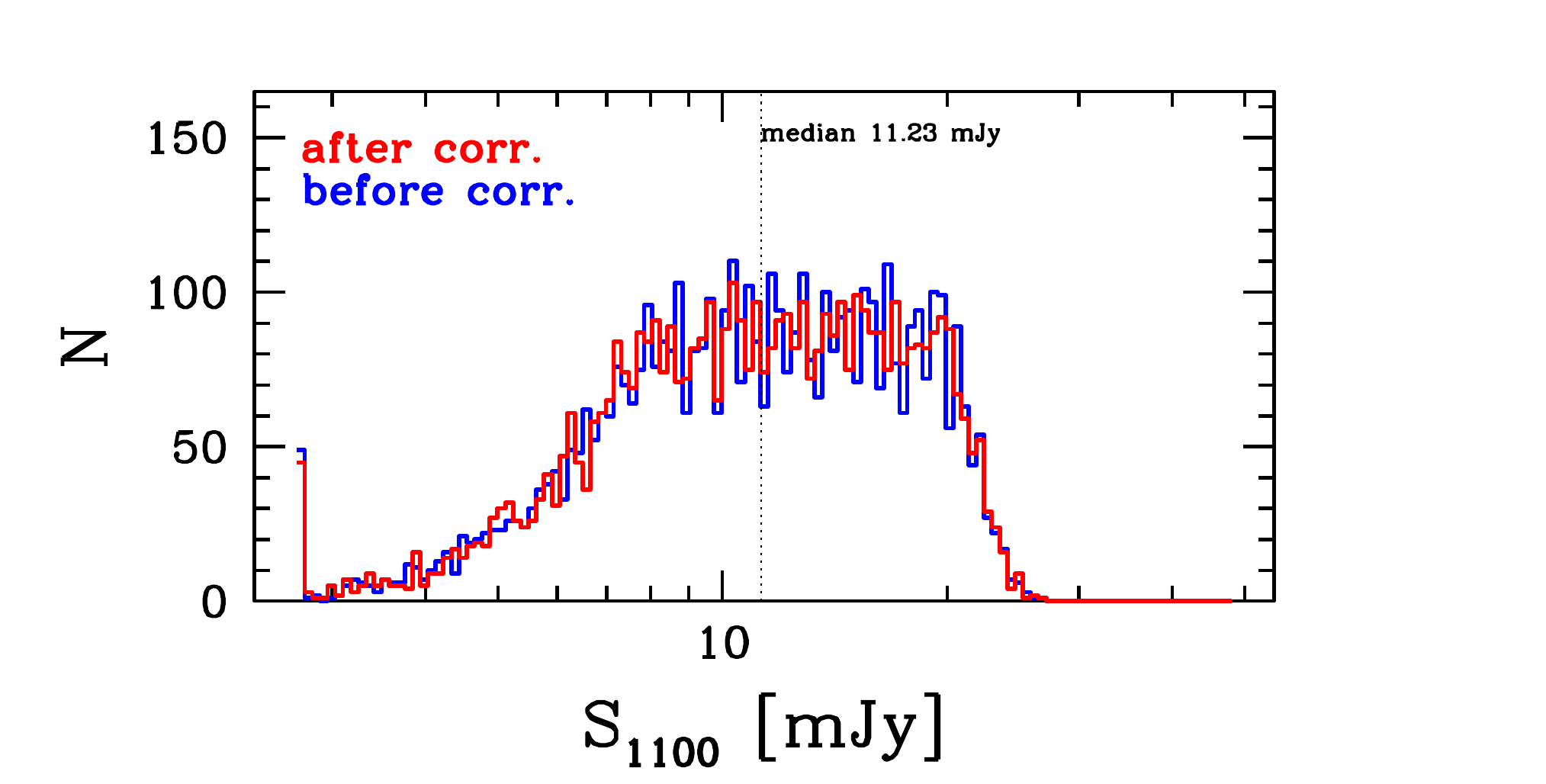}
	\includegraphics[width=0.23\textwidth, trim={1cm 15cm 0cm 2.5cm}, clip]{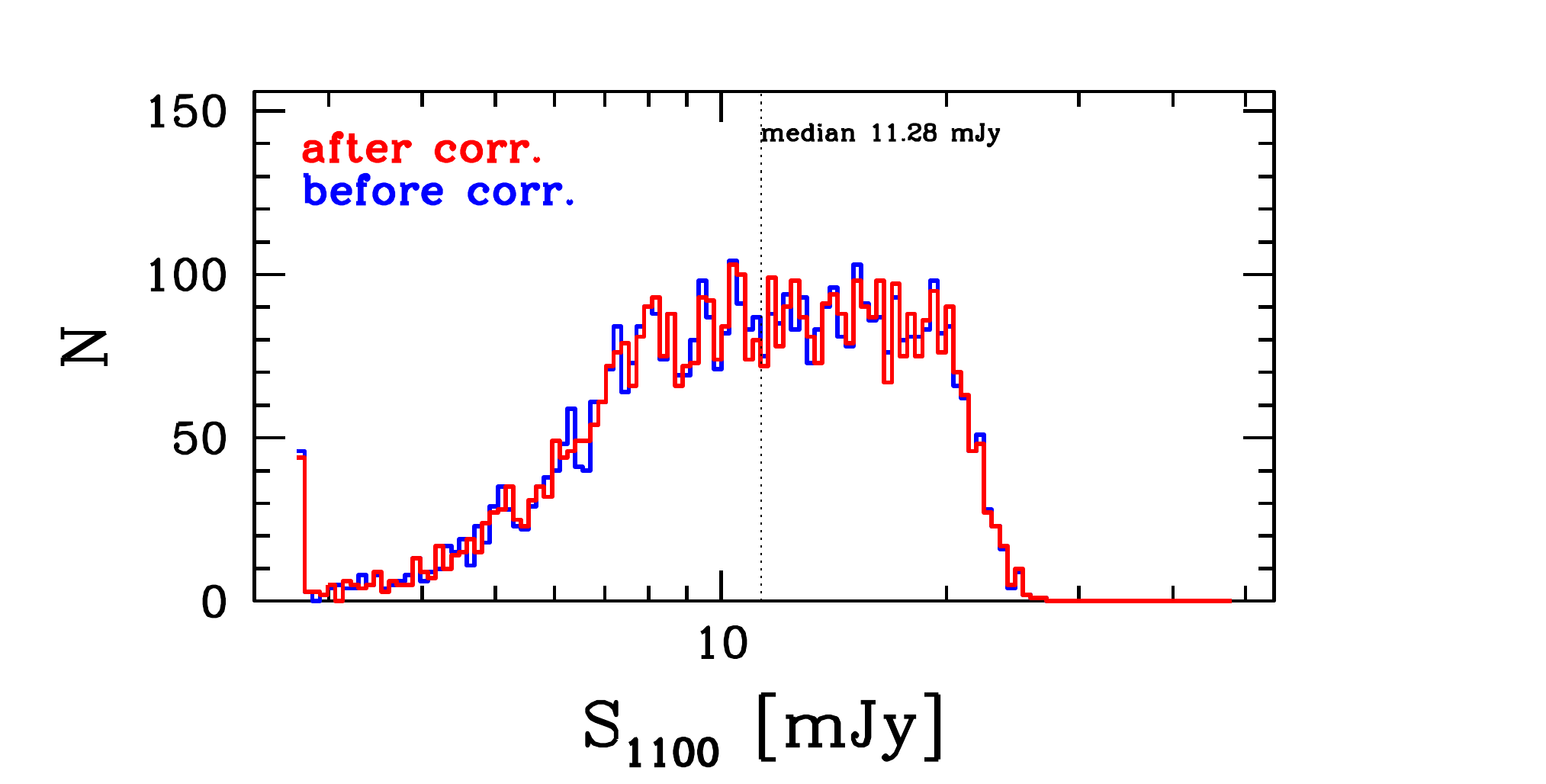}
	\includegraphics[width=0.23\textwidth, trim={1cm 15cm 0cm 2.5cm}, clip]{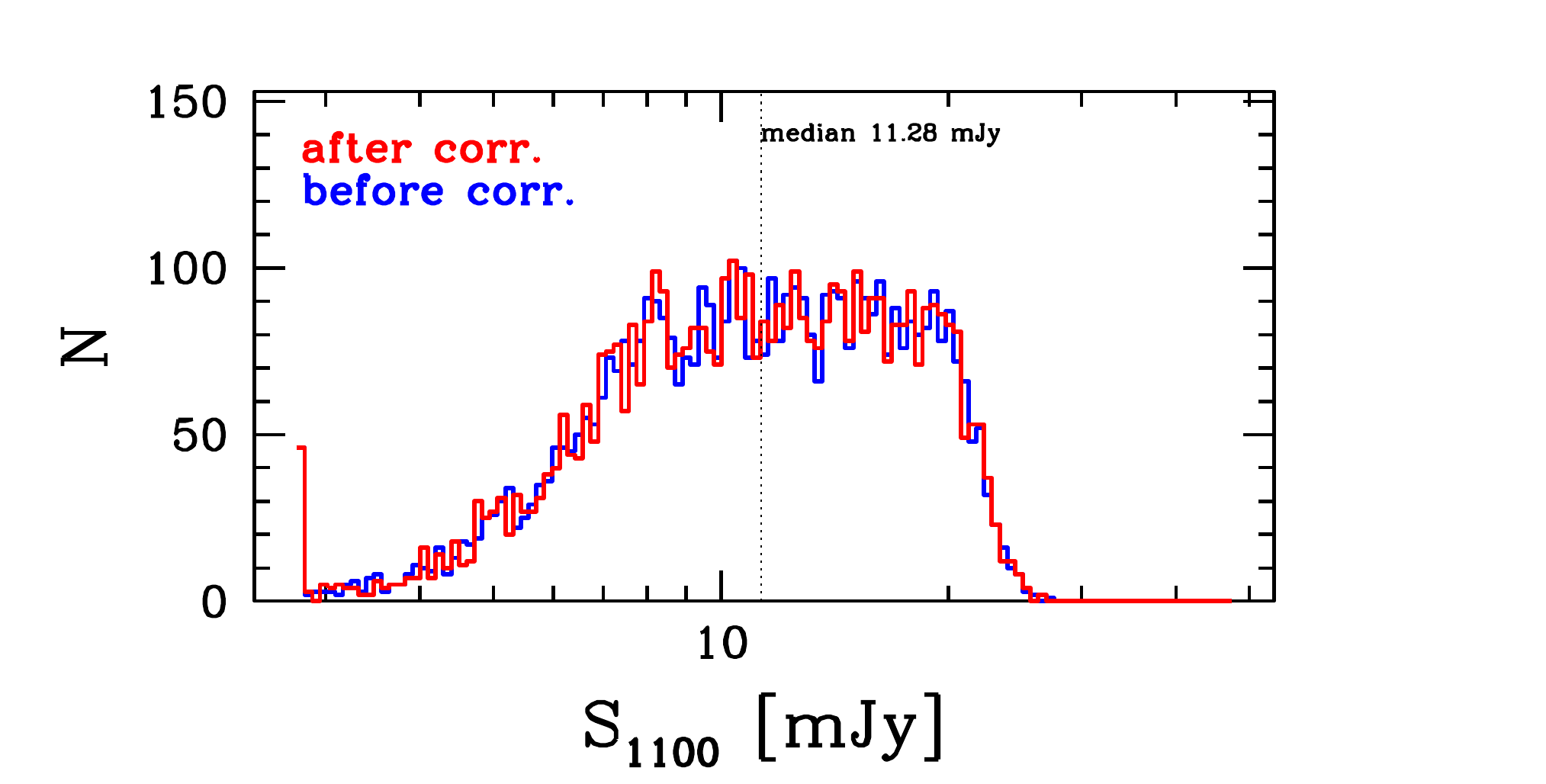}
    \end{subfigure}
    
    \begin{subfigure}[b]{\textwidth}\centering
	\includegraphics[width=0.23\textwidth, trim={1cm 15cm 0cm 2.5cm}, clip]{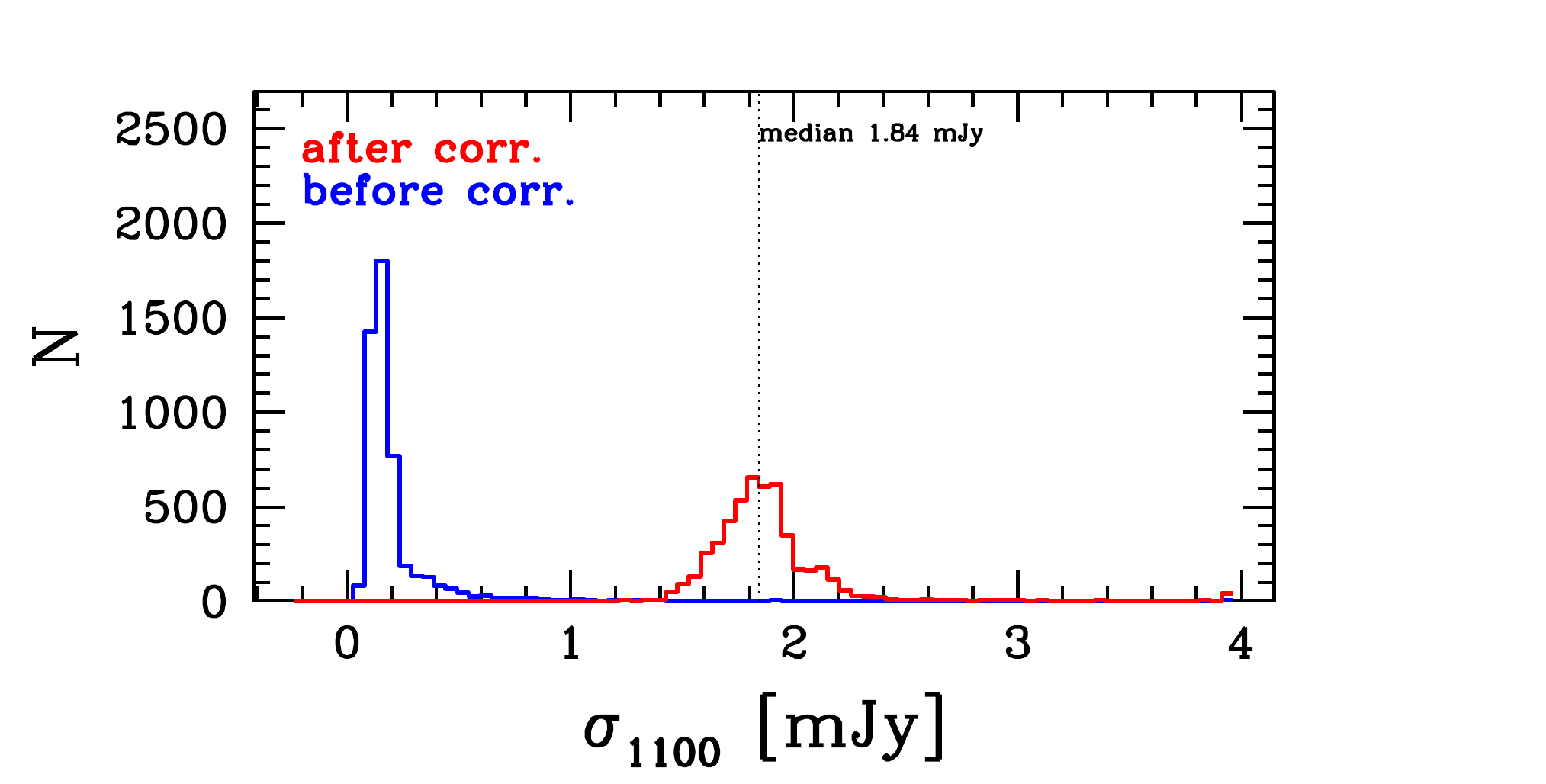}
	\includegraphics[width=0.23\textwidth, trim={1cm 15cm 0cm 2.5cm}, clip]{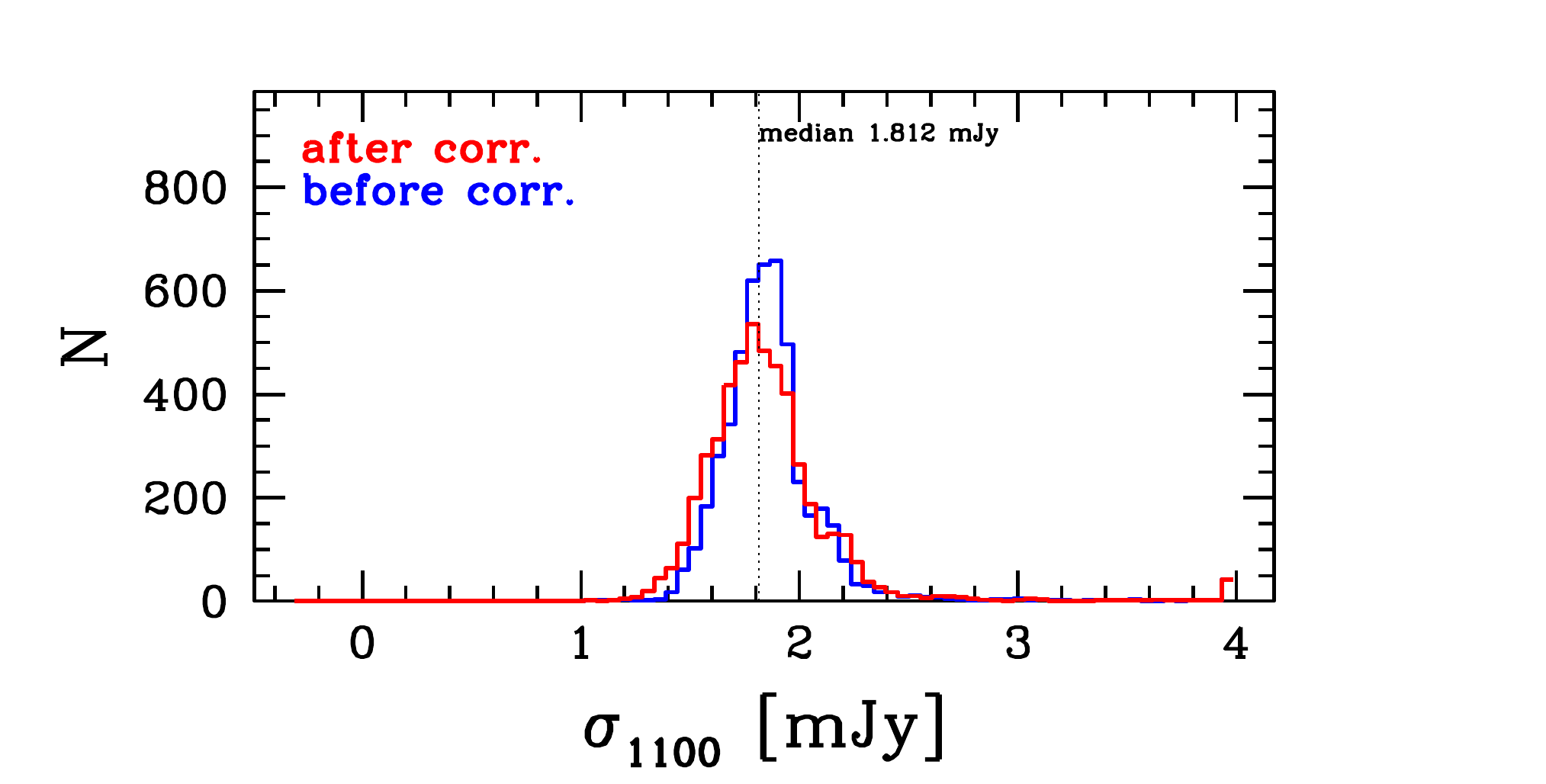}
	\includegraphics[width=0.23\textwidth, trim={1cm 15cm 0cm 2.5cm}, clip]{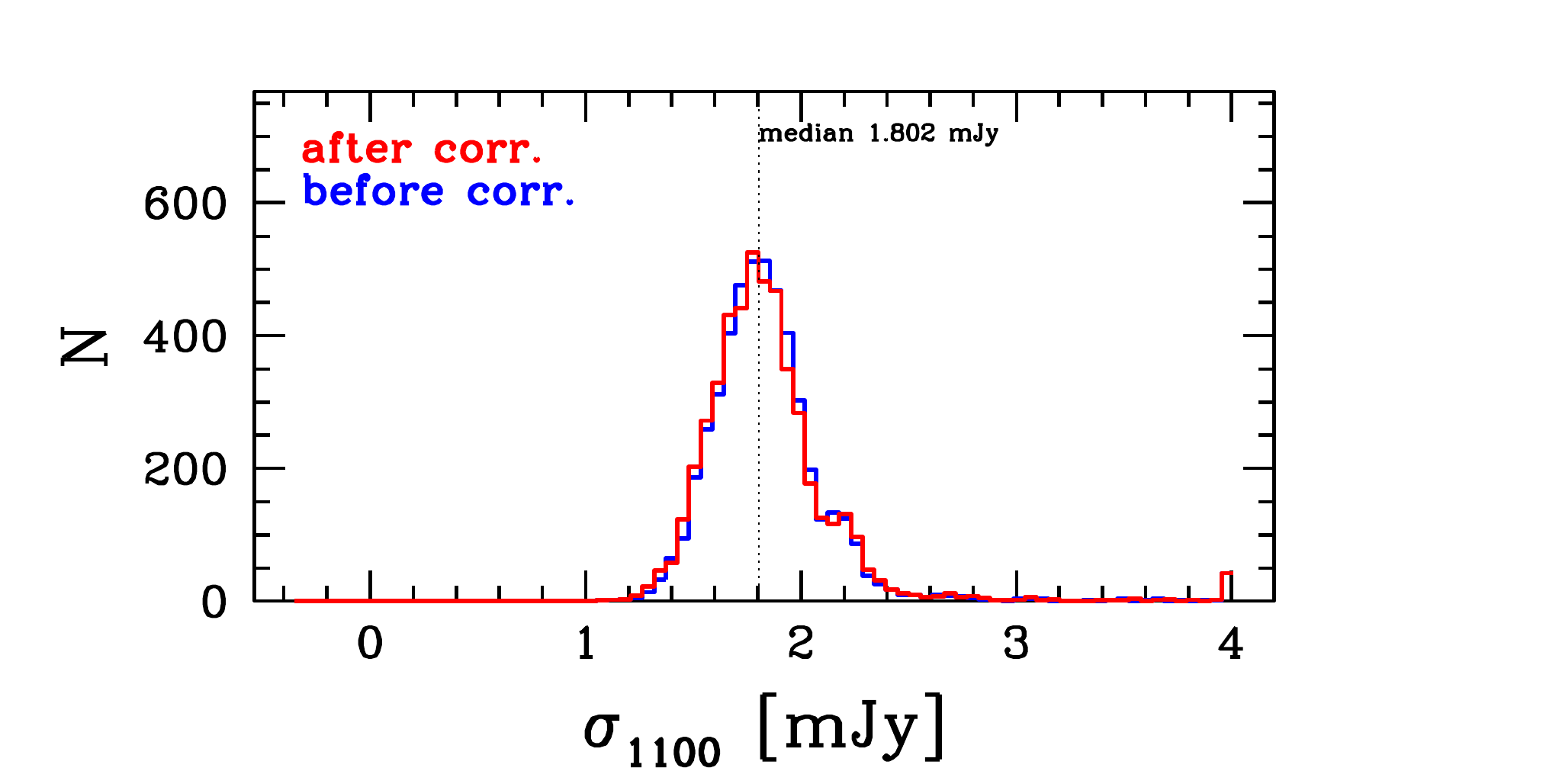}
	\includegraphics[width=0.23\textwidth, trim={1cm 15cm 0cm 2.5cm}, clip]{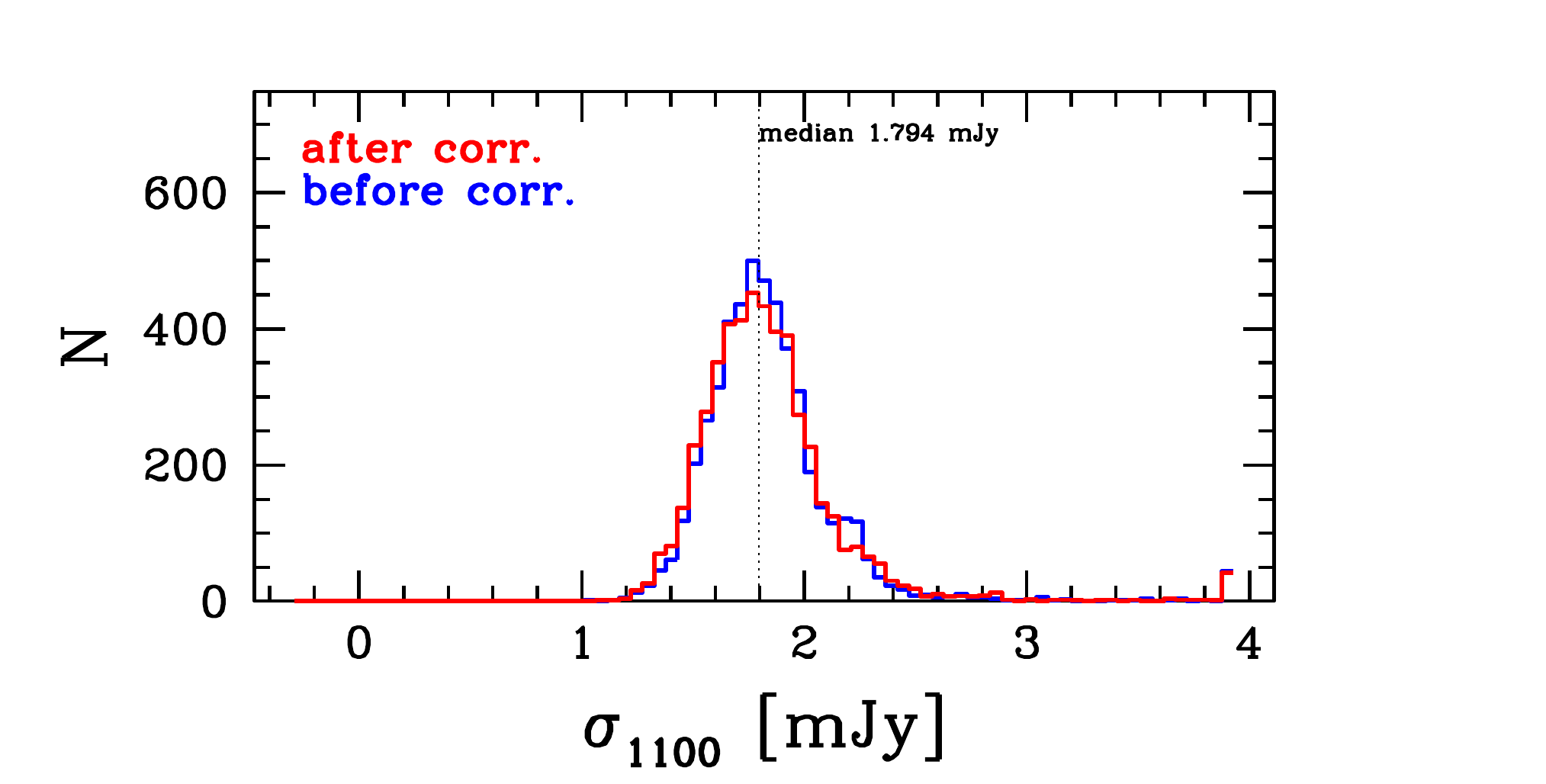}
    \end{subfigure}

\caption{%
	Simulation correction analyses at AzTEC 1.1~mm. See descriptions in text. 
}
\end{figure}


\begin{figure}
	\centering
    
    \begin{subfigure}[b]{\textwidth}\centering
	\includegraphics[width=0.3\textwidth, trim={1cm 15cm 0cm 2.5cm}, clip]{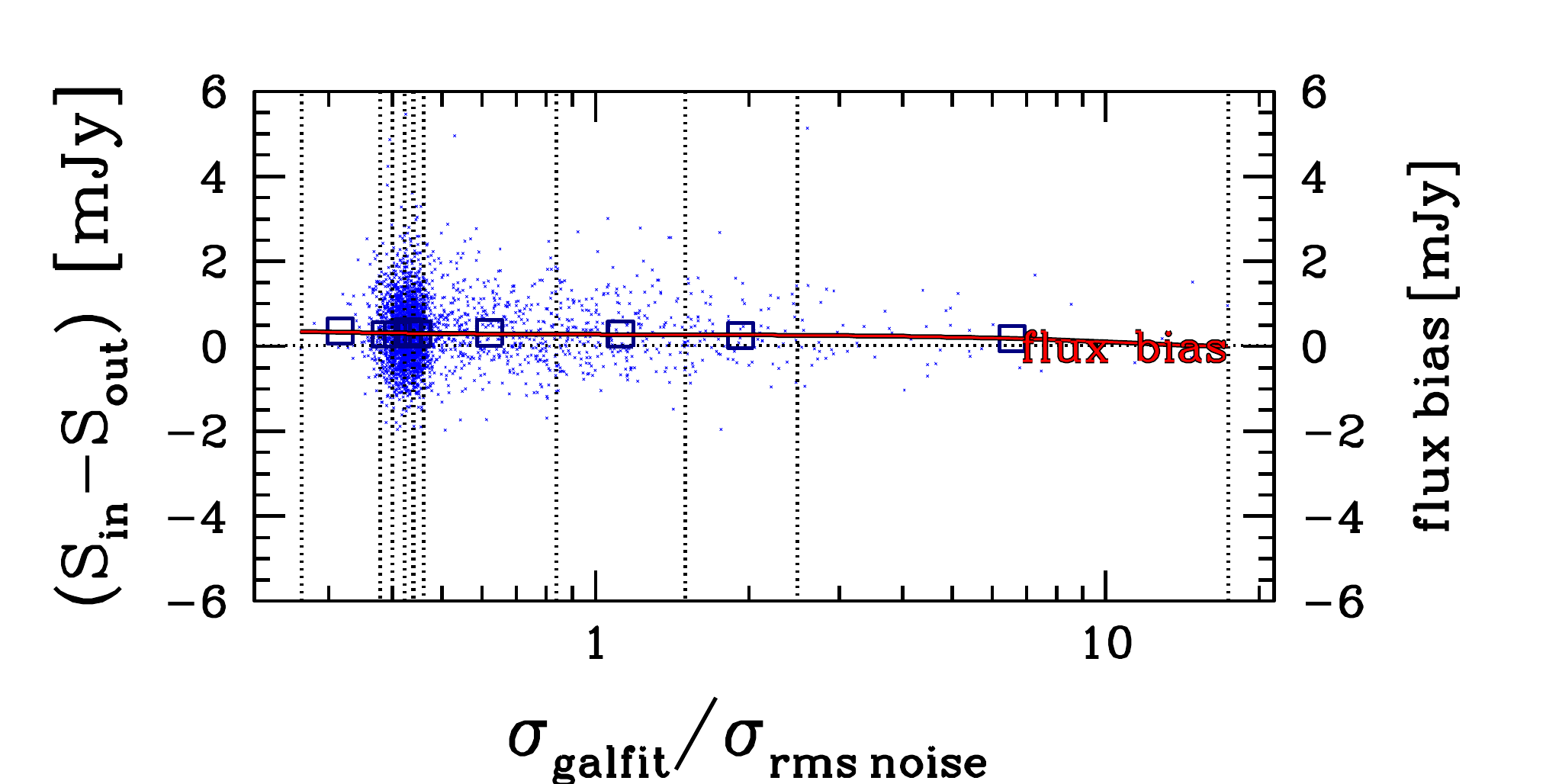}
	\includegraphics[width=0.3\textwidth, trim={1cm 15cm 0cm 2.5cm}, clip]{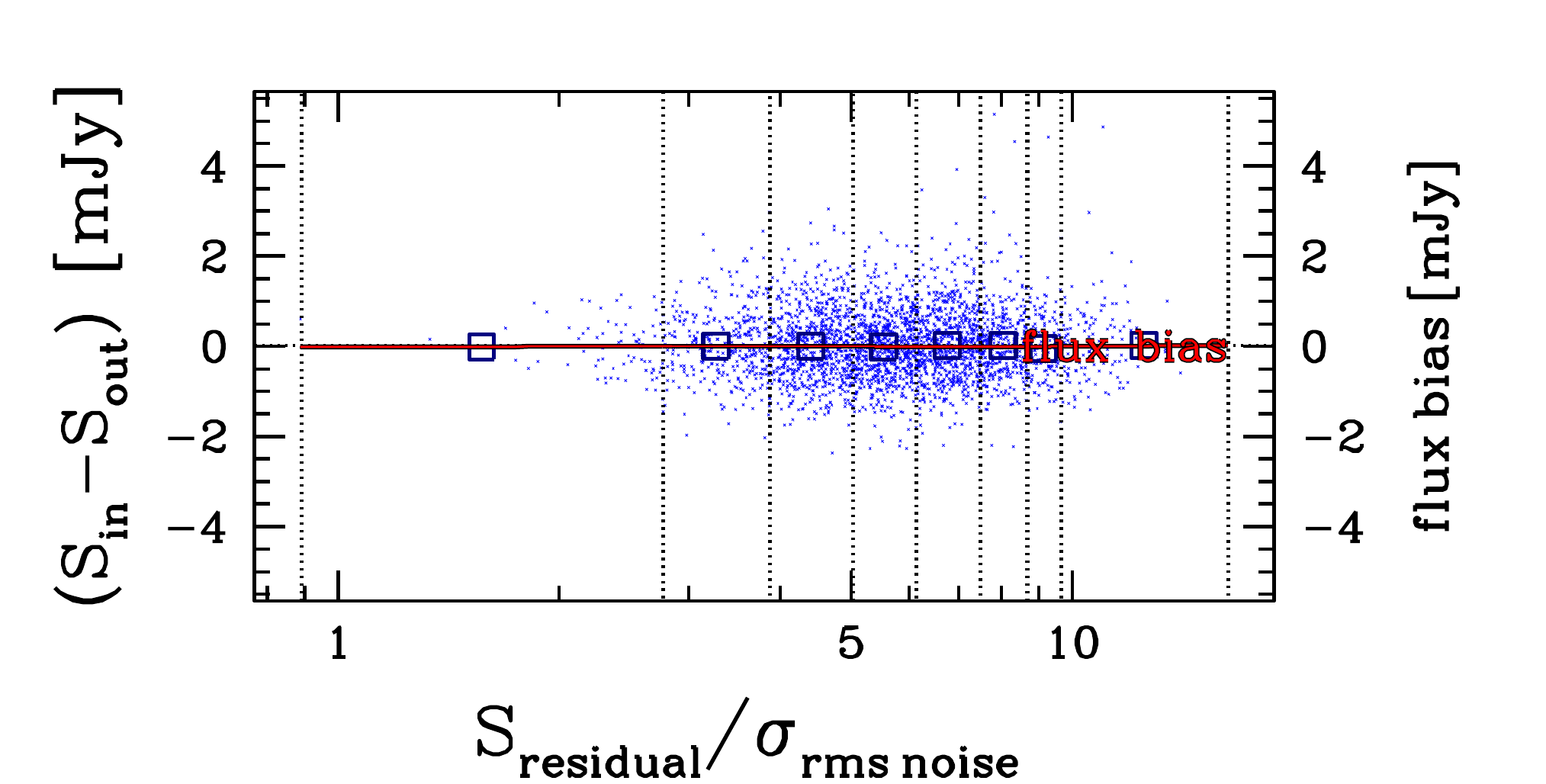}
	\includegraphics[width=0.3\textwidth, trim={1cm 15cm 0cm 2.5cm}, clip]{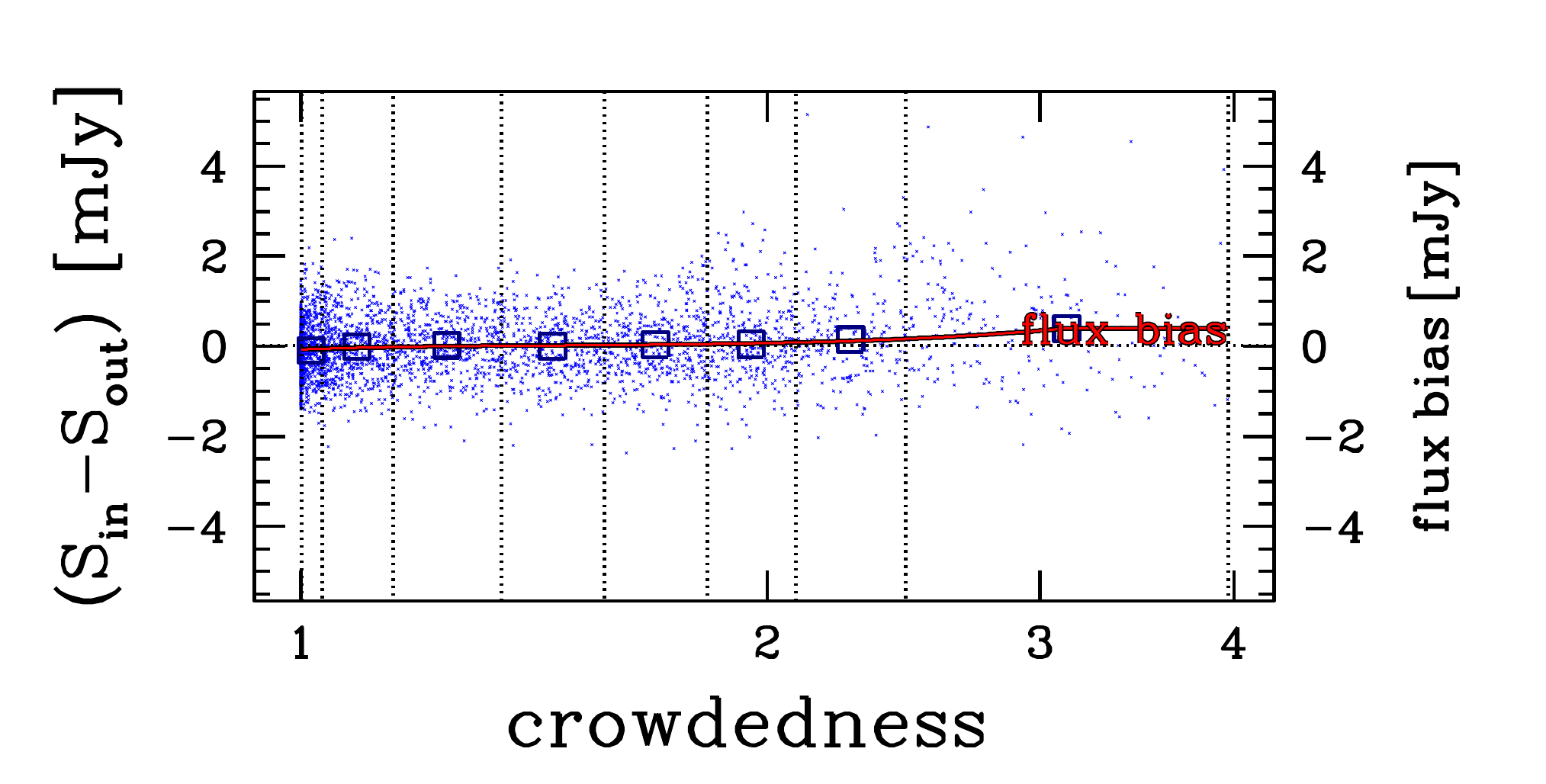}
    \end{subfigure}
    
    \begin{subfigure}[b]{\textwidth}\centering
	\includegraphics[width=0.3\textwidth, trim={1cm 15cm 0cm 2.5cm}, clip]{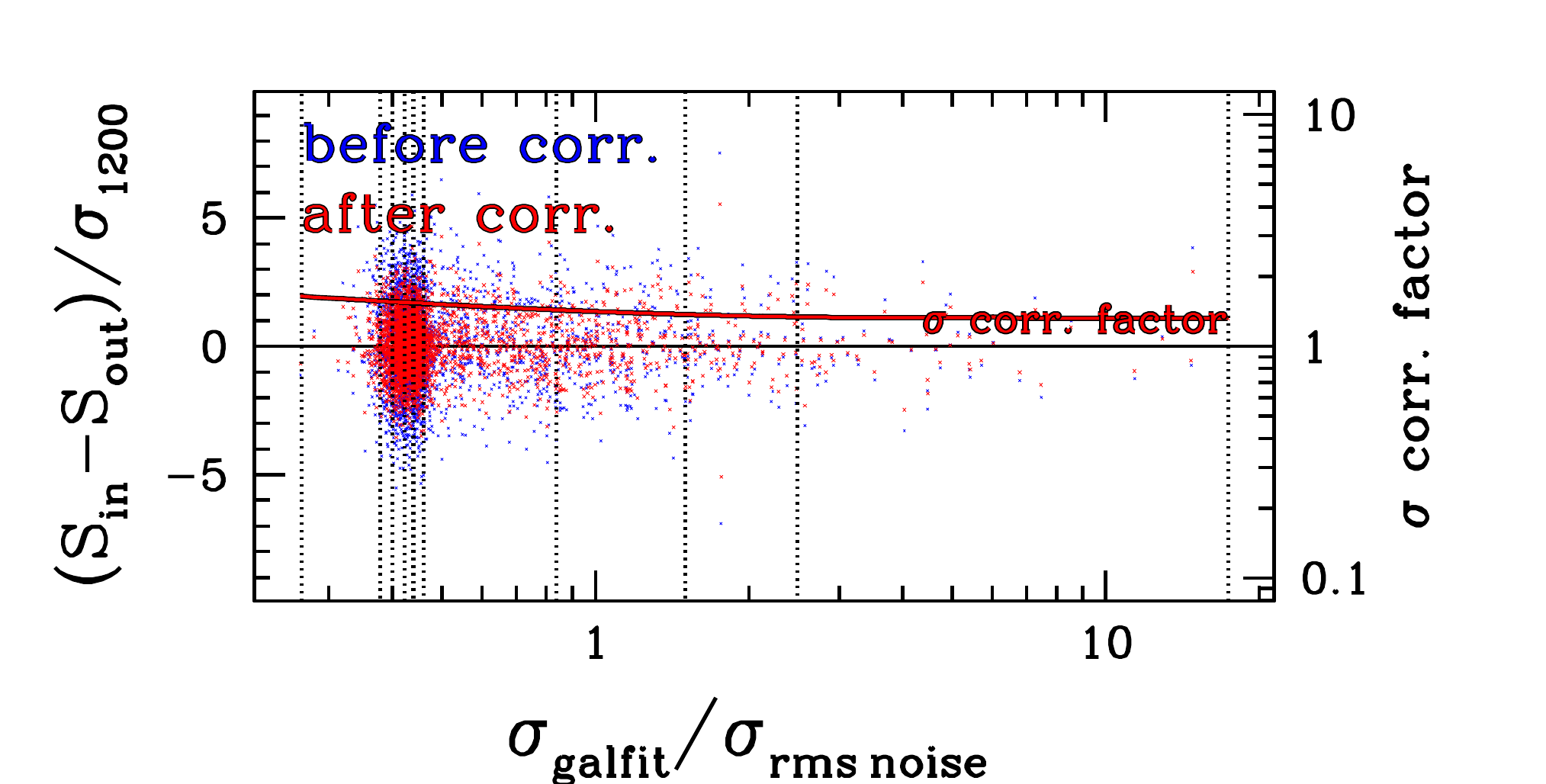}
	\includegraphics[width=0.3\textwidth, trim={1cm 15cm 0cm 2.5cm}, clip]{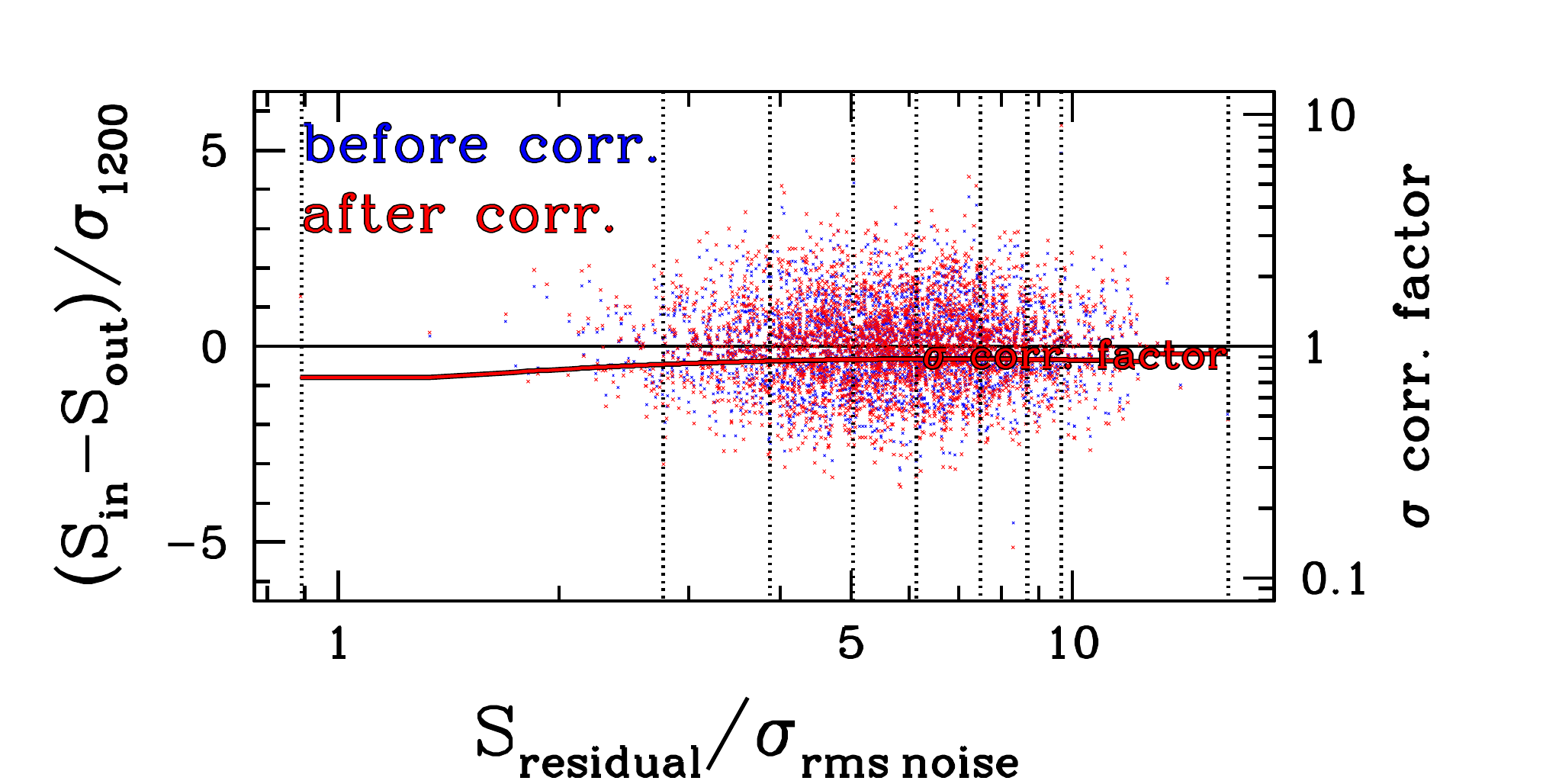}
	\includegraphics[width=0.3\textwidth, trim={1cm 15cm 0cm 2.5cm}, clip]{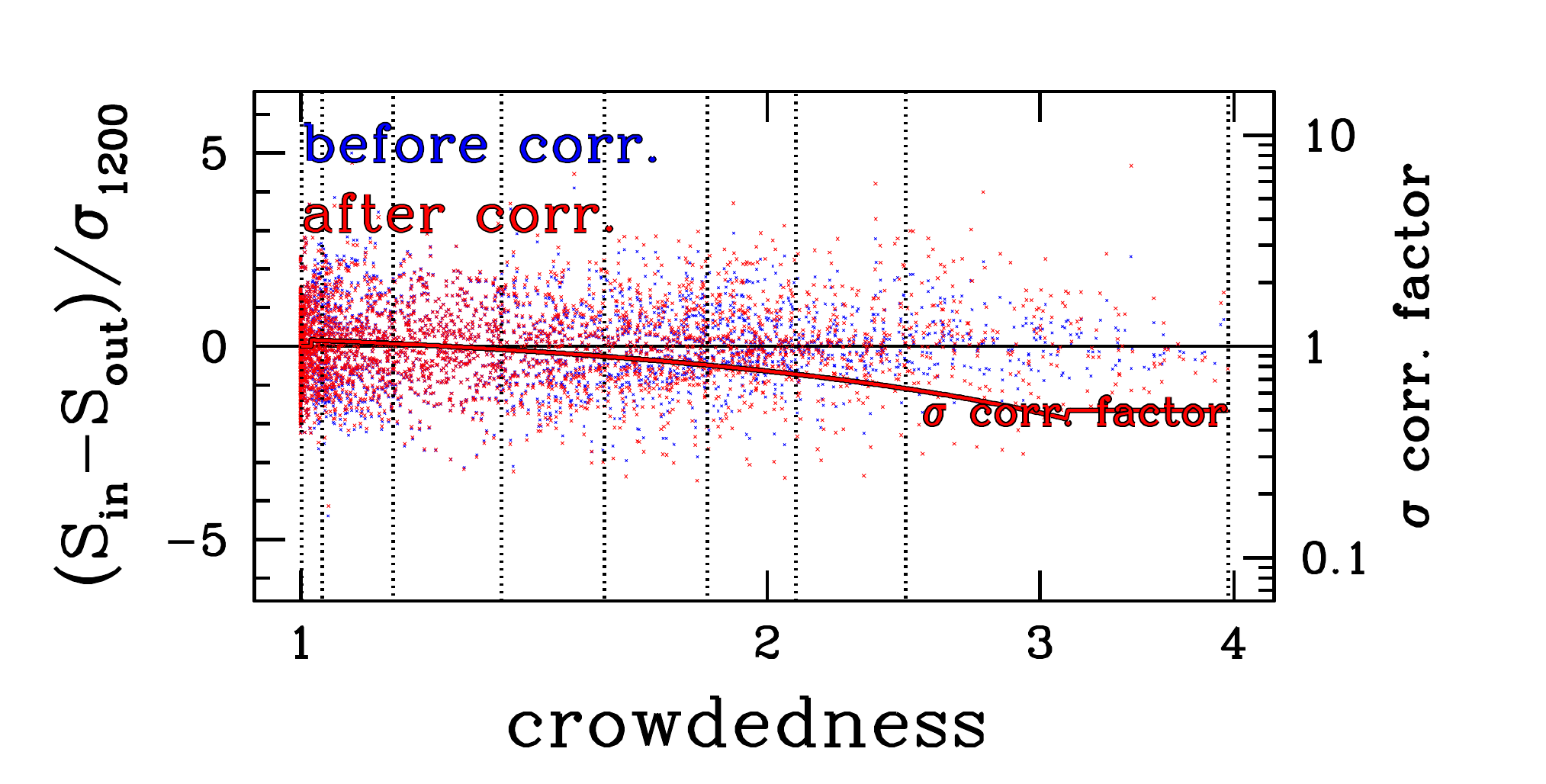}
    \end{subfigure}
    
    \begin{subfigure}[b]{\textwidth}\centering
	\includegraphics[width=0.3\textwidth, trim={1cm 13cm 0cm 2.5cm}, clip]{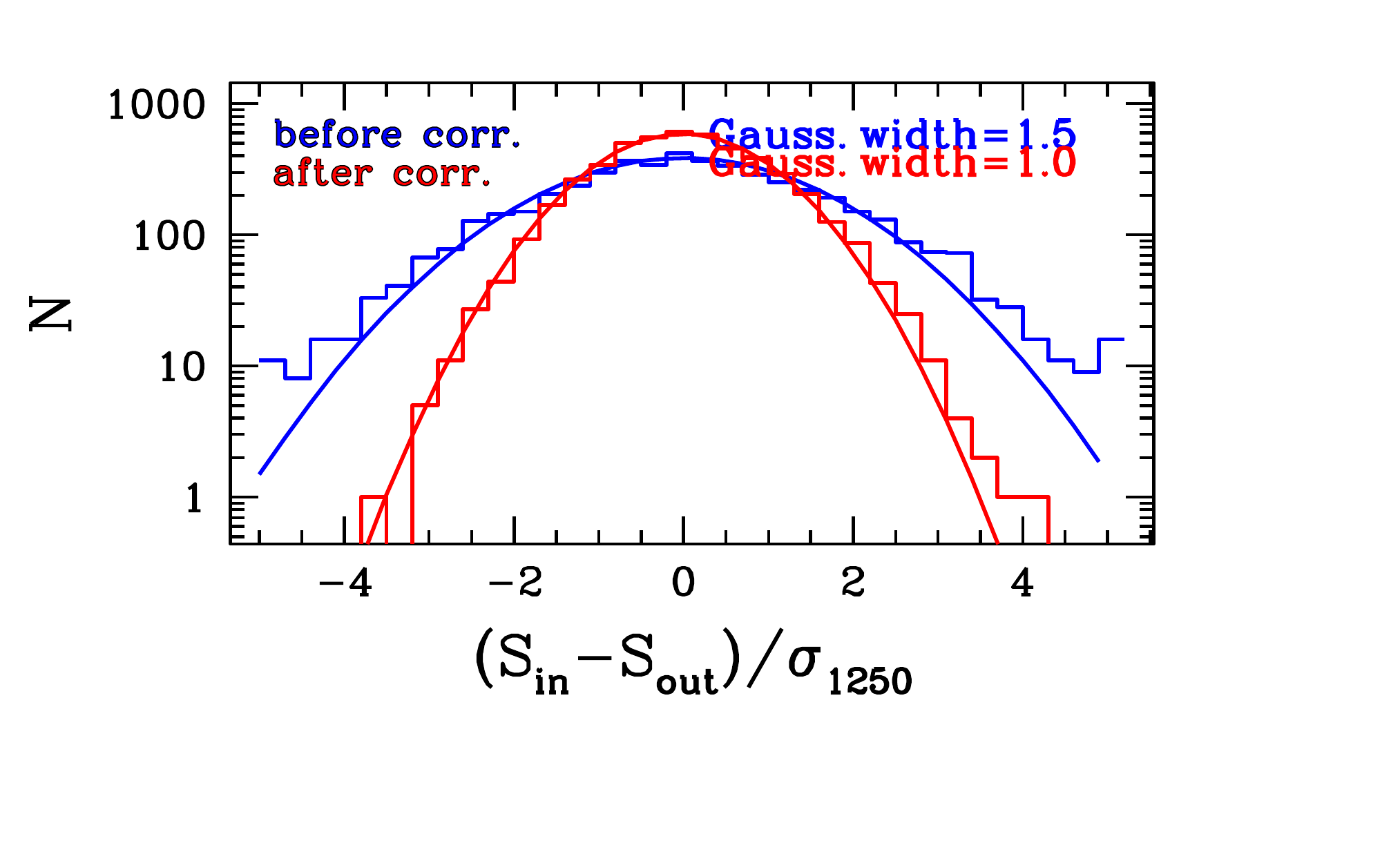}
	\includegraphics[width=0.3\textwidth, trim={1cm 13cm 0cm 2.5cm}, clip]{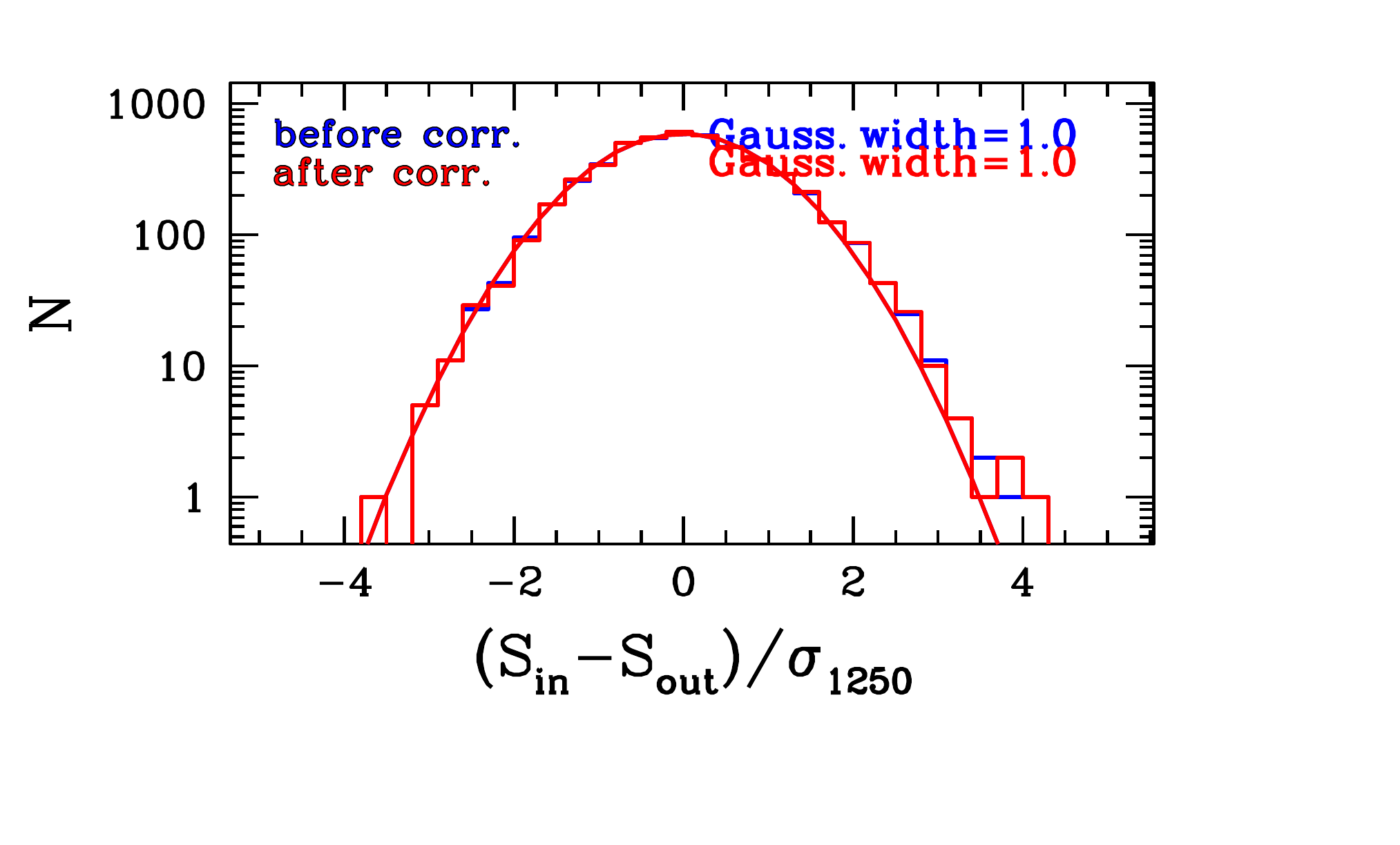}
	\includegraphics[width=0.3\textwidth, trim={1cm 13cm 0cm 2.5cm}, clip]{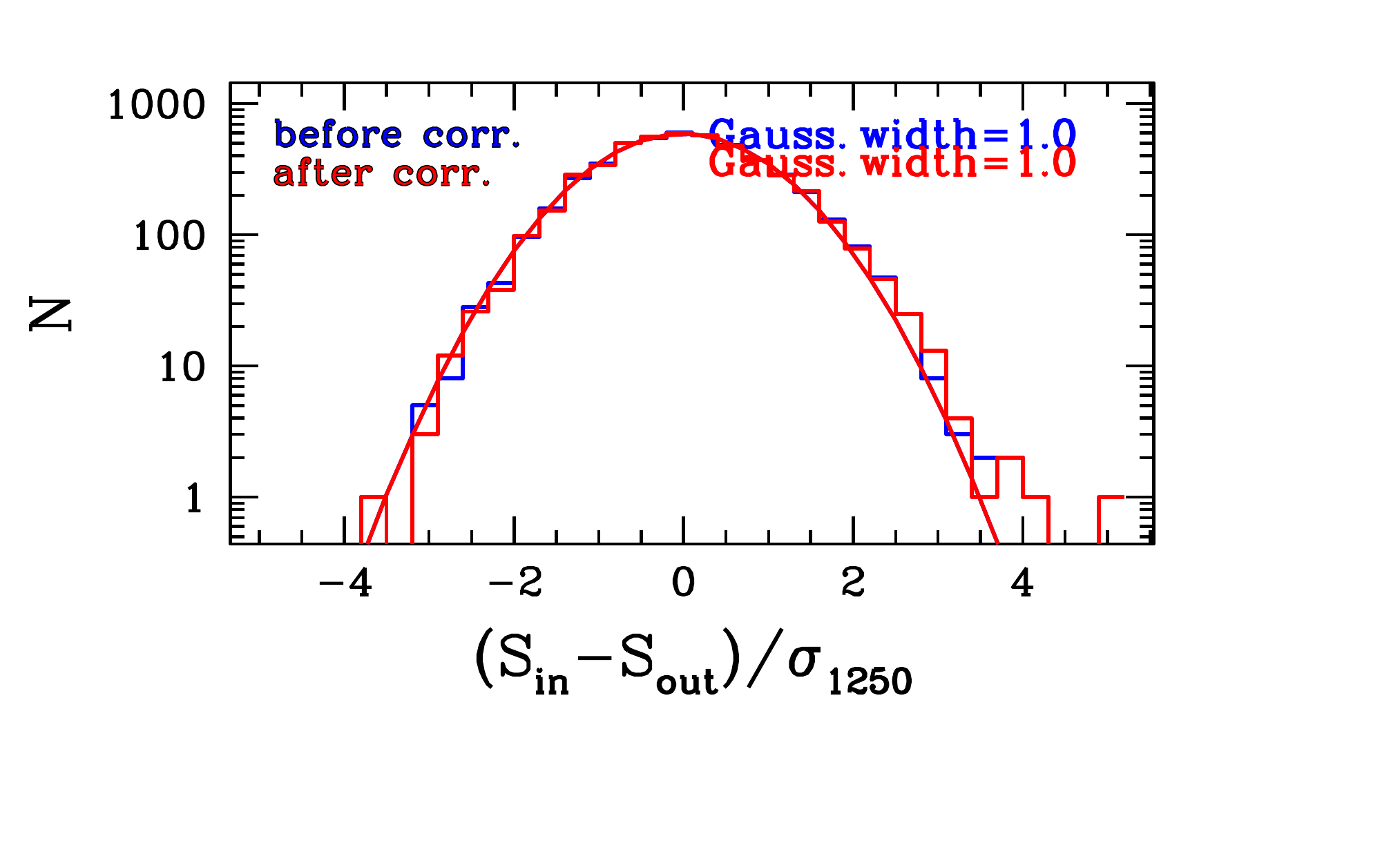}
    \end{subfigure}
    
    \begin{subfigure}[b]{\textwidth}\centering
	\includegraphics[width=0.3\textwidth, trim={1cm 13cm 0cm 2.5cm}, clip]{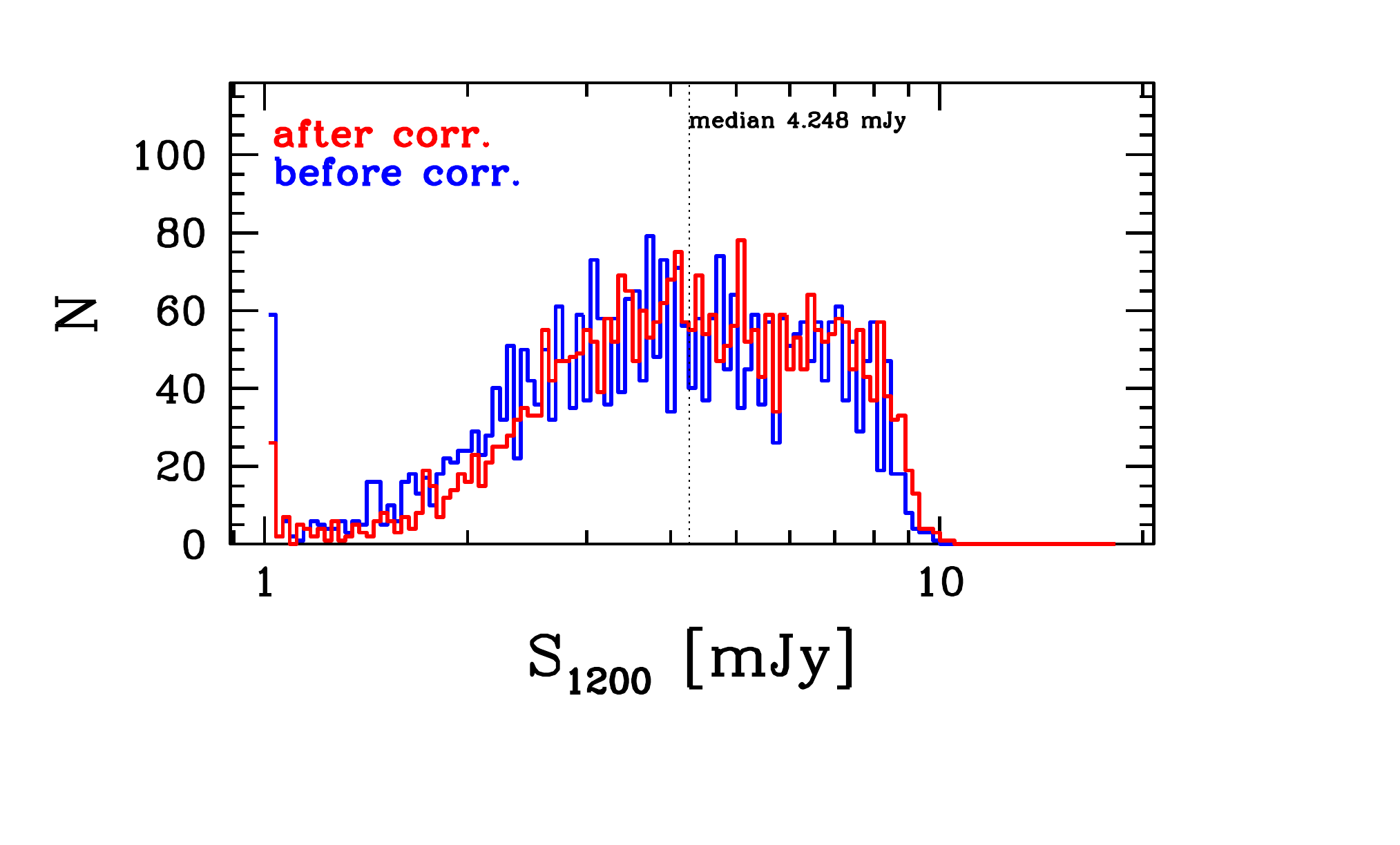}
	\includegraphics[width=0.3\textwidth, trim={1cm 13cm 0cm 2.5cm}, clip]{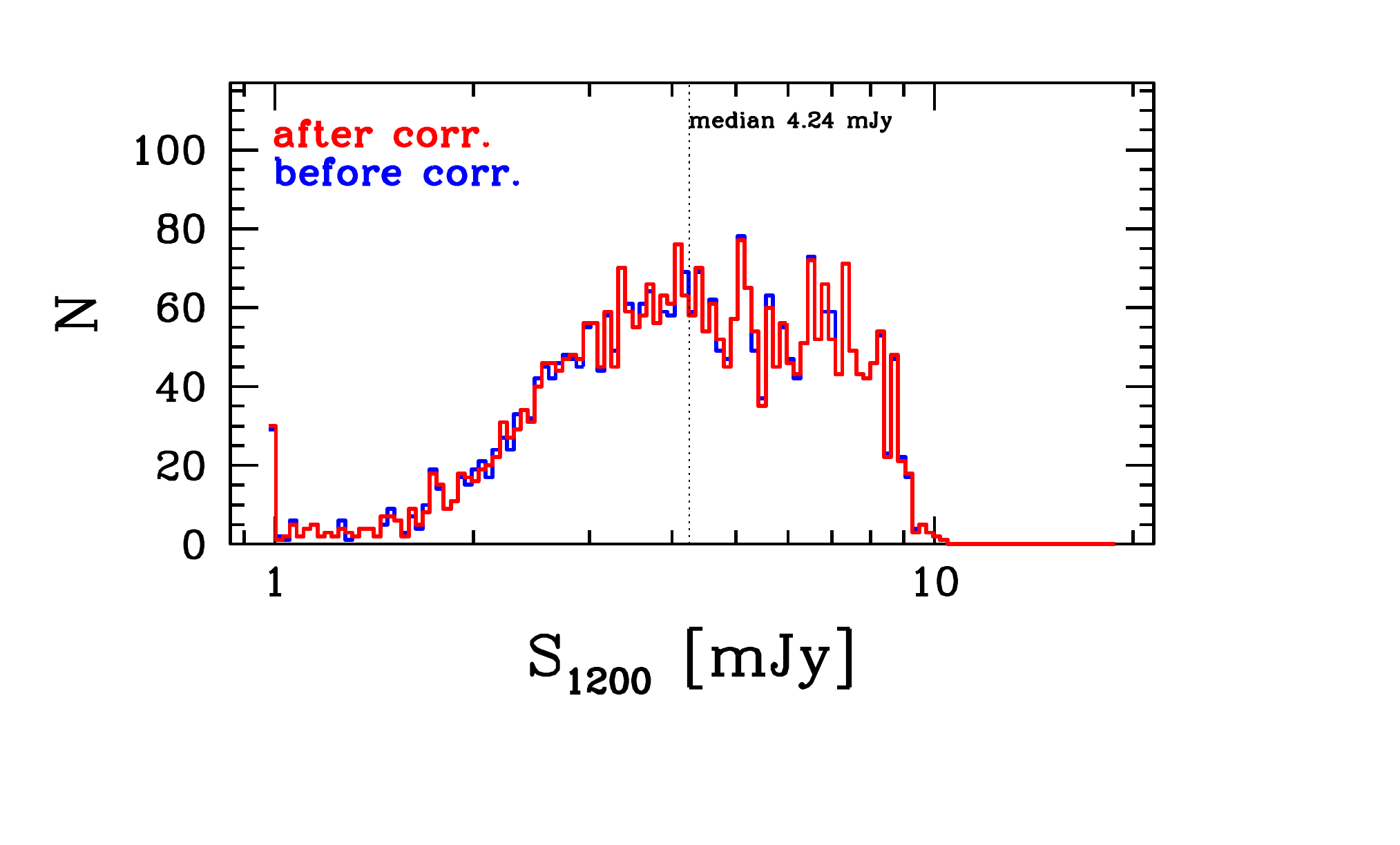}
	\includegraphics[width=0.3\textwidth, trim={1cm 13cm 0cm 2.5cm}, clip]{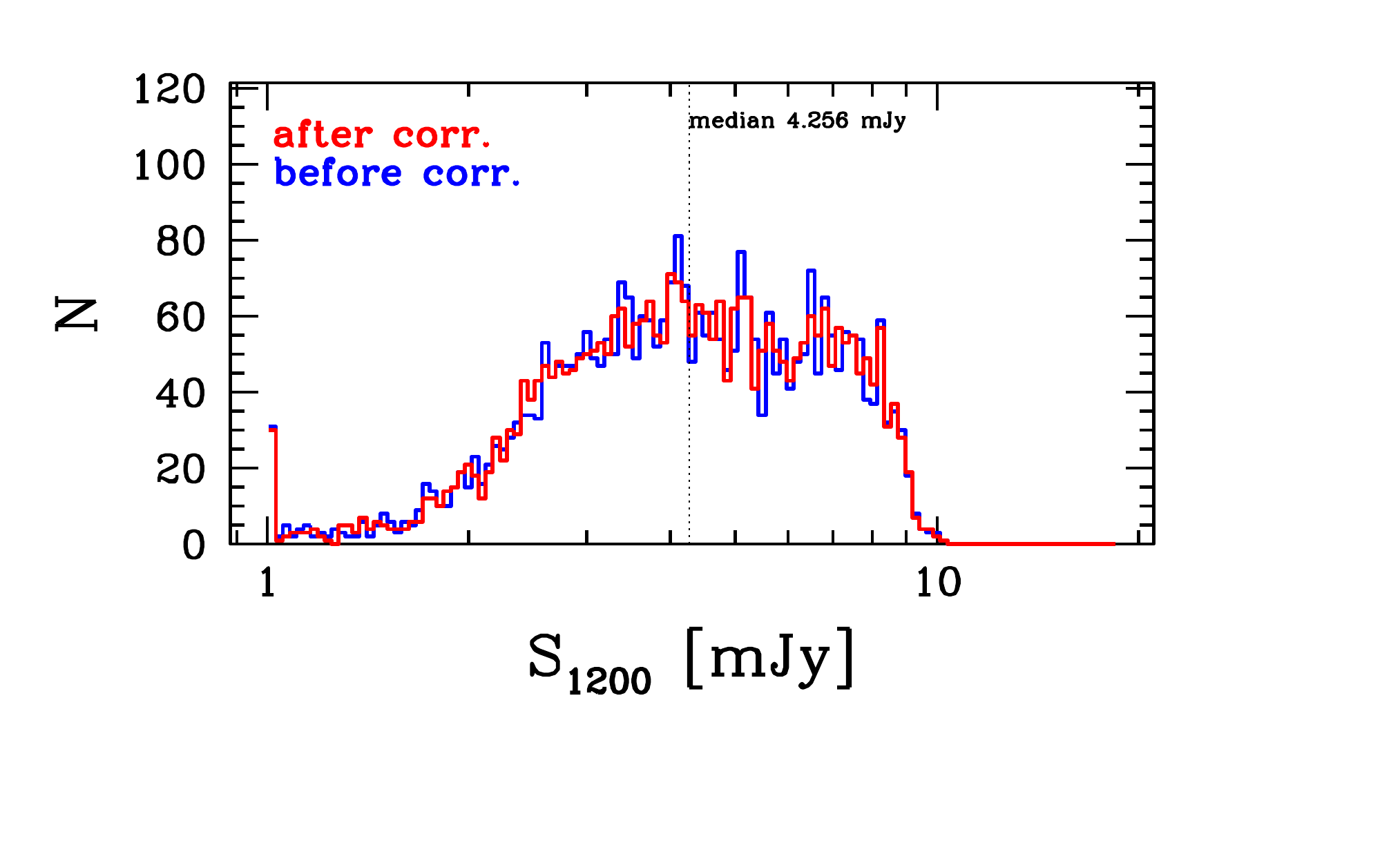}
    \end{subfigure}
    
    \begin{subfigure}[b]{\textwidth}\centering
	\includegraphics[width=0.3\textwidth, trim={1cm 13cm 0cm 2.5cm}, clip]{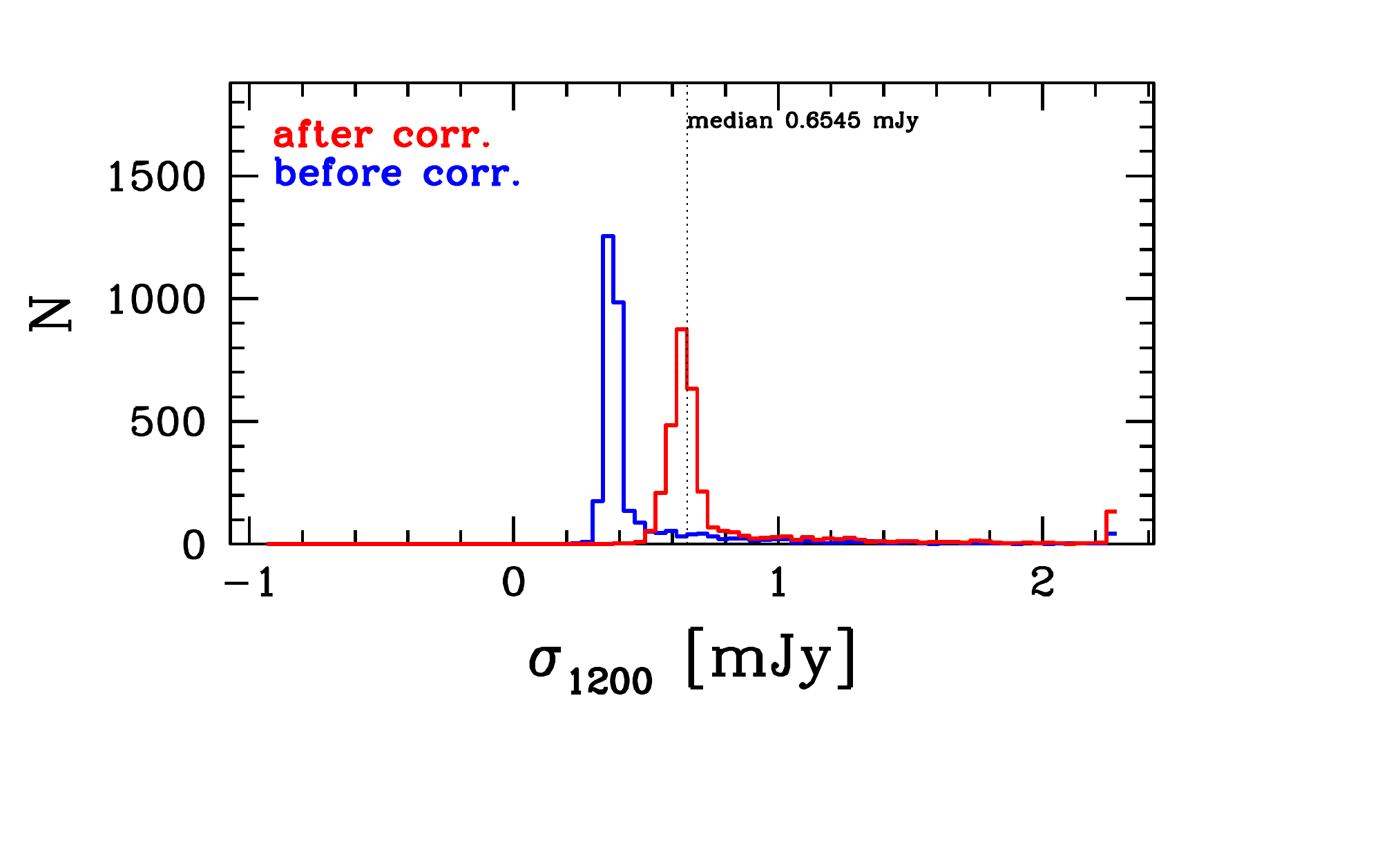}
	\includegraphics[width=0.3\textwidth, trim={1cm 13cm 0cm 2.5cm}, clip]{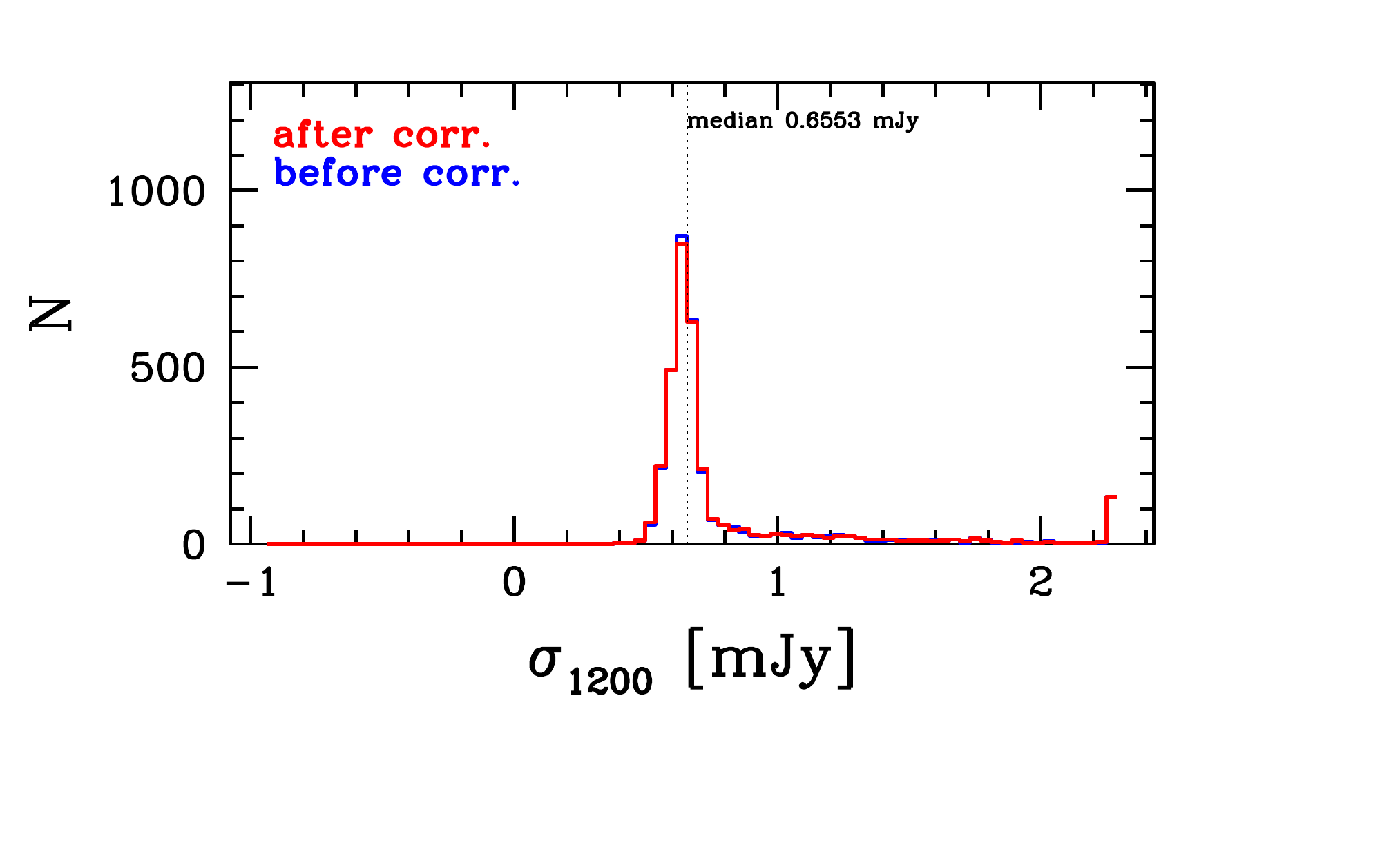}
	\includegraphics[width=0.3\textwidth, trim={1cm 13cm 0cm 2.5cm}, clip]{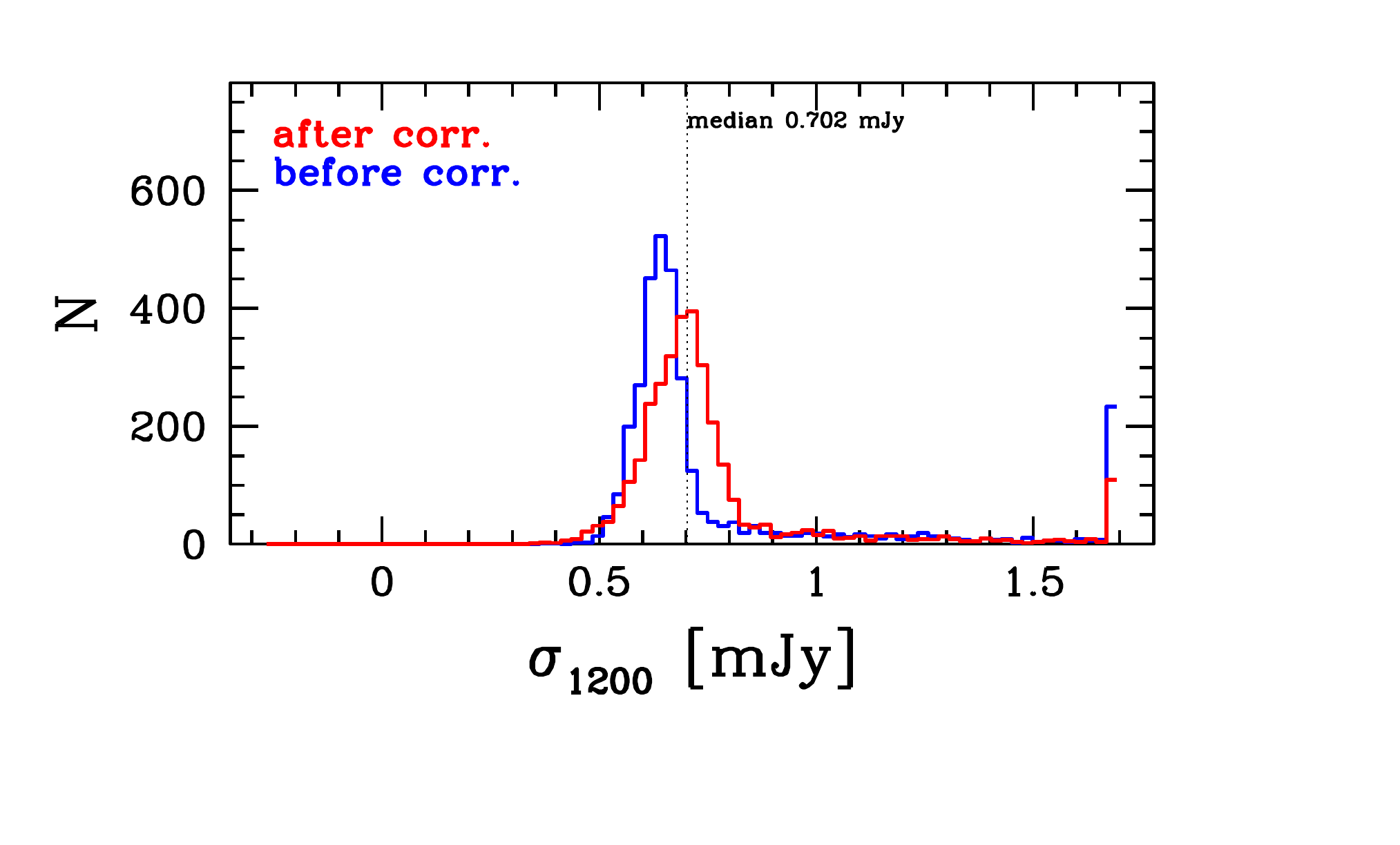}
    \end{subfigure}

\caption{%
	Simulation correction analyses at MAMBO 1.2~mm. See descriptions in text. 
}
\end{figure}



\begin{figure}
	\centering
    
    \begin{subfigure}[b]{\textwidth}\centering
	\includegraphics[width=0.42\textwidth, trim={0.8cm 15cm 0cm 2.5cm}, clip]{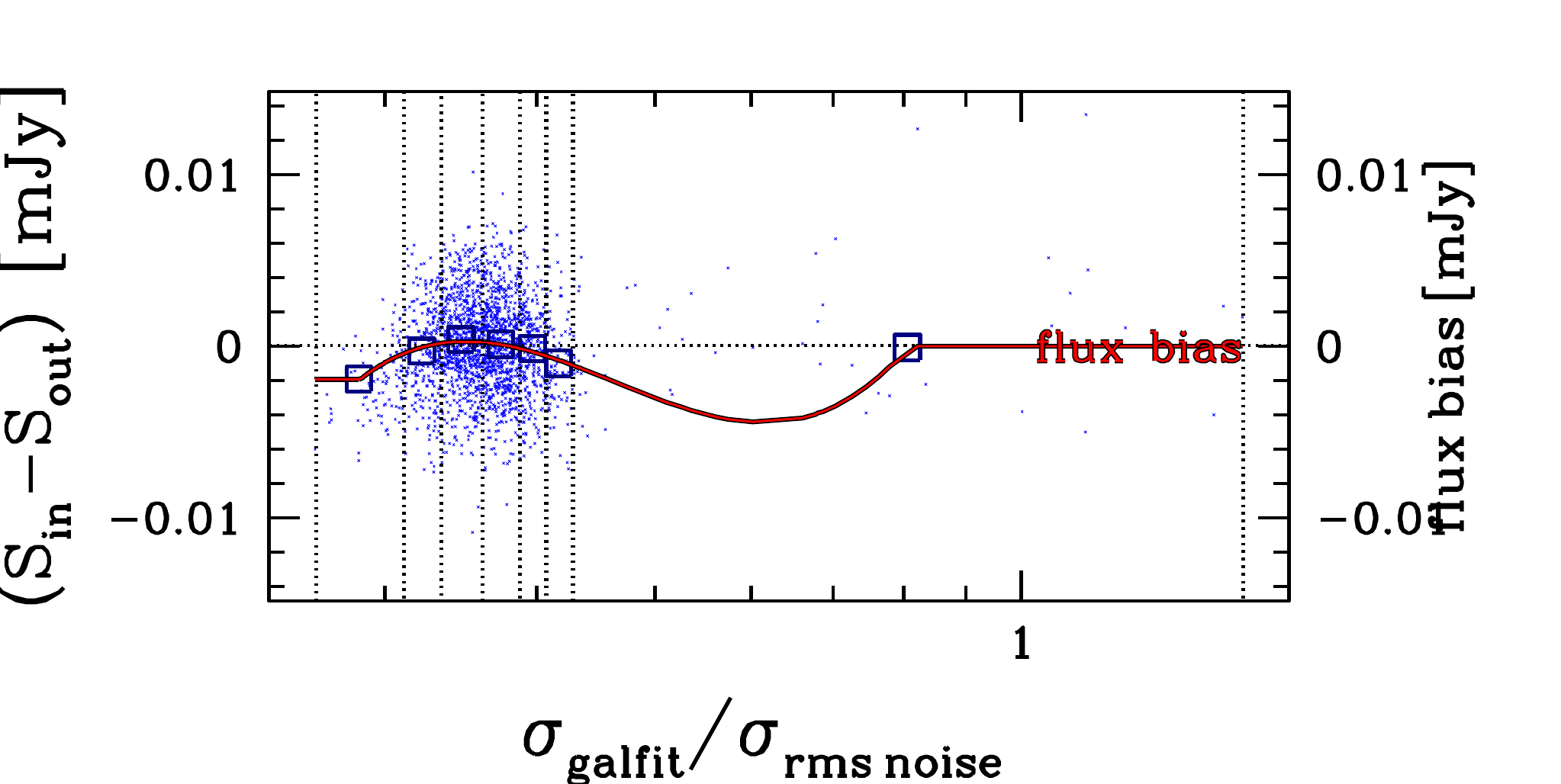}
	\includegraphics[width=0.42\textwidth, trim={0.8cm 15cm 0cm 2.5cm}, clip]{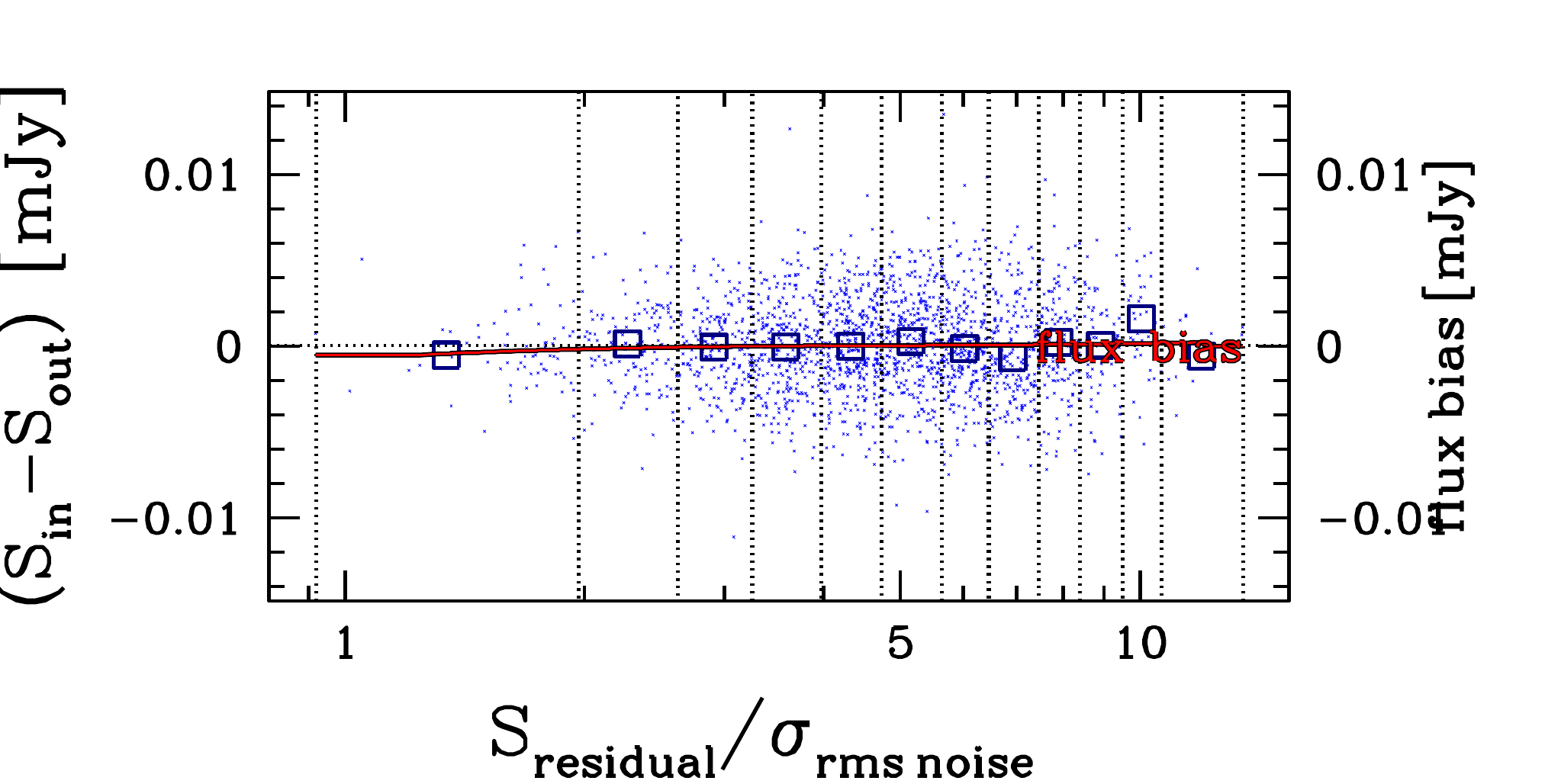}
    \end{subfigure}
    
    \begin{subfigure}[b]{\textwidth}\centering
	\includegraphics[width=0.42\textwidth, trim={0.8cm 15cm 0cm 2.5cm}, clip]{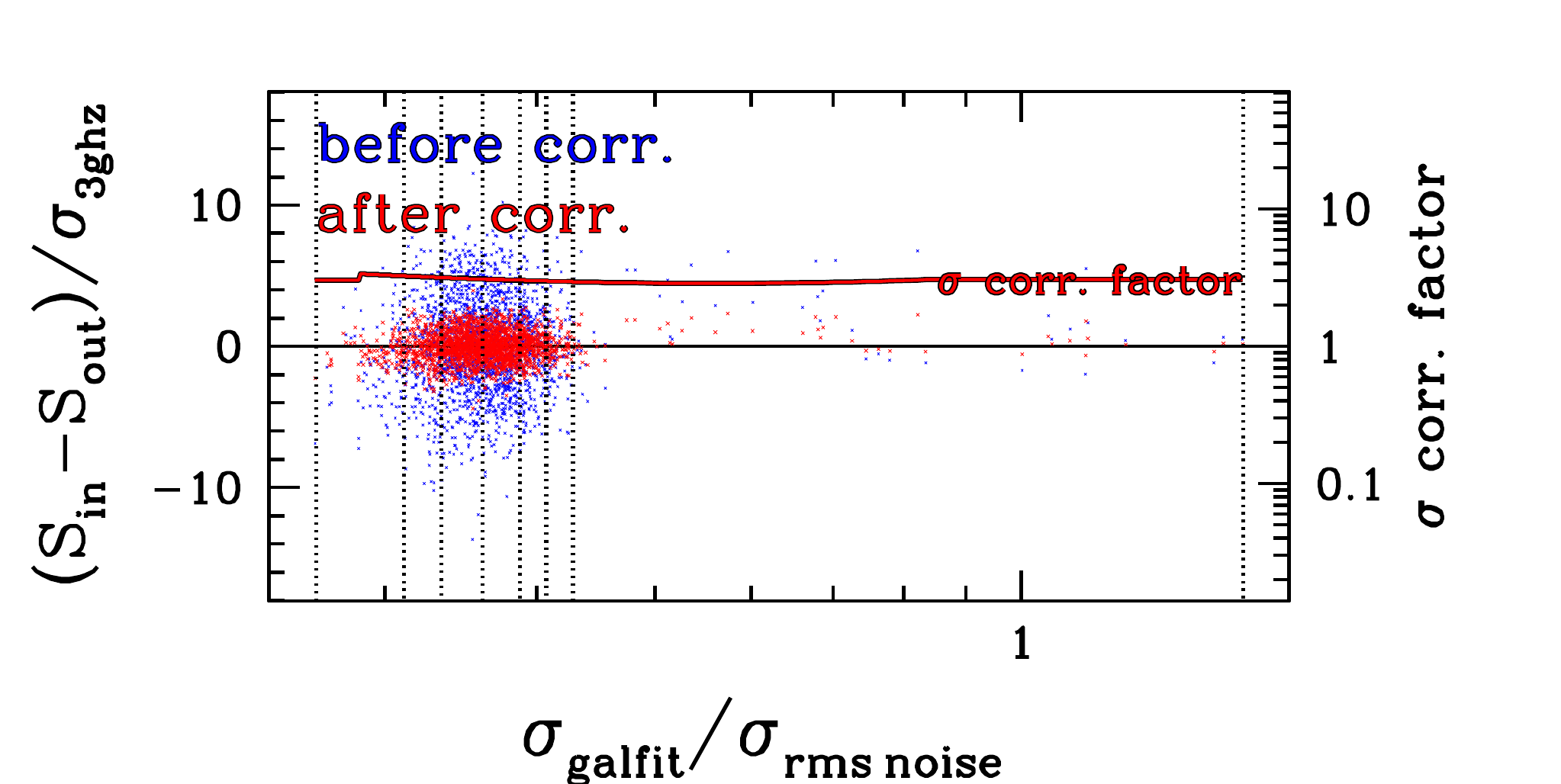}
	\includegraphics[width=0.42\textwidth, trim={1cm 15cm 0cm 2.5cm}, clip]{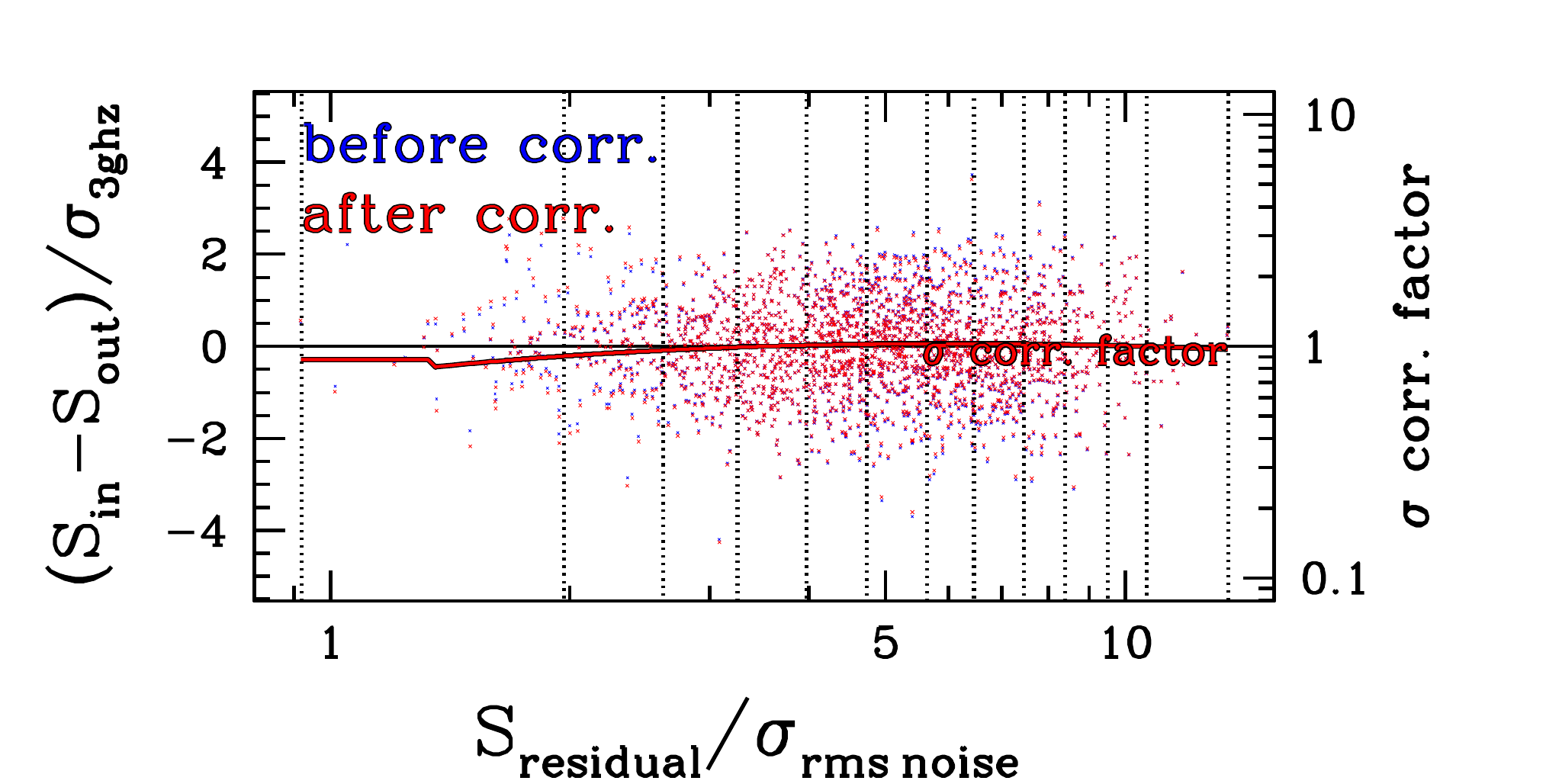}
    \end{subfigure}
    
    \begin{subfigure}[b]{\textwidth}\centering
	\includegraphics[width=0.42\textwidth, trim={0.8cm 15cm 0cm 2.5cm}, clip]{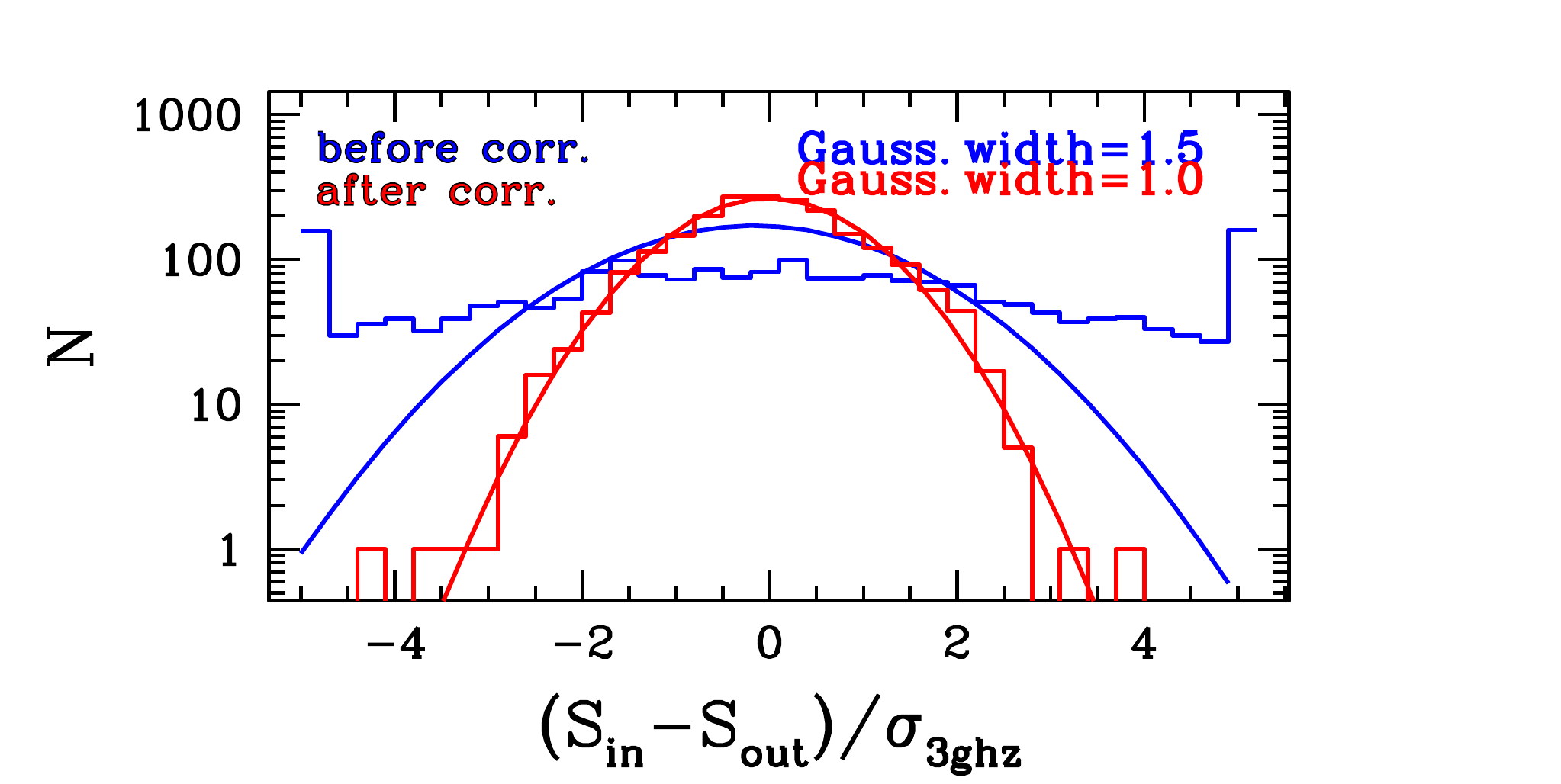}
	\includegraphics[width=0.42\textwidth, trim={0.8cm 15cm 0cm 2.5cm}, clip]{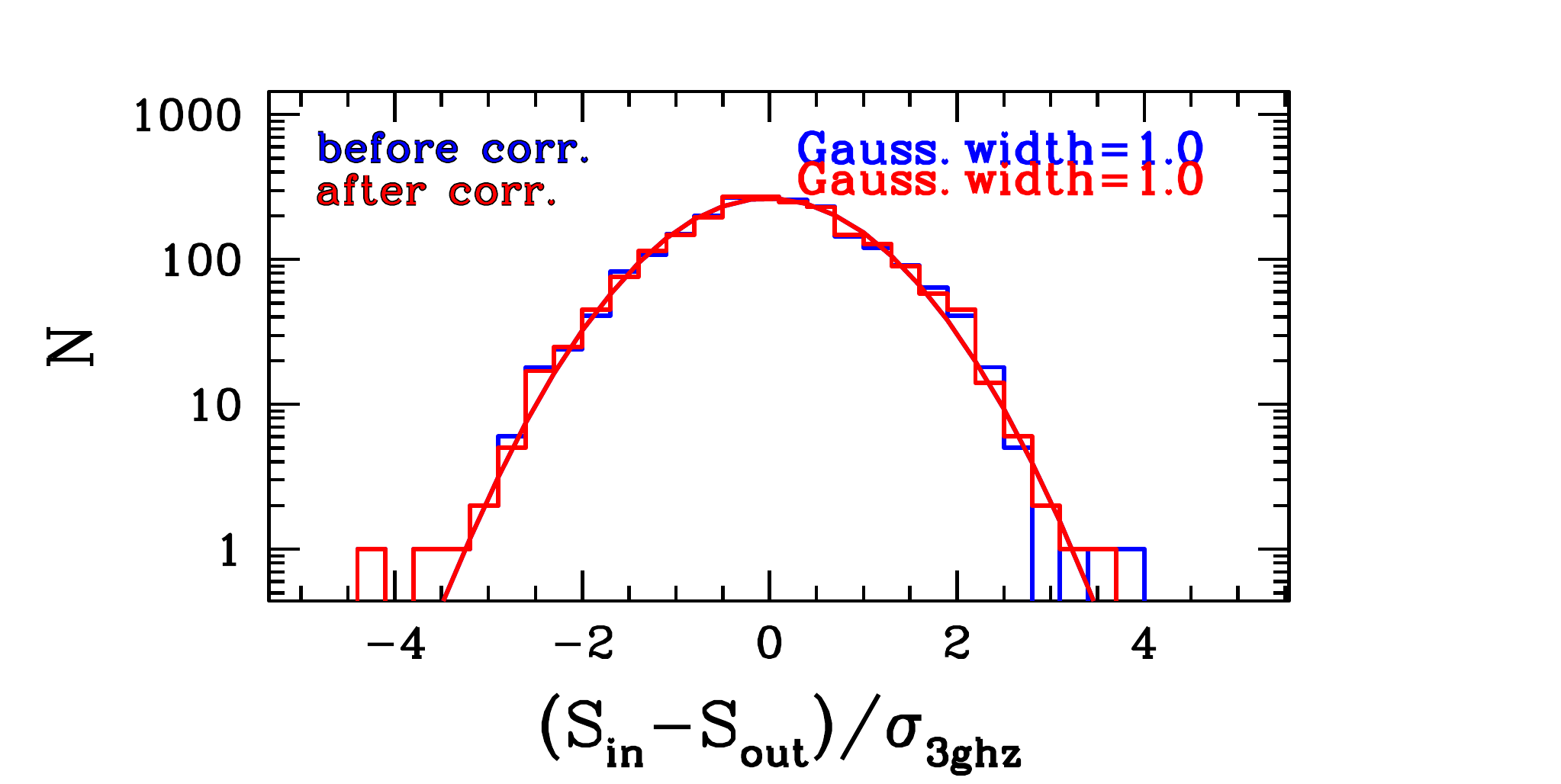}
    \end{subfigure}
    
    
    \begin{subfigure}[b]{\textwidth}\centering
	\includegraphics[width=0.42\textwidth, trim={0.8cm 15cm 0cm 2.5cm}, clip]{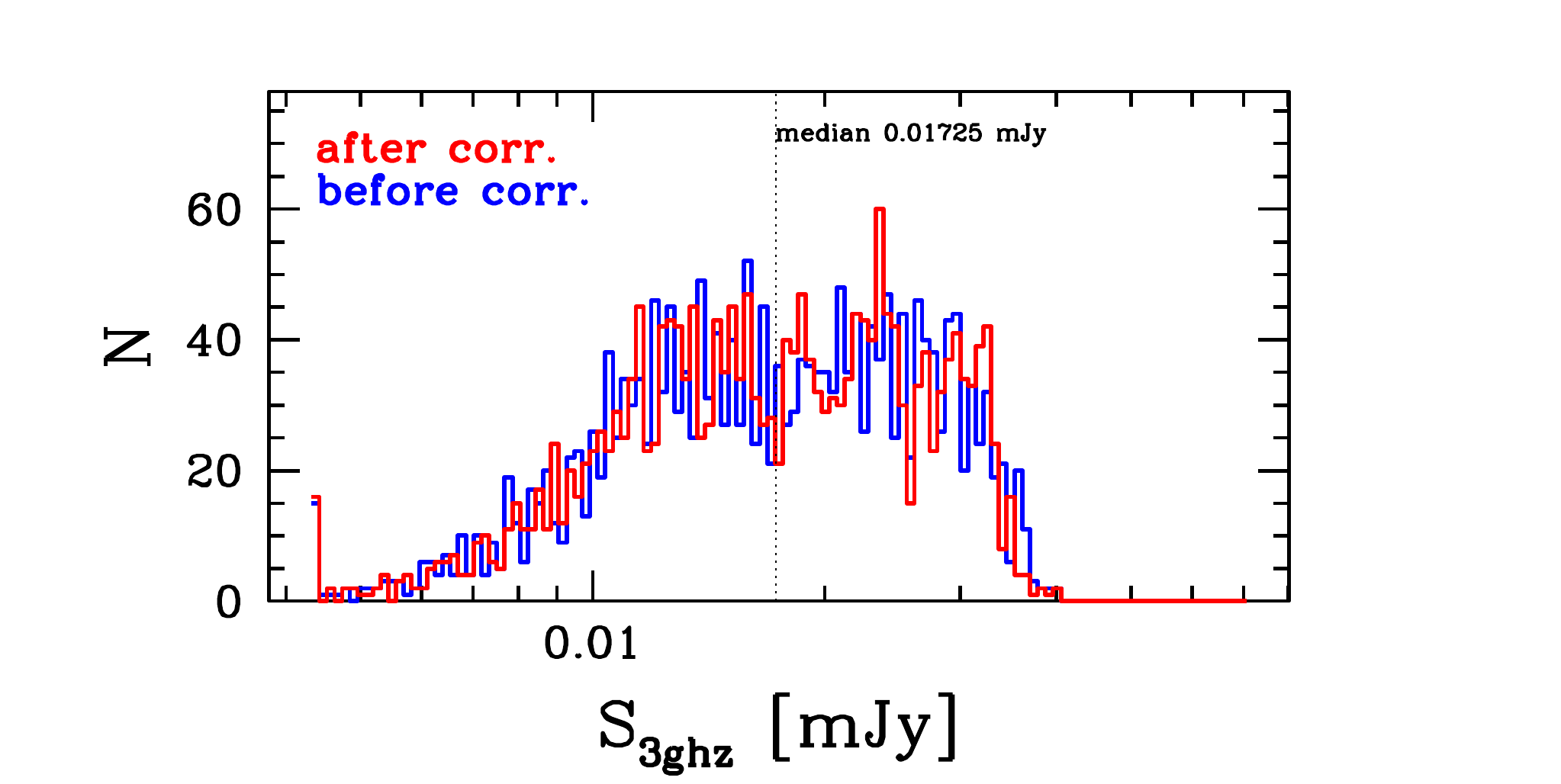}
	\includegraphics[width=0.42\textwidth, trim={0.8cm 15cm 0cm 2.5cm}, clip]{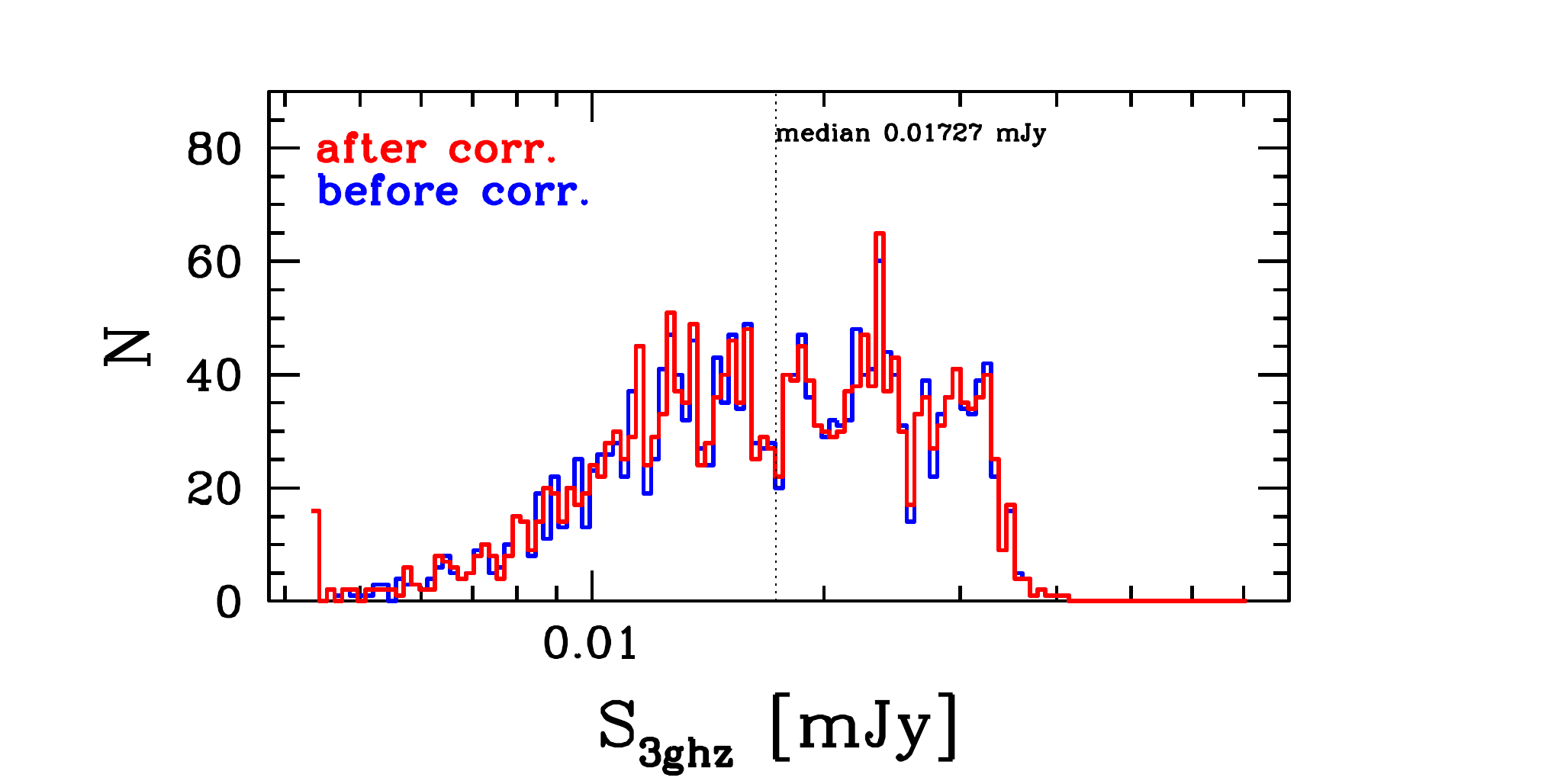}
    \end{subfigure}
    
    \begin{subfigure}[b]{\textwidth}\centering
	\includegraphics[width=0.42\textwidth, trim={0.8cm 15cm 0cm 2.5cm}, clip]{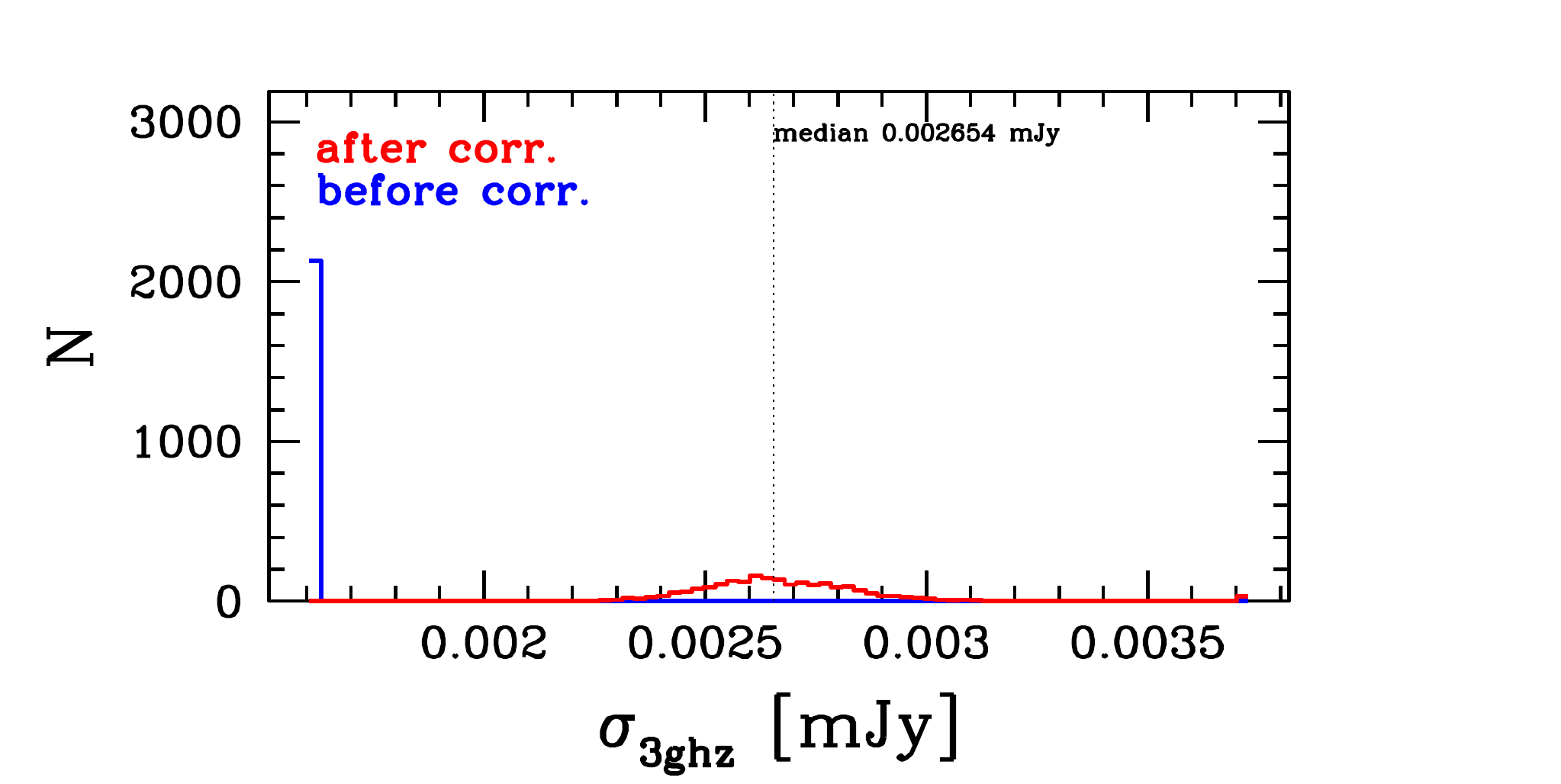}
	\includegraphics[width=0.42\textwidth, trim={0.8cm 15cm 0cm 2.5cm}, clip]{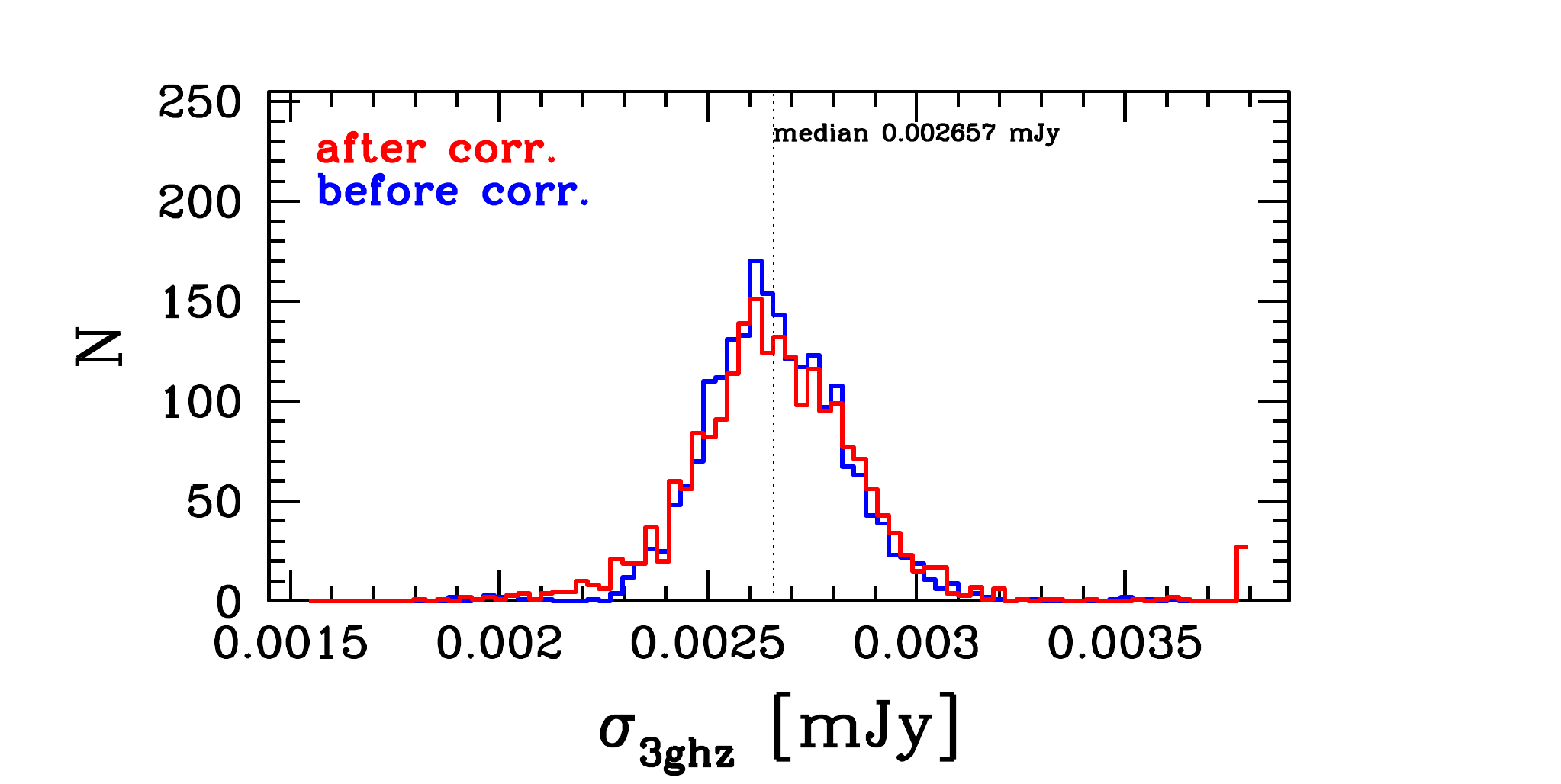}
    \end{subfigure}
    
	\caption{%
		Simulation correction analyses at VLA 3~GHz. See descriptions in text. 
	}
    \label{Simu_fig_3ghz}
\end{figure}

\begin{figure}
	\centering
    
    \begin{subfigure}[b]{\textwidth}\centering
	\includegraphics[width=0.42\textwidth, trim={0.8cm 15cm 0cm 2.5cm}, clip]{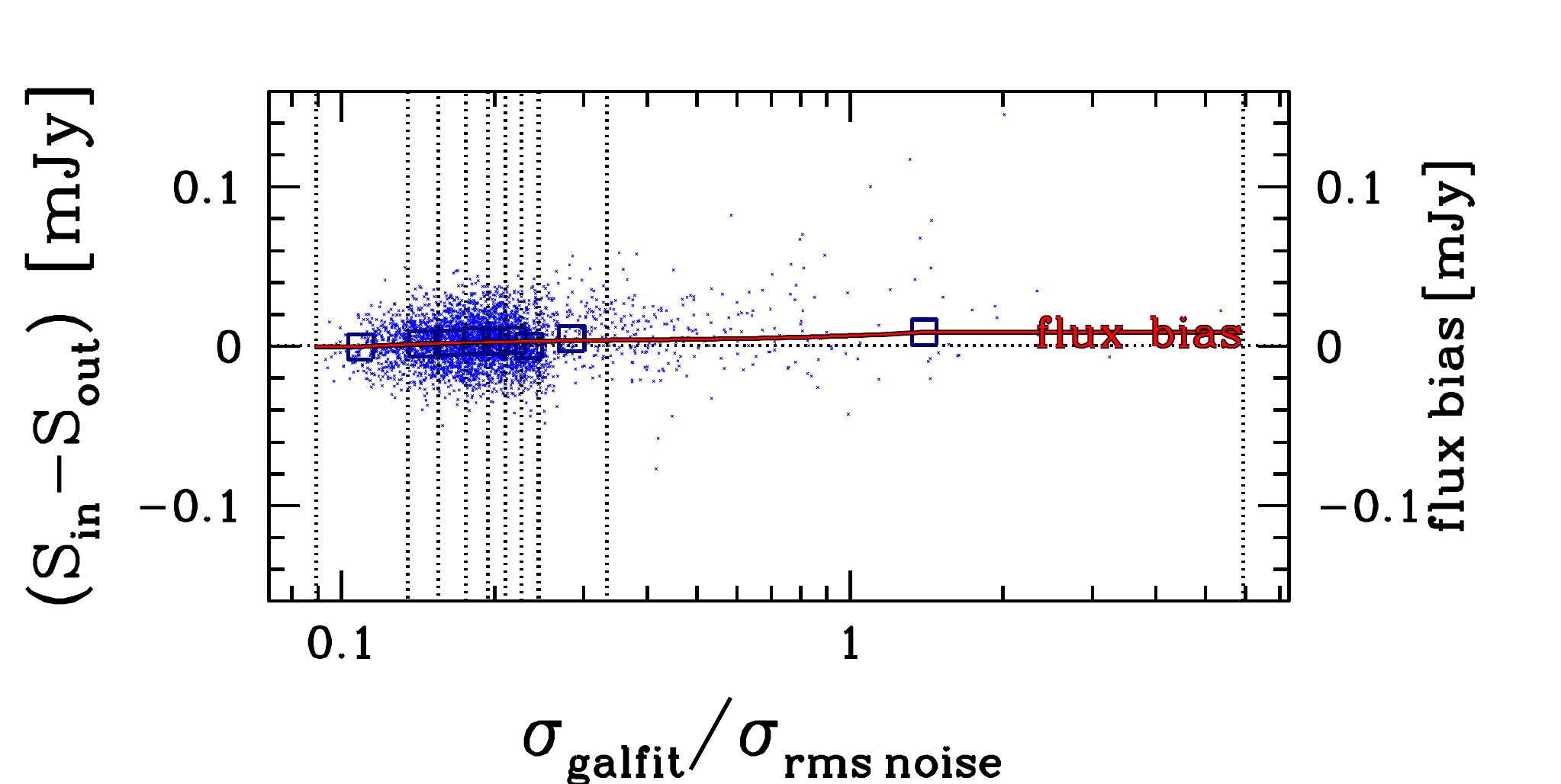}
	\includegraphics[width=0.42\textwidth, trim={0.8cm 15cm 0cm 2.5cm}, clip]{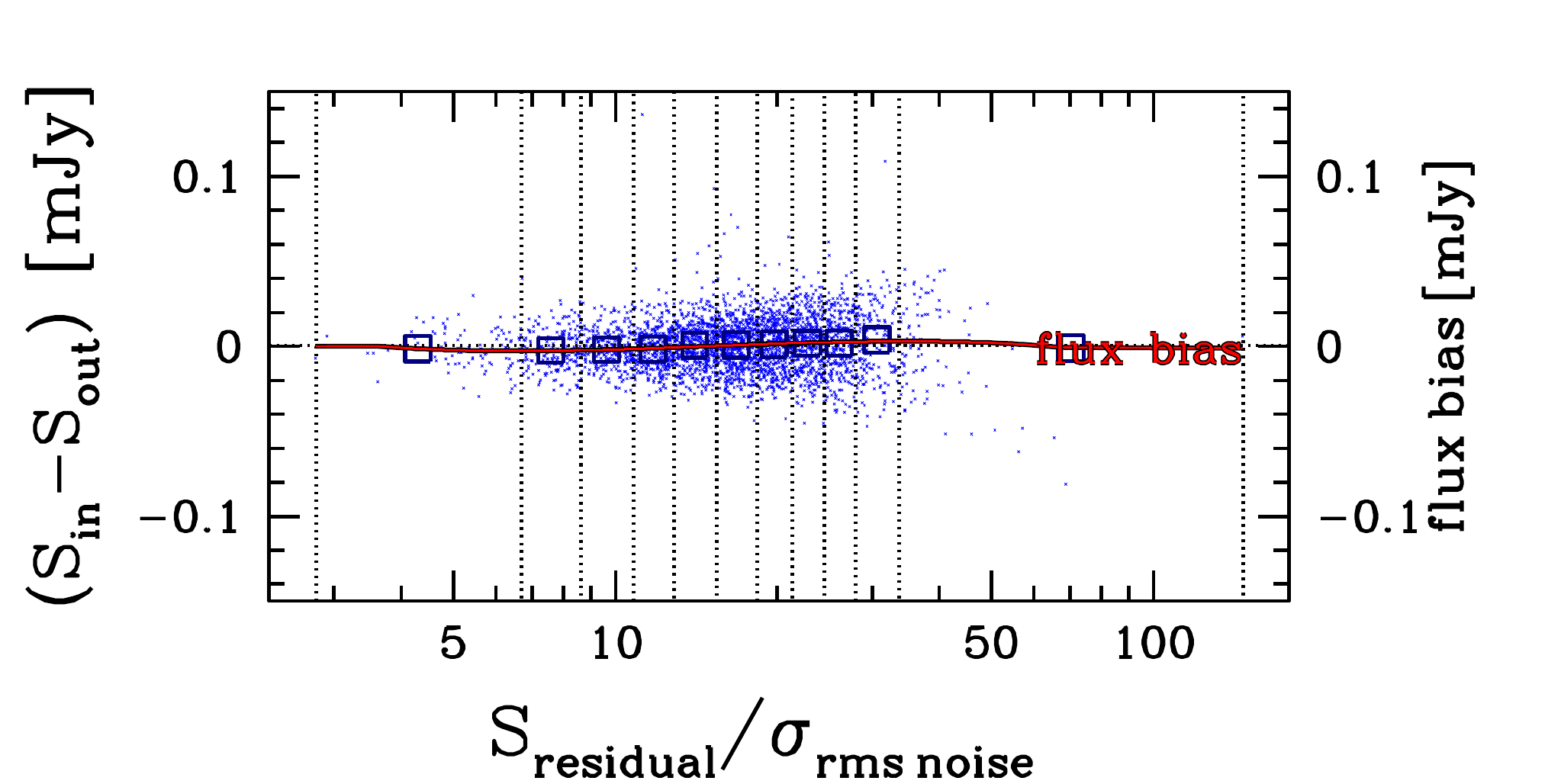}
    \end{subfigure}
    
    \begin{subfigure}[b]{\textwidth}\centering
	\includegraphics[width=0.42\textwidth, trim={0.8cm 15cm 0cm 2.5cm}, clip]{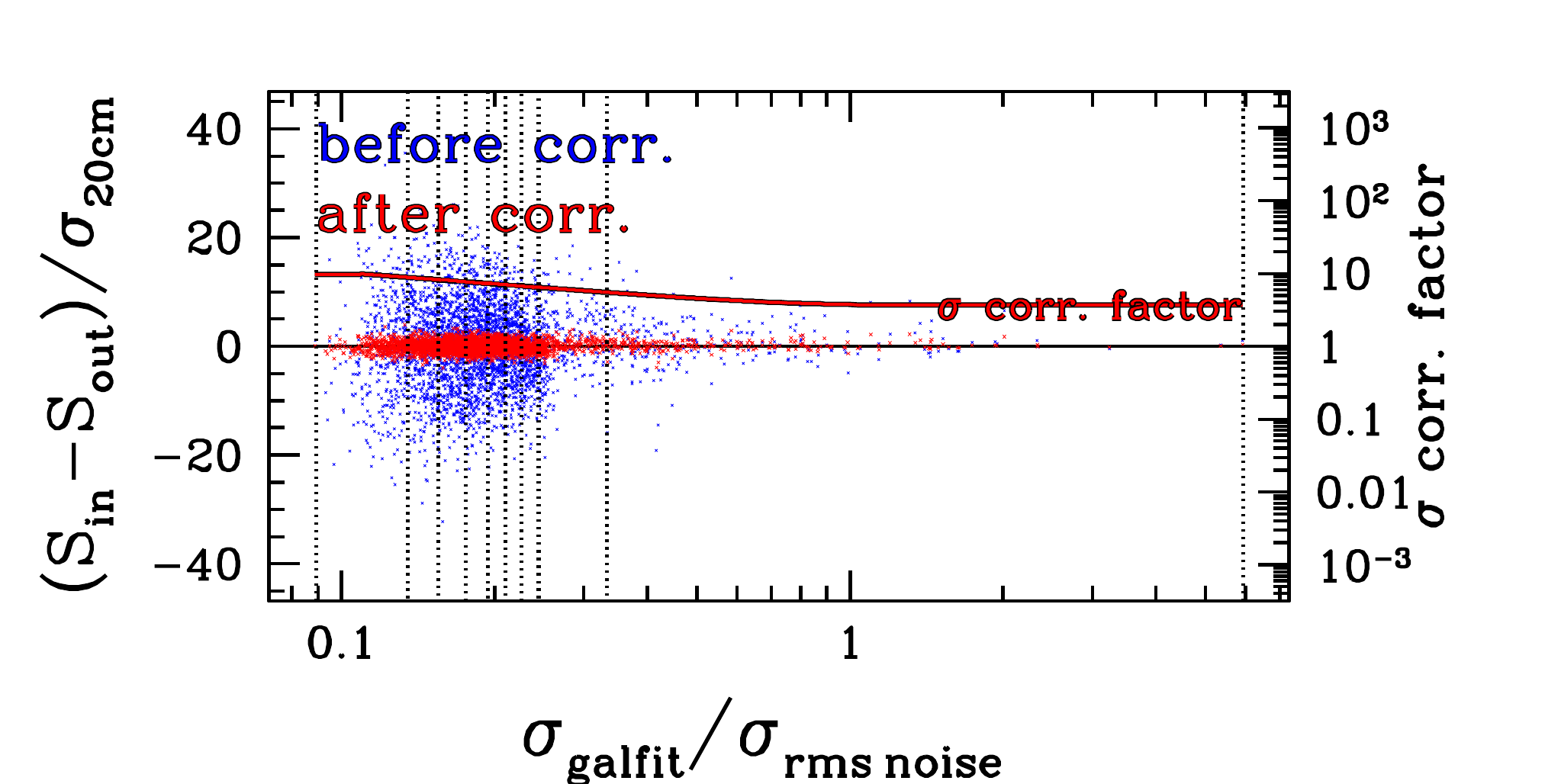}
	\includegraphics[width=0.42\textwidth, trim={0.8cm 15cm 0cm 2.5cm}, clip]{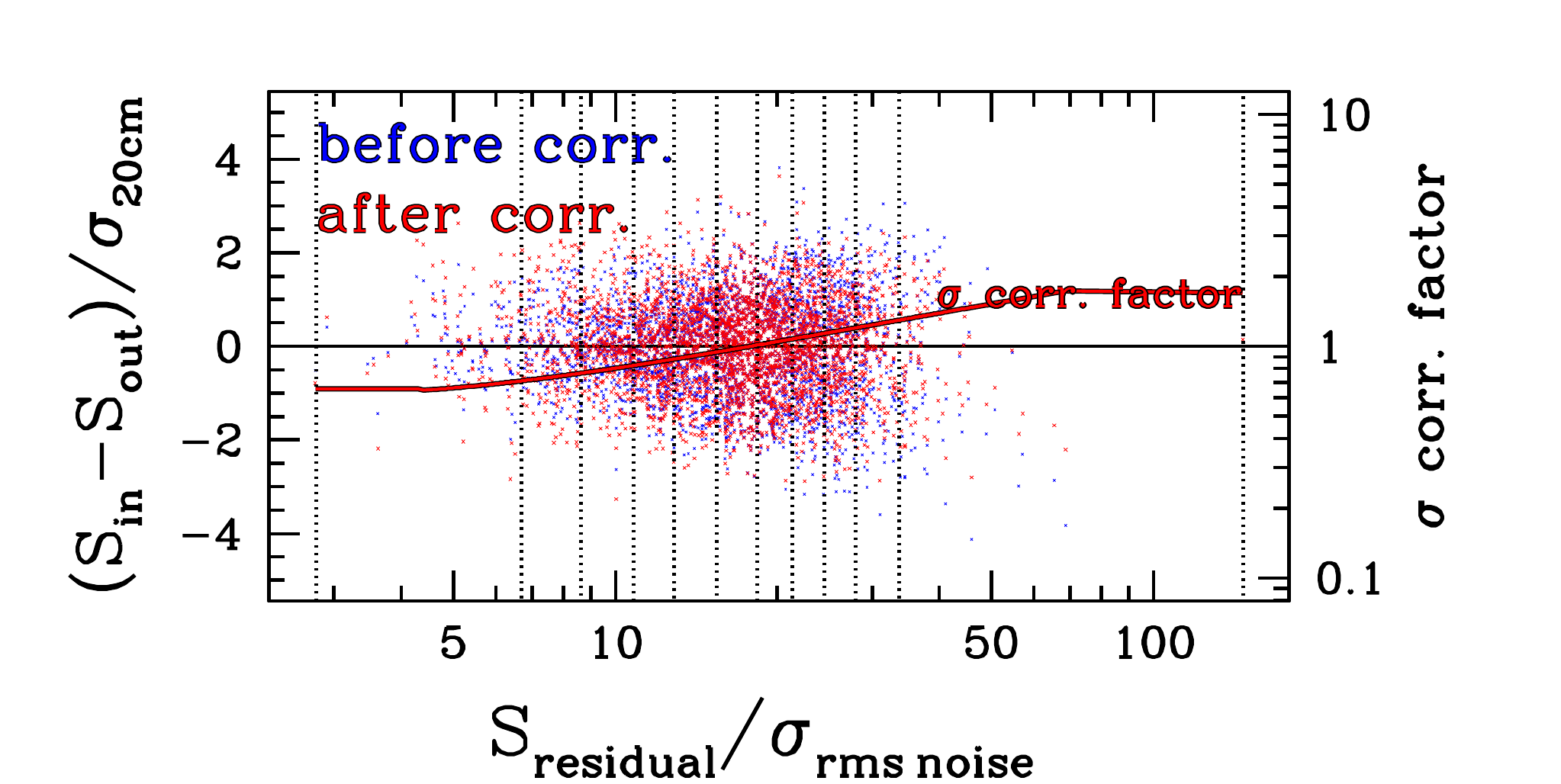}
    \end{subfigure}
    
    \begin{subfigure}[b]{\textwidth}\centering
	\includegraphics[width=0.42\textwidth, trim={0.8cm 15cm 0cm 2.5cm}, clip]{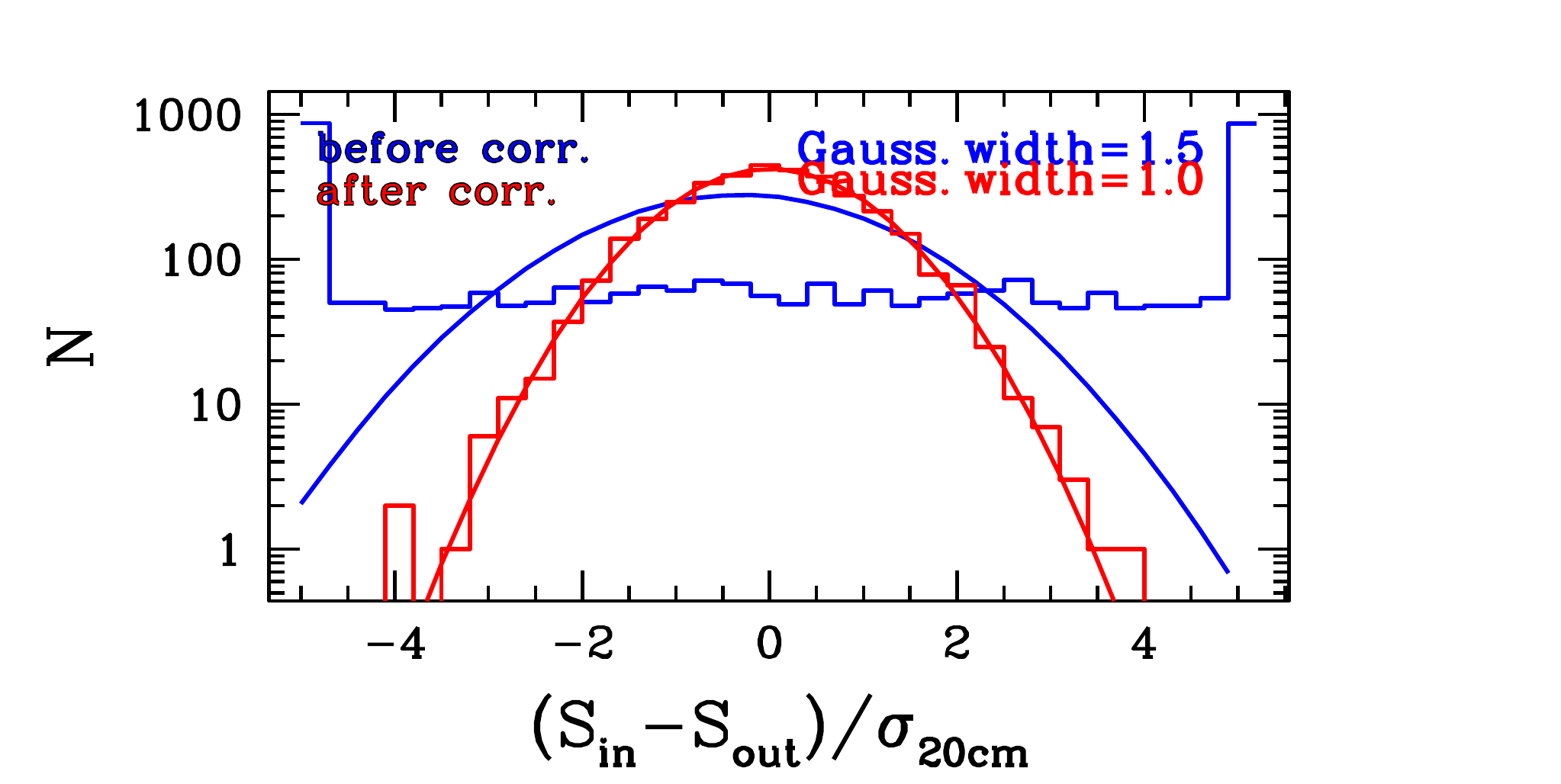}
	\includegraphics[width=0.42\textwidth, trim={0.8cm 15cm 0cm 2.5cm}, clip]{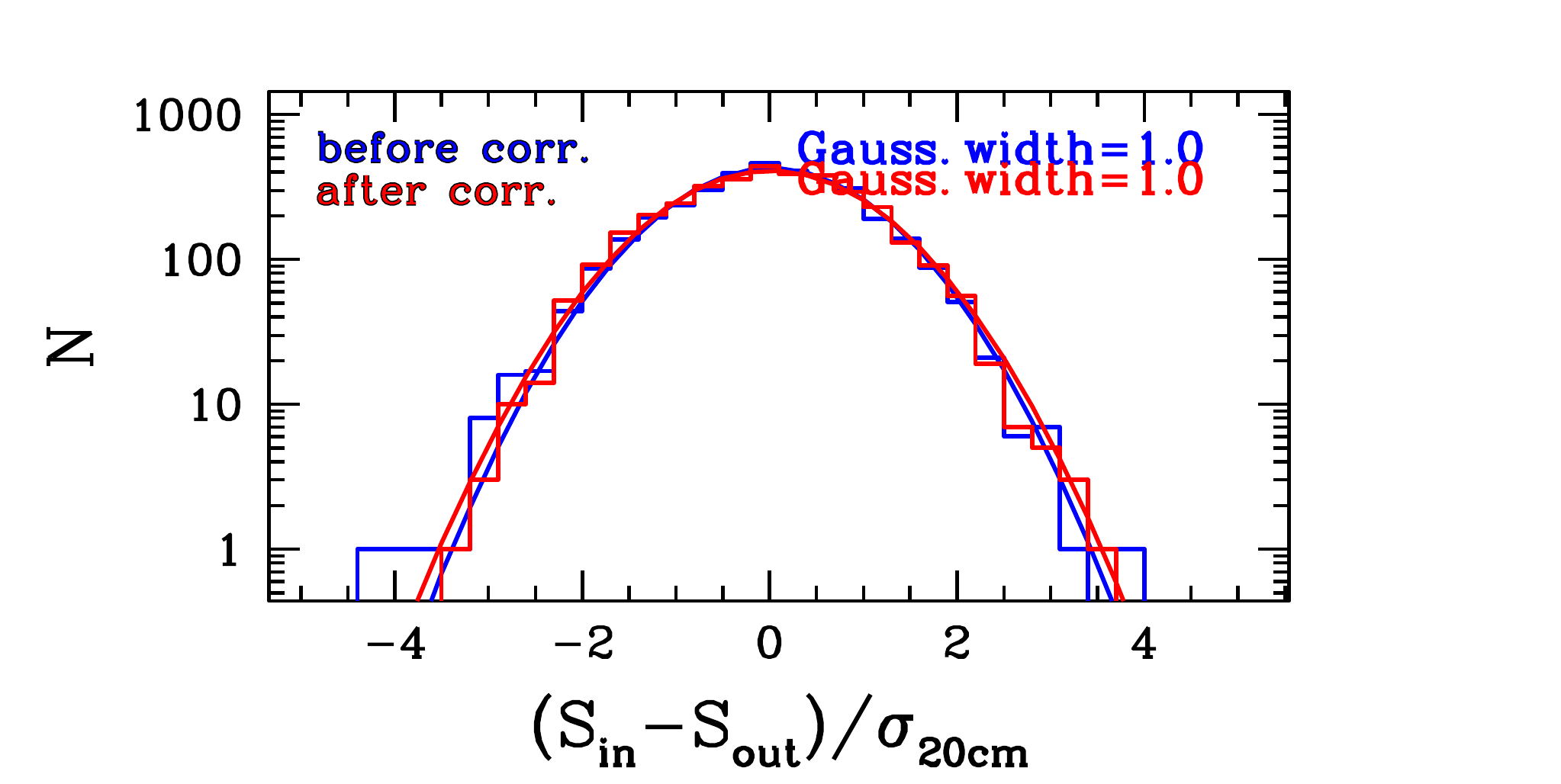}
    \end{subfigure}
    
    
    \begin{subfigure}[b]{\textwidth}\centering
	\includegraphics[width=0.42\textwidth, trim={0.8cm 15cm 0cm 2.5cm}, clip]{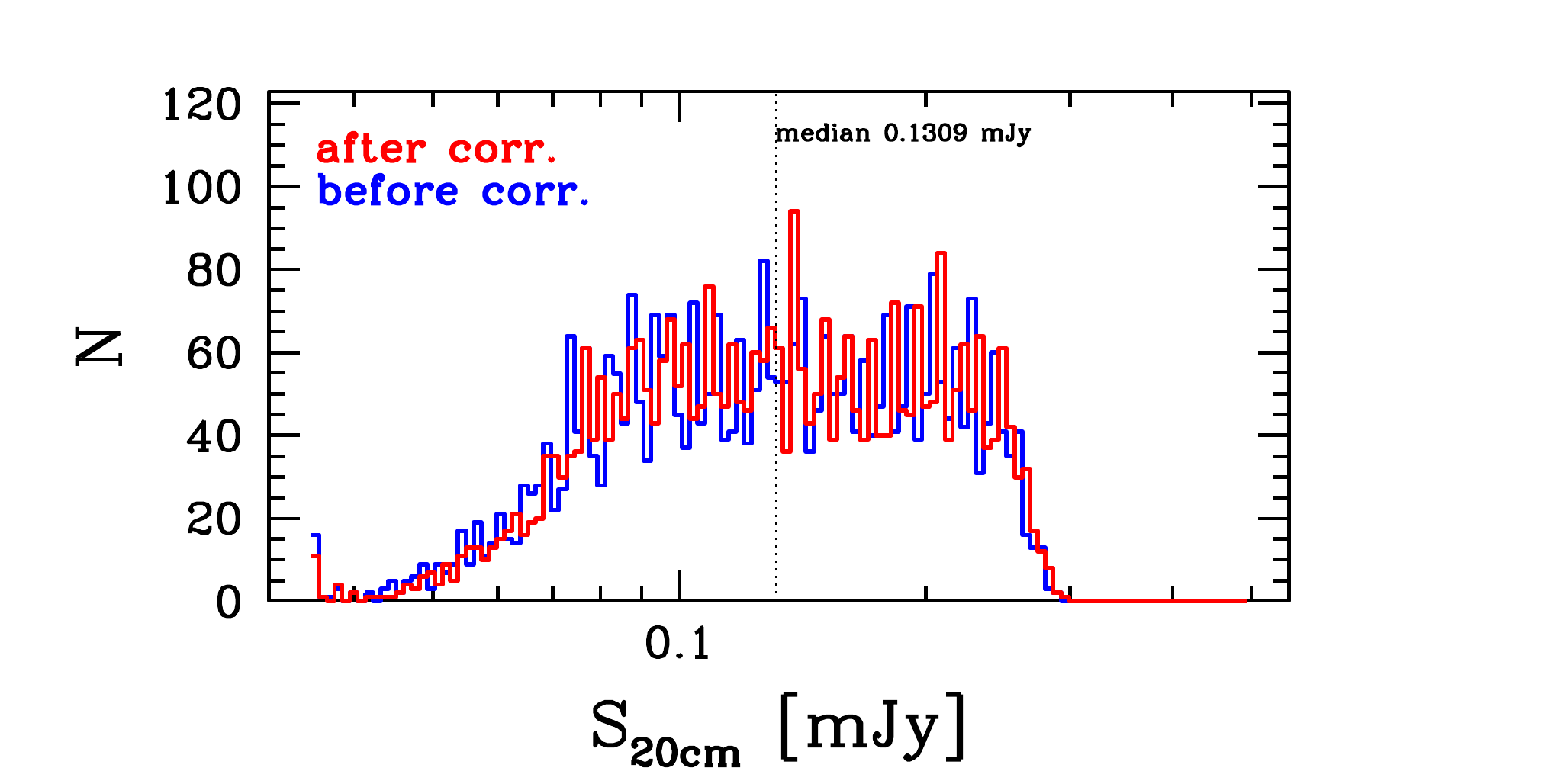}
	\includegraphics[width=0.42\textwidth, trim={0.8cm 15cm 0cm 2.5cm}, clip]{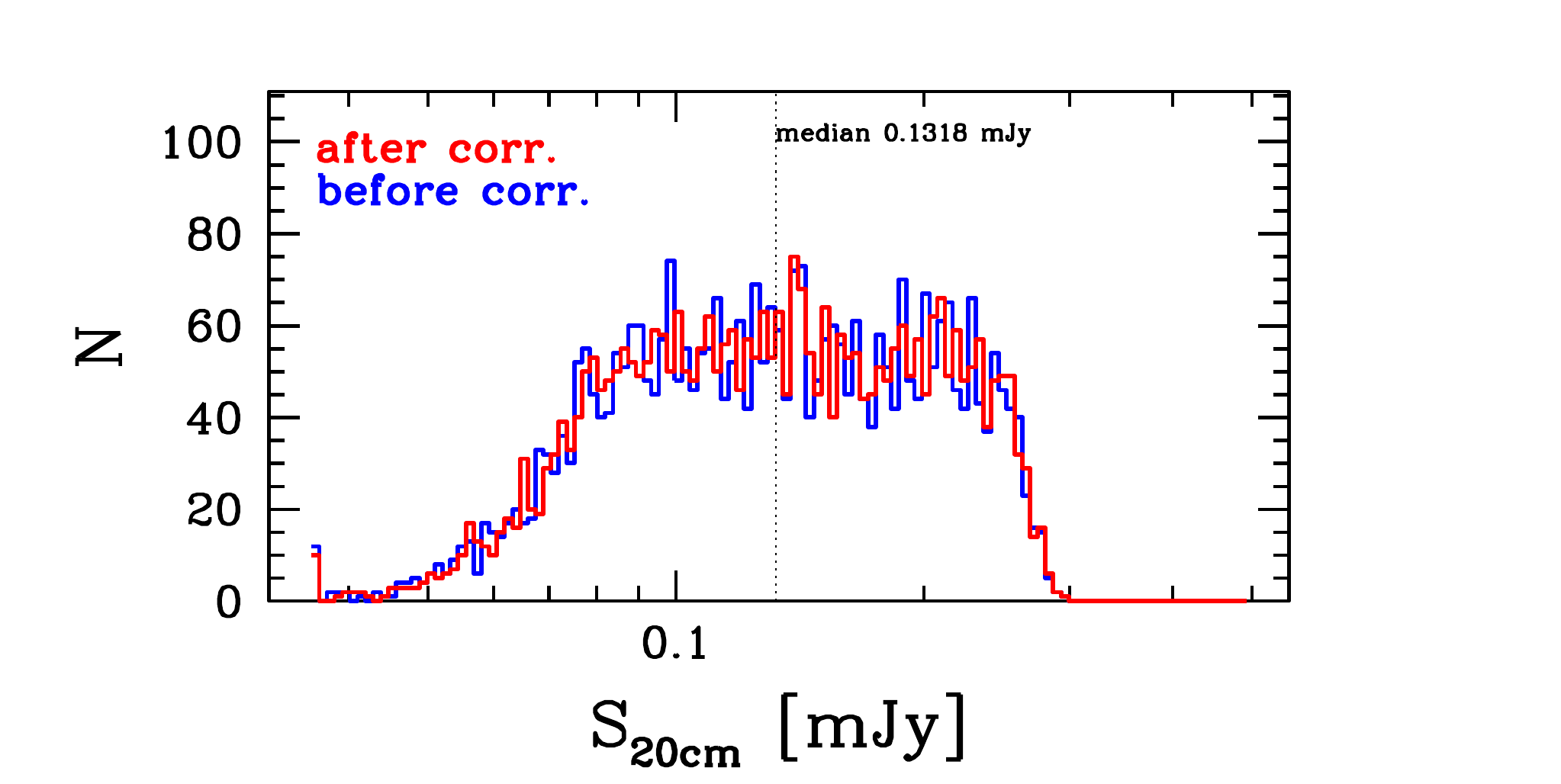}
    \end{subfigure}
    
    \begin{subfigure}[b]{\textwidth}\centering
	\includegraphics[width=0.42\textwidth, trim={0.8cm 15cm 0cm 2.5cm}, clip]{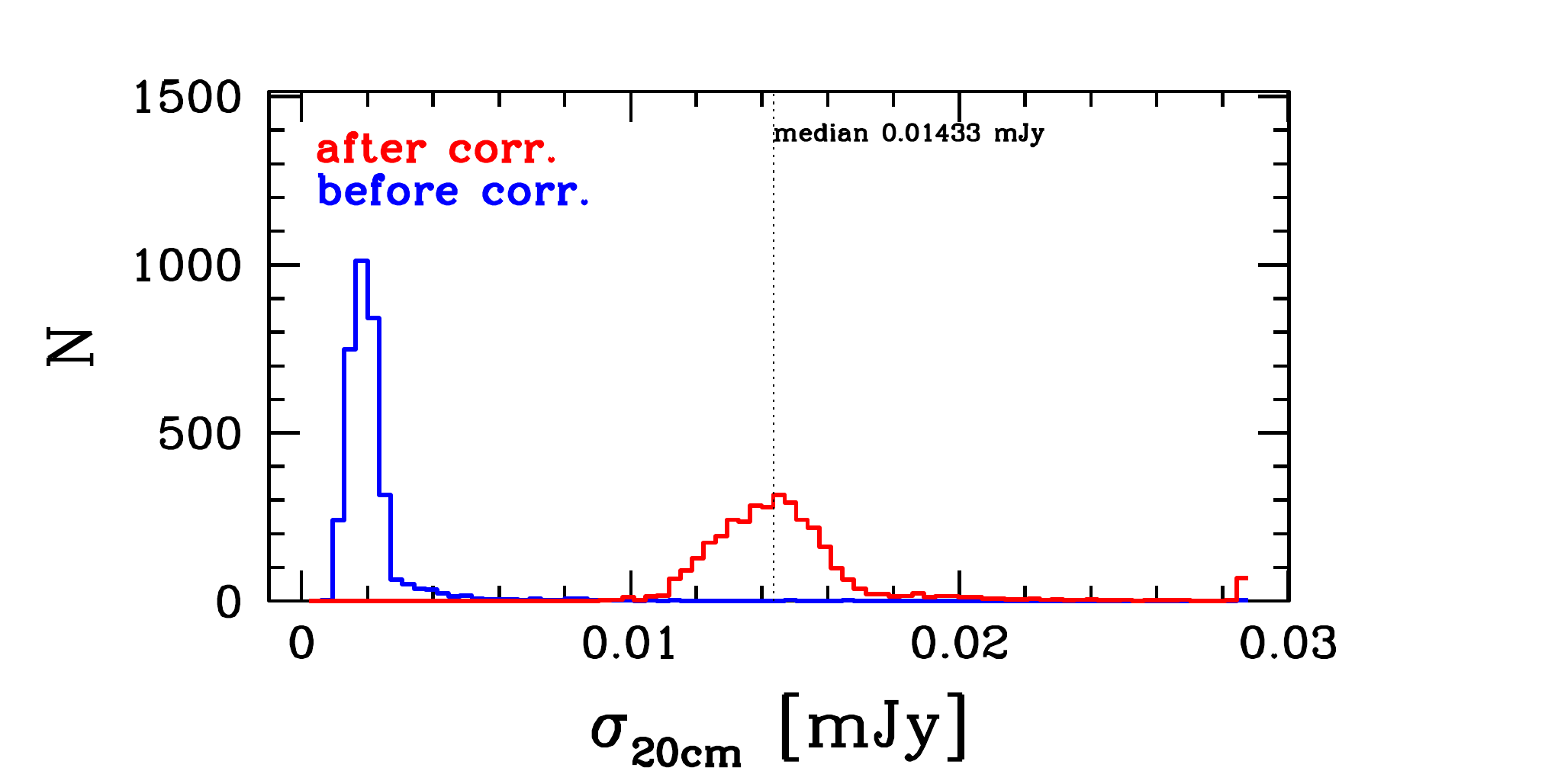}
	\includegraphics[width=0.42\textwidth, trim={0.8cm 15cm 0cm 2.5cm}, clip]{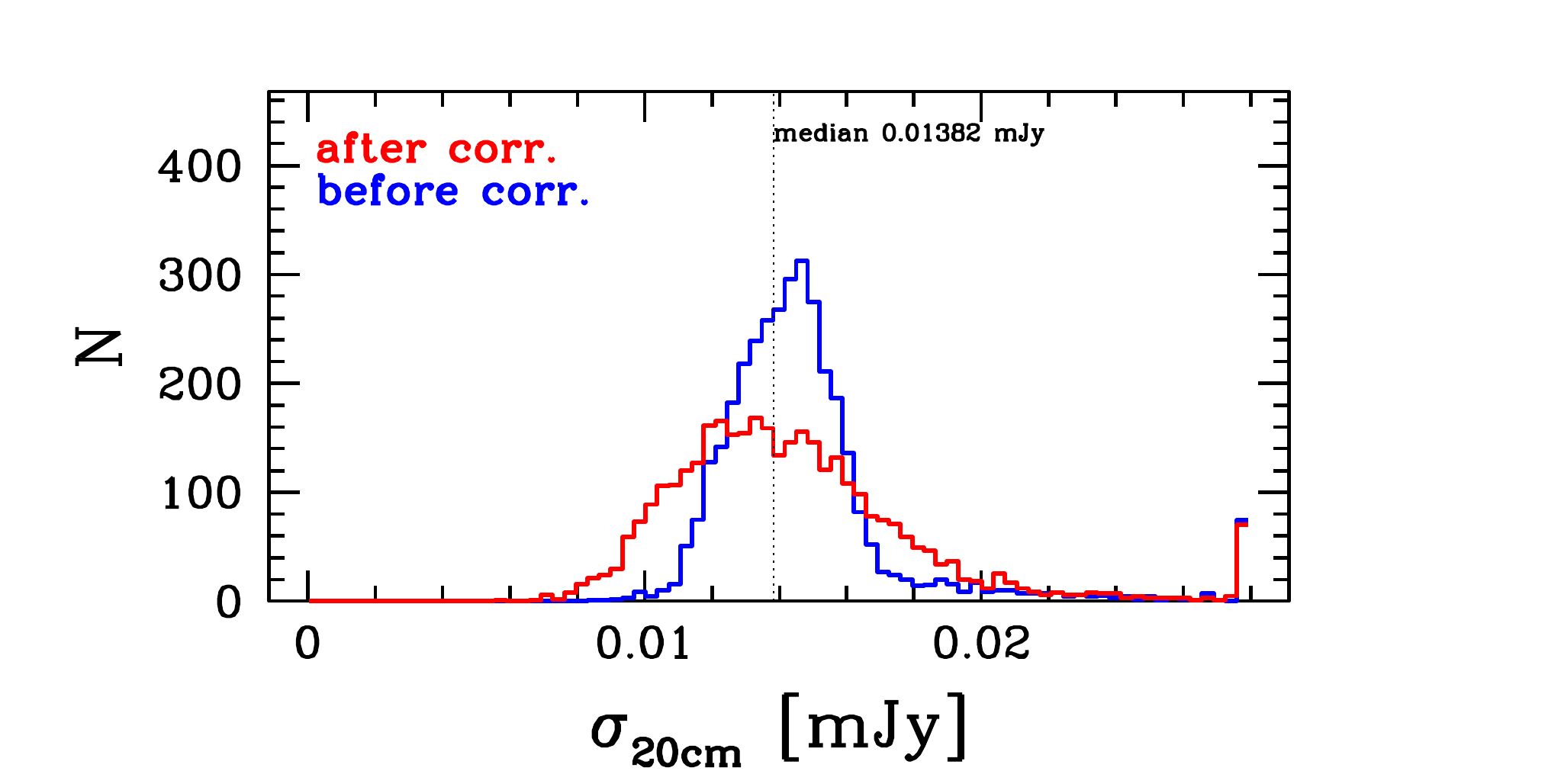}
    \end{subfigure}
    
	\caption{%
		Simulation correction analyses at VLA 1.4~GHz. See descriptions in text. 
	}
    \label{Simu_fig_20cm}

\end{figure}



\begin{figure}
	\centering
    
    \begin{subfigure}[b]{\textwidth}\centering
	\includegraphics[width=0.3\textwidth, trim={1cm 15cm 0cm 2.5cm}, clip]{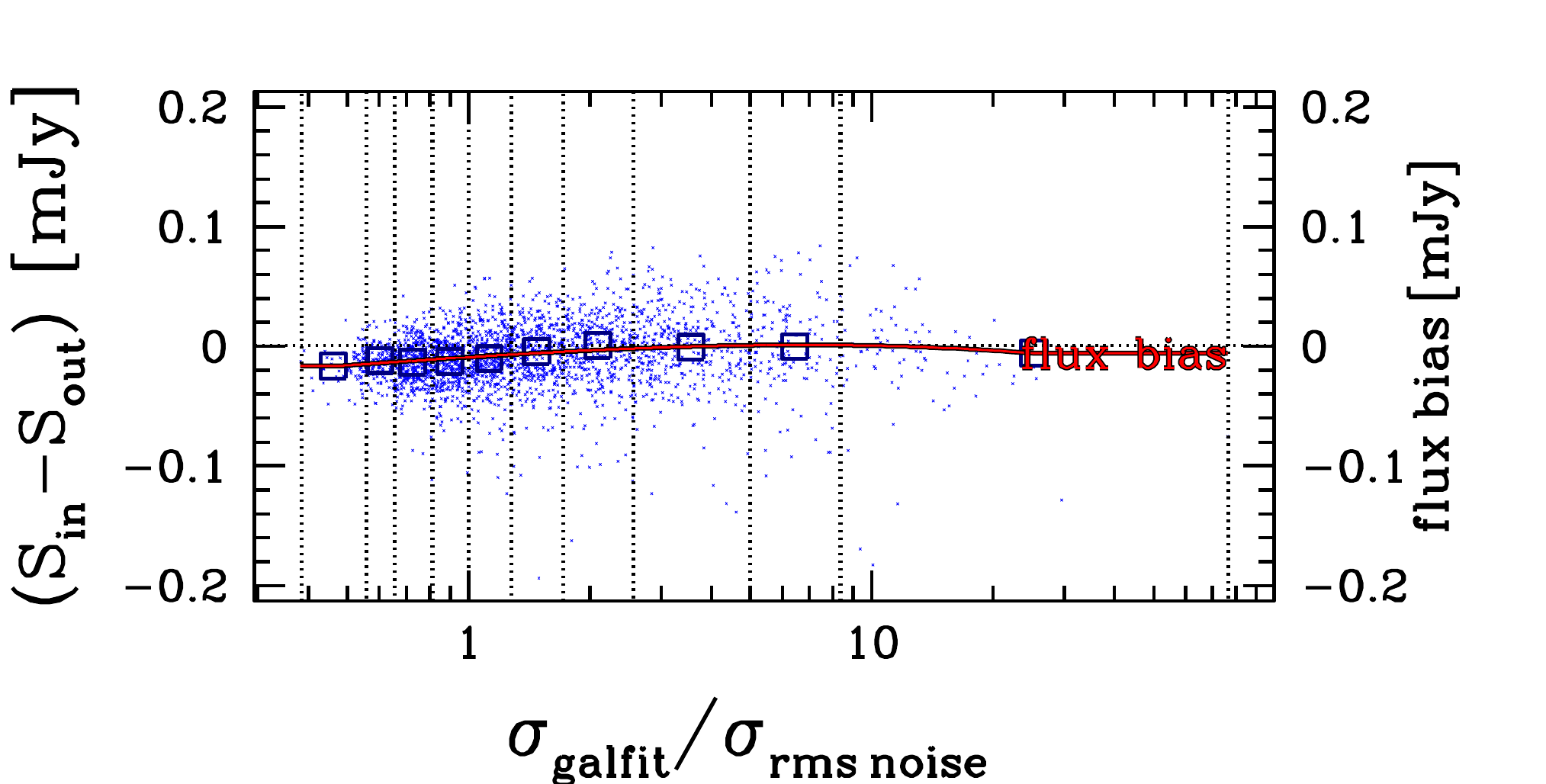}
	\includegraphics[width=0.3\textwidth, trim={1cm 15cm 0cm 2.5cm}, clip]{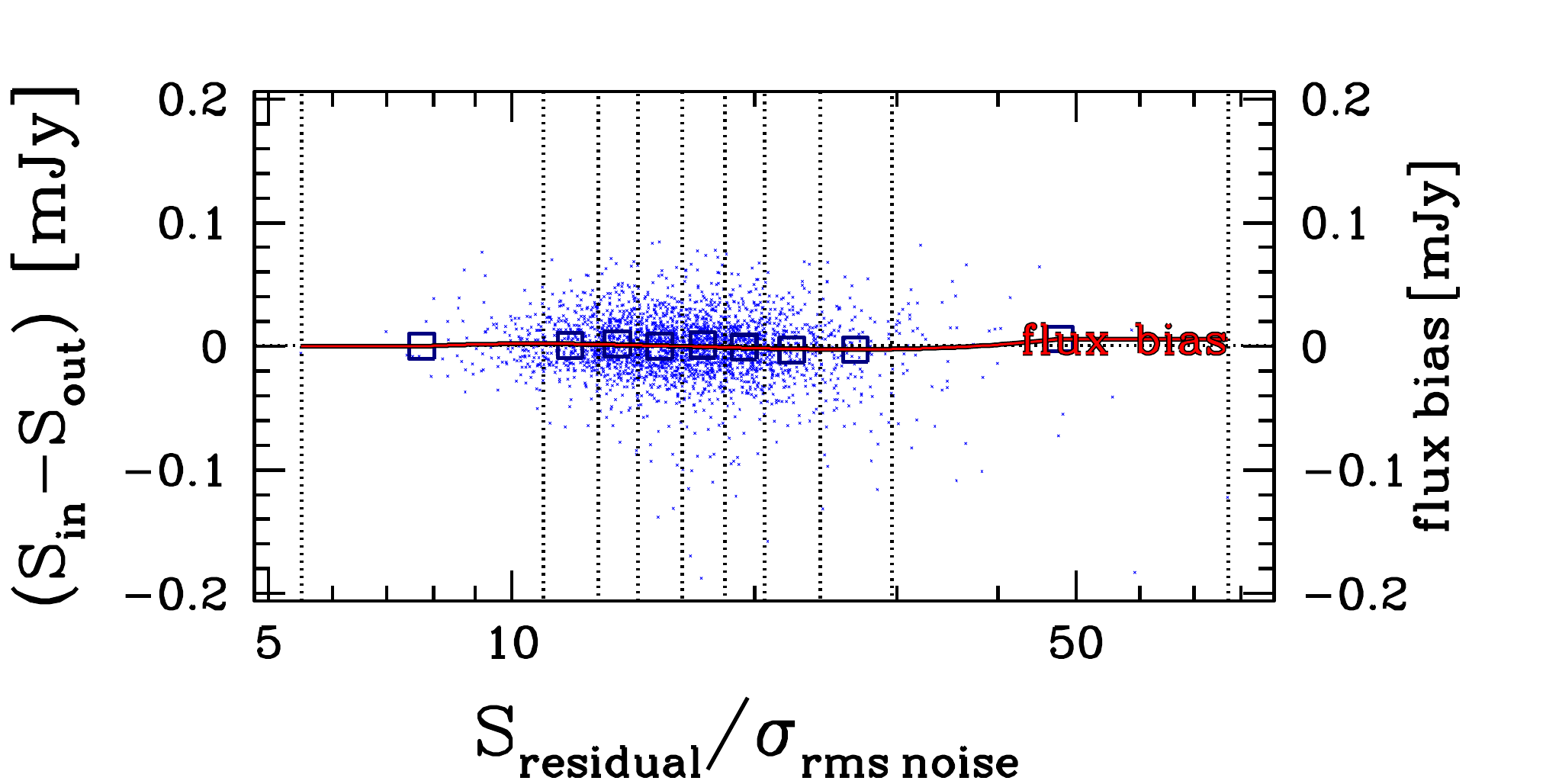}
	\includegraphics[width=0.3\textwidth, trim={1cm 15cm 0cm 2.5cm}, clip]{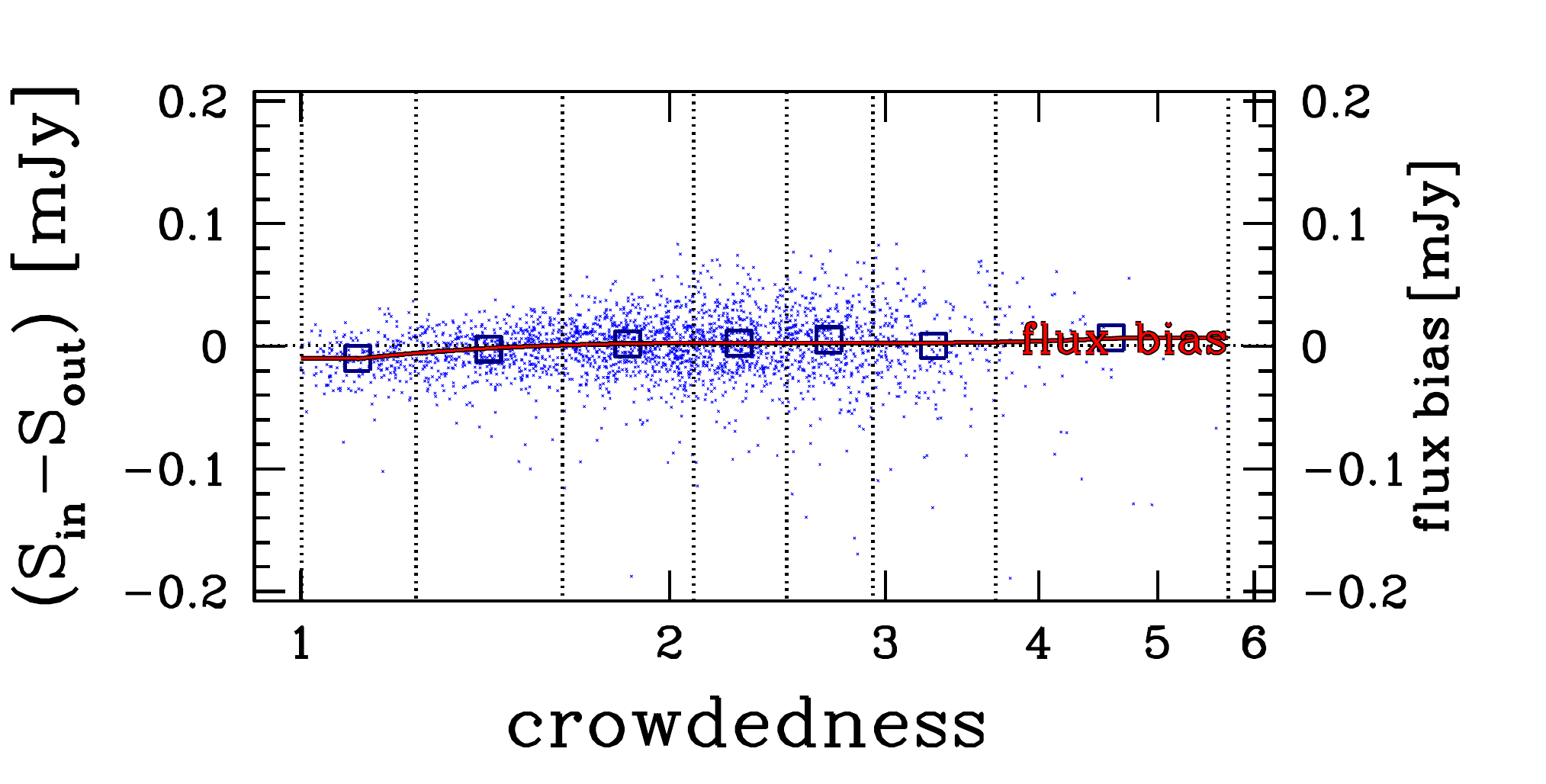}
    \end{subfigure}
    
    \begin{subfigure}[b]{\textwidth}\centering
	\includegraphics[width=0.3\textwidth, trim={1cm 15cm 0cm 2.5cm}, clip]{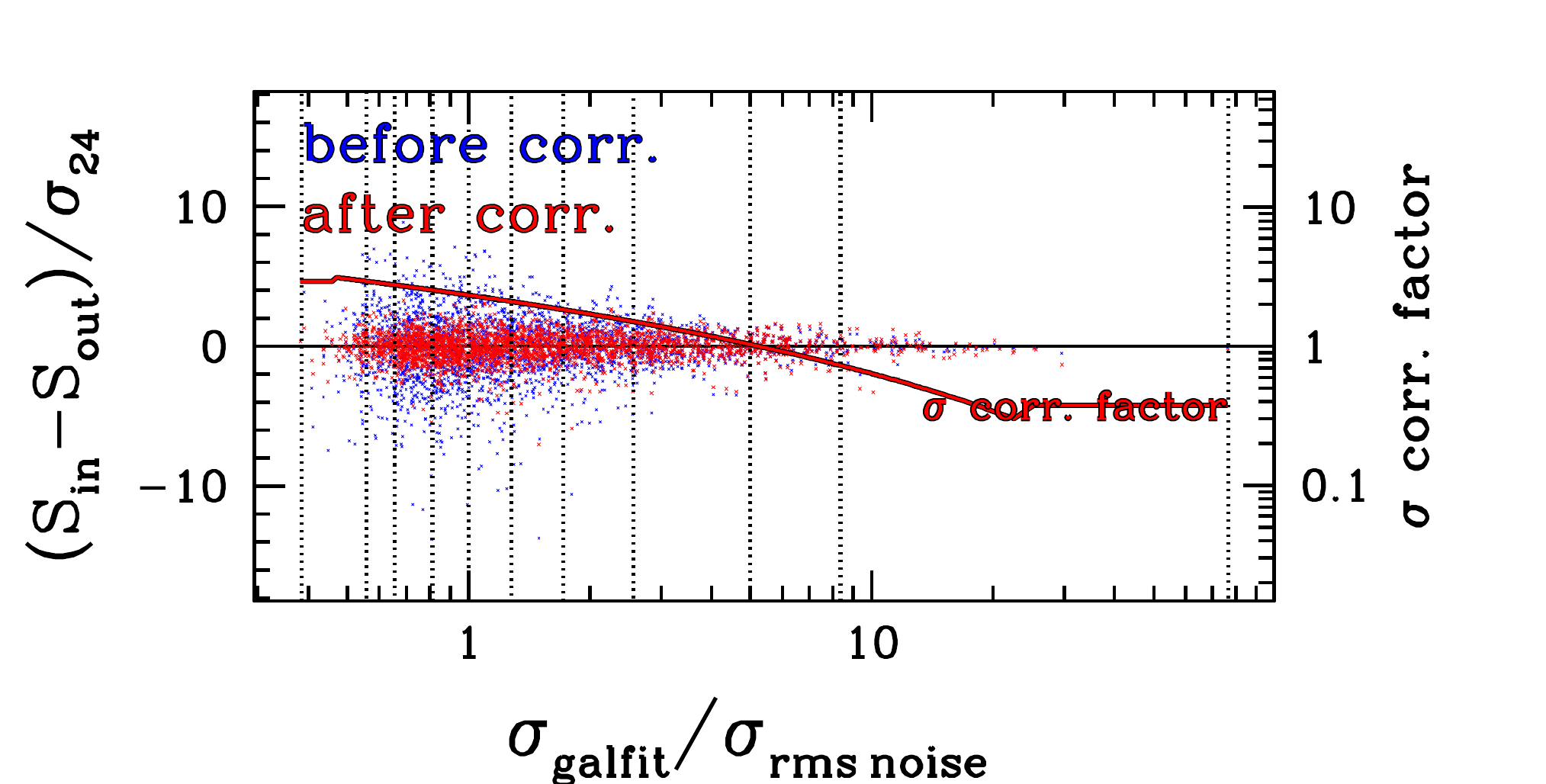}
	\includegraphics[width=0.3\textwidth, trim={1cm 15cm 0cm 2.5cm}, clip]{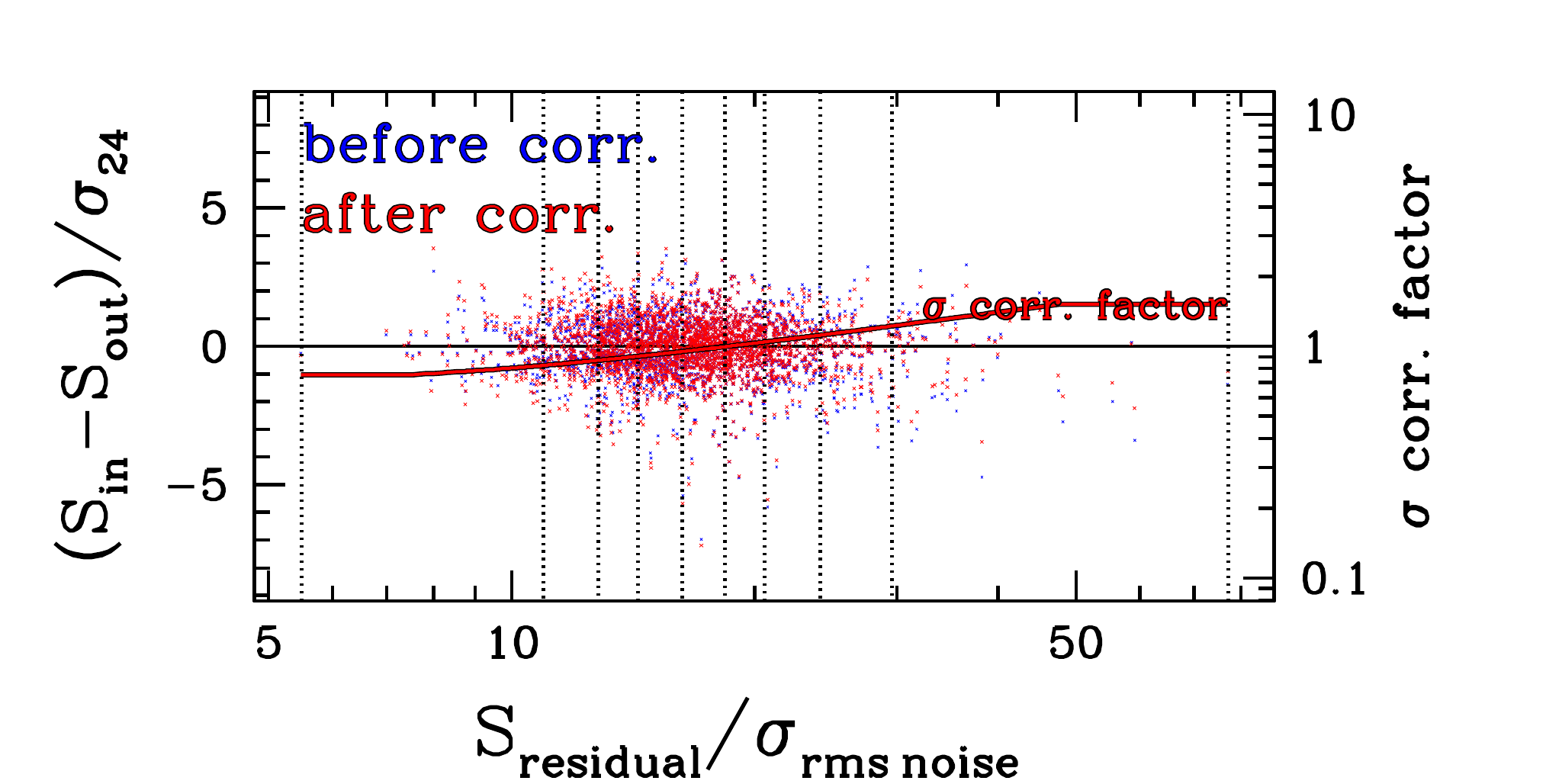}
	\includegraphics[width=0.3\textwidth, trim={1cm 15cm 0cm 2.5cm}, clip]{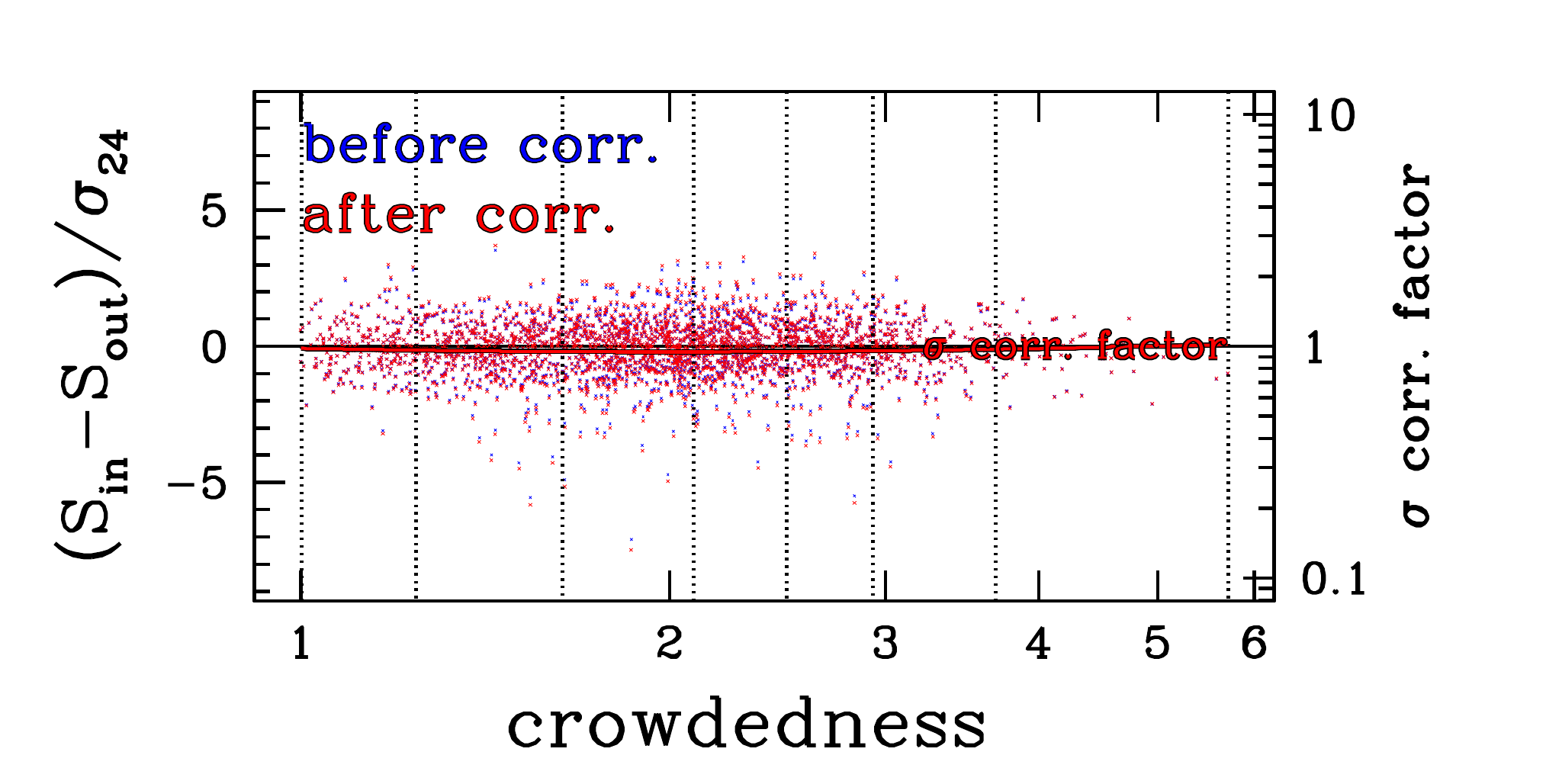}
    \end{subfigure}
    
    \begin{subfigure}[b]{\textwidth}\centering
	\includegraphics[width=0.3\textwidth, trim={1cm 15cm 0cm 2.5cm}, clip]{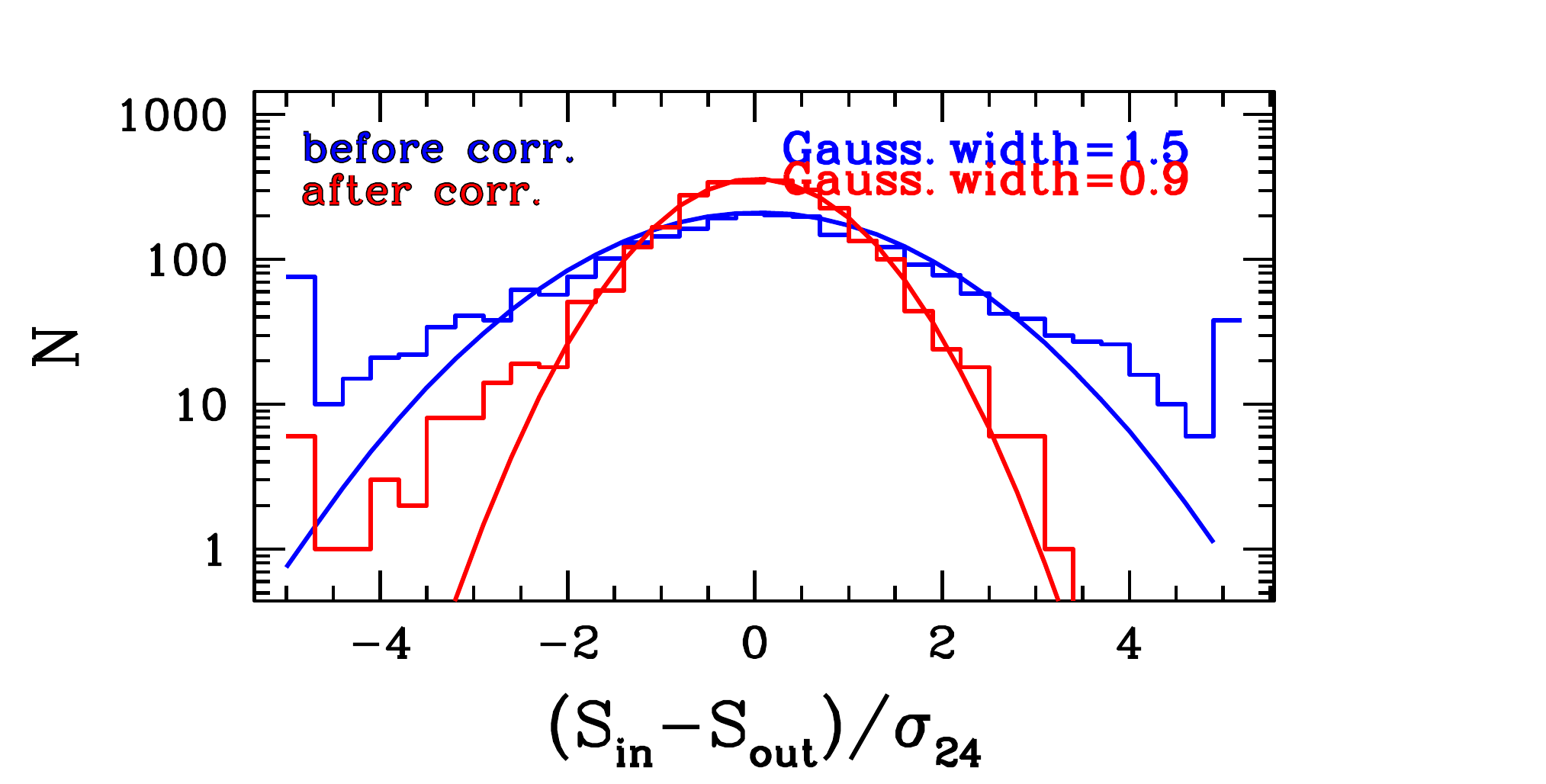}
	\includegraphics[width=0.3\textwidth, trim={1cm 15cm 0cm 2.5cm}, clip]{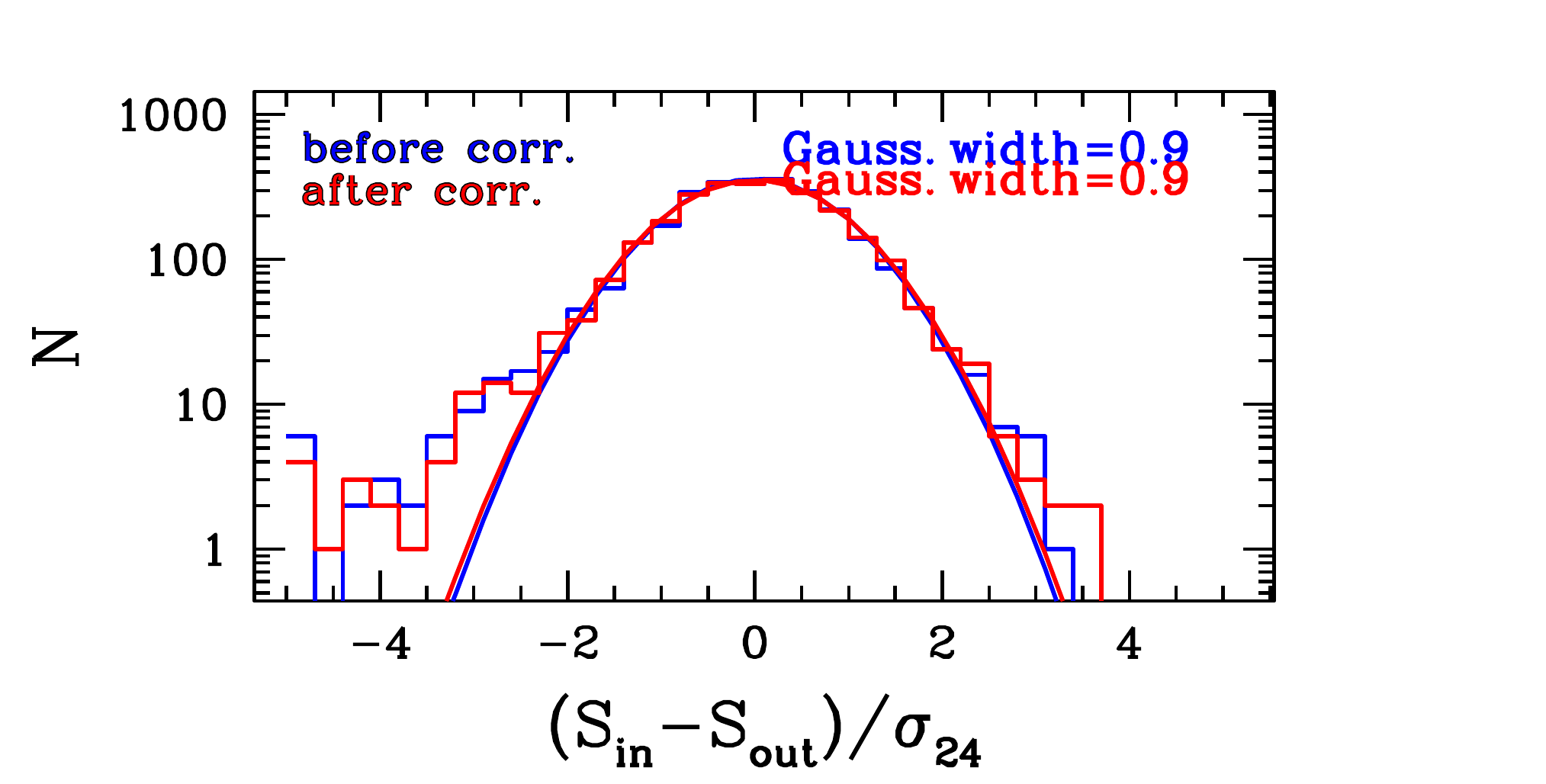}
	\includegraphics[width=0.3\textwidth, trim={1cm 15cm 0cm 2.5cm}, clip]{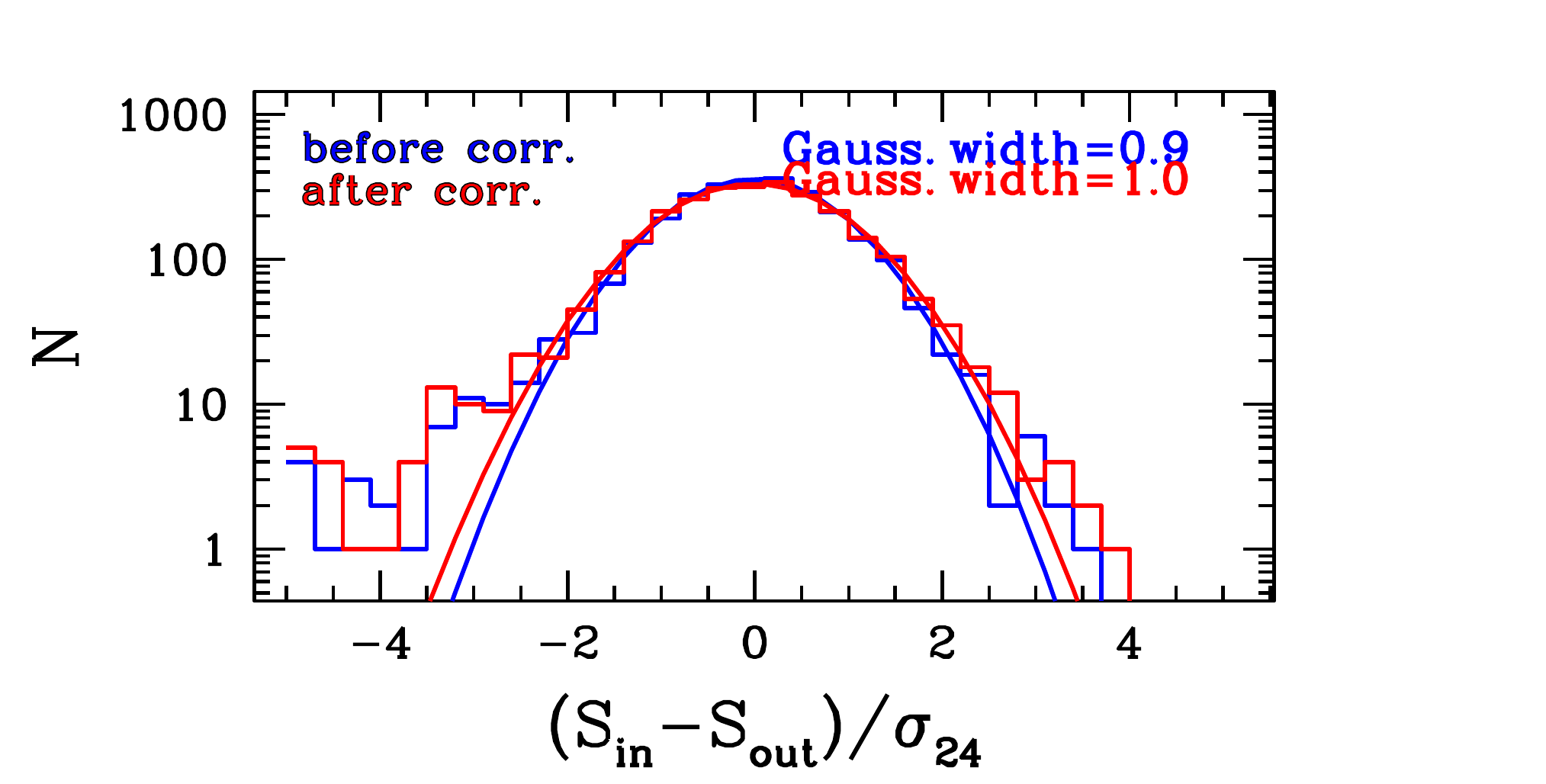}
    \end{subfigure}
    
    \begin{subfigure}[b]{\textwidth}\centering
	\includegraphics[width=0.3\textwidth, trim={1cm 15cm 0cm 2.5cm}, clip]{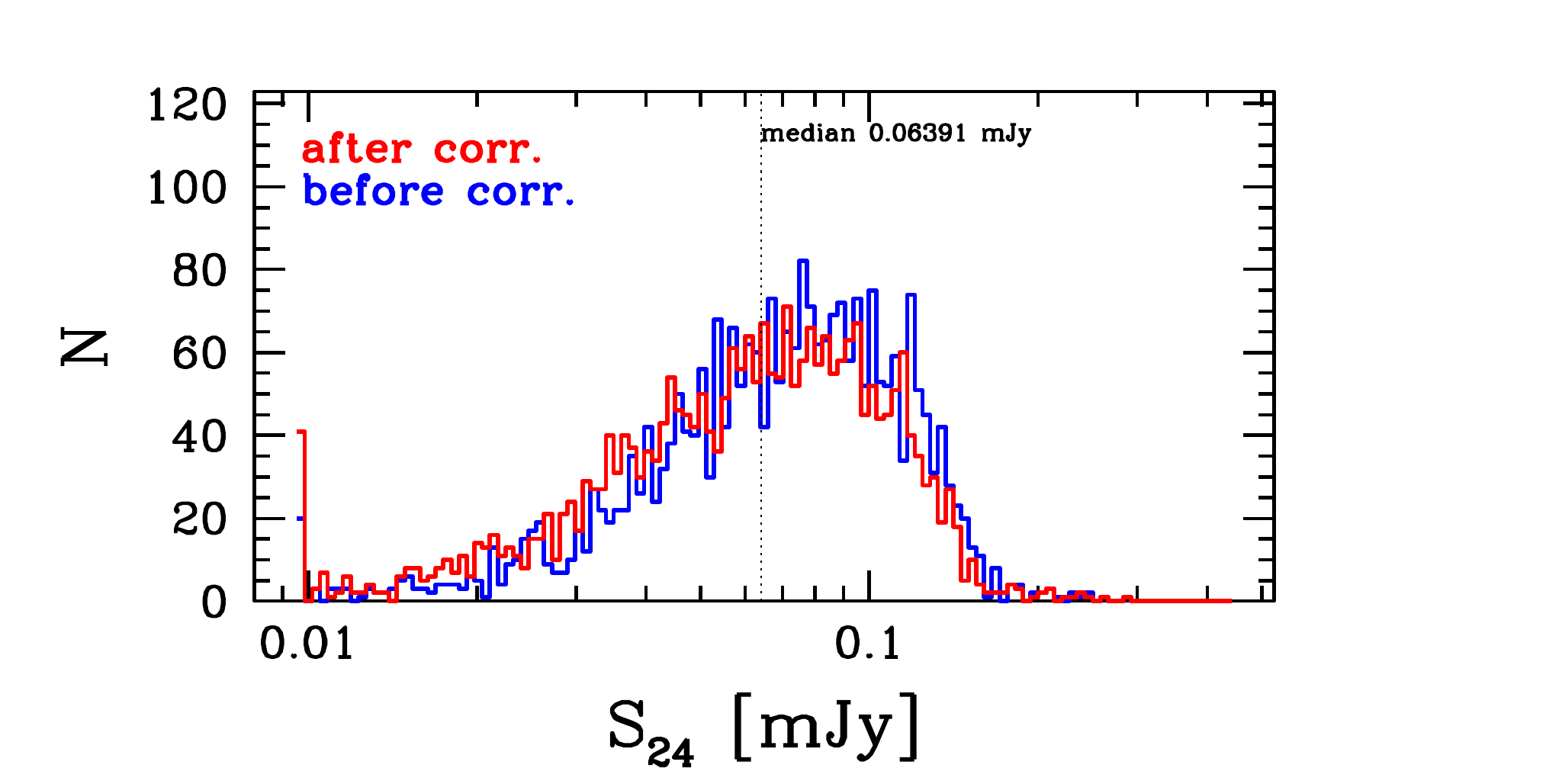}
	\includegraphics[width=0.3\textwidth, trim={1cm 15cm 0cm 2.5cm}, clip]{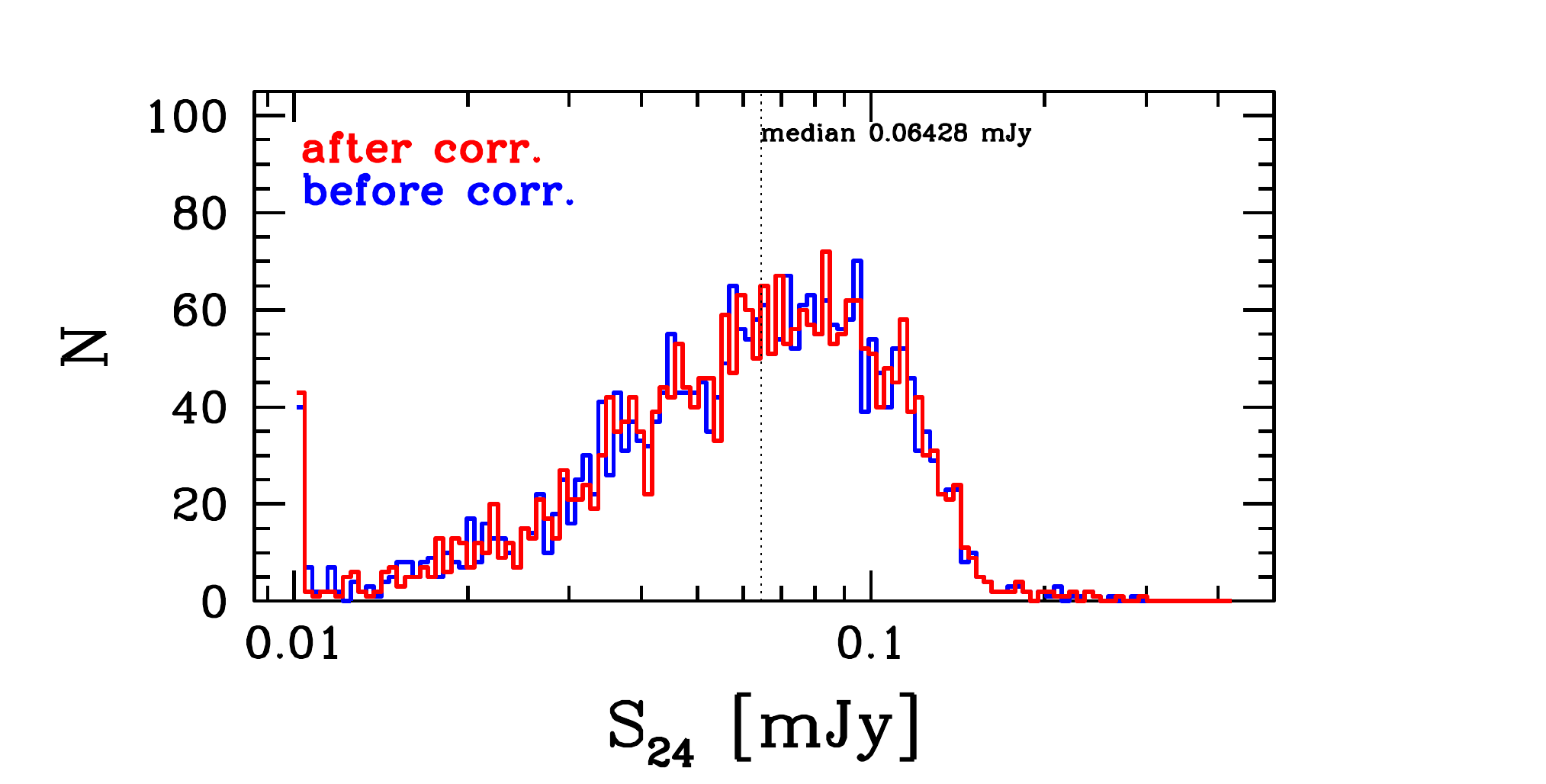}
	\includegraphics[width=0.3\textwidth, trim={1cm 15cm 0cm 2.5cm}, clip]{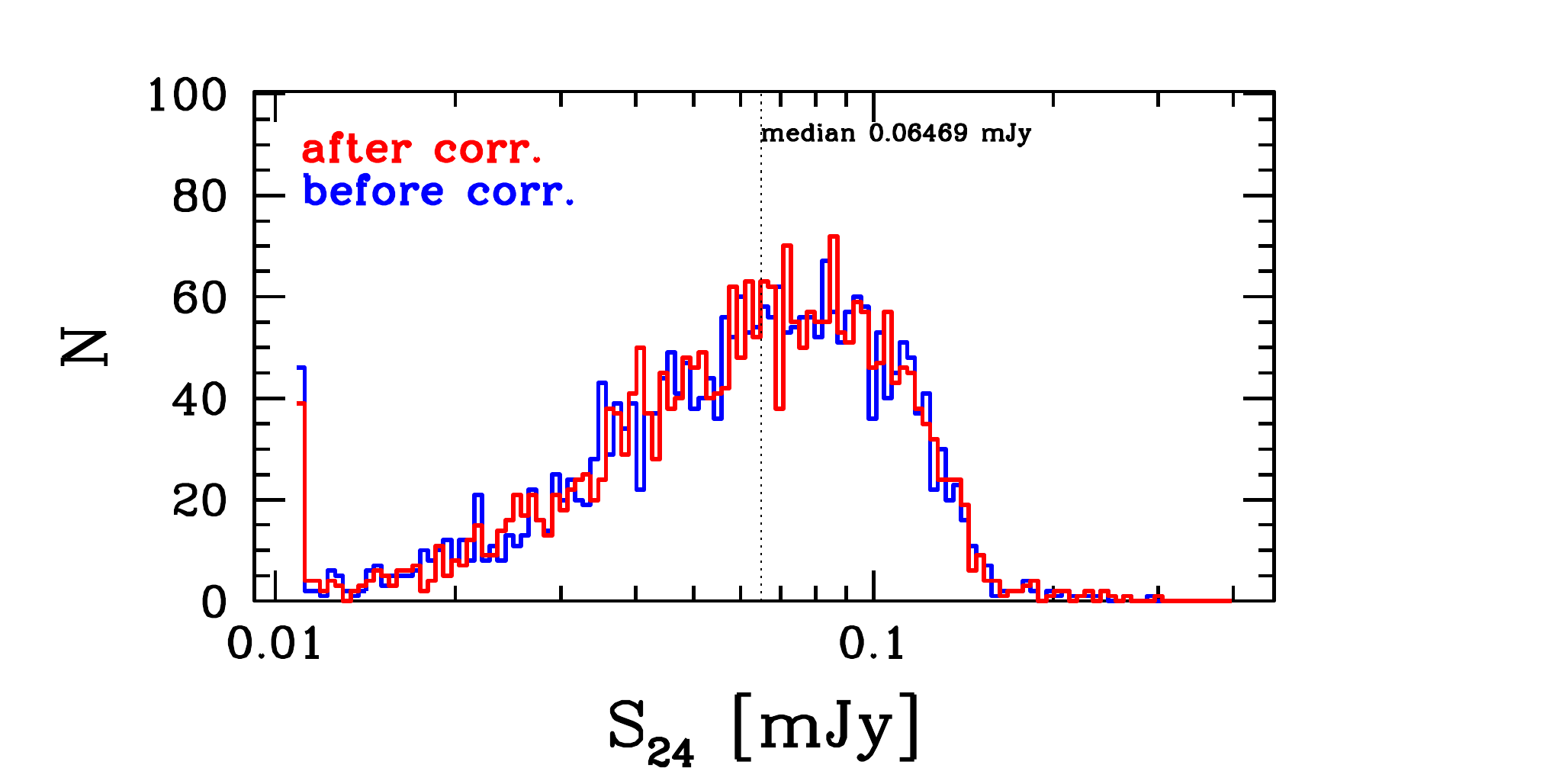}
    \end{subfigure}
    
    \begin{subfigure}[b]{\textwidth}\centering
	\includegraphics[width=0.3\textwidth, trim={1cm 15cm 0cm 2.5cm}, clip]{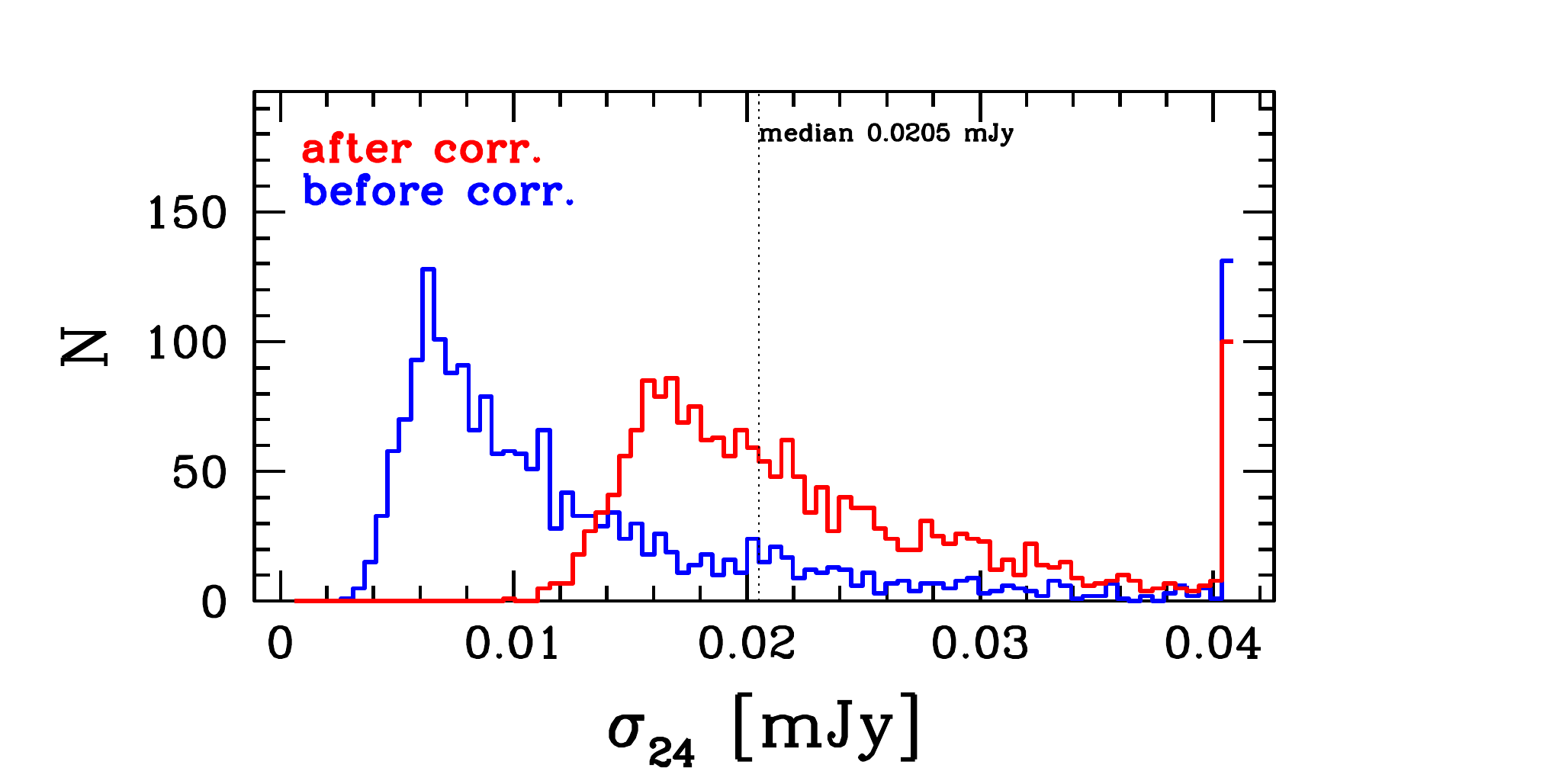}
	\includegraphics[width=0.3\textwidth, trim={1cm 15cm 0cm 2.5cm}, clip]{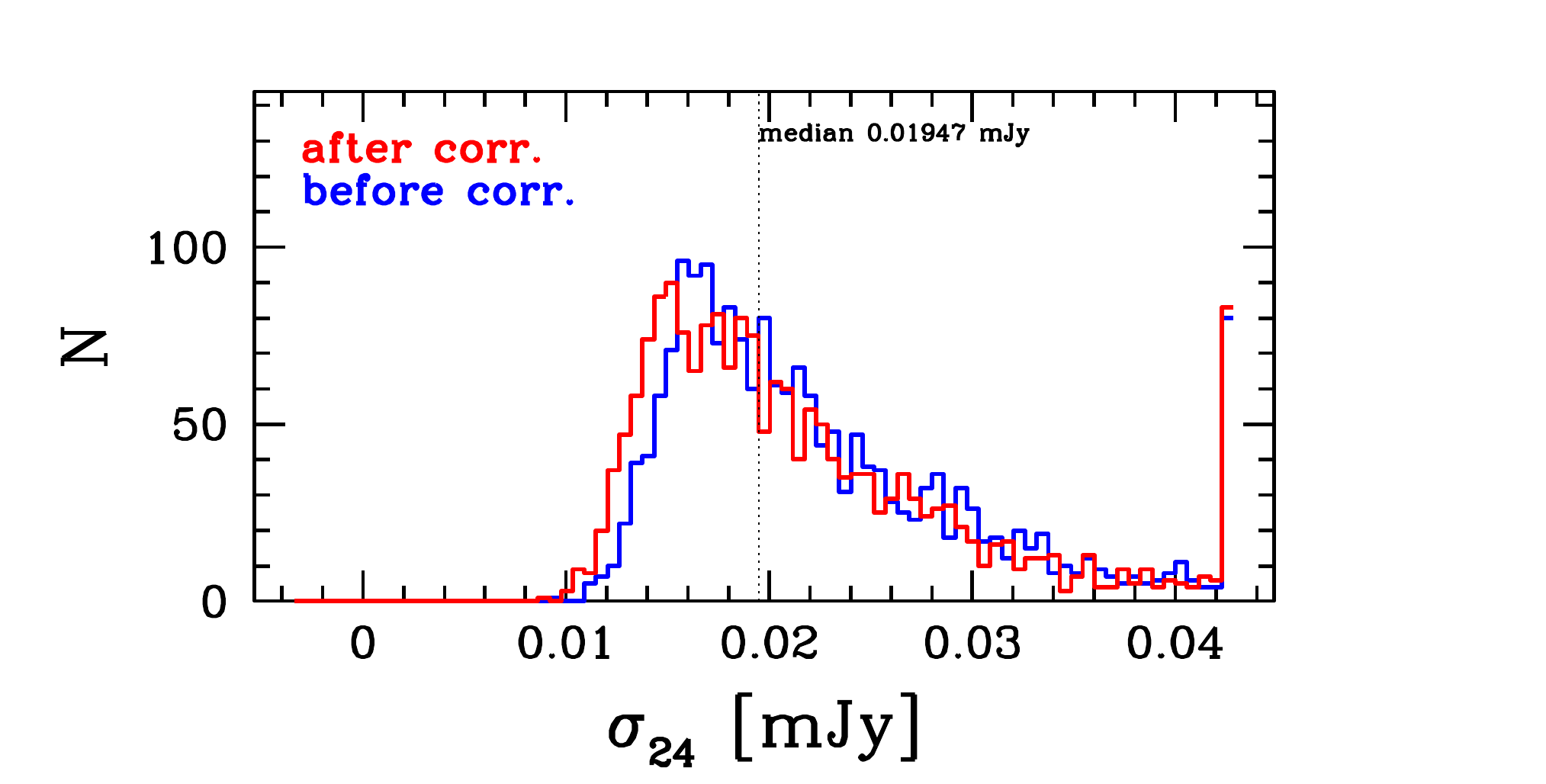}
	\includegraphics[width=0.3\textwidth, trim={1cm 15cm 0cm 2.5cm}, clip]{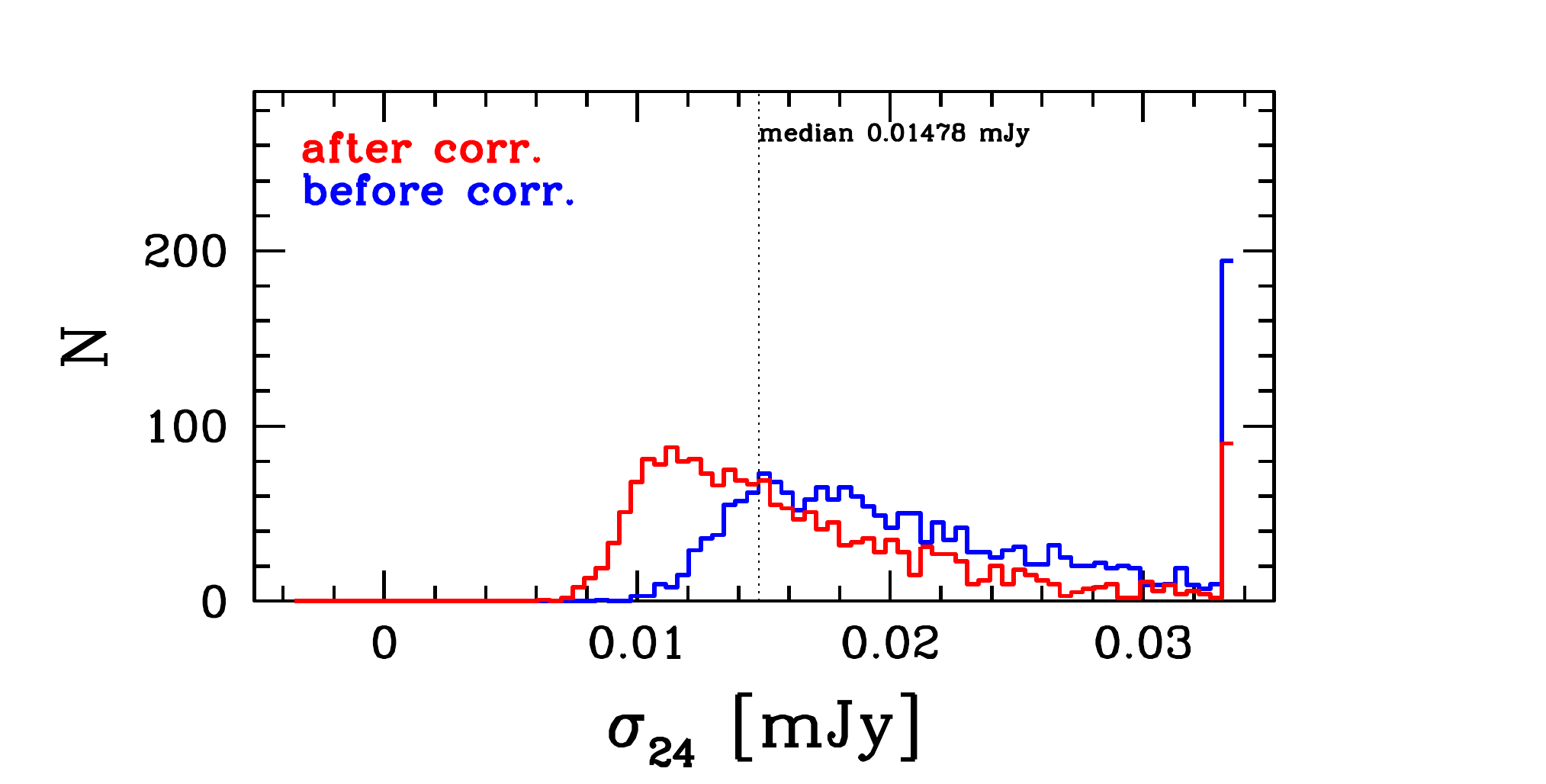}
    \end{subfigure}
    
\caption{%
	Simulation correction analyses at MIPS 24~$\mu$m. See descriptions in text. 
}
\label{Simu_fig_24um}
\end{figure}

\clearpage

\section{D. High redshift candidates}
\label{Section_highz}

We present cutouts and SEDs of high redshift candidates here. For each candidate, we show cutouts of UltraVISTA $K_s$, SPLASH 3.6~$\mu$m \& 4.5~$\mu$m, VLA 3~GHz, MIPS 24~$\mu$m, $Herschel$, SCUBA2 and MAMBO images on the left, and accompany its SED on the right. 
The instrument, wavelength (in unit of $\mu$m) and field of view (FoV) are shown in green text in each cutout. 
{The scheme of the symbols in the SED panels are identical to those in Fig.~\ref{example_SED}}.

\begin{figure}
	\centering
	\includegraphics[width=0.28\textwidth]{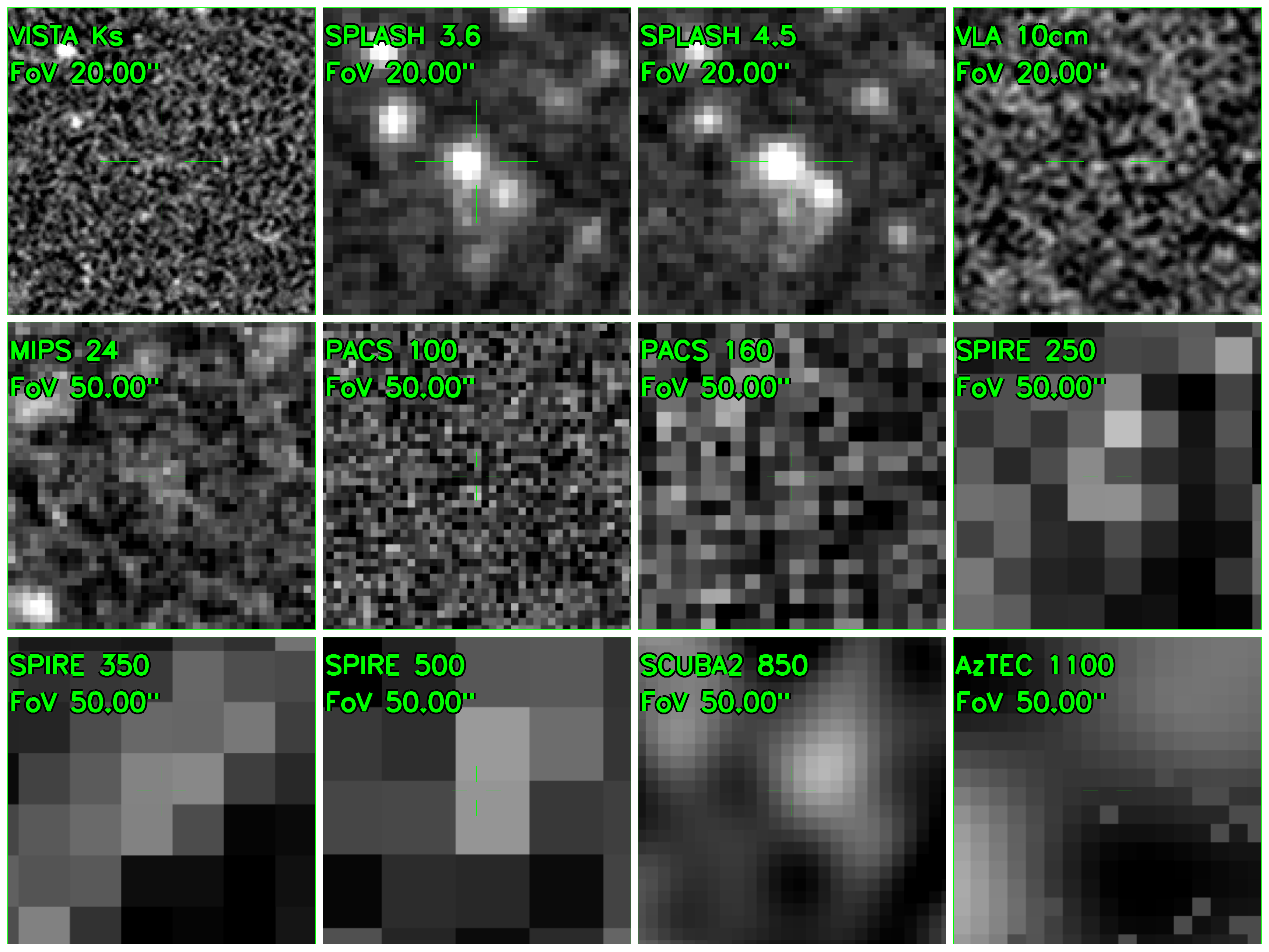}
	\includegraphics[width=0.21\textwidth, trim={0.6cm 5cm 1cm 3.5cm}, clip]{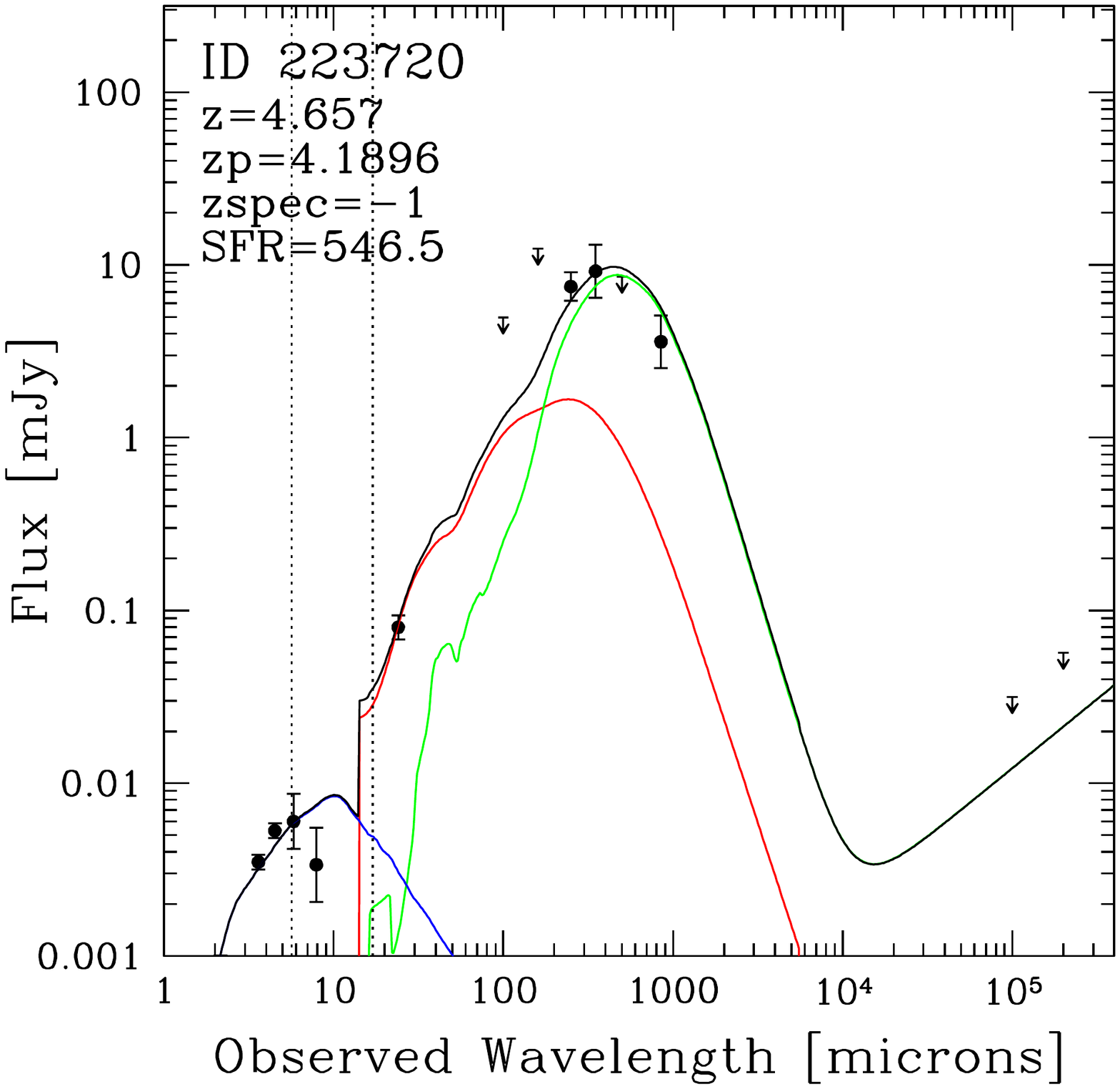}
    \includegraphics[width=0.28\textwidth]{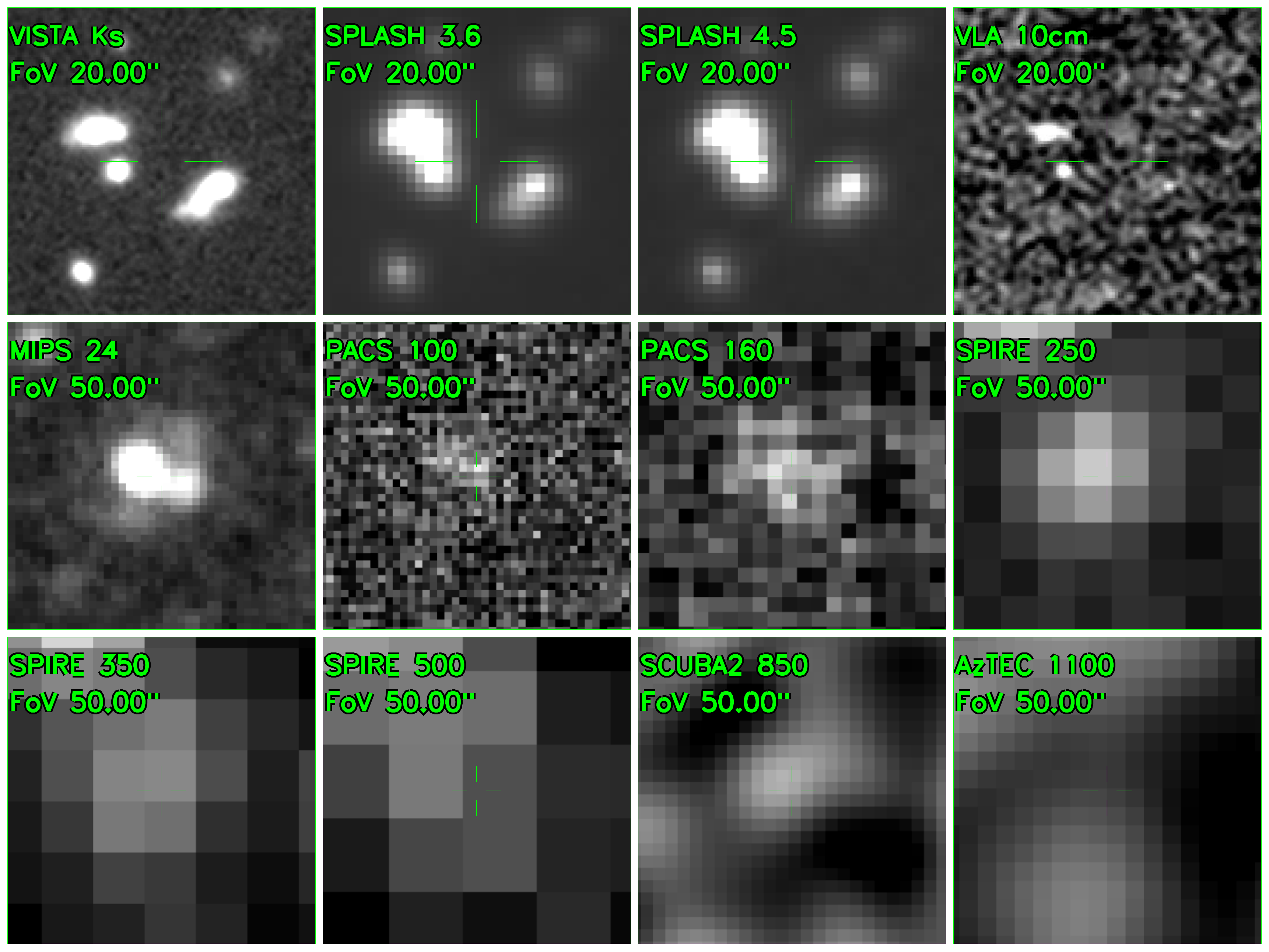}
	\includegraphics[width=0.21\textwidth, trim={0.6cm 5cm 1cm 3.5cm}, clip]{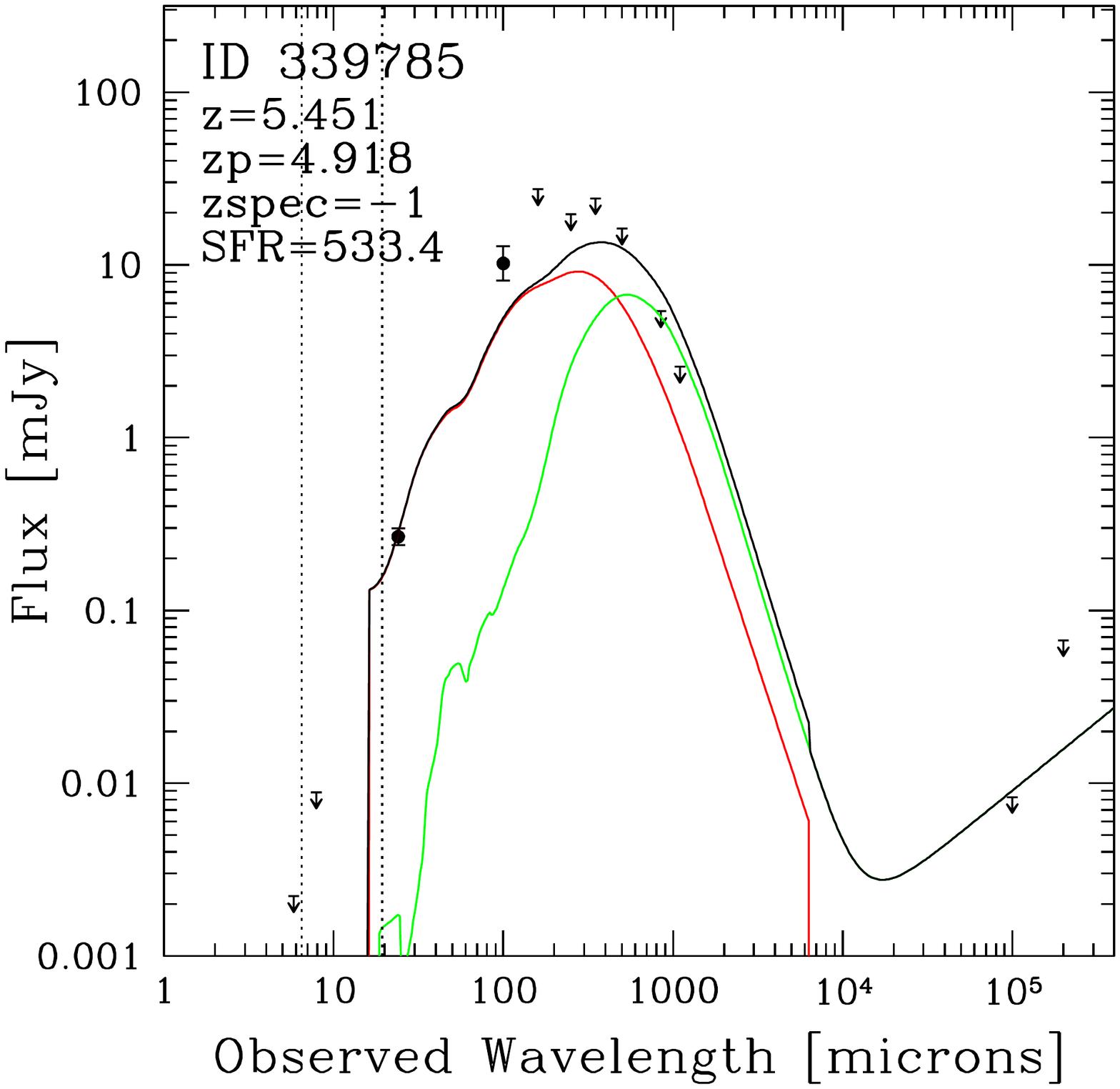}
	\includegraphics[width=0.28\textwidth]{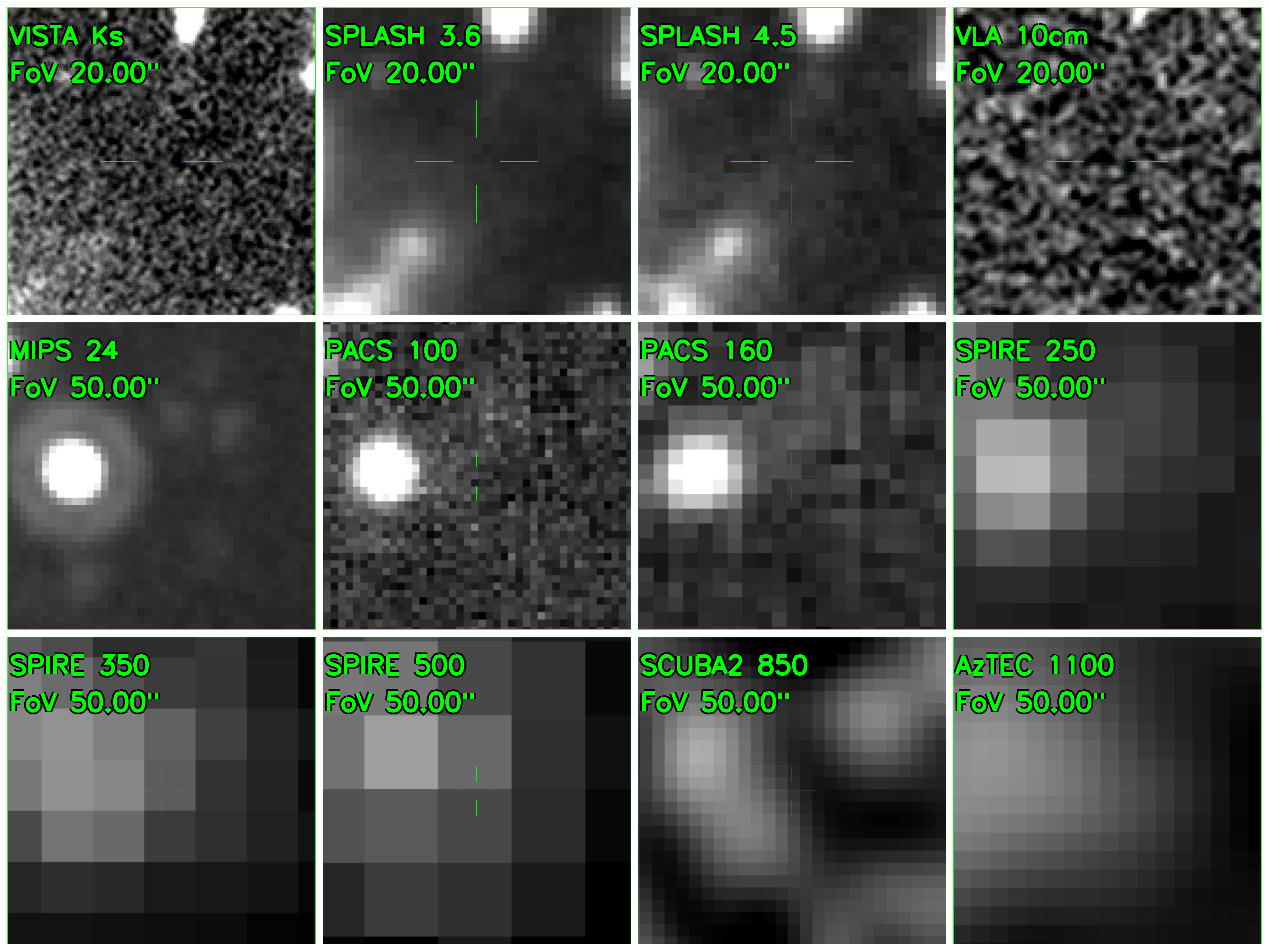}
	\includegraphics[width=0.21\textwidth, trim={0.6cm 5cm 1cm 3.5cm}, clip]{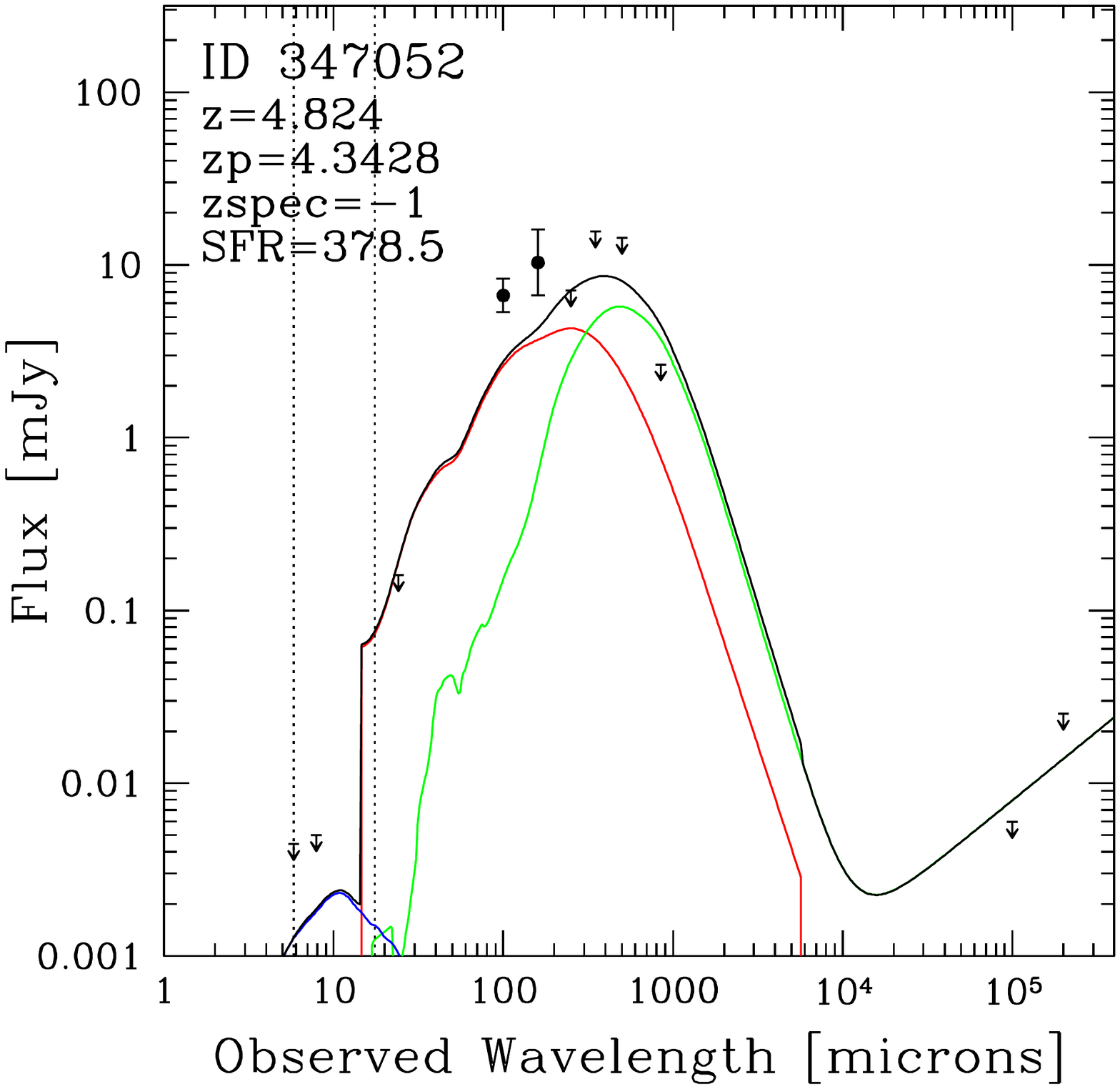}
    \includegraphics[width=0.28\textwidth]{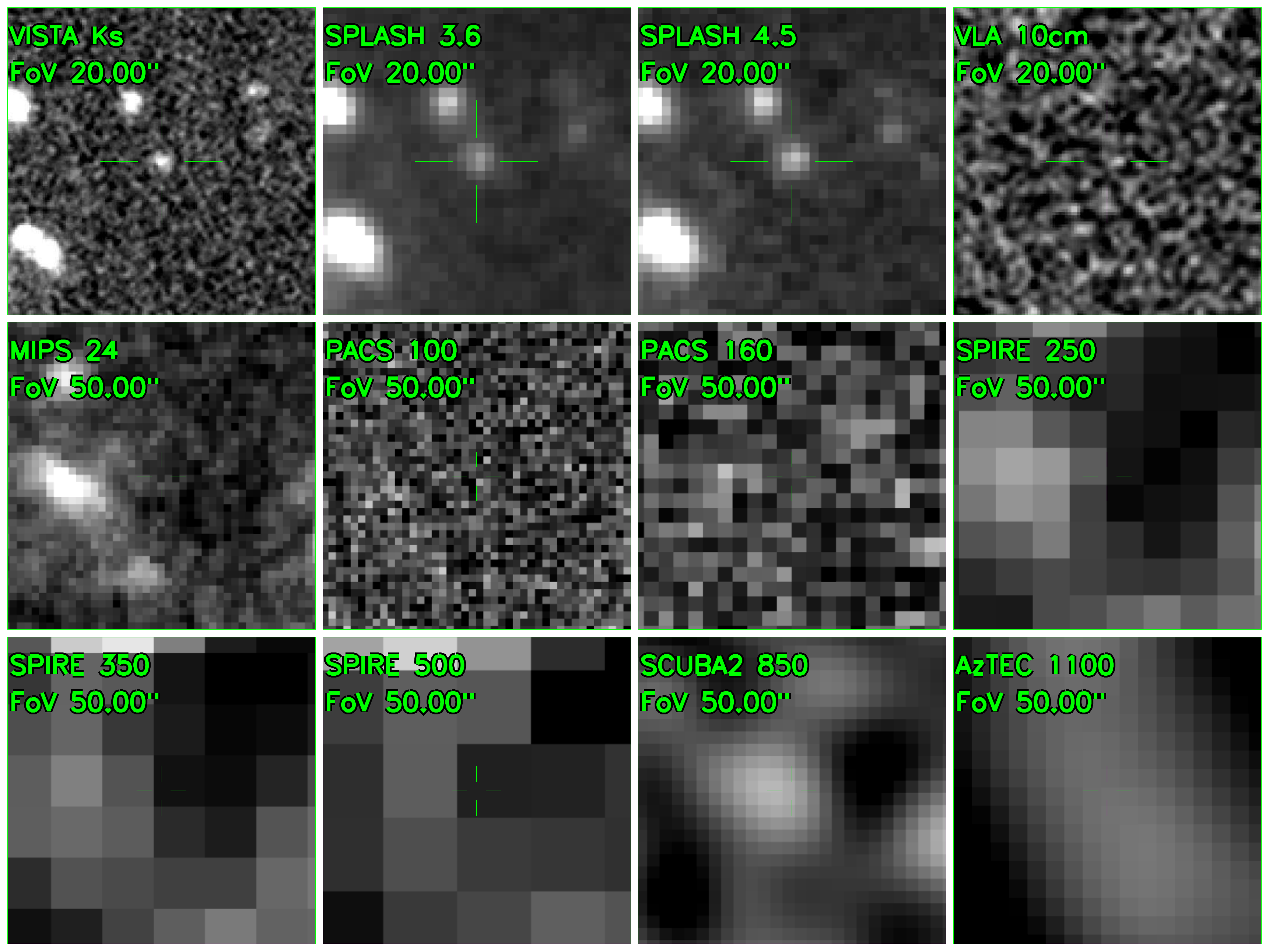}
	\includegraphics[width=0.21\textwidth, trim={0.6cm 5cm 1cm 3.5cm}, clip]{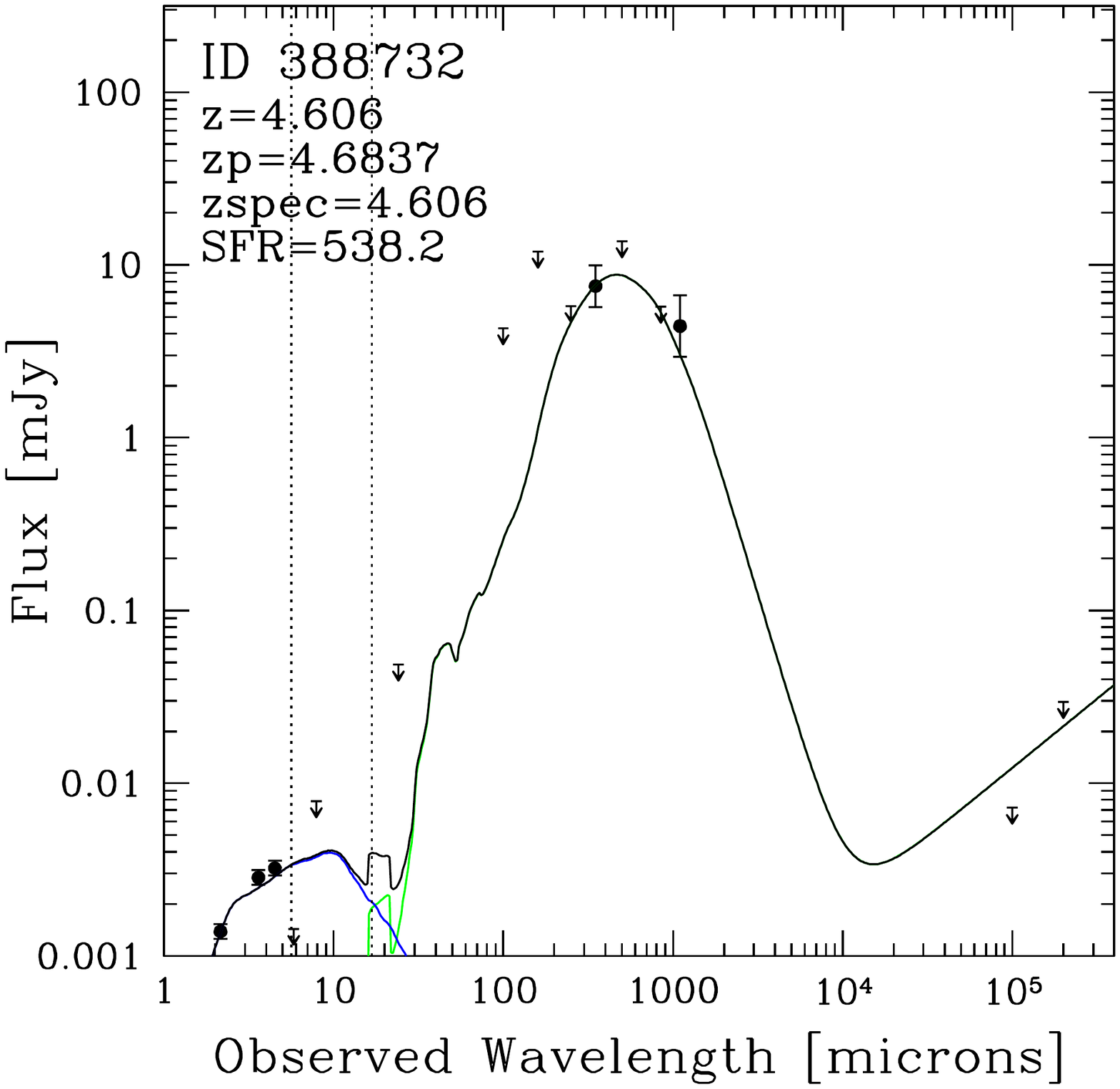}
    \includegraphics[width=0.28\textwidth]{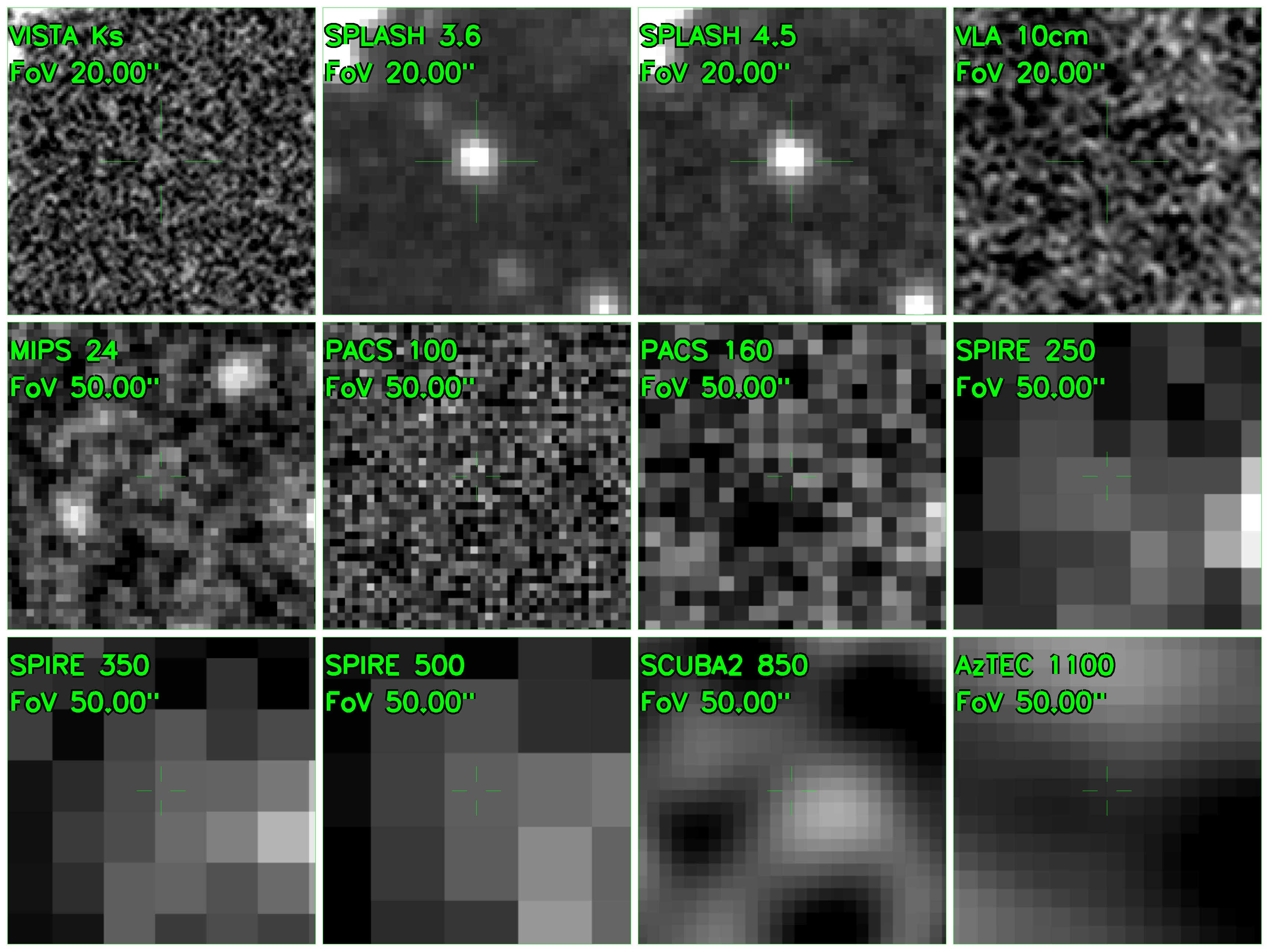}
	\includegraphics[width=0.21\textwidth, trim={0.6cm 5cm 1cm 3.5cm}, clip]{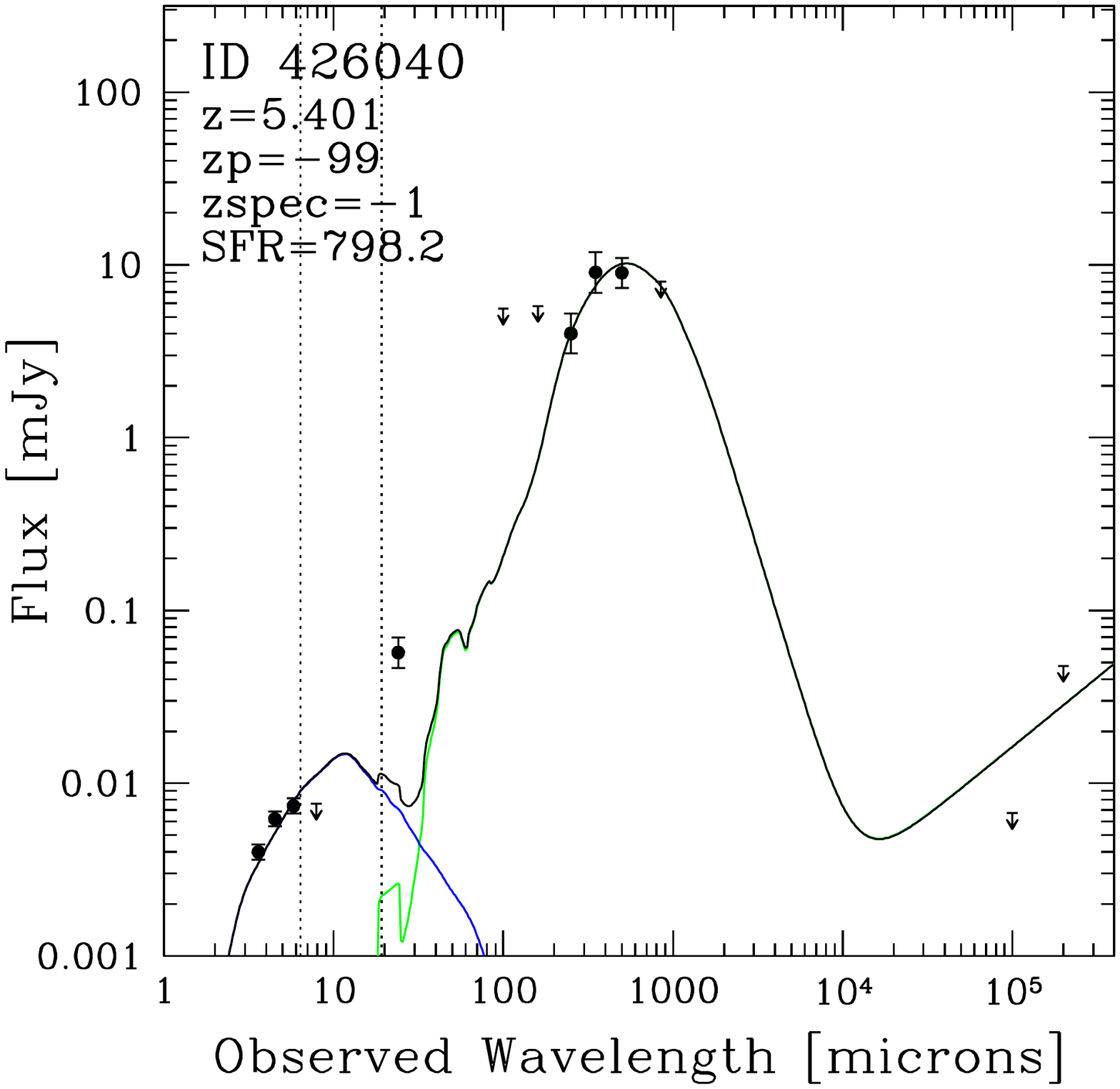}
    \includegraphics[width=0.28\textwidth]{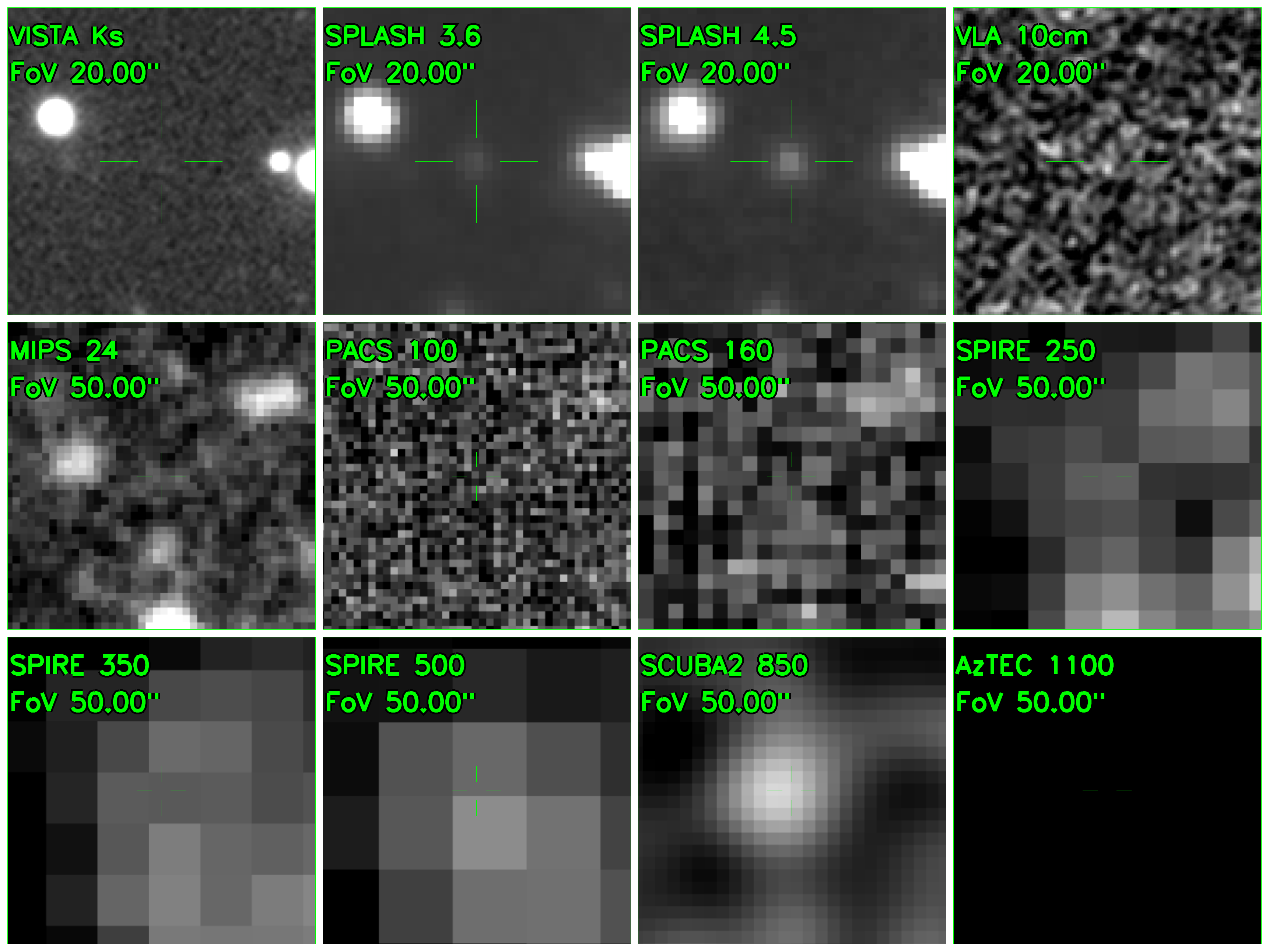}
	\includegraphics[width=0.21\textwidth, trim={0.6cm 5cm 1cm 3.5cm}, clip]{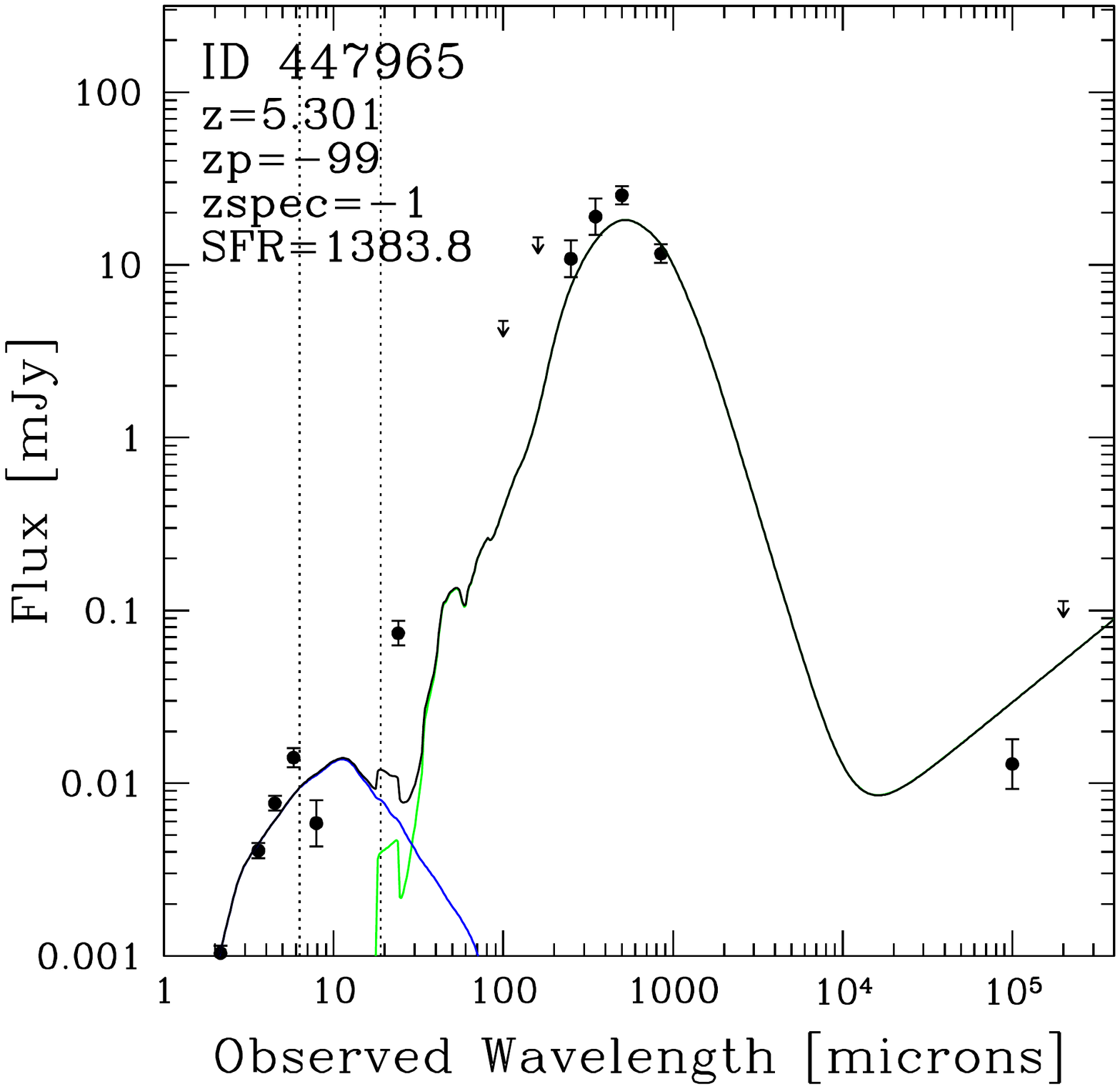}
    \includegraphics[width=0.28\textwidth]{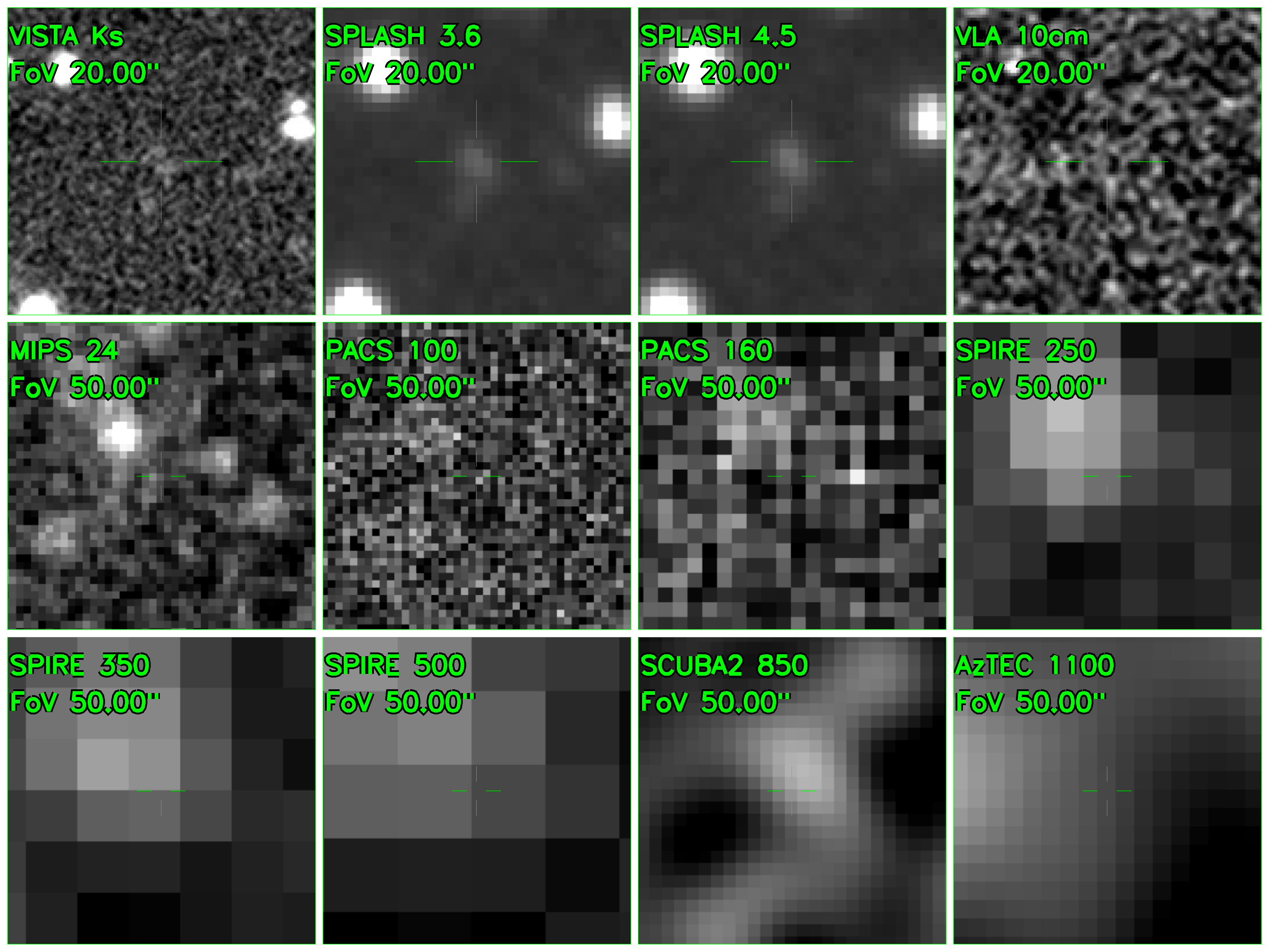}
	\includegraphics[width=0.21\textwidth, trim={0.6cm 5cm 1cm 3.5cm}, clip]{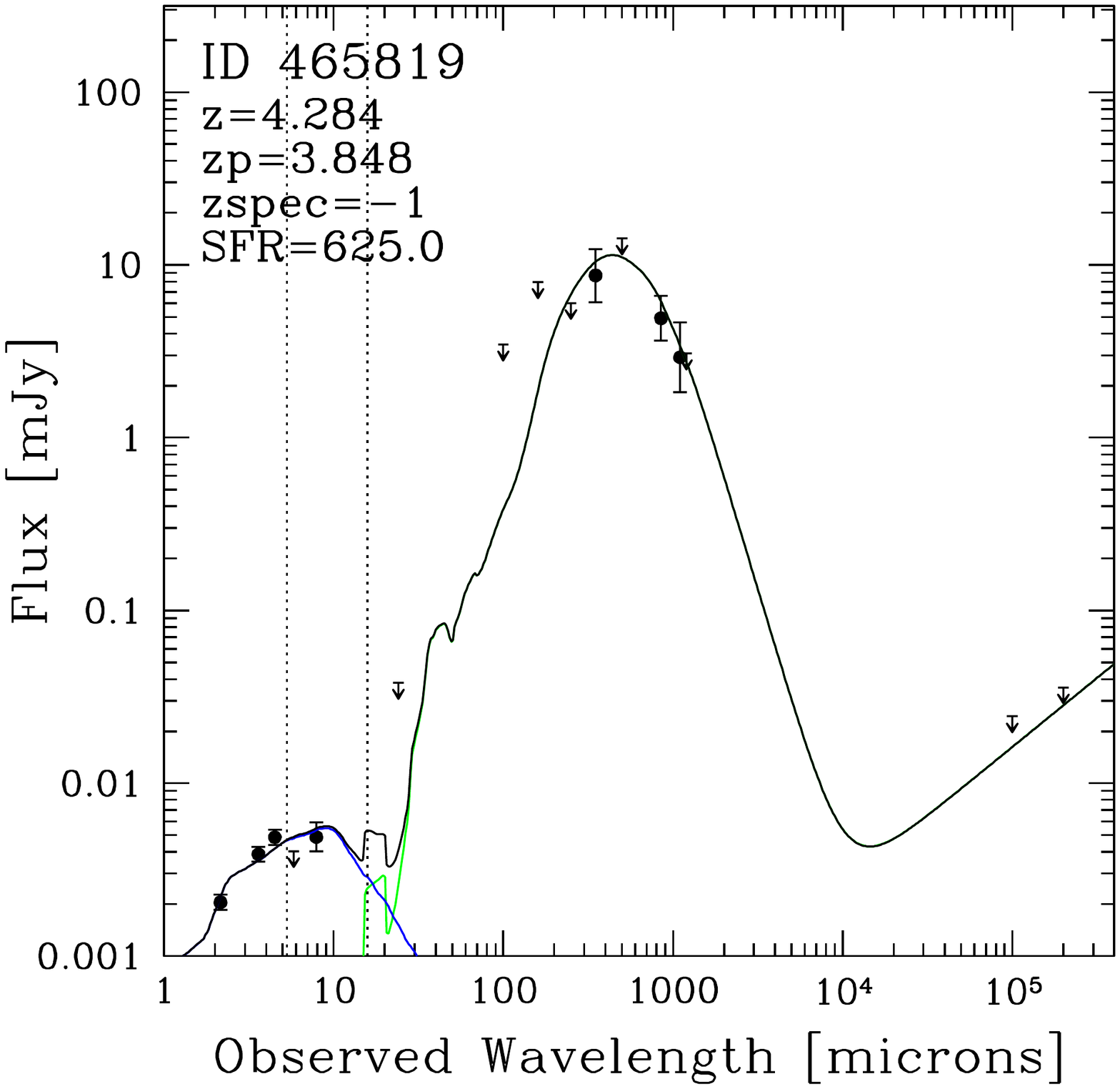}
    \includegraphics[width=0.28\textwidth]{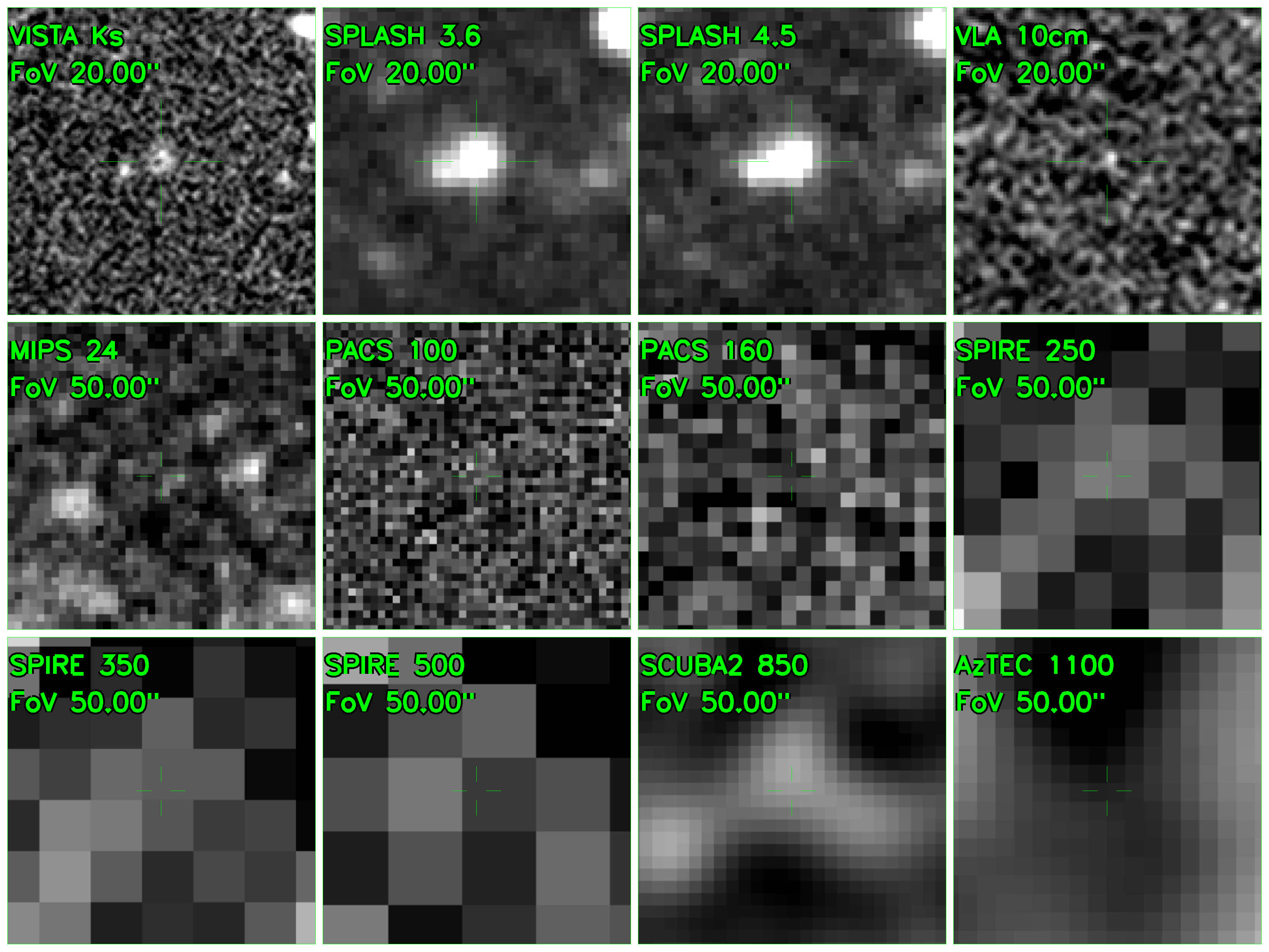}
	\includegraphics[width=0.21\textwidth, trim={0.6cm 5cm 1cm 3.5cm}, clip]{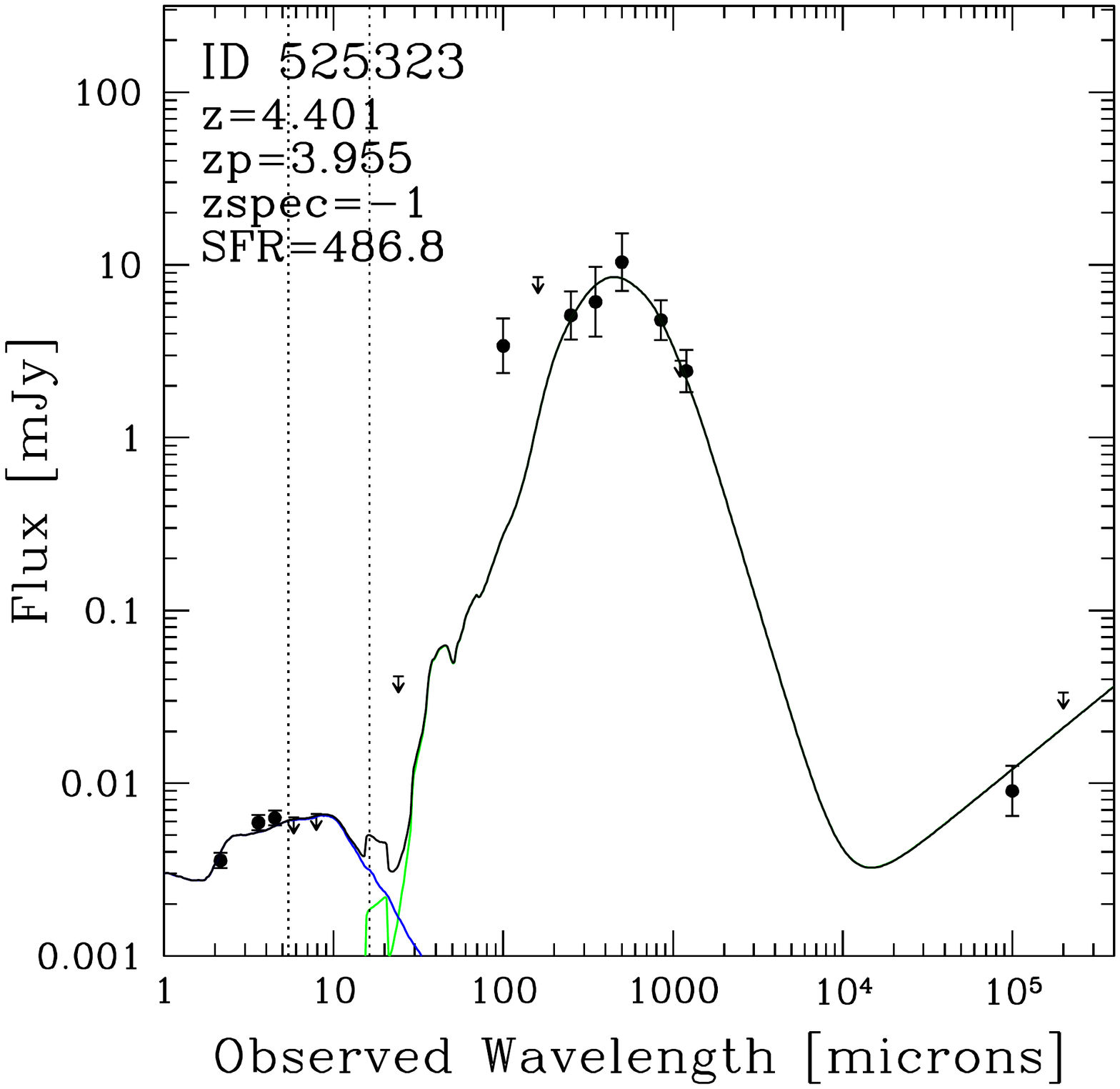}
    \includegraphics[width=0.28\textwidth]{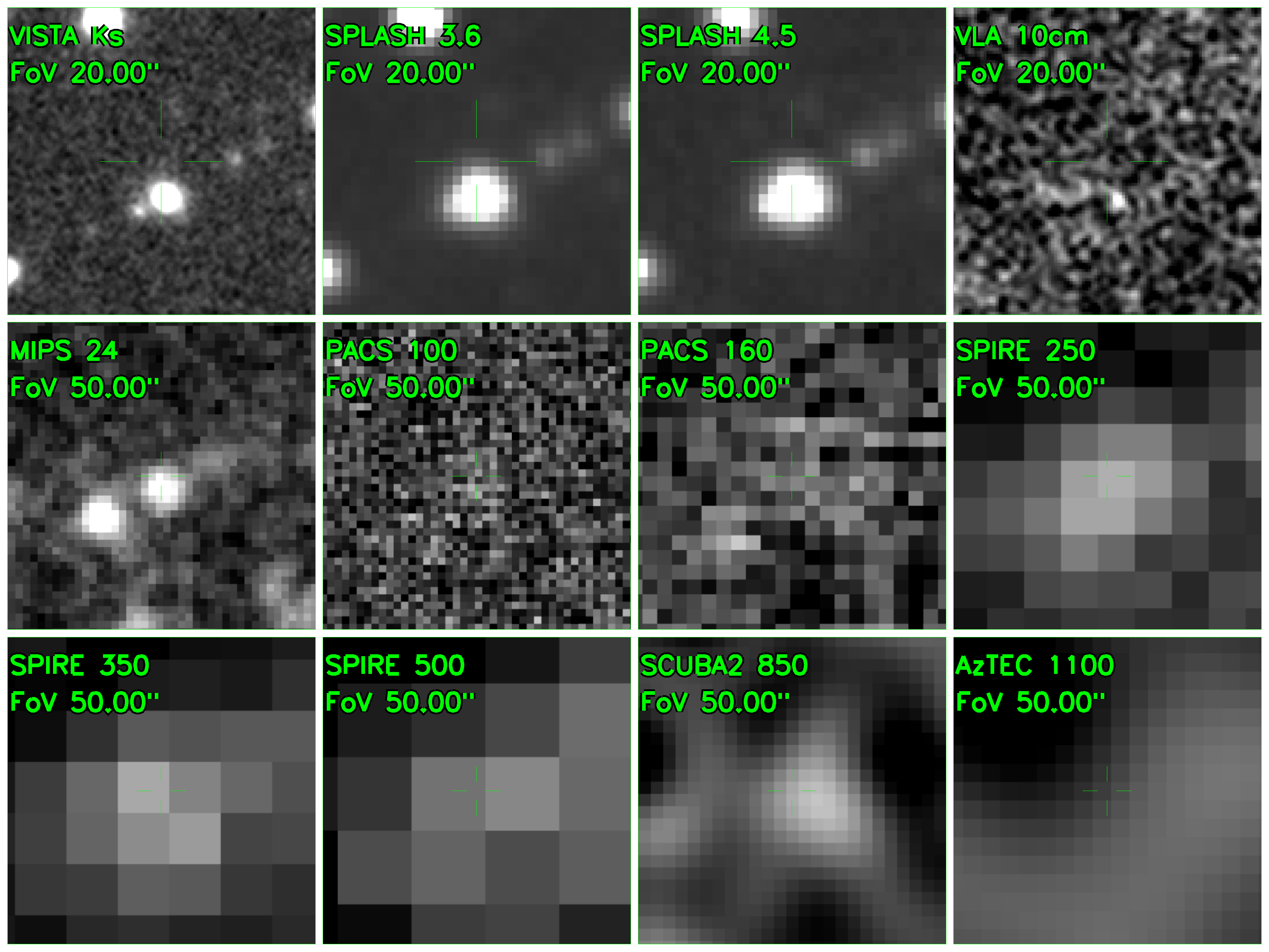}
	\includegraphics[width=0.21\textwidth, trim={0.6cm 5cm 1cm 3.5cm}, clip]{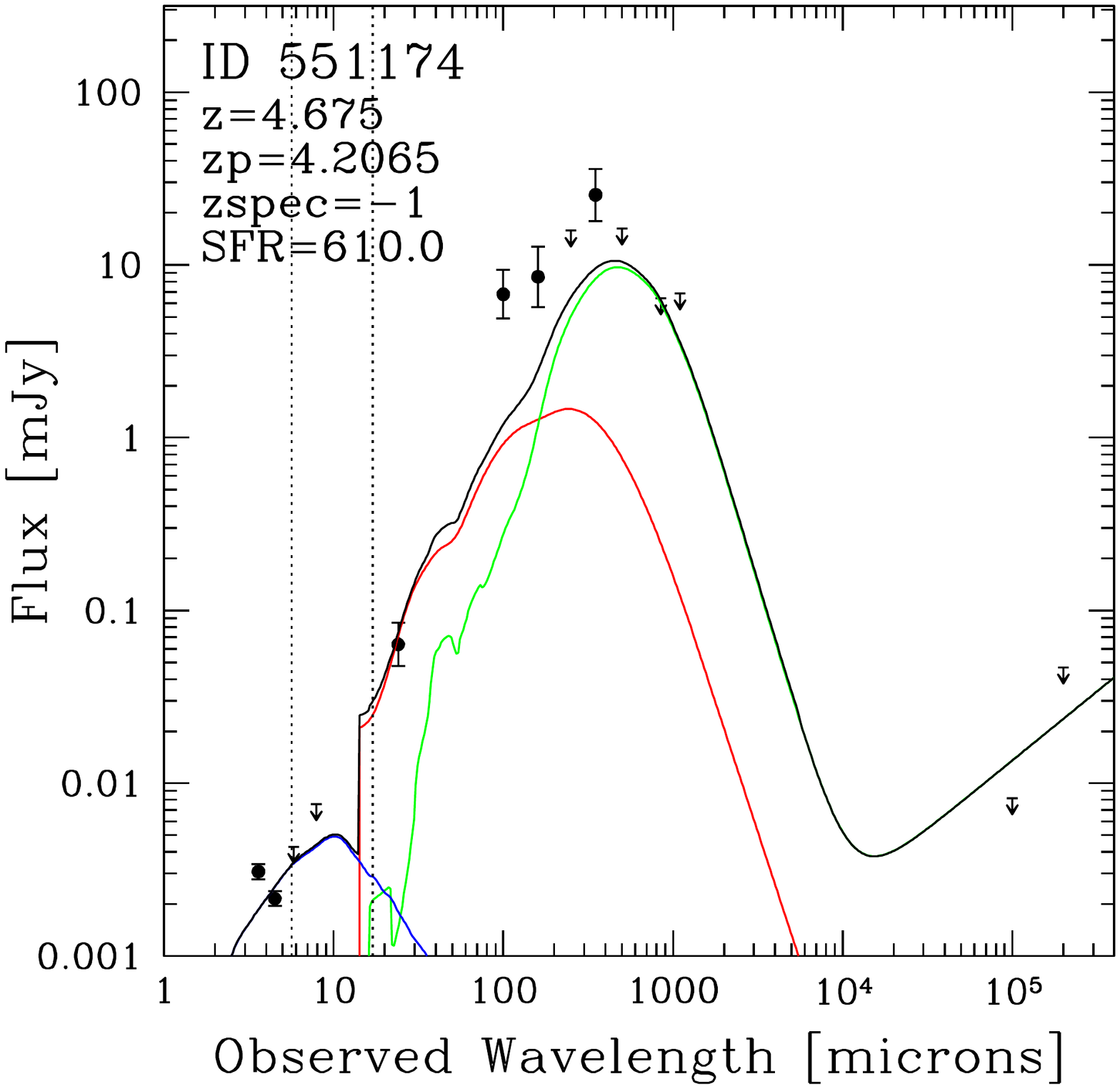}
    \includegraphics[width=0.28\textwidth]{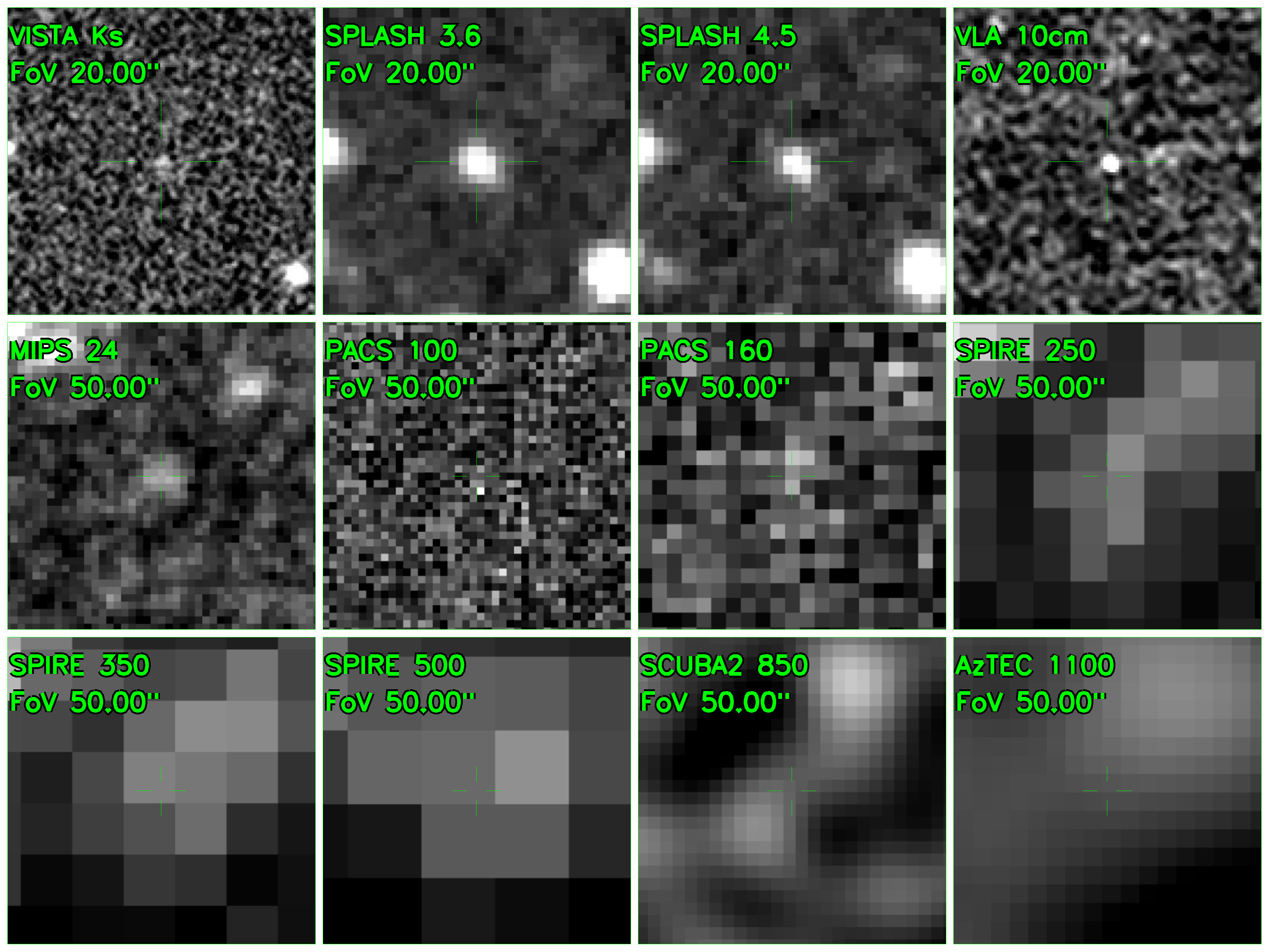}
	\includegraphics[width=0.21\textwidth, trim={0.6cm 5cm 1cm 3.5cm}, clip]{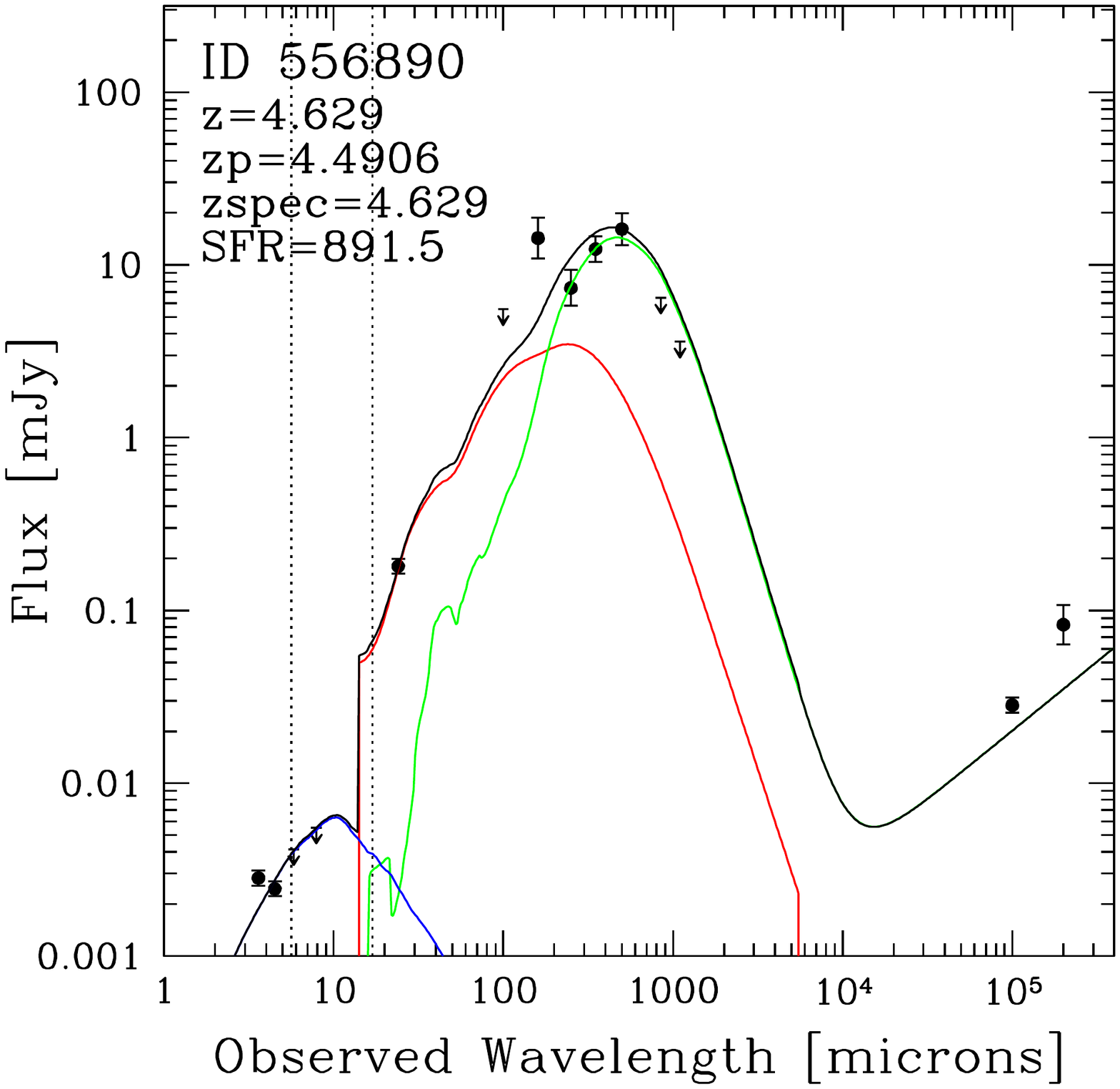}
	\caption{%
		Multi-band cutouts and SEDs of high redshift candidates. {}
		\label{highz_cutouts1}
		}
\end{figure}

\begin{figure}
	\centering
    \includegraphics[width=0.28\textwidth]{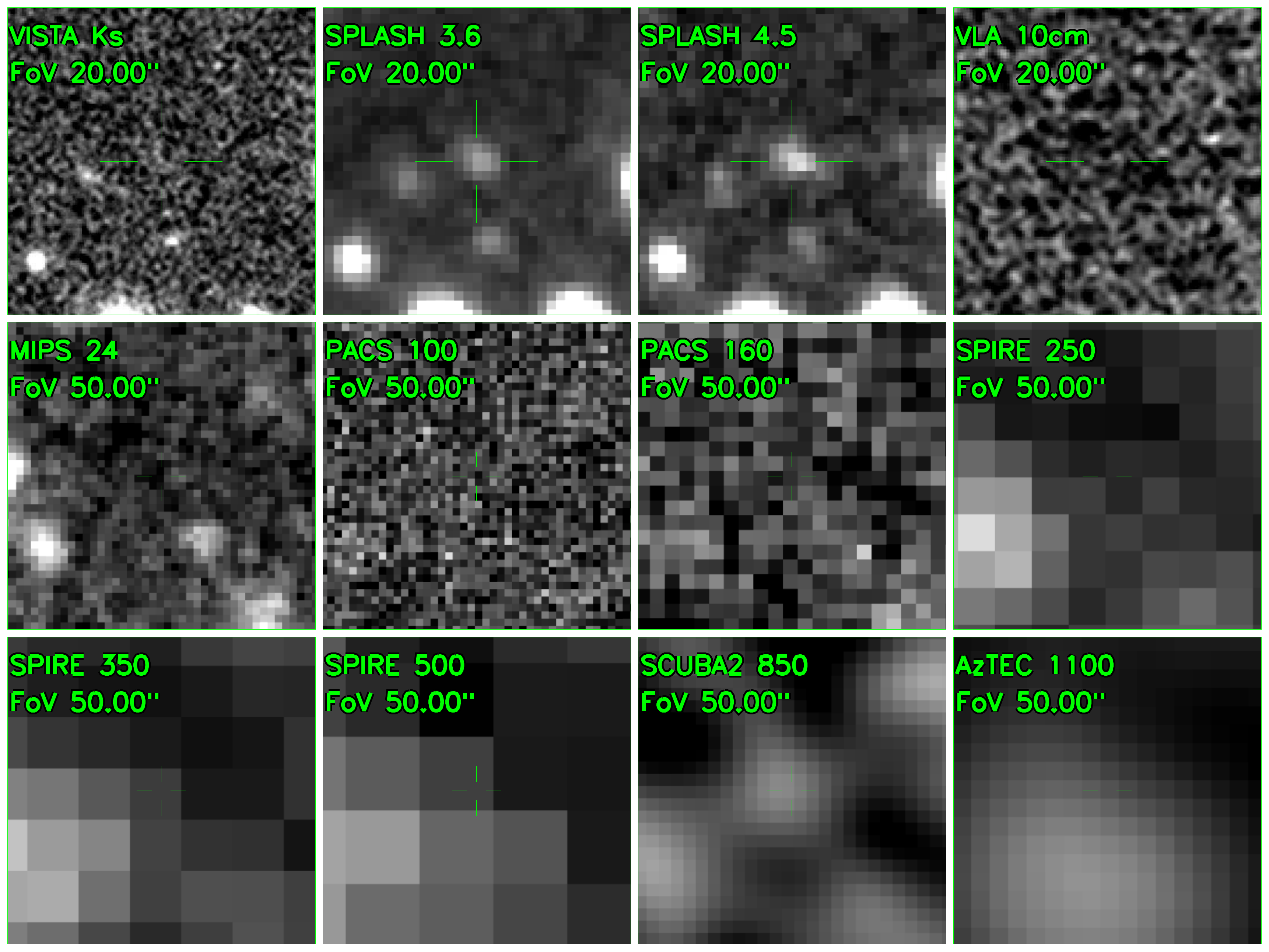}
	\includegraphics[width=0.21\textwidth, trim={0.6cm 5cm 1cm 3.5cm}, clip]{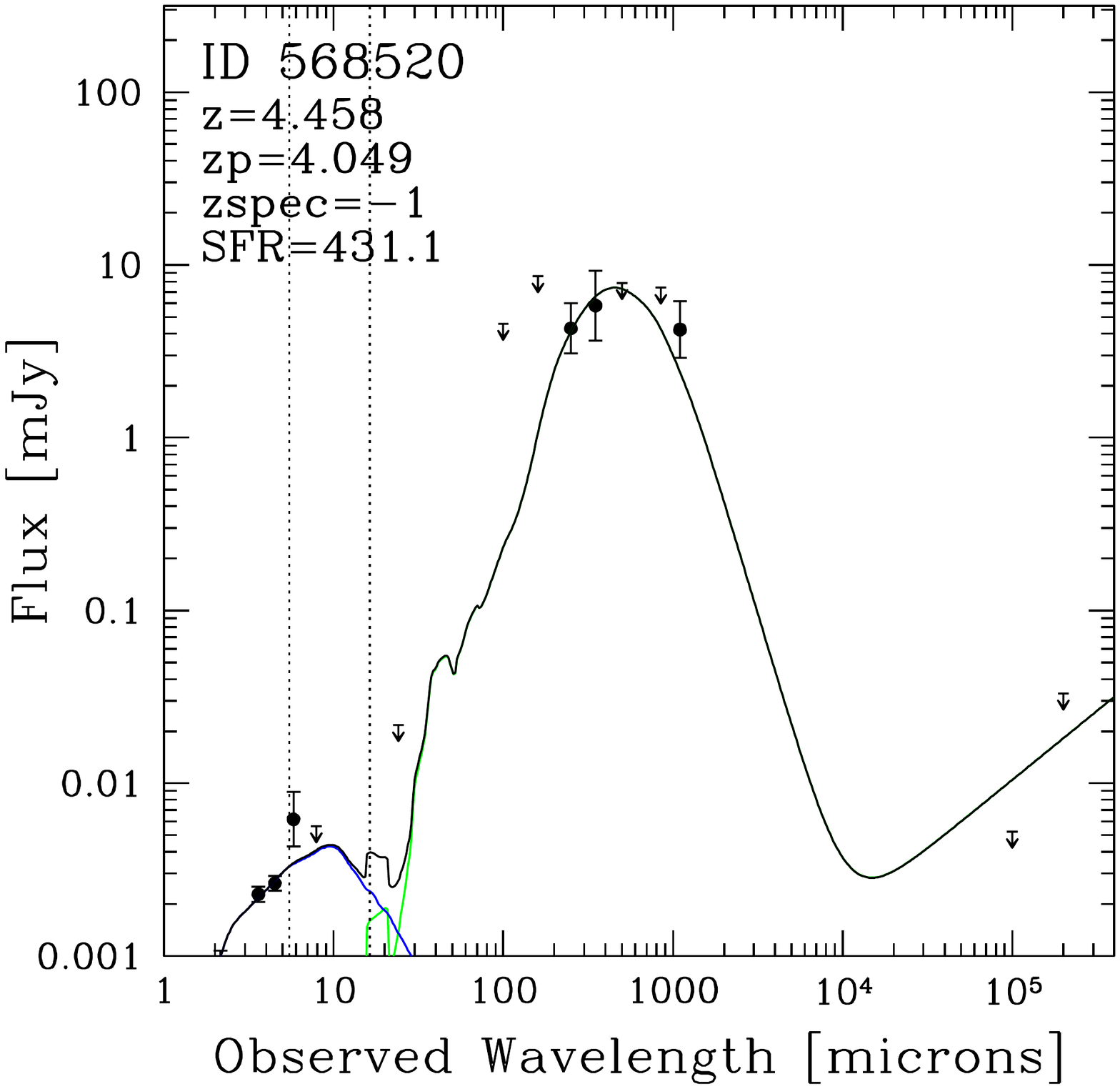}
    \includegraphics[width=0.28\textwidth]{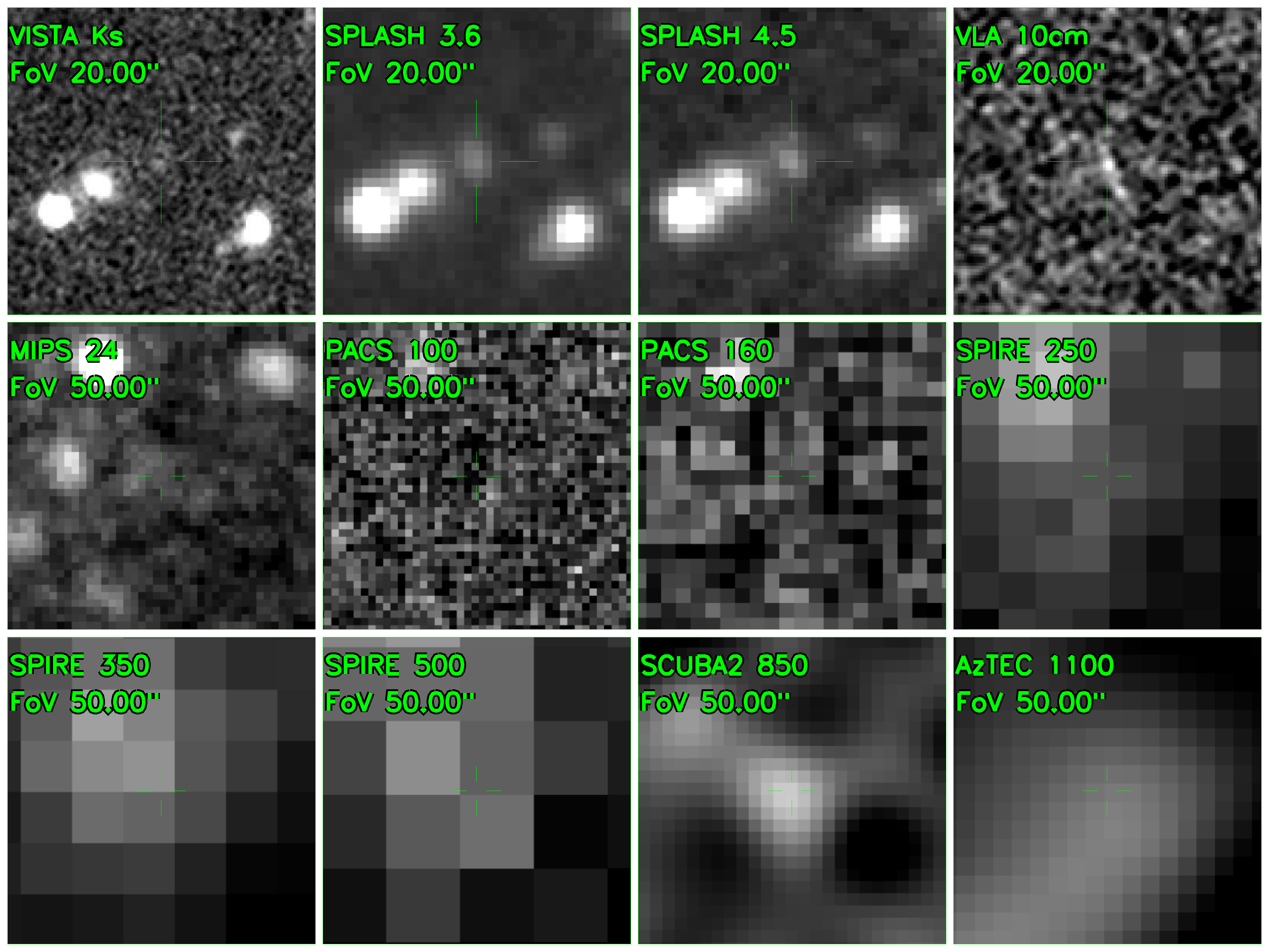}
	\includegraphics[width=0.21\textwidth, trim={0.6cm 5cm 1cm 3.5cm}, clip]{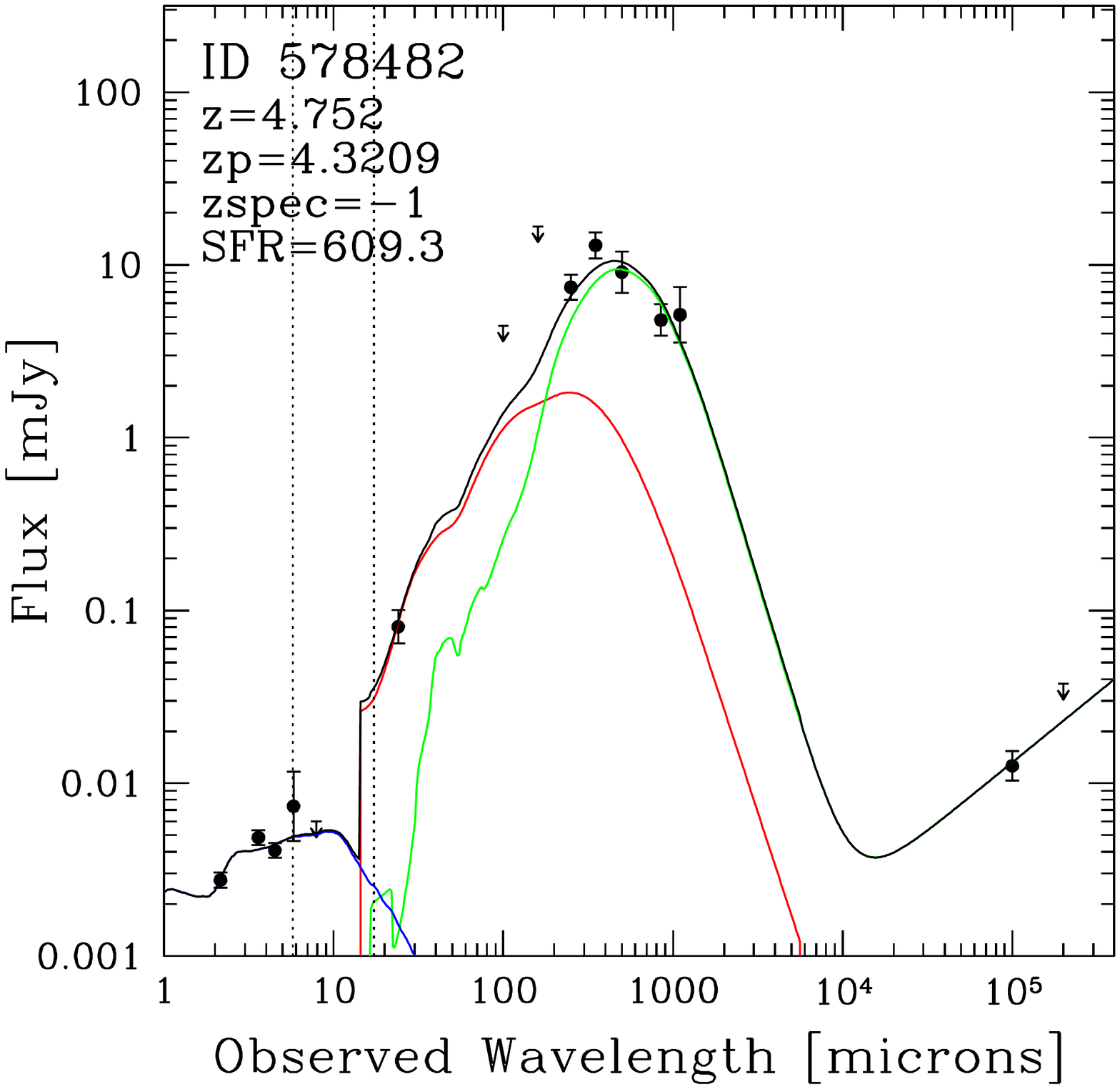}
    \includegraphics[width=0.28\textwidth]{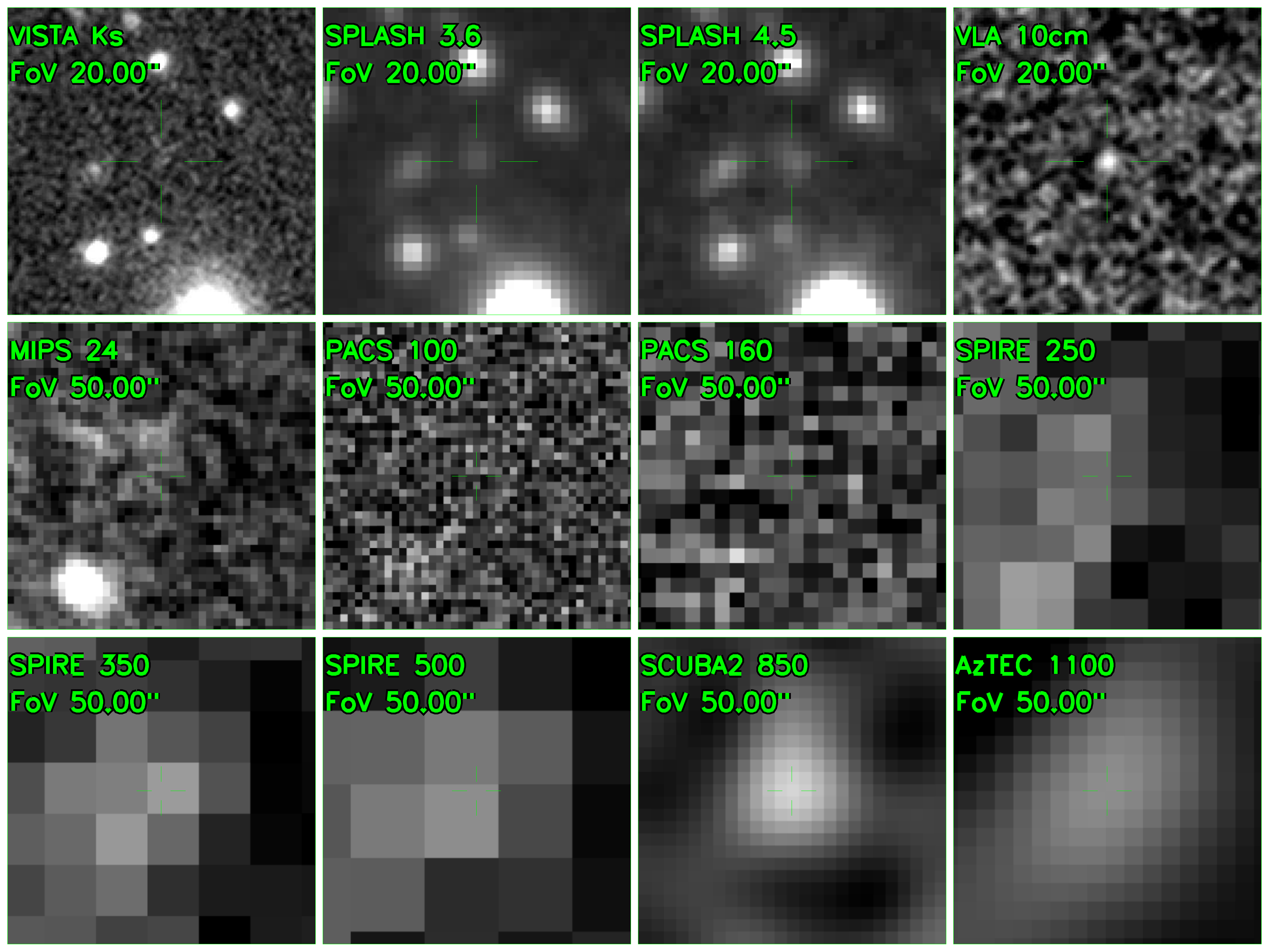}
	\includegraphics[width=0.21\textwidth, trim={0.6cm 5cm 1cm 3.5cm}, clip]{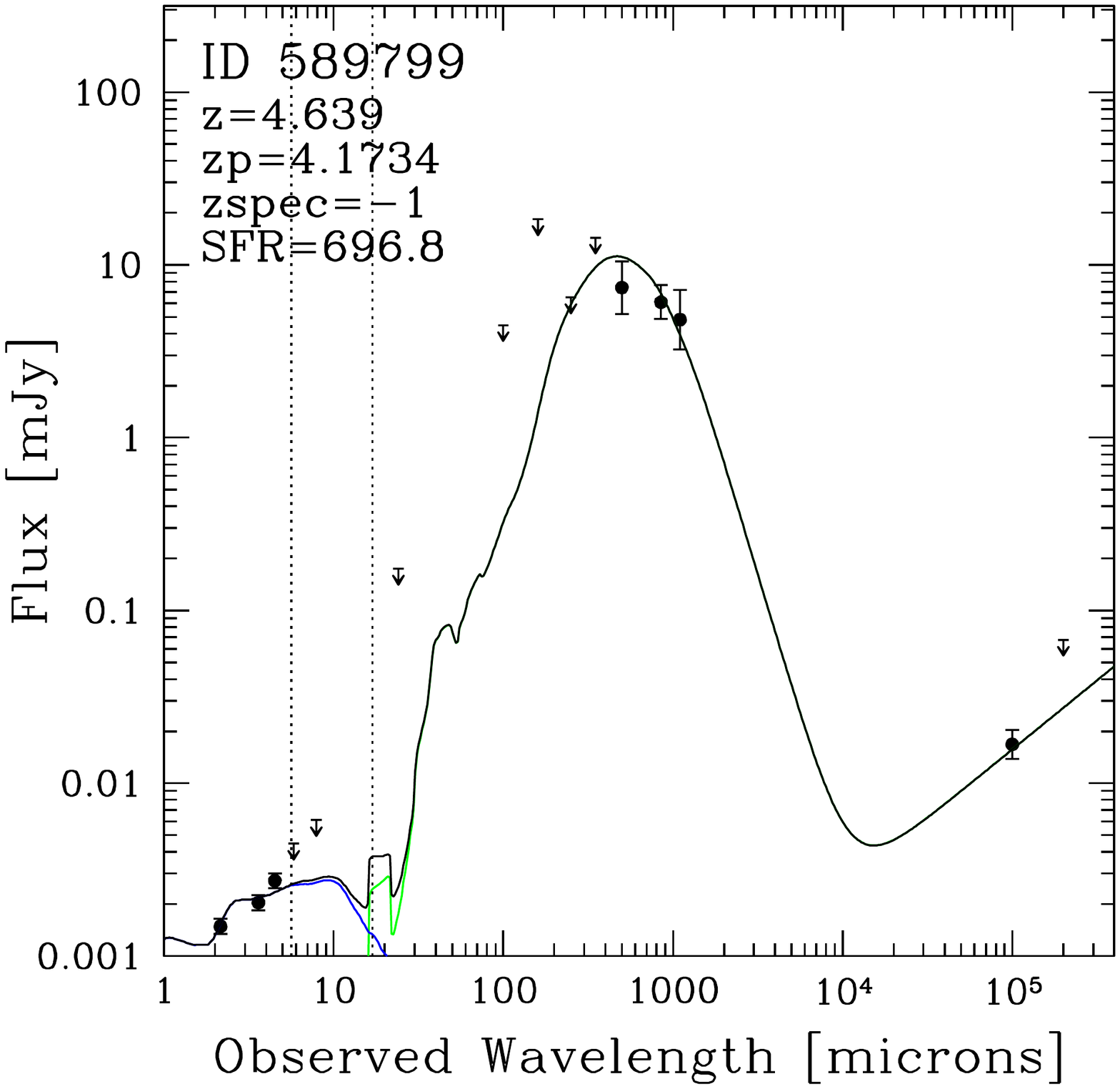}
    \includegraphics[width=0.28\textwidth]{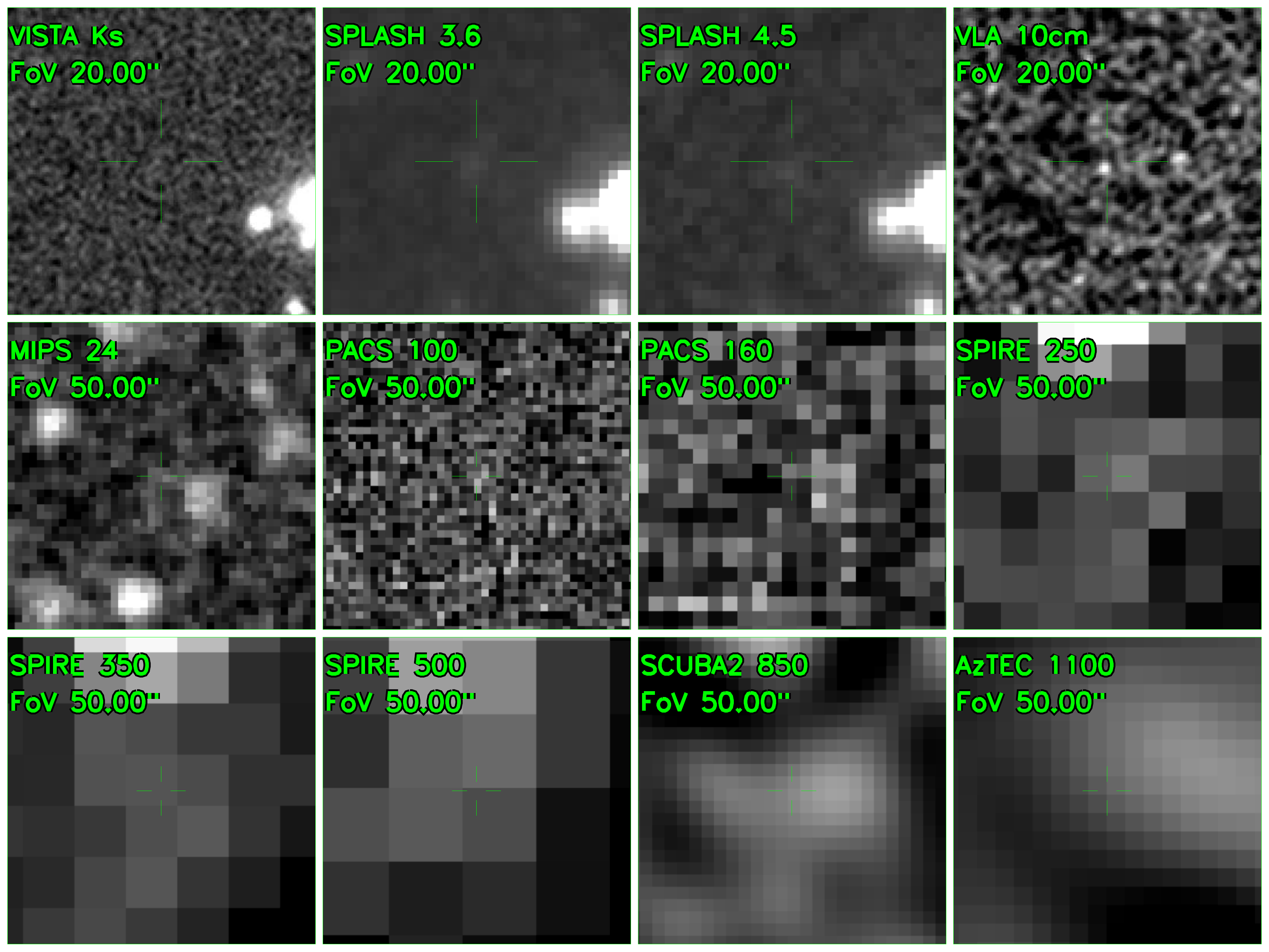}
	\includegraphics[width=0.21\textwidth, trim={0.6cm 5cm 1cm 3.5cm}, clip]{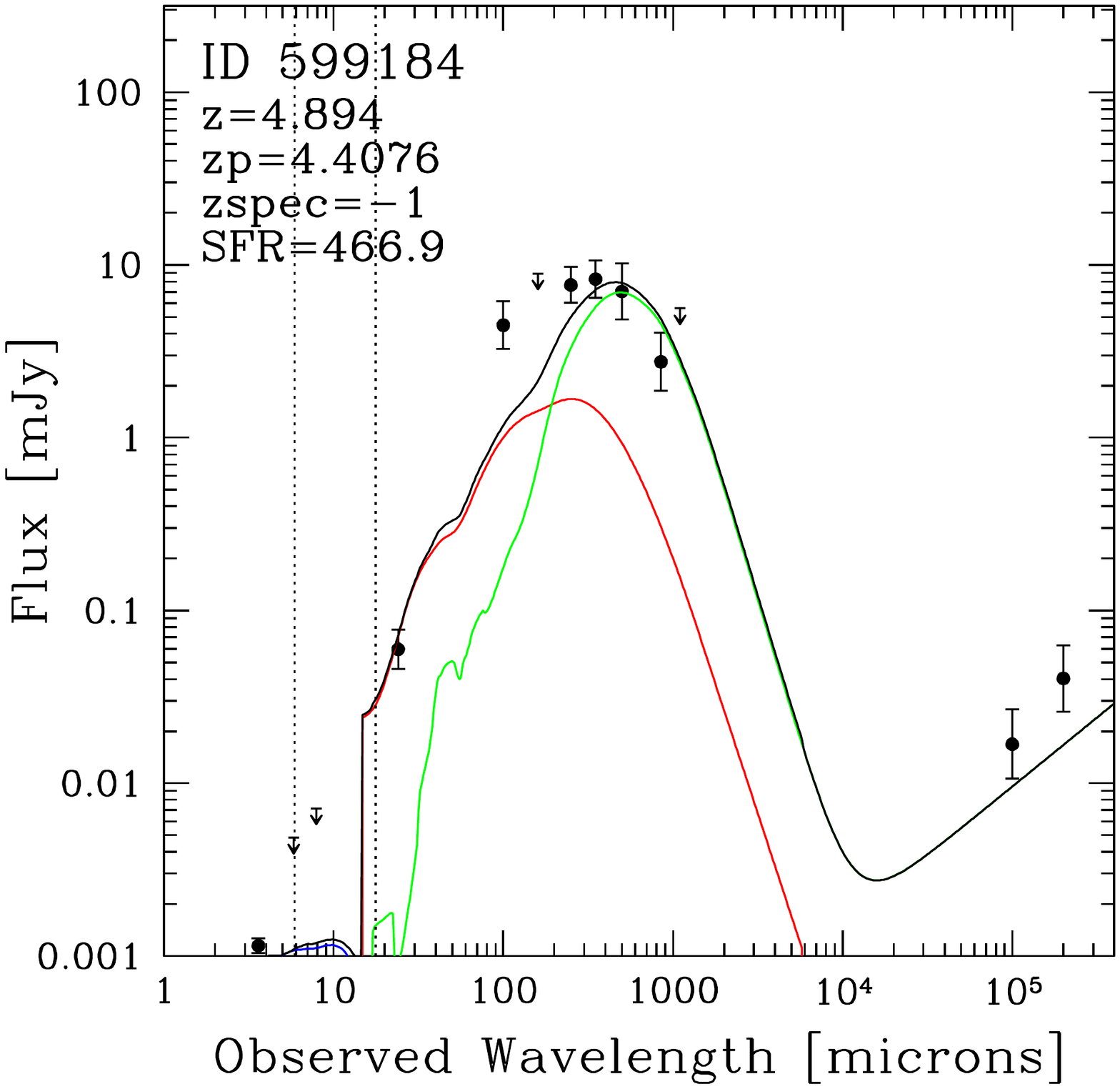}
    \includegraphics[width=0.28\textwidth]{599521.pdf}
	\includegraphics[width=0.21\textwidth, trim={0.6cm 5cm 1cm 3.5cm}, clip]{Plot_SED_599521.pdf}
    \includegraphics[width=0.28\textwidth]{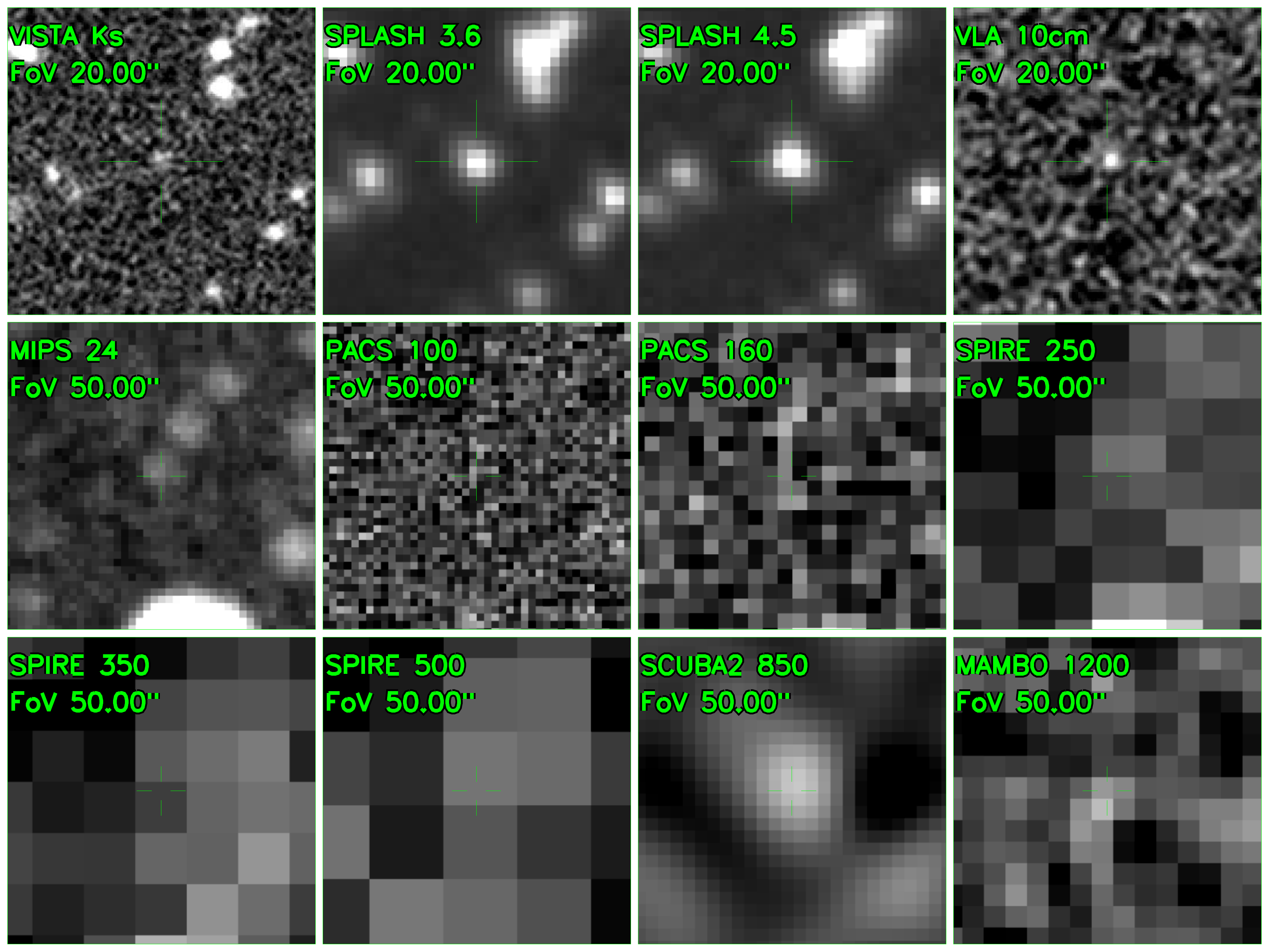}
	\includegraphics[width=0.21\textwidth, trim={0.6cm 5cm 1cm 3.5cm}, clip]{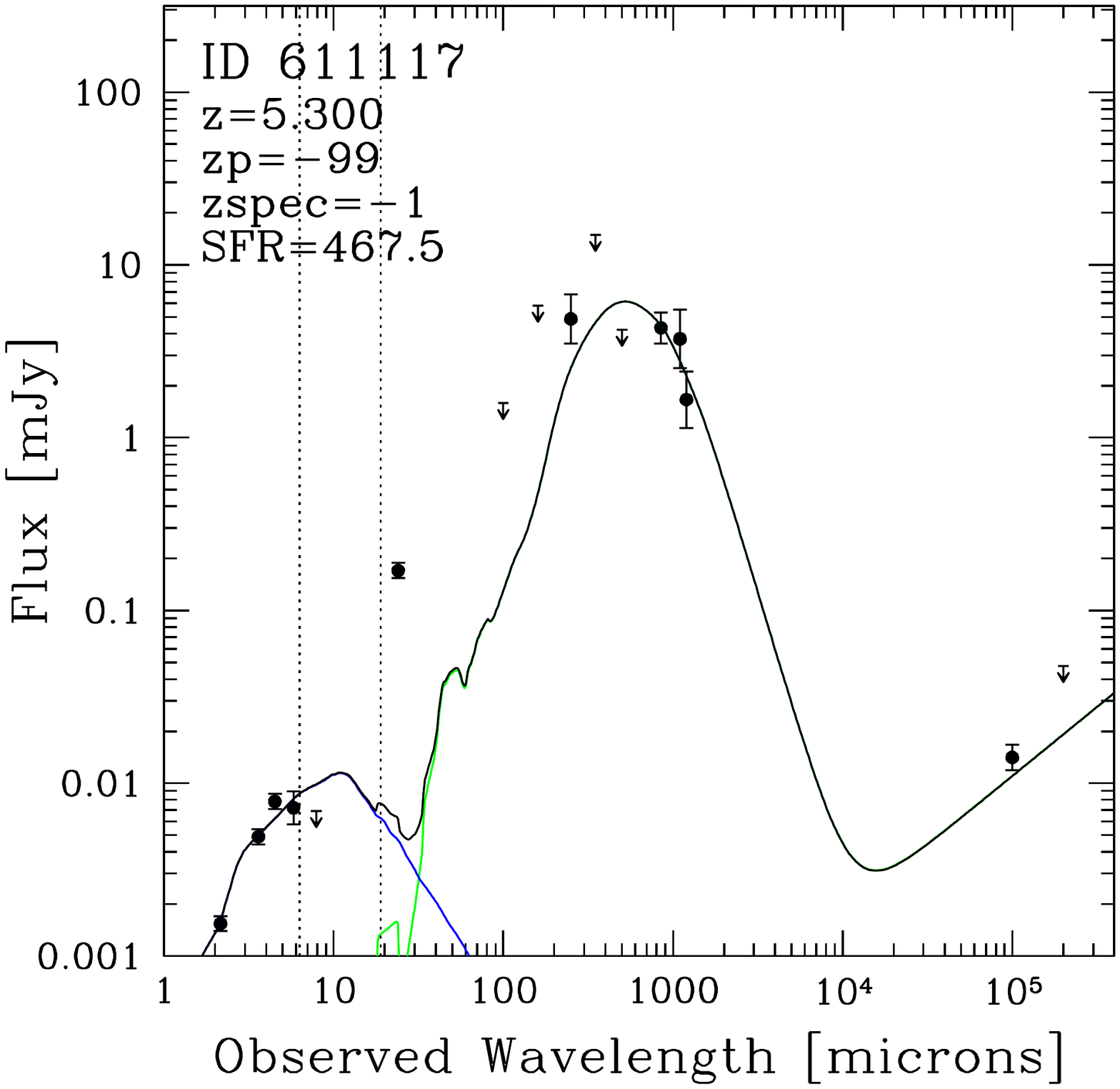}
    \includegraphics[width=0.28\textwidth]{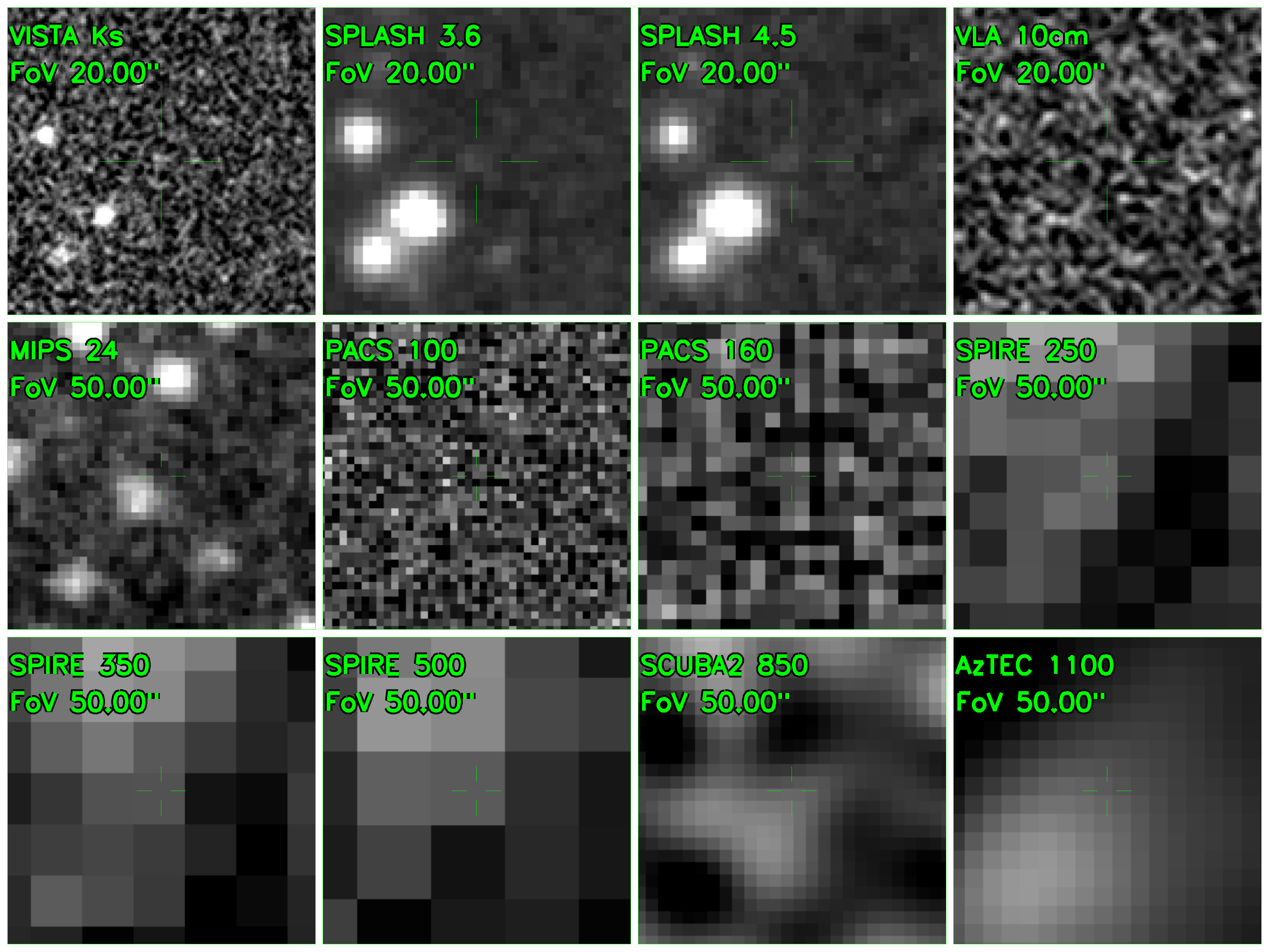}
	\includegraphics[width=0.21\textwidth, trim={0.6cm 5cm 1cm 3.5cm}, clip]{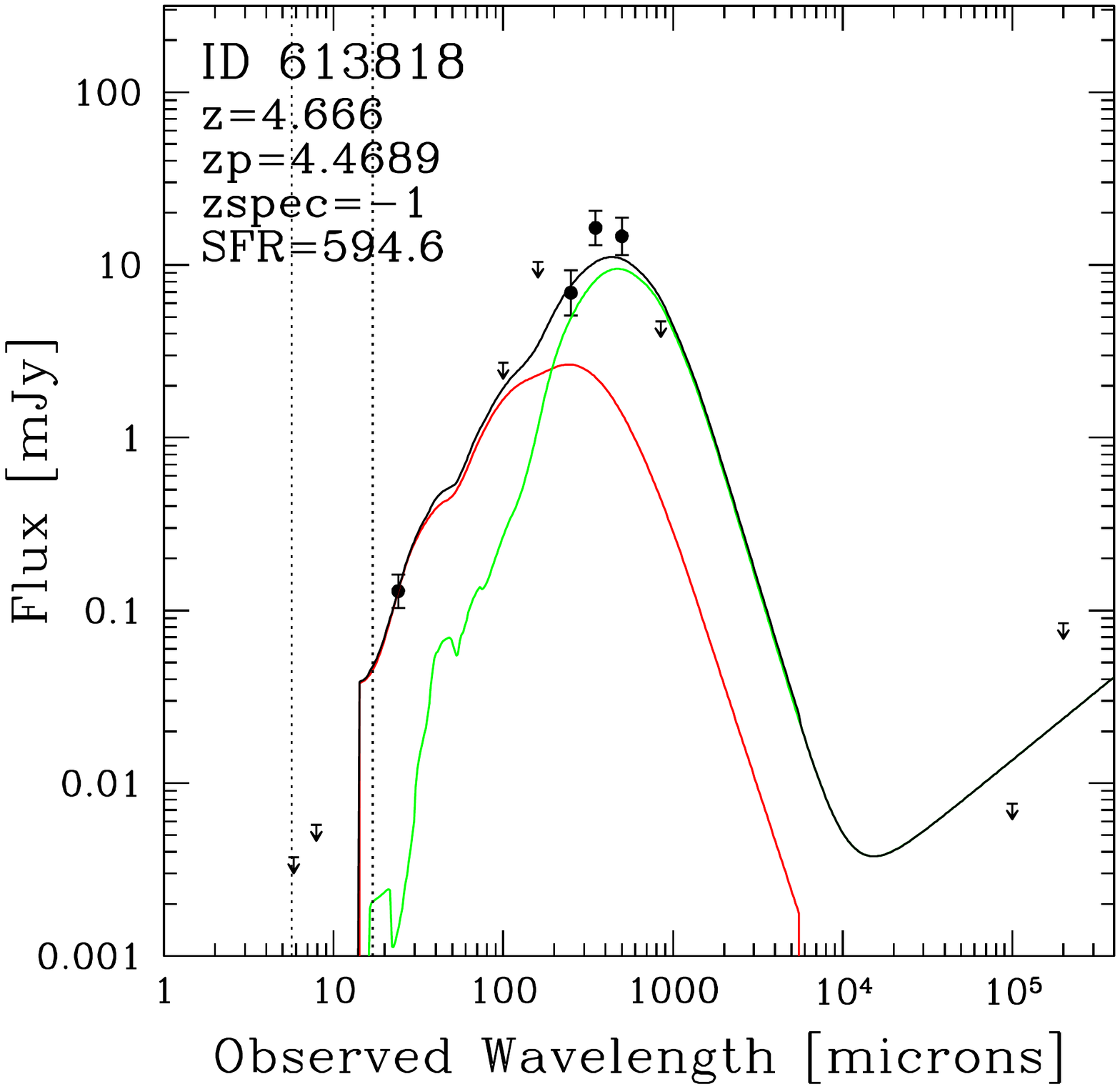}
    \includegraphics[width=0.28\textwidth]{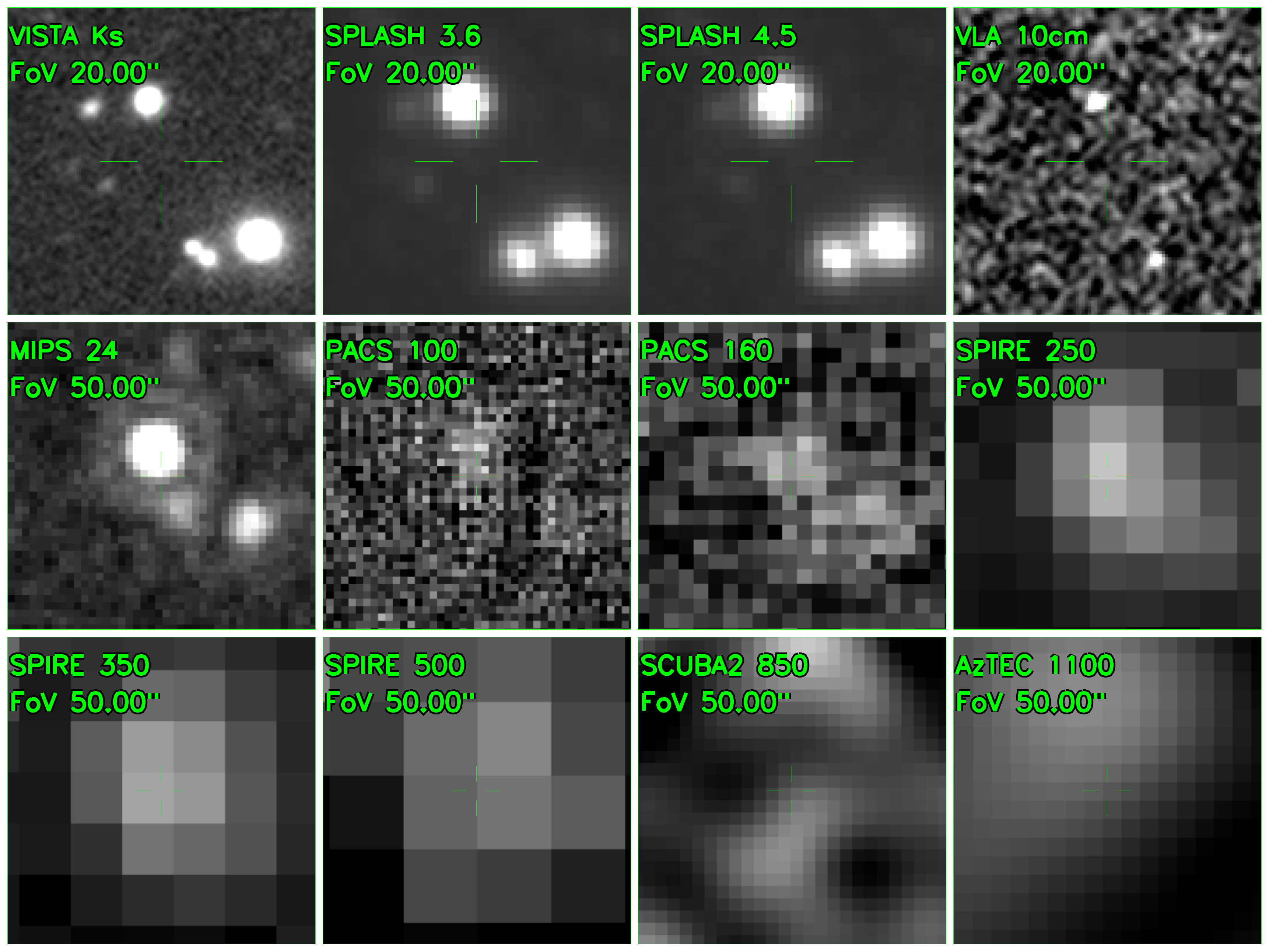}
	\includegraphics[width=0.21\textwidth, trim={0.6cm 5cm 1cm 3.5cm}, clip]{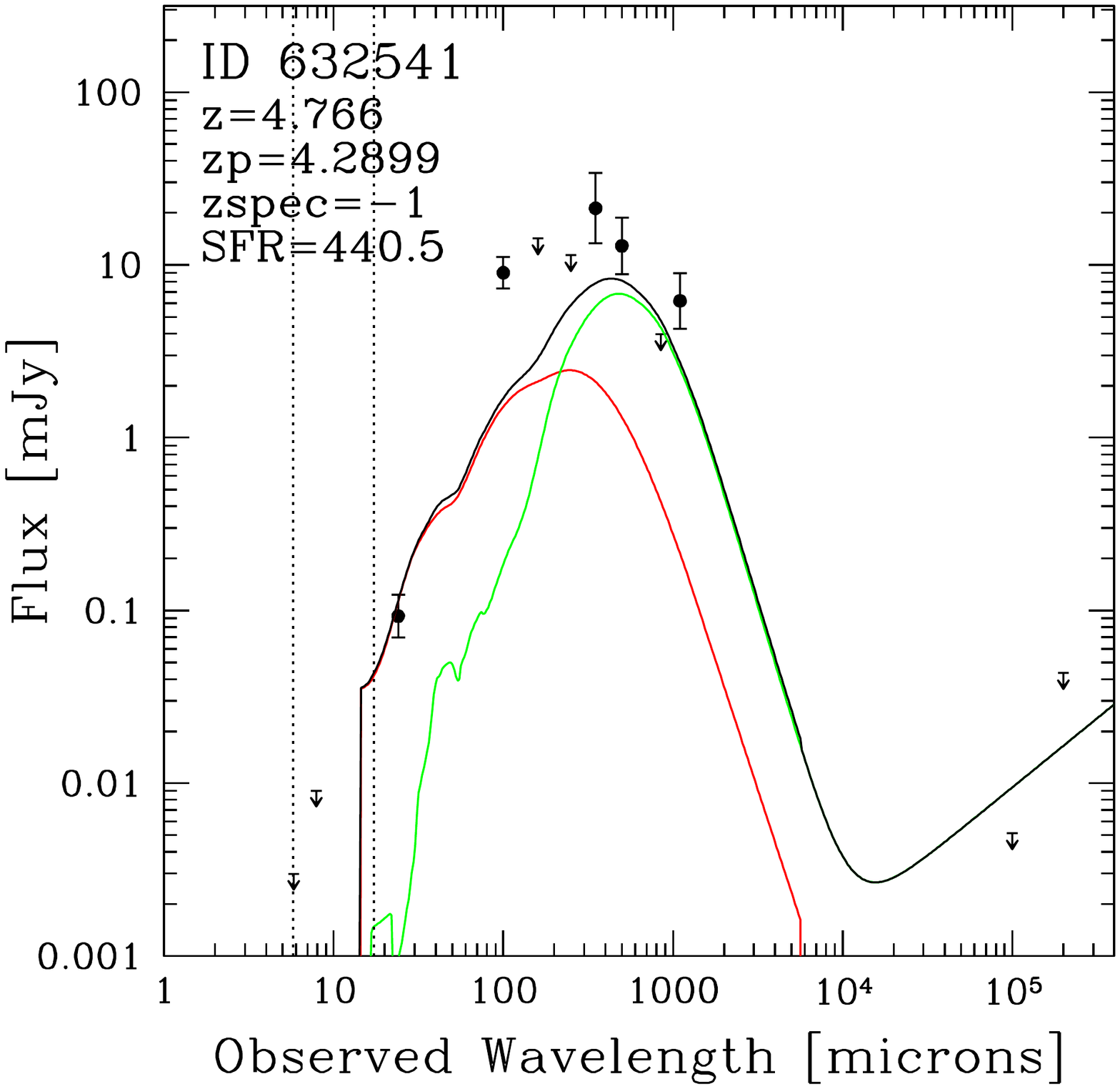}
    \includegraphics[width=0.28\textwidth]{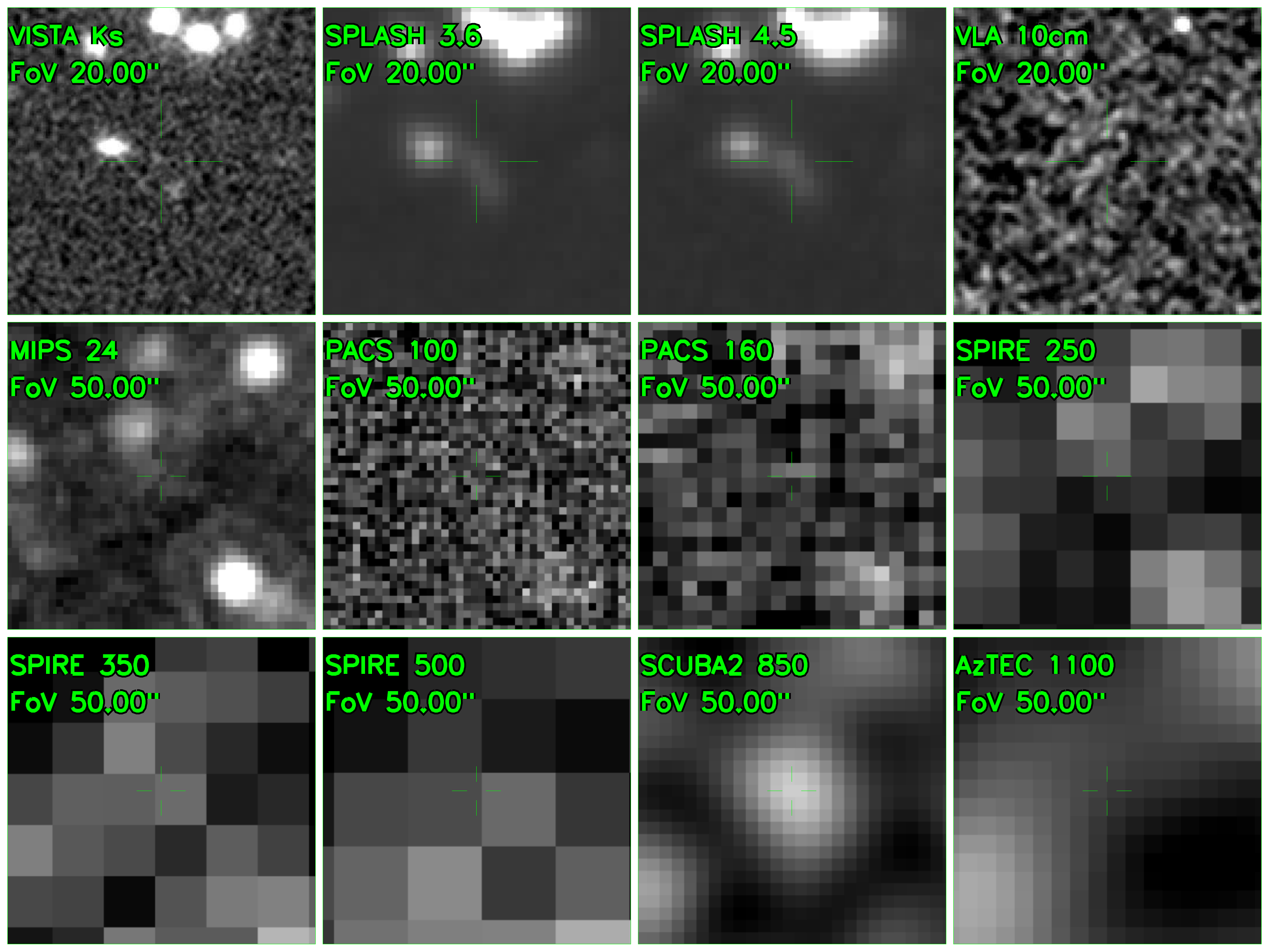}
	\includegraphics[width=0.21\textwidth, trim={0.6cm 5cm 1cm 3.5cm}, clip]{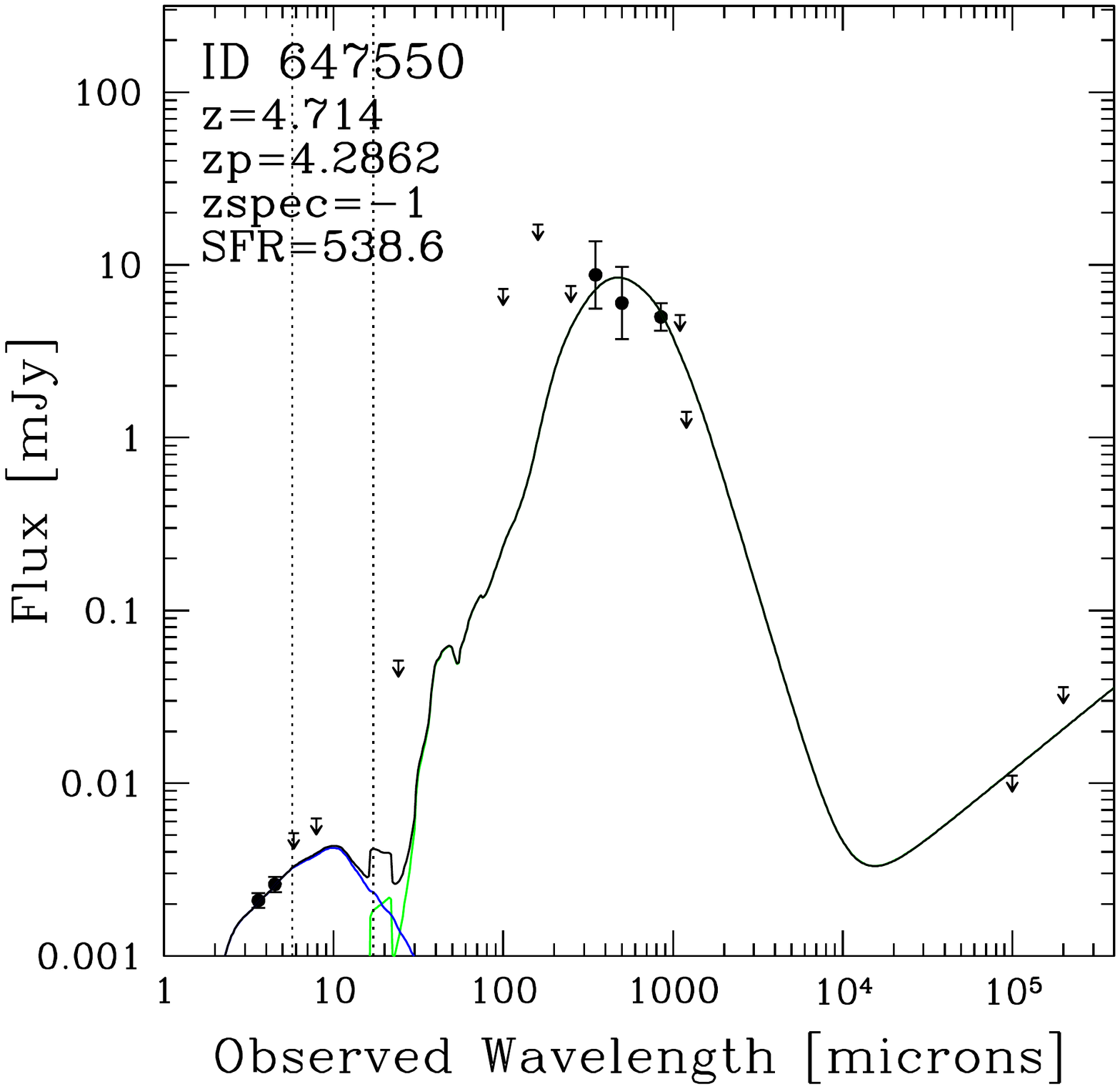}
    \includegraphics[width=0.28\textwidth]{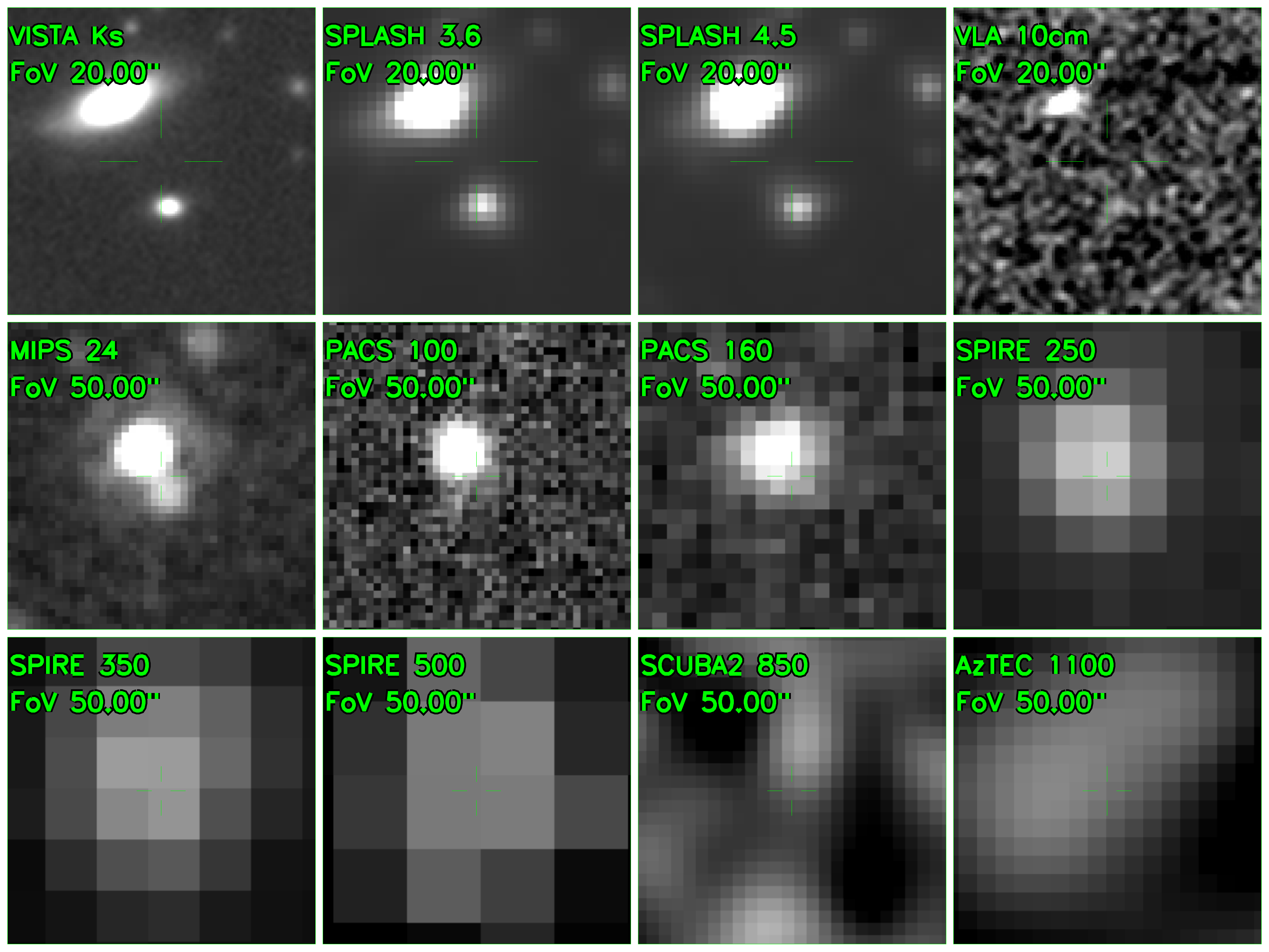}
	\includegraphics[width=0.21\textwidth, trim={0.6cm 5cm 1cm 3.5cm}, clip]{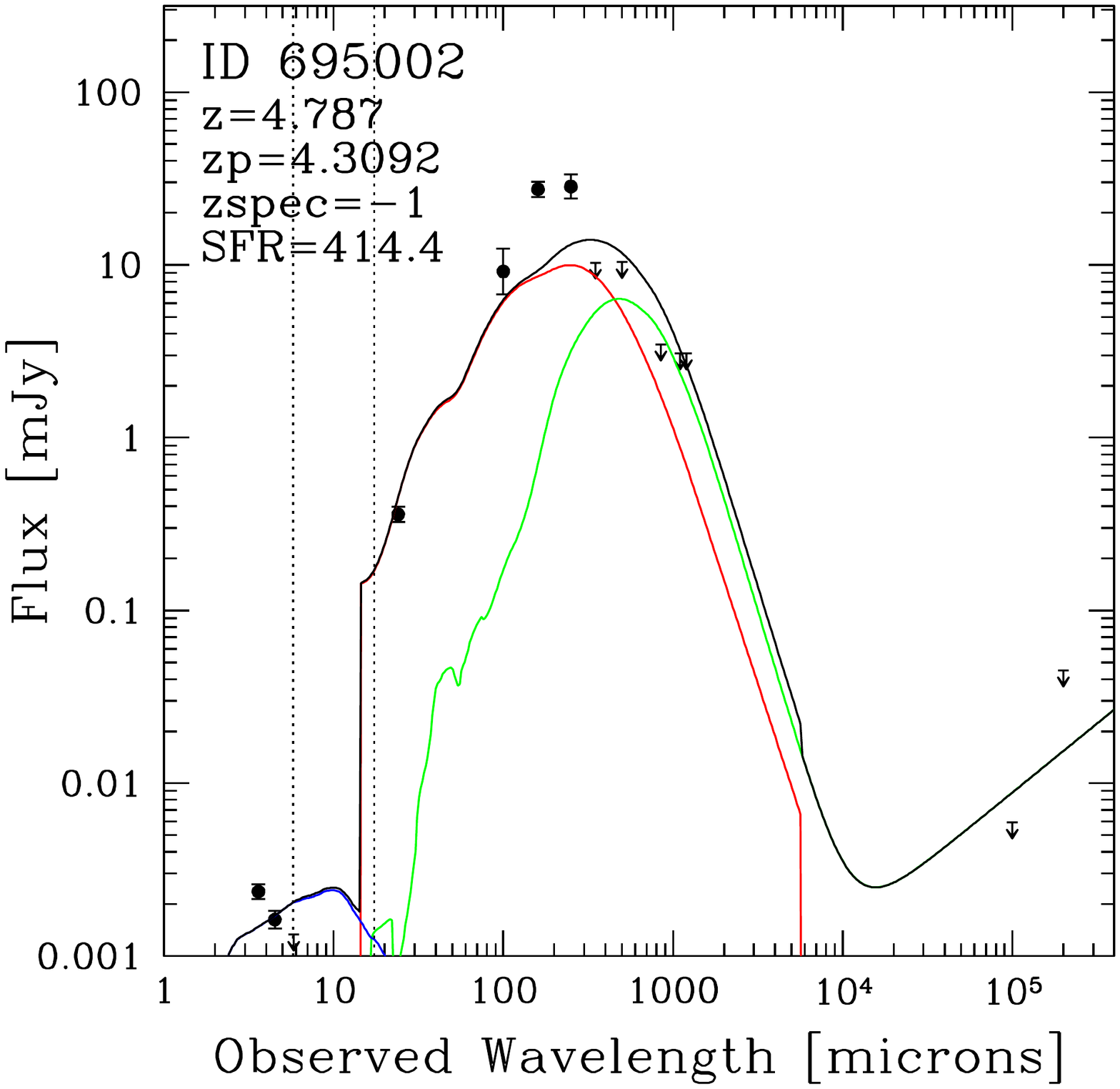}
    
	\caption{%
		Multi-band cutouts and SEDs of high redshift candidates. {}
		\label{highz_cutouts1}
		}
\end{figure}

\begin{figure}
	\centering
    \includegraphics[width=0.28\textwidth]{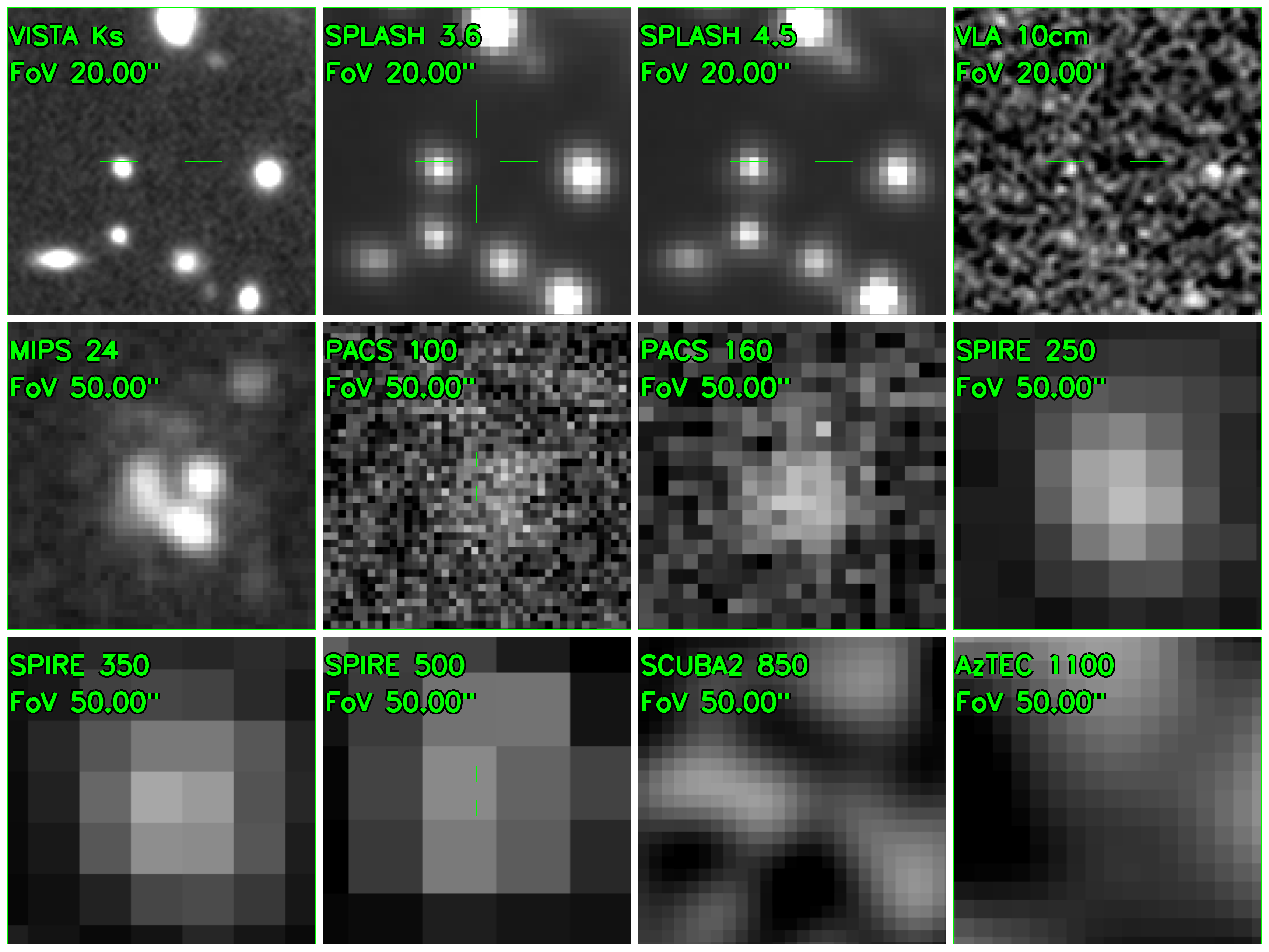}
	\includegraphics[width=0.21\textwidth, trim={0.6cm 5cm 1cm 3.5cm}, clip]{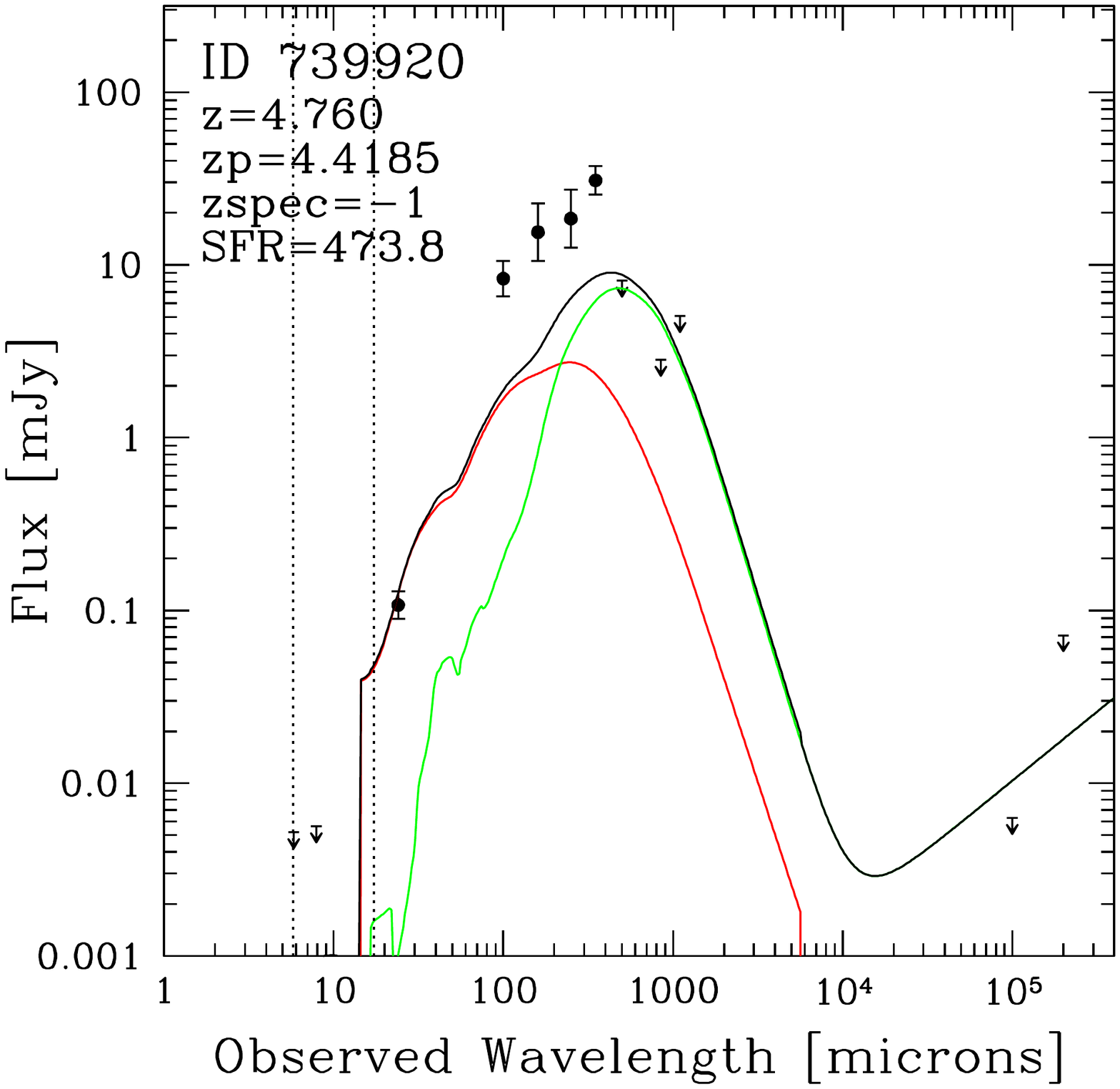}
    \includegraphics[width=0.28\textwidth]{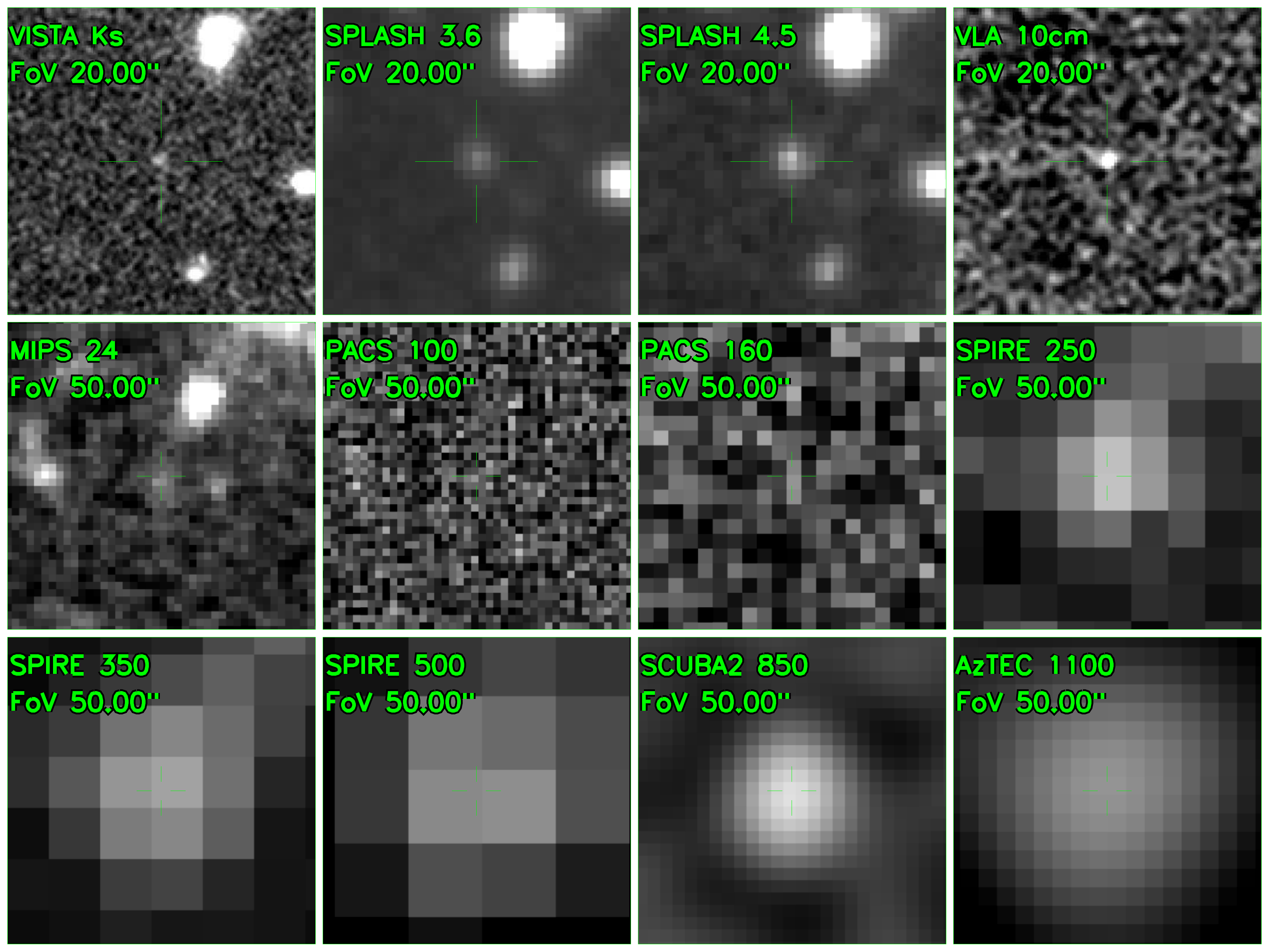}
	\includegraphics[width=0.21\textwidth, trim={0.6cm 5cm 1cm 3.5cm}, clip]{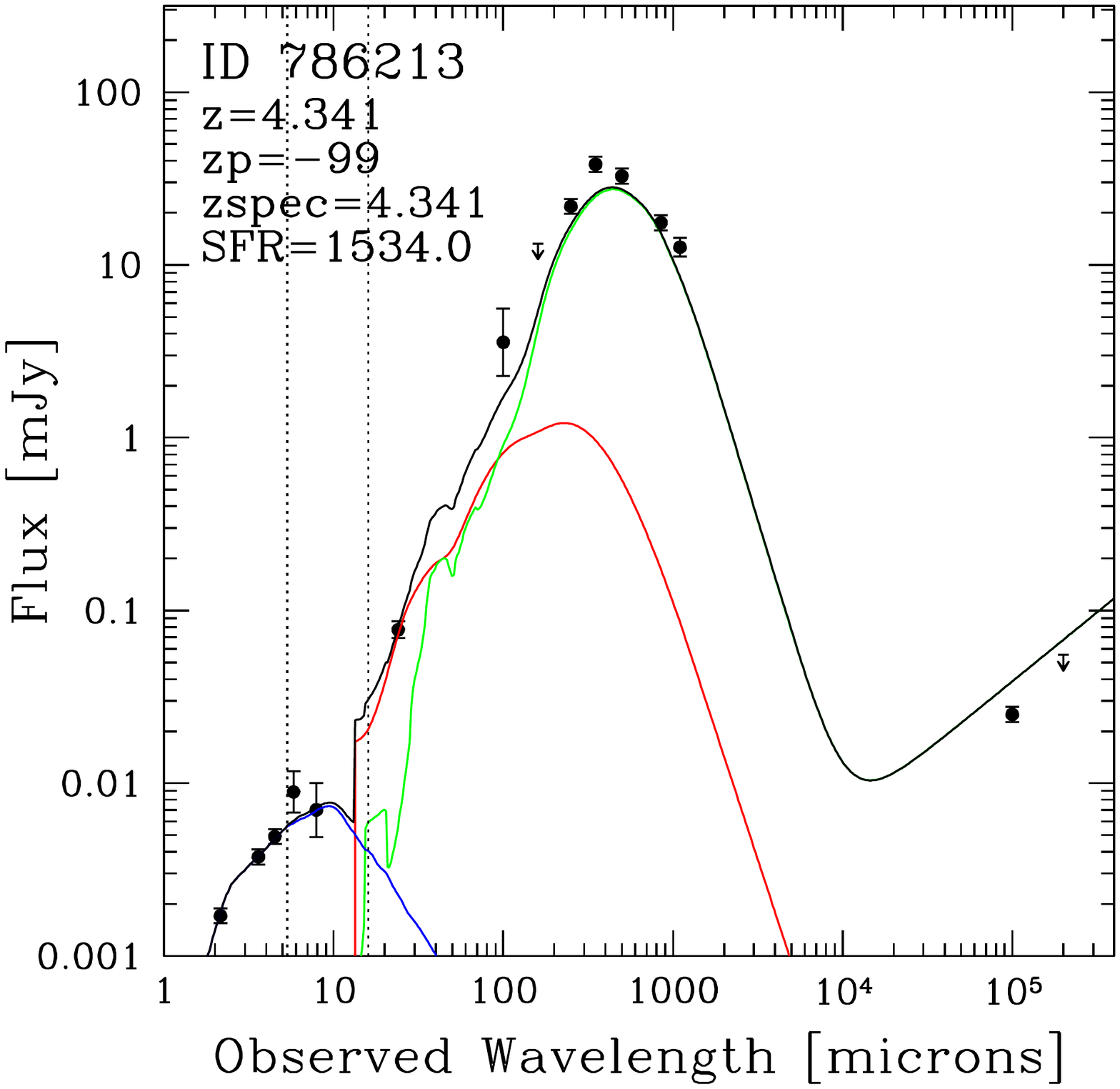}
    \includegraphics[width=0.28\textwidth]{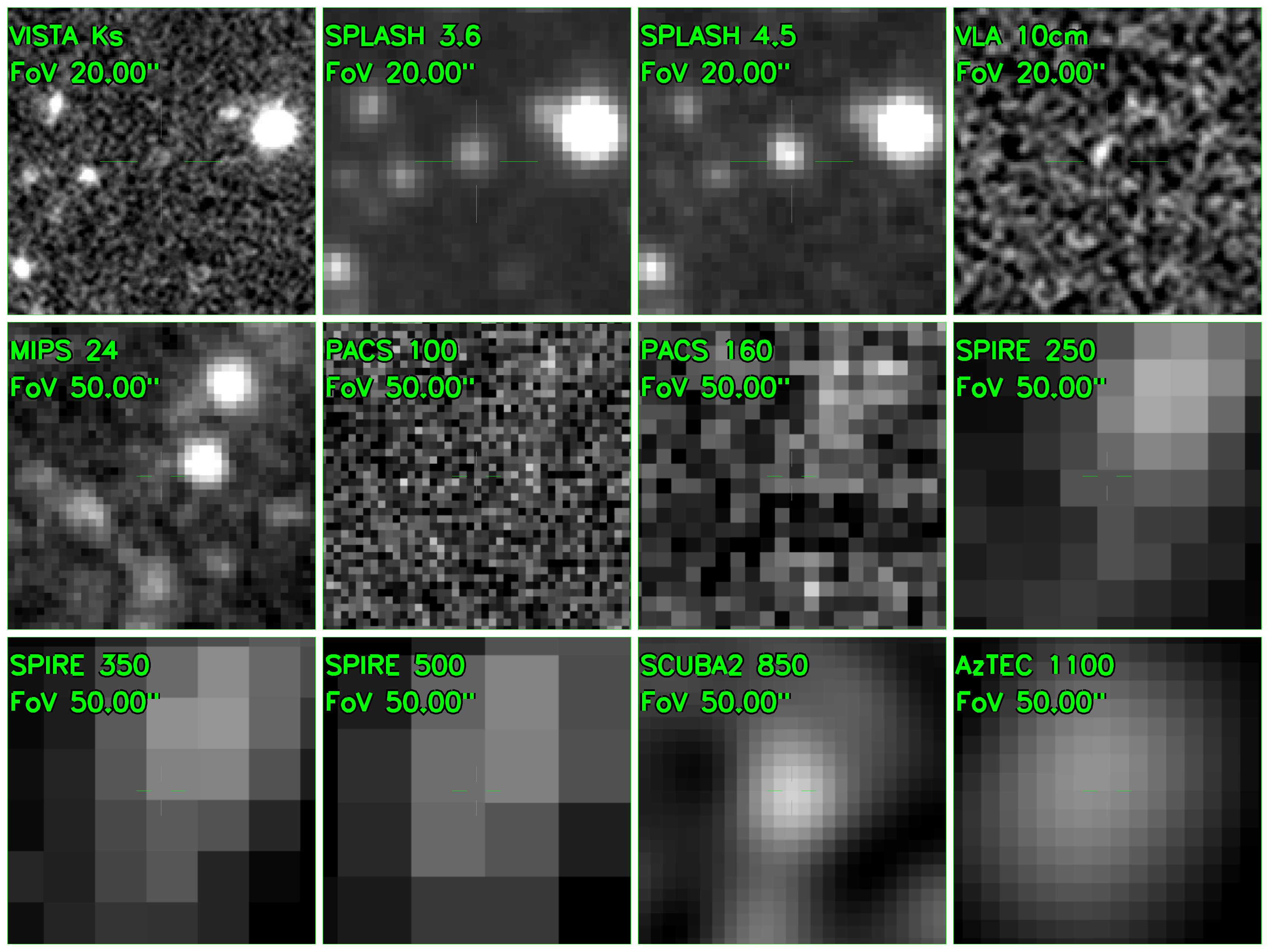}
	\includegraphics[width=0.21\textwidth, trim={0.6cm 5cm 1cm 3.5cm}, clip]{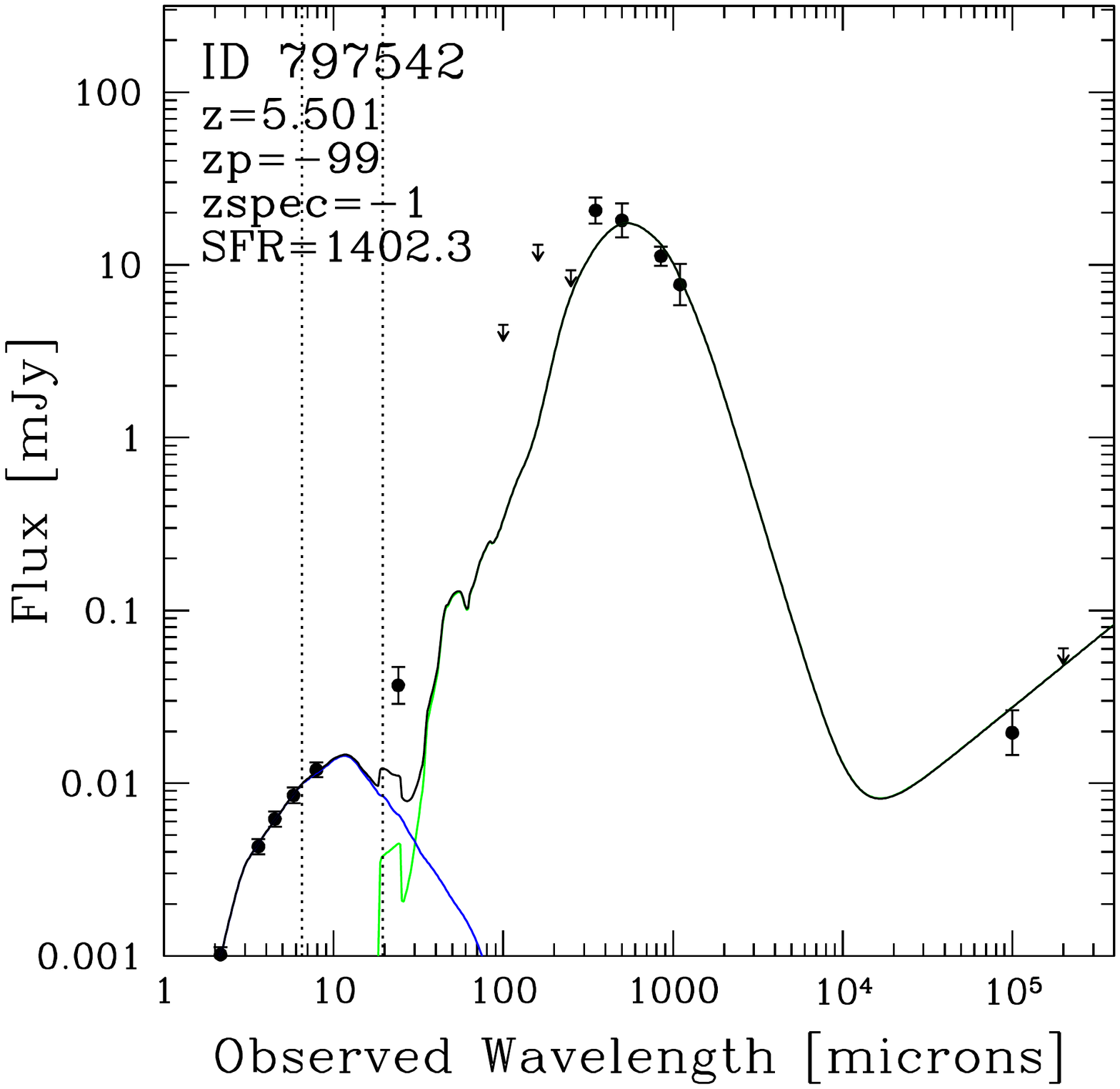}
    \includegraphics[width=0.28\textwidth]{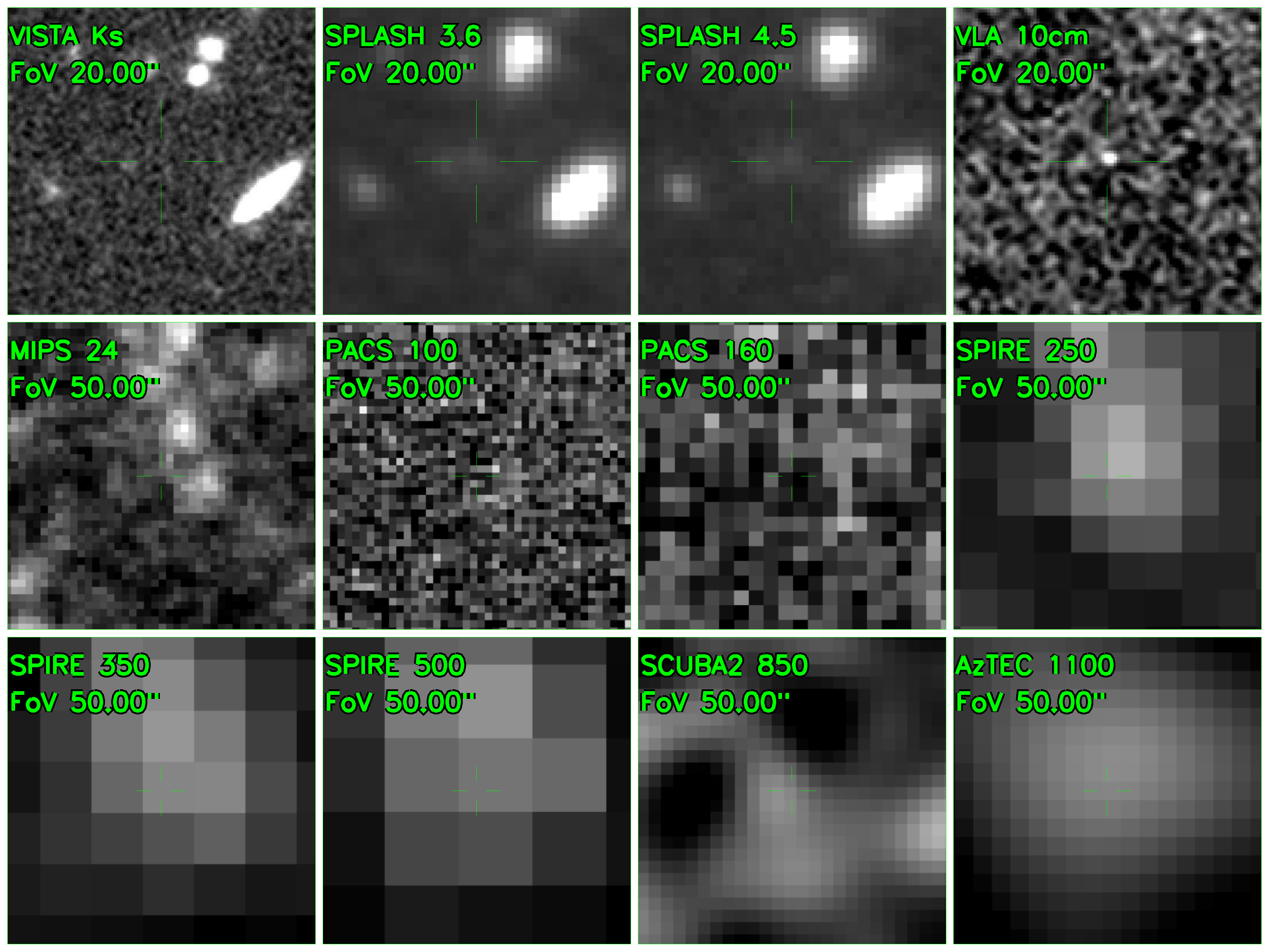}
	\includegraphics[width=0.21\textwidth, trim={0.6cm 5cm 1cm 3.5cm}, clip]{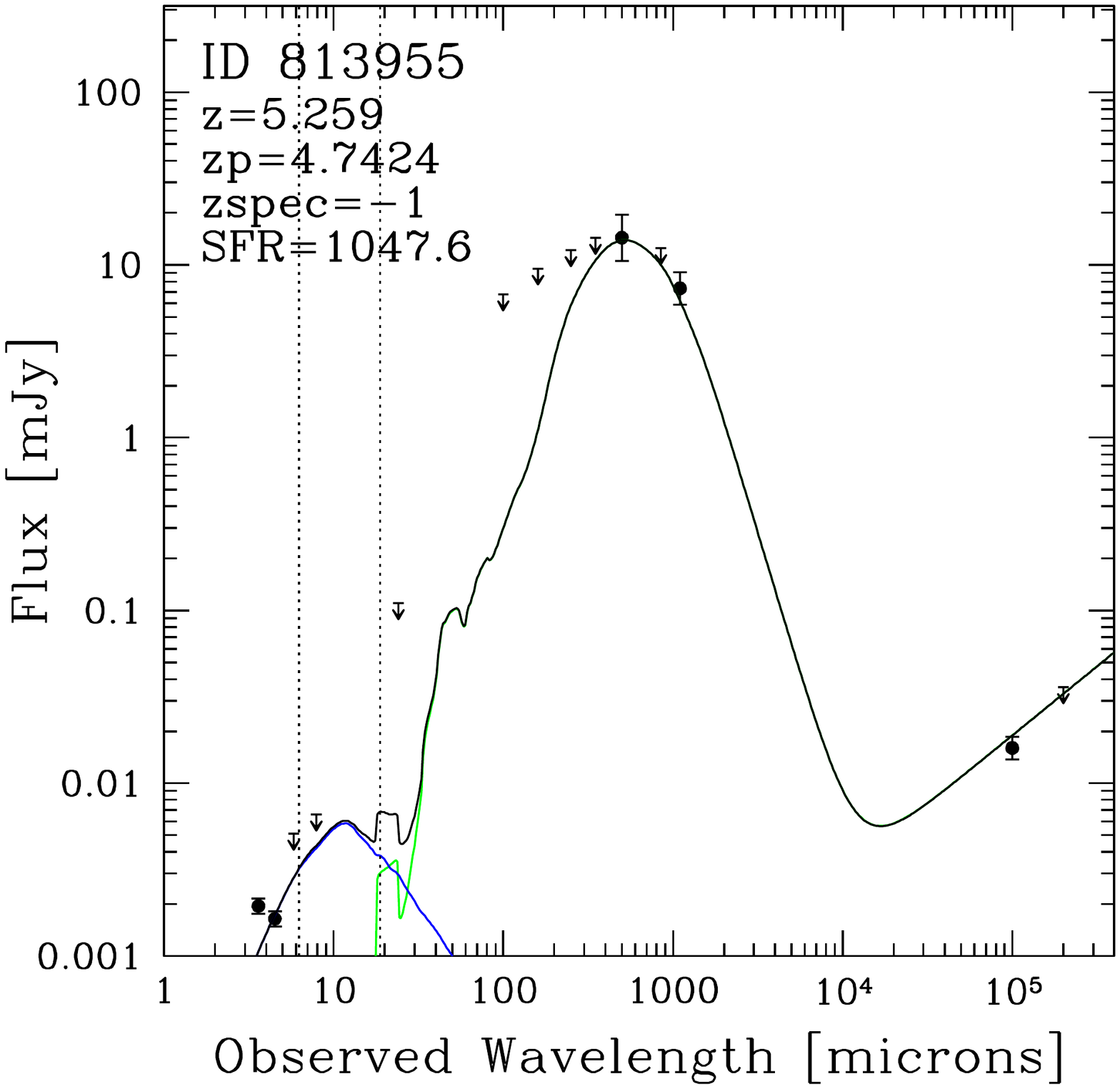}
    \includegraphics[width=0.28\textwidth]{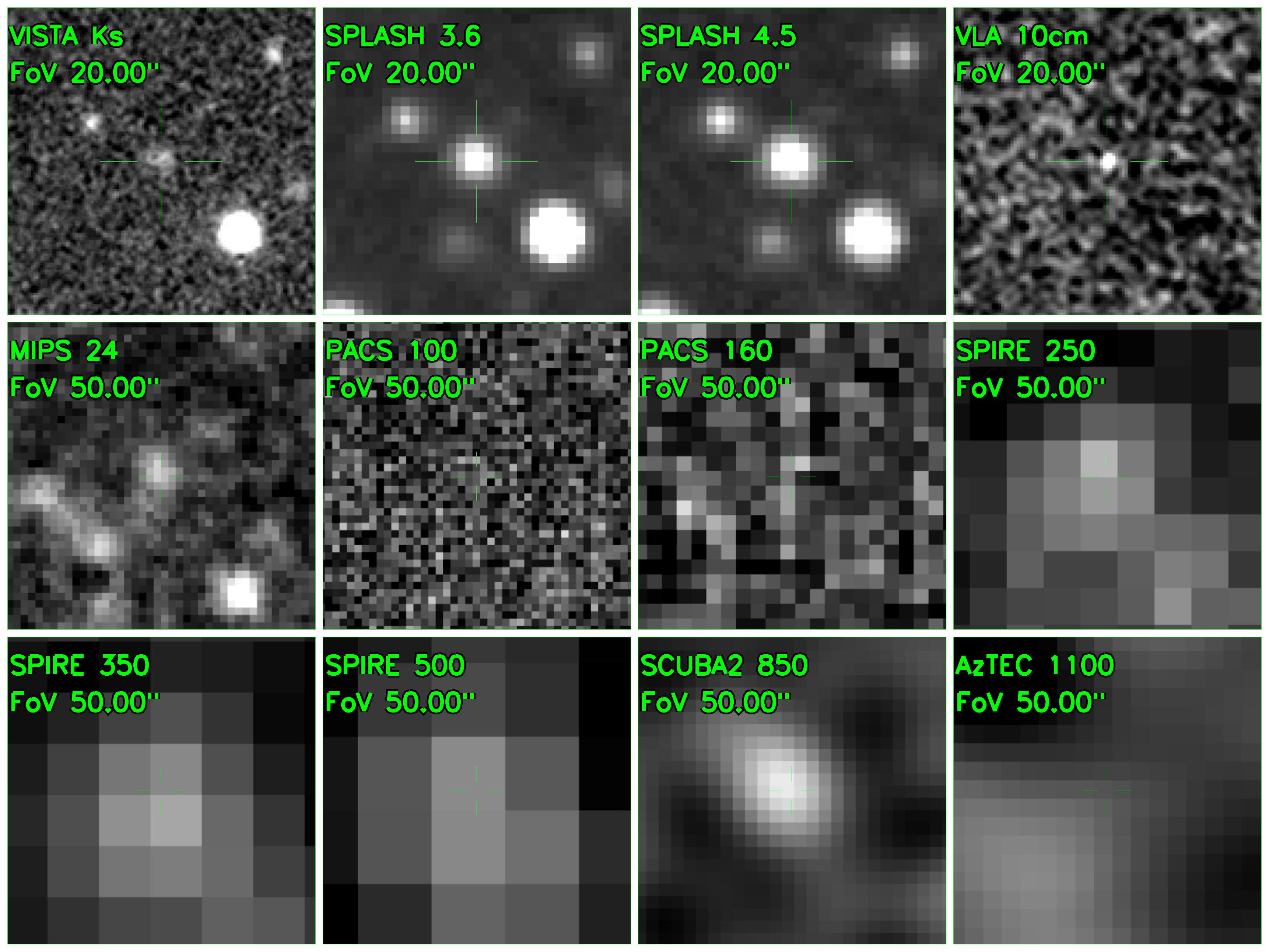}
	\includegraphics[width=0.21\textwidth, trim={0.6cm 5cm 1cm 3.5cm}, clip]{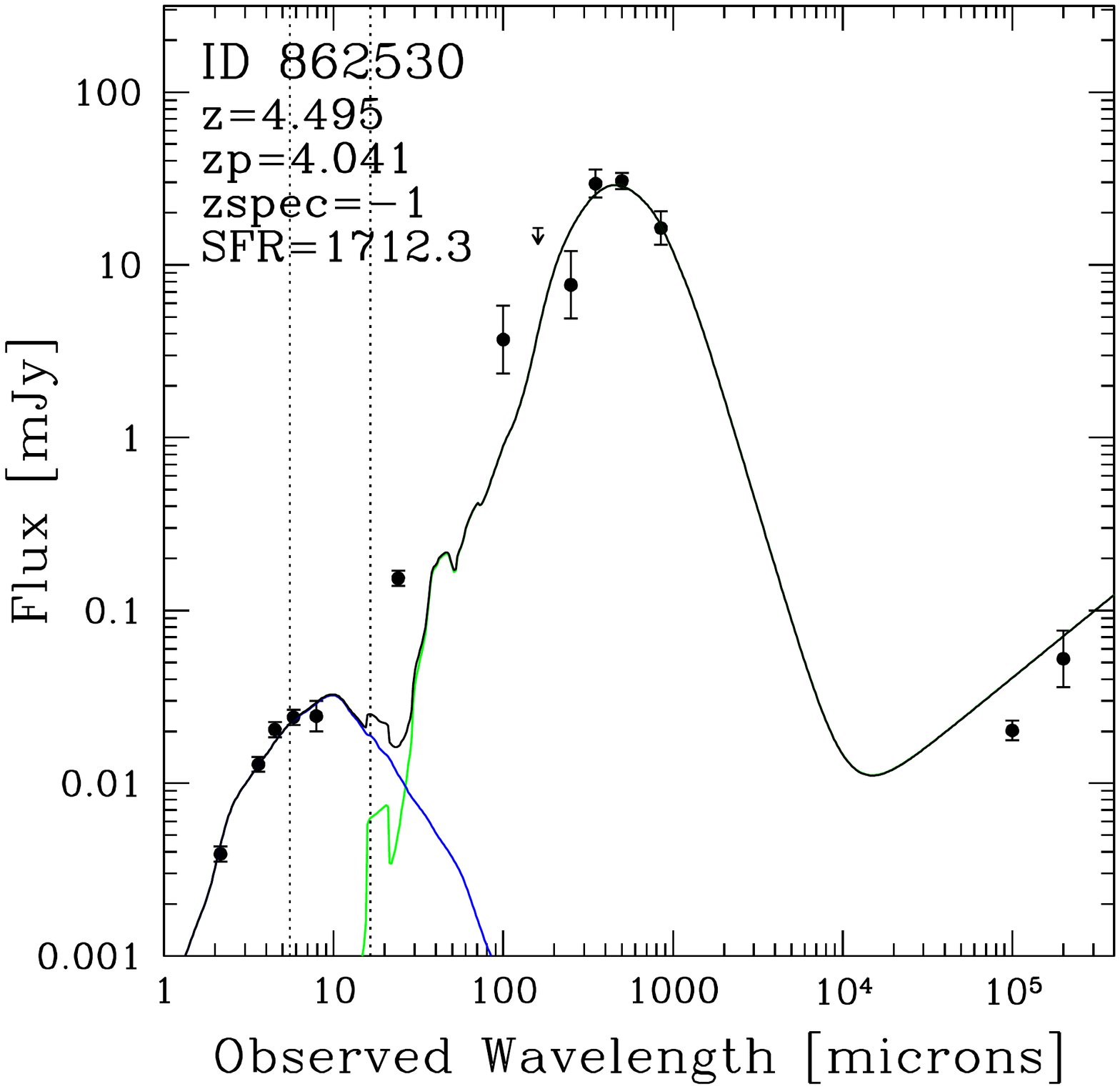}
    \includegraphics[width=0.28\textwidth]{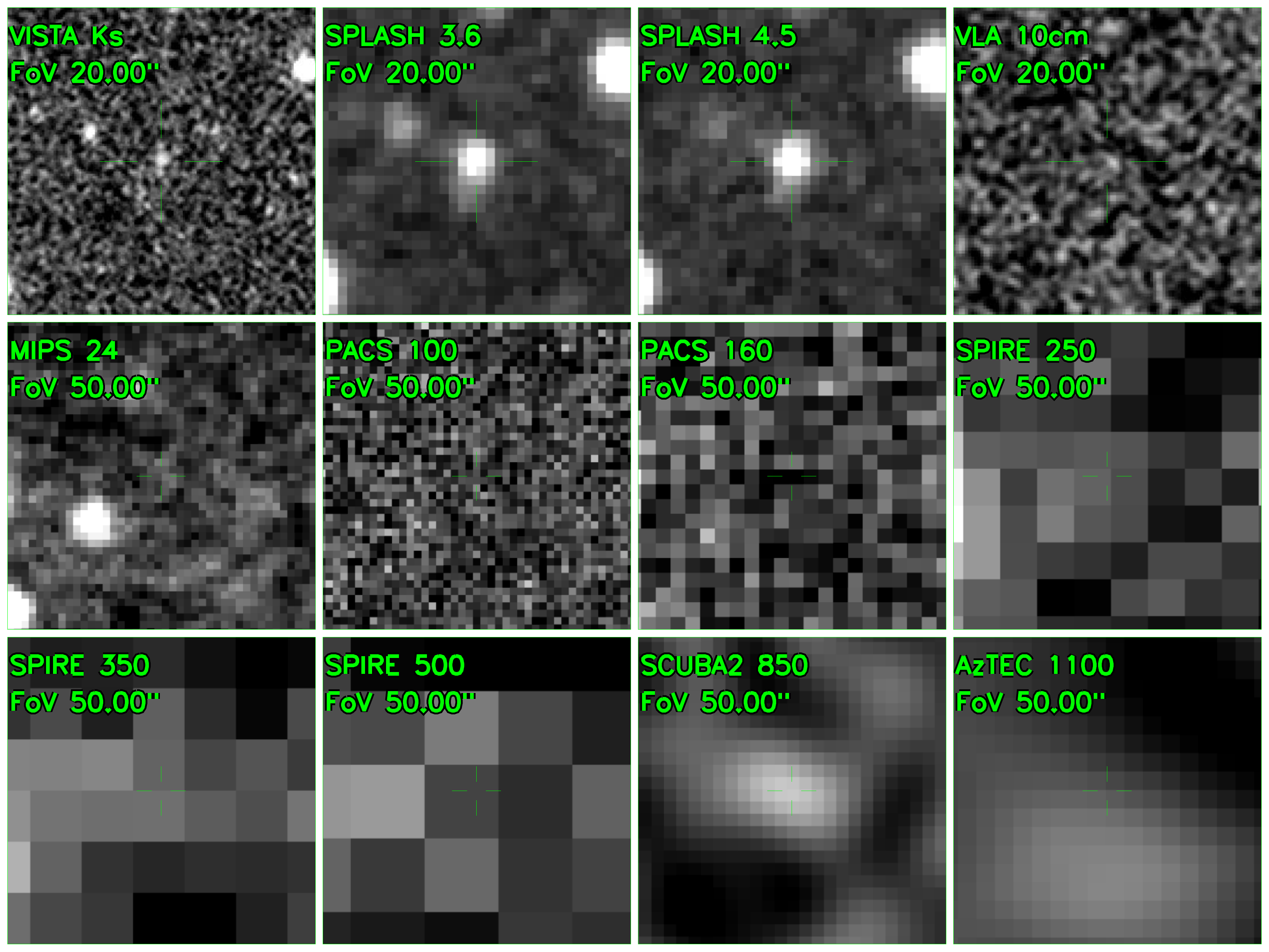}
	\includegraphics[width=0.21\textwidth, trim={0.6cm 5cm 1cm 3.5cm}, clip]{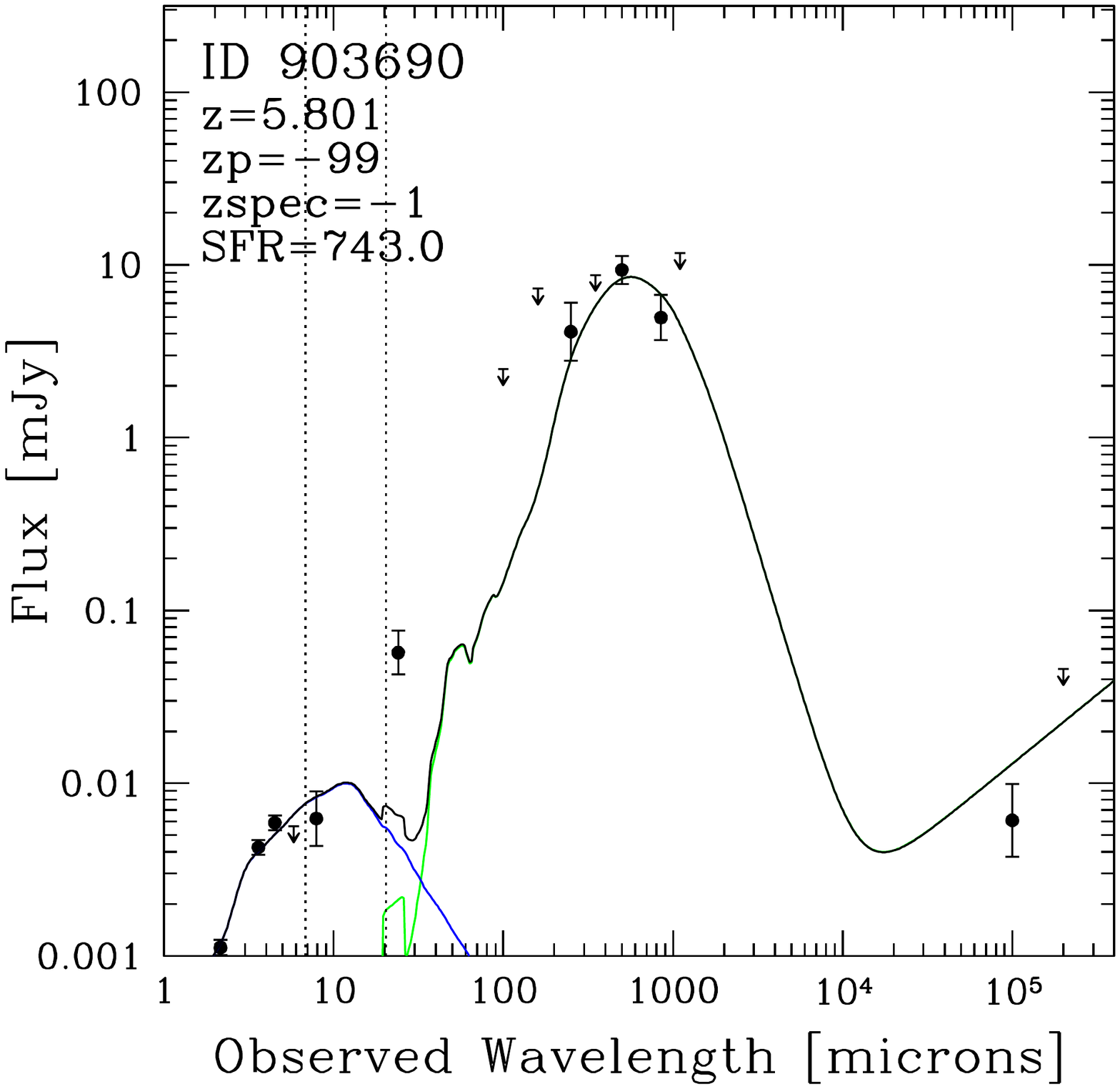}
    \includegraphics[width=0.28\textwidth]{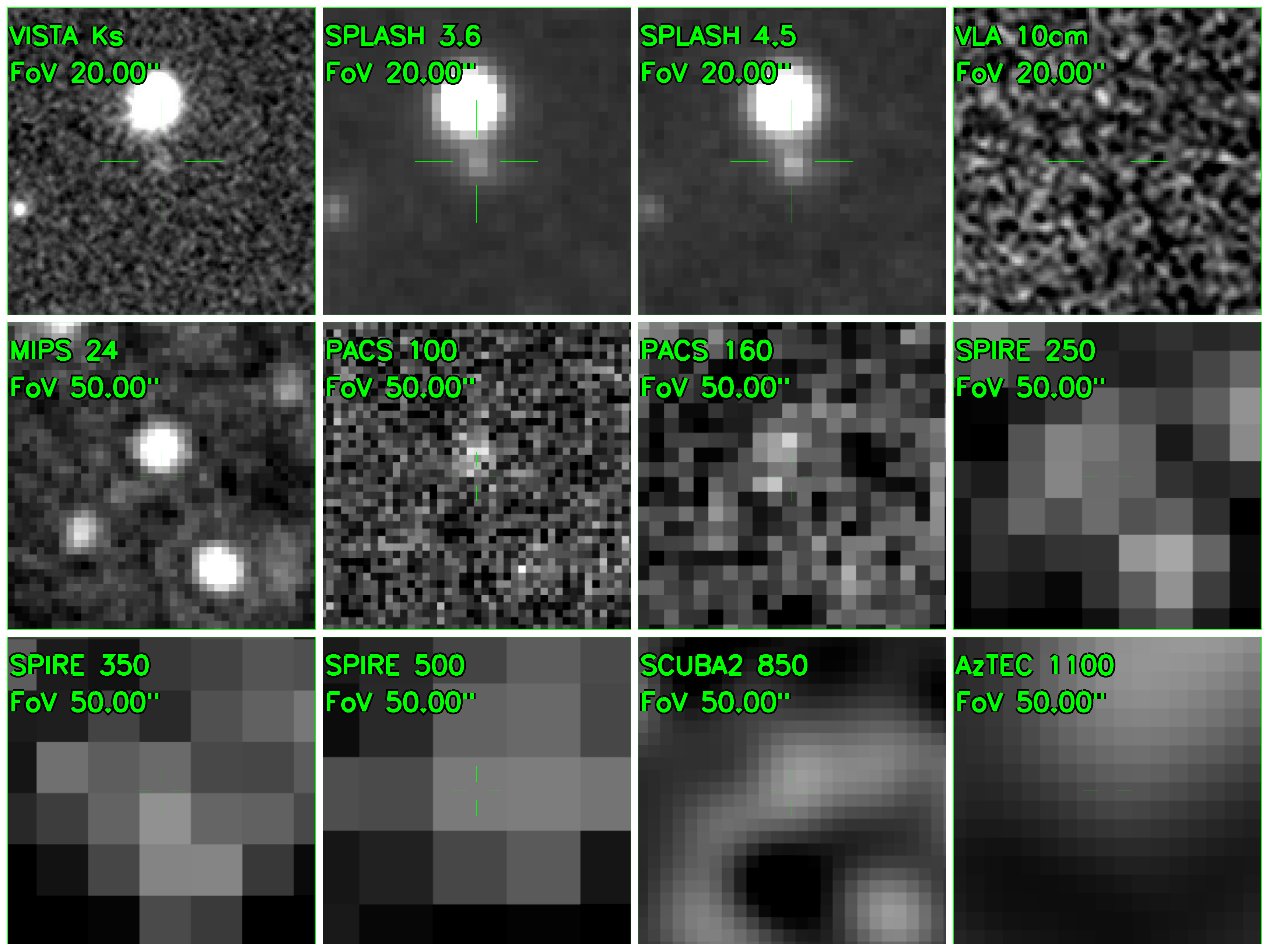}
	\includegraphics[width=0.21\textwidth, trim={0.6cm 5cm 1cm 3.5cm}, clip]{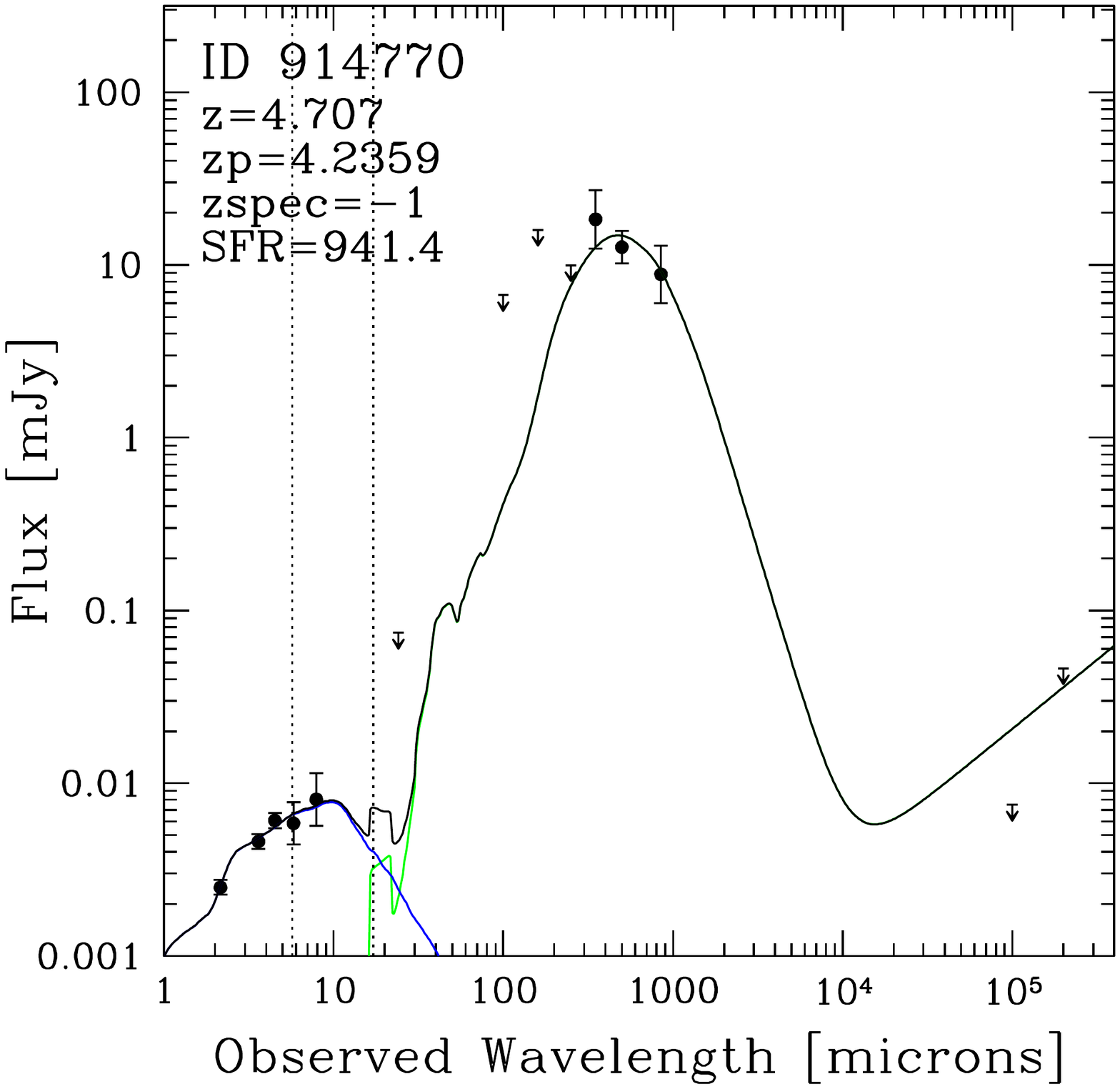}
    \includegraphics[width=0.28\textwidth]{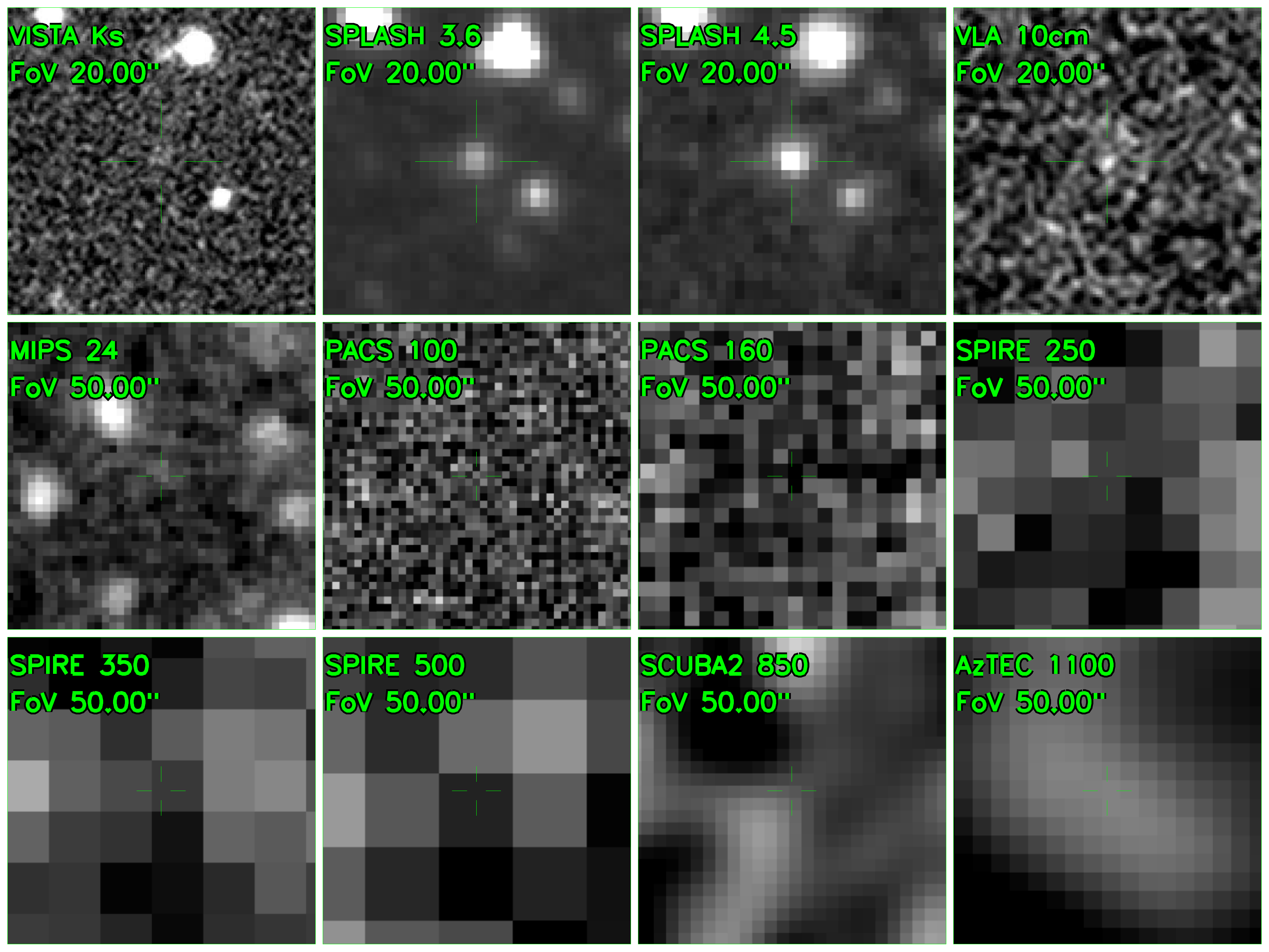}
	\includegraphics[width=0.21\textwidth, trim={0.6cm 5cm 1cm 3.5cm}, clip]{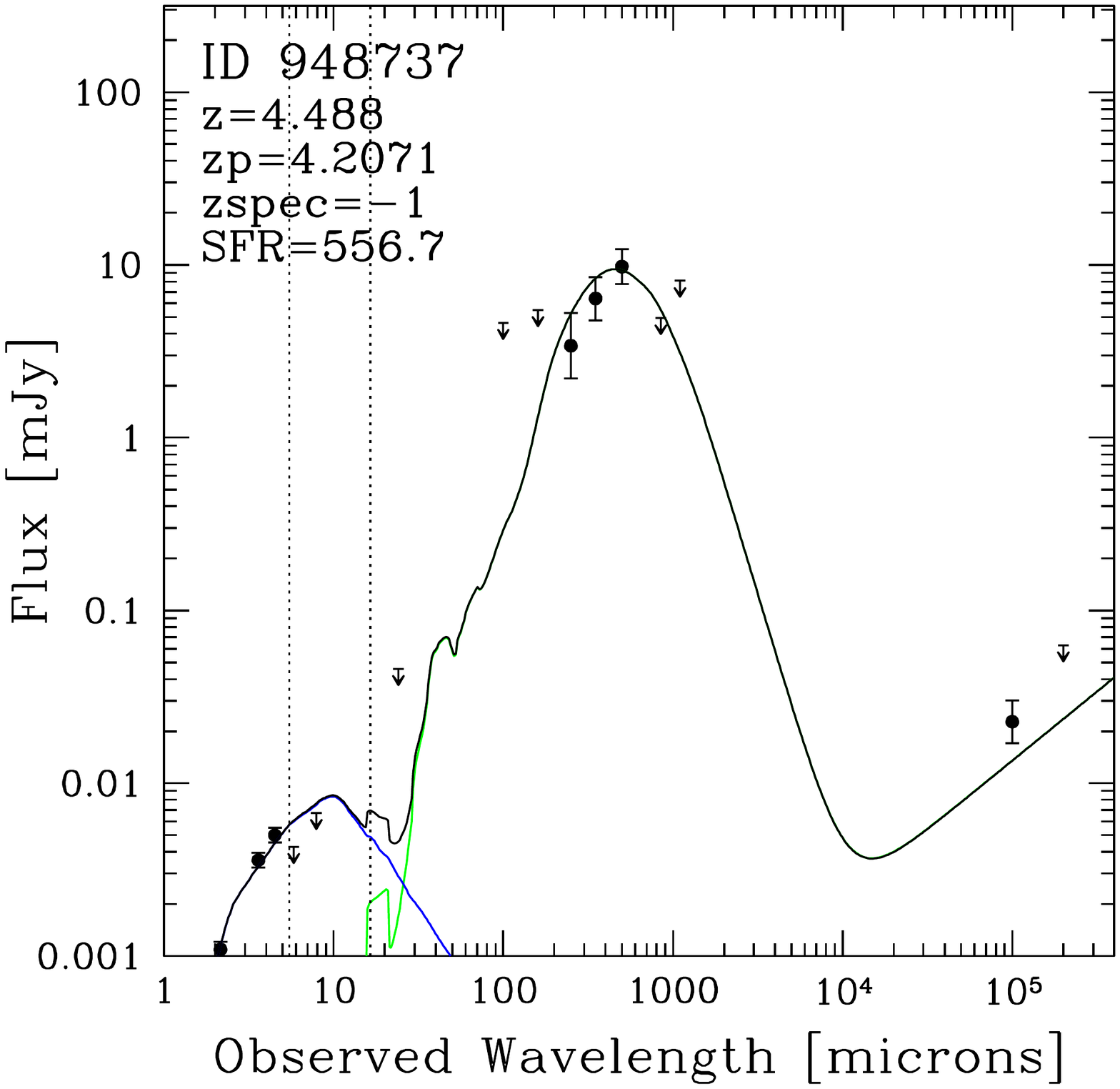}
    \includegraphics[width=0.28\textwidth]{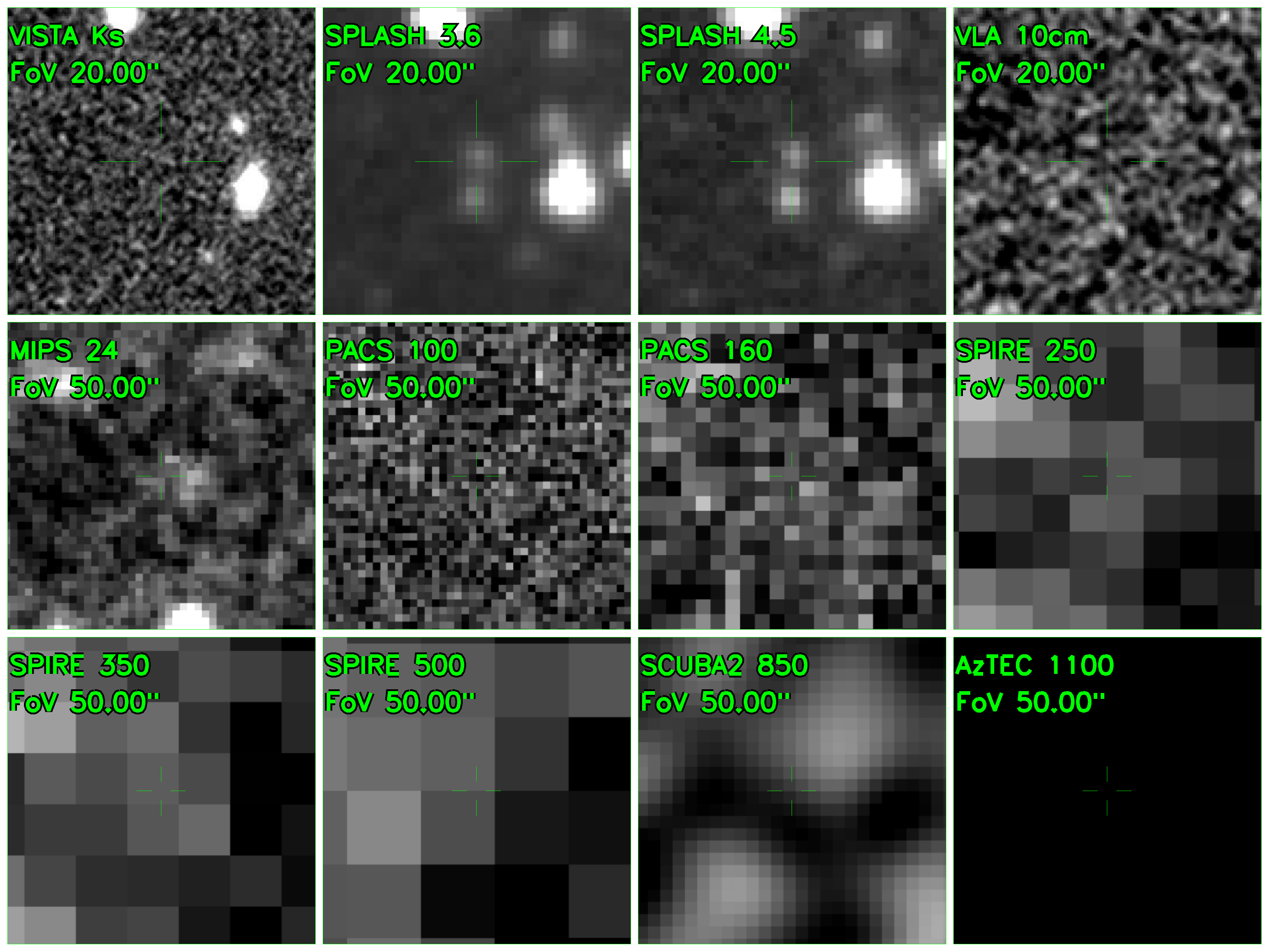}
	\includegraphics[width=0.21\textwidth, trim={0.6cm 5cm 1cm 3.5cm}, clip]{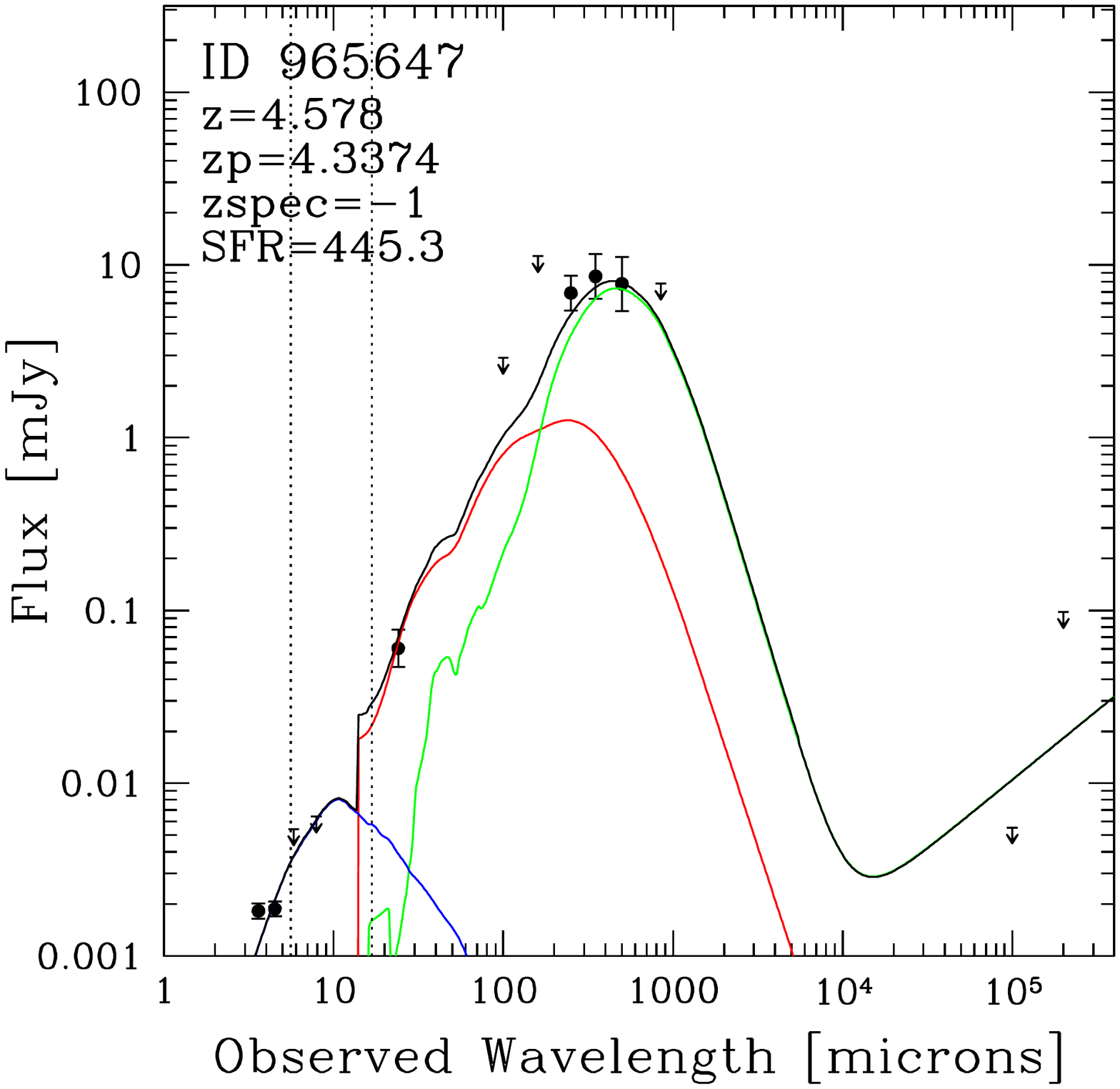}
    \includegraphics[width=0.28\textwidth]{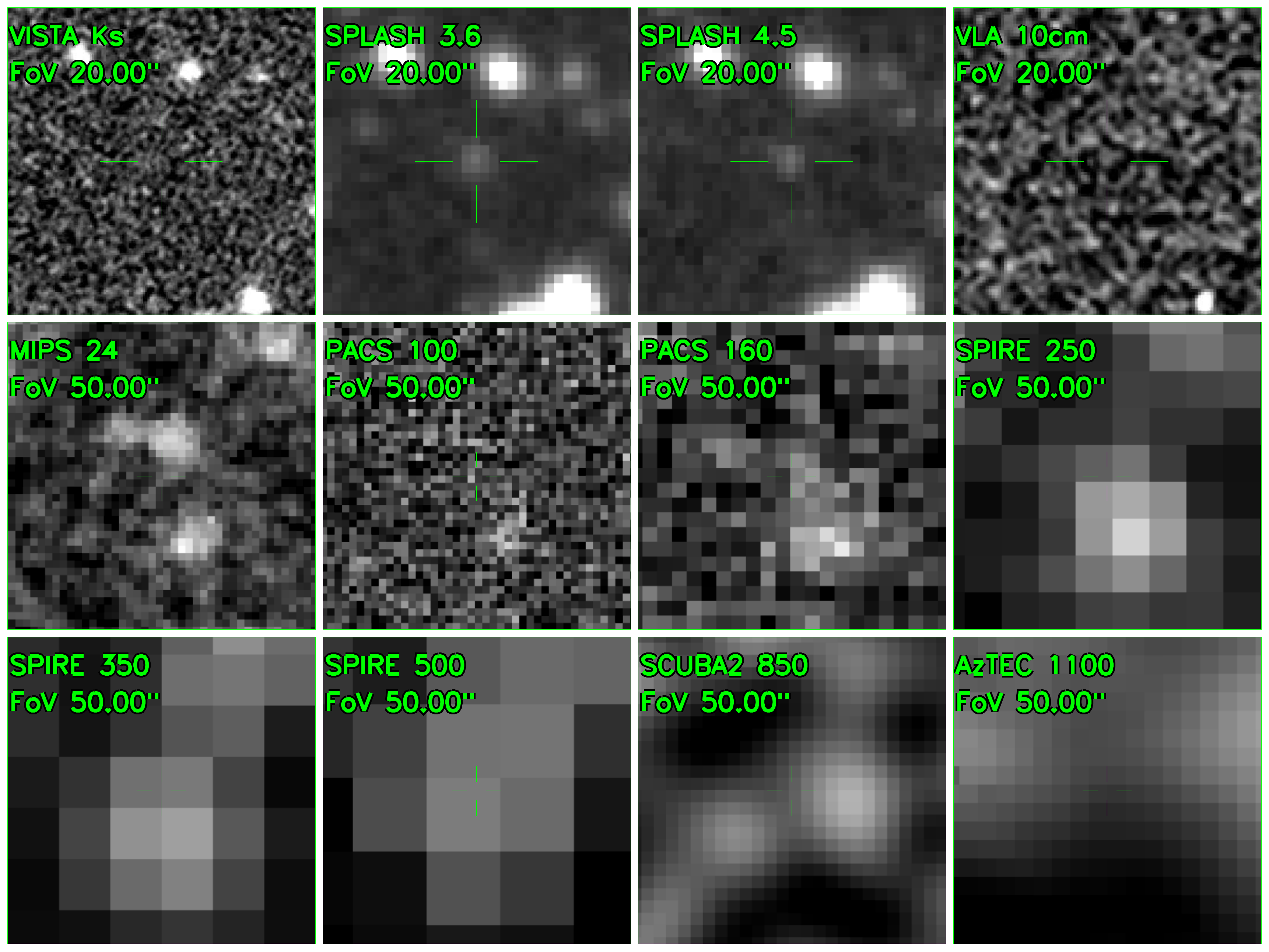}
	\includegraphics[width=0.21\textwidth, trim={0.6cm 5cm 1cm 3.5cm}, clip]{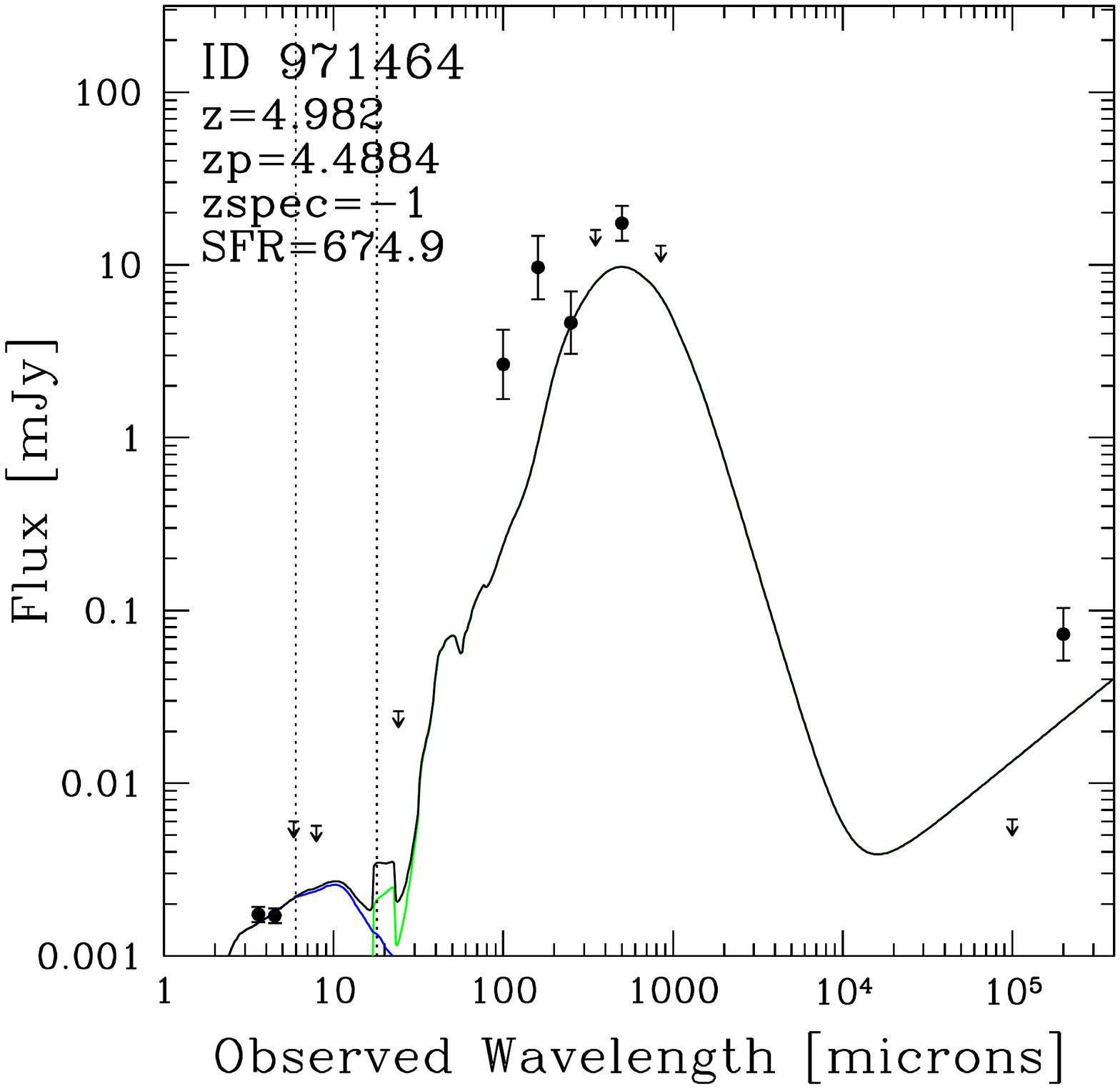}
	\caption{%
		Multi-band cutouts and SEDs of high redshift candidates.
		\label{highz_cutouts1}
		}
\end{figure}

\begin{figure}
	\centering%
    \includegraphics[width=0.28\textwidth]{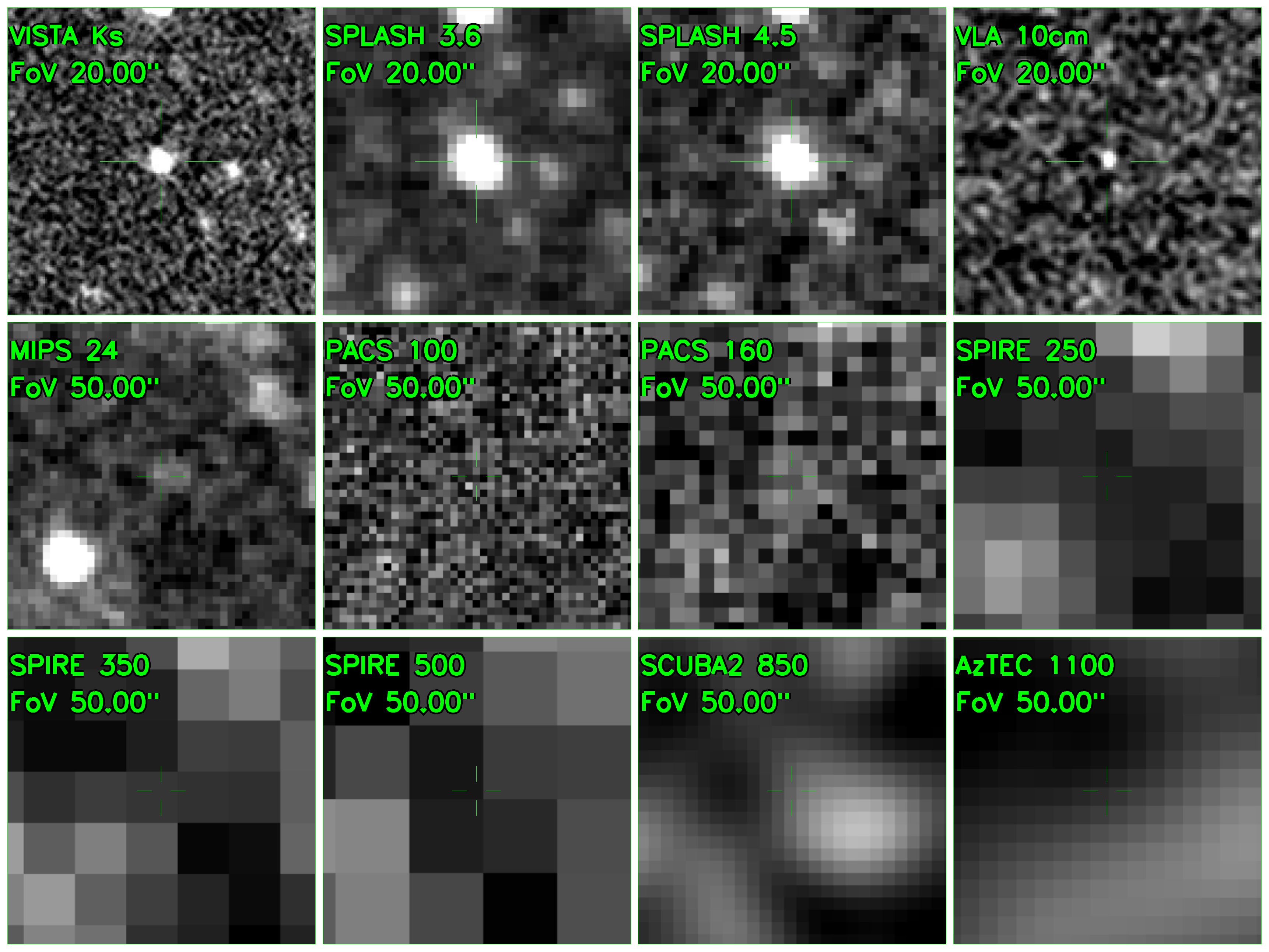}
	\includegraphics[width=0.21\textwidth, trim={0.6cm 5cm 1cm 3.5cm}, clip]{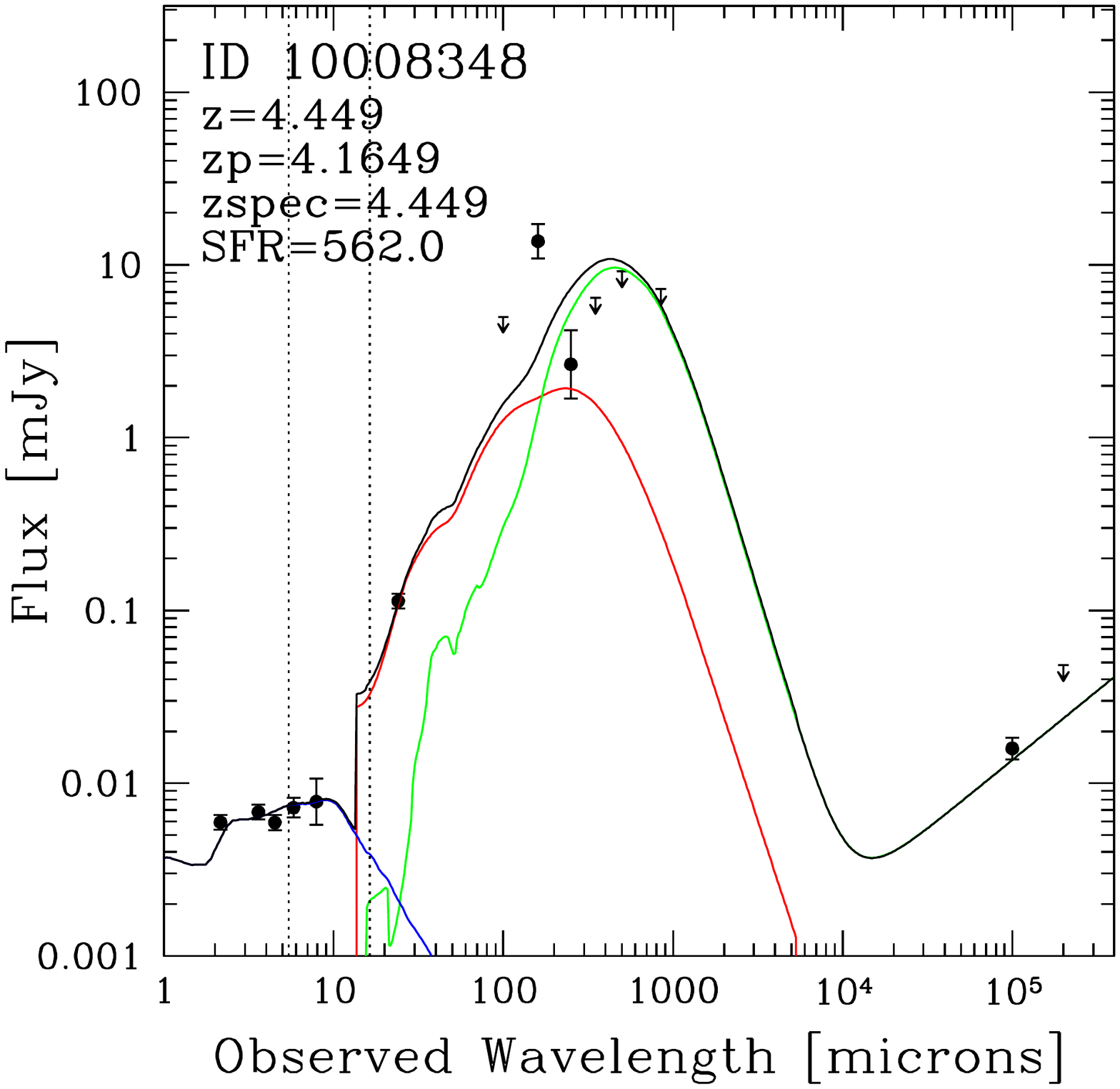}
    \includegraphics[width=0.28\textwidth]{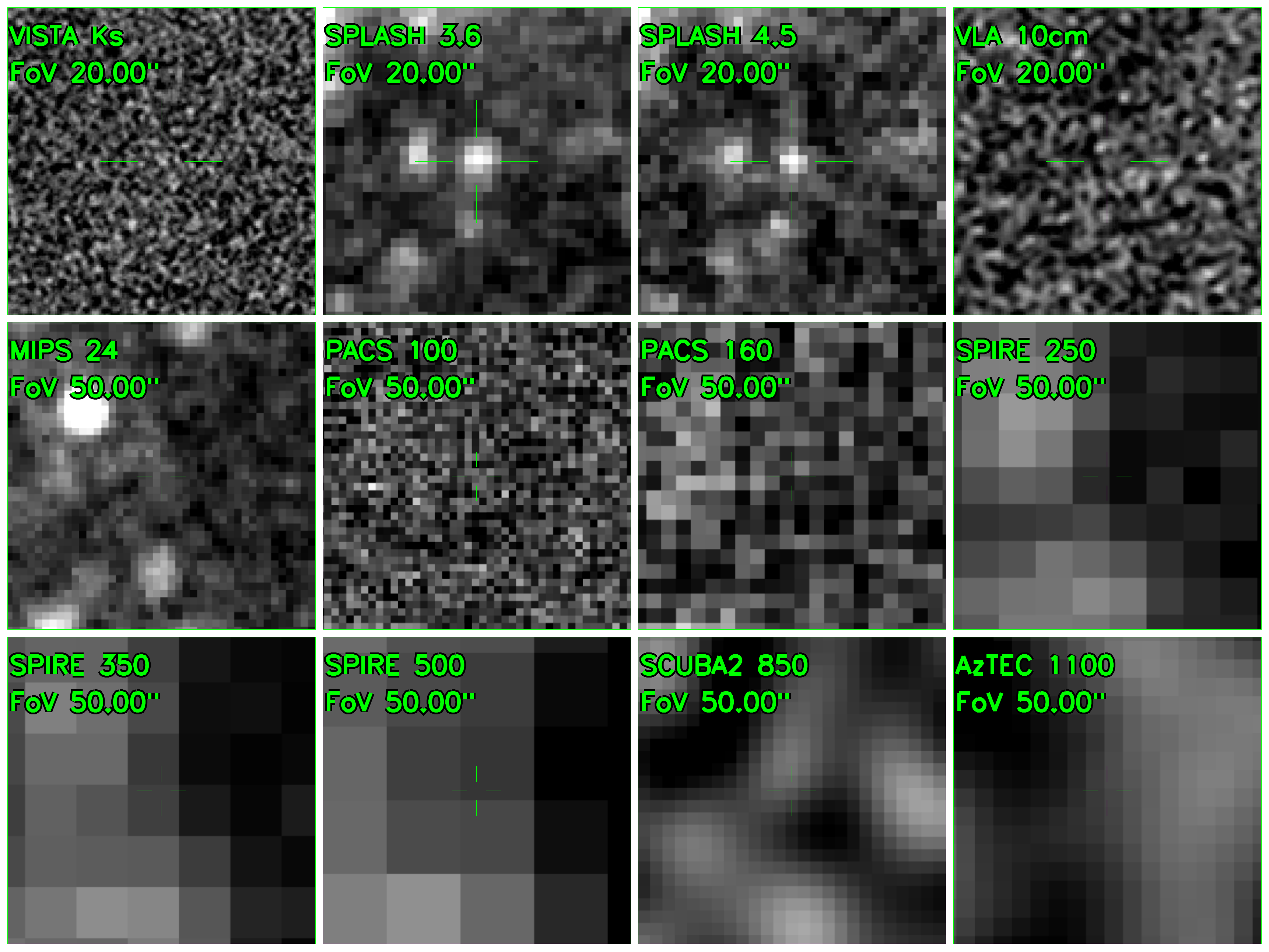}
	\includegraphics[width=0.21\textwidth, trim={0.6cm 5cm 1cm 3.5cm}, clip]{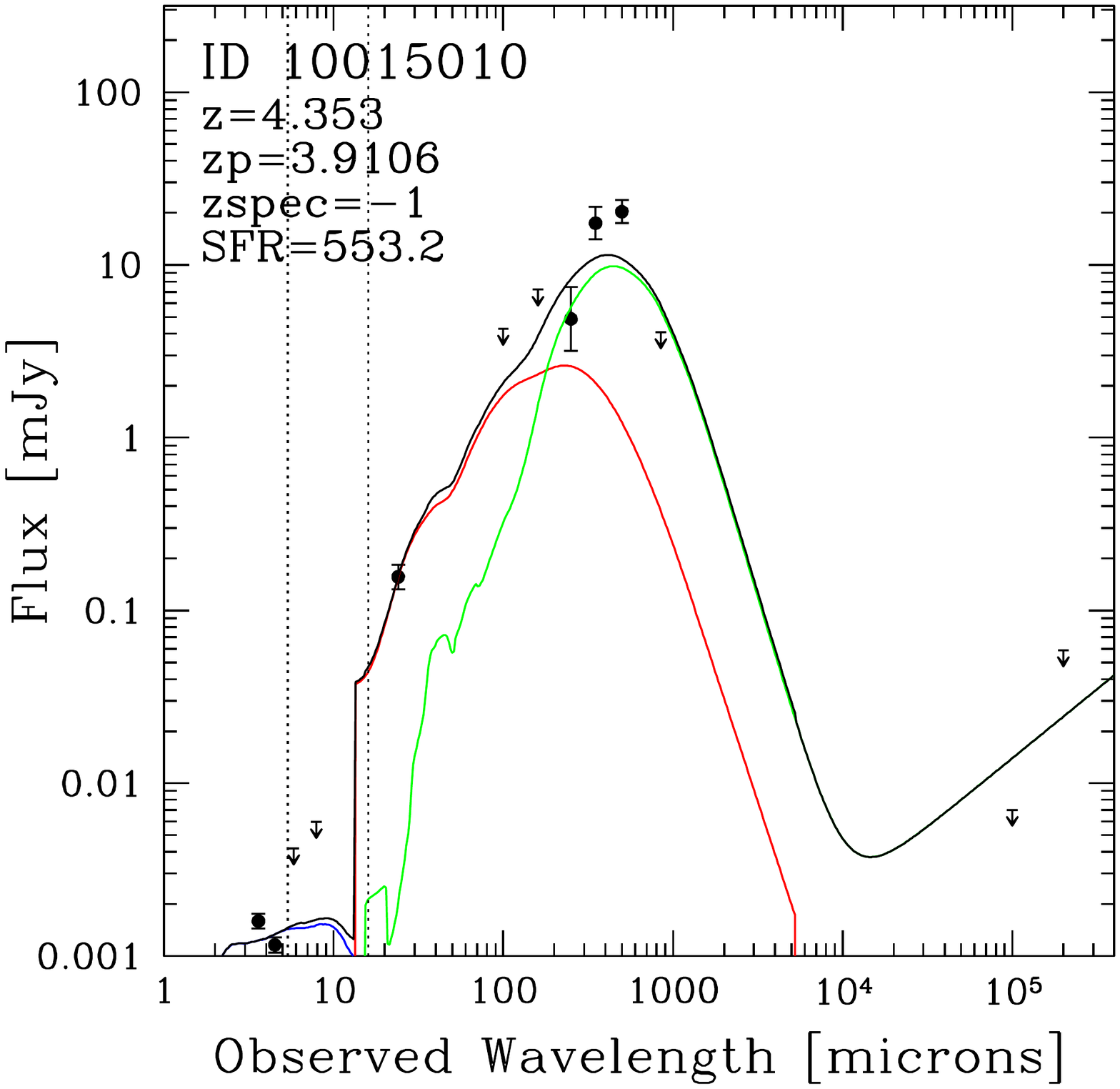}
    \includegraphics[width=0.28\textwidth]{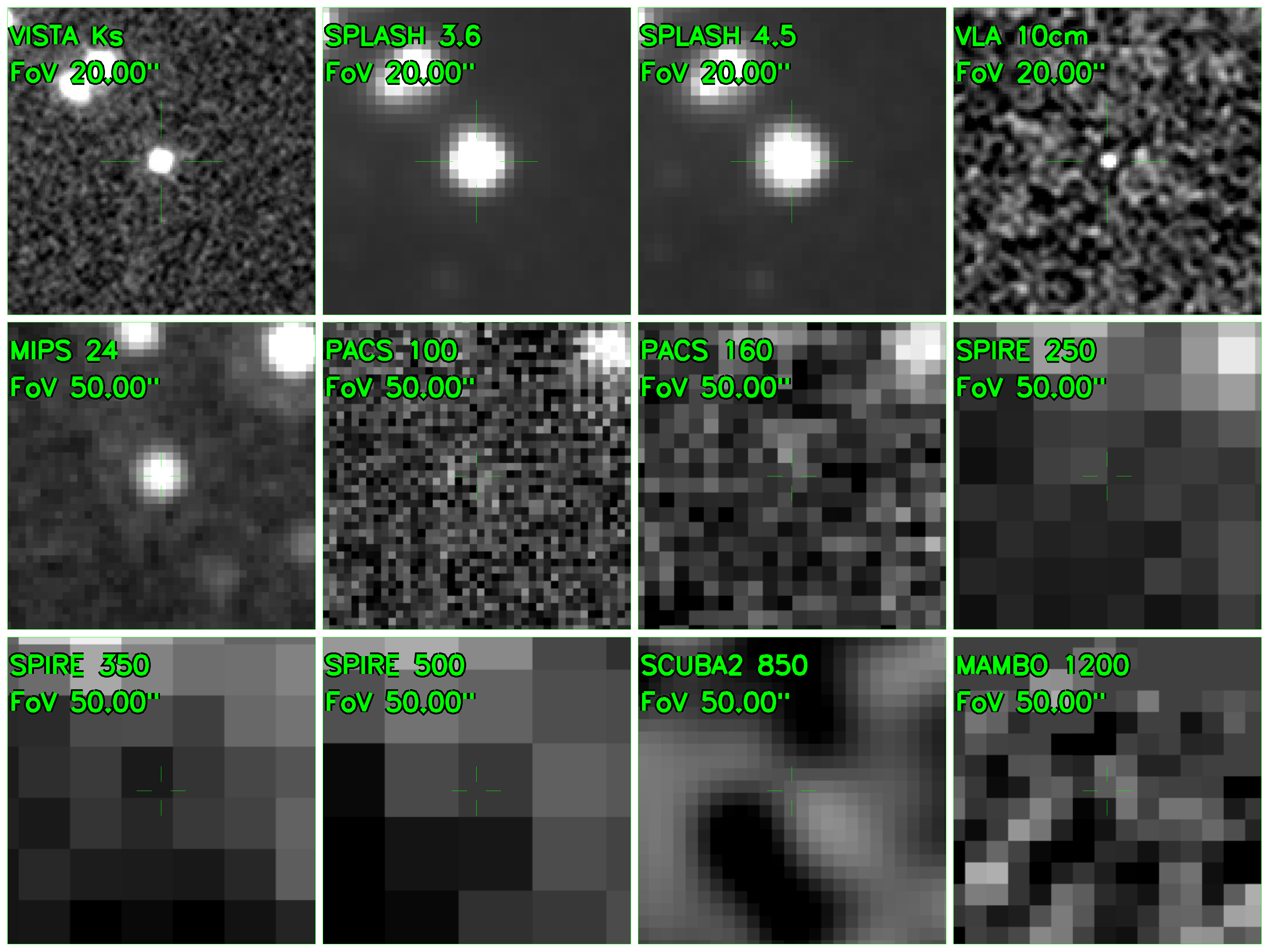}
	\includegraphics[width=0.21\textwidth, trim={0.6cm 5cm 1cm 3.5cm}, clip]{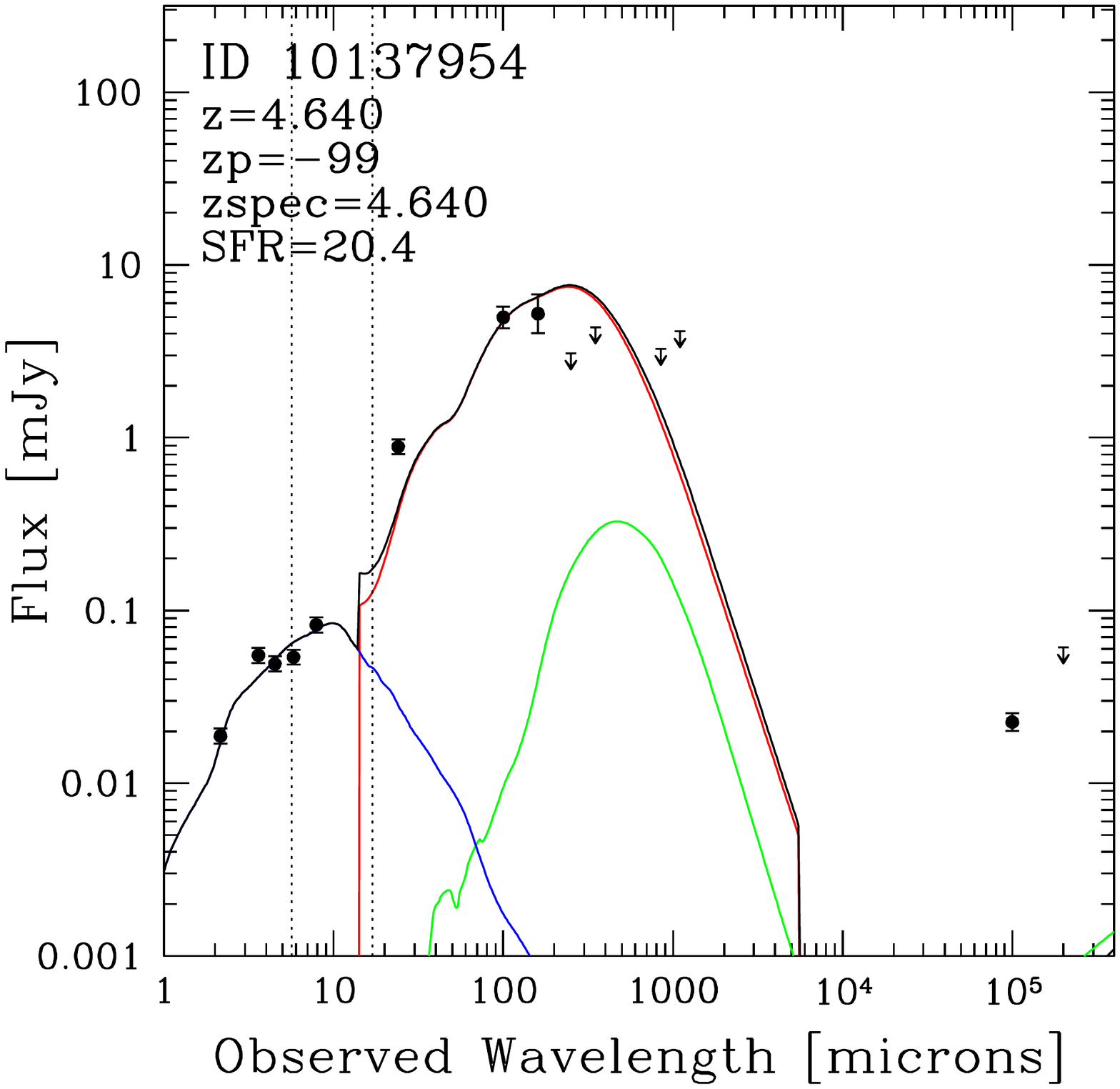}
    \includegraphics[width=0.28\textwidth]{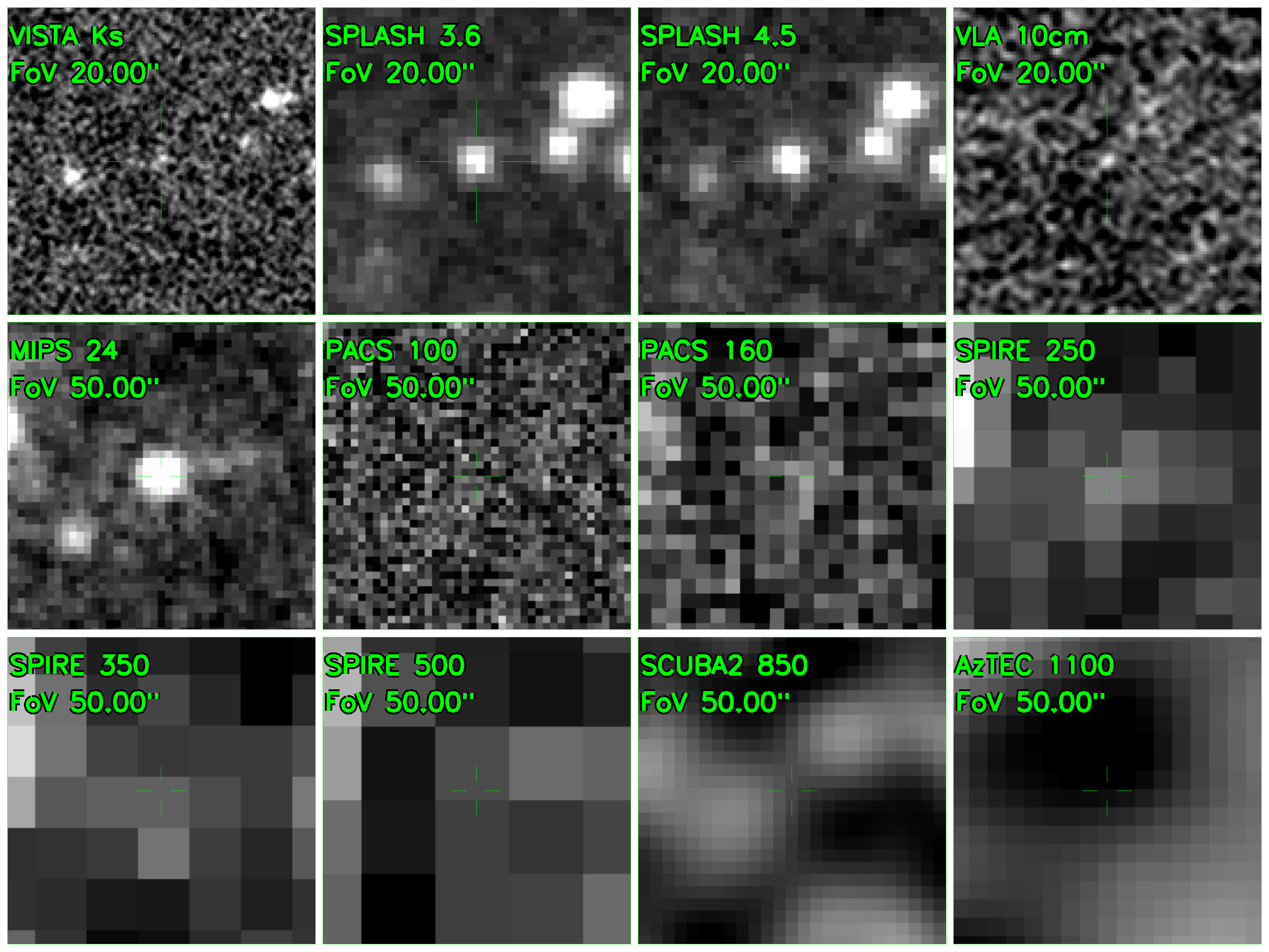}
	\includegraphics[width=0.21\textwidth, trim={0.6cm 5cm 1cm 3.5cm}, clip]{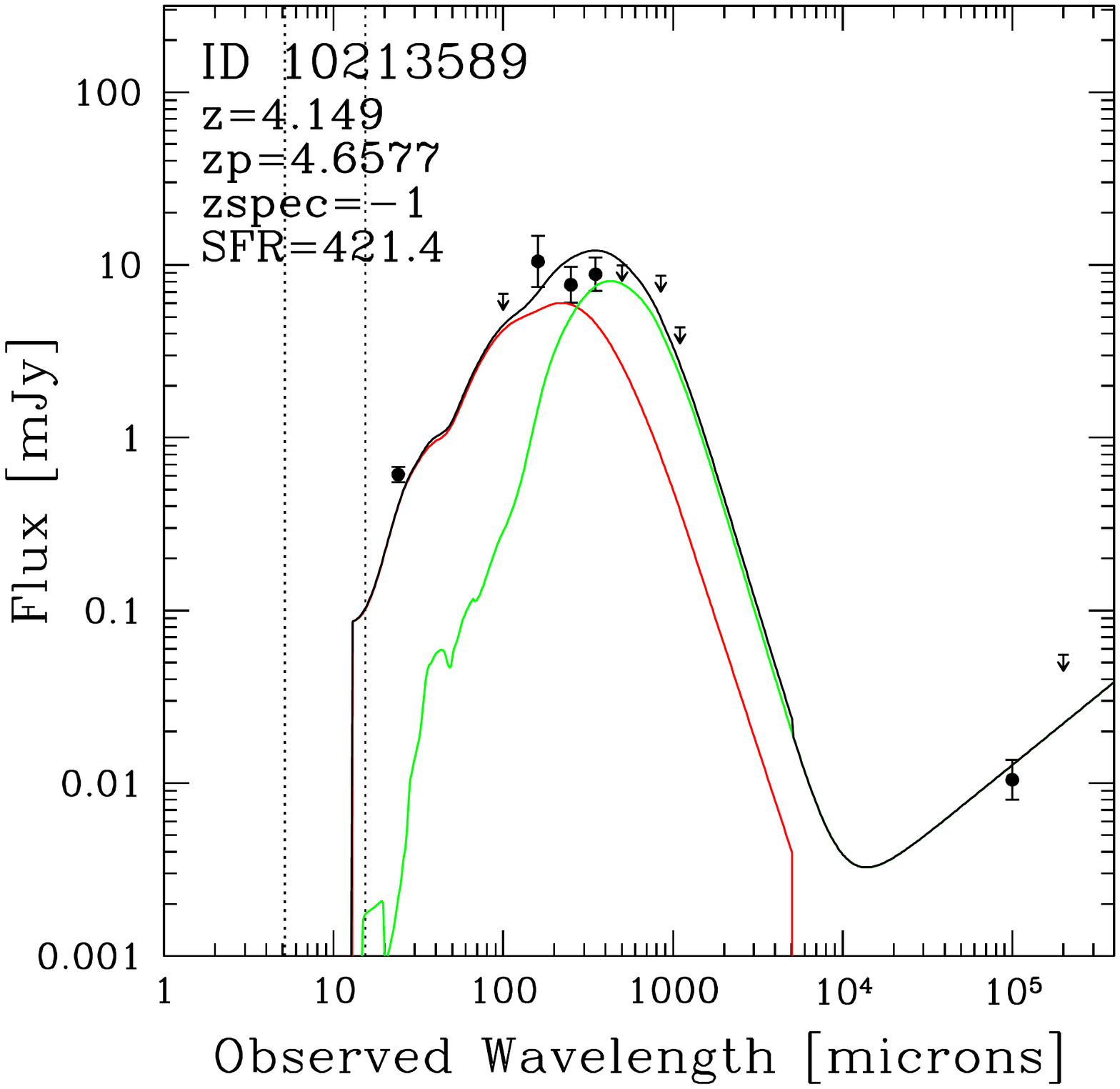}
    \includegraphics[width=0.28\textwidth]{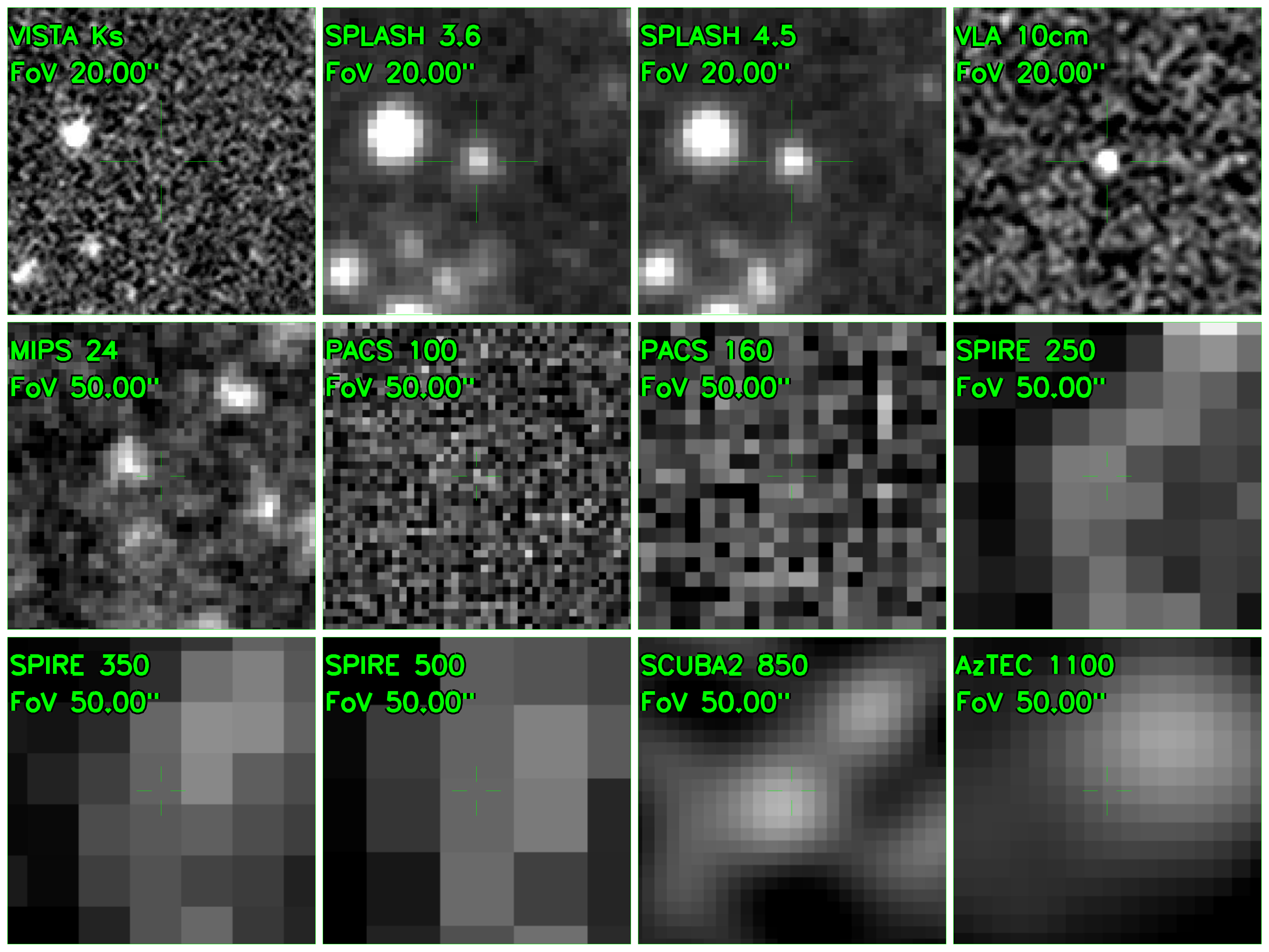}
	\includegraphics[width=0.21\textwidth, trim={0.6cm 5cm 1cm 3.5cm}, clip]{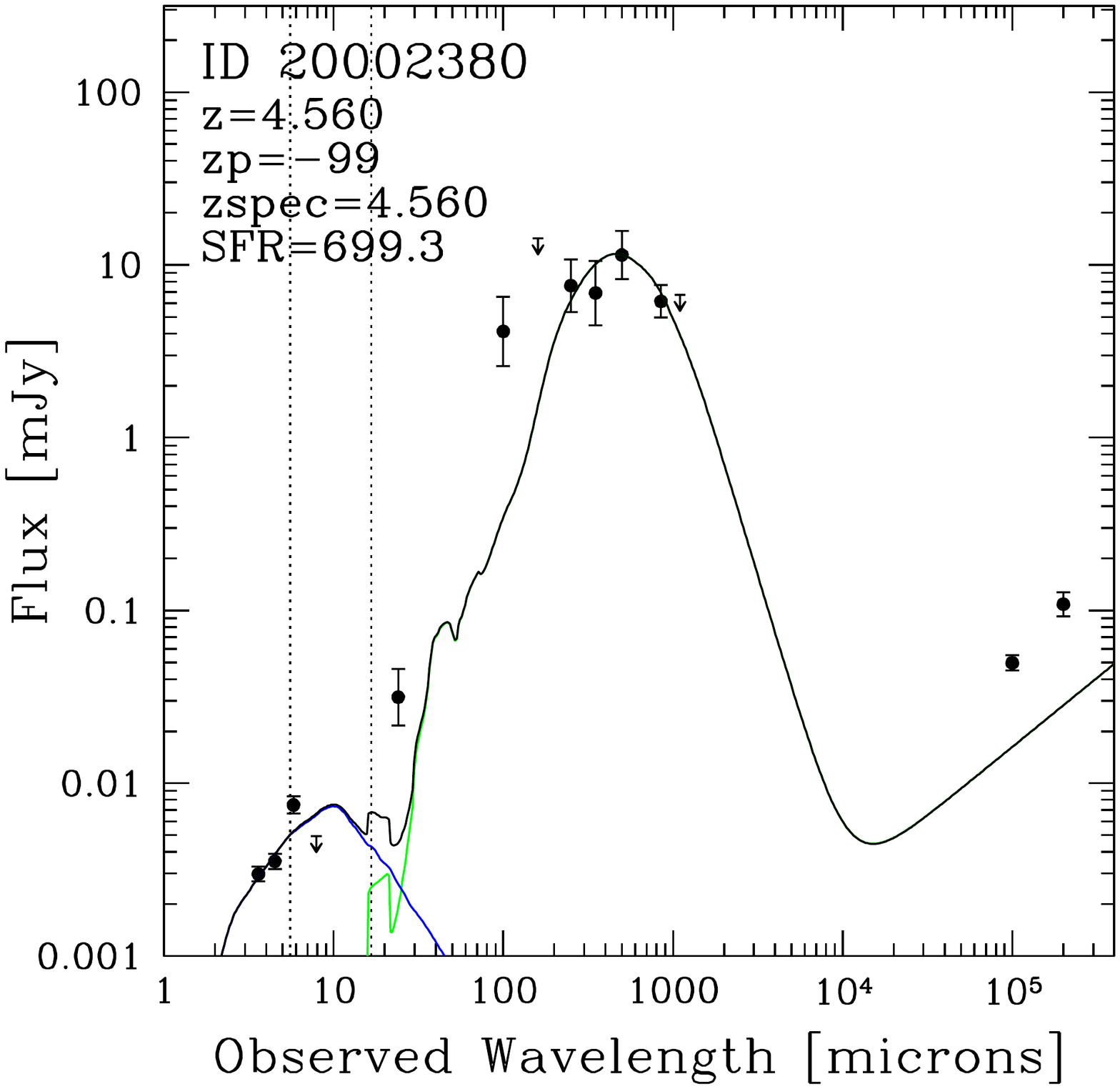}
    \includegraphics[width=0.28\textwidth]{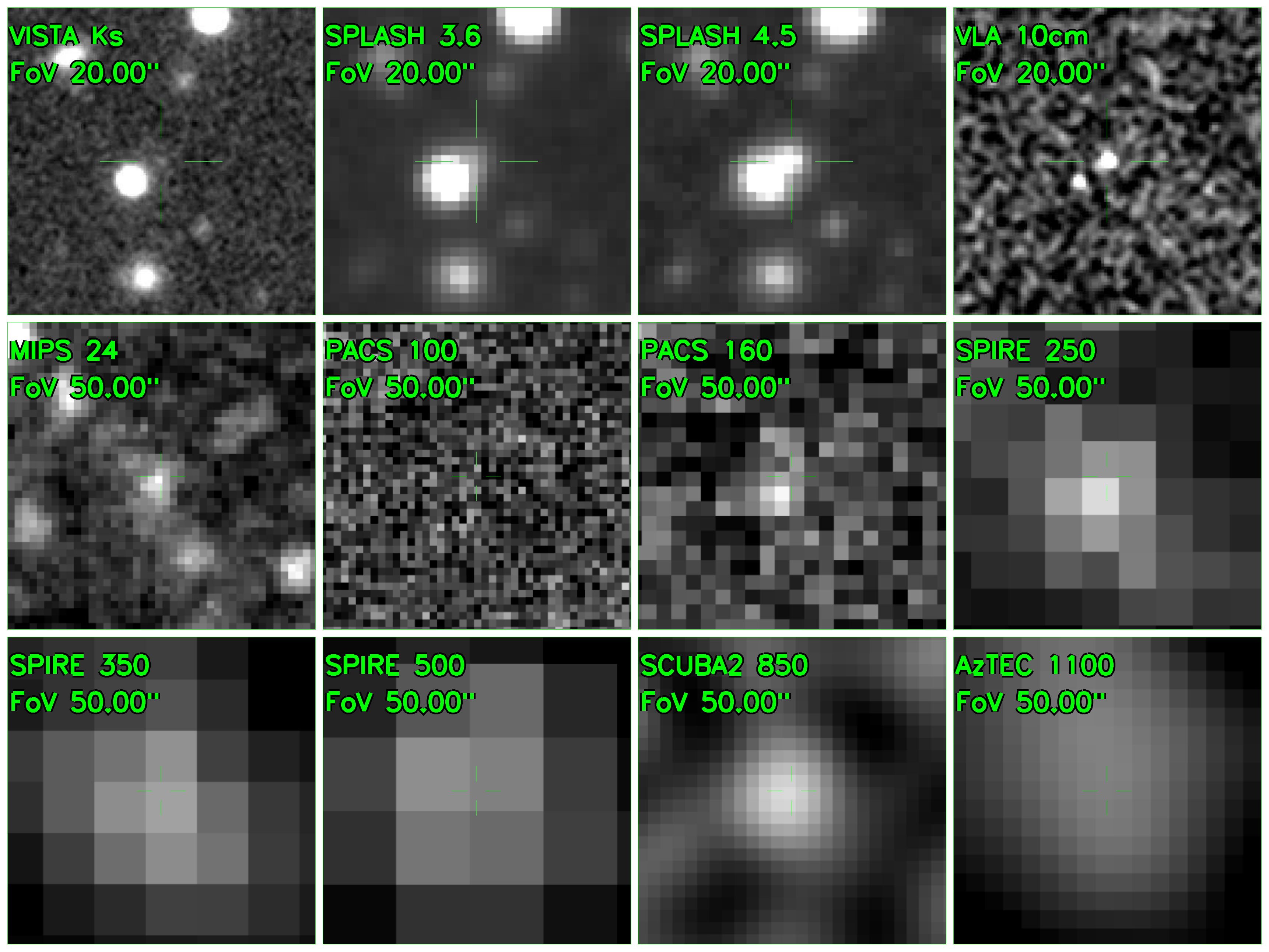}
	\includegraphics[width=0.21\textwidth, trim={0.6cm 5cm 1cm 3.5cm}, clip]{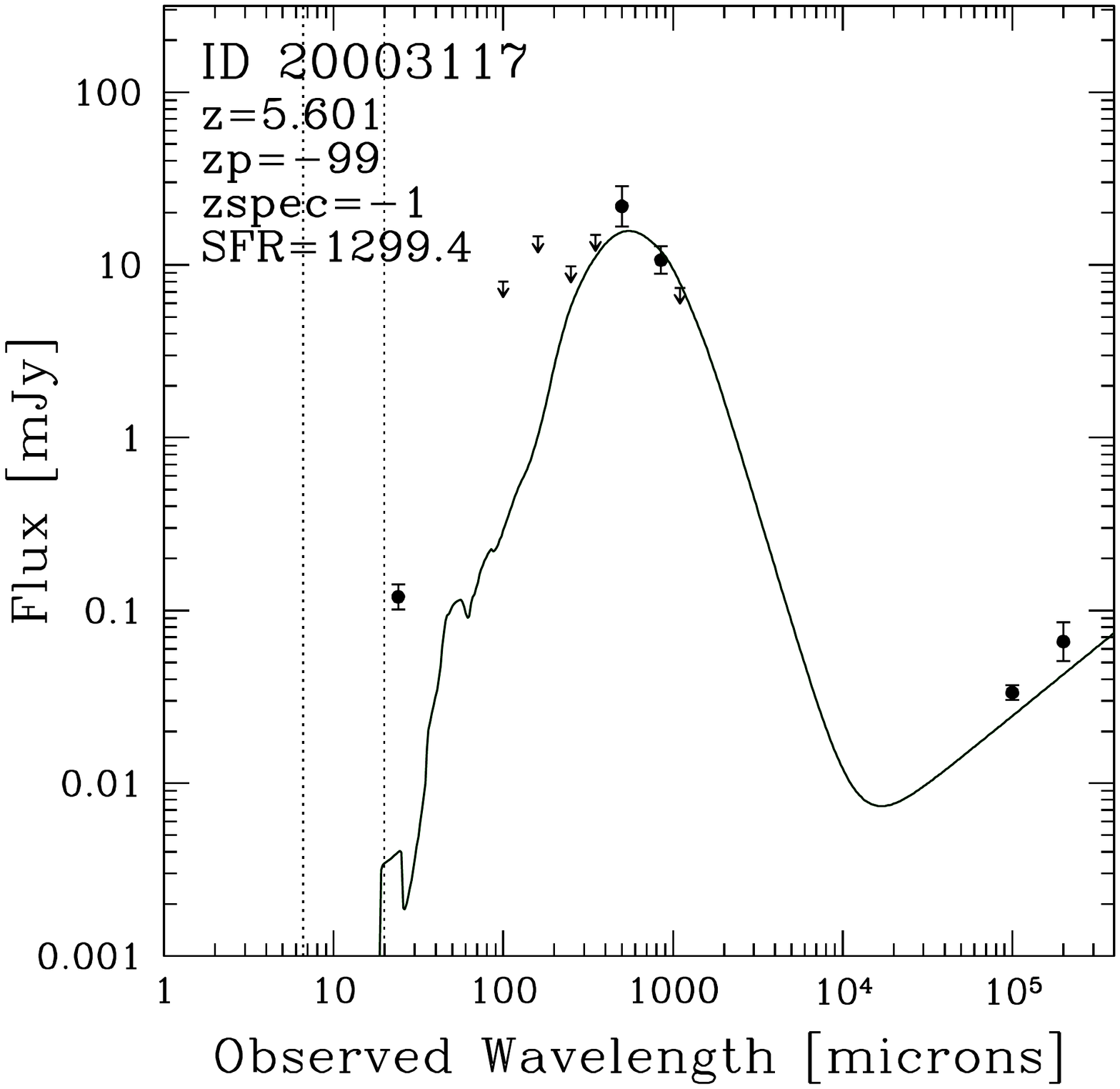}
    \includegraphics[width=0.28\textwidth]{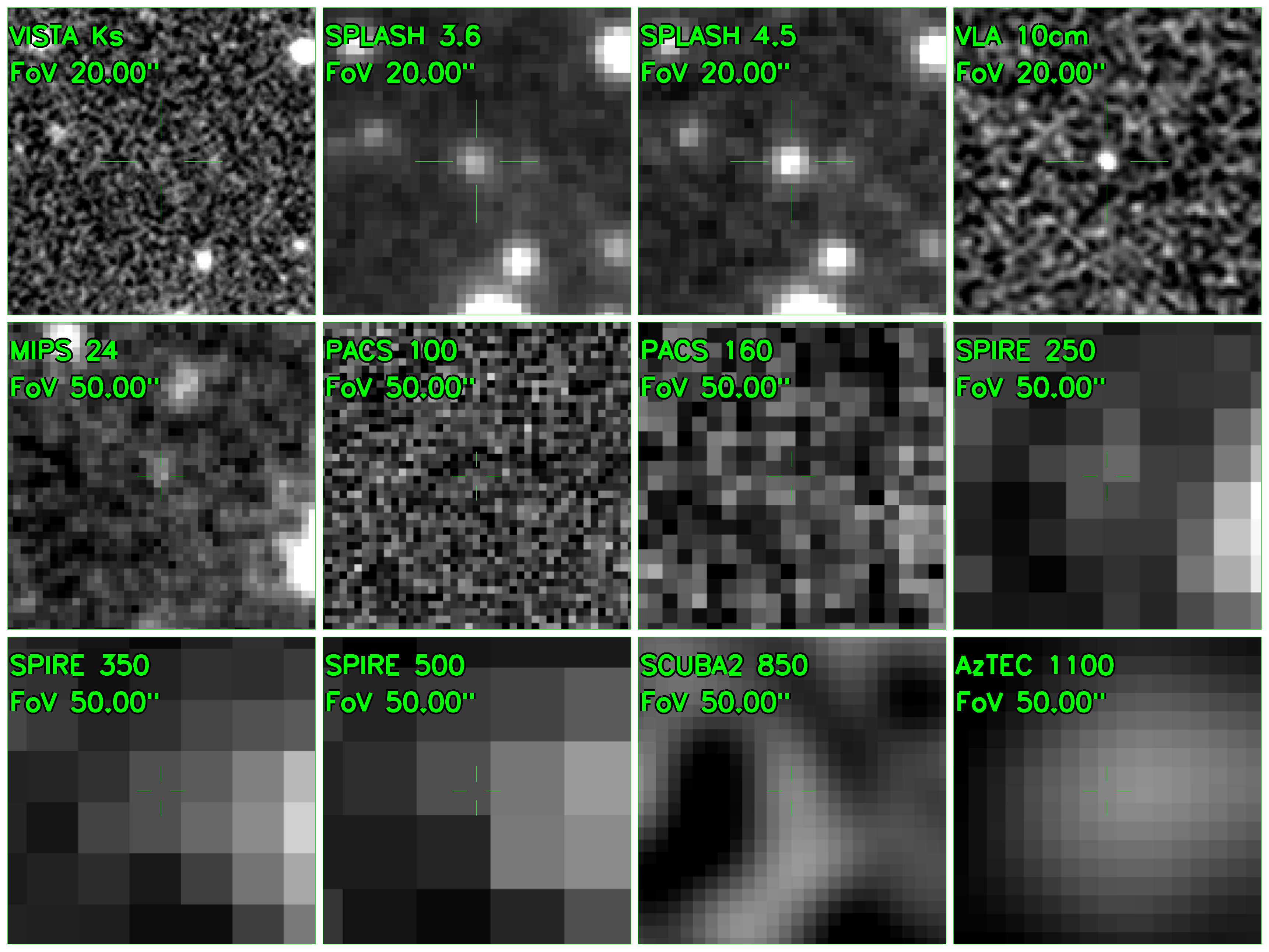}
	\includegraphics[width=0.21\textwidth, trim={0.6cm 5cm 1cm 3.5cm}, clip]{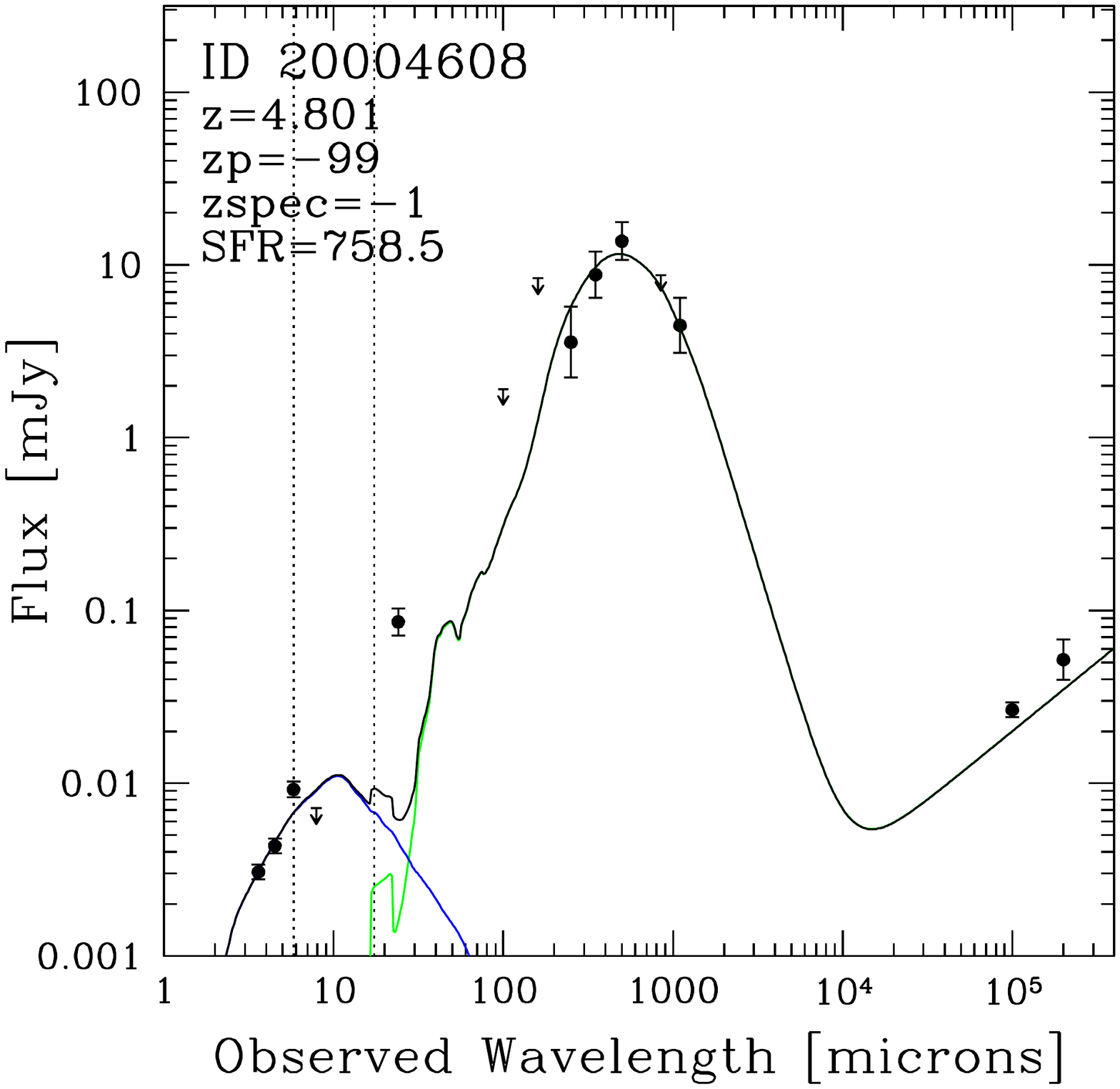}
    \includegraphics[width=0.28\textwidth]{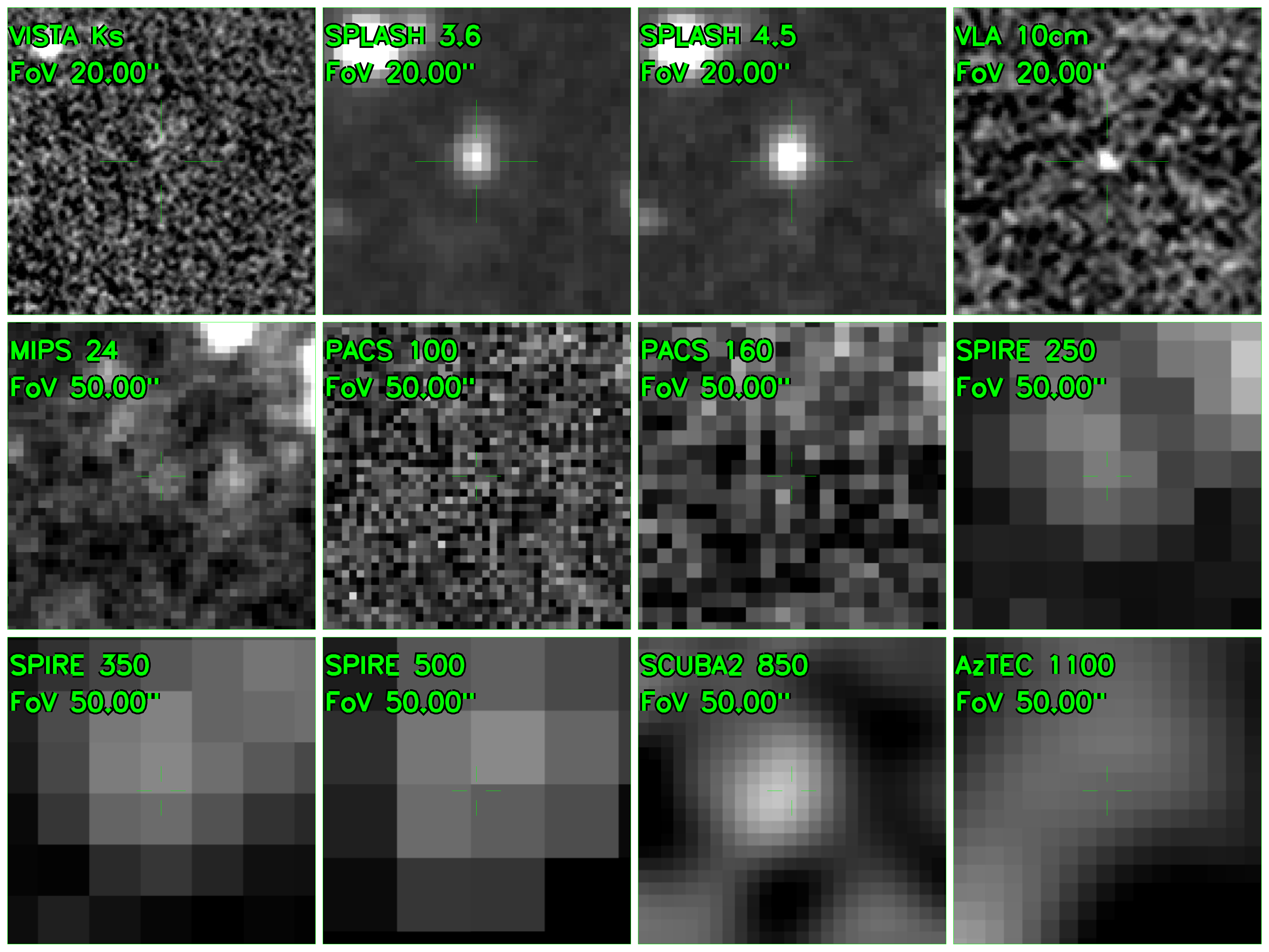}
	\includegraphics[width=0.21\textwidth, trim={0.6cm 5cm 1cm 3.5cm}, clip]{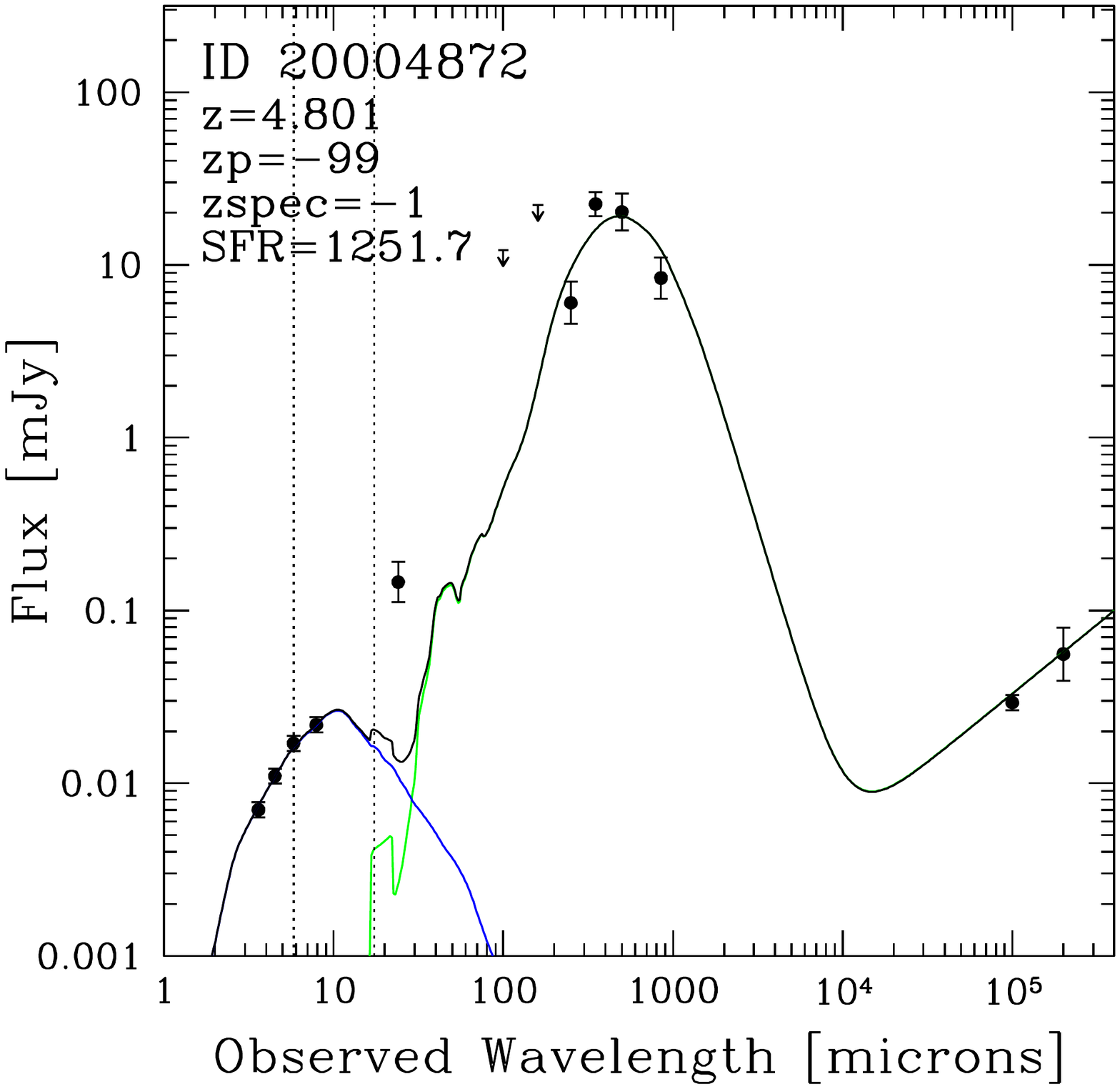}
    \includegraphics[width=0.28\textwidth]{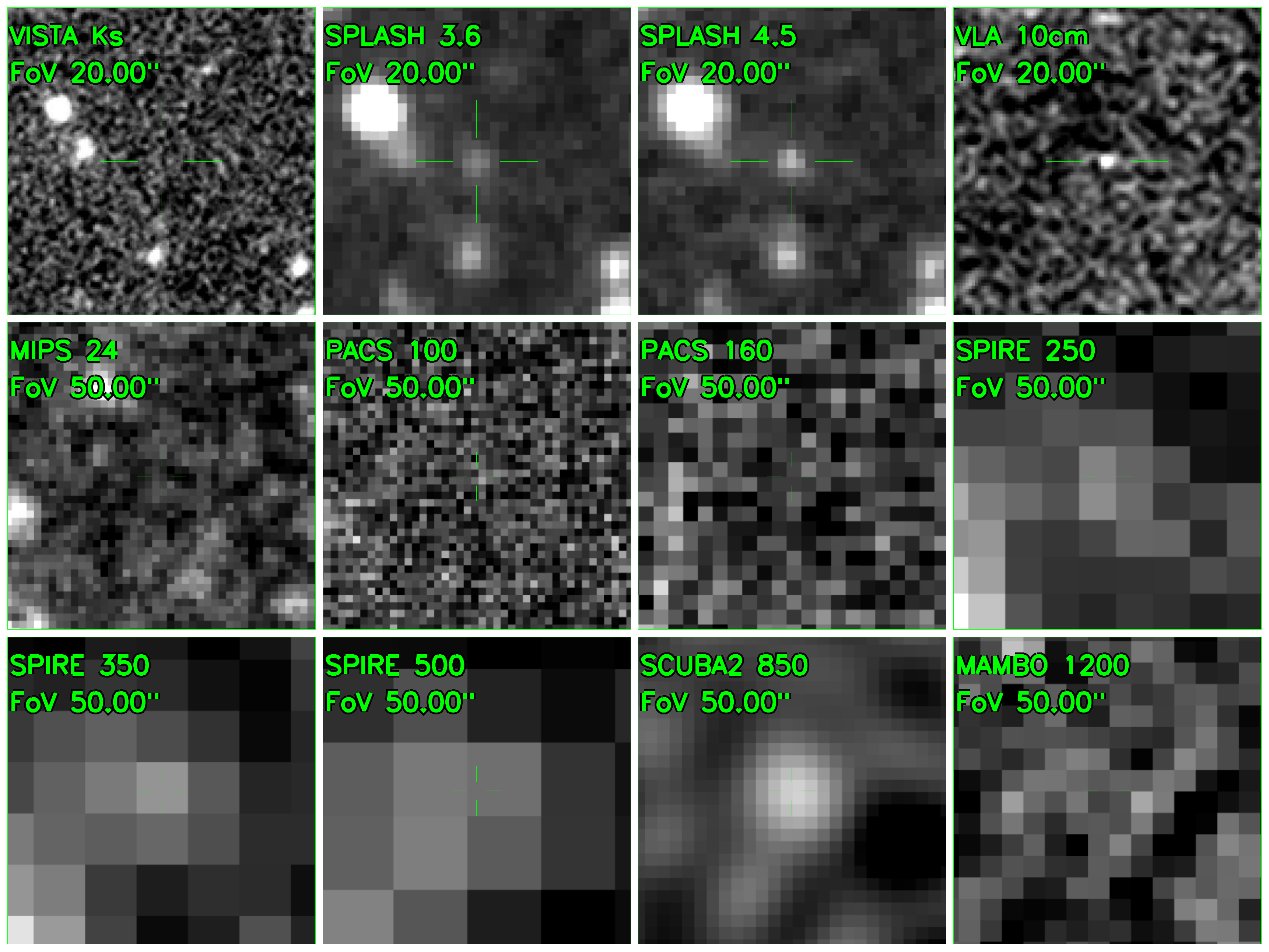}
	\includegraphics[width=0.21\textwidth, trim={0.6cm 5cm 1cm 3.5cm}, clip]{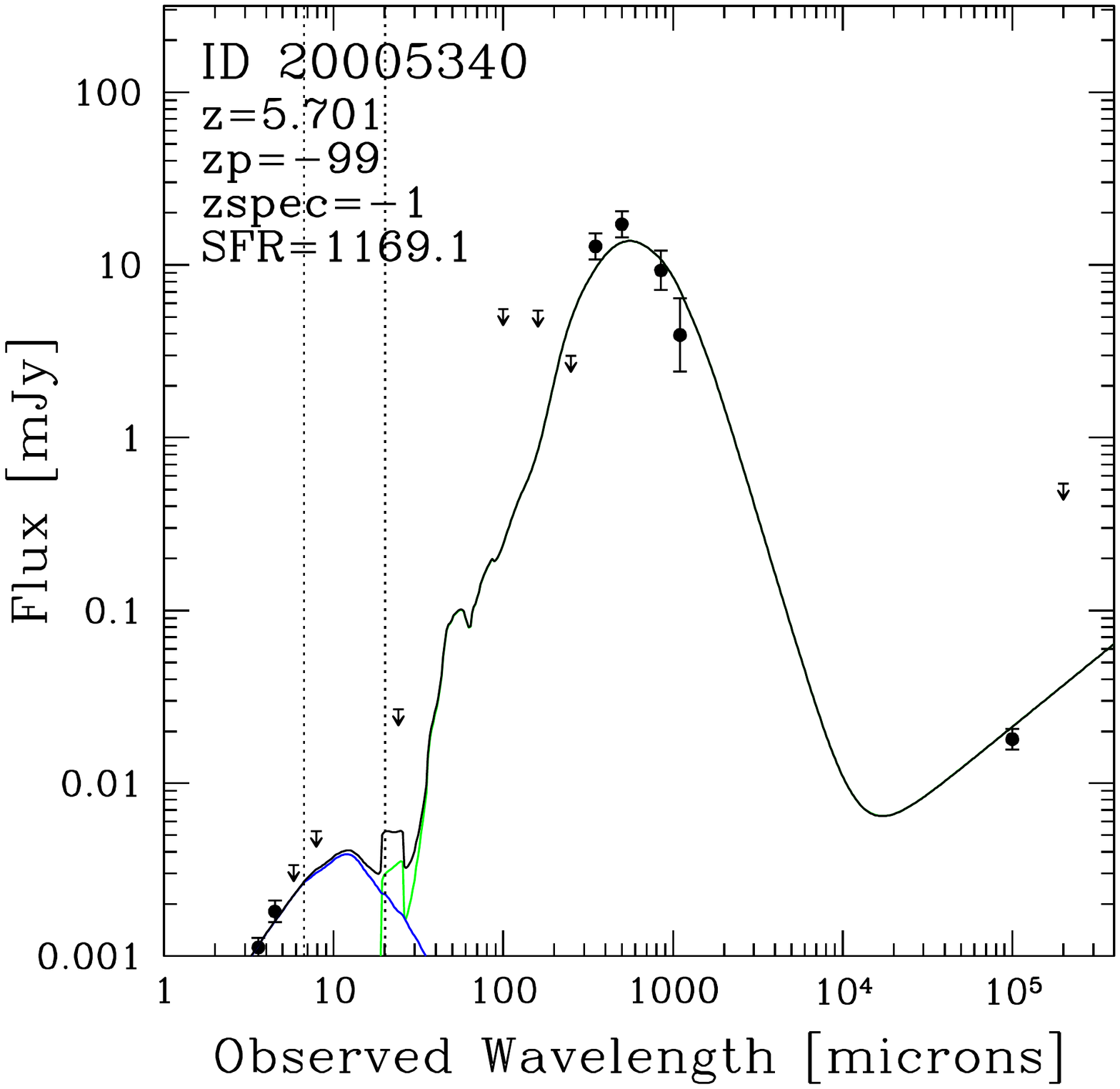}
    \includegraphics[width=0.28\textwidth]{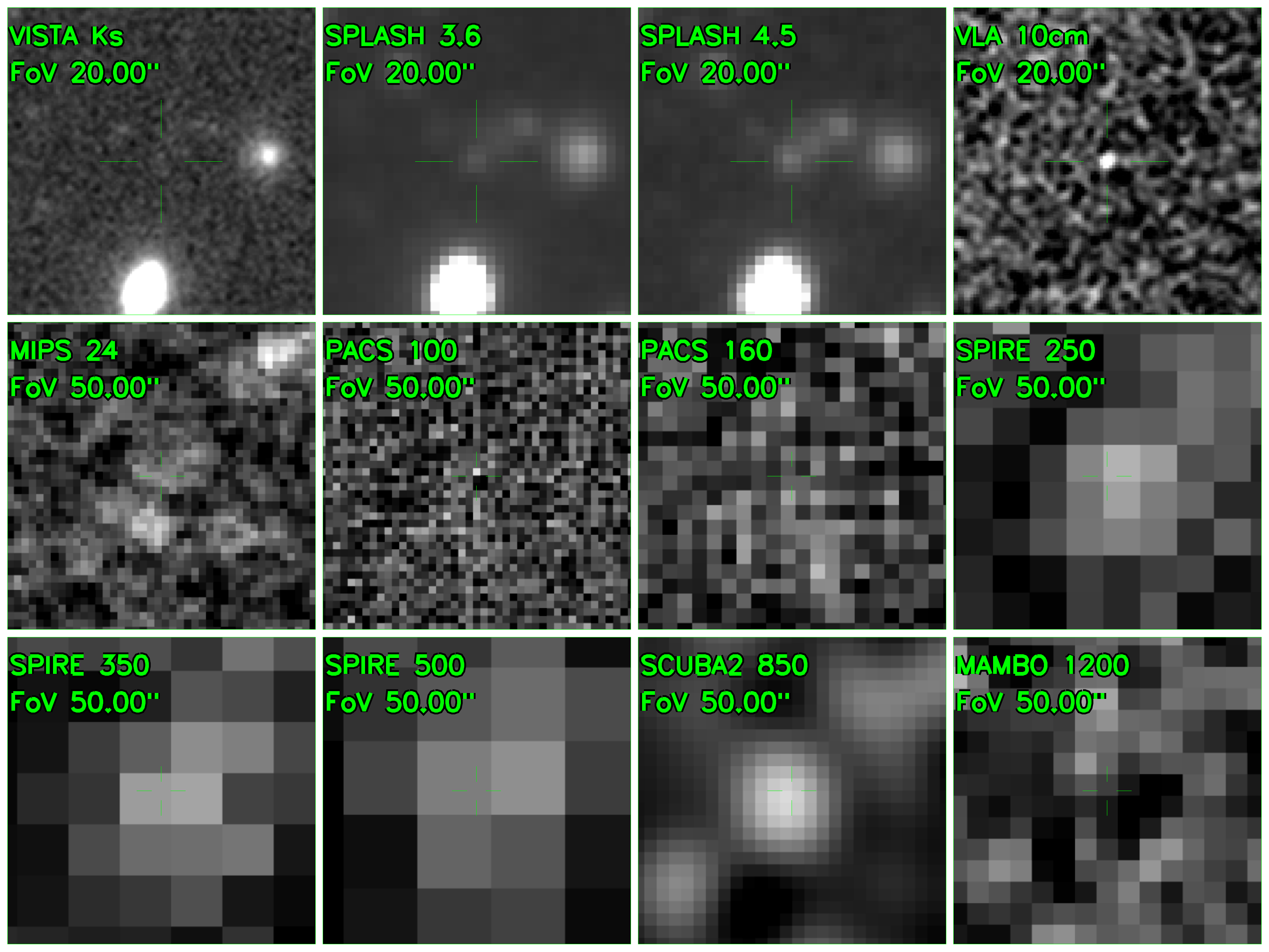}
	\includegraphics[width=0.21\textwidth, trim={0.6cm 5cm 1cm 3.5cm}, clip]{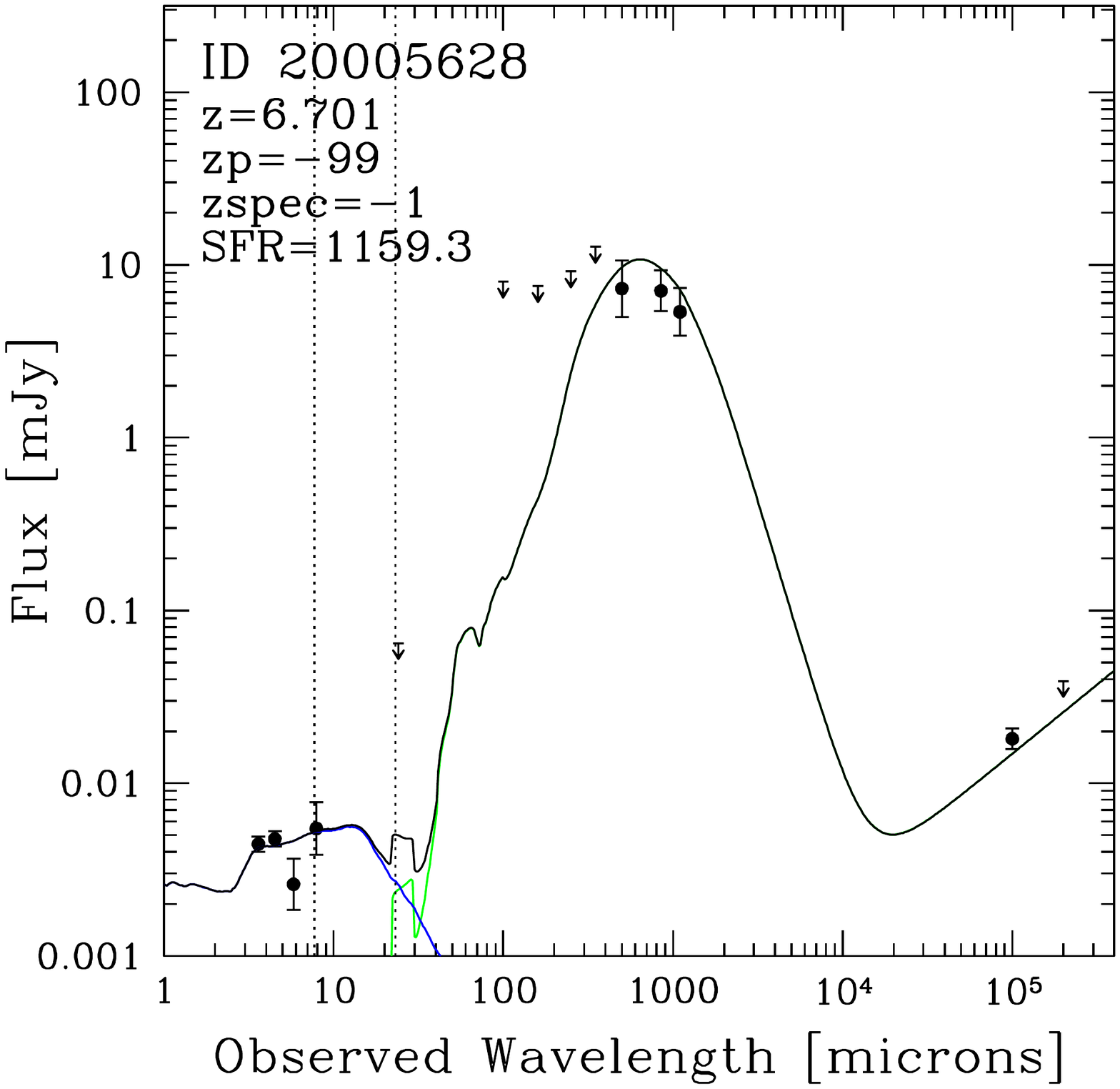}
    \caption{%
		Multi-band cutouts and SEDs of high redshift candidates, continued.  
		\label{highz_cutouts2}
		}
\end{figure}

\begin{figure}
	\centering%
    \includegraphics[width=0.28\textwidth]{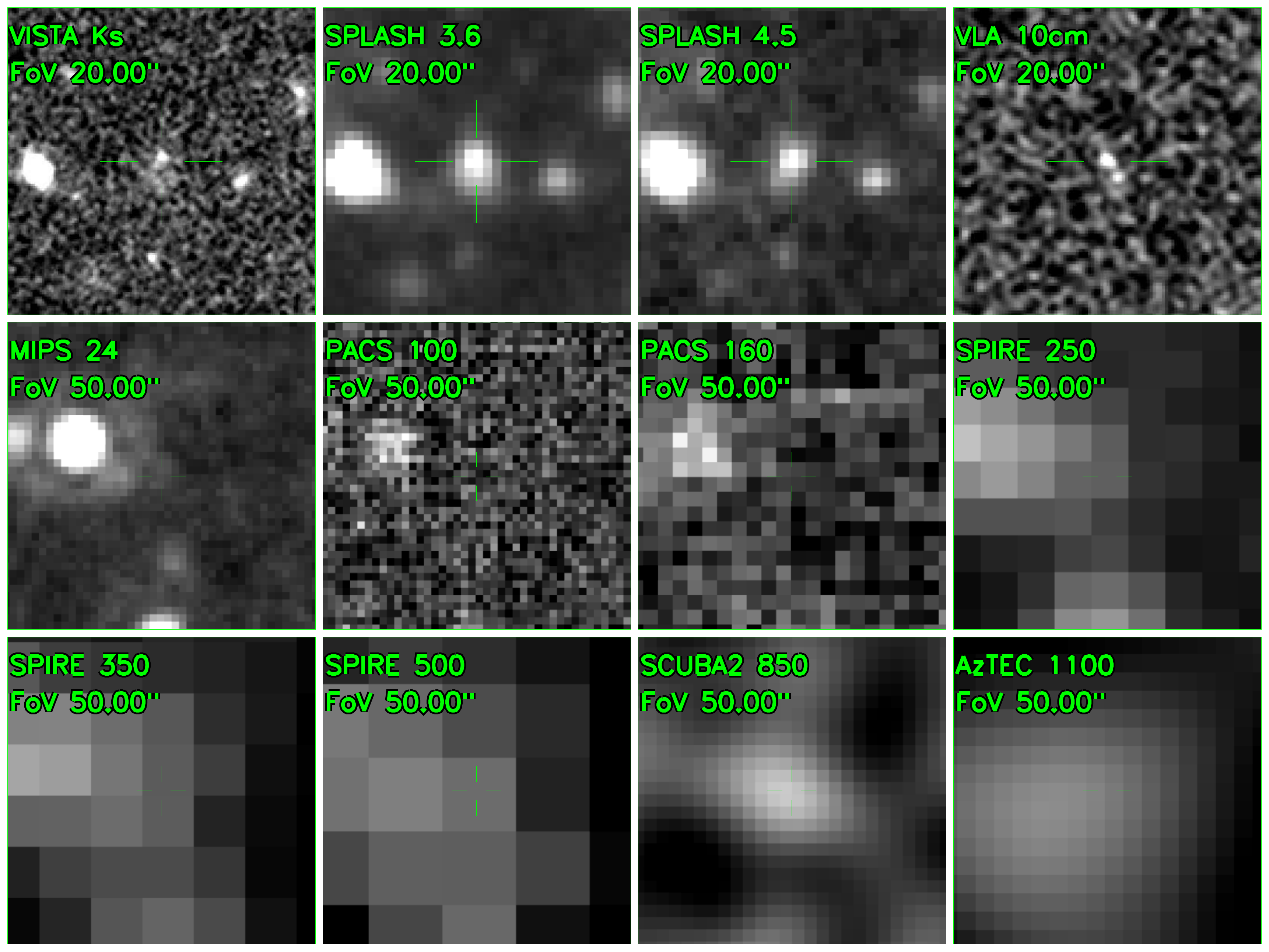}
	\includegraphics[width=0.21\textwidth, trim={0.6cm 5cm 1cm 3.5cm}, clip]{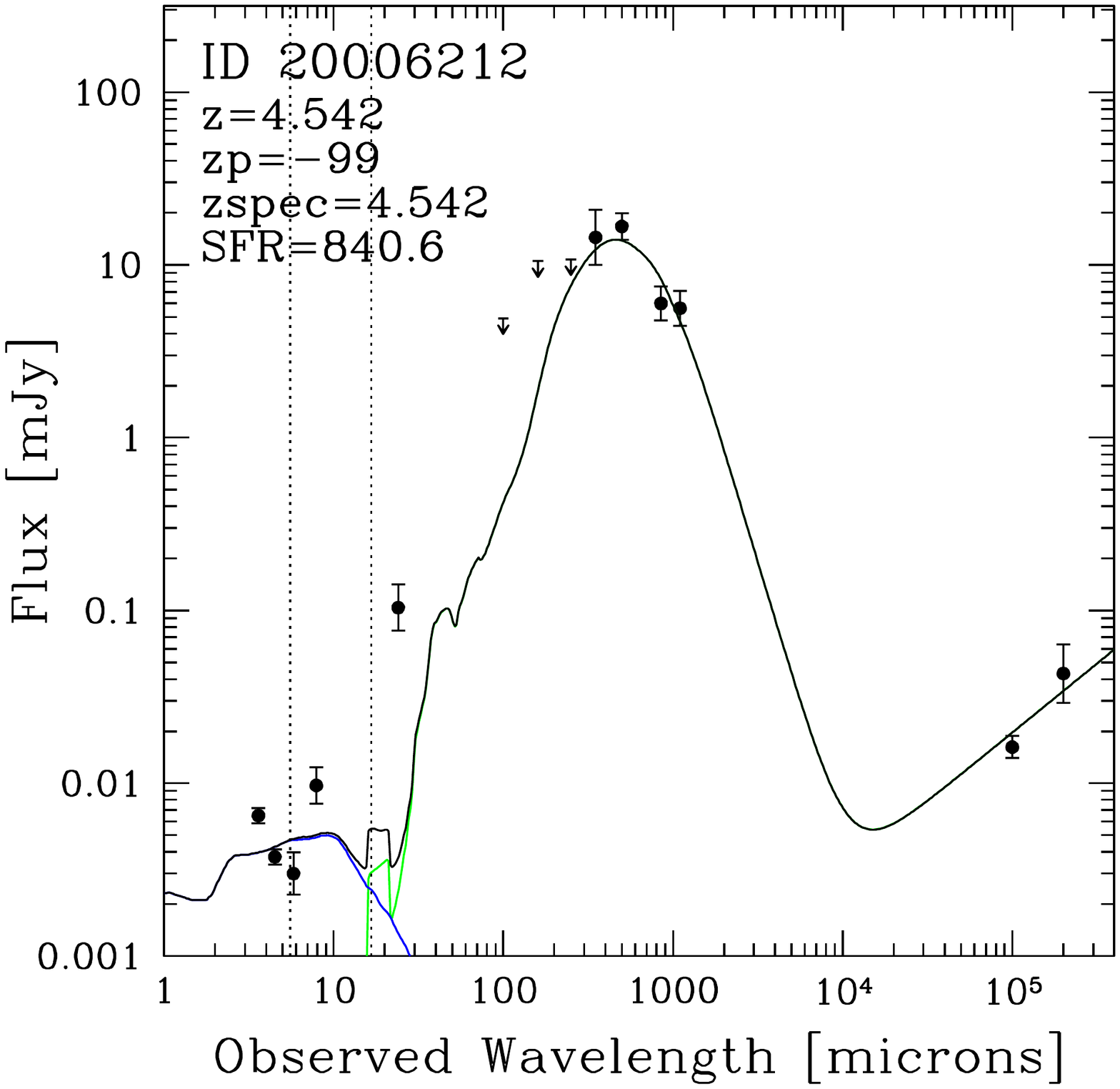}
	\includegraphics[width=0.28\textwidth]{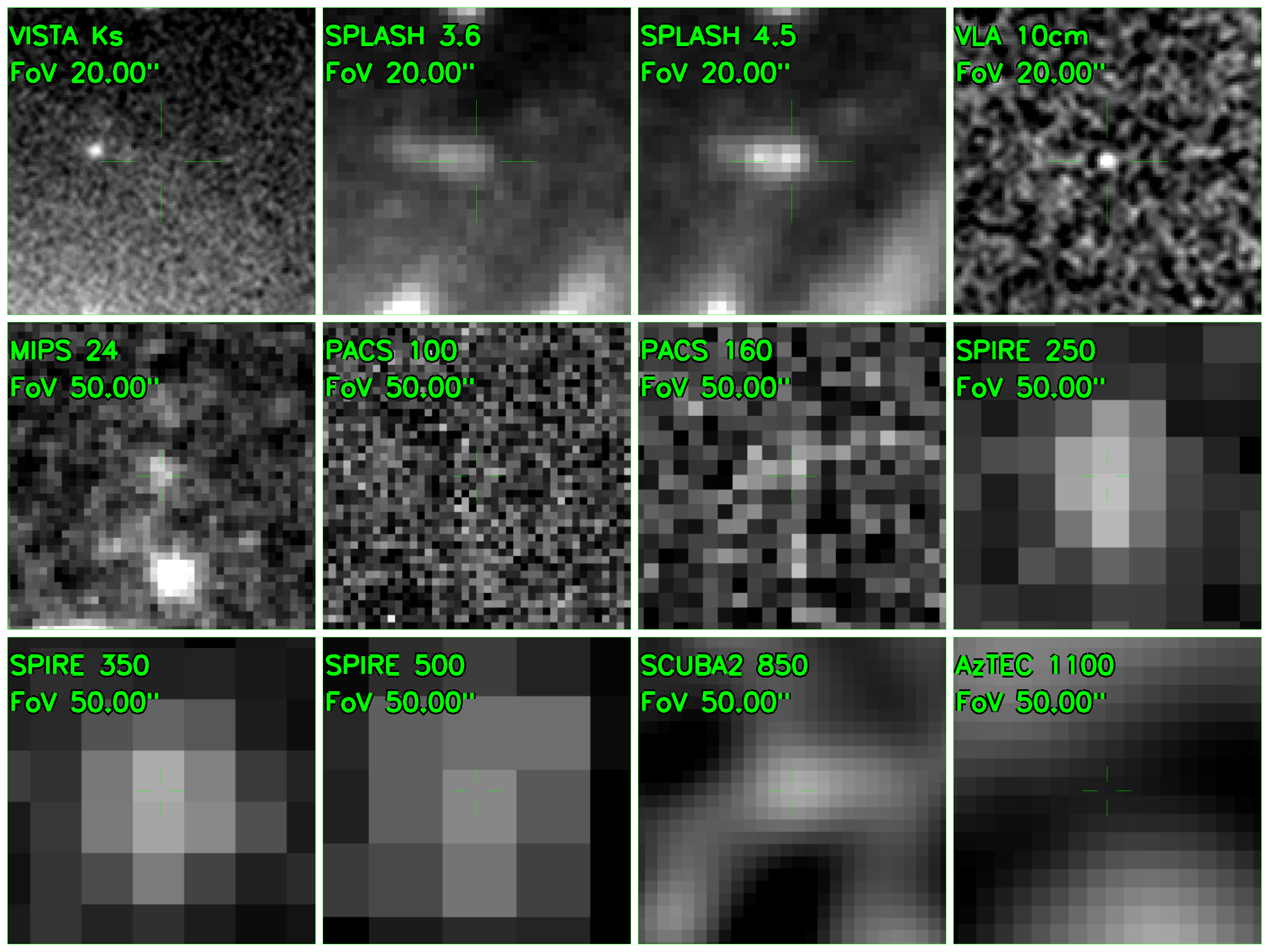}
	\includegraphics[width=0.21\textwidth, trim={0.6cm 5cm 1cm 3.5cm}, clip]{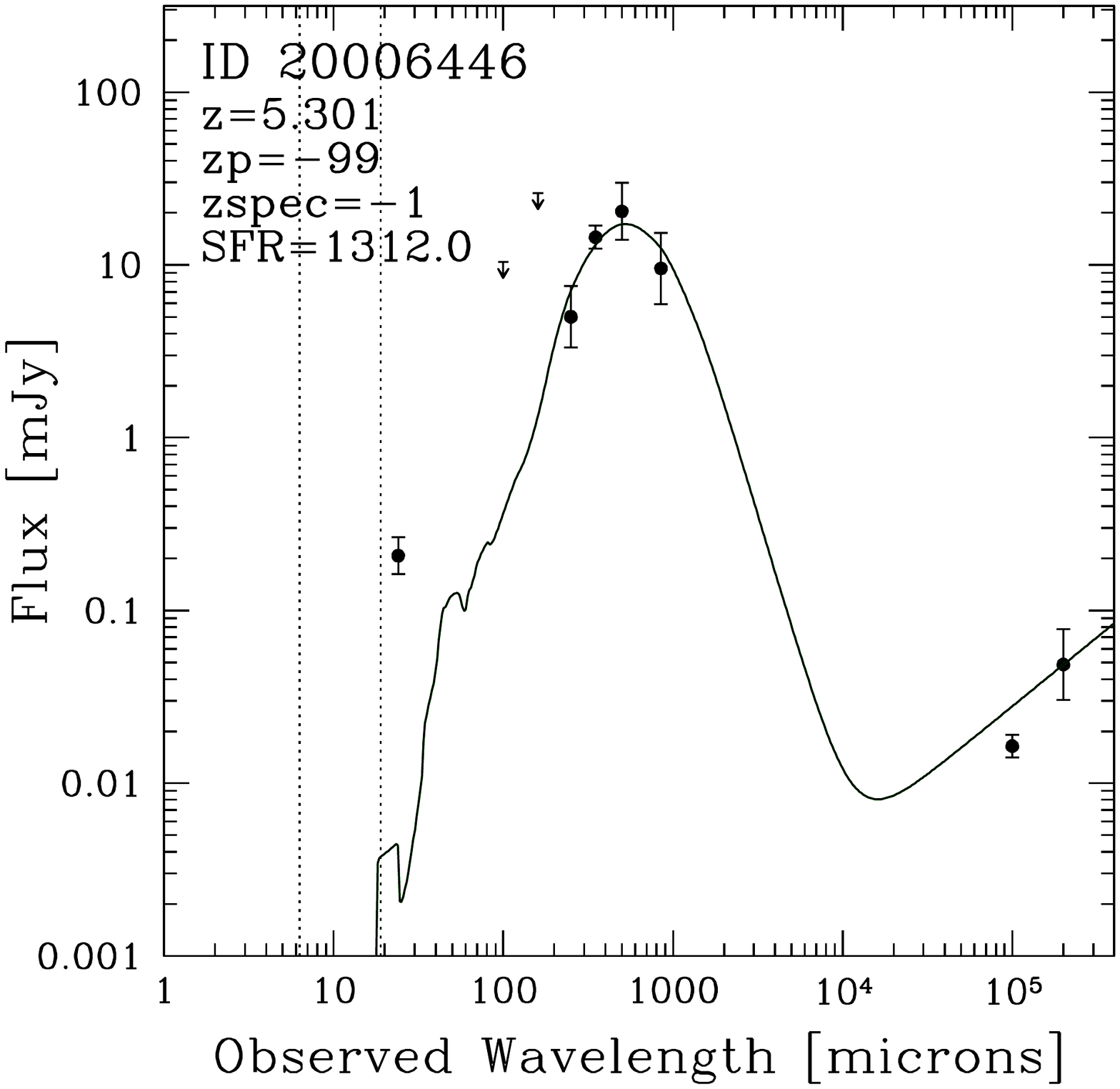}
    \includegraphics[width=0.28\textwidth]{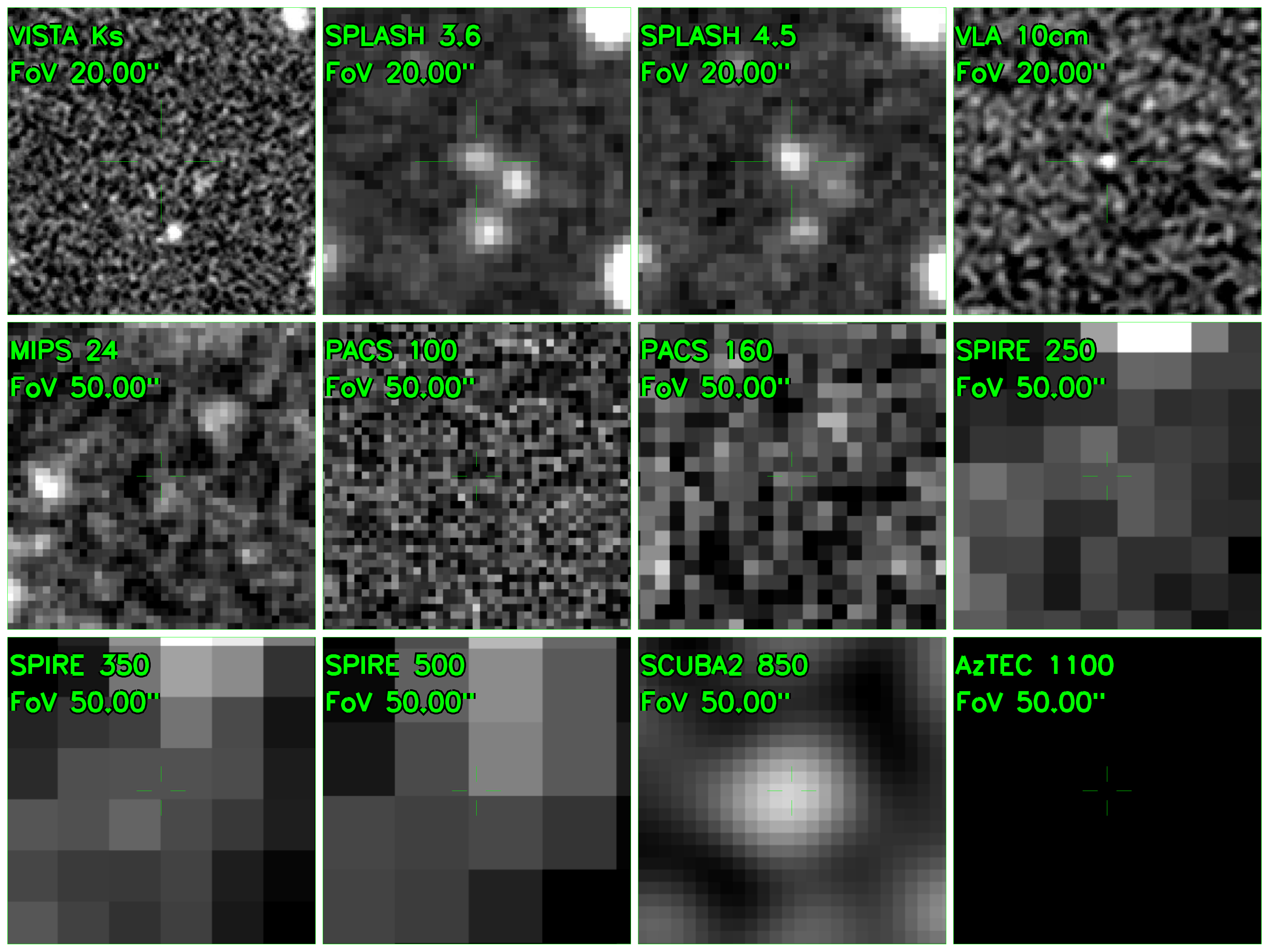}
	\includegraphics[width=0.21\textwidth, trim={0.6cm 5cm 1cm 3.5cm}, clip]{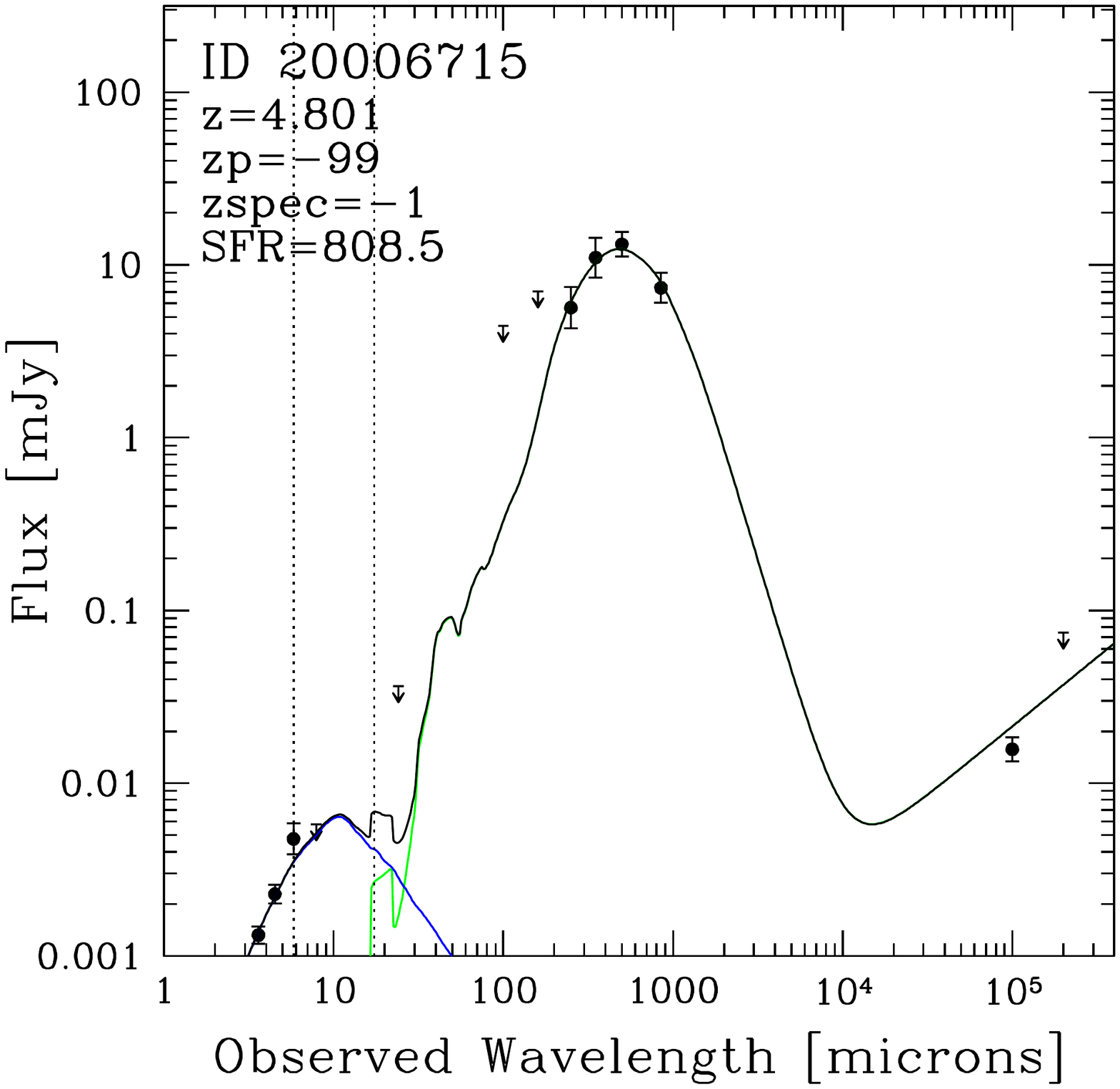}
    \includegraphics[width=0.28\textwidth]{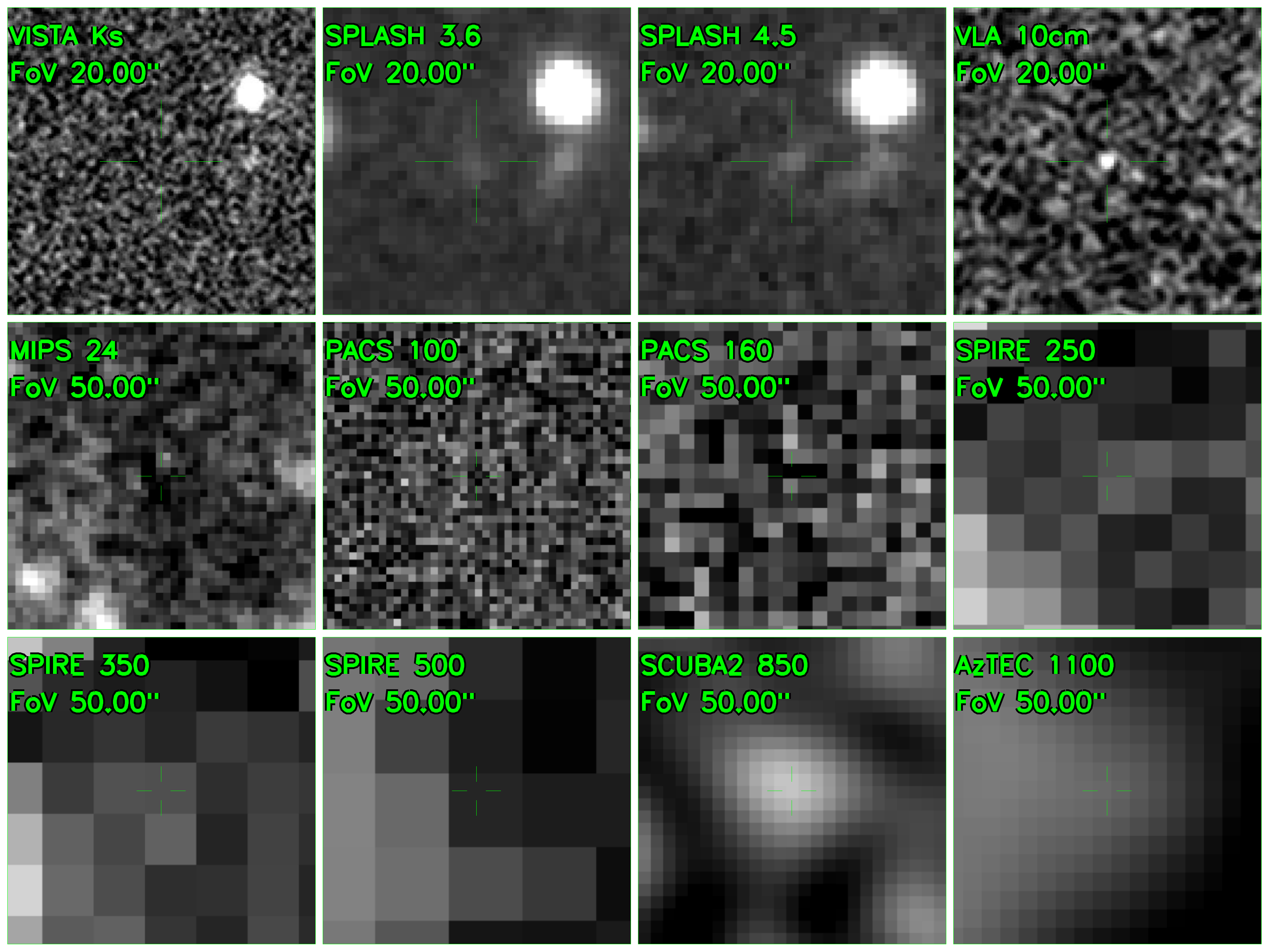}
	\includegraphics[width=0.21\textwidth, trim={0.6cm 5cm 1cm 3.5cm}, clip]{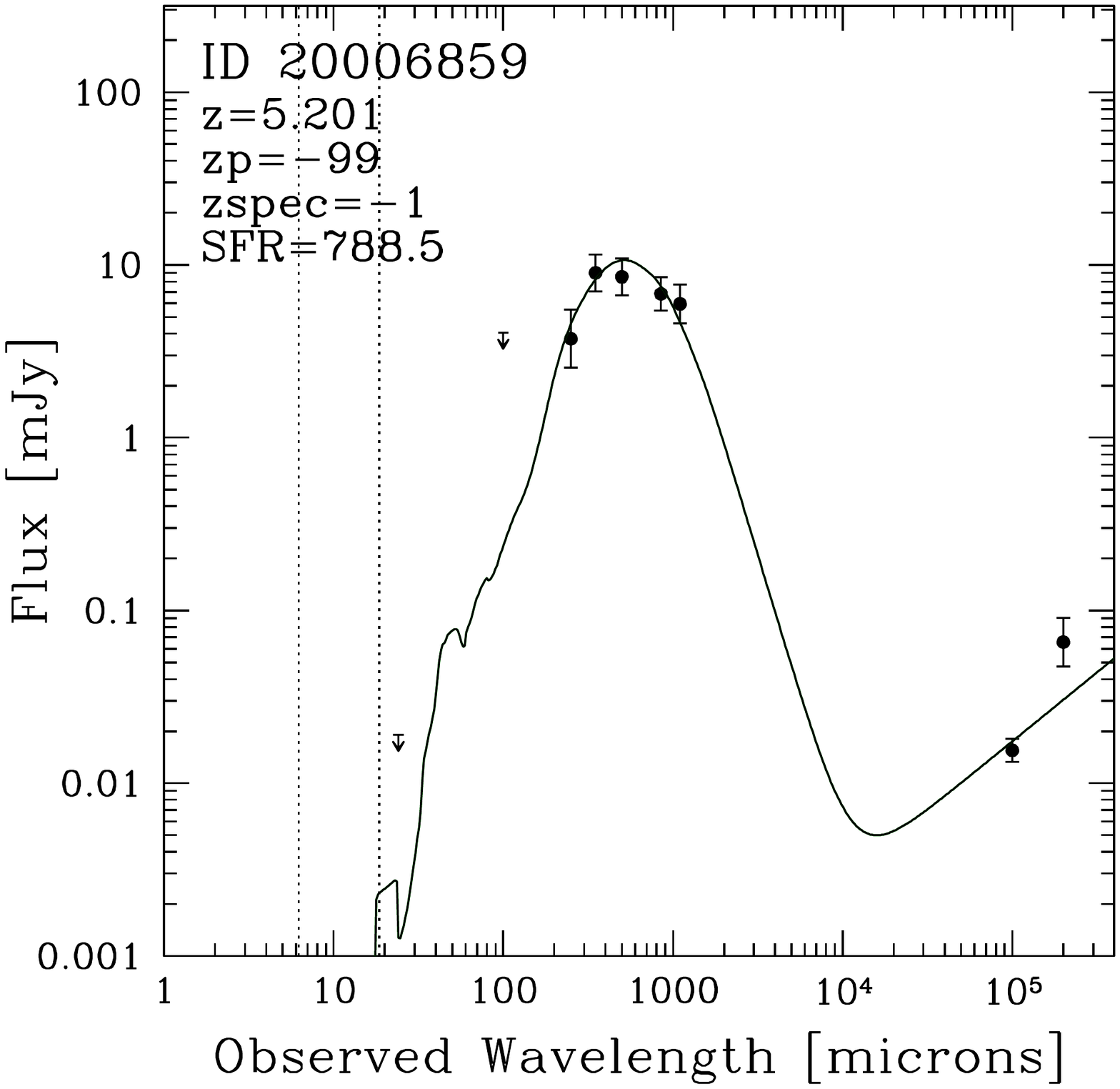}
    \includegraphics[width=0.28\textwidth]{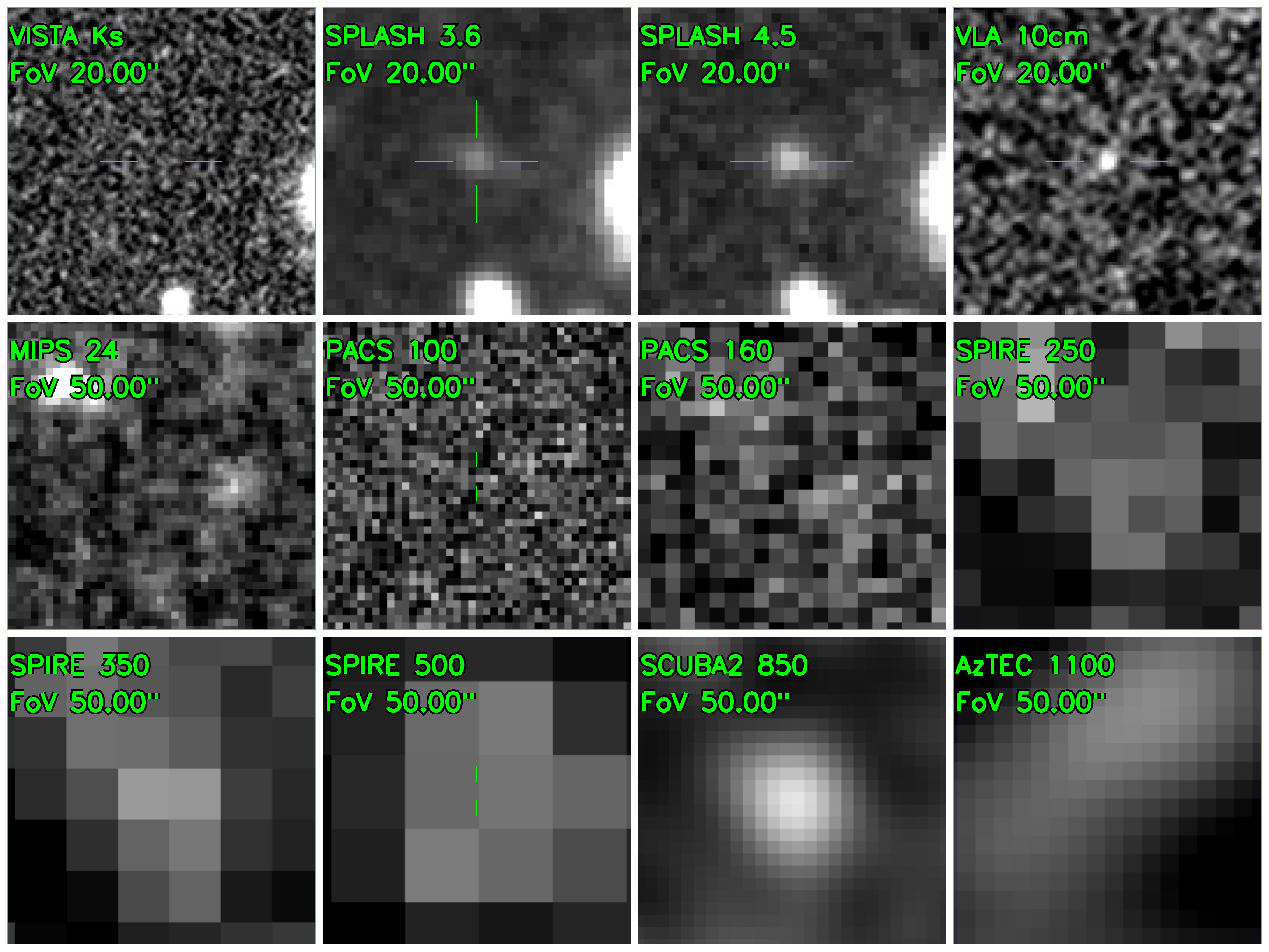}
	\includegraphics[width=0.21\textwidth, trim={0.6cm 5cm 1cm 3.5cm}, clip]{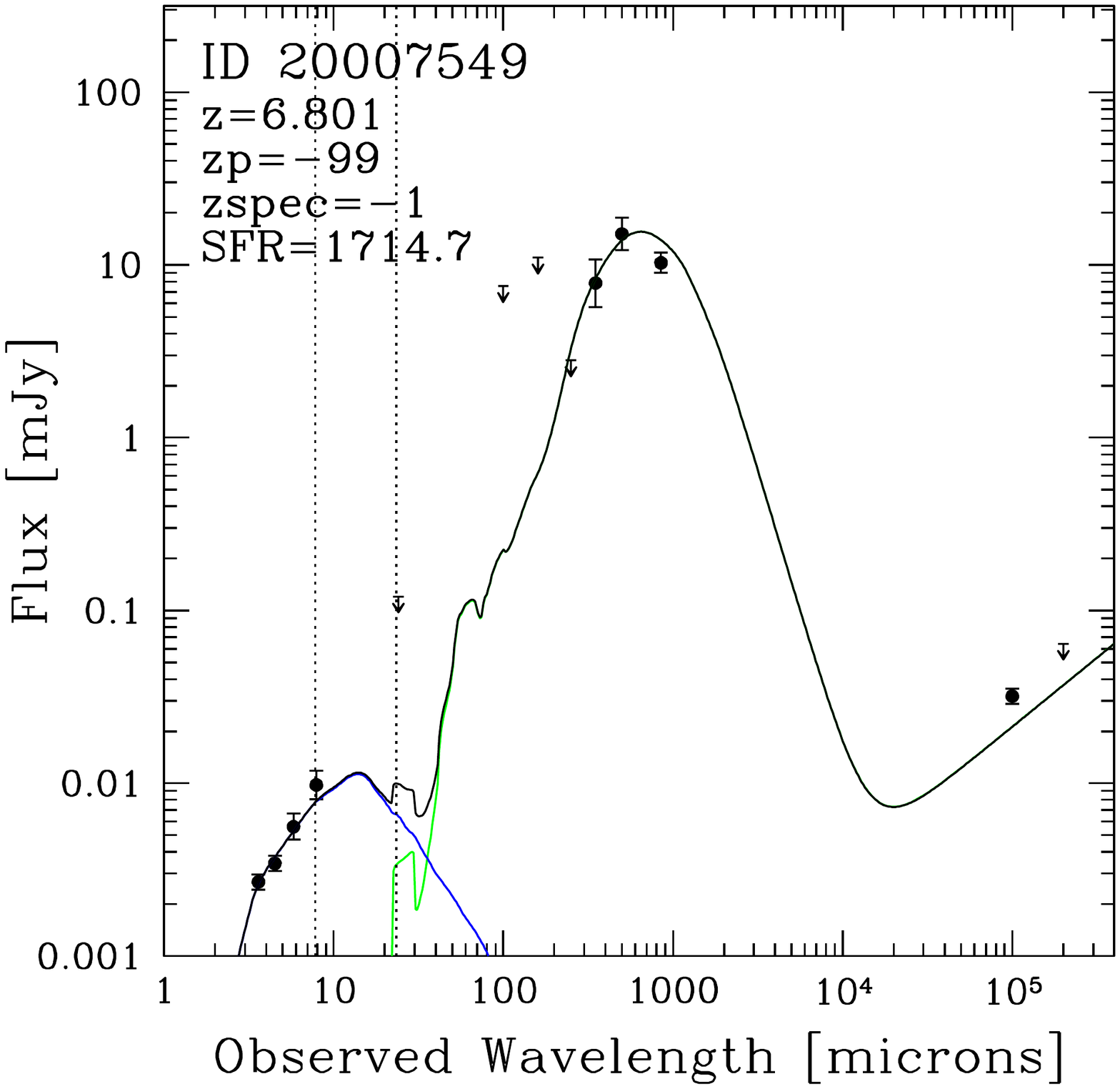}
    \includegraphics[width=0.28\textwidth]{20007898.pdf}
	\includegraphics[width=0.21\textwidth, trim={0.6cm 5cm 1cm 3.5cm}, clip]{Plot_SED_20007898.pdf}
    \includegraphics[width=0.28\textwidth]{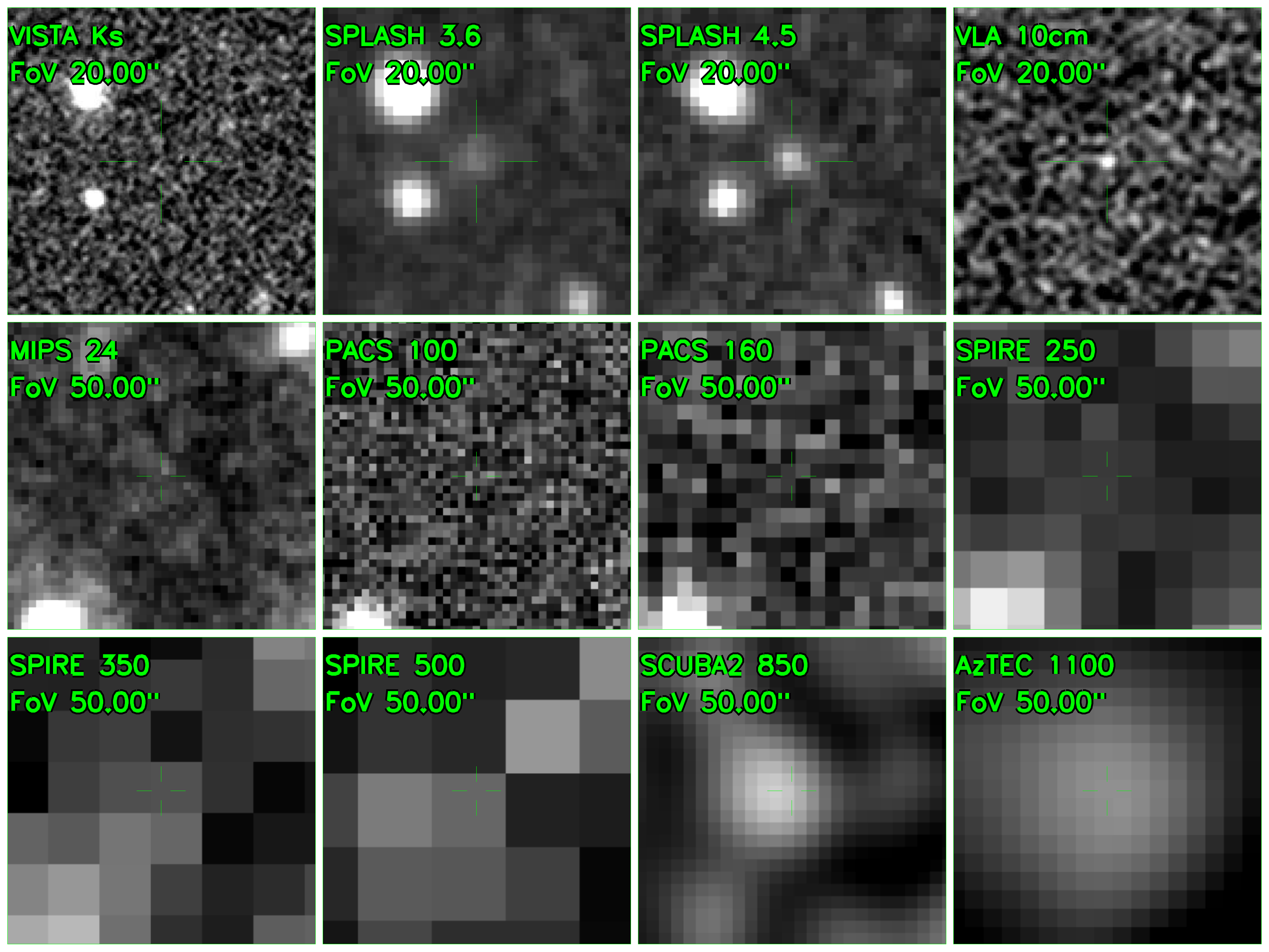}
	\includegraphics[width=0.21\textwidth, trim={0.6cm 5cm 1cm 3.5cm}, clip]{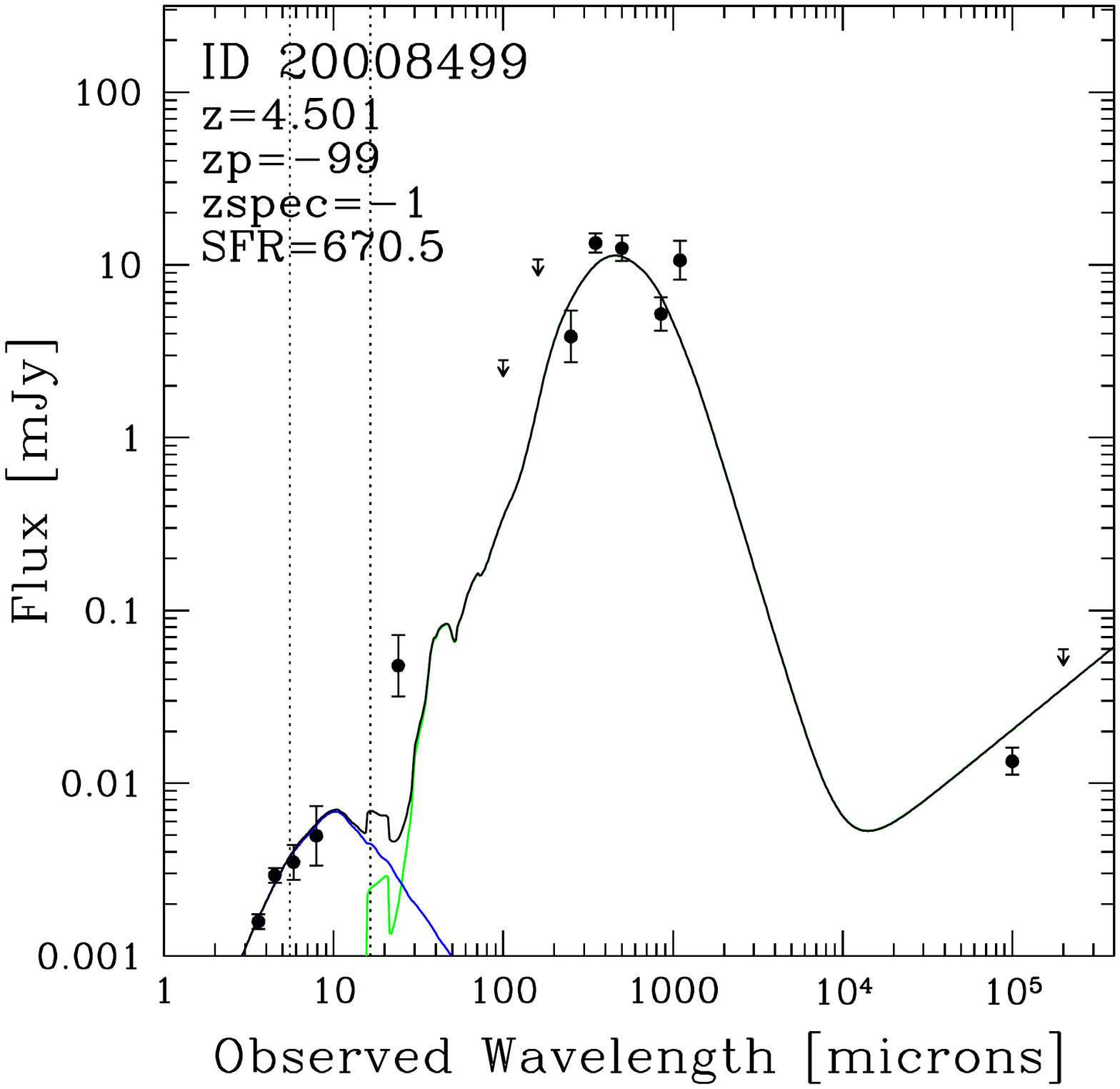}
    \includegraphics[width=0.28\textwidth]{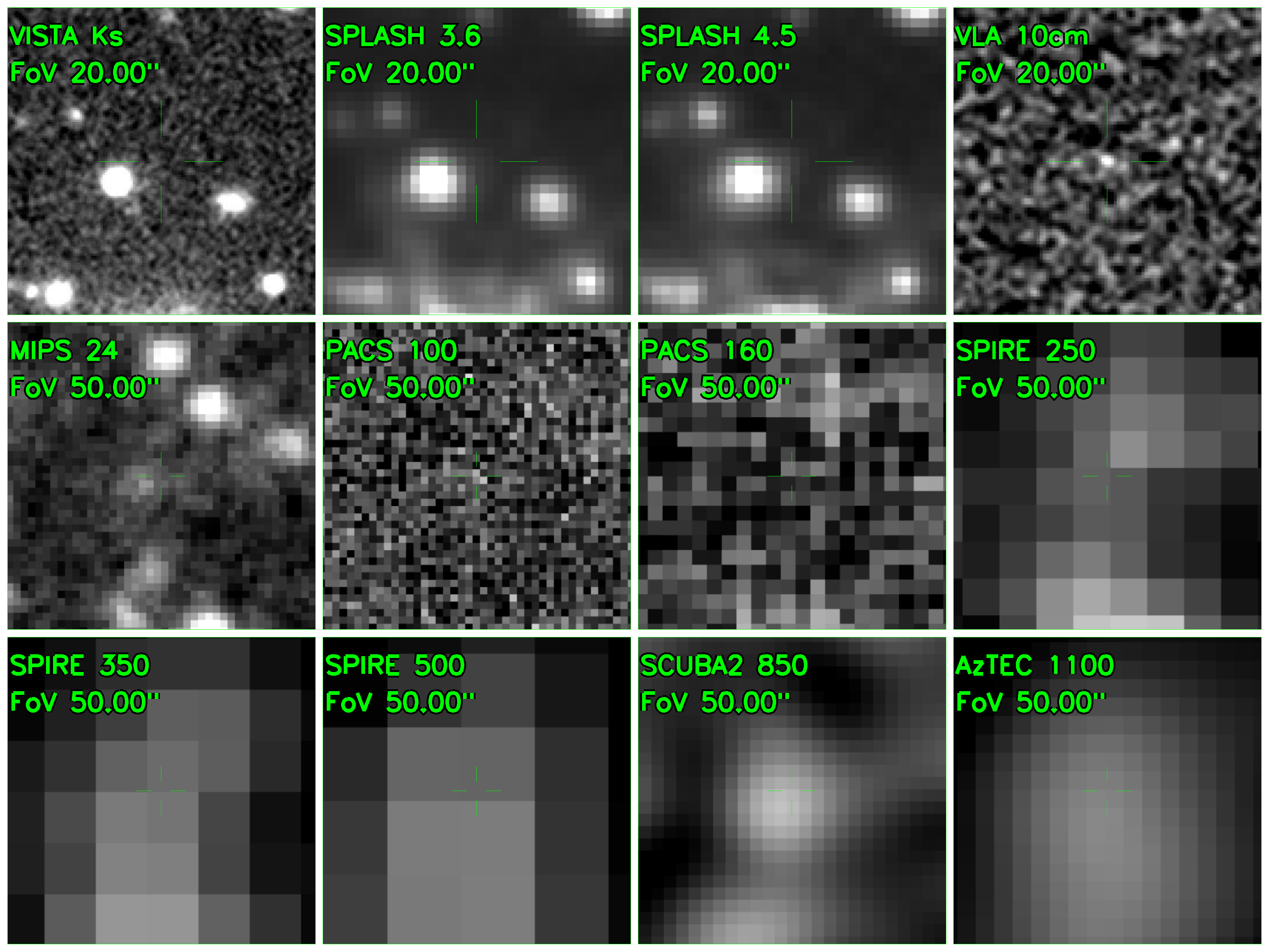}
	\includegraphics[width=0.21\textwidth, trim={0.6cm 5cm 1cm 3.5cm}, clip]{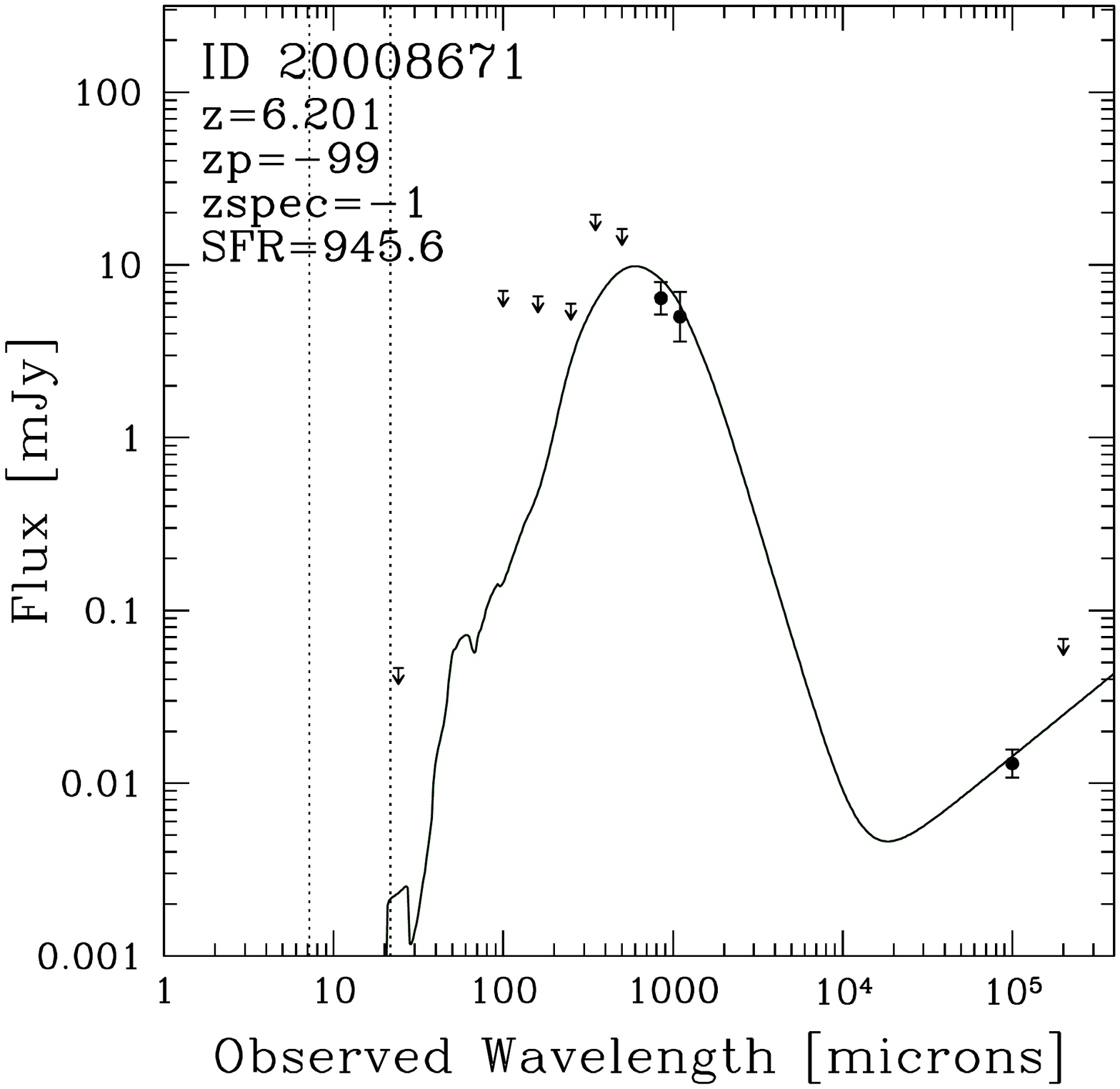}
     \includegraphics[width=0.28\textwidth]{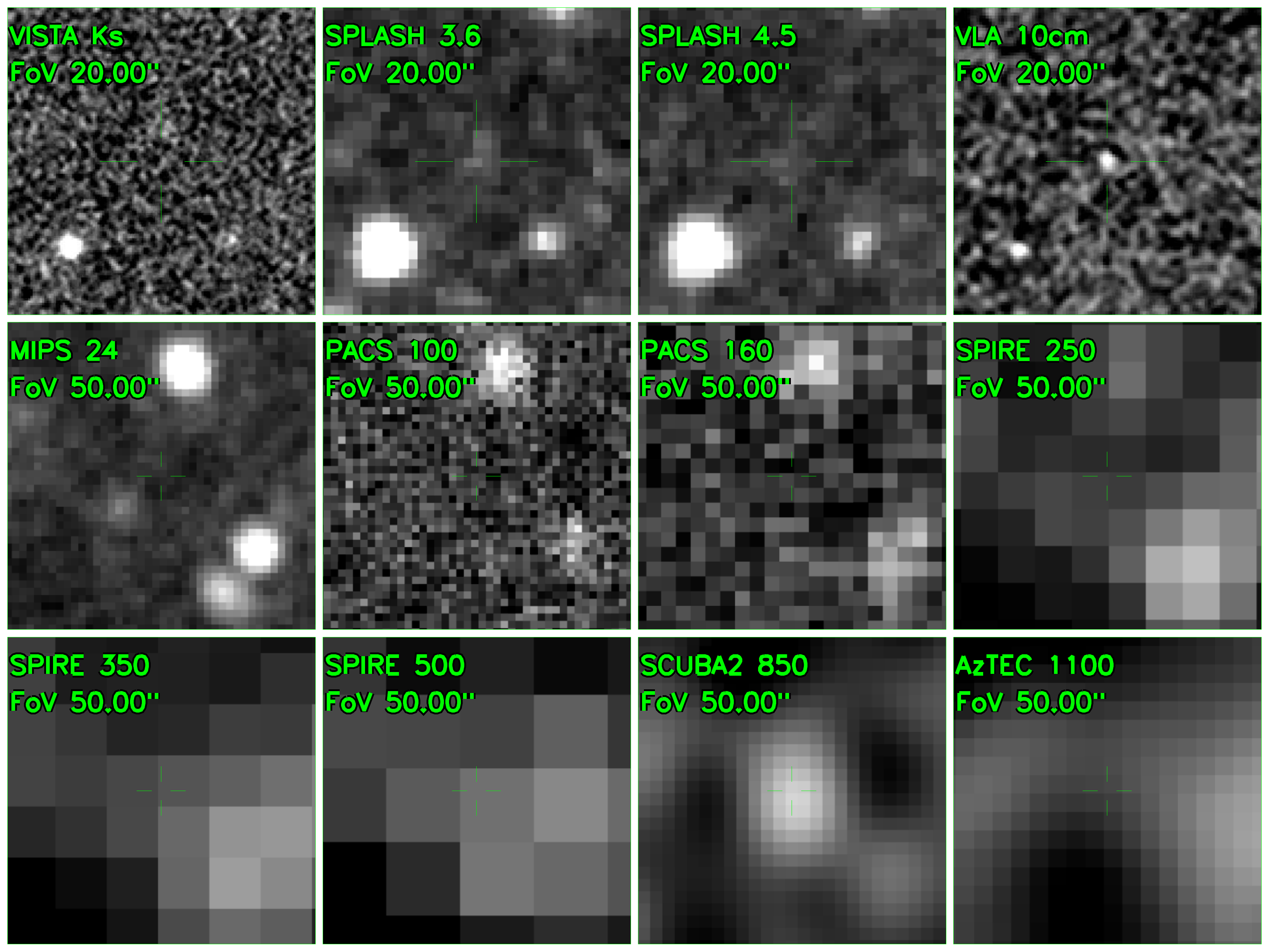}
	\includegraphics[width=0.21\textwidth, trim={0.6cm 5cm 1cm 3.5cm}, clip]{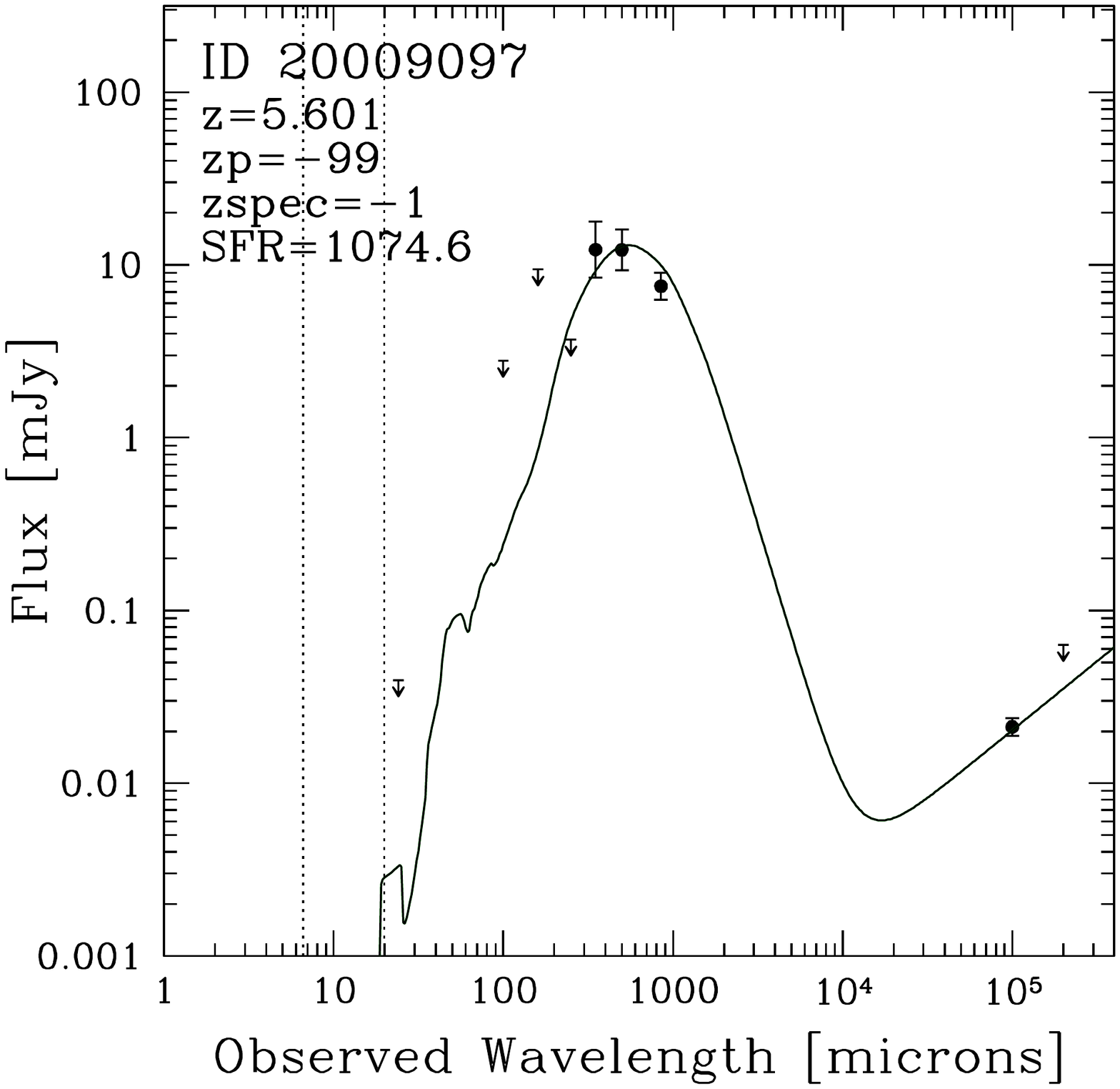}
     \includegraphics[width=0.28\textwidth]{20010161.pdf}
	\includegraphics[width=0.21\textwidth, trim={0.6cm 5cm 1cm 3.5cm}, clip]{Plot_SED_20010161.pdf}
    \caption{%
		Multi-band cutouts and SEDs of high redshift candidates, continued.  
		\label{highz_cutouts2}
		}
\end{figure}

 \begin{figure}
	\centering%
    \includegraphics[width=0.28\textwidth]{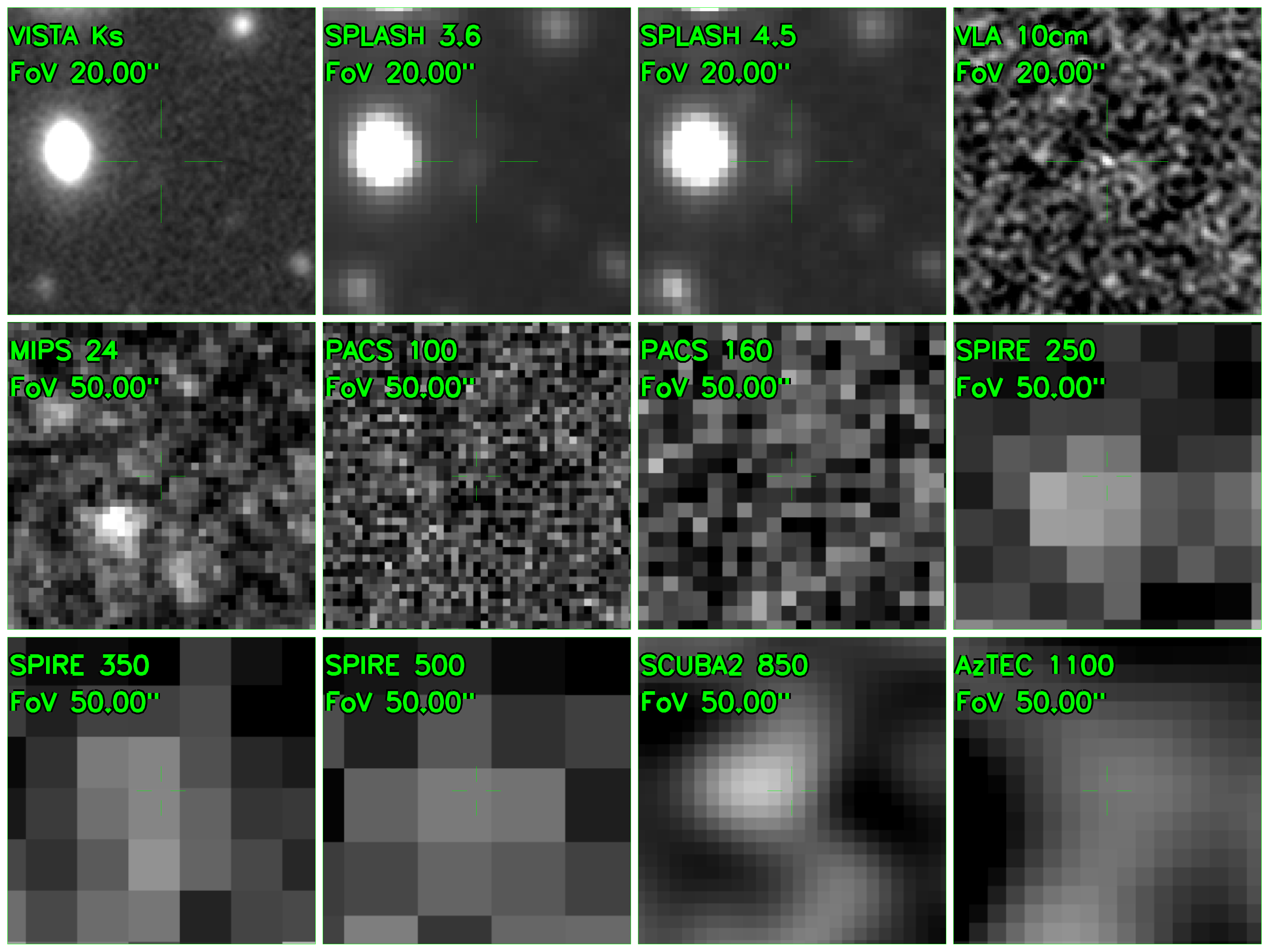}
	\includegraphics[width=0.21\textwidth, trim={0.6cm 5cm 1cm 3.5cm}, clip]{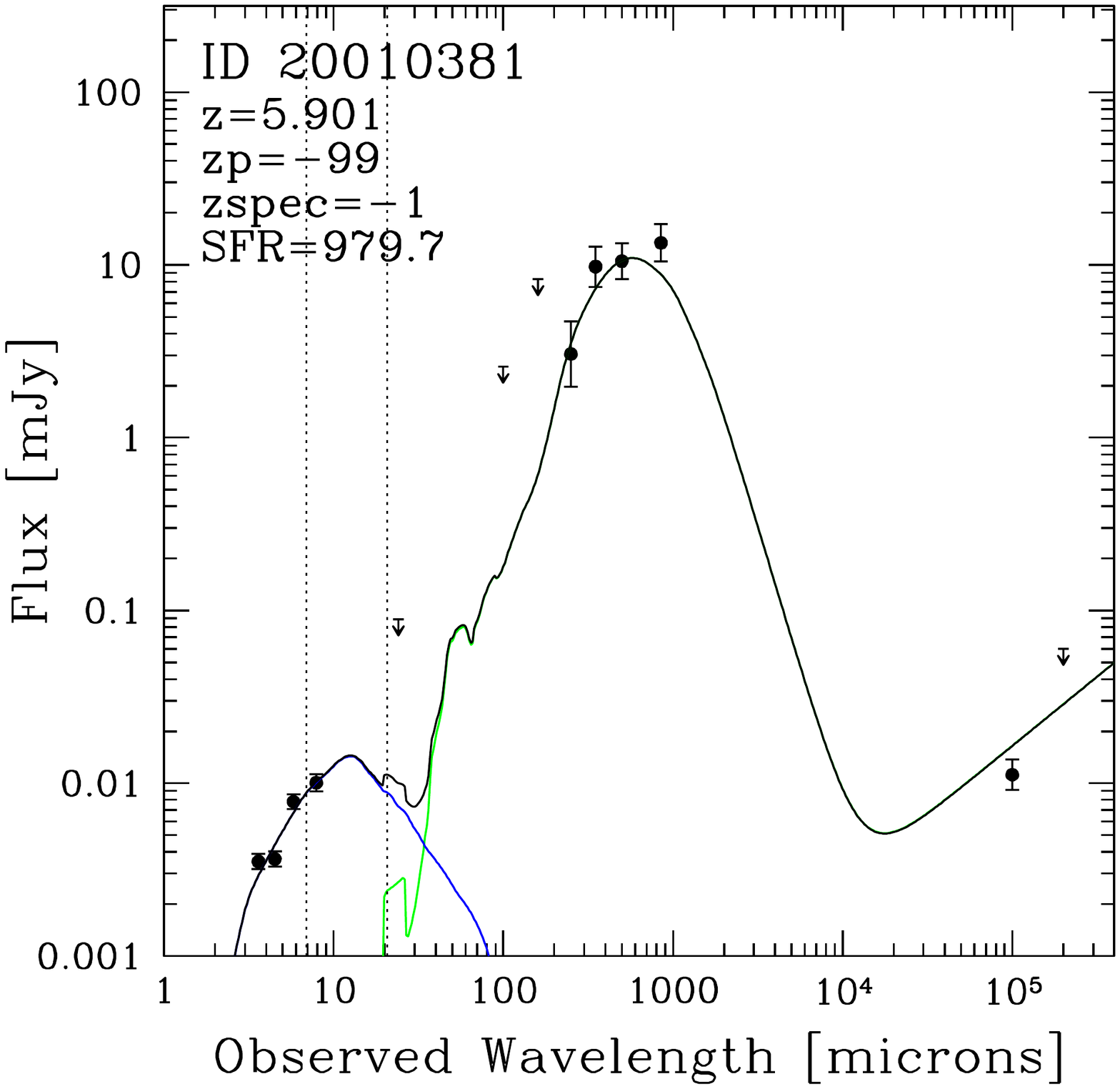}
    	\includegraphics[width=0.28\textwidth]{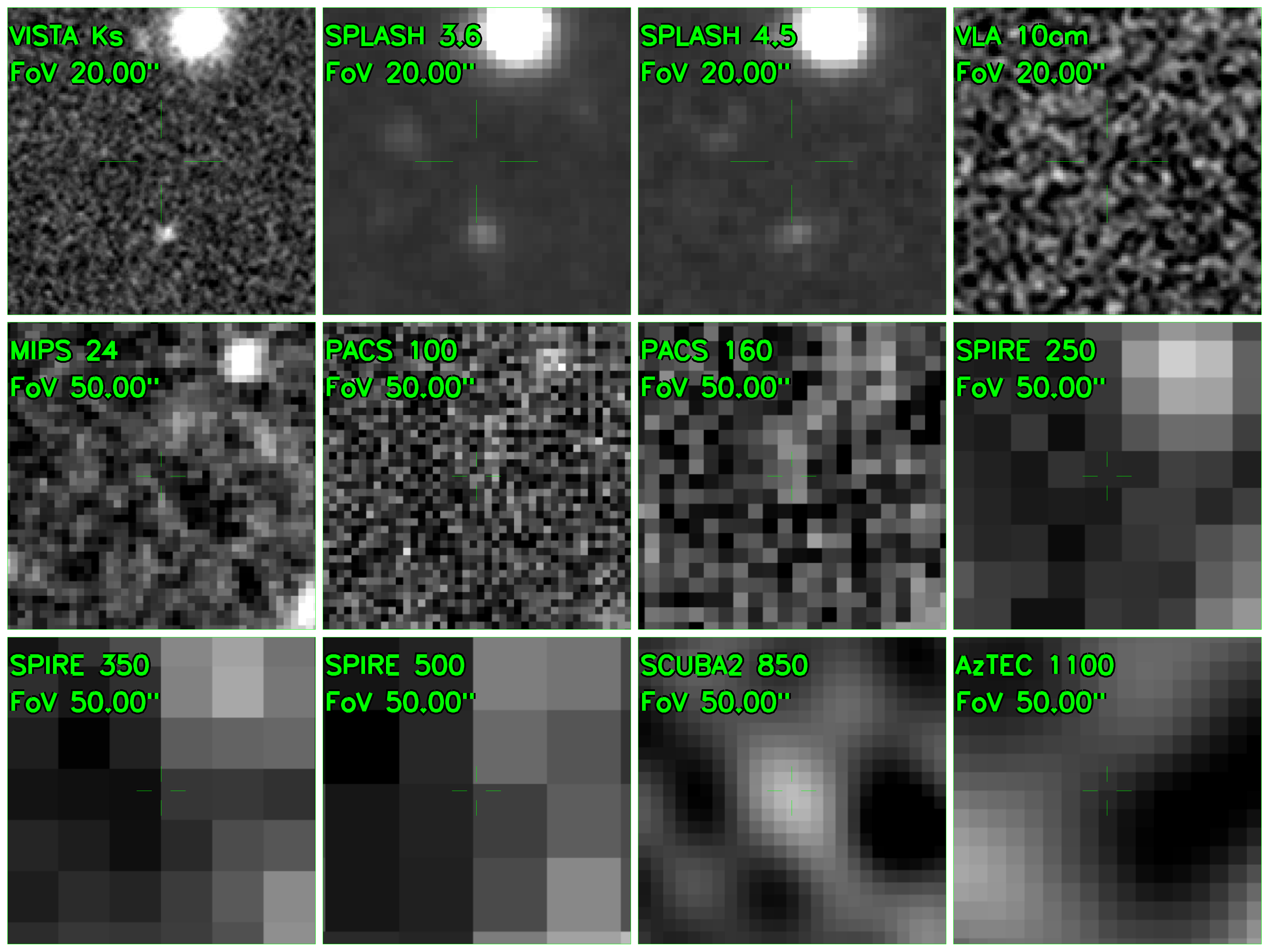}
	\includegraphics[width=0.21\textwidth, trim={0.6cm 5cm 1cm 3.5cm}, clip]{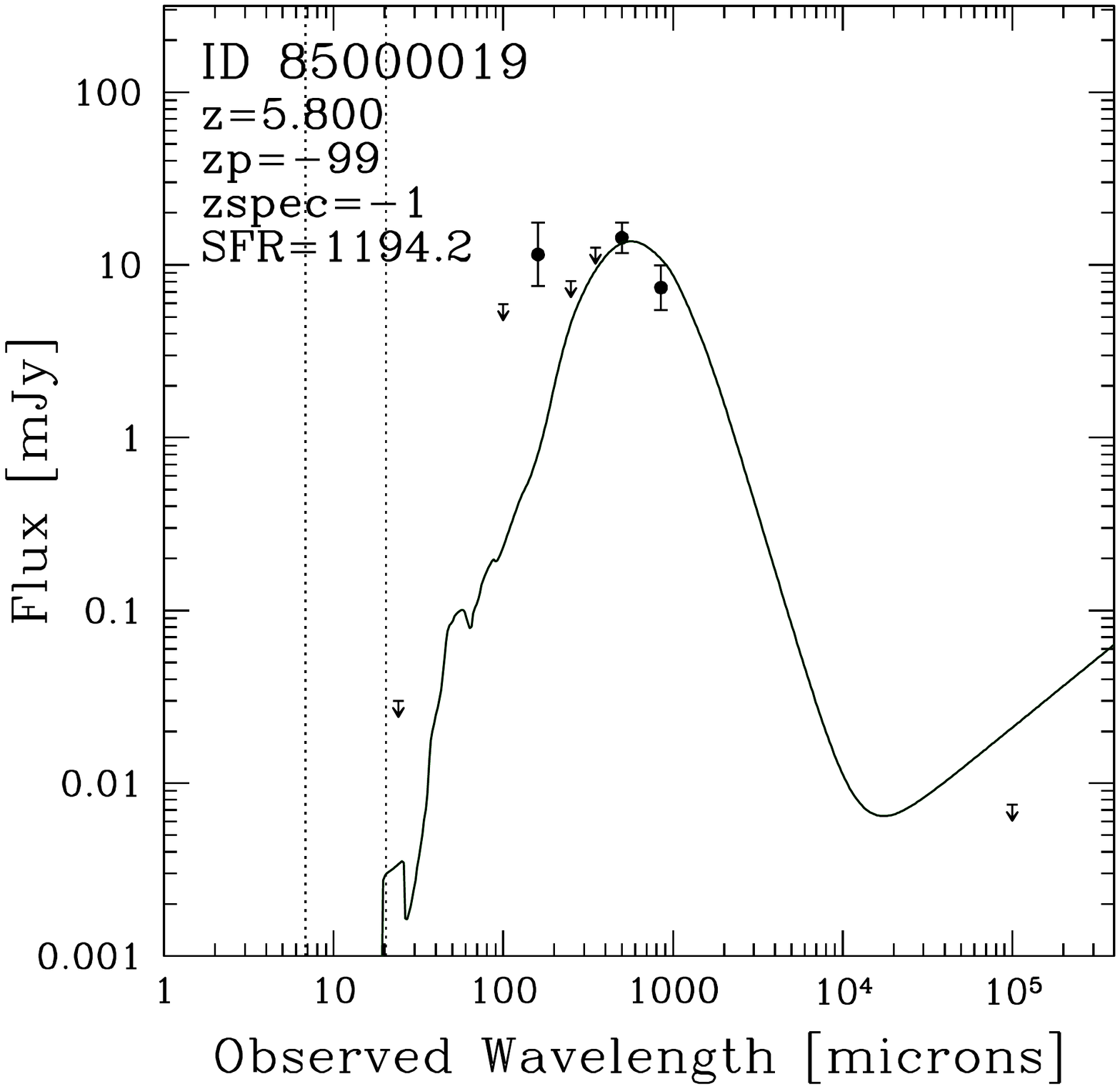}
    \includegraphics[width=0.28\textwidth]{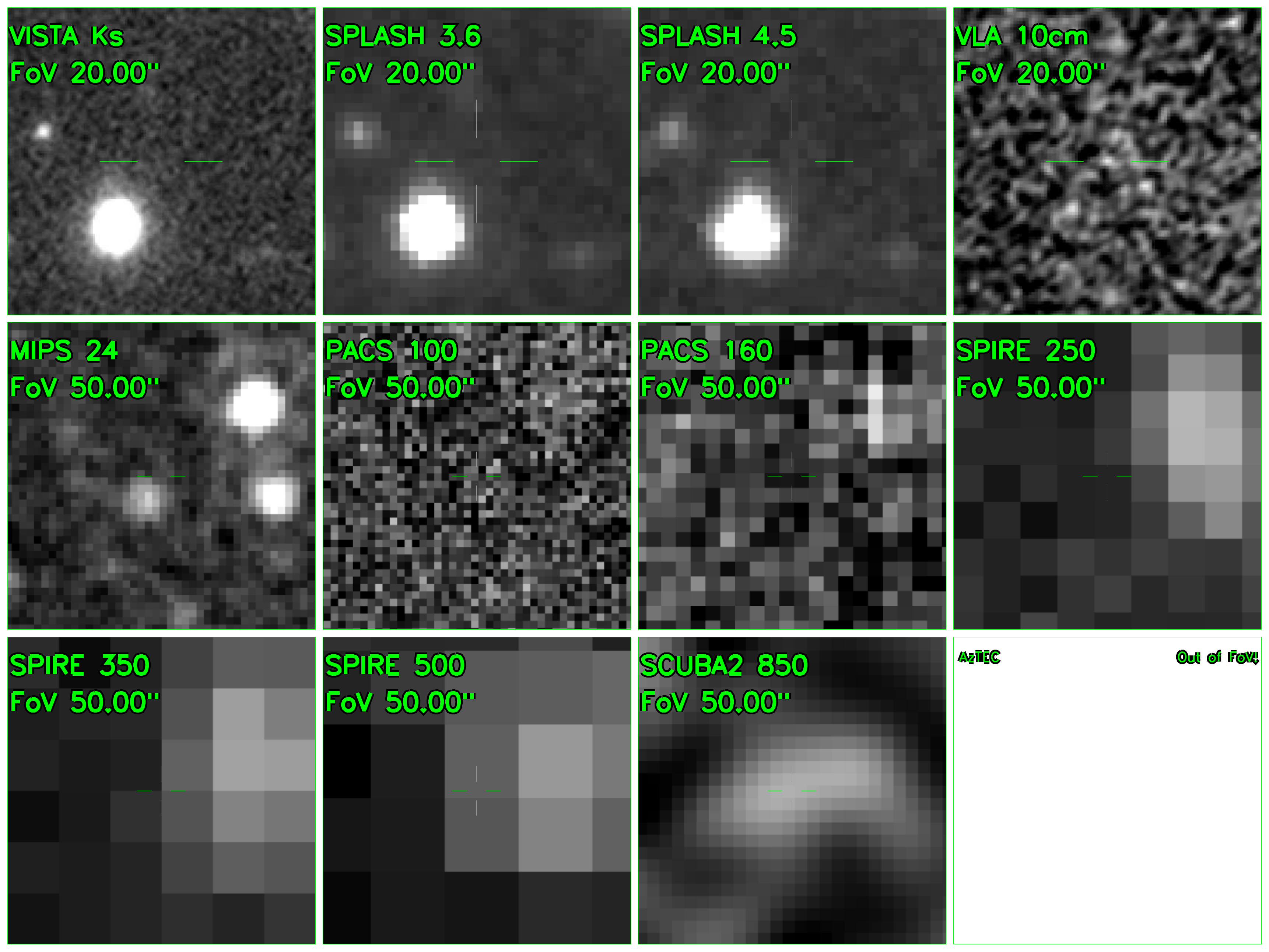}
	\includegraphics[width=0.21\textwidth, trim={0.6cm 5cm 1cm 3.5cm}, clip]{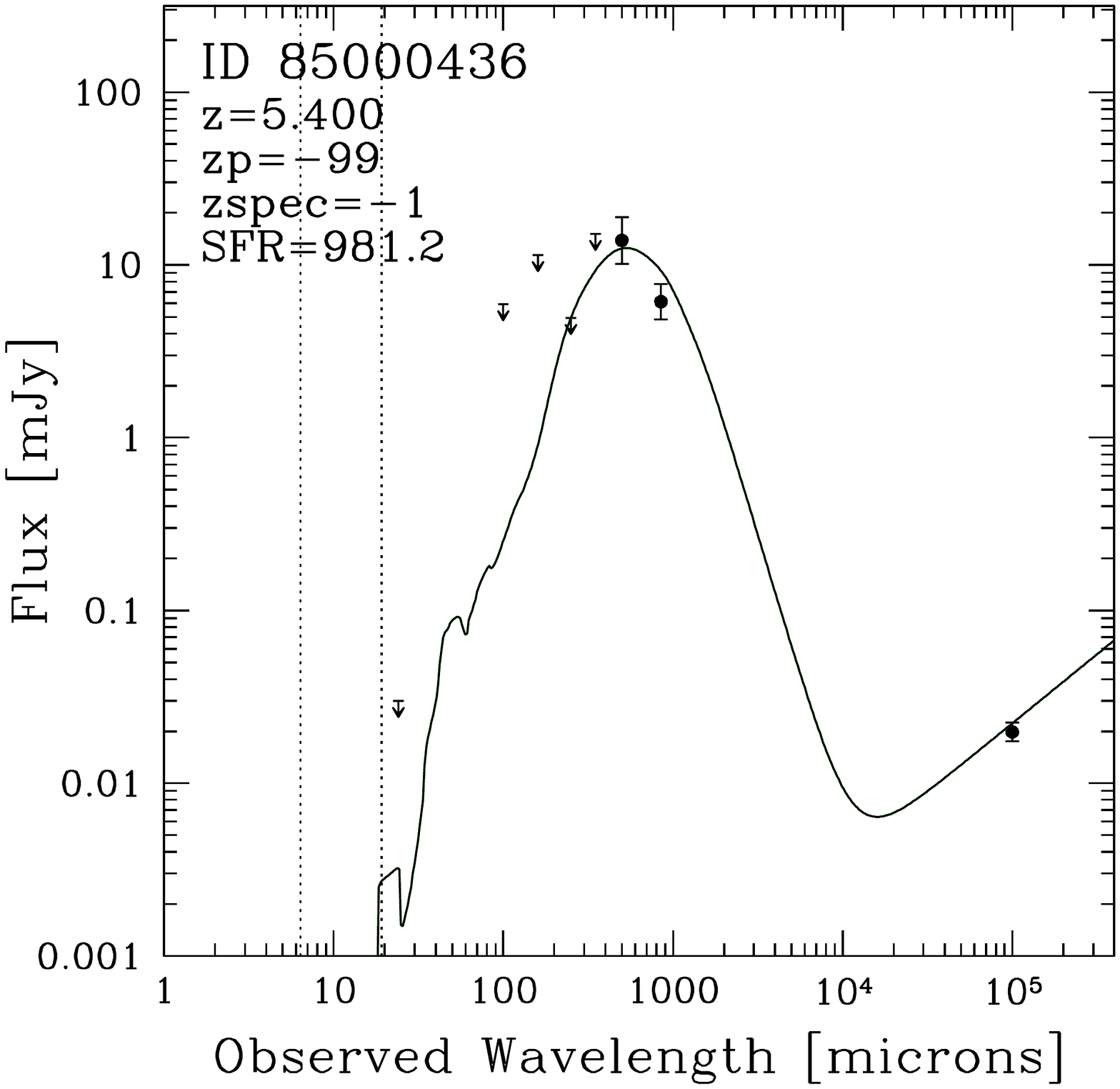}
    \includegraphics[width=0.28\textwidth]{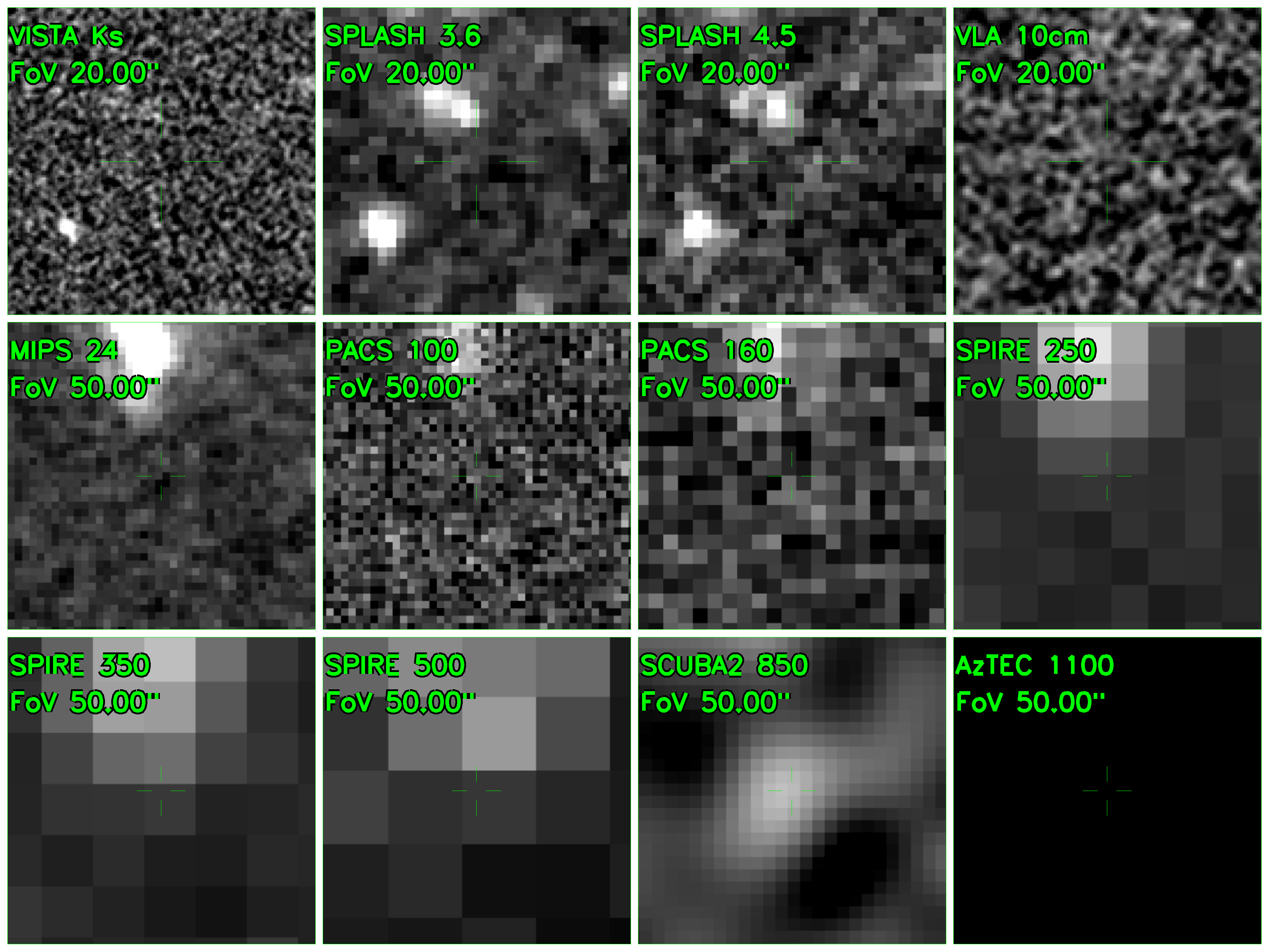}
	\includegraphics[width=0.21\textwidth, trim={0.6cm 5cm 1cm 3.5cm}, clip]{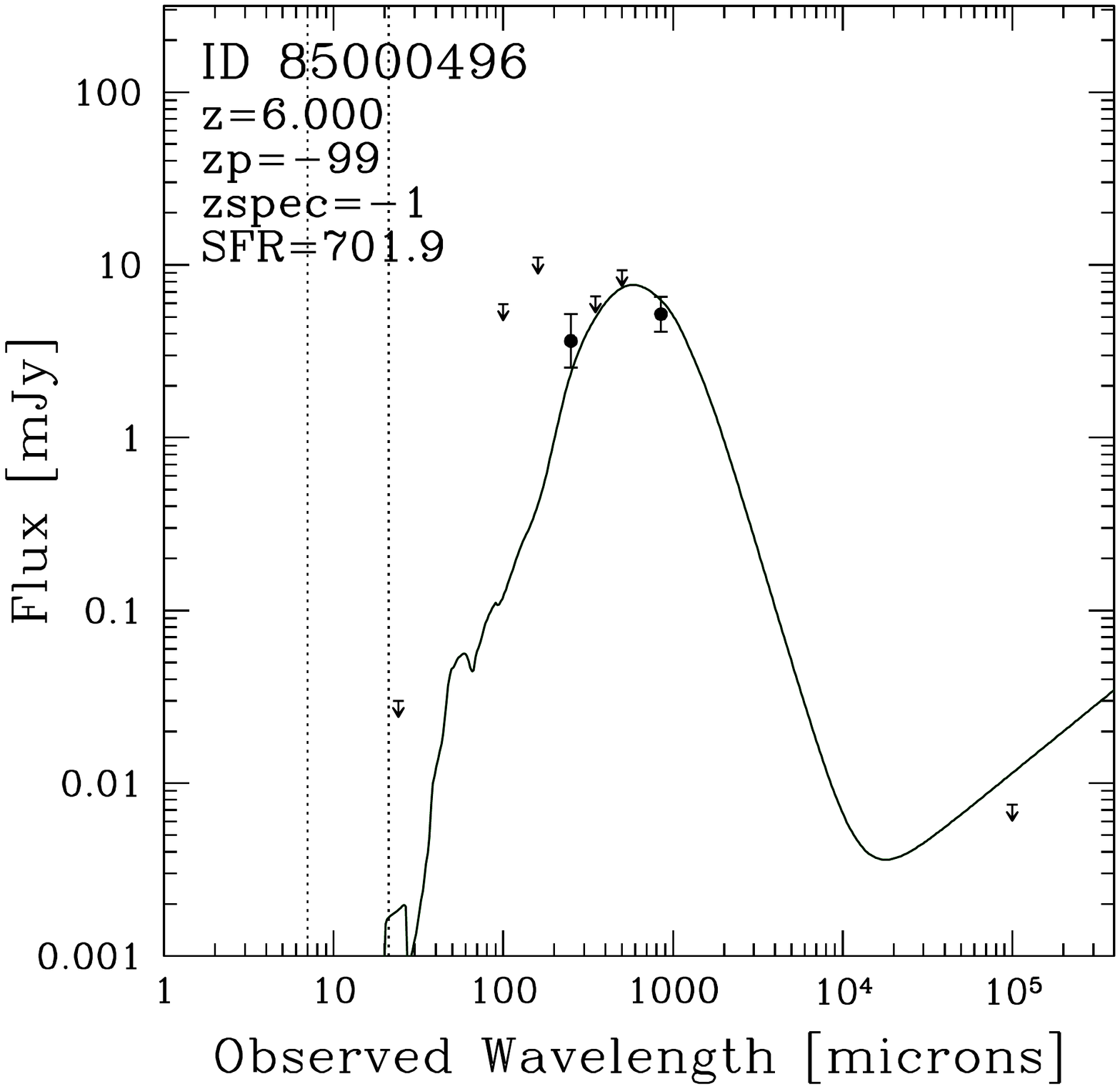}
    \includegraphics[width=0.28\textwidth]{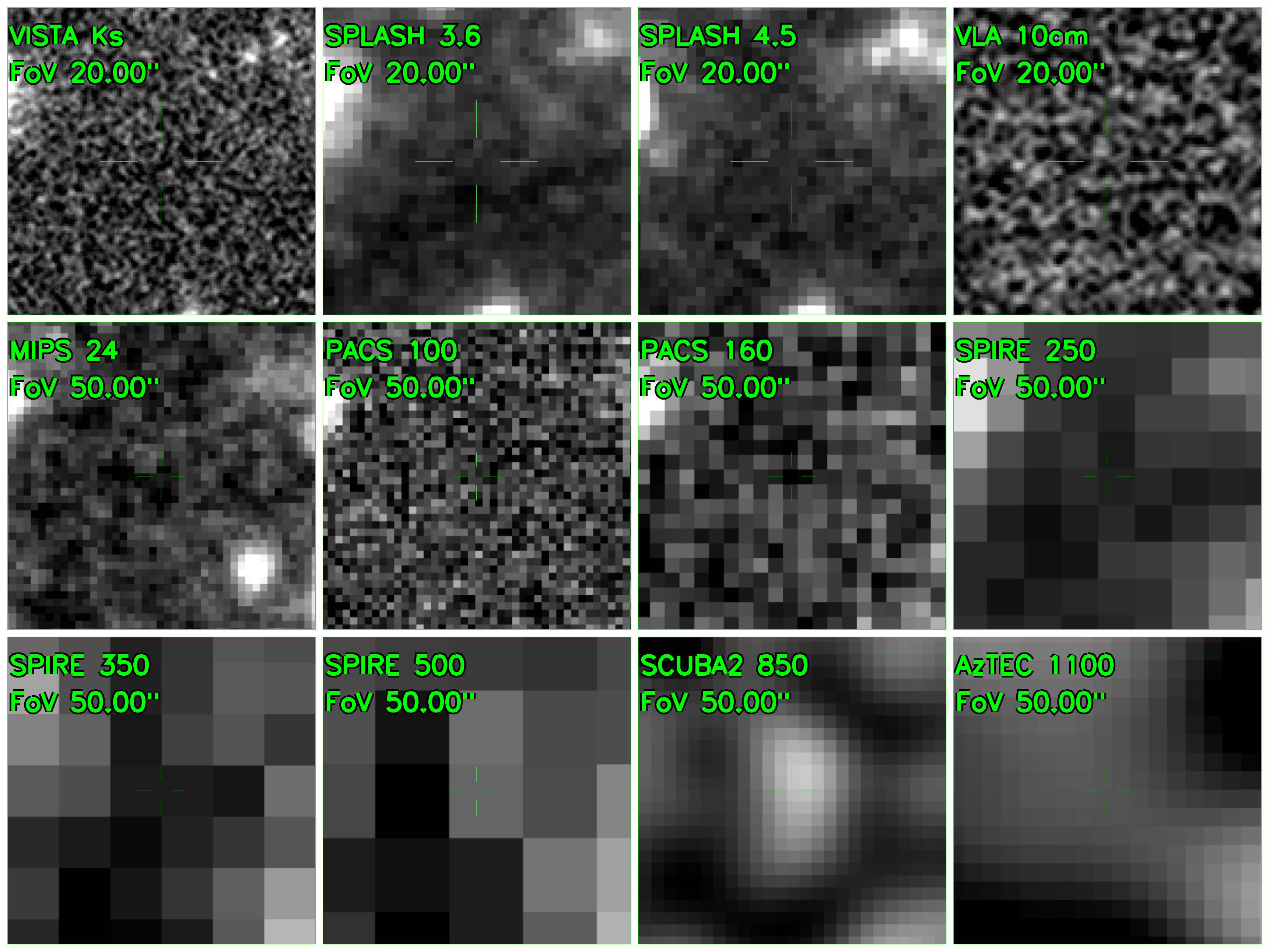}
	\includegraphics[width=0.21\textwidth, trim={0.6cm 5cm 1cm 3.5cm}, clip]{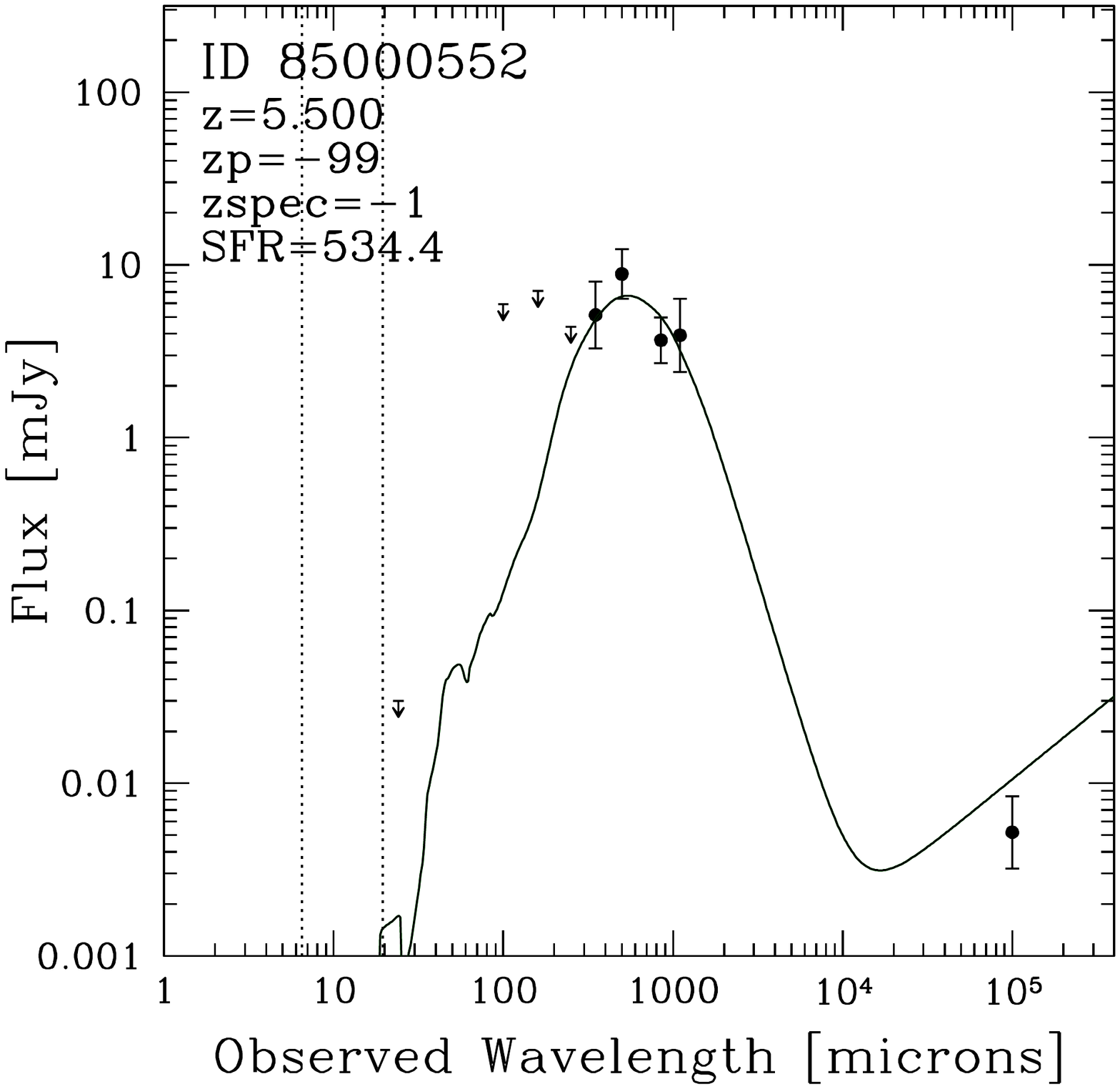}
    \includegraphics[width=0.28\textwidth]{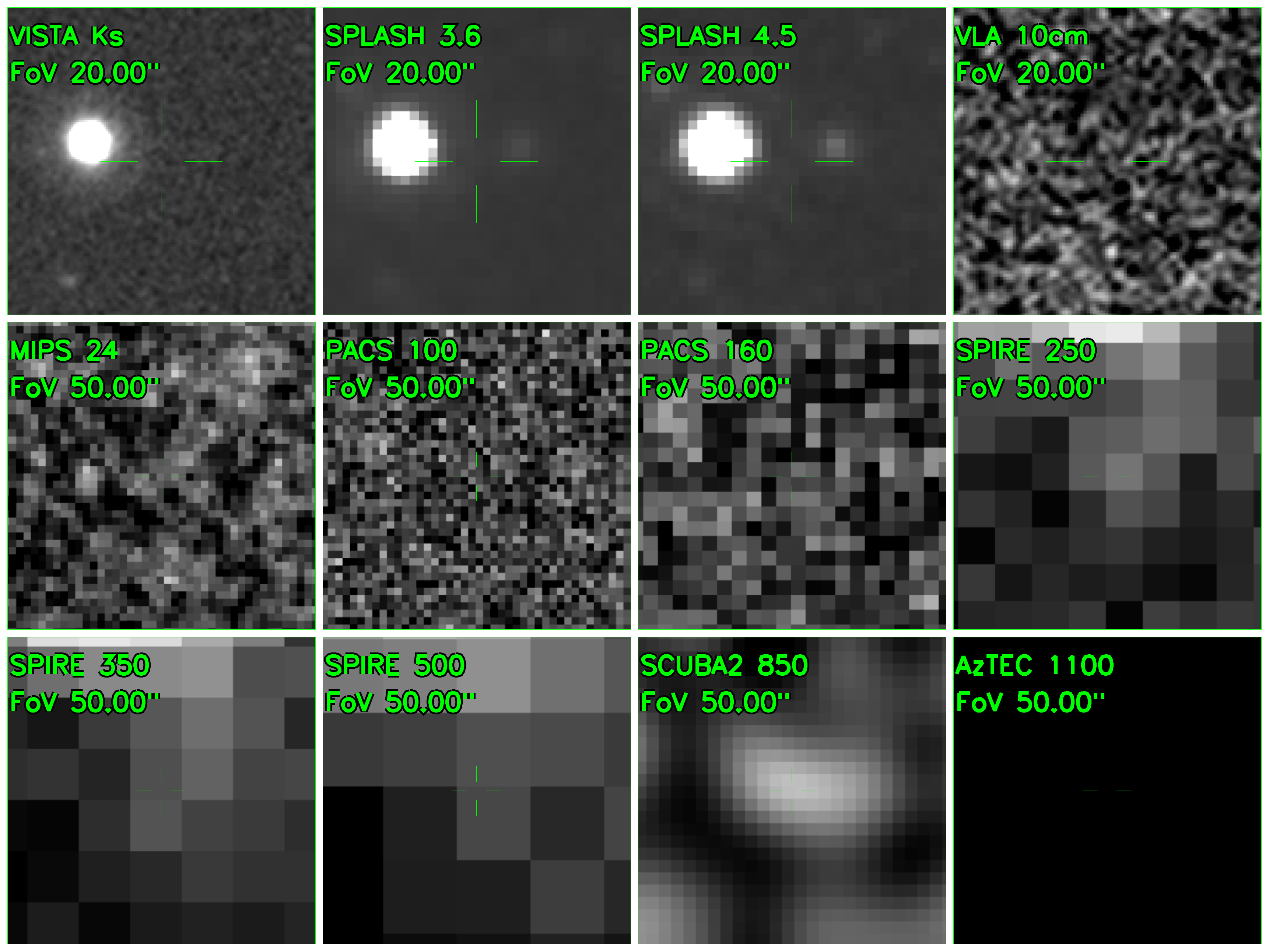}
	\includegraphics[width=0.21\textwidth, trim={0.6cm 5cm 1cm 3.5cm}, clip]{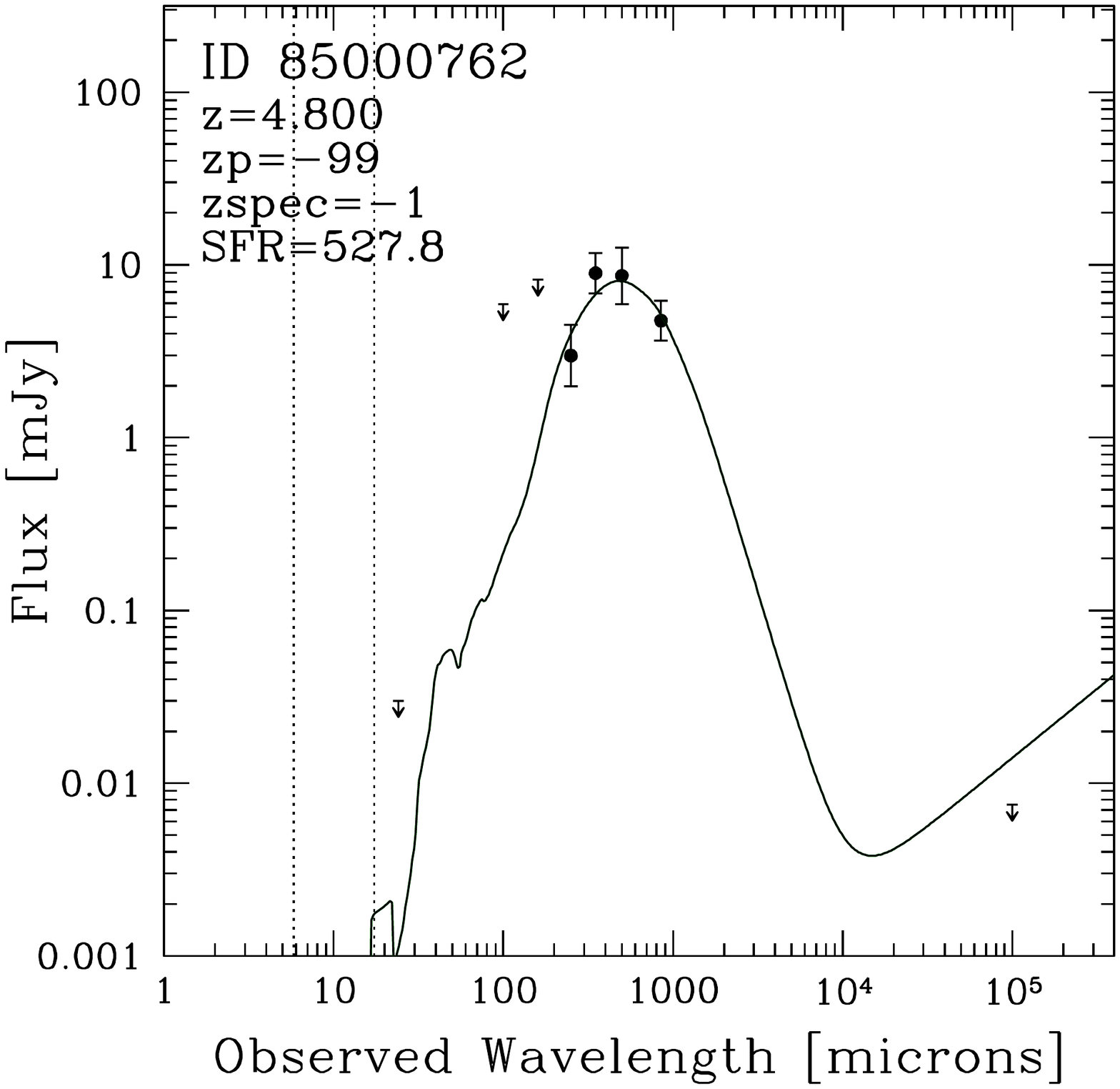}
    \includegraphics[width=0.28\textwidth]{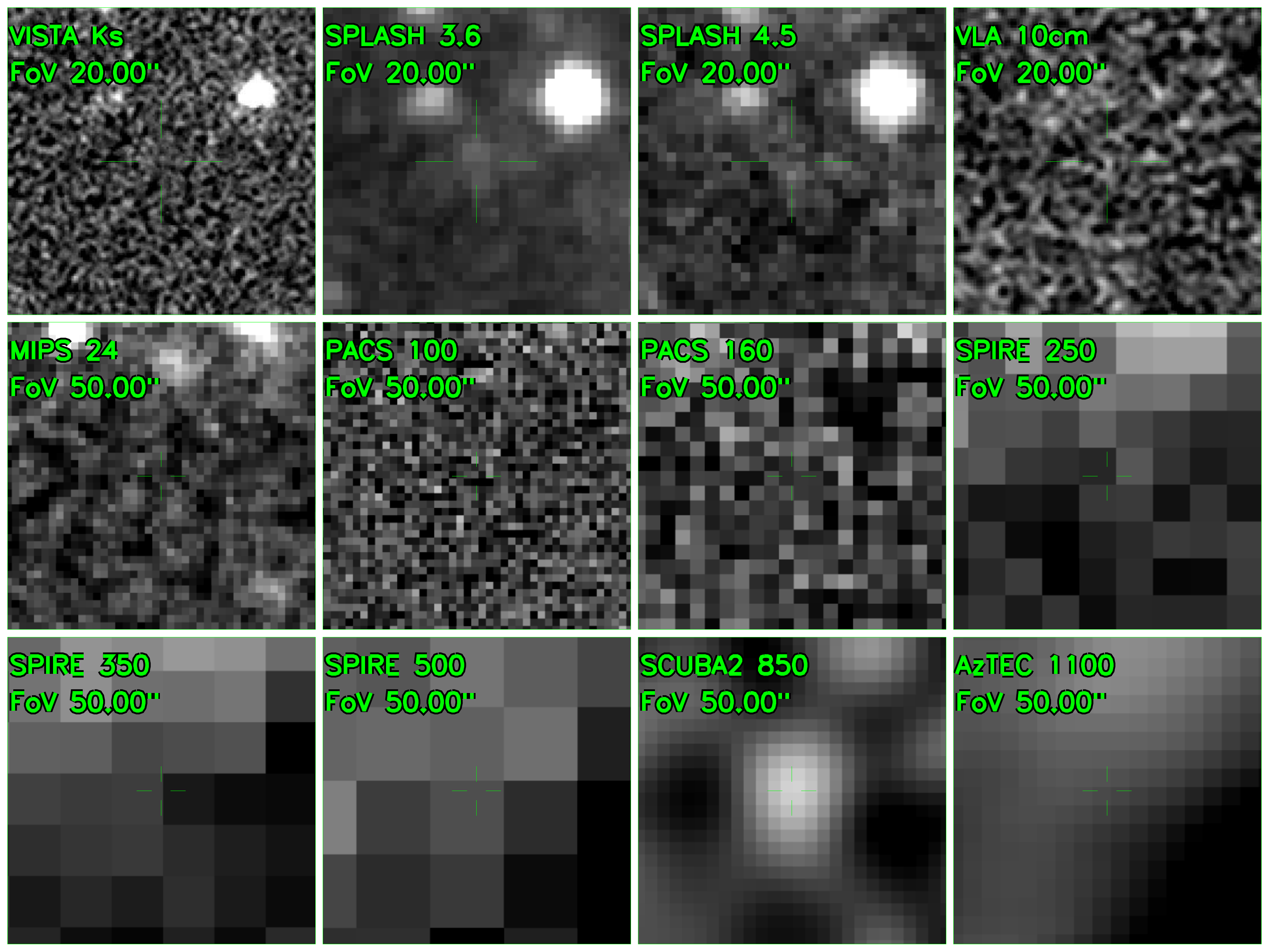}
	\includegraphics[width=0.21\textwidth, trim={0.6cm 5cm 1cm 3.5cm}, clip]{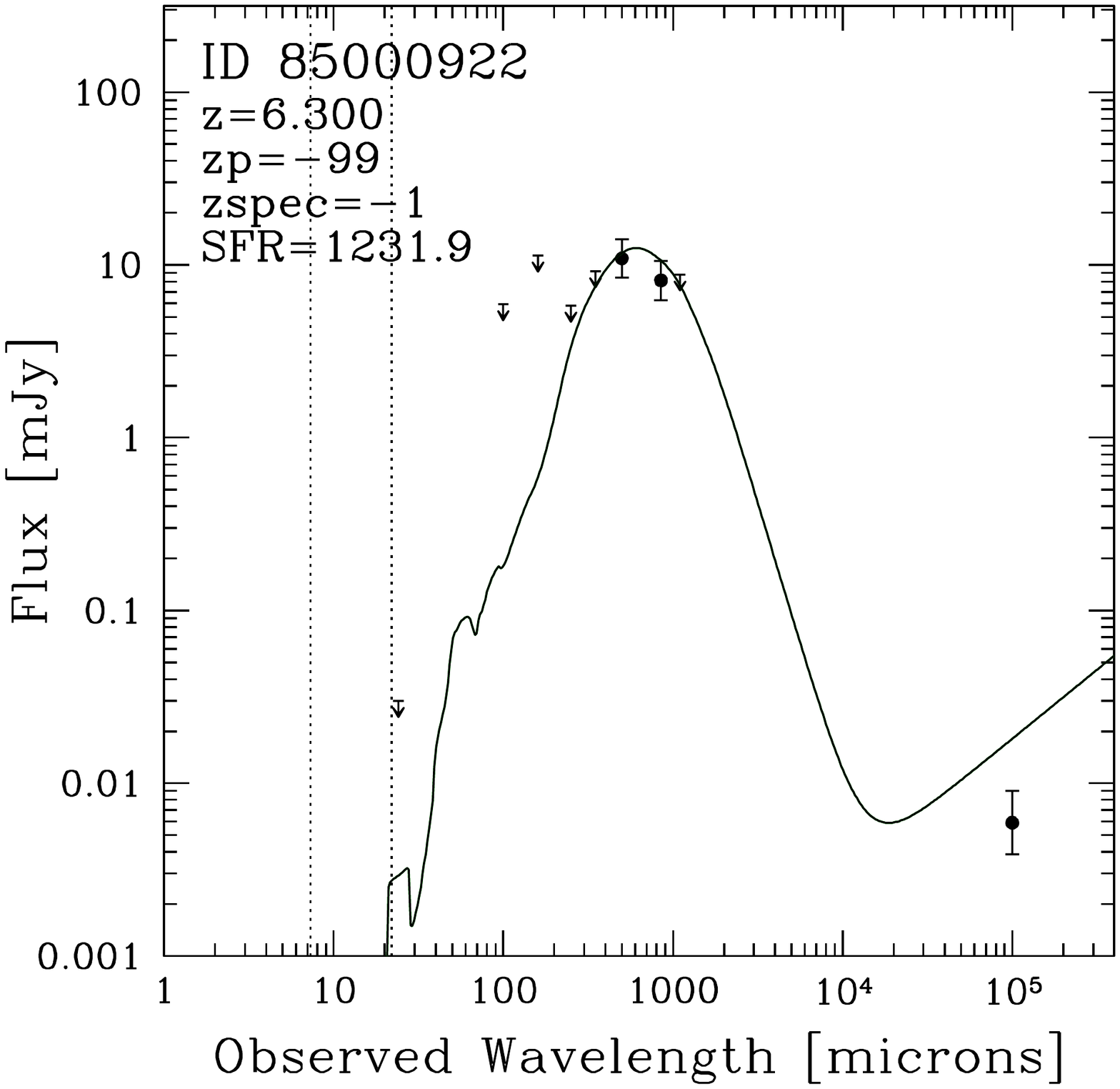}
    \includegraphics[width=0.28\textwidth]{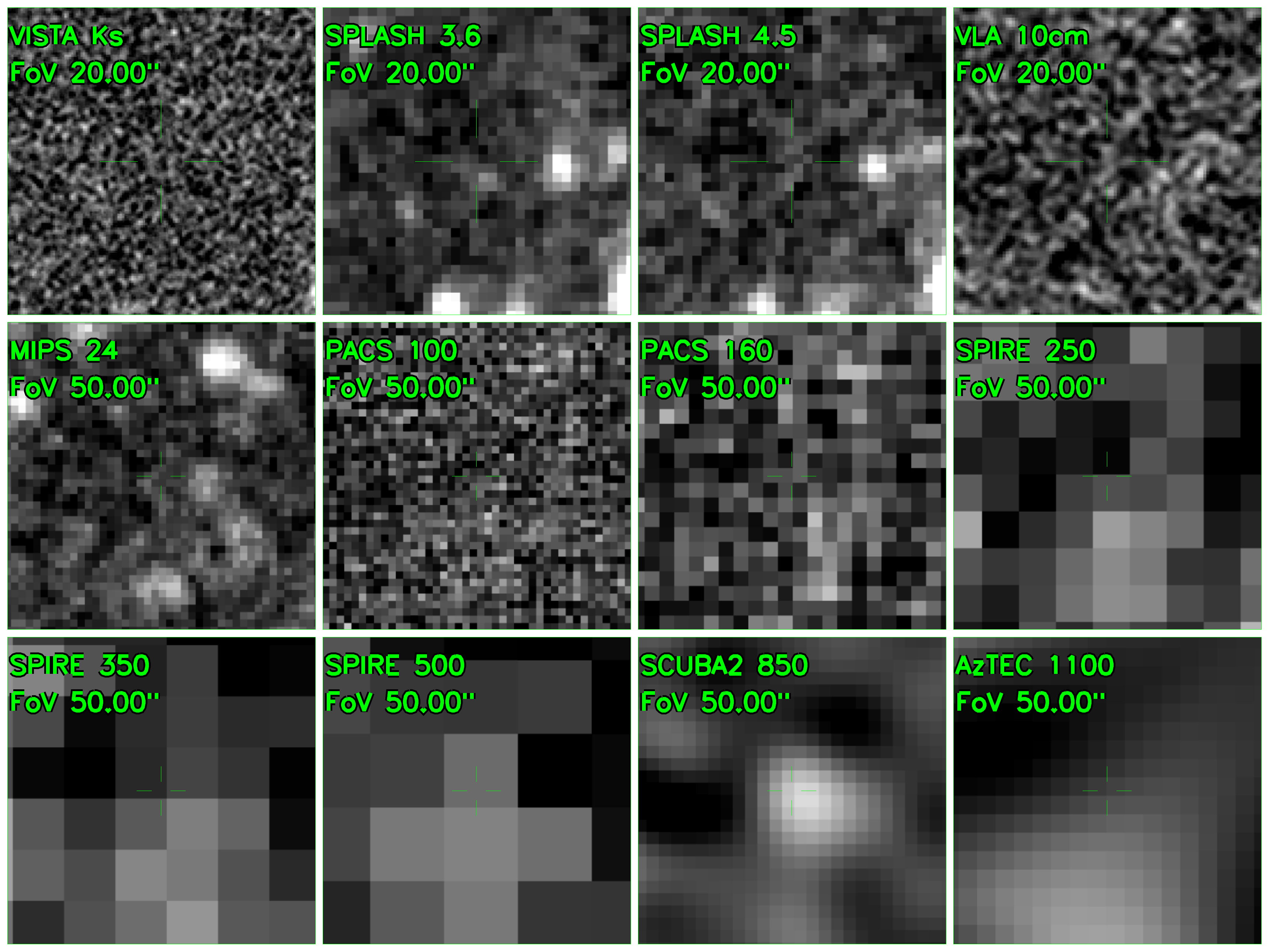}
	\includegraphics[width=0.21\textwidth, trim={0.6cm 5cm 1cm 3.5cm}, clip]{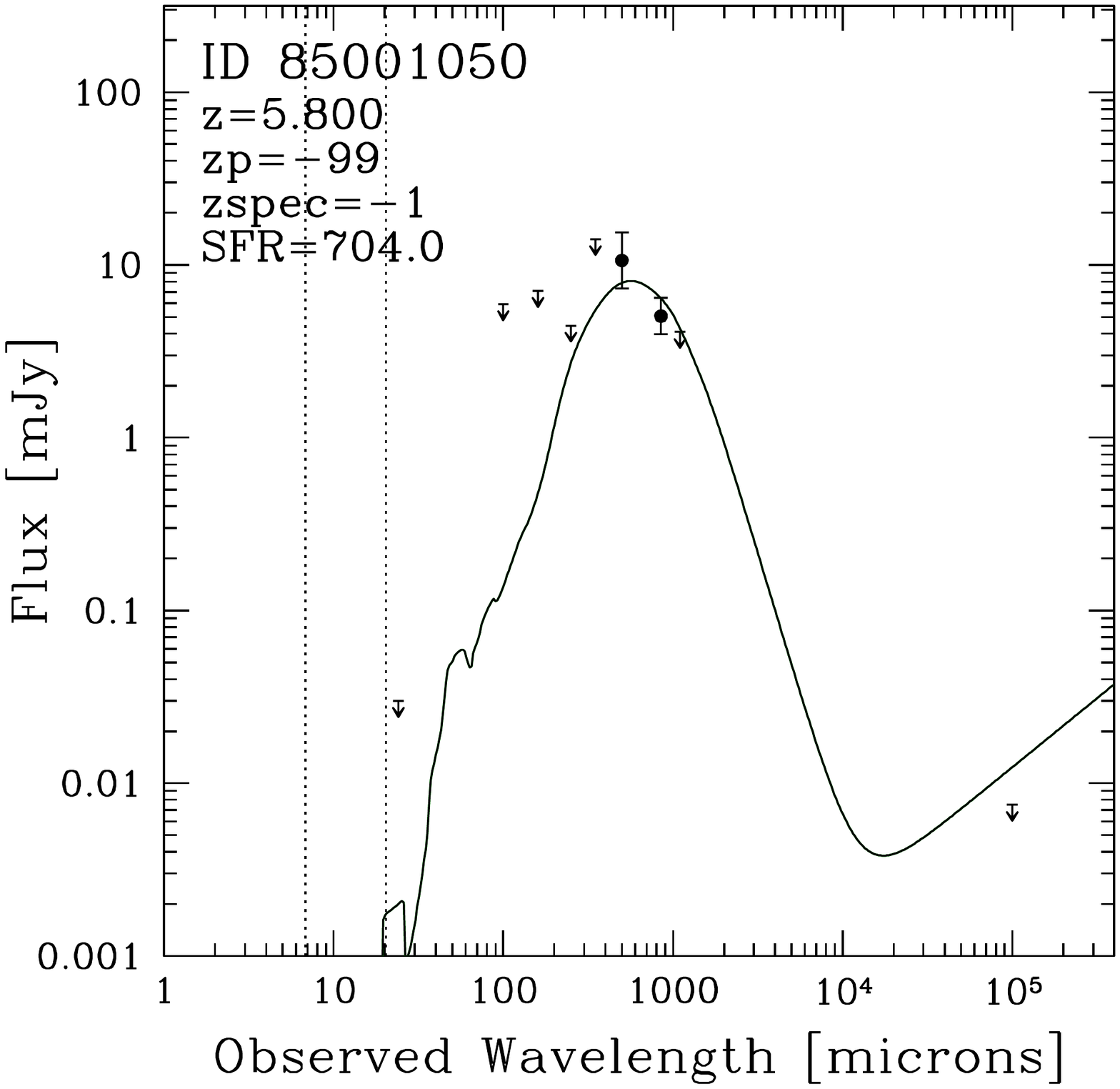}
    \includegraphics[width=0.28\textwidth]{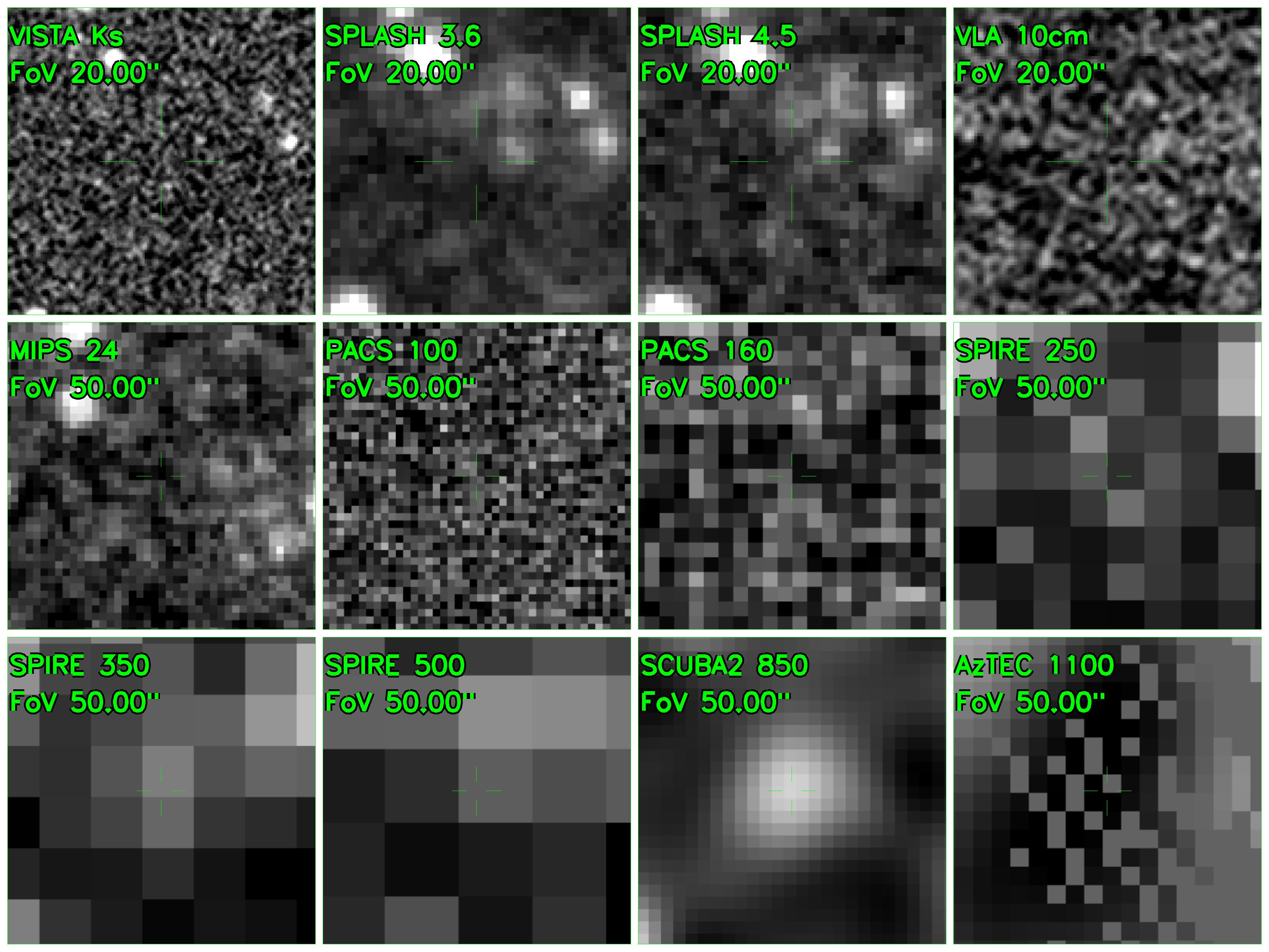}
	\includegraphics[width=0.21\textwidth, trim={0.6cm 5cm 1cm 3.5cm}, clip]{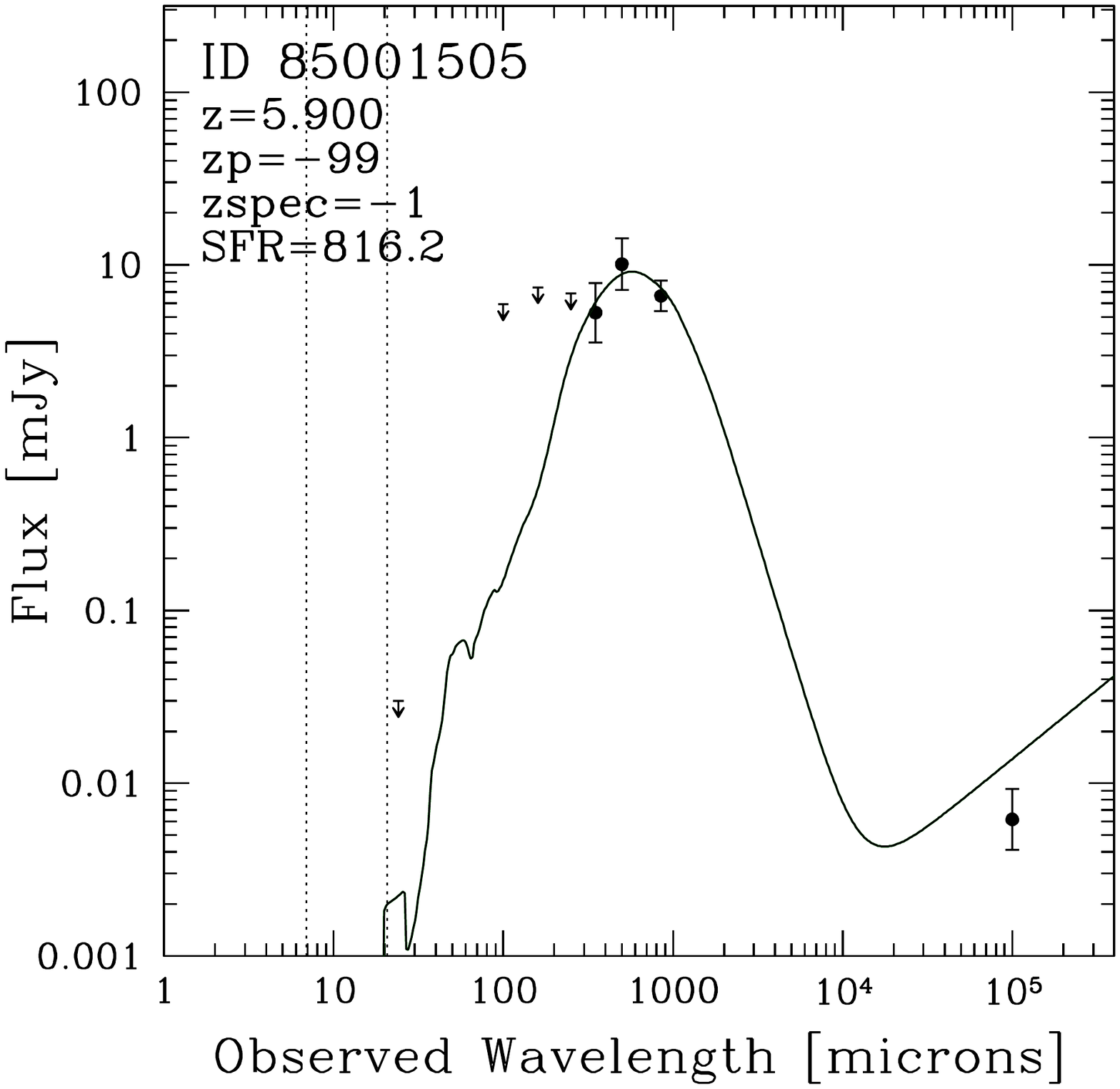}
    \includegraphics[width=0.28\textwidth]{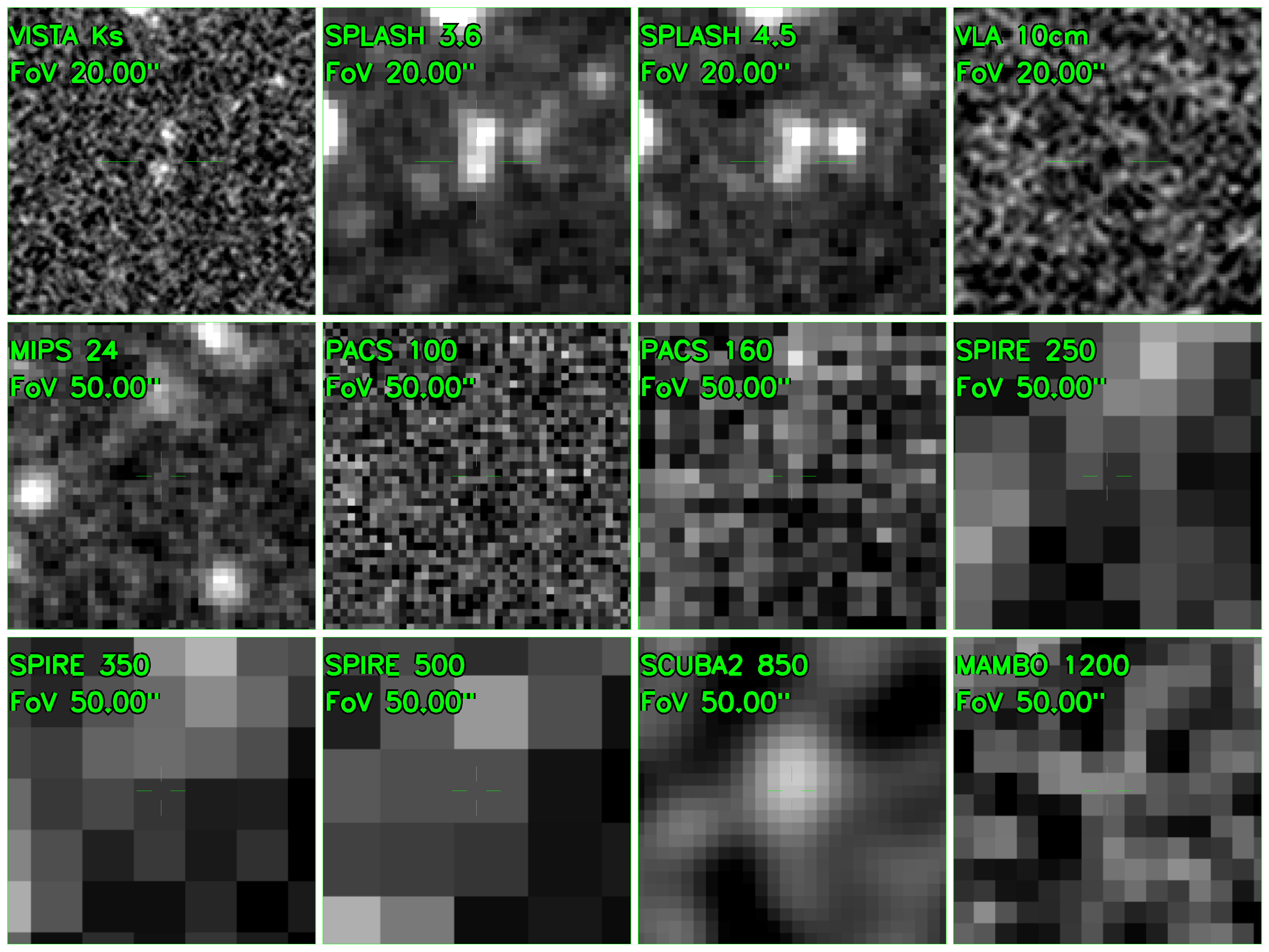}
	\includegraphics[width=0.21\textwidth, trim={0.6cm 5cm 1cm 3.5cm}, clip]{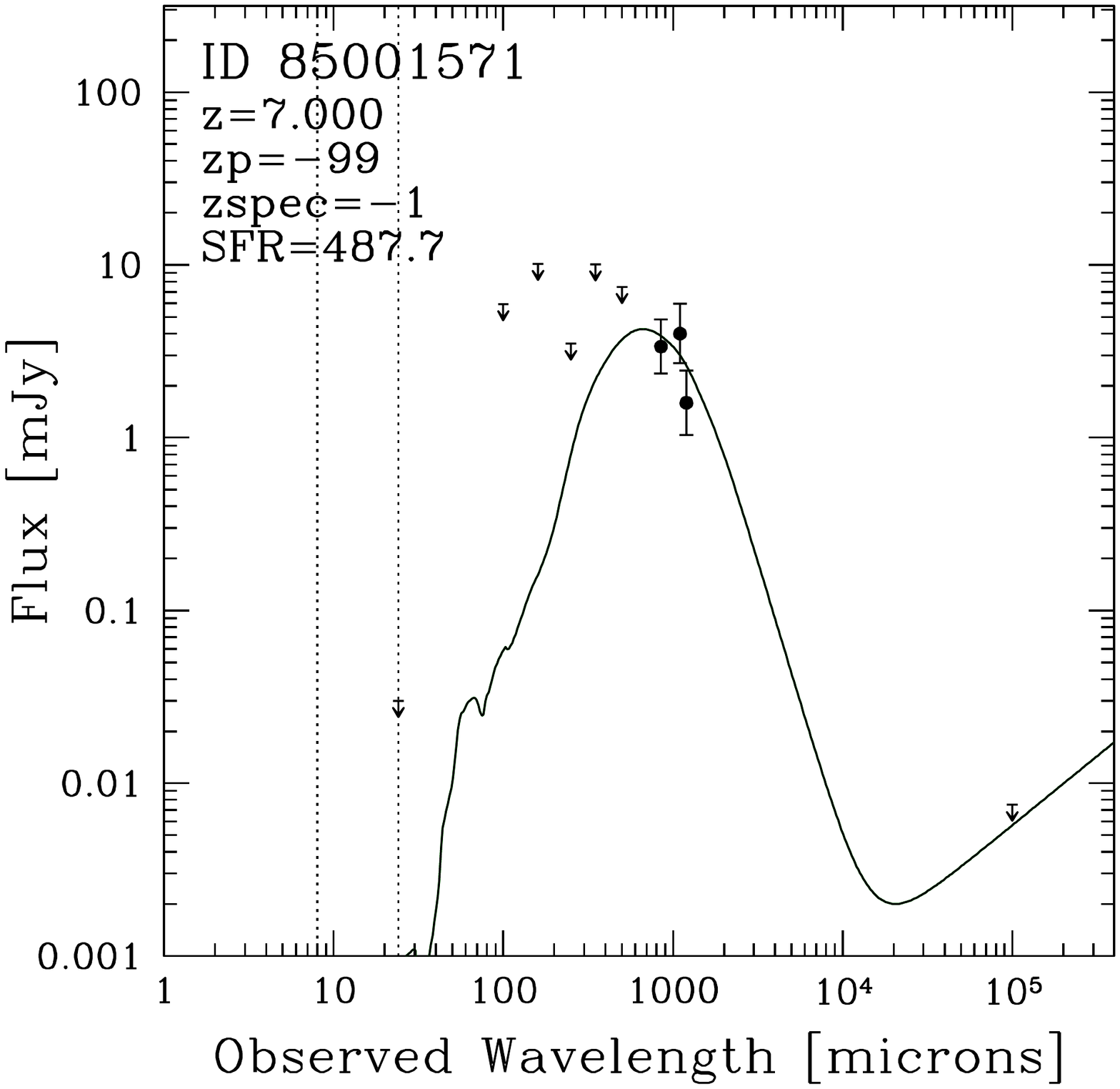}
	\caption{%
		Multi-band cutouts and SEDs of high redshift candidates, continued.  
		\label{highz_cutouts2}
		}
\end{figure}

\begin{figure}
	\centering
    \includegraphics[width=0.28\textwidth]{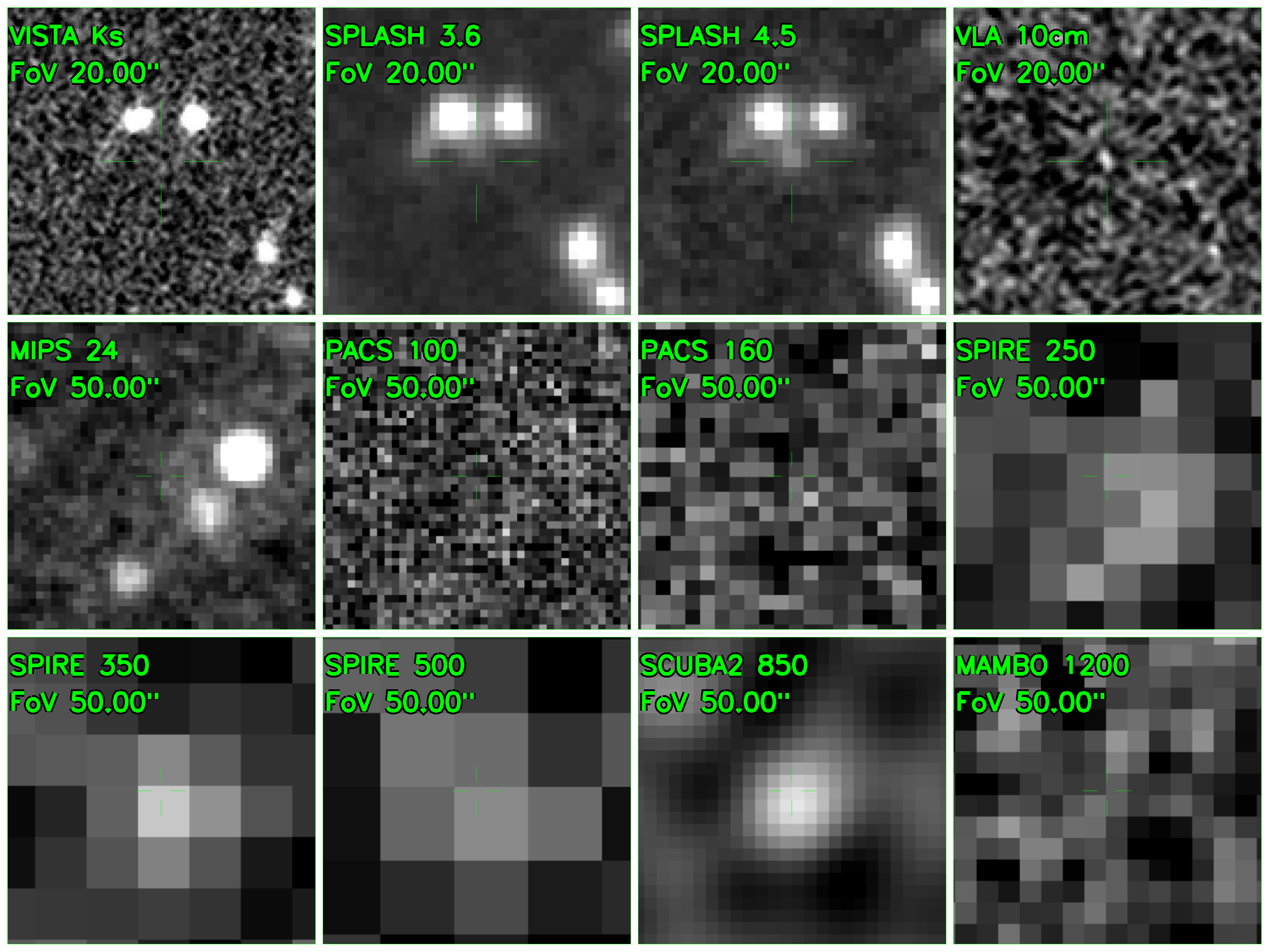}
	\includegraphics[width=0.21\textwidth, trim={0.6cm 5cm 1cm 3.5cm}, clip]{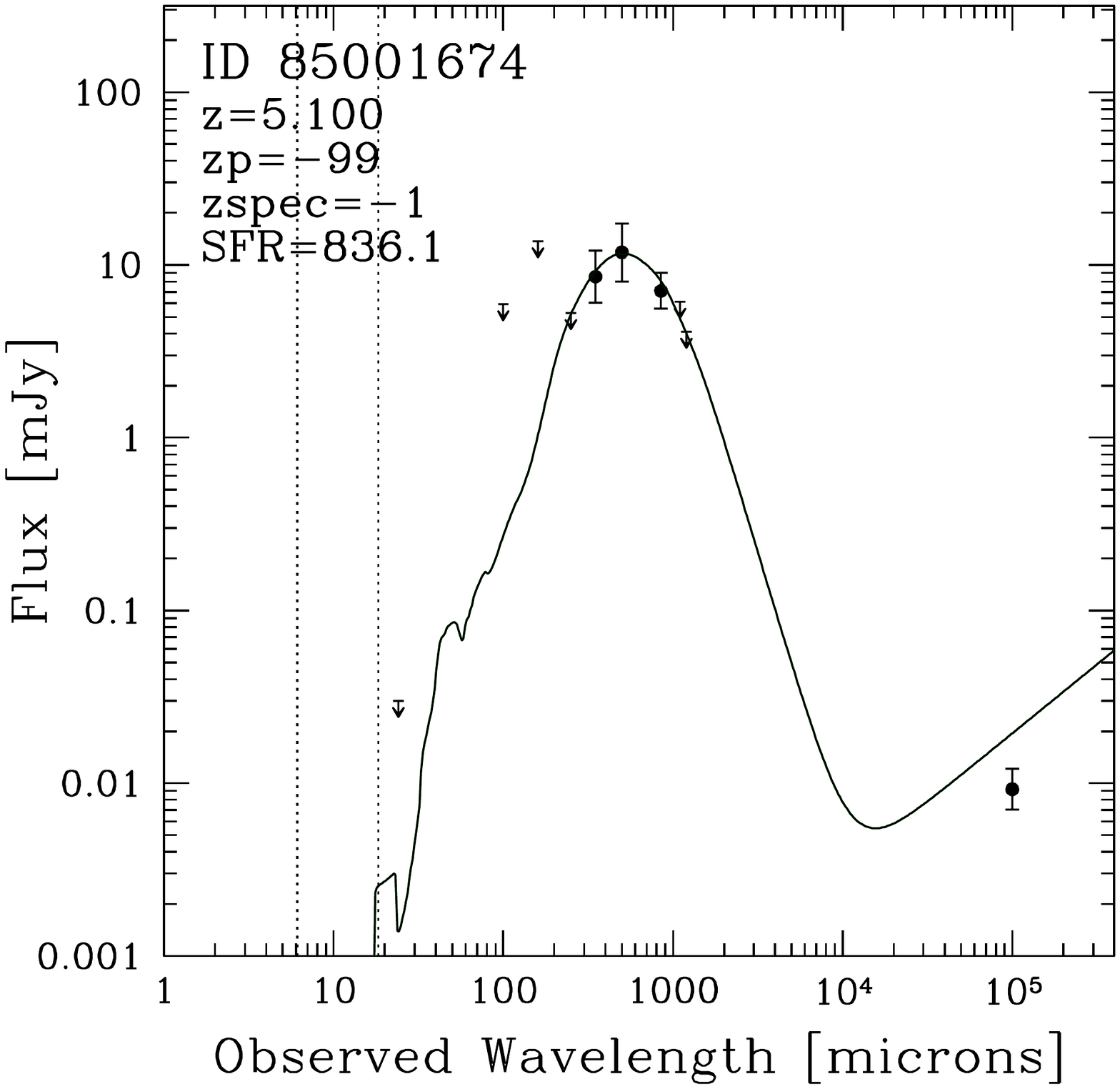}
    \includegraphics[width=0.28\textwidth]{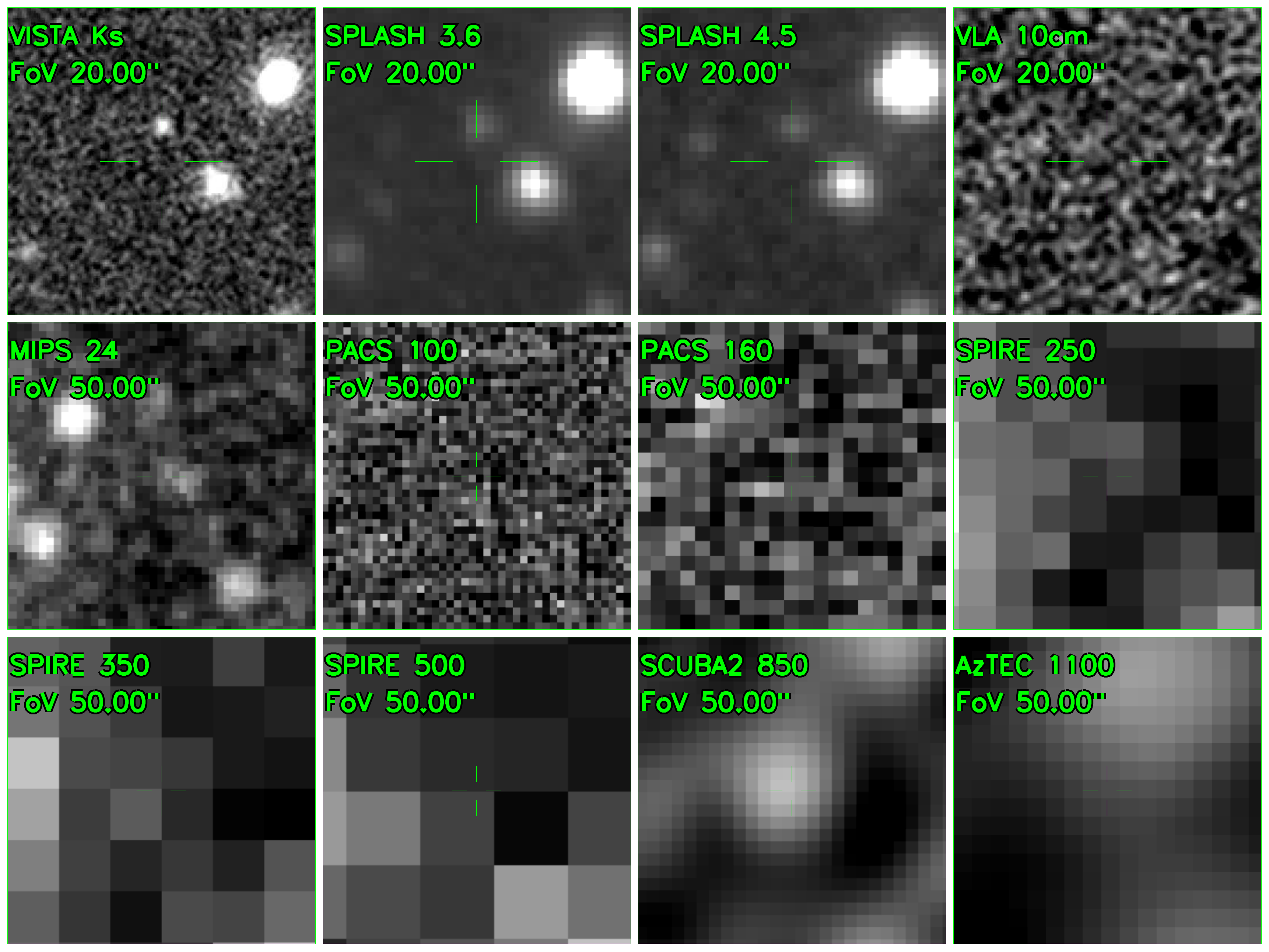}
	\includegraphics[width=0.21\textwidth, trim={0.6cm 5cm 1cm 3.5cm}, clip]{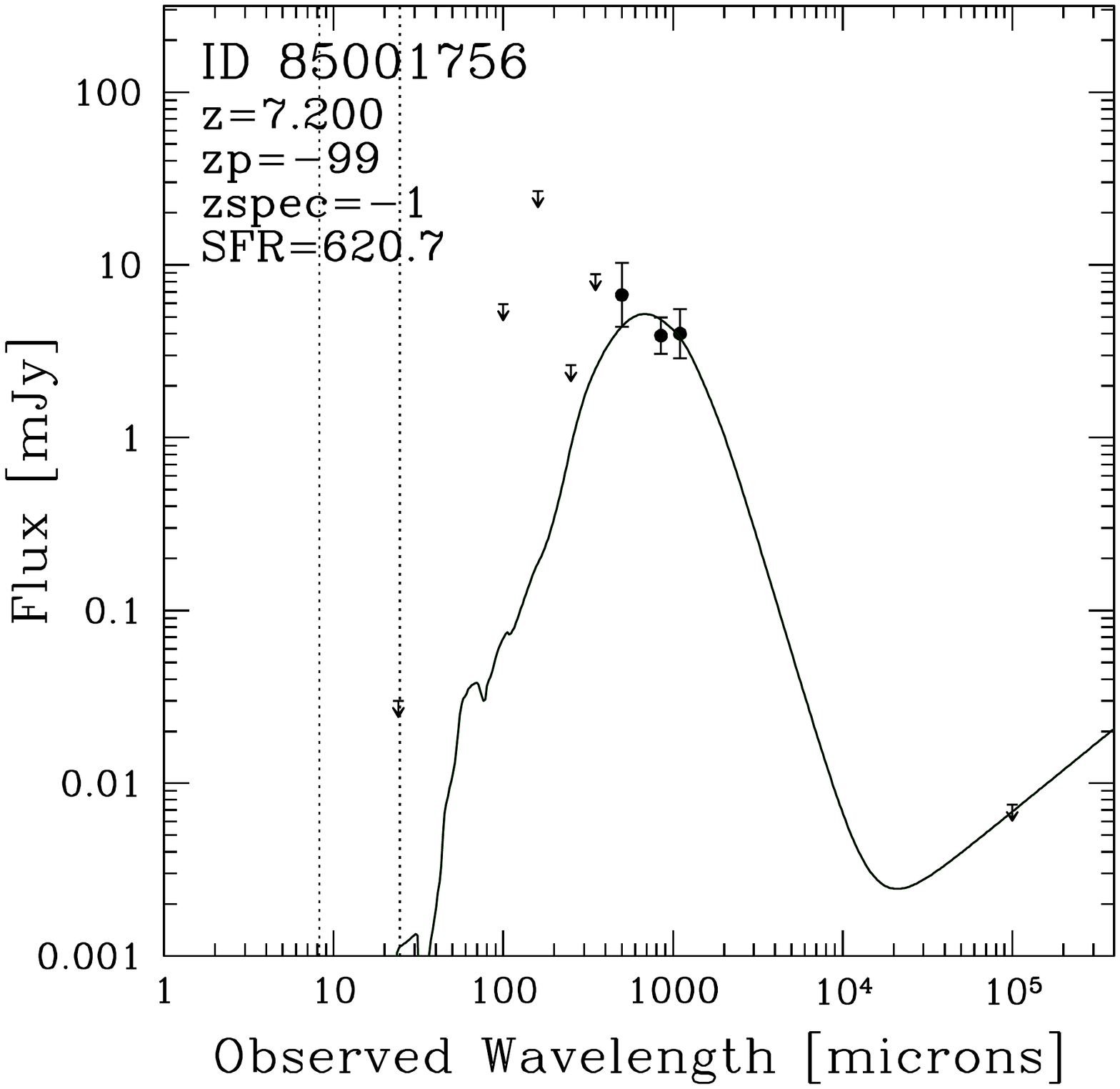}
    \includegraphics[width=0.28\textwidth]{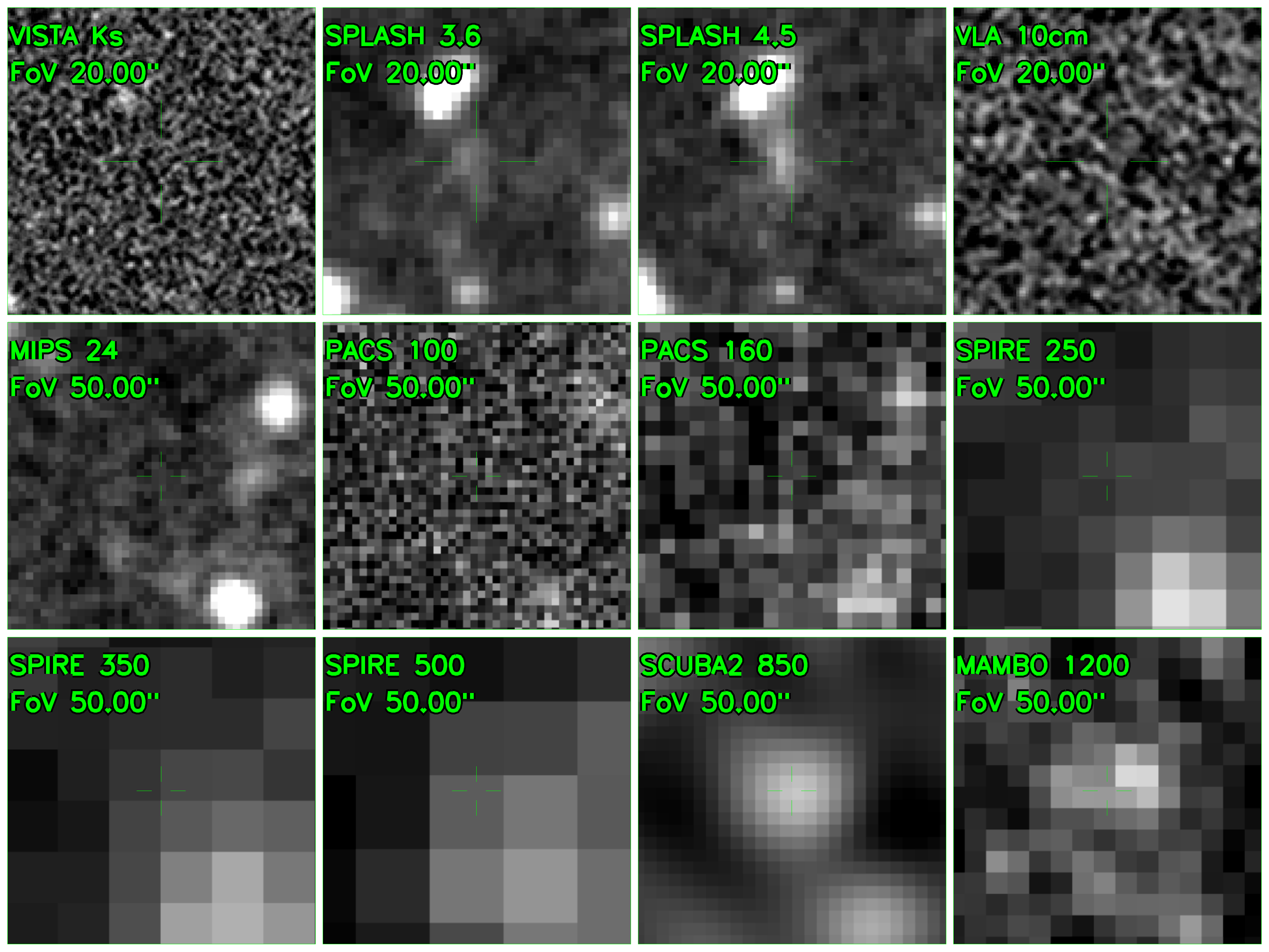}
	\includegraphics[width=0.21\textwidth, trim={0.6cm 5cm 1cm 3.5cm}, clip]{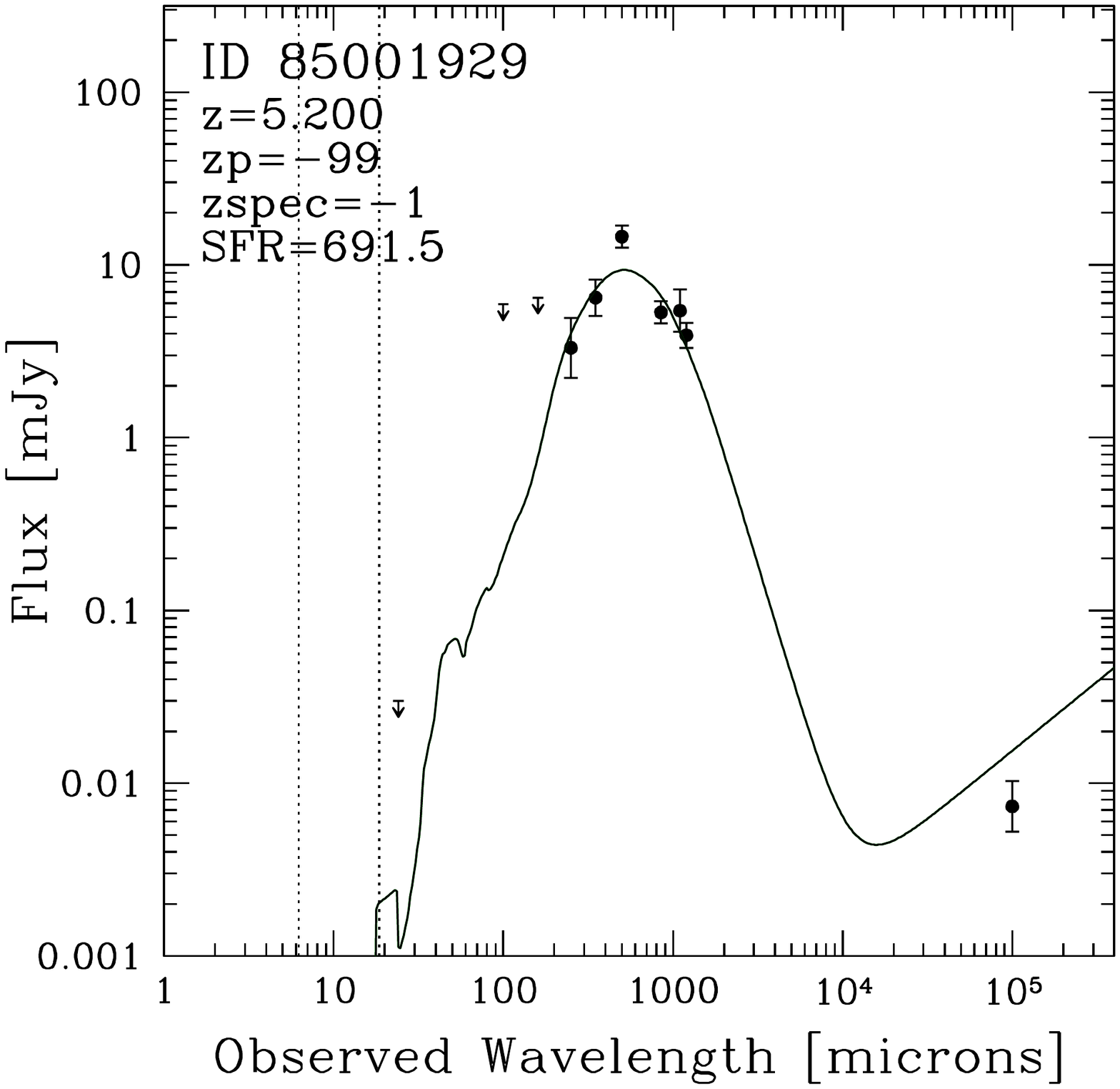}
    \includegraphics[width=0.28\textwidth]{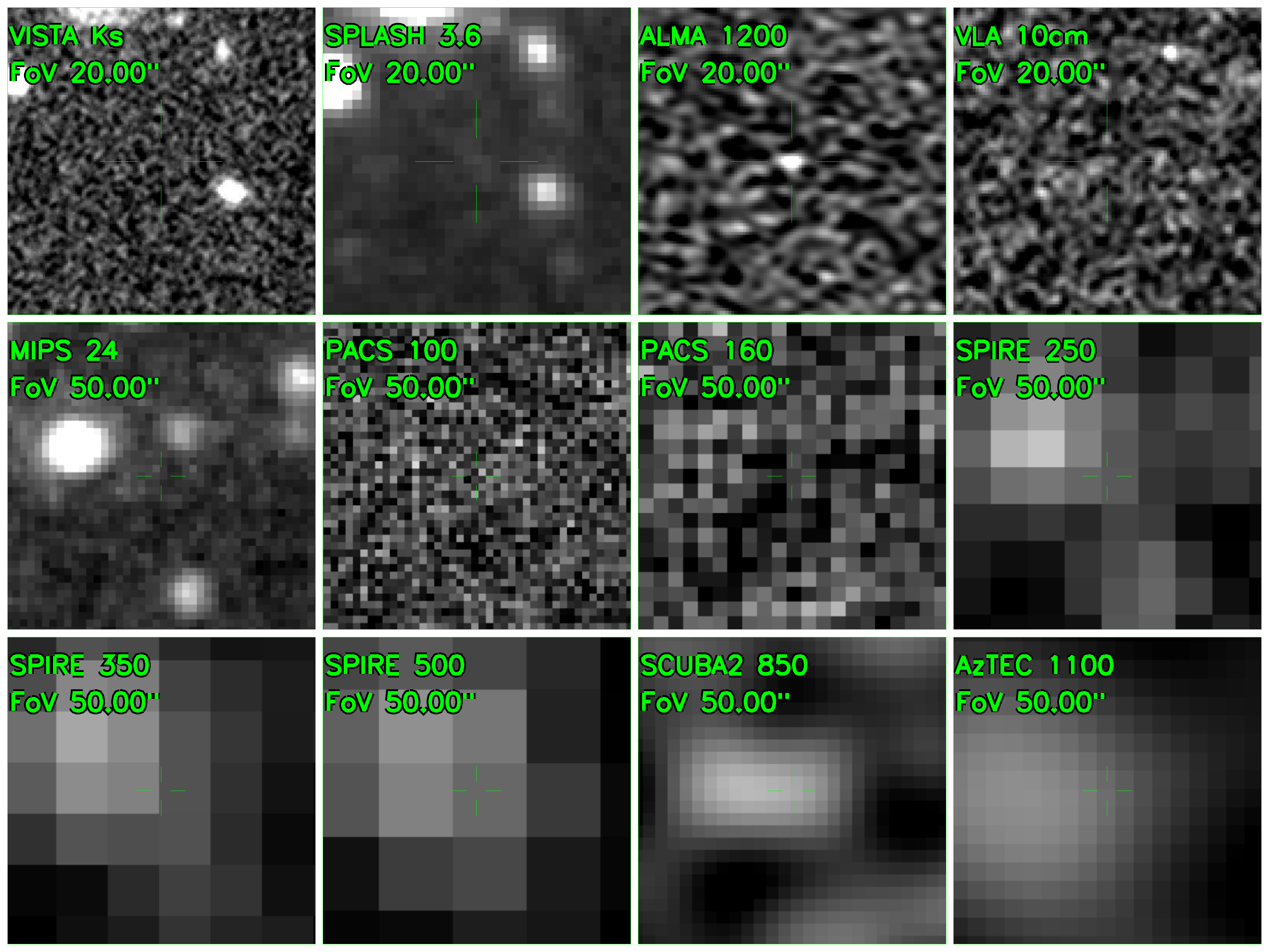}
	\includegraphics[width=0.21\textwidth, trim={0.6cm 5cm 1cm 3.5cm}, clip]{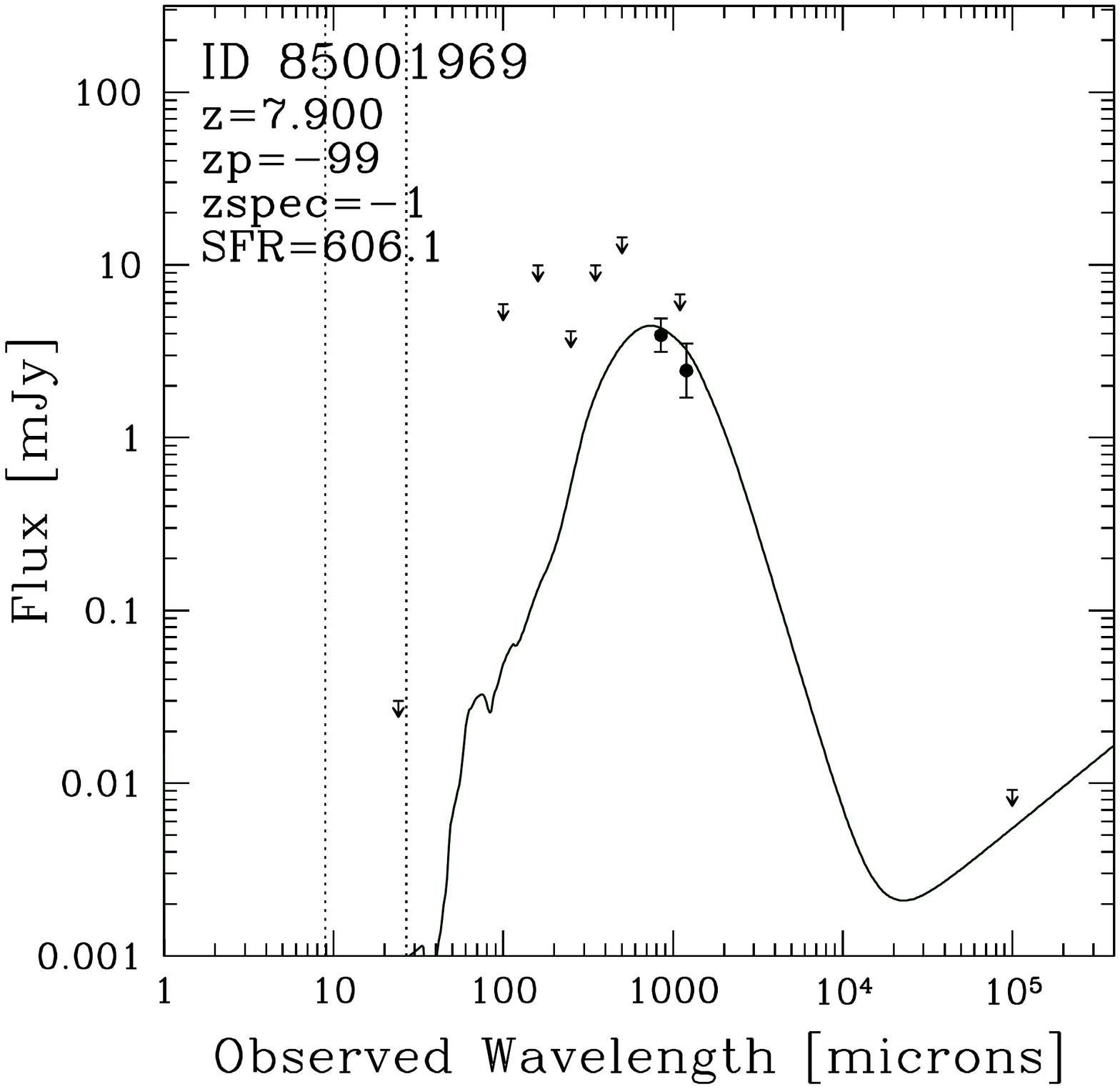}
    \includegraphics[width=0.28\textwidth]{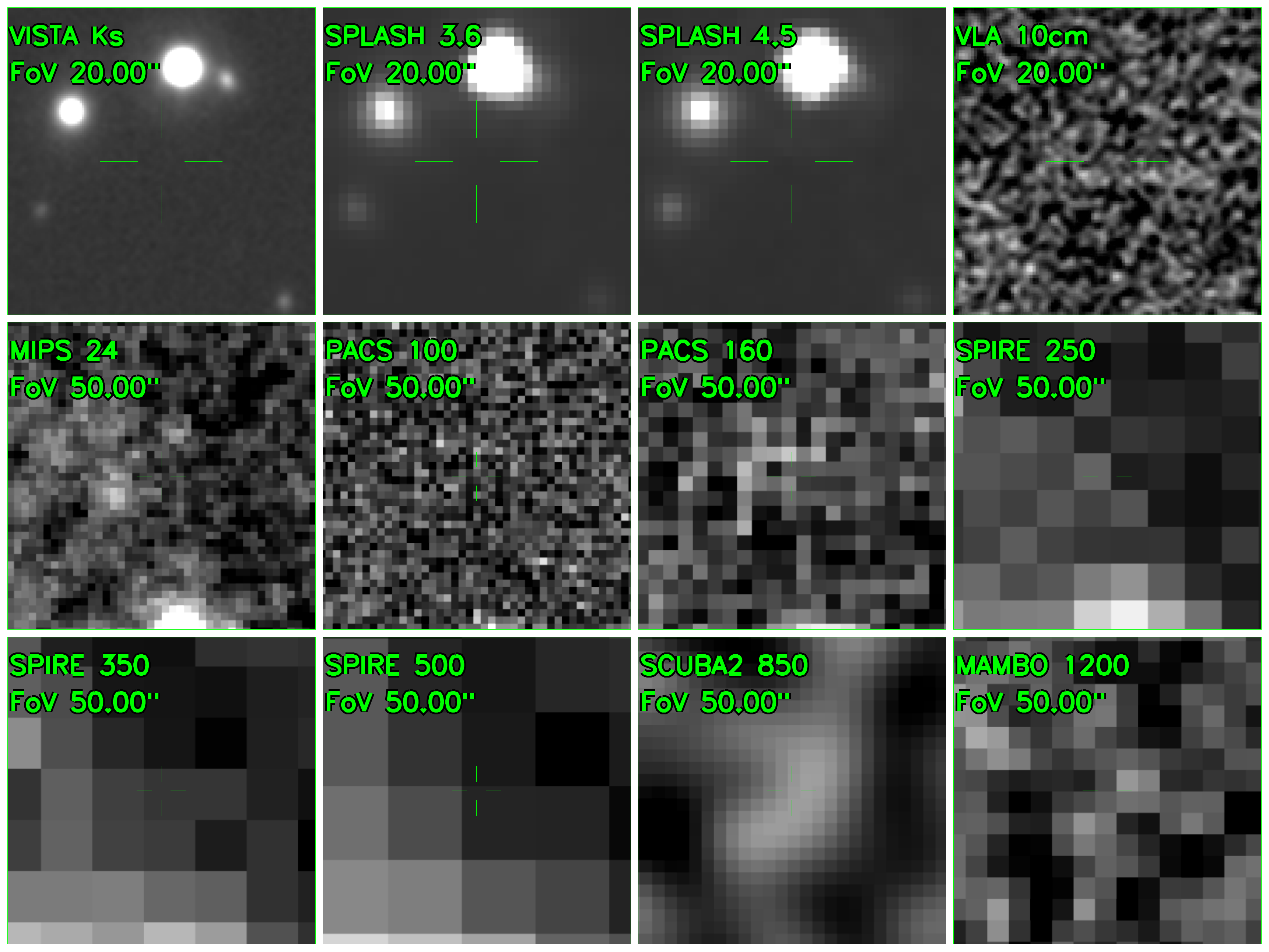}
	\includegraphics[width=0.21\textwidth, trim={0.6cm 5cm 1cm 3.5cm}, clip]{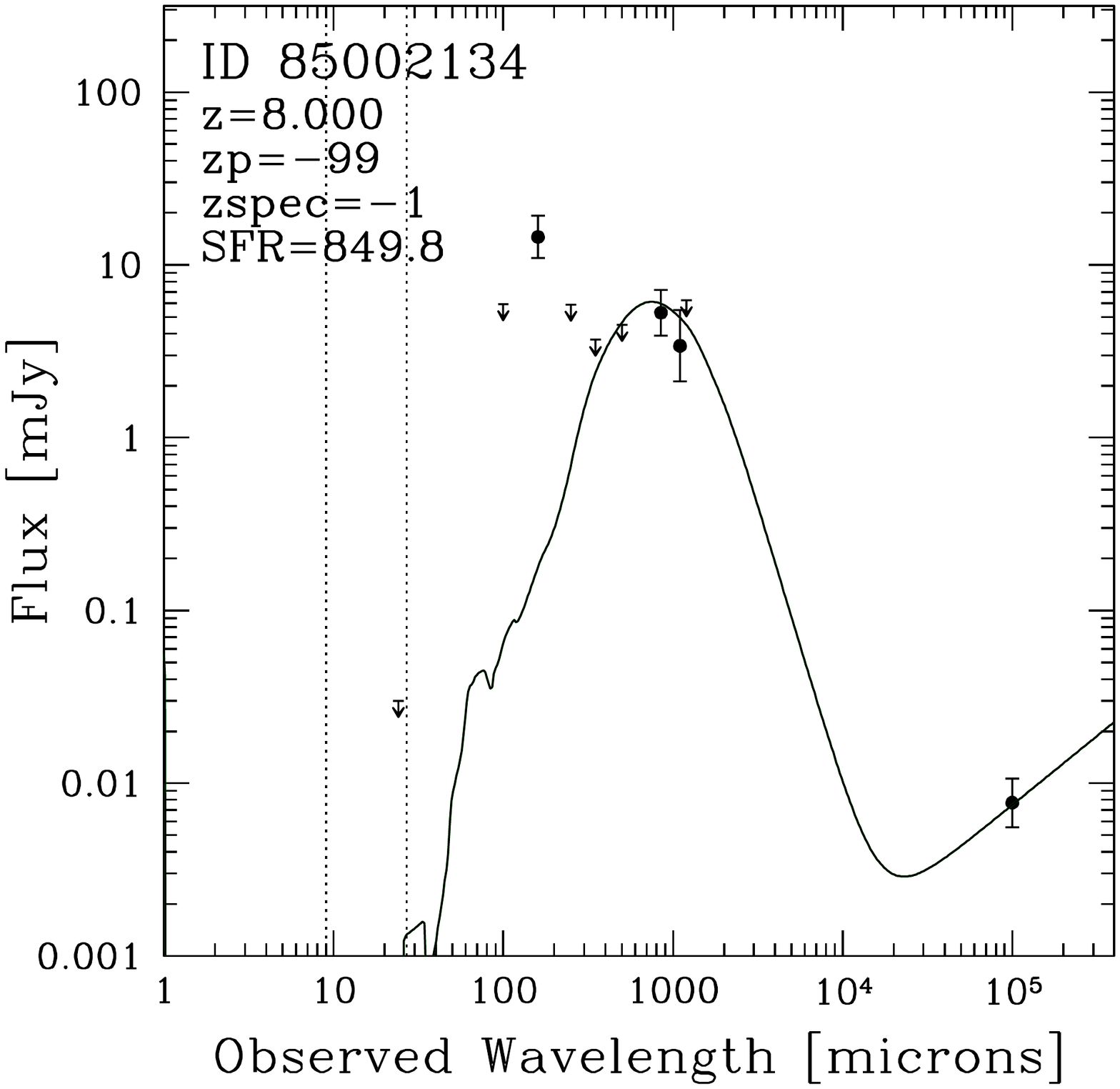}
    \includegraphics[width=0.28\textwidth]{85002215.pdf}
	\includegraphics[width=0.21\textwidth, trim={0.6cm 5cm 1cm 3.5cm}, clip]{Plot_SED_85002215.pdf}
    \includegraphics[width=0.28\textwidth]{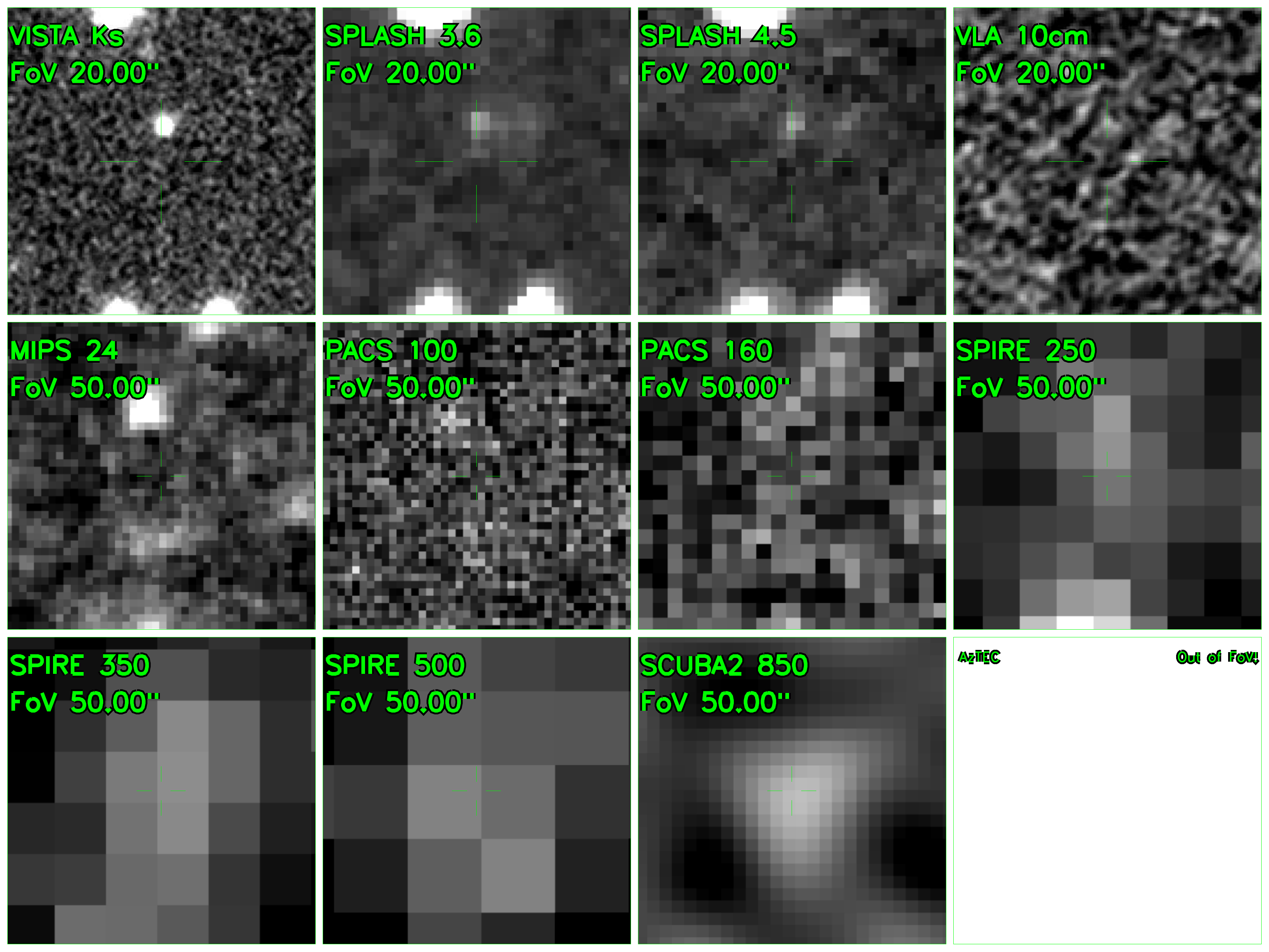}
	\includegraphics[width=0.21\textwidth, trim={0.6cm 5cm 1cm 3.5cm}, clip]{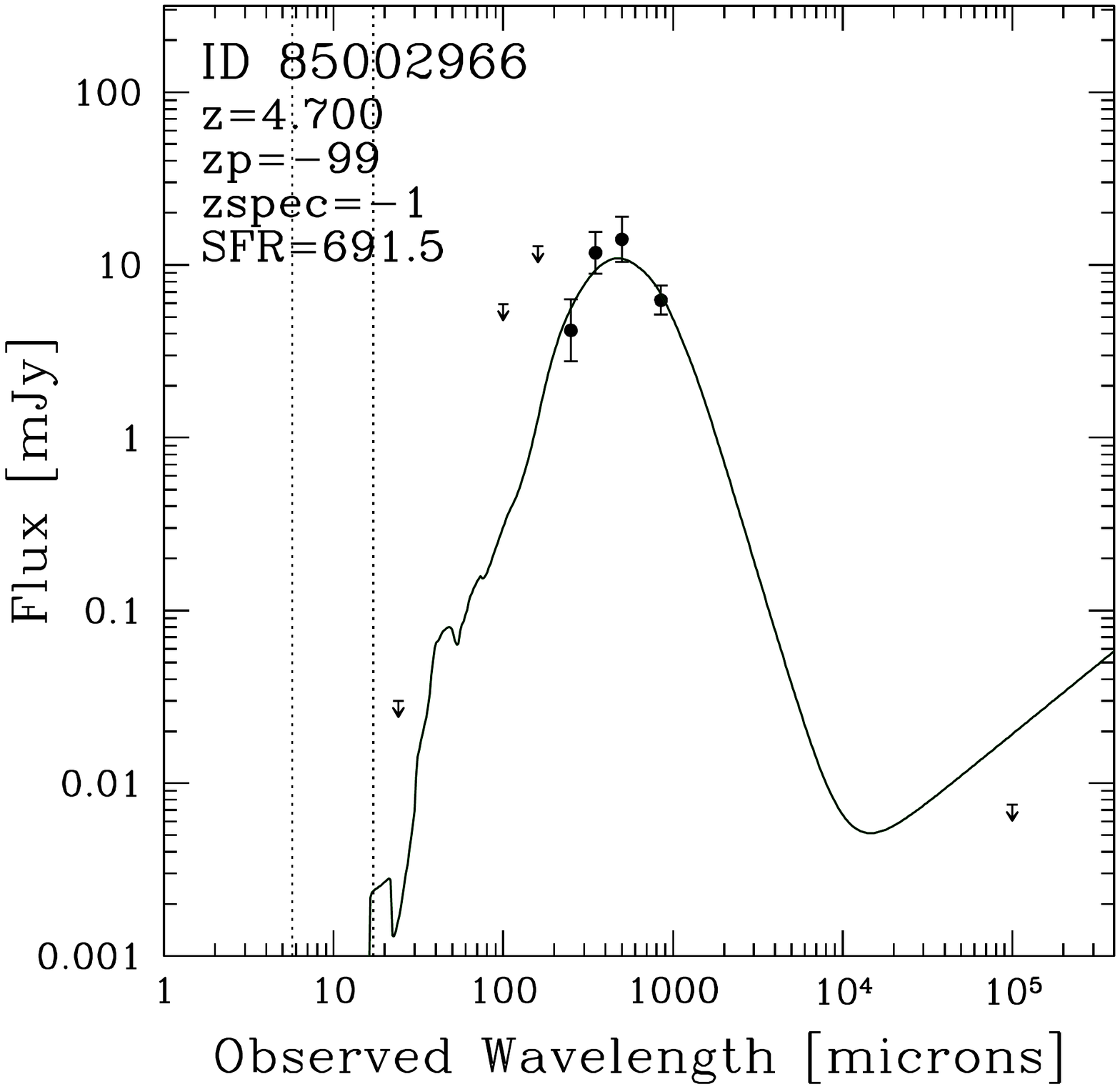}
    \includegraphics[width=0.28\textwidth]{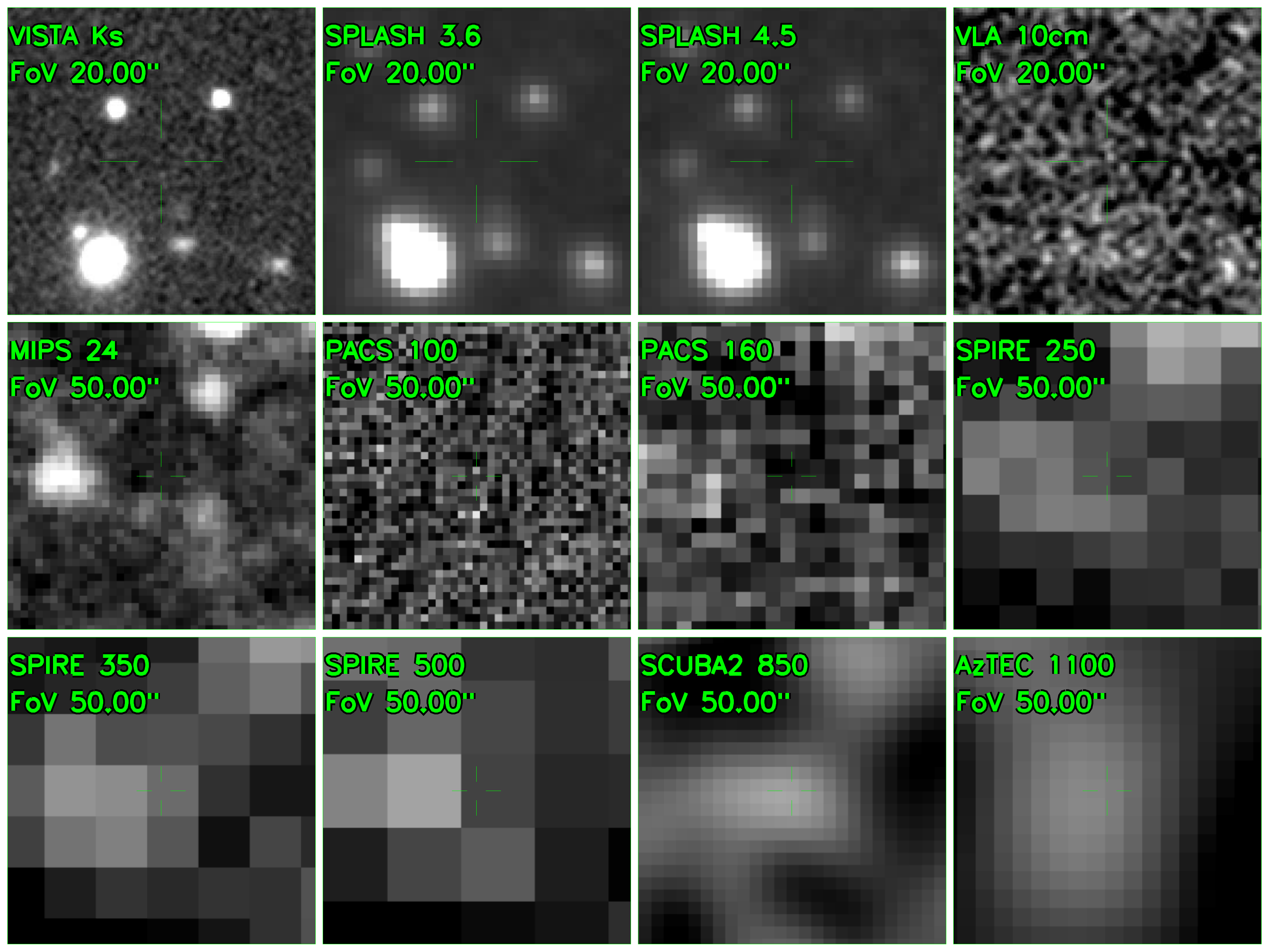}
	\includegraphics[width=0.21\textwidth, trim={0.6cm 5cm 1cm 3.5cm}, clip]{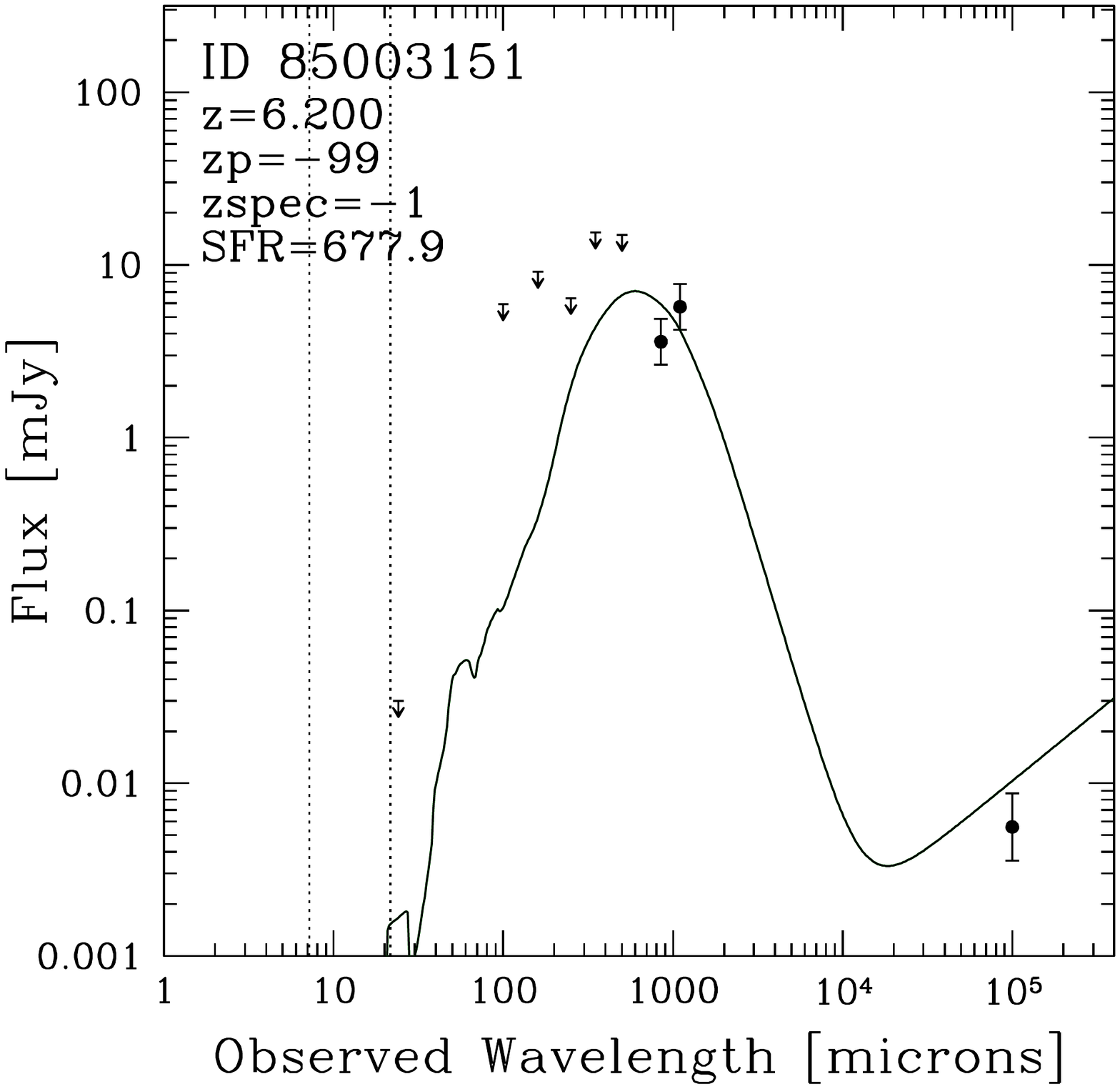}
    \includegraphics[width=0.28\textwidth]{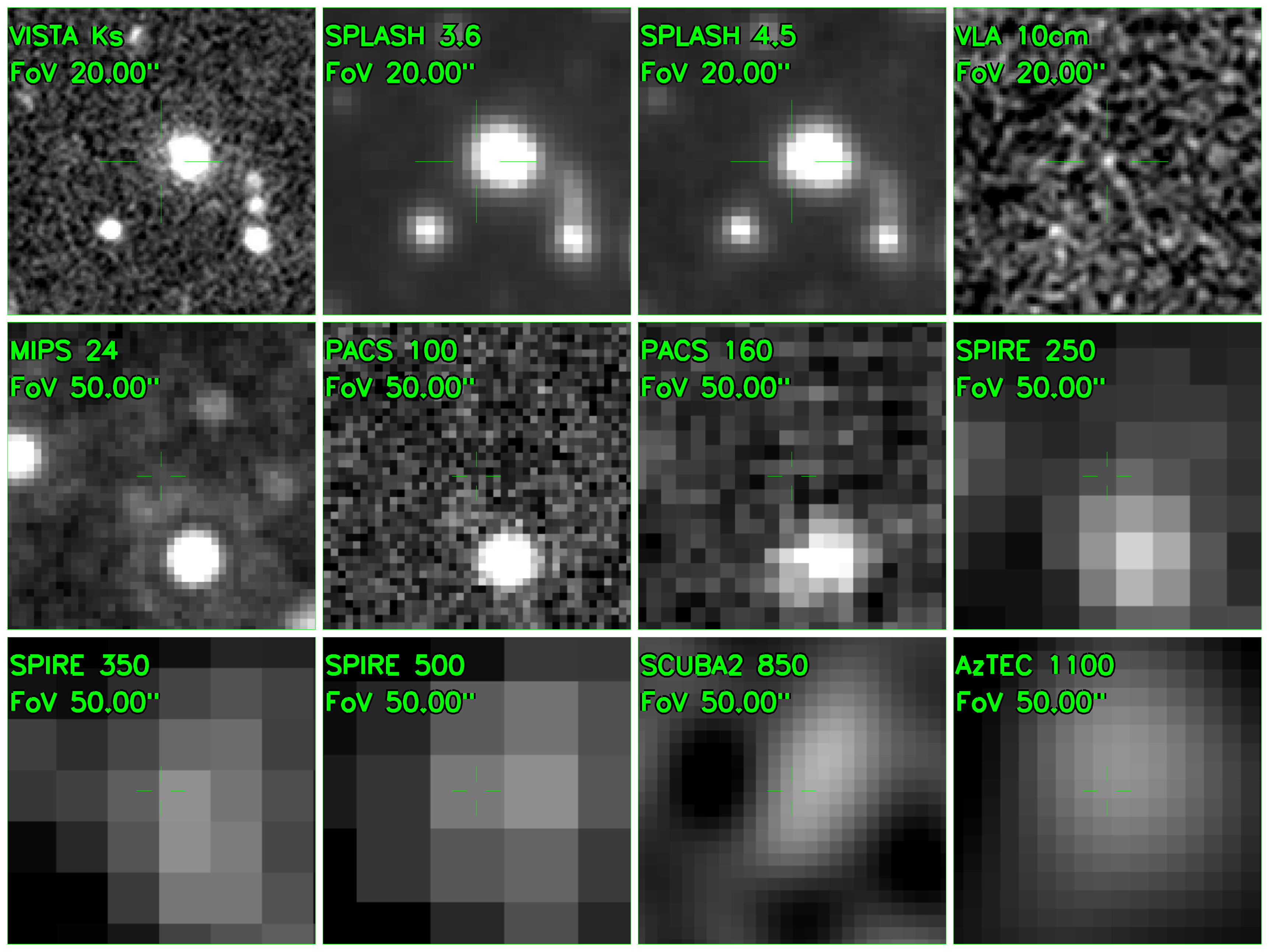}
	\includegraphics[width=0.21\textwidth, trim={0.6cm 5cm 1cm 3.5cm}, clip]{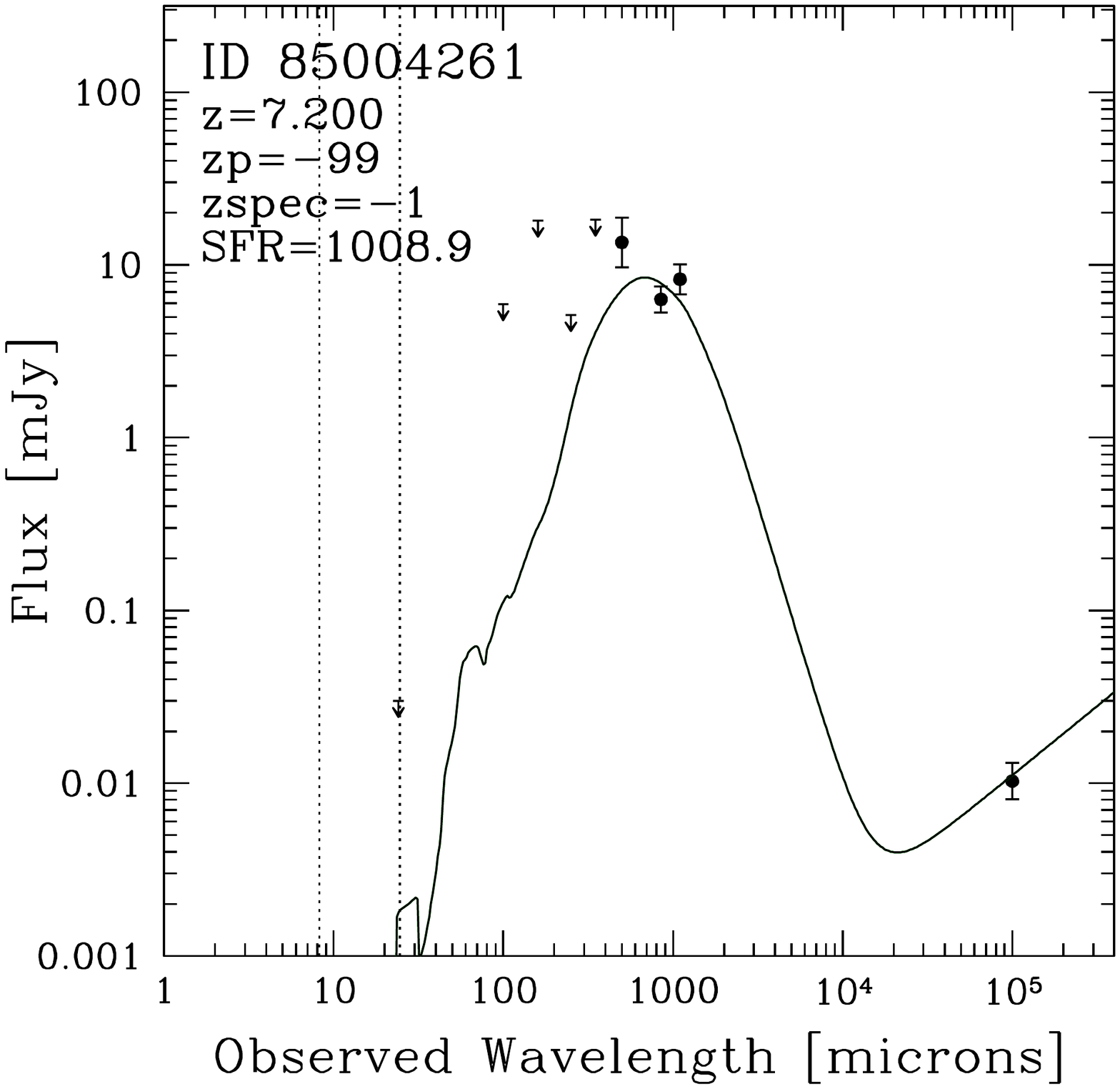}
    \includegraphics[width=0.28\textwidth]{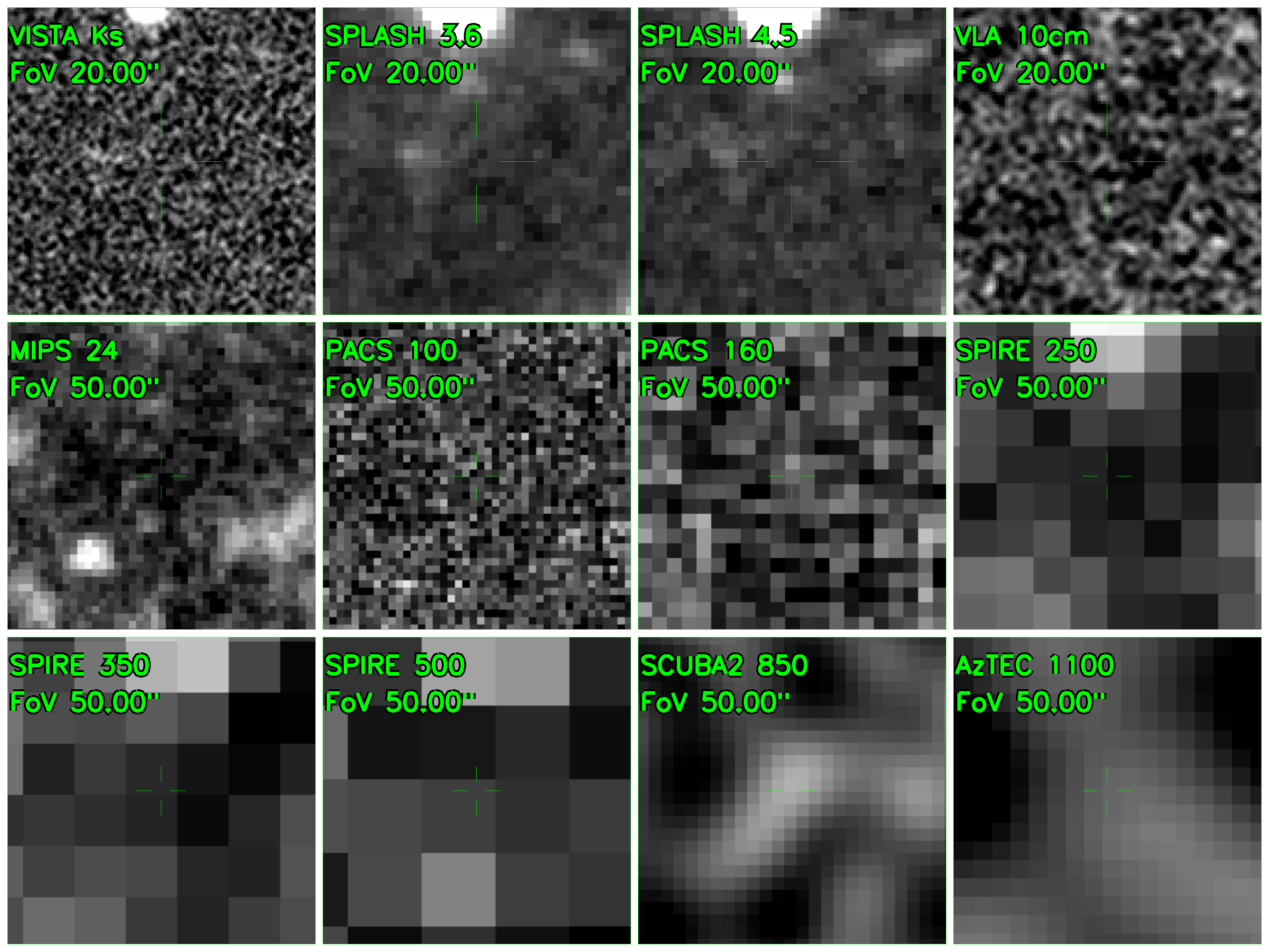}
	\includegraphics[width=0.21\textwidth, trim={0.6cm 5cm 1cm 3.5cm}, clip]{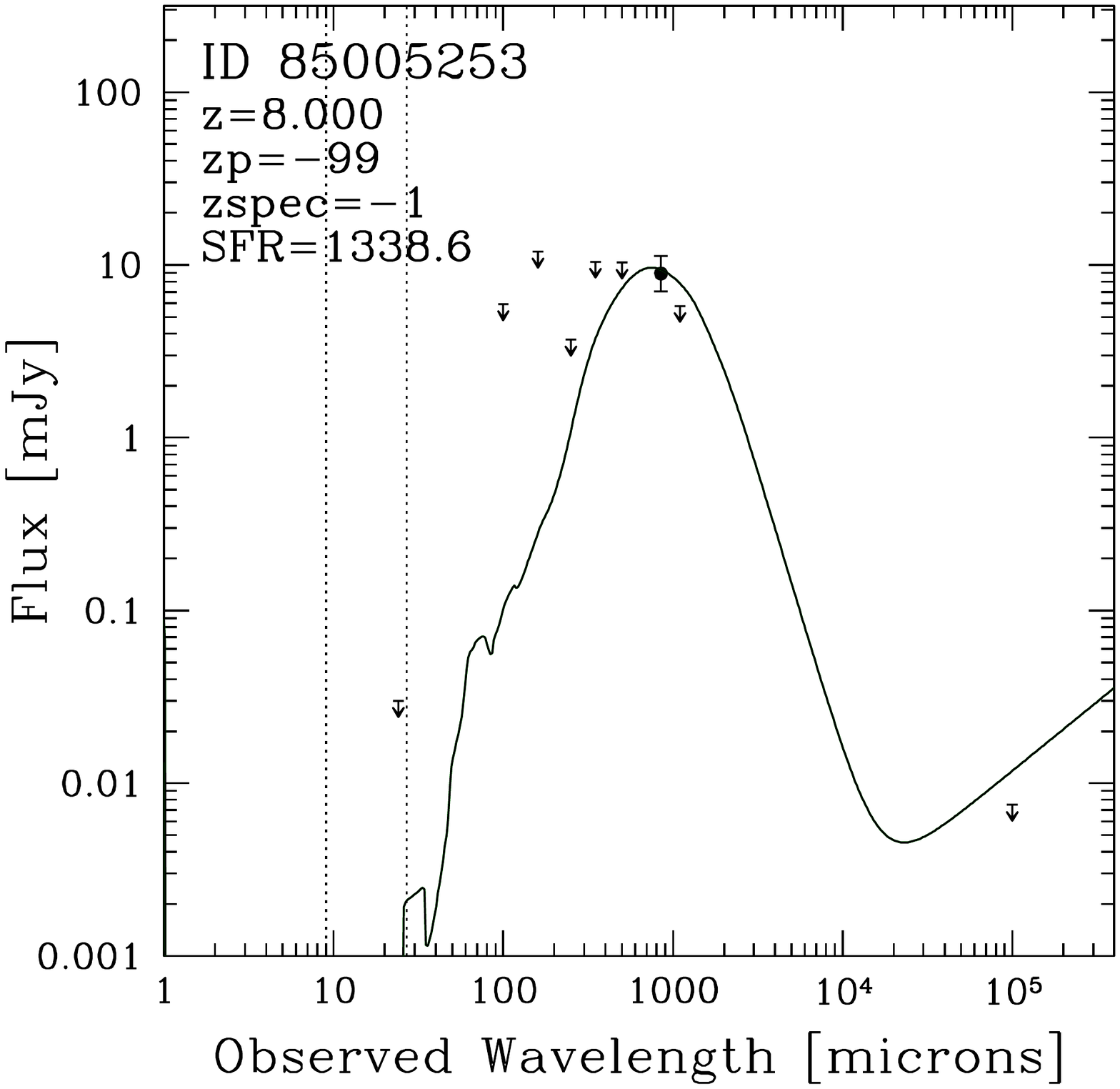}
    \caption{%
		Multi-band cutouts and SEDs of high redshift candidates, continued.  
		\label{highz_cutouts2}
		}
\end{figure}

\begin{figure}
	\centering
    \includegraphics[width=0.28\textwidth]{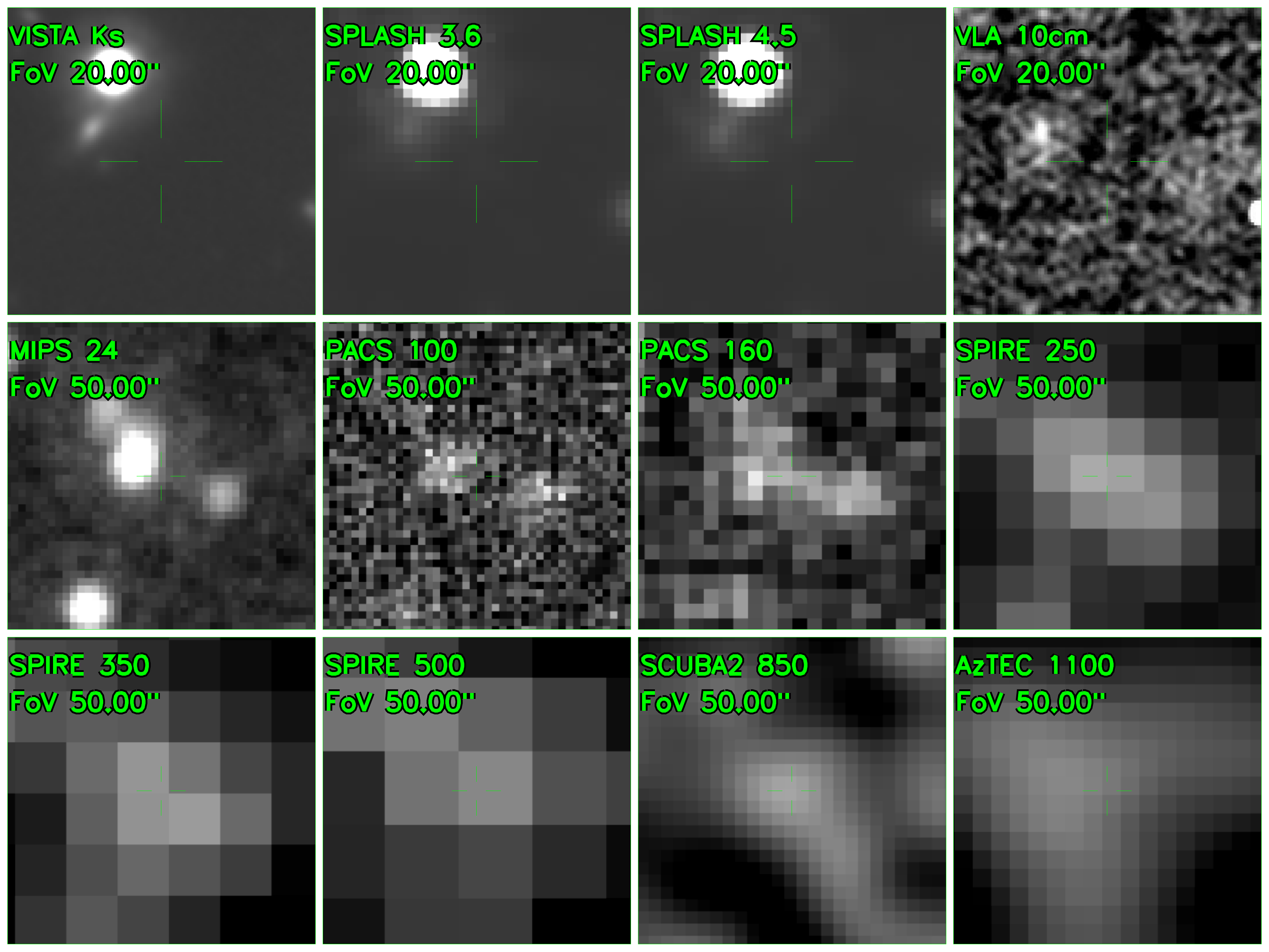}
	\includegraphics[width=0.21\textwidth, trim={0.6cm 5cm 1cm 3.5cm}, clip]{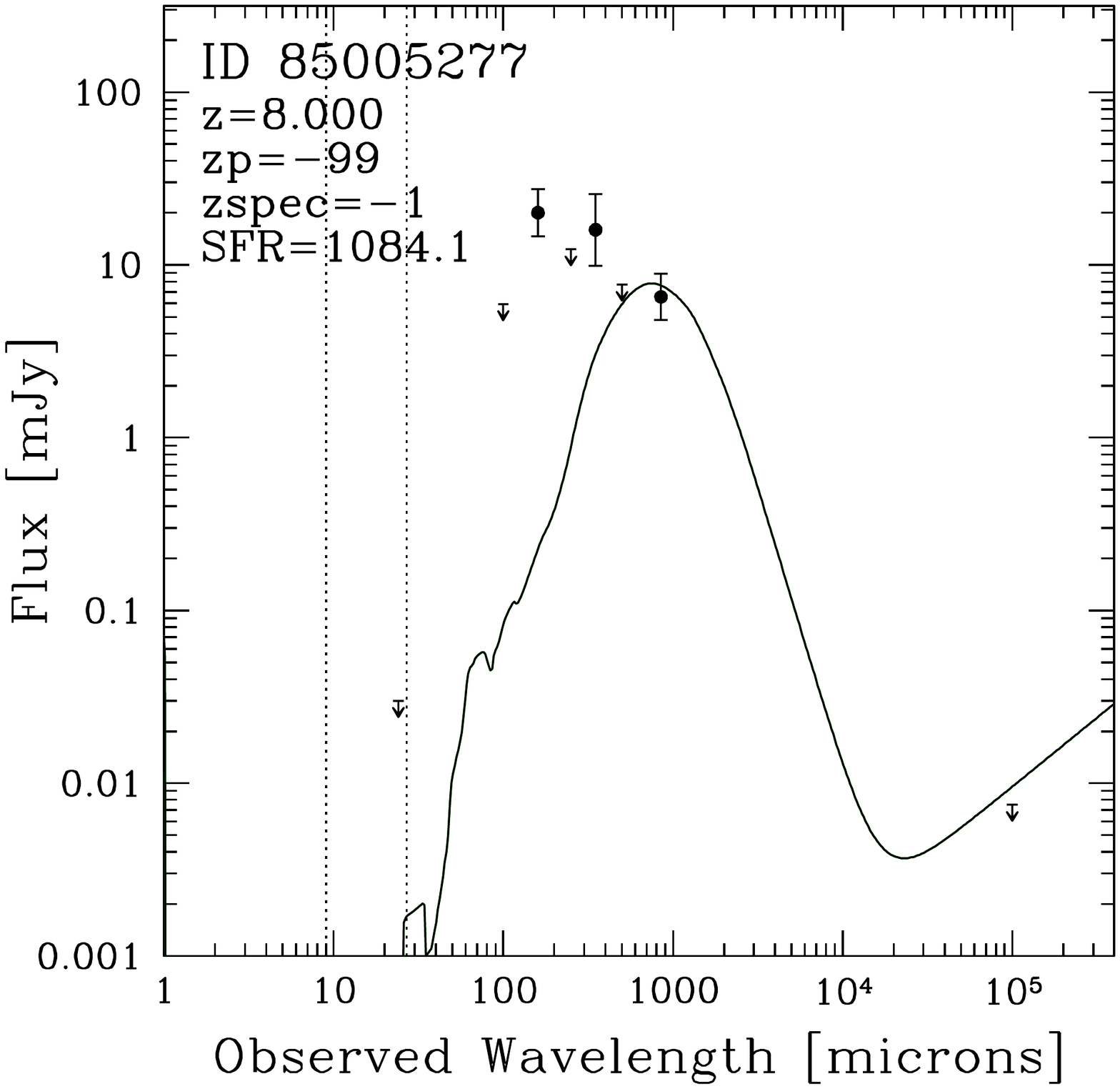}
    \includegraphics[width=0.28\textwidth]{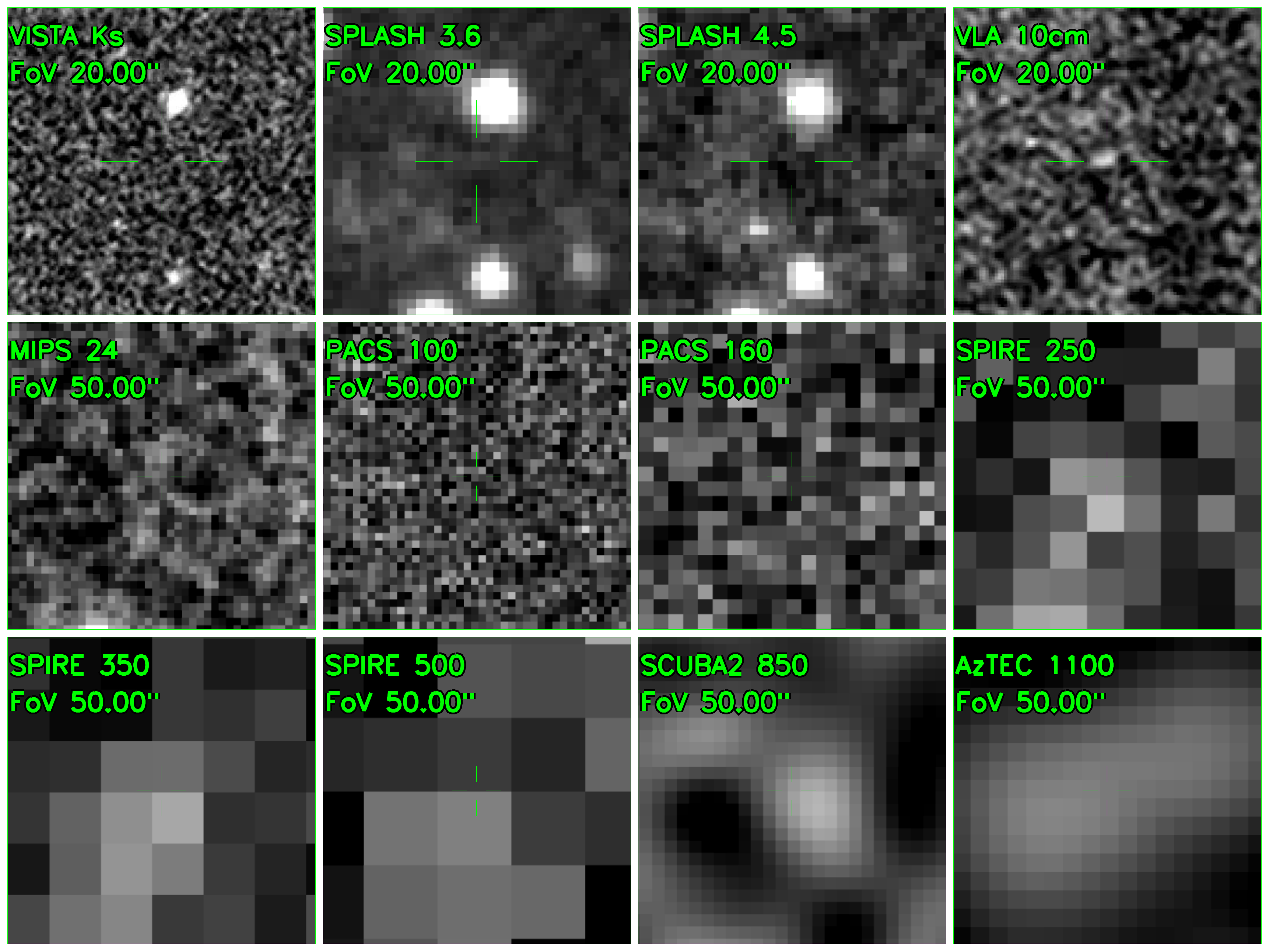}
	\includegraphics[width=0.21\textwidth, trim={0.6cm 5cm 1cm 3.5cm}, clip]{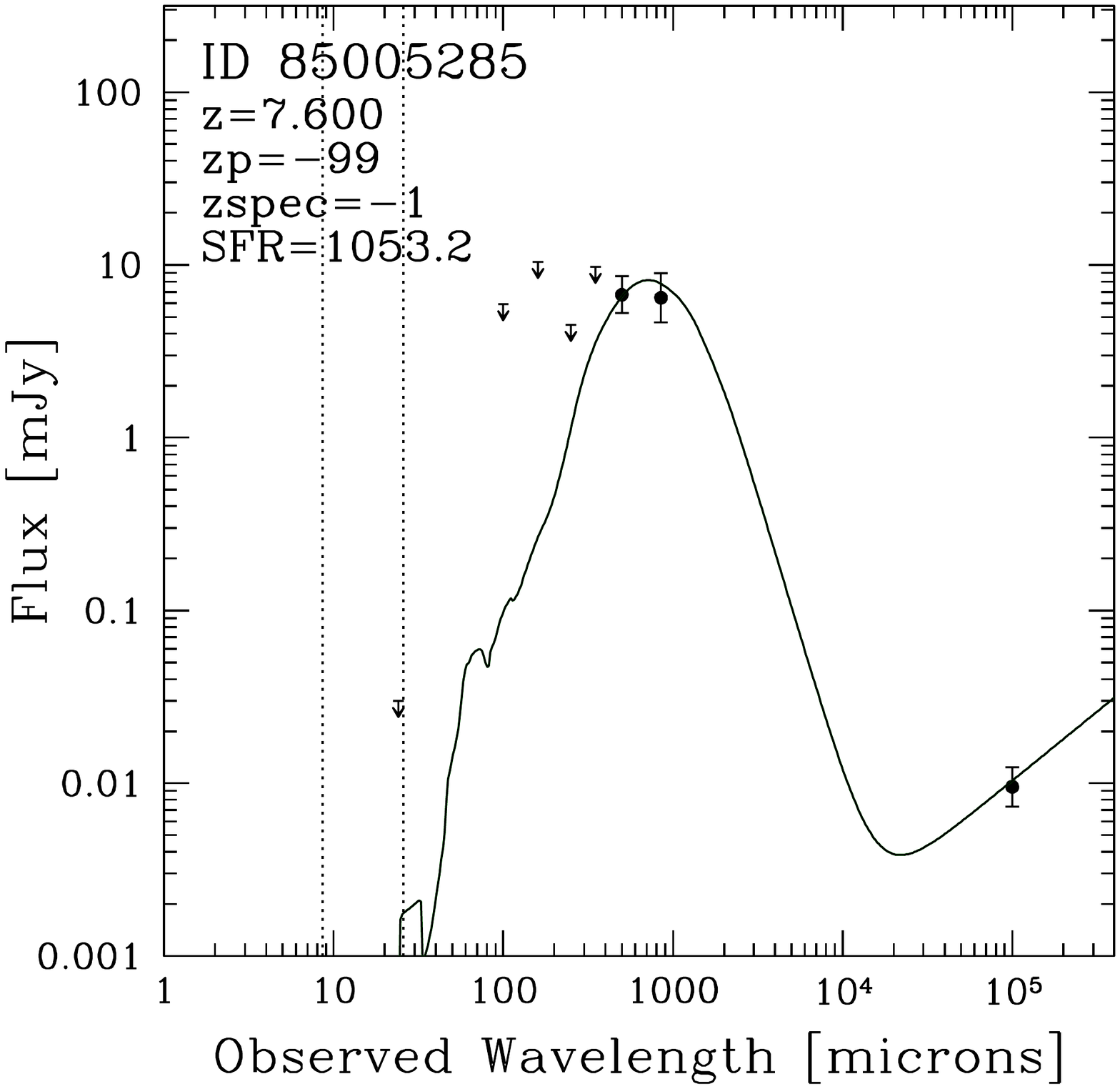}
    \includegraphics[width=0.28\textwidth]{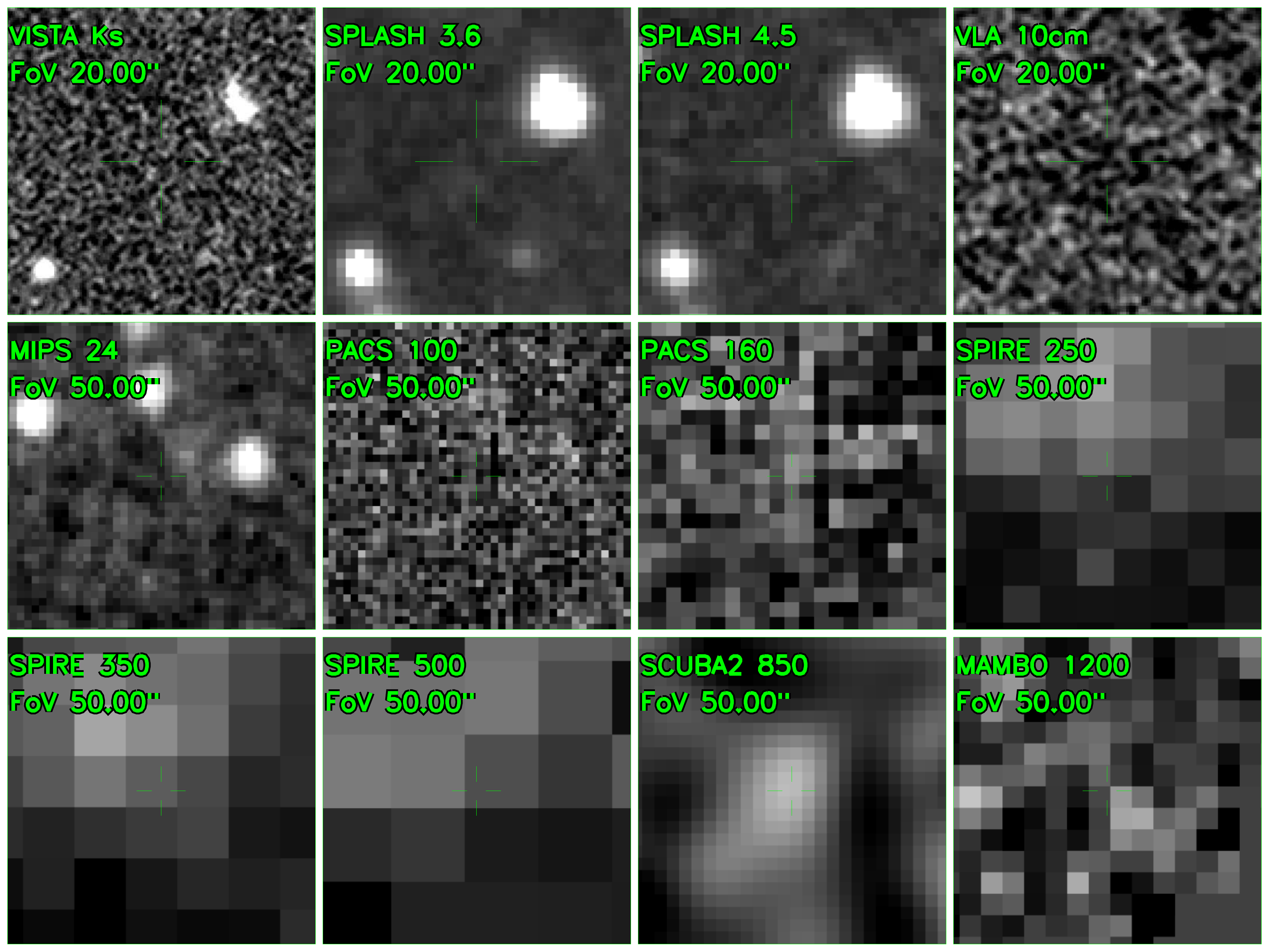}
	\includegraphics[width=0.21\textwidth, trim={0.6cm 5cm 1cm 3.5cm}, clip]{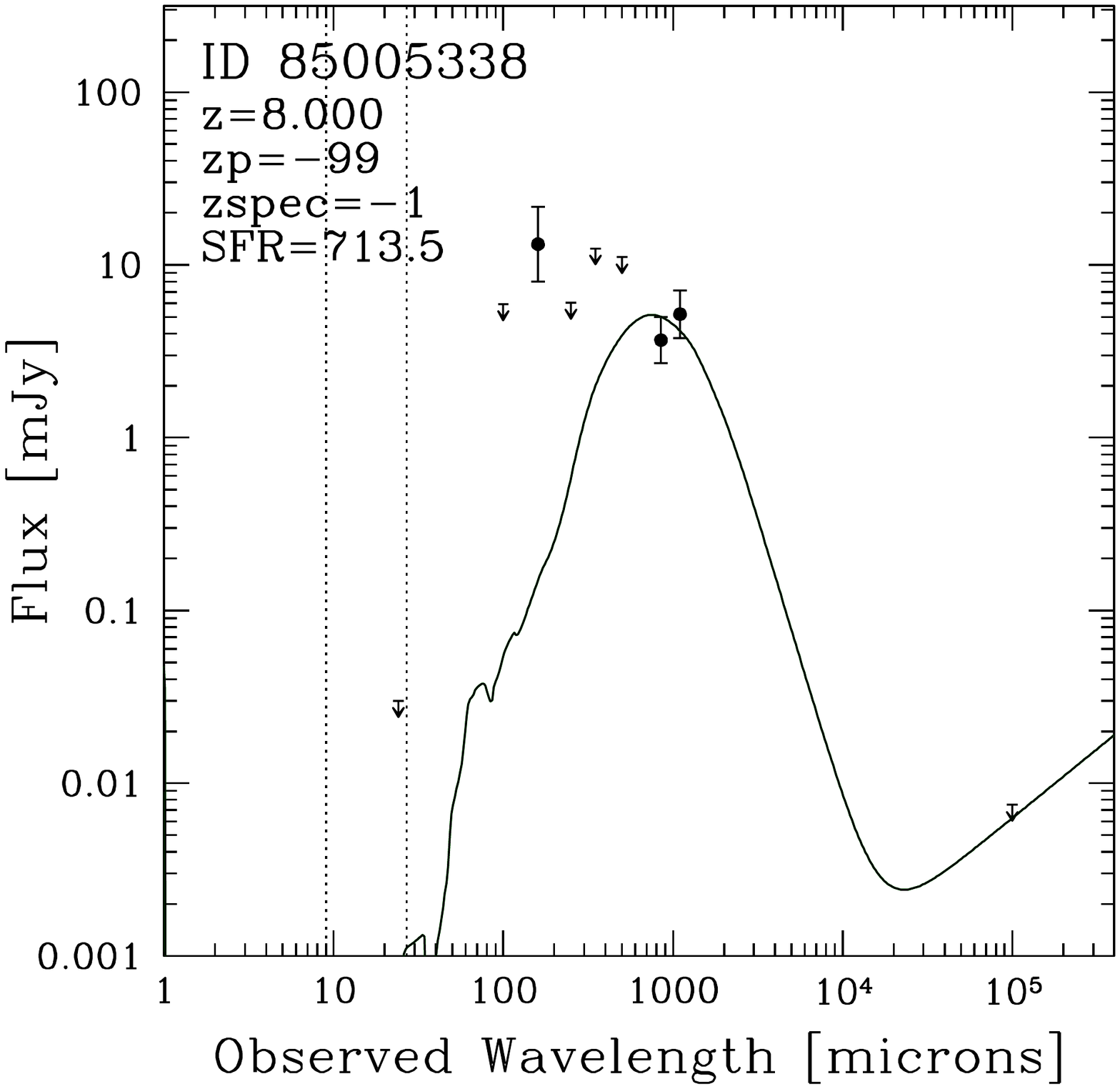}
    \includegraphics[width=0.28\textwidth]{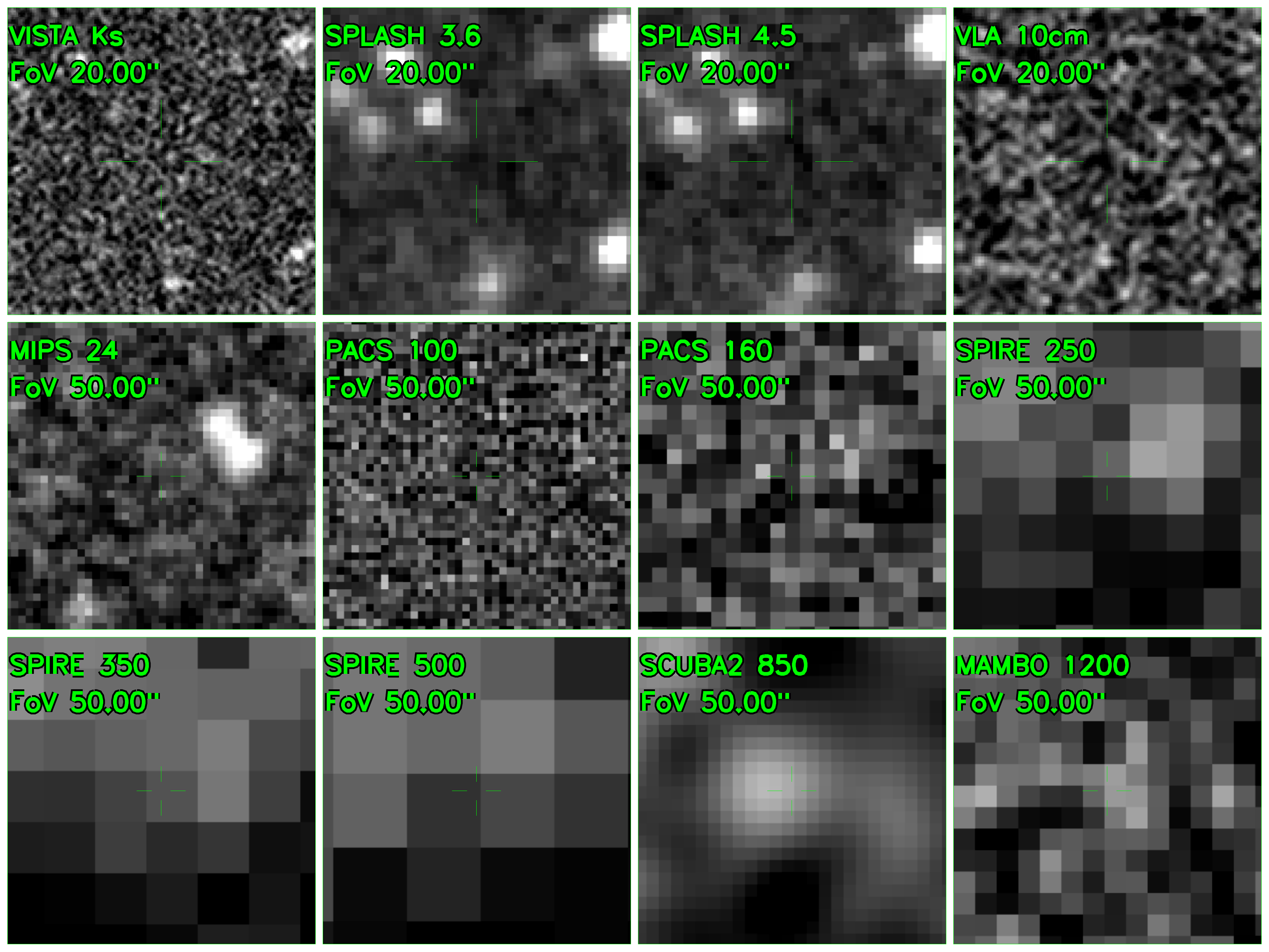}
	\includegraphics[width=0.21\textwidth, trim={0.6cm 5cm 1cm 3.5cm}, clip]{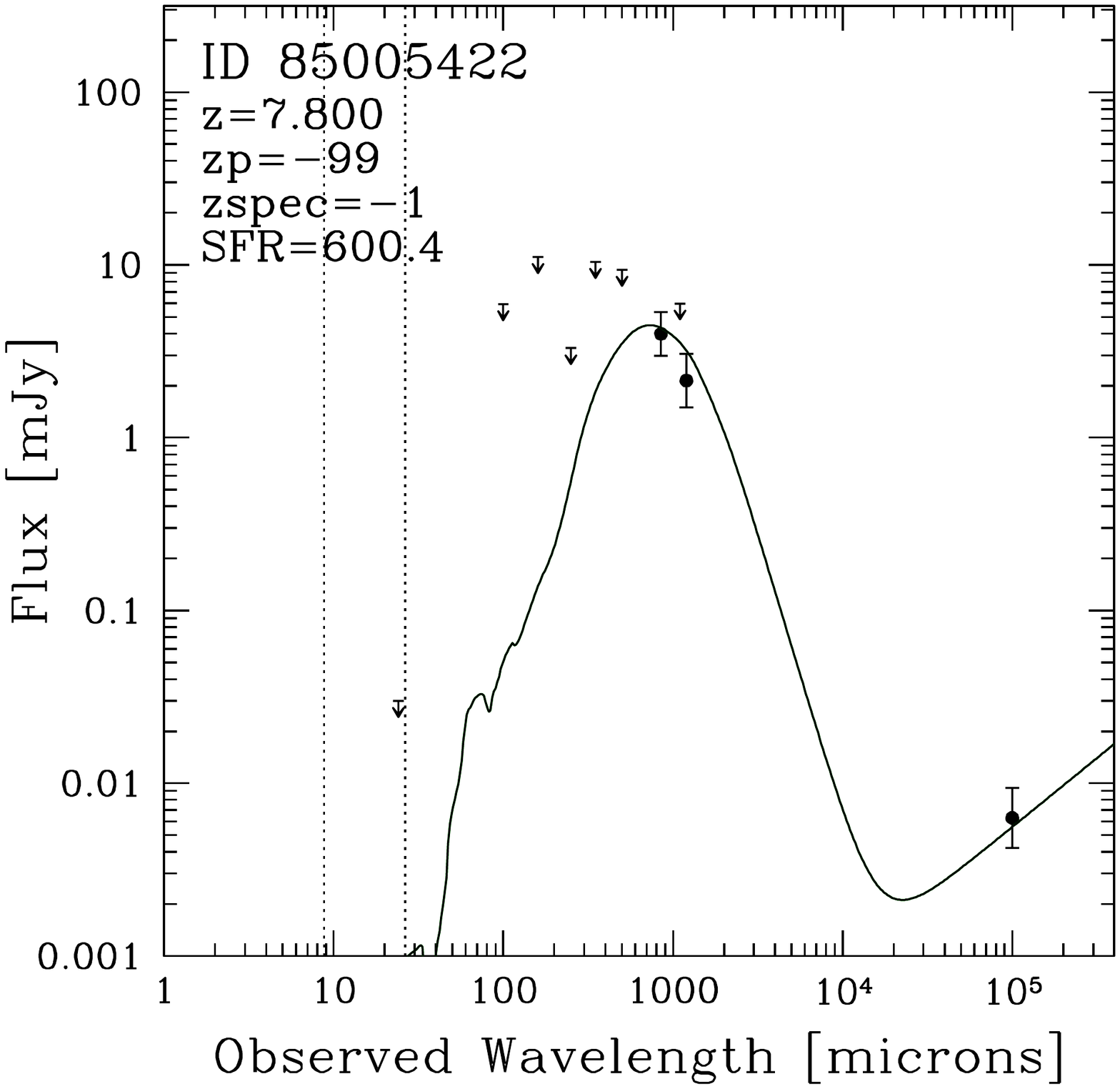}
    \includegraphics[width=0.28\textwidth]{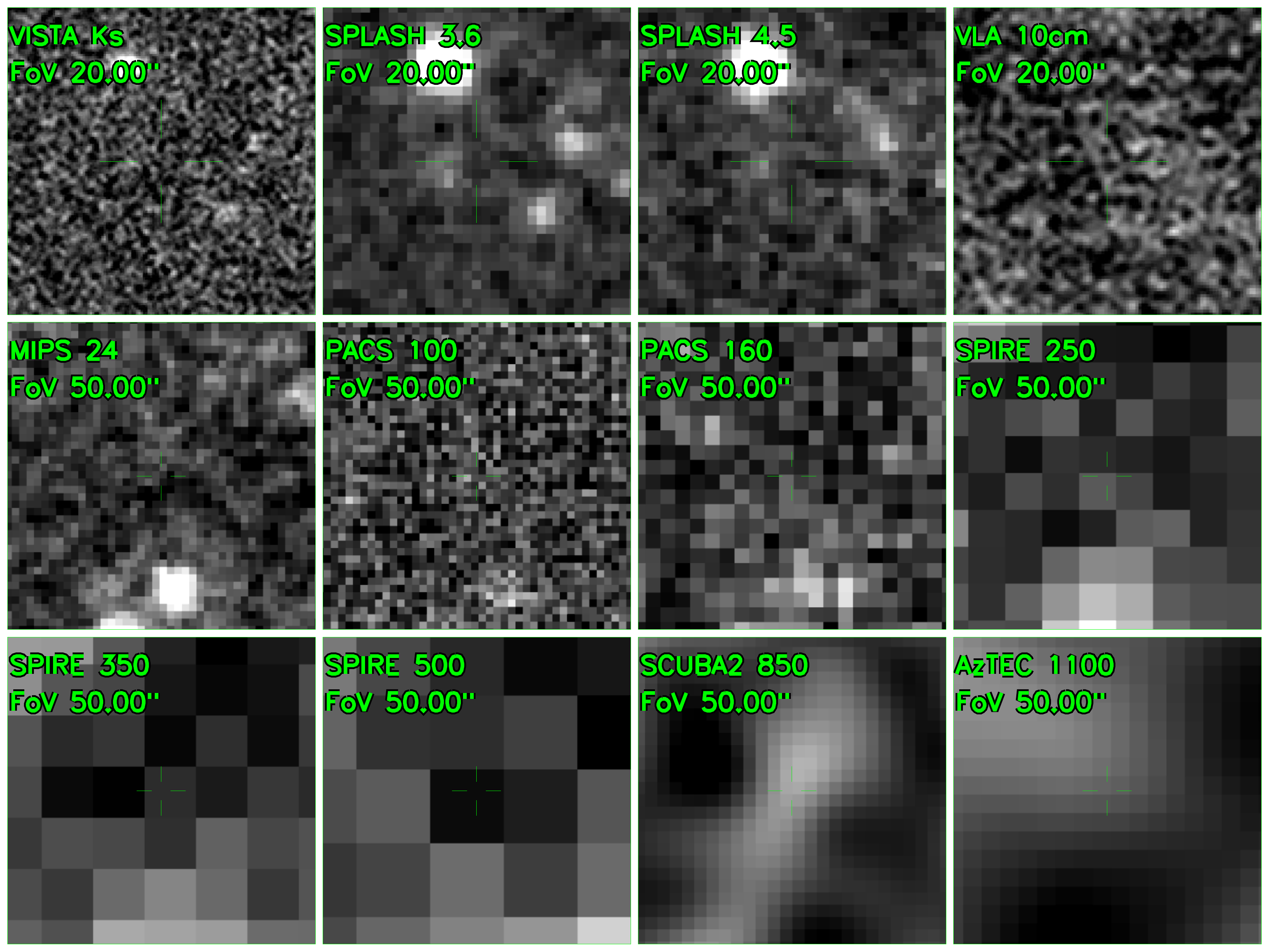}
	\includegraphics[width=0.21\textwidth, trim={0.6cm 5cm 1cm 3.5cm}, clip]{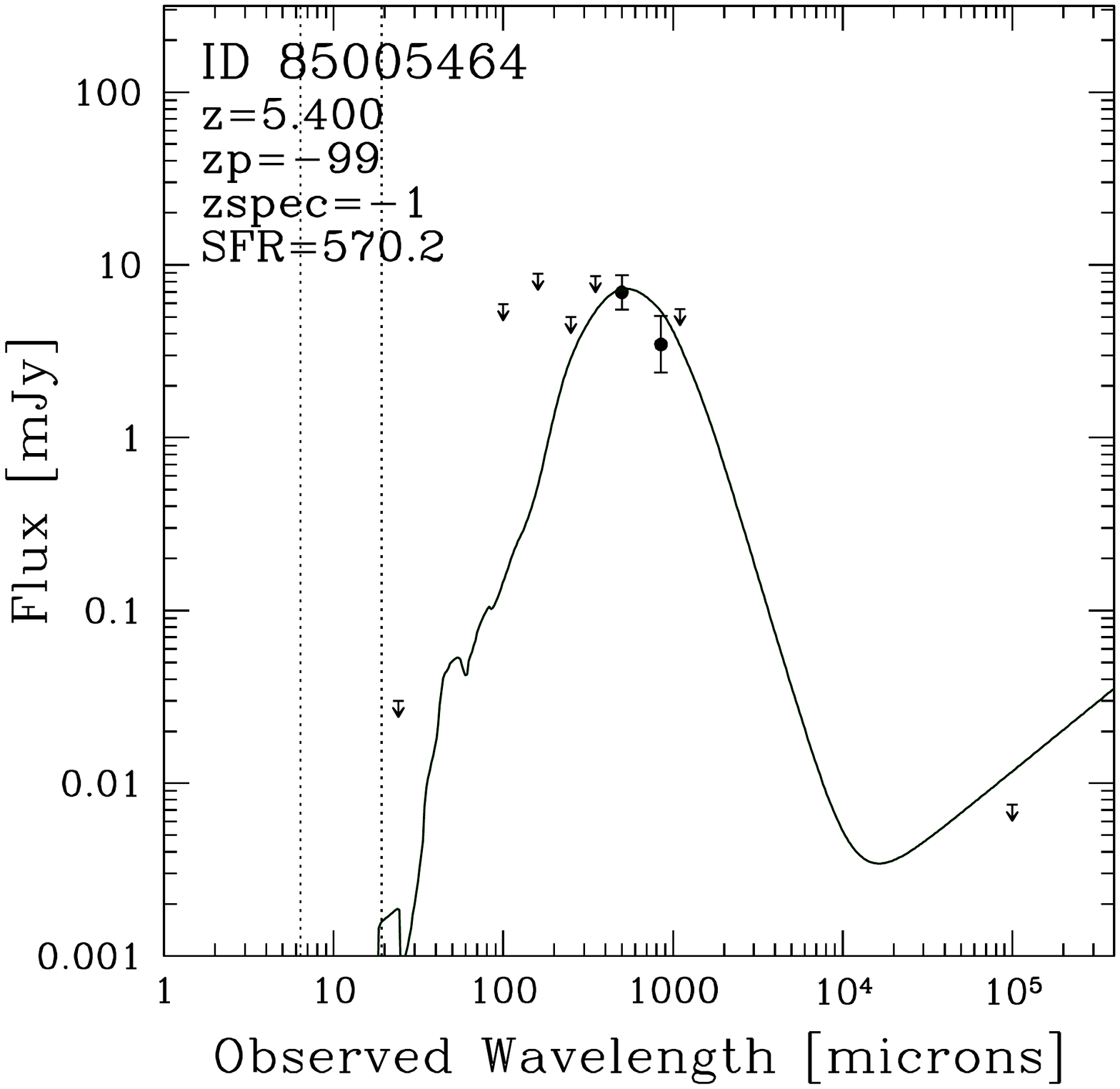}
	\includegraphics[width=0.28\textwidth]{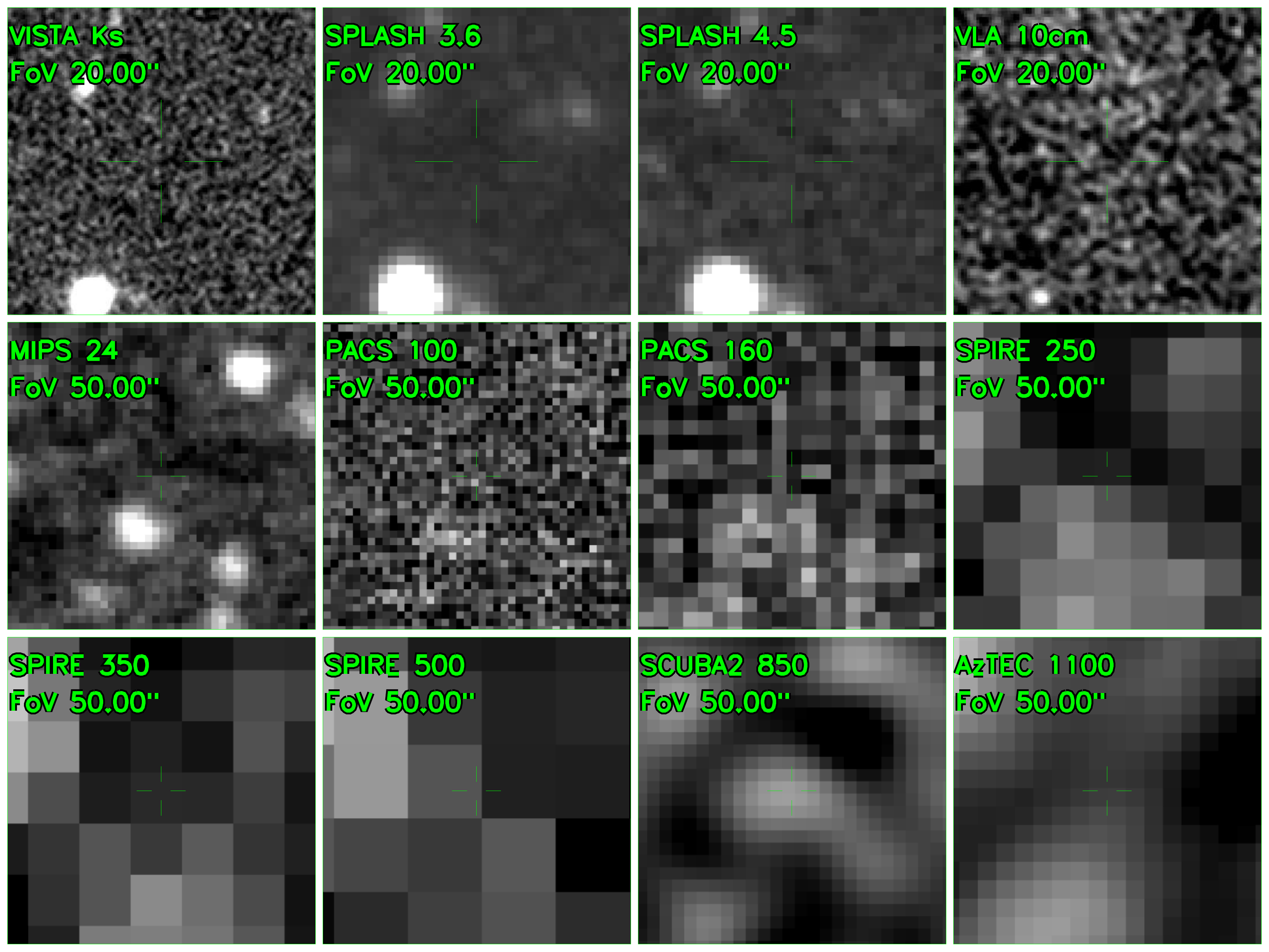}
	\includegraphics[width=0.21\textwidth, trim={0.6cm 5cm 1cm 3.5cm}, clip]{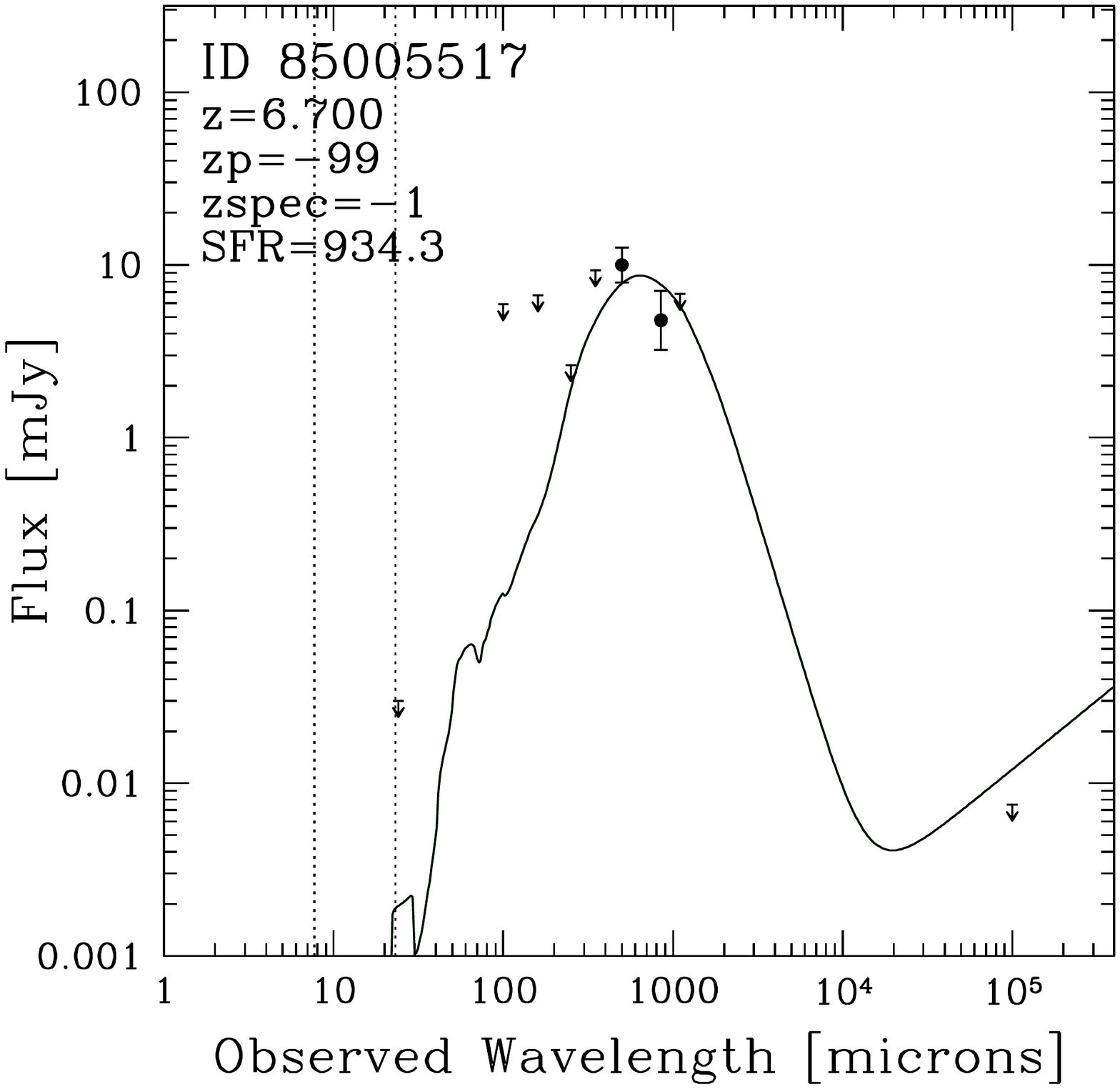}
    \includegraphics[width=0.28\textwidth]{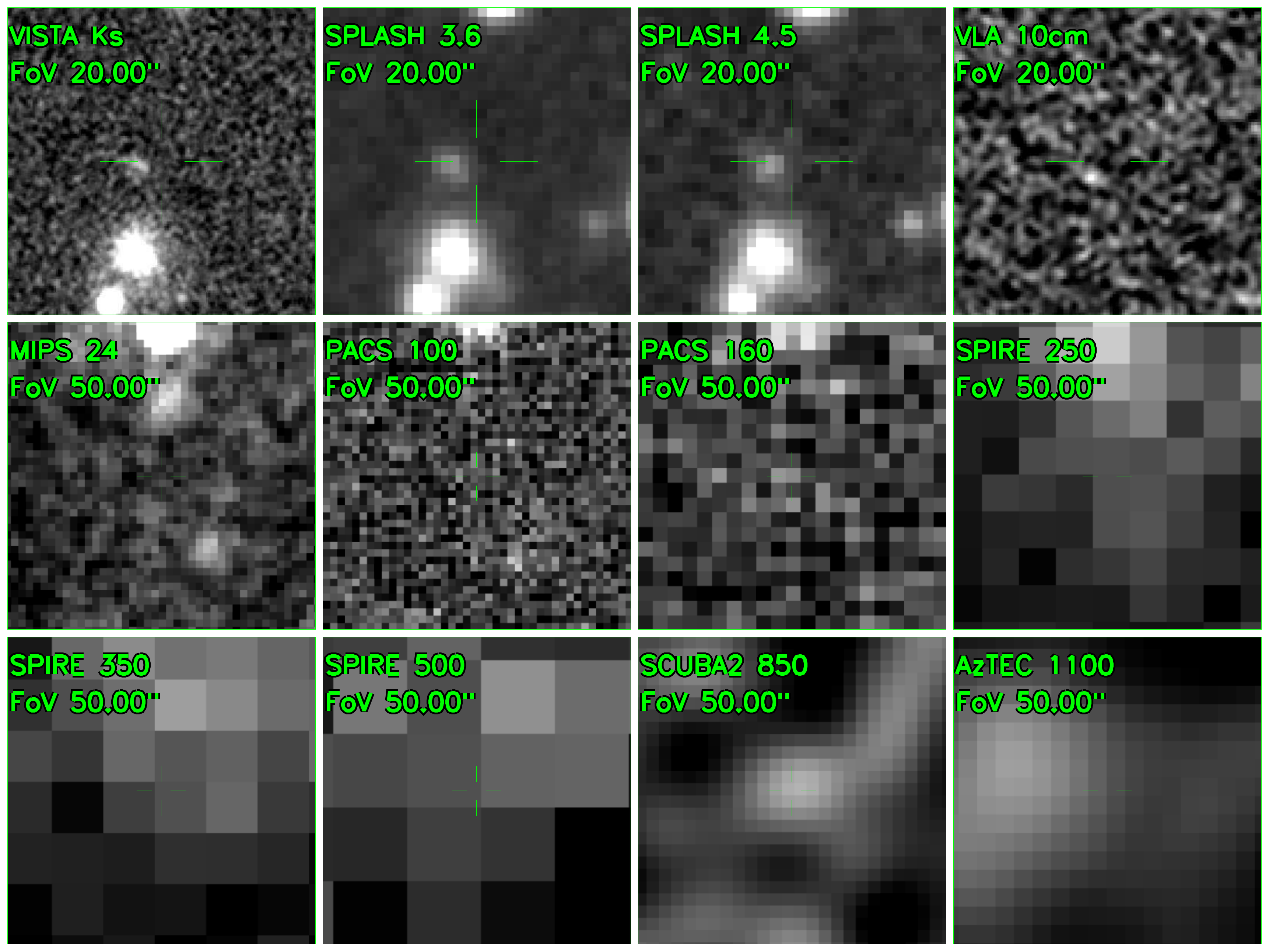}
	\includegraphics[width=0.21\textwidth, trim={0.6cm 5cm 1cm 3.5cm}, clip]{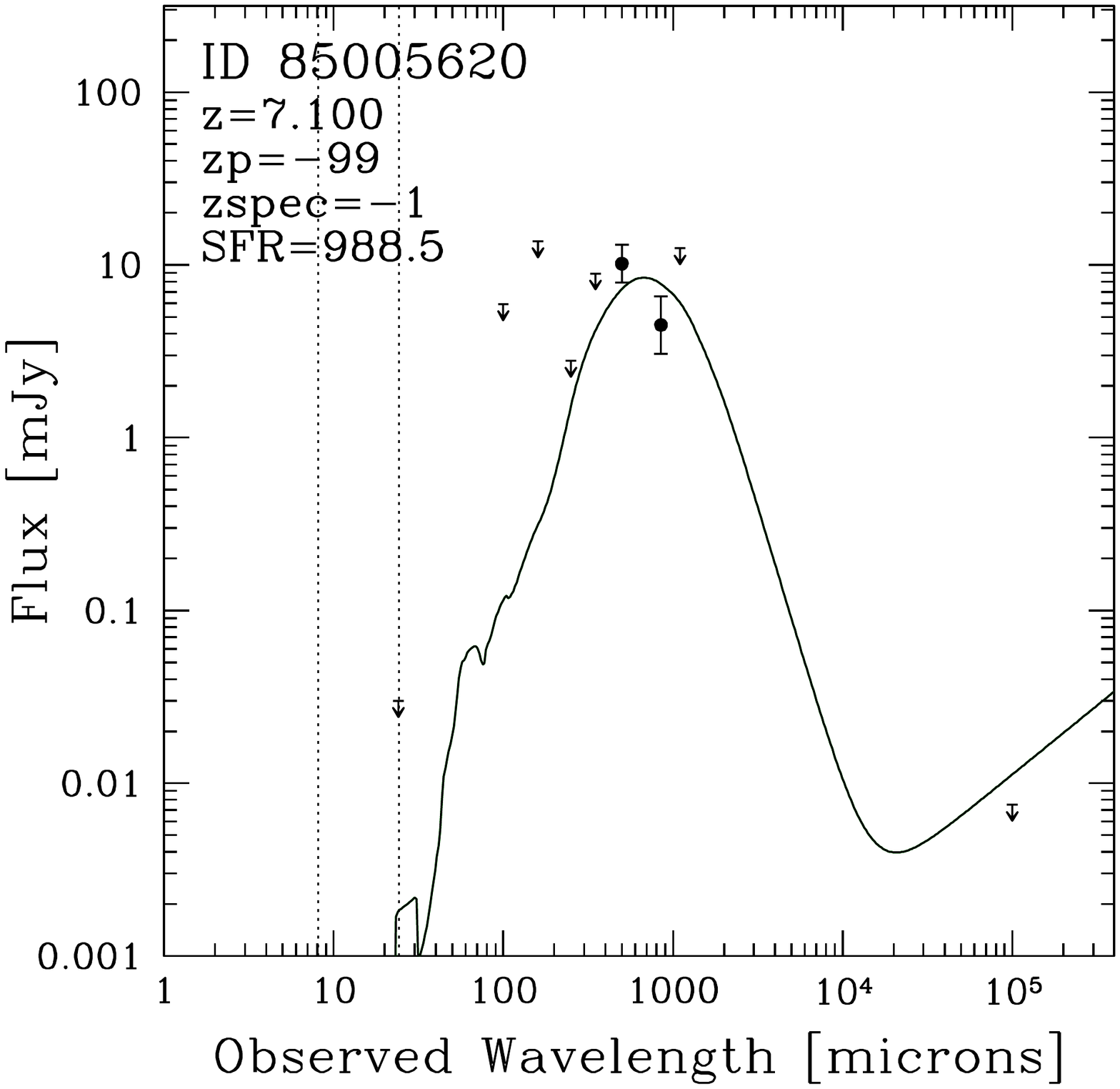}
    \includegraphics[width=0.28\textwidth]{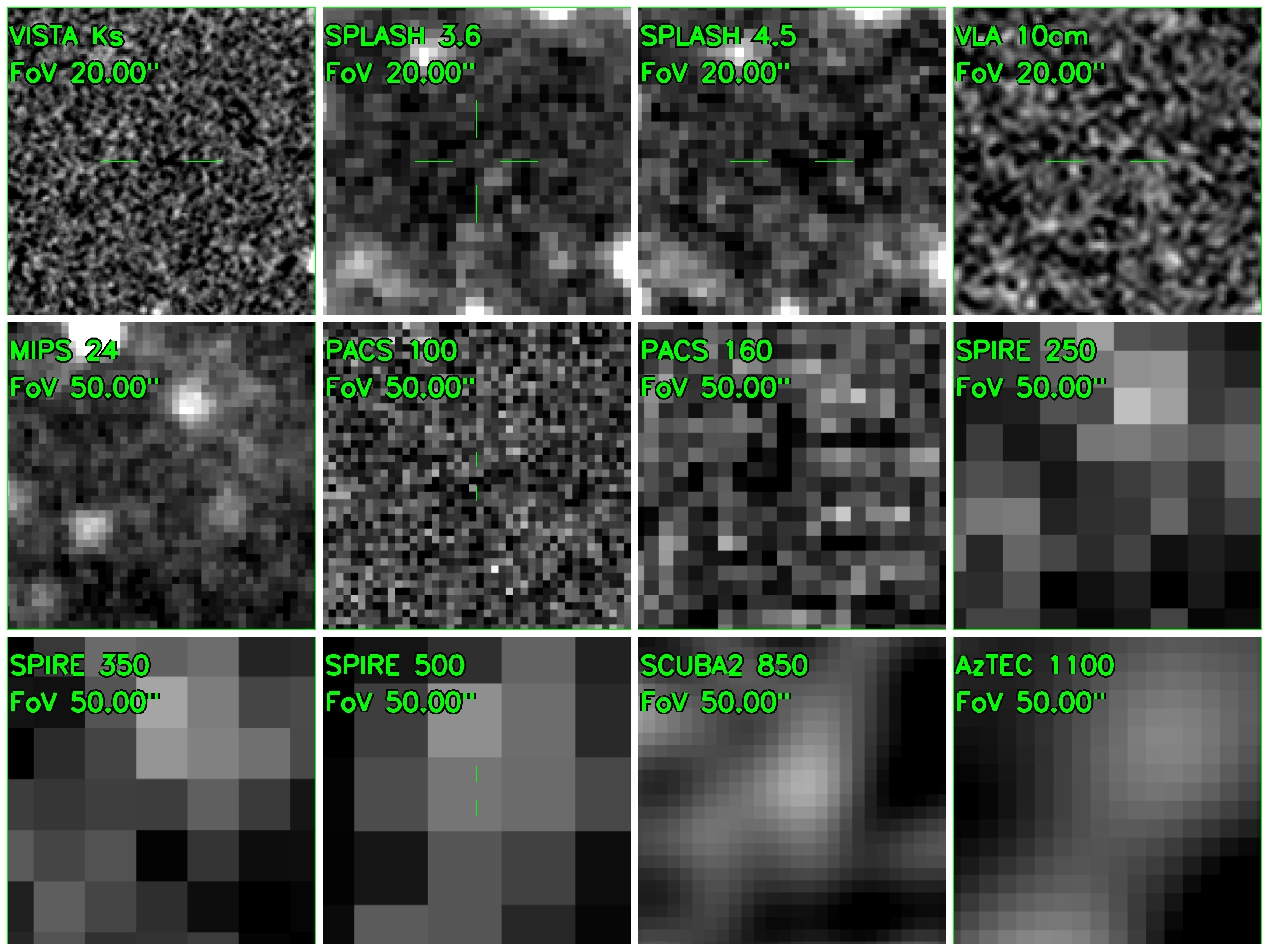}
	\includegraphics[width=0.21\textwidth, trim={0.6cm 5cm 1cm 3.5cm}, clip]{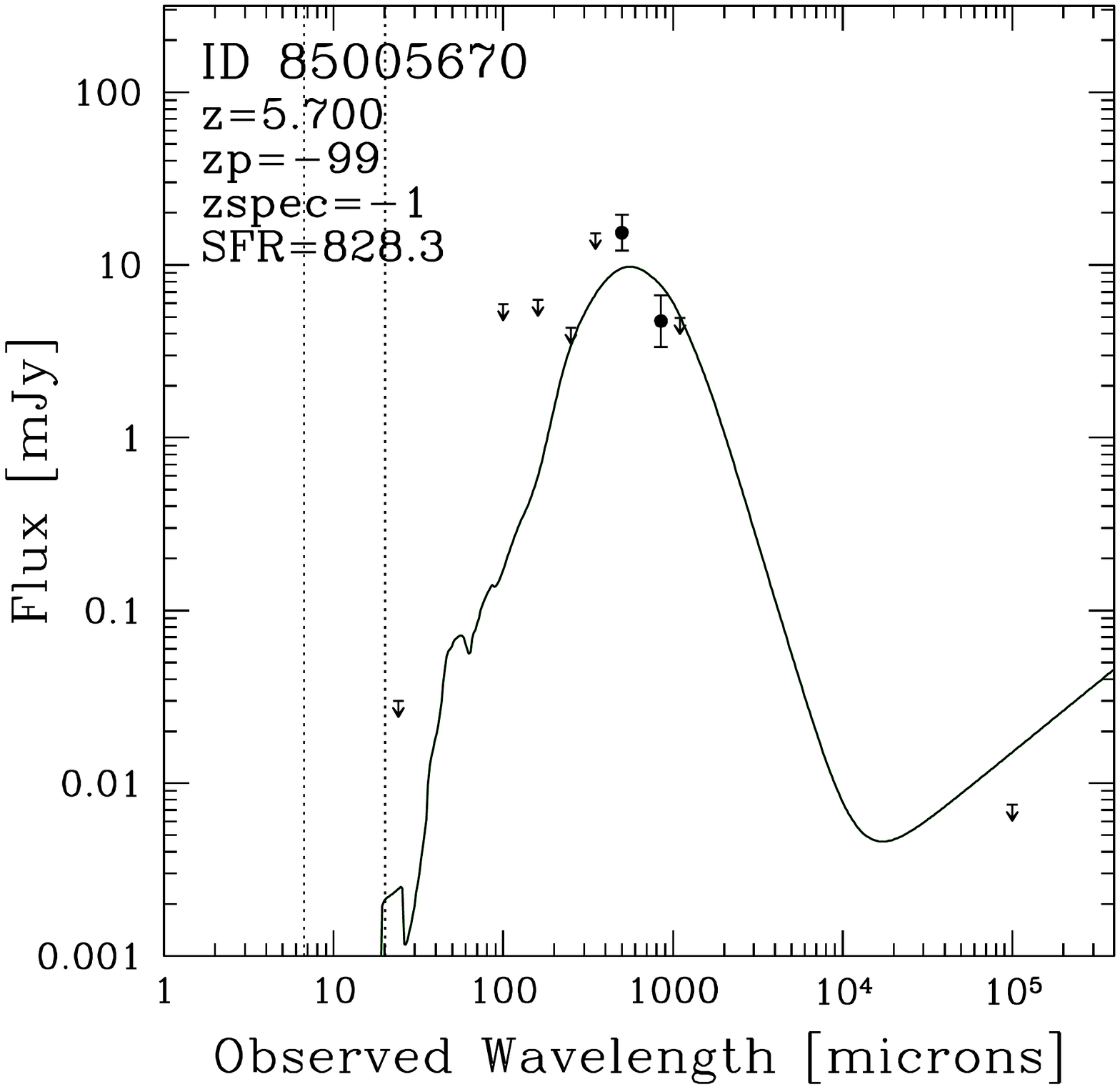}
    \includegraphics[width=0.28\textwidth]{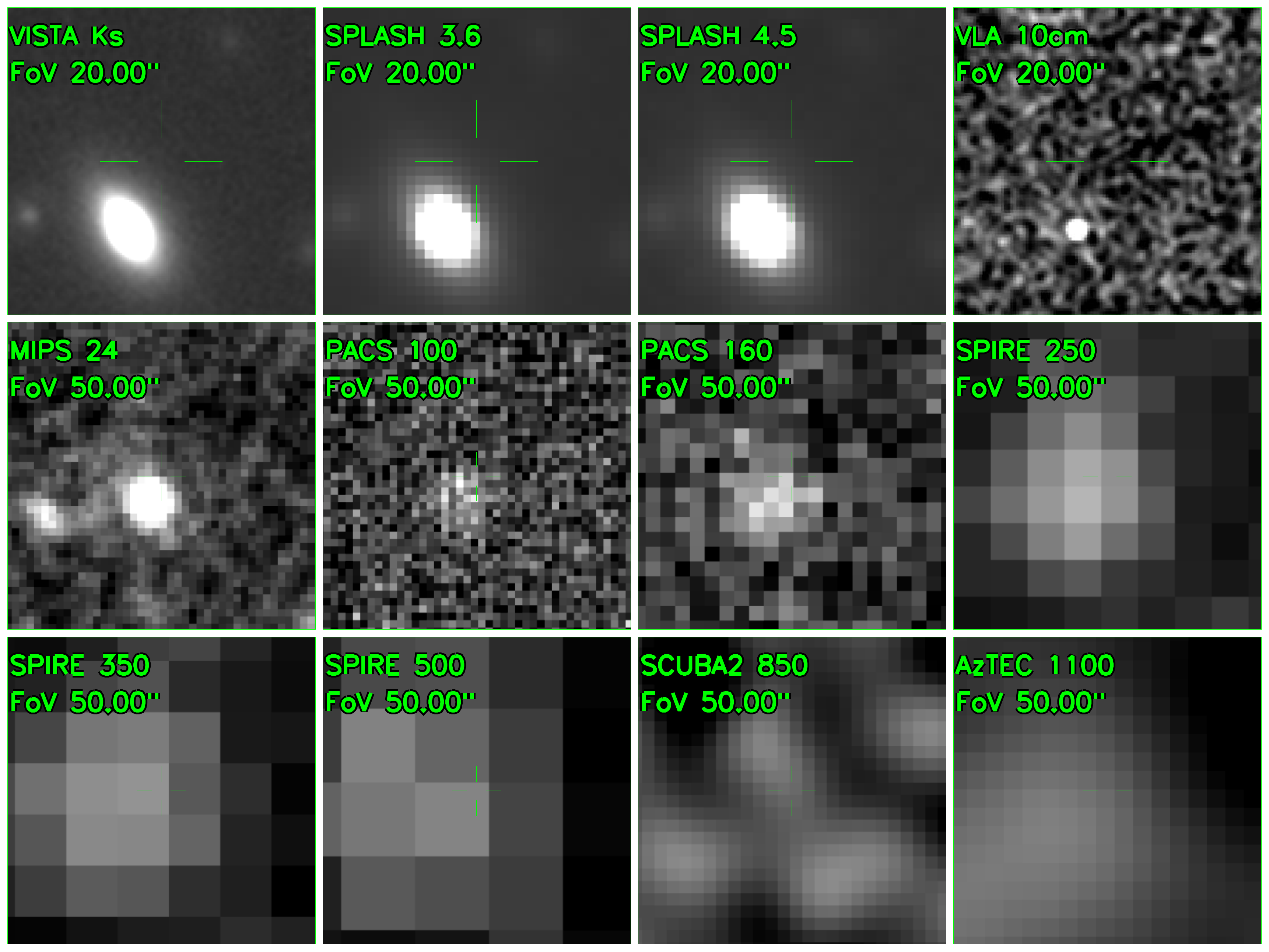}
	\includegraphics[width=0.21\textwidth, trim={0.6cm 5cm 1cm 3.5cm}, clip]{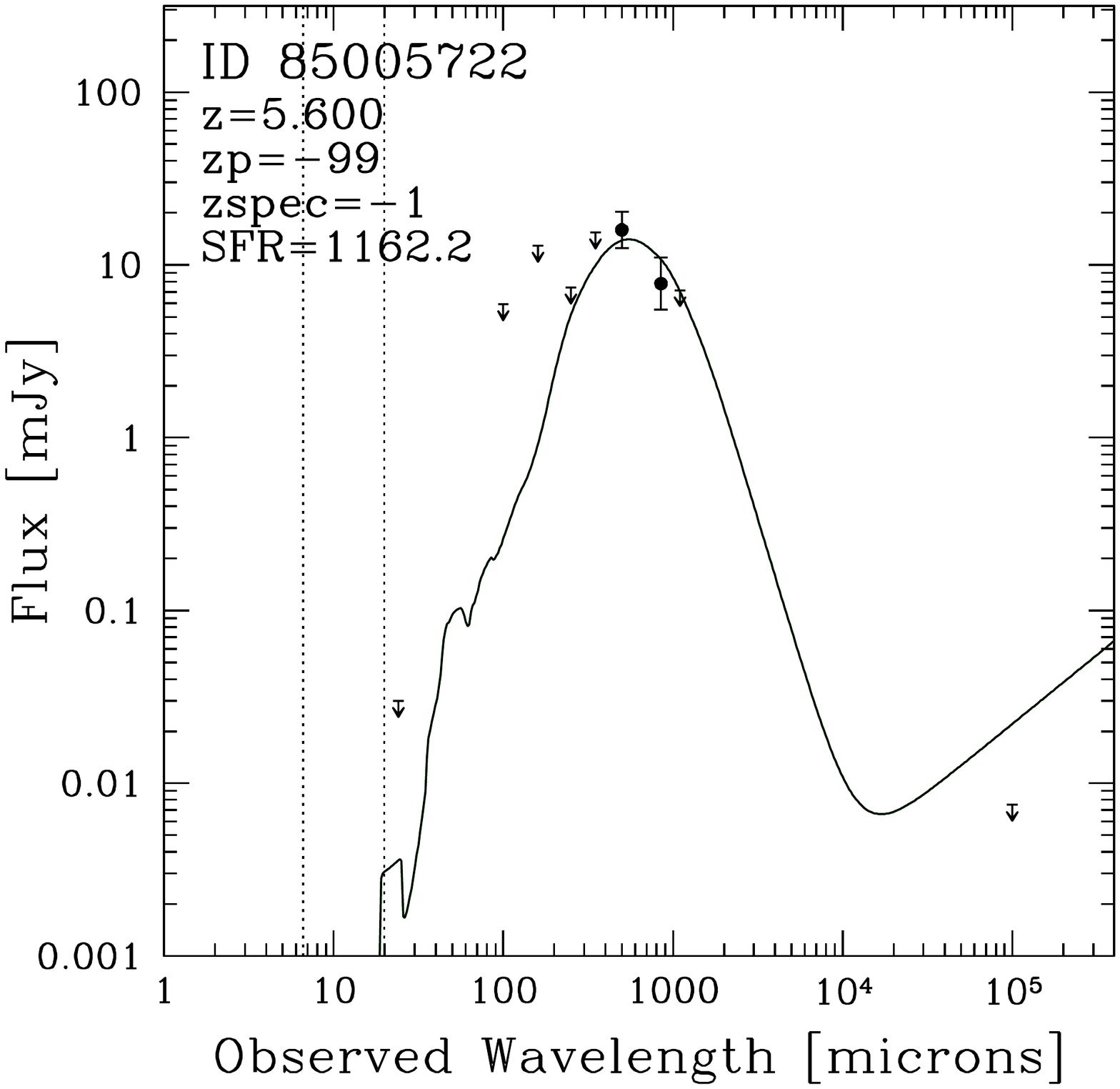}
    \includegraphics[width=0.28\textwidth]{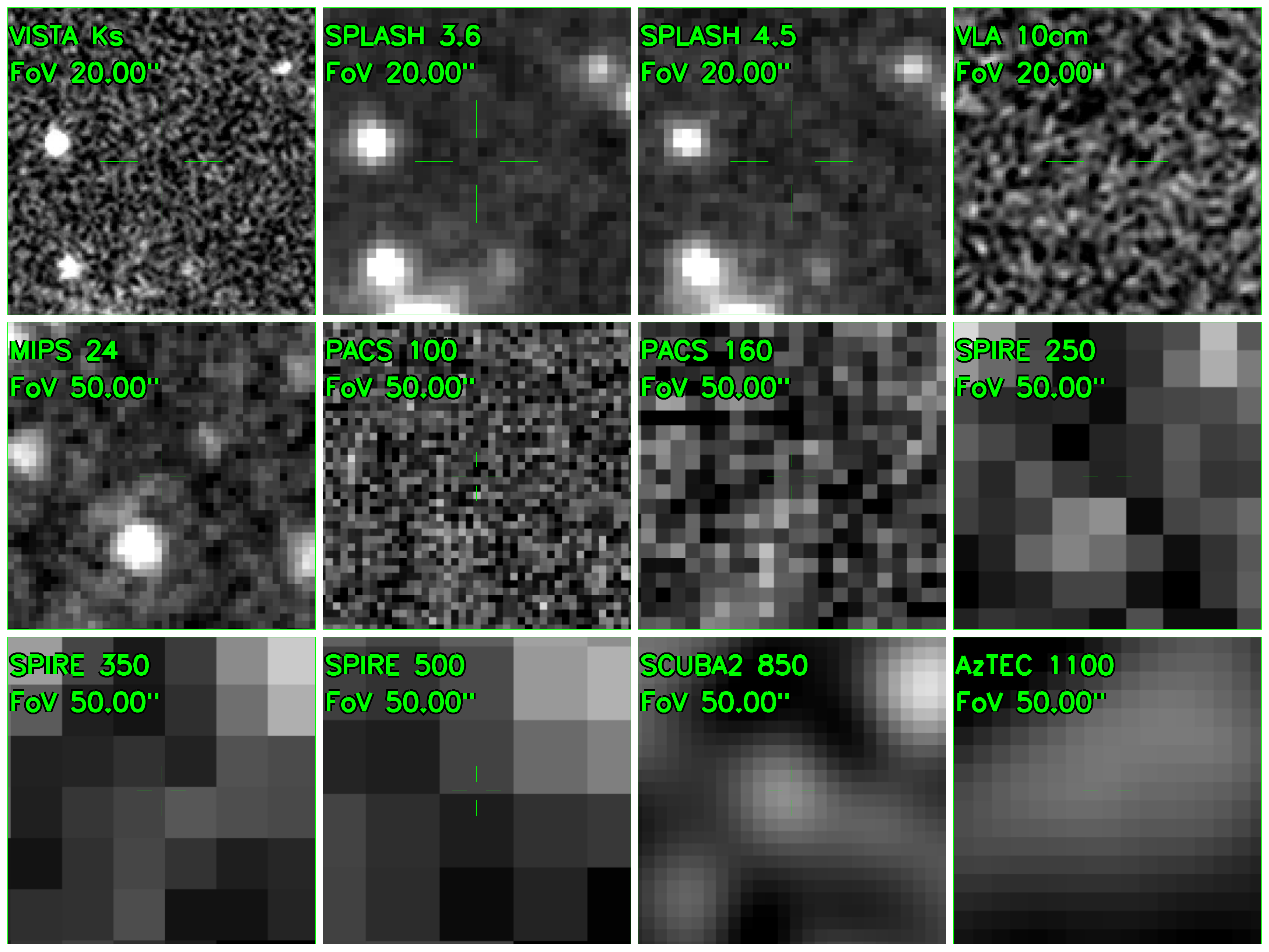}
	\includegraphics[width=0.21\textwidth, trim={0.6cm 5cm 1cm 3.5cm}, clip]{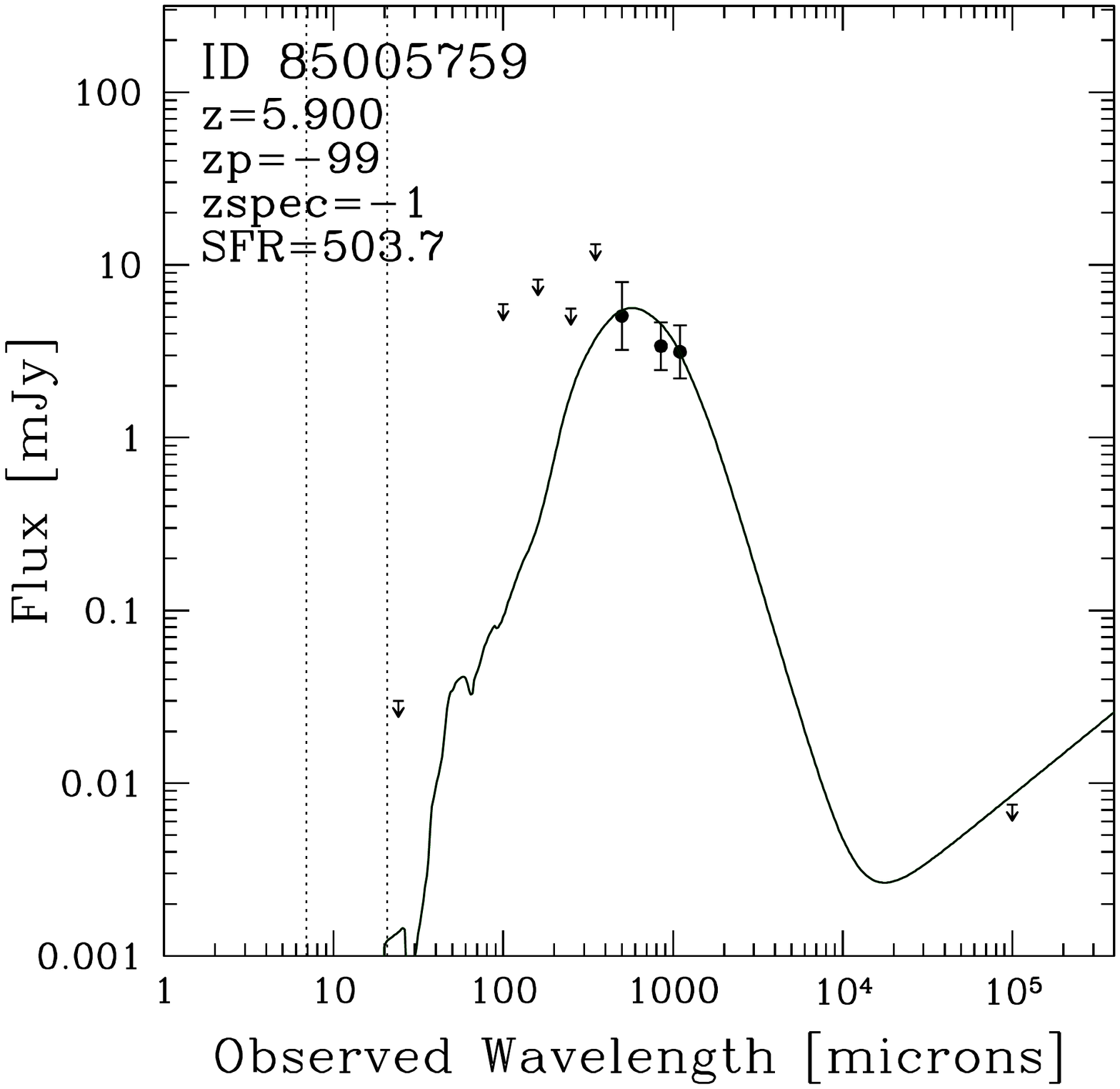}
    \caption{%
		Multi-band cutouts and SEDs of high redshift candidates, continued.  
		\label{highz_cutouts2}
		}
\end{figure}

\begin{figure}
	\centering
    \includegraphics[width=0.28\textwidth]{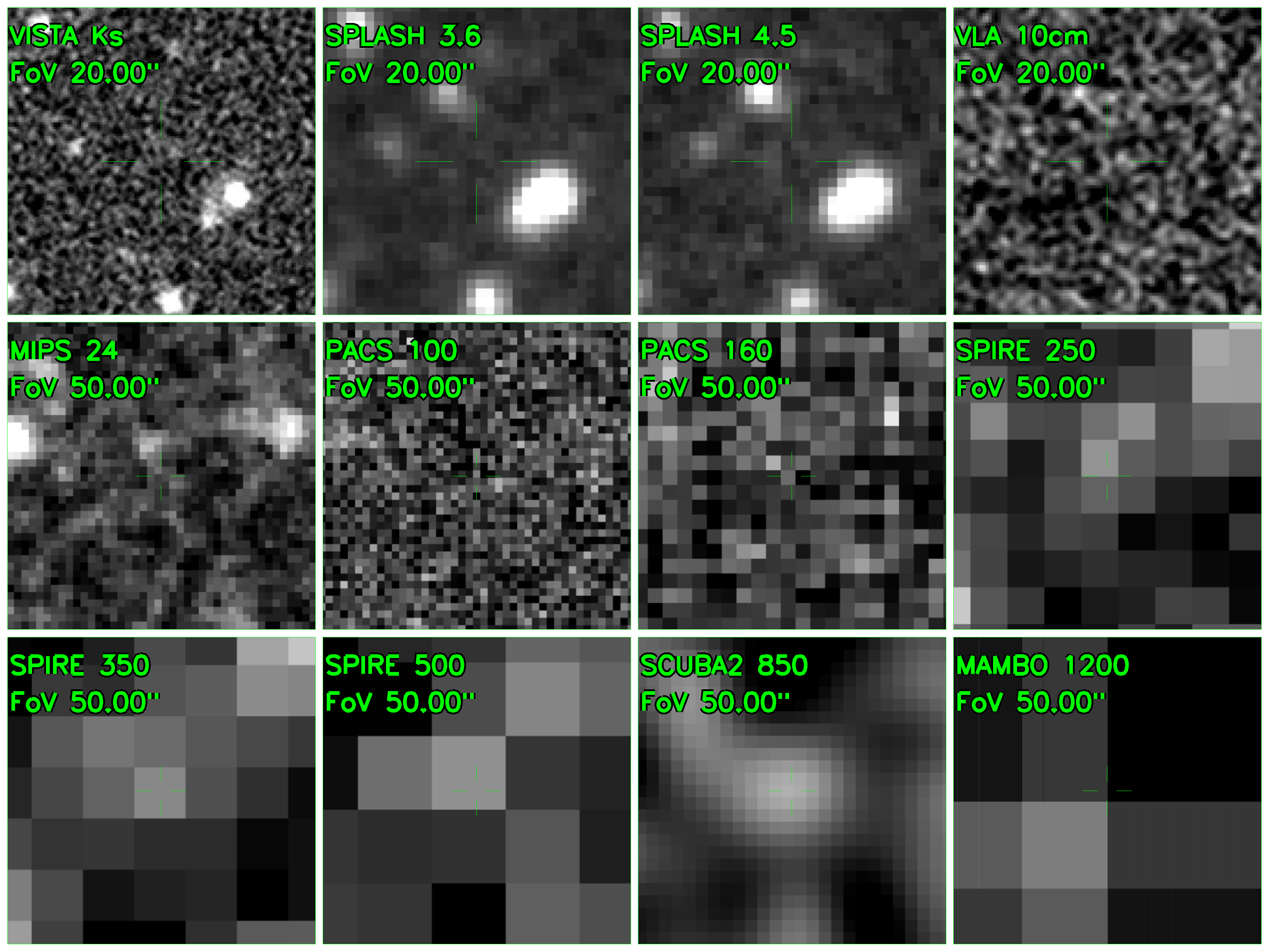}
	\includegraphics[width=0.21\textwidth, trim={0.6cm 5cm 1cm 3.5cm}, clip]{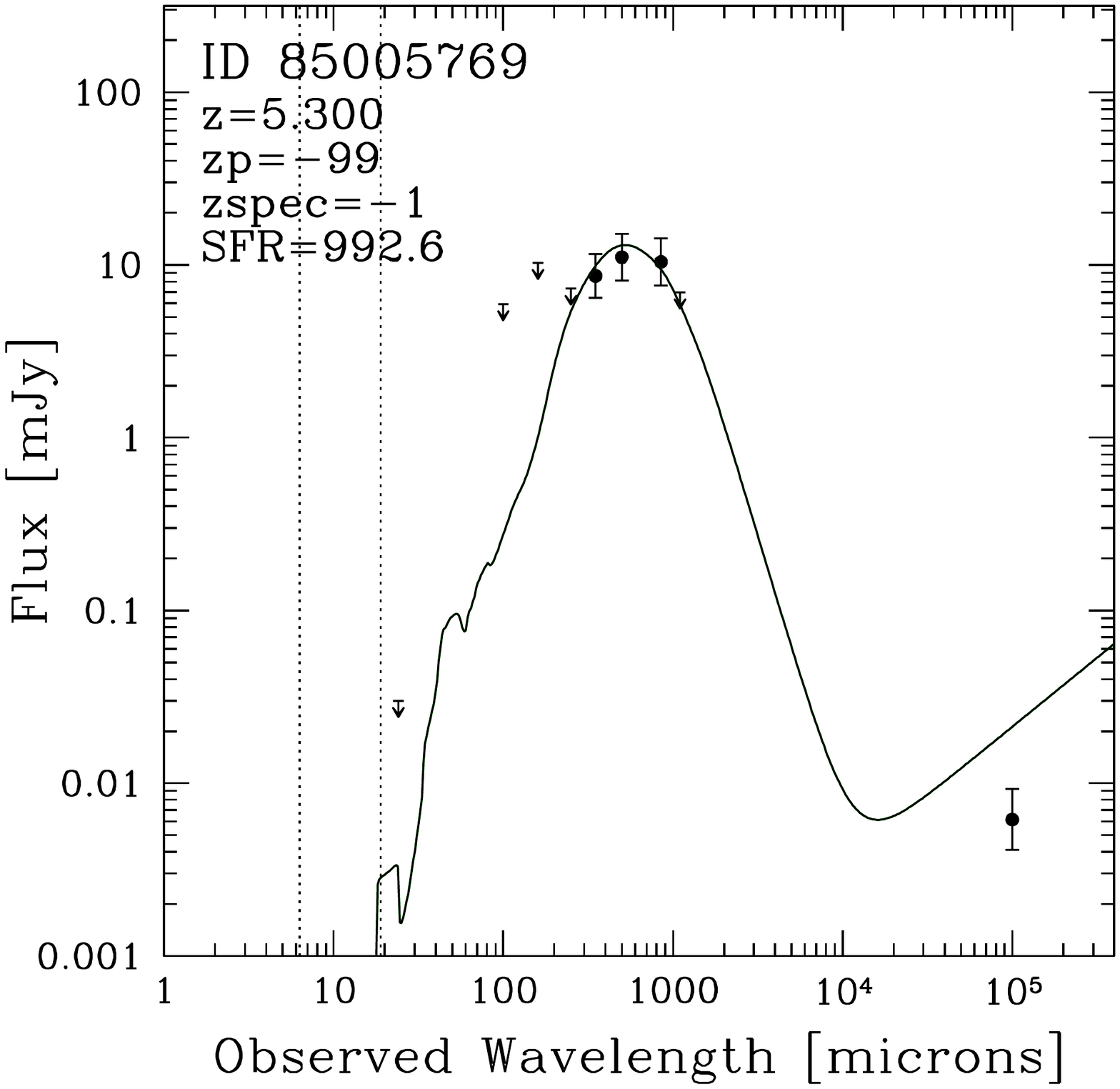}
    \includegraphics[width=0.28\textwidth]{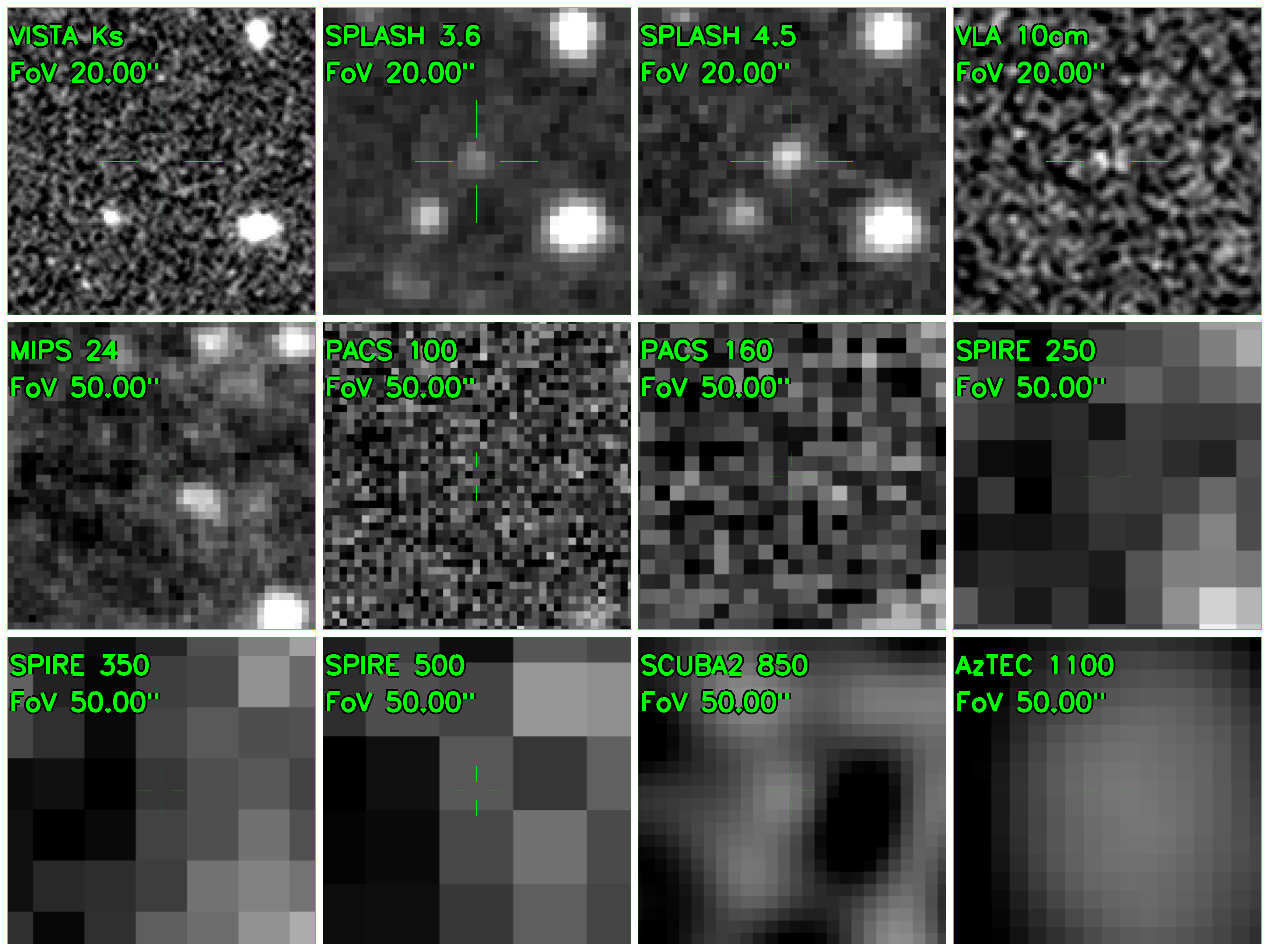}
	\includegraphics[width=0.21\textwidth, trim={0.6cm 5cm 1cm 3.5cm}, clip]{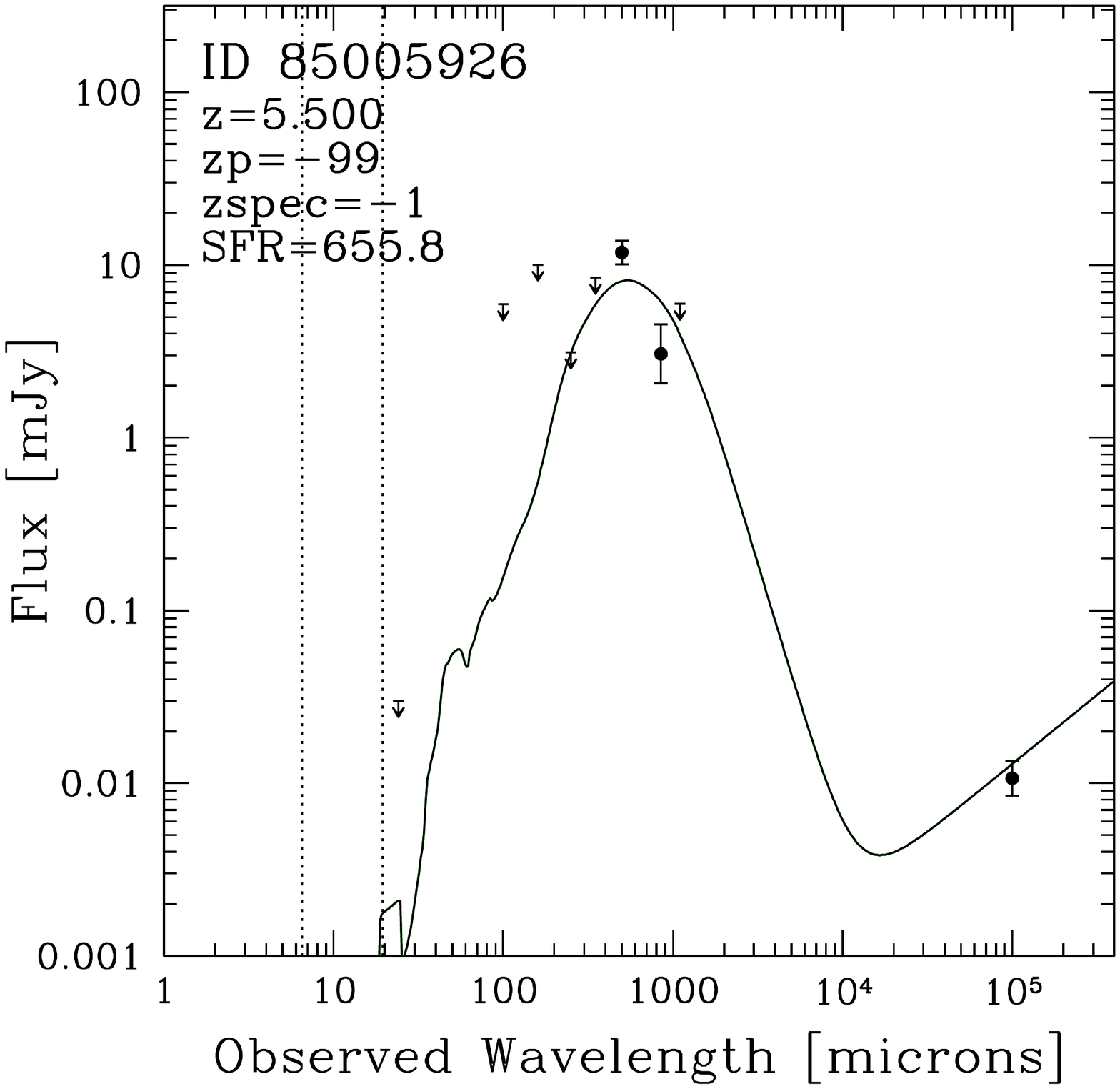}
    \includegraphics[width=0.28\textwidth]{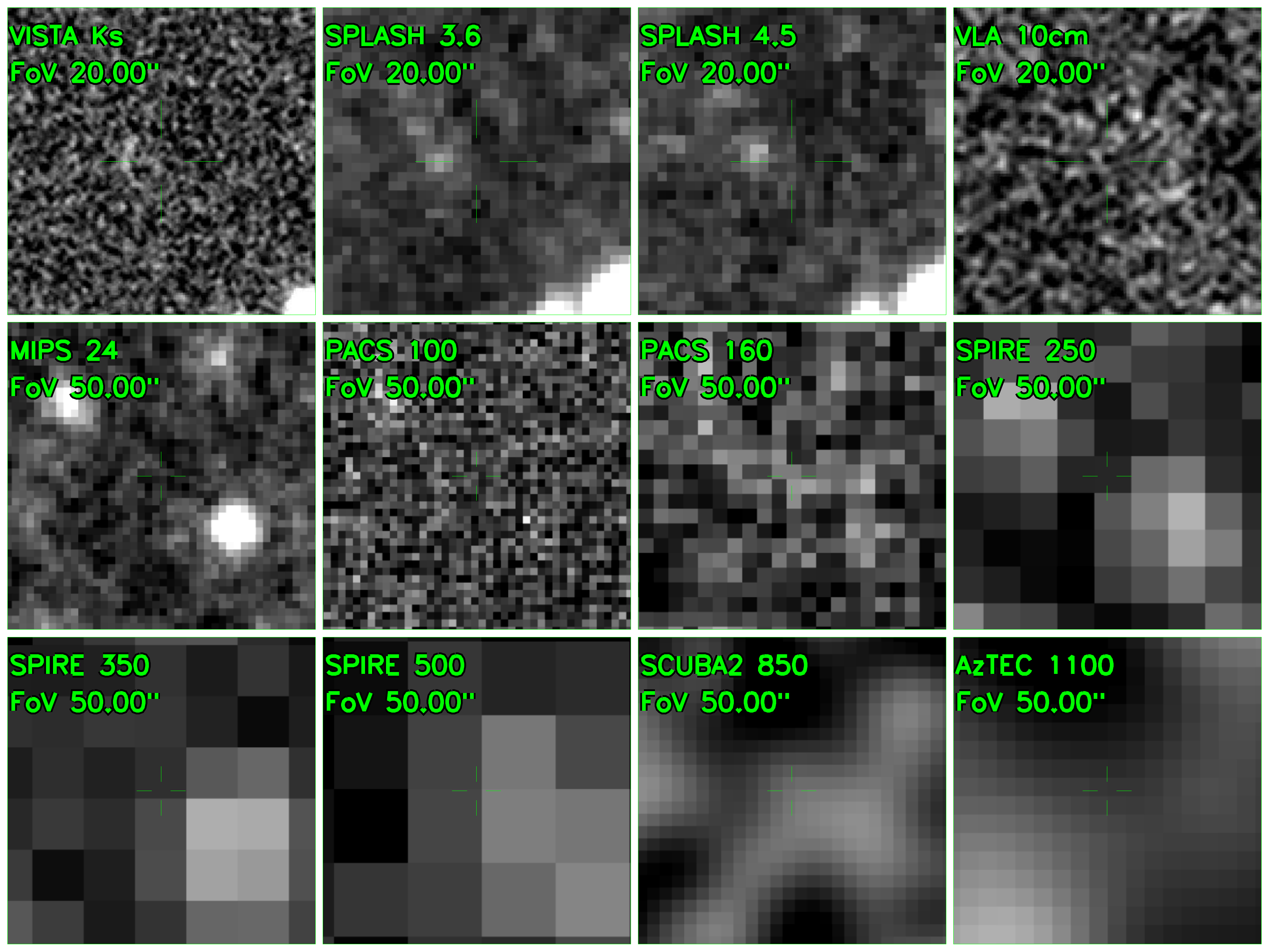}
	\includegraphics[width=0.21\textwidth, trim={0.6cm 5cm 1cm 3.5cm}, clip]{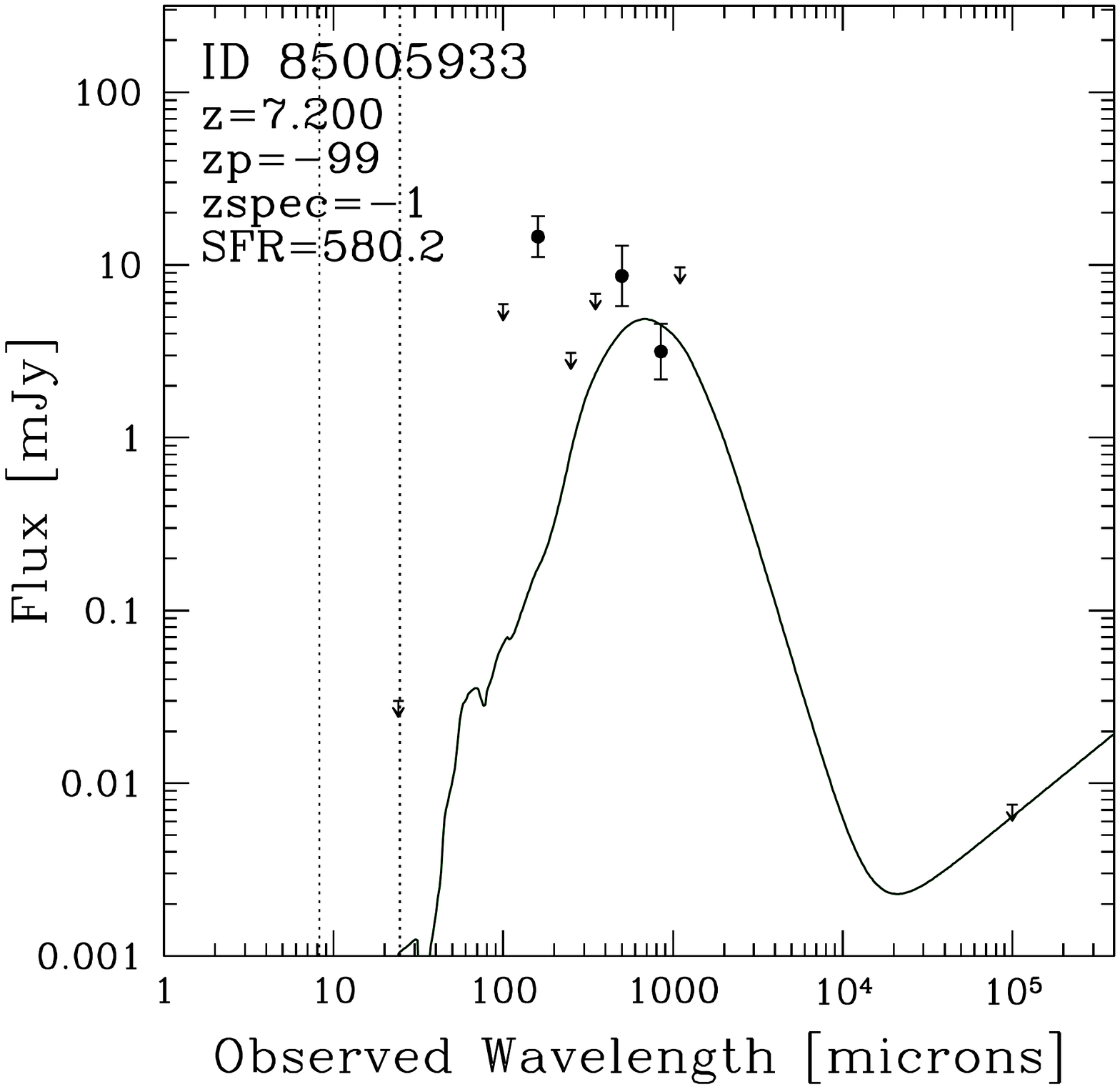}
    \includegraphics[width=0.28\textwidth]{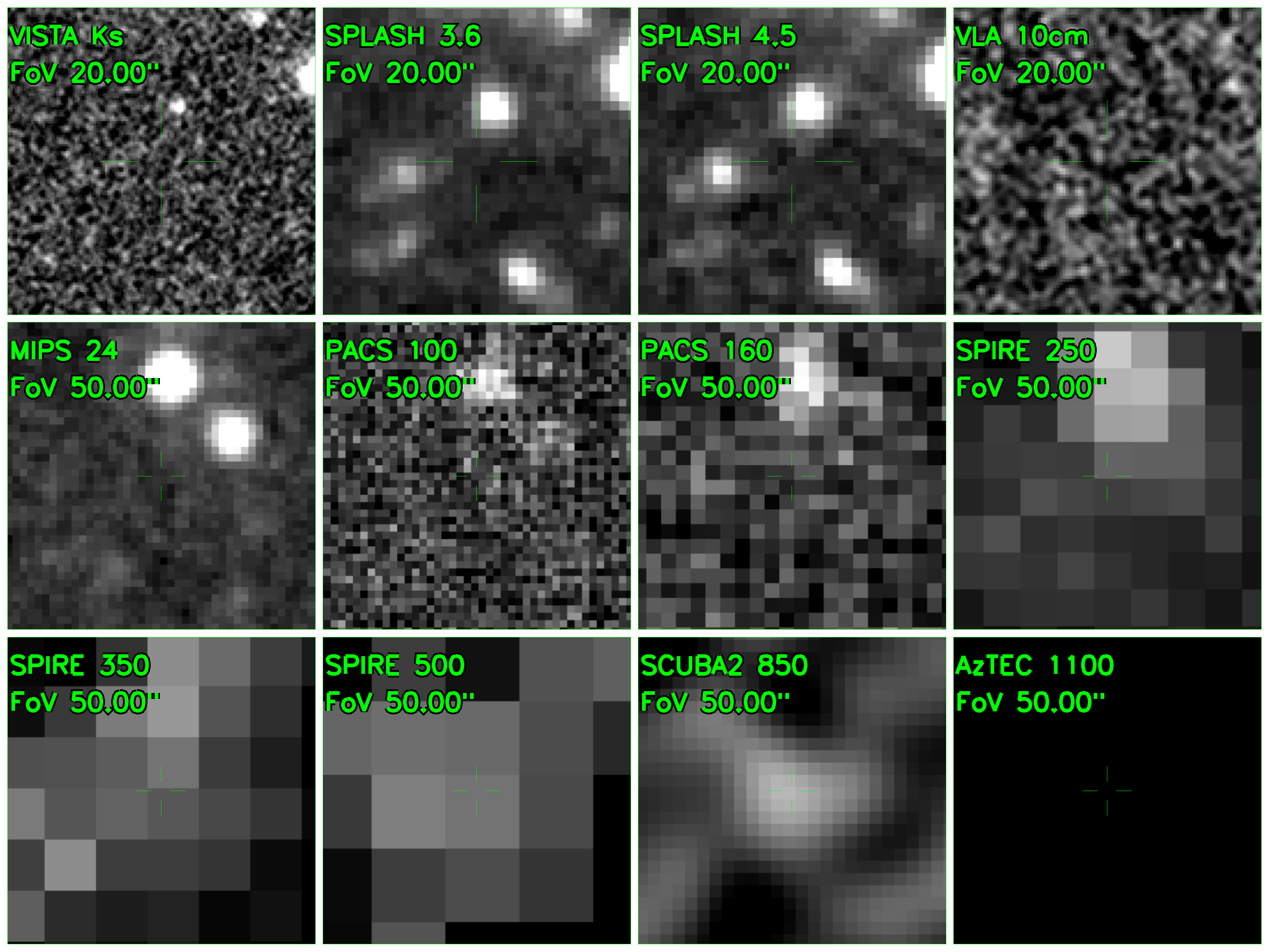}
	\includegraphics[width=0.21\textwidth, trim={0.6cm 5cm 1cm 3.5cm}, clip]{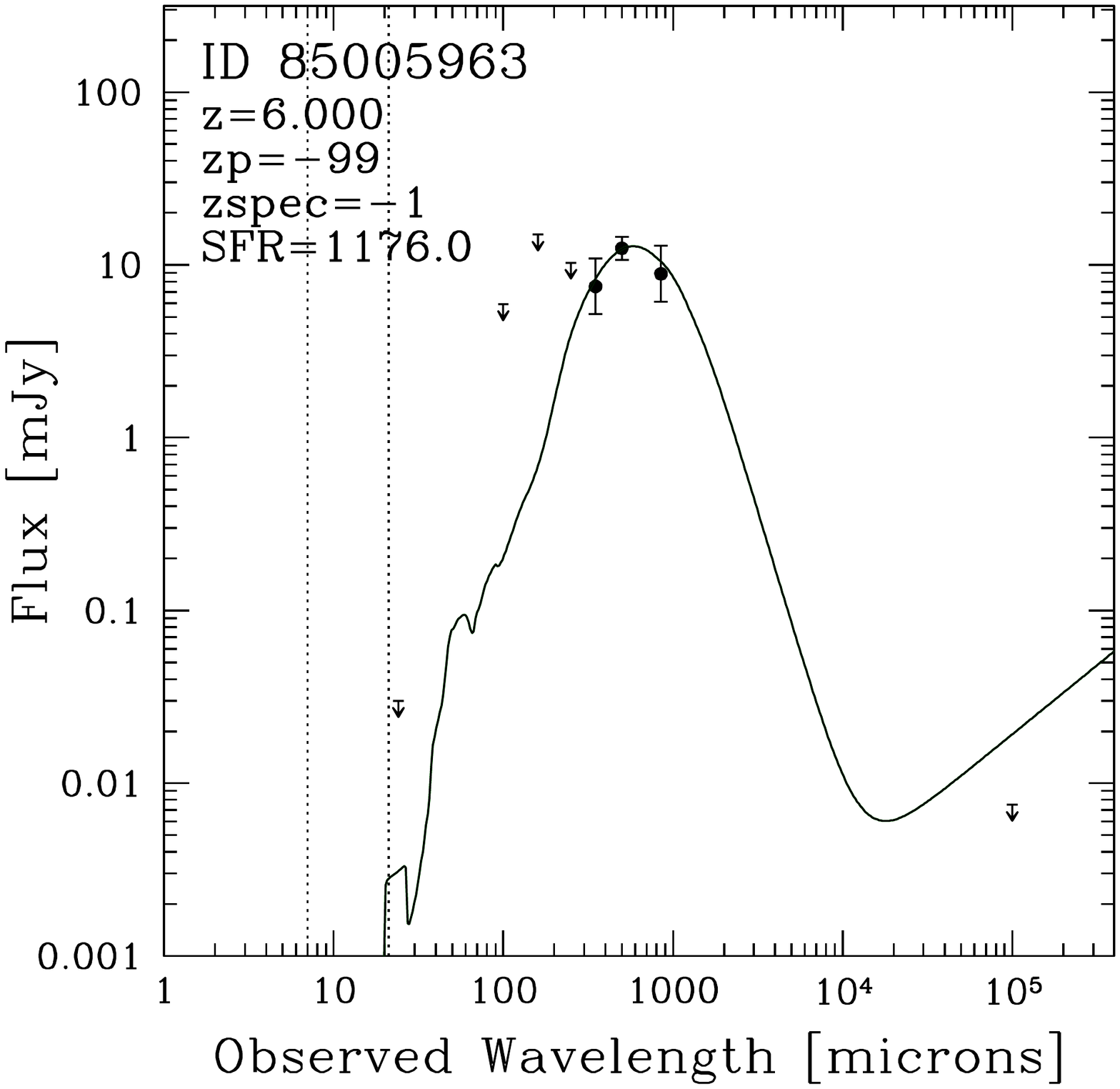}
    	\includegraphics[width=0.28\textwidth]{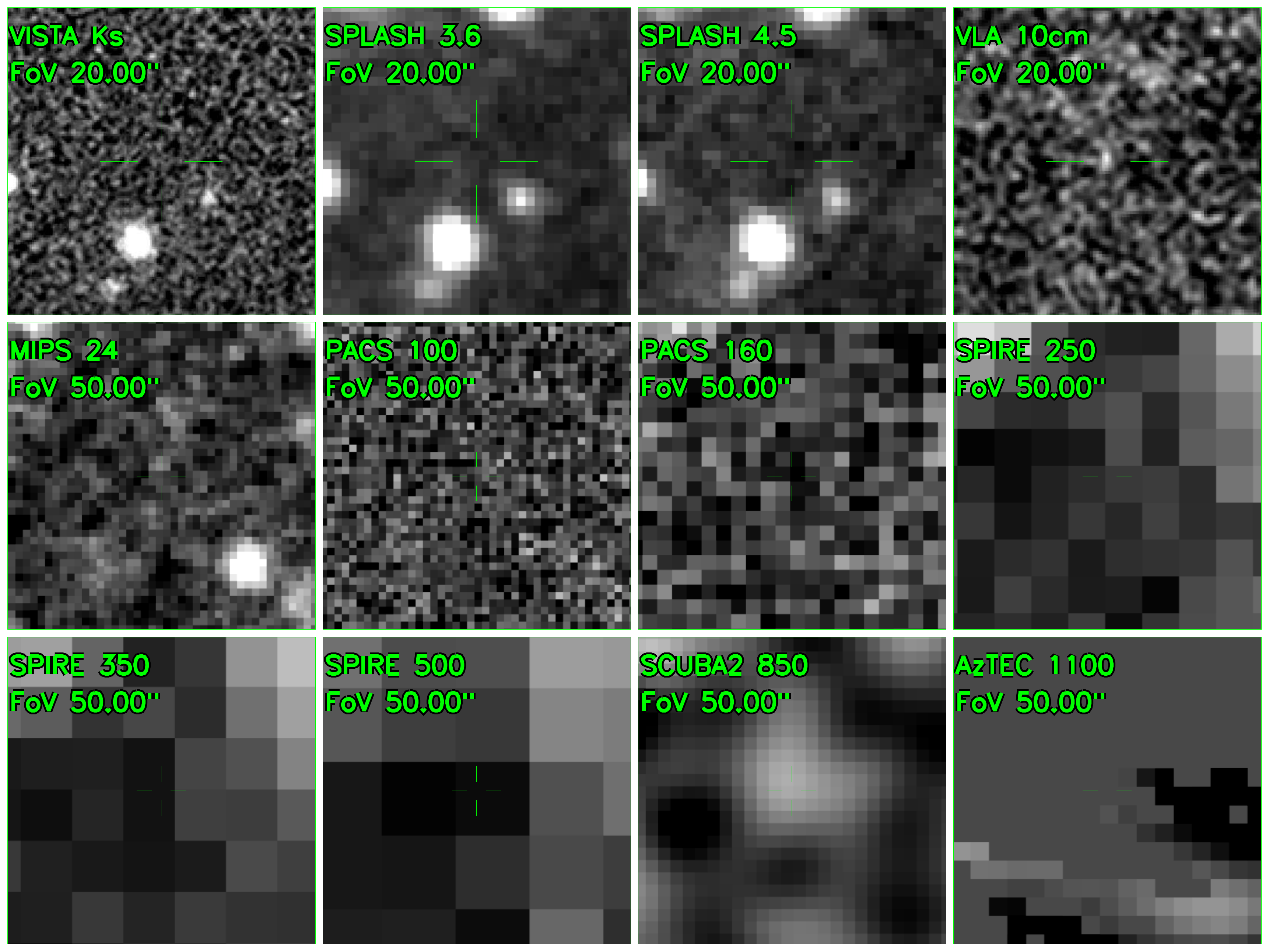}
	\includegraphics[width=0.21\textwidth, trim={0.6cm 5cm 1cm 3.5cm}, clip]{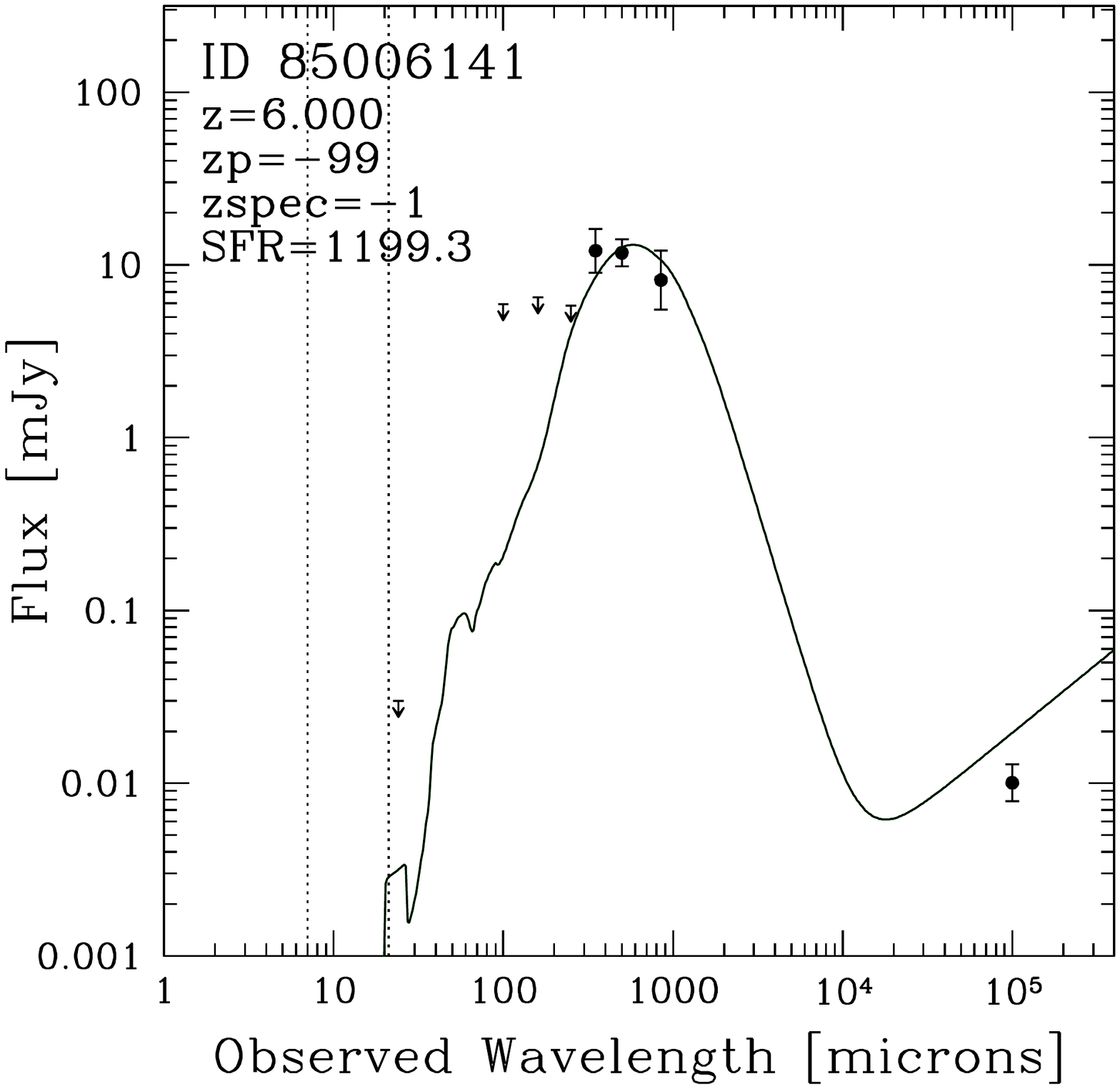}
 \caption{%
		Multi-band cutouts and SEDs of high redshift candidates, continued. 
		\label{highz_cutouts3}
		}
\end{figure}


\clearpage



\end{document}